\documentclass[twoside,12pt]{Classes/CUEDthesisPSnPDF}

\ifpdf
    \pdfcatalog { /PageMode (/UseOutlines)
                  /OpenAction (fitbh)  }
\fi

\title{Medida de la secci\'on eficaz\\[2pt]
	de producci\'on de dibosones WZ\\[2pt]
	a 7 TeV y 8 TeV\\[2pt]
	de energ\'ia del centro de masas\\[2pt]
        en el experimento\\[2pt]
        CMS\\[2pt]}
\titleplane{WZ production cross-section measurement
at 7 TeV and 8 TeV center of mass energies
within the Compact Muon Solenoid experiment}
\titleEN{WZ production cross-section\\[2pt]
	measurement at\\[2pt]
	7 TeV and 8 TeV\\[2pt]
	center of mass energies within\\[2pt]
	the Compact Muon Solenoid\\[2pt] 
	experiment\\[2pt]}

\ifpdf
  \author{\href{mailto:duarte@ifca.unican.es}{Jordi Duarte Campderr\'os}}
  \collegeordept{\href{http://www.ifca.es}{Departamento de F\'isica Moderna}}
  \university{\href{http://www.unican.es}{Instituto de F\'isica de Cantabria (CSIC-UC)}}
  \crest{\includegraphics[width=140mm,height=30mm]{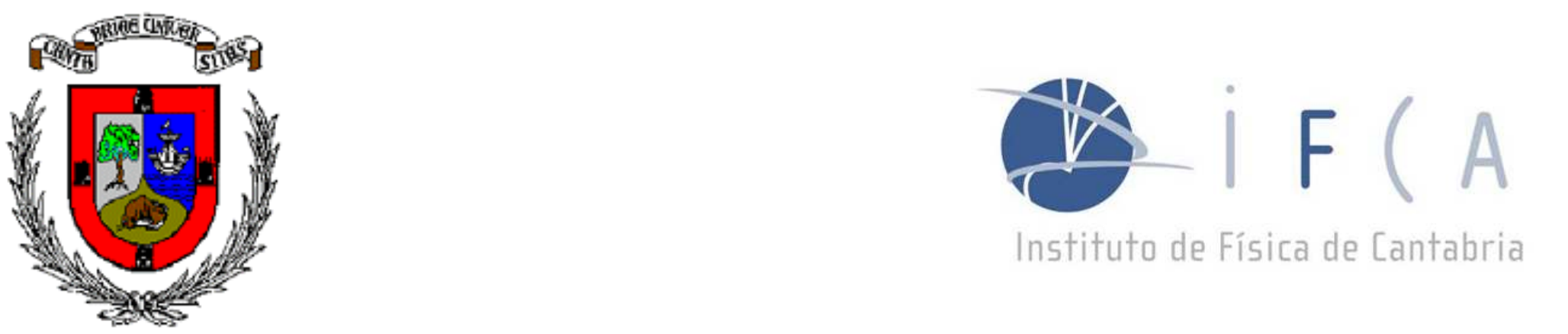}}
  \badgeUniv{\includegraphics[width=30mm]{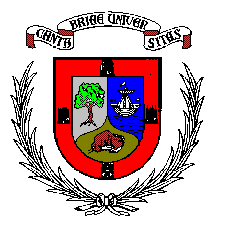}}
  \badgeDep{\includegraphics[width=30mm]{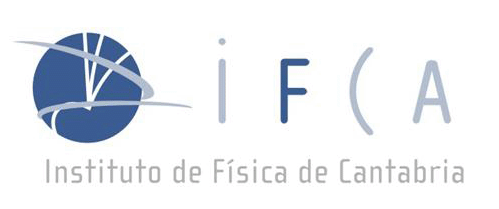}}
\else
  \author{Jordi Duarte Campderr\'os}
  \collegeordept{Departamento de F\'isica Moderna}
  \university{{Instituto de F\'isica de Cantabria (CSIC-UC)}
  \crest{\includegraphics[bb = 0 0 292 336, width=30mm]{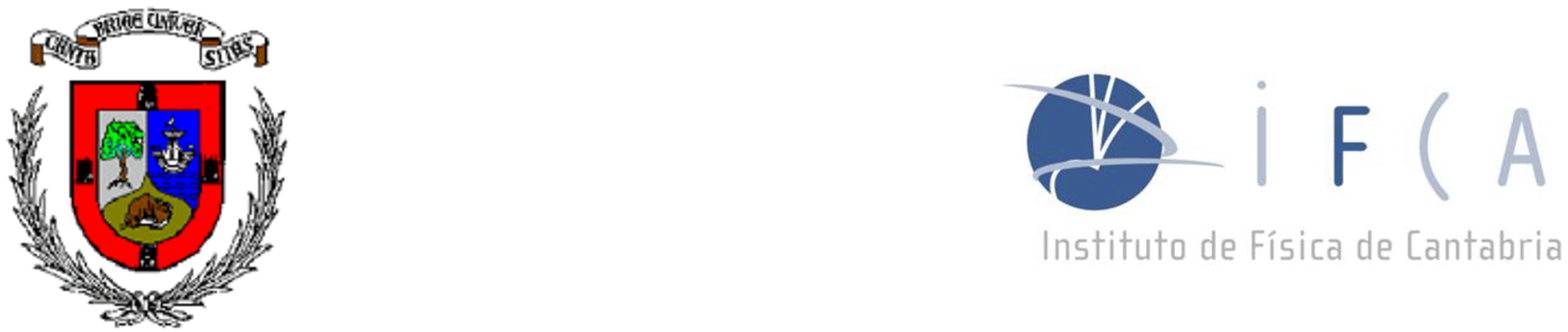}}
  \badgeUniv{\includegraphics[width=30mm]{UNICANLogo}}
  \badgeDep{\includegraphics[width=30mm]{IFCALogo}}
\fi
\degreedate{Santander, Diciembre de 2013}
\supervisor{Iv\'an Vila \'Alvarez}
\supervisortelf{942202083}
\supervisoremail{vila@ifca.unican.es}
\codirector{Teresa Rodrigo Anoro}

\hbadness=10000
\hfuzz=50pt

\usepackage{multirow}

\usepackage[protrusion=true]{microtype}

\usepackage{feynmp}
\DeclareGraphicsRule{*}{mps}{*}{}

\makeatletter
\def\endfmffile{%
  \fmfcmd{\p@rcent\space the end.^^J%
          end.^^J%
          endinput;}%
  \if@fmfio
    \immediate\closeout\@outfmf
  \fi
  \ifnum\pdfshellescape>\z@
    \immediate\write18{mpost \thefmffile}%
  \fi}
\makeatother

\onehalfspacing

\begin{document}
\selectlanguage{english}

\maketitle


\setcounter{secnumdepth}{3}
\setcounter{tocdepth}{1}

\frontmatter 
\pagenumbering{roman}
\begin{dedication}
A la meva cara N\'uria i al nostre preci\'os fill, Nuc.
\end{dedication}

\begin{abstractslong}
	The \WZ associated diboson production is studied by measuring both inclusive cross section
	and, for the first time, the ratio between the \wzm and the \wzp cross sections. The 
	measurements are performed using data samples of proton-proton collisions collected during 
	the years 2011 and 2012, at 7 and 8~\TeV of centre-of-mass energies, respectively, by the CMS
	experiment at the LHC, updating the 7~\TeV cross section measurement available in CMS, and
	presenting the new cross section measurement in CMS at 8~\TeV. The data sample used for the
	7~\TeV measurements correspond to an integrated luminosity of 4.9~\fbinv, whence the data 
	for the 8~\TeV correspond to $\lumi_{int}=19.6~\fbinv$. 

\end{abstractslong}

\begin{acknowledgementslong}
  Quisiera expresar mi sincera gratitud a mi director de tesis, Iv\'an, por su constante 
  procrastinaci\'on, entendido en su sentido etimol\'ogico y no patol\'ogico, que nos ha llevado a 
  un sinf\'in de interesantes caminos que resolvieron dirigirse hacia este trabajo de tesis. En el 
  proceso, he aprendido a argumentar cuidadosamente mis enunciados gracias a su, ahora s\'i, 
  patol\'ogica obsesi\'on de controlar y entender todos los pasos que esconde cualquier 
  razonamiento; el ardor de la discusi\'on se incrementa cuando la idea preconcebida de Iv\'an es 
  opuesta a la que t\'u intentas argumentar. En su defensa dir\'e, que si la raz\'on est\'a de tu 
  parte, finalmente aceptar\'a tus argumentos y ceder\'a. Pero es un \'arduo trabajo. Tambi\'en 
  dir\'e, que en muchas ocasiones, la raz\'on no estaba de la m\'ia... Ha sido un verdadero placer
  trabajar con \'el. 
  
  Mi agradecimiento al programa de becas predoctorales de la Universidad de Cantabria, a trav\'es 
  del cual este trabajo de investigaci\'on ha sido financiado en su mayor parte.

  The author thankfully acknowledges the computer resources, technical expertise and assistance 
  provided by the Advanced Computing and e-Science team at IFCA.

  A mi compa\~nera de fatigas Clara Jord\'a, o mejor, la Dra. Clara Jord\'a, me gustar\'ia 
  dedicarle un agradecimiento especial. Por tantas horas de discusiones, tags, probes,
  c\'odigos, tesis, ROOTs y dem\'as sucesiones de historias breves, que llenaron muchos de los 
  momentos de fatiga, sobretodo en el CERN, y otros de relajaci\'on, como los caf\'es, ya en el IFCA,
  ya en el 40. 

  Son muchas las personas que he conocido y que me han ayudado de una forma u otra, 
  indirecta o directamente, en la realizaci\'on de este trabajo. En el IFCA, a Teresa Rodrigo, 
  tambi\'en directora de tesis, me gustar\'ia agradecerle su constante energ\'ia y motivaci\'on,
  y sus siempre acertadas aportaciones a los m\'as diversos problemas. A Roc\'io Vilar su 
  grand\'isima ayuda y disposici\'on, adem\'as de las instructivas charlas en las que nos hemos 
  enfrascado. A Celso, Pupi y Alberto agradecerles sus comentarios y puntuales contribuciones. A 
  Sven, sus acertadas correcciones. A Ana Yaiza, le agradezco que me proporcionase el c\'odigo del 
  BLUE method en C++ que posteriormente modifiqu\'e a python. Y a la gente de Oviedo, Javi 
  Fern\'andez, por su infatigable trabajo en el procesadado de datos, Chiqui, Santi y Lara, por sus
  instructivas discusiones en las reuniones semanales y, por supuesto, Javier Cuevas, cuyos caf\'es 
  compartidos en el 40, o las cenas en la piscina frente al CERN, me ayudaron en muchos casos a 
  avanzar el proceso de an\'alisis. A J\'onatan Piedra me gustar\'ia reconocerle su disposici\'on a 
  la charla y a la discusi\'on sana, agradeci\'endole los meses de despacho compartido en el CERN y,
  por supuesto, el an\'alisis de datos a 8~\TeV. 
  
  Alicia Calder\'on merece menci\'on aparte, su excelente trabajo en el m\'etodo data-driven es 
  indispensable en esta tesis, donde adem\'as de proporcionarme los fakes rates, hemos tenido 
  instructivas e interminables discusiones del m\'etodo y sus aplicaciones. Y por supuesto, su 
  infatigable capacidad de trabajo le ha permitido, a pesar de sus heterog\'eneas obligaciones, 
  estar ah\'i en cualquier momento que la necesitase, y han sido unos cuantos momentos.

  I am also indebted with the IRB-BU group in the \WZ analysis. In particular, I am grateful to 
  Sre\'cko Morovi\'c the shared spent hours checking and analyzing our analysis differences, and 
  of course, the fruitful exchange of emails making possible the outstanding analysis convergence
  we obtained. My gratitude is also to Vuko Brigljevi\'c who always help me to understand the 
  analysis deeply. The analysis effort it would not be possible without his leadership, being a 
  pleasure to work with him. And finally Cory Fantasia, for the productive discussions in the 
  meetings.

  I would like to express my gratitude to the Upsilon team, specially to Nuno Leonardo and Zoltan
  Gecse, for their patience and ease to help, during the exciting days of the first collision data,
  and Ian Shipsey for his kindly presence.

  En cuanto a la gente que indirectamente me han ayudado a la re\-a\-li\-za\-ci\'on de esta tesis, me 
  gustar\'ia agradecer especialmente a David Moya los caf\'es compartidos, las discusiones, su
  generosa predisposici\'on y, en conclusi\'on, su amistad. A Gerva su estupendo sentido del 
  humor, las pachangas futboleras y los omnipresentes caf\'es. A Enol y Ana, y por 
  supuesto, a Ico, su solidaridad y experiencias compartidas. A Pablo y \'Alvaro, a parte de su
  trabajo en el cluster que ha sido indispensable para la realizaci\'on de esta tesis, les 
  agradezco las discusiones absurdas y las cervezas tomadas (cuando a\'un exist\'ia vida m\'as 
  all\'a de las nueve de la noche). A Ib\'an, su black hole supermasivo. A mis compa\~neros o ya 
  ex-compa\~neros del IFCA, Paqui, Javier Brochero, P\'arbol, Rebeca, Flori, Rafa, Amparo, Javi, 
  Richard, Miguel \'Angel, Lu\'is Lanz, Ra\'ul, Airam les agradezco los caf\'es y ratos pasados con ellos, no 
  siempre fruct\'iferos pero  siempre agradables. 

  Tamb\'e voldria agrair a la Biuse la seva amistat. A la meva tieta Emilia, pel seu suport que em va
  permetre torna a agafar el llibres, a m\'es de les vegades que ha
  vingut a ajudar-nos quan les coses es possaven complicades, aix\'i com als meus pares. A mes
  germanes, nebot, neboda, cunyats, simplement per que em surt. I, suposso, que no es necessari
  i, tot i que sigui evident, a la meva d\'ona i al meu fillet per, b\`asicament, tot.
\end{acknowledgementslong}

\newterm[nonumberlist=false,name={pileup}]{ind:pileup}
\newterm[nonumberlist=false,name={Monte Carlo}]{ind:mc}
\newterm[nonumberlist=false,name={trigger}]{ind:gltrigger}
\newterm[nonumberlist=false,name={tag and probe}]{ind:tap}
\newterm[nonumberlist=false,name={data-driven}]{ind:data-driven}
\newterm[nonumberlist=false,name={underlying event}]{ind:ue}
\newterm[nonumberlist=false,name={hard scattering}]{ind:hardscattering}
\newterm[nonumberlist=false,name={signature}]{ind:signature}
\newterm[nonumberlist=false,name={initial state radiation}]{ind:isr}
\newterm[nonumberlist=false,name={final state radiation}]{ind:fsr}
\newterm[nonumberlist=false,name={multivariate analysis}]{ind:mva}
\newterm[nonumberlist=false,name={parton distribution function}]{ind:pdf}
\newterm[nonumberlist=false,name={CMS}]{ind:cms}
\newterm[nonumberlist=false,name={LHC}]{ind:lhc}
\newterm[nonumberlist=false,name={gauge}]{ind:gauge}

\newterm[nonumberlist=false,name={standard model}]{ind:sm}
\newterm[nonumberlist=false,name={quantum field theory}]{ind:qft}
\newterm[nonumberlist=false,name={leading order}]{ind:lo}
\newterm[nonumberlist=false,name={next to leading order}]{ind:nlo}
\newterm[nonumberlist=false,name={next to next to leading order}]{ind:nnlo}
\newterm[nonumberlist=false,name={quantum cromodynamics}]{ind:qcd}
\newterm[nonumberlist=false,name={quantum electrodynamics}]{ind:qed}
\newterm[nonumberlist=false,name={Yang-Mills}]{ind:ym}
\newterm[nonumberlist=false,name={electroweak theory}]{ind:ew}
\newterm[nonumberlist=false,name={CKM (Cabibbo--Kobayashi--Maskawa) matrix}]{ind:ckm}
\newterm[nonumberlist=false,name={grand unification theories}]{ind:gut}
\newterm[nonumberlist=false,name={MSTW8 PDFs set}]{ind:mstw8}
\newterm[nonumberlist=false,name={CT-10 PDFs set}]{ind:ct10}

\newterm[nonumberlist=false,name={LEP}]{ind:lep}
\newterm[nonumberlist=false,name={electromagnetic calorimeter}]{ind:ecal}
\newterm[nonumberlist=false,name={hadronic calorimeter}]{ind:hcal}

\newterm[nonumberlist=false,name={interaction point}]{ind:ip}
\newterm[nonumberlist=false,name={event data model}]{ind:edm}
\newterm[nonumberlist=false,name={stand-alone muon track}]{ind:sta}
\newterm[nonumberlist=false,name={global muon track}]{ind:global}
\newterm[nonumberlist=false,name={gaussian sum filter}]{ind:gsf}
\newterm[nonumberlist=false,name={jet energy scale}]{ind:jec}

\newterm[nonumberlist=false,name={high level trigger (HLT)}]{ind:hlt}
\newterm[nonumberlist=false,name={primary dataset}]{ind:pd}
\newterm[nonumberlist=false,name={data acquisition system}]{ind:daq}

\newterm[nonumberlist=false,name={boosted decision tree}]{ind:bdt}
\newterm[nonumberlist=false,name={particle object group (POG)}]{ind:pog}
\newterm[nonumberlist=false,name={working point}]{ind:wp}

\newterm[nonumberlist=false,name={scale factor}]{ind:sf}

\newterm[nonumberlist=false,name={fom method}]{ind:fom}
\newterm[nonumberlist=false,name={fakeable, \emph{see also} \gls{ind:loose}}]{ind:fakeable}
\newterm[nonumberlist=false,name={loose}]{ind:loose}

\newterm[nonumberlist=false,name={prompt}]{ind:prompt}
\newterm[nonumberlist=false,name={fake}]{ind:fake}

\newterm[nonumberlist=false,name={blue method}]{ind:blue}

\newterm[nonumberlist=false,name={anomalous triple gauge boson couplings}]{ind:atgc}

\newterm[nonumberlist=false,name={event reconstruction}]{ind:reco}

\newterm[nonumberlist=false,name={trigger path}]{ind:triggerpath}
\newterm[nonumberlist=false,name={event}]{ind:event}
\newterm[nonumberlist=false,name={signal}]{ind:signal}
\newterm[nonumberlist=false,name={background}]{ind:background}
\newterm[nonumberlist=false,name={decay channel}]{ind:decaychannel}
\newterm[nonumberlist=false,name={channel, \emph{see also} \gls{ind:decaychannel}}]{ind:decay}

\newglossaryentry{gltrigger}
{
   name={Trigger},
   description={Data-filtering system used to keep a low rate of stored events
                among the billions per second produced at the LHC},
   text={trigger}
}

\newglossaryentry{tap}
{
   name={Tag and Probe},
   description={Technique used in high energy physics to calculate efficiency
               using well defined events containing resonances},
   text={tag and probe}
}

\newglossaryentry{data-driven}
{
   name={Data-driven},
   description={Technical term referring to the use of experimental data to obtain an
                estimation, avoiding the use of simulated data},
   text={data-driven\glsadd{ind:data-driven}}
}

\newglossaryentry{pileup}
{
   name={Pileup},
   description={Collisions accompanying the hard scattering produced at the same bunch crossing and 
               recorded along the hard scattering event},
   text={pileup}
}

\newglossaryentry{mc}
{
   name={Monte Carlo},
   description={Wide set of techniques used to simulate events relying on repeated random sampling 
                based in theoretical probabilities. Actually, the term Monte Carlo is broader, the 
		definition given here is within the context of this dissertation}
}

\newglossaryentry{ue}
{
   name={Underlying Event},
   description={Additional parton interactions occurring in a hadron-hadron collision, apart from 
                the hard scattering process},
   text={underlaying event}
}

\newglossaryentry{hardscattering}
{
   name={Hard Scattering},
   description={Process involving a large momentum transfer in a hadron-hadron collision},
   text={hard scattering}
}

\newglossaryentry{signature}
{
   name={Signature},
   description={Experimental hallmark left by the particles in the detector, which characterises
                an event},
   text={signature}
}

\newglossaryentry{glo:isr}
{
   name={Initial State Radiation},
   description={Radiation produced by a particle in its way to a scattering or annihilating 
                process, usually referred to the hard scattering}
}

\newglossaryentry{glo:fsr}
{
   name={Final State Radiation},
   description={Radiation produced by a particle coming from a scattering or annihilating process,
		usually referred to the hard scattering}
}

\newglossaryentry{glo:mva}
{
   name={Multivariate Analysis},
   description={Statistical technique to analyse simultaneously more than one outcome variable}
}

\newglossaryentry{glo:pdf}
{
   name={Parton Distribution Functions},
   description={Phenomenological functions fitted from experimental data, describing the 
               probability to find a parton inside a hadron carrying a given fraction of the
               hadron total momentum}
}

\newglossaryentry{gauge}
{
   name={Gauge Invariance},
   description={Local invariance of a Lagrangian under some kind of transformations}
}

\newglossaryentry{glo:ip}
{
   name={Interaction Point},
   description={Physical point where the collision between two particles occurs}
}

\newglossaryentry{glo:reco}
{
   name={Event Reconstruction},
   description={Software algorithms applied to the raw data acquired in order to obtain the high
   	        level objects produced in the event, \emph{see also \gls{glo:event}}}
}

\newglossaryentry{glo:lo}
{
   name={Leading Order},
   description={Calculation of the matrix element up to first order of $\alpha_s$. Called also 
               Born or tree level}
}

\newglossaryentry{glo:pd}
{
   name={Primary Dataset},
   description={Set of experimental data recorded by CMS sorted by trigger paths, \emph{see also
                \gls{glo:triggerpath}}}
}

\newglossaryentry{glo:triggerpath}
{
   name={Trigger Path},
   description={Sequence of hardware and software algorithms applied to the raw experimental data 
                outcome used to reject or accept an event. Each path is optimised to select events
                with specific characteristics (muon, electron, hadronic content, ...)}
}

\newglossaryentry{glo:event}
{
   name={Event},
   description={Data obtained in a particles collision},
}

\newglossaryentry{glo:signal}
{
   name={Signal},
   description={Data events containing the theoretical expected process under study},
}

\newglossaryentry{glo:background}
{
   name={Background},
   description={Data events containing the theoretical expected processes which are not defined as 
                signal, \emph{see also \gls{glo:signal}}}
}

\newglossaryentry{glo:decaychannel}
{
   name={Decay Channel},
   description={Each combination of particles that a unstable particle can disintegrate}
}
\newglossaryentry{glo:channel}
{
   name={Channel},
   description={\emph{See \gls{glo:decaychannel}}}
}


\newacronym{lhc}{LHC}{Large Hadron Collider}
\newacronym{psb}{PSB}{Proton Synchrotron Booster}
\newacronym{ps}{PS}{Proton Synchrotron}
\newacronym{cms}{CMS}{Compact Muon Solenoid}
\newacronym{atlas}{ATLAS}{A Toroidal Large ApparatuS}
\newacronym{alice}{ALICE}{A Large Ion Collider Experiment}
\newacronym{lhcb}{LHCb}{Large Hadron Collider beauty}
\newacronym{lhcf}{LHCf}{Large Hadron Collider forward}
\newacronym{totem}{TOTEM}{TOTal Elastic and diffractive cross section Measurement}
\newacronym{sm}{SM}{Standard Model}
\newacronym{qft}{QFT}{Quantum Field Theory}
\newacronym{lo}{LO}{Leading Order}
\newacronym{nlo}{NLO}{Next to Leading Order}
\newacronym{nnlo}{NNLO}{Next to Next to Leading Order}
\newacronym{qcd}{QCD}{Quantum Chromodynamics}
\newacronym{qed}{QED}{Quantum Electrodynamics}
\newacronym{gws}{GWS}{Glashow, Weinberg and Salam}
\newacronym{ym}{YM}{Yang-Mills}
\newacronym{ew}{EW}{Electro-Weak}
\newacronym{ckm}{CKM}{Cabibbo--Kobayashi--Maskawa}
\newacronym{gut}{GUT}{Grand Unification Theory}
\newacronym{pdf}{PDF}{Parton Distribution Functions}
\newacronym{mstw8}{MSTW8}{Martin--Stirling--Thorne--Watt}

\newacronym{lep}{LEP}{Large Electron--Positron Collider}
\newacronym{ecal}{ECAL}{Electromagnetic CALorimeter}
\newacronym{hcal}{HCAL}{Hadronic CALorimeter}

\newacronym{ip}{IP}{Interaction Point}
\newacronym{edm}{EDM}{Event Data Model}
\newacronym{sta}{STA}{Stand-alone muon track}
\newacronym{gsf}{GSF}{Gaussian Sum Filter}
\newacronym{jec}{JEC}{Jet energy correction}

\newacronym{hlt}{HLT}{High Level Trigger}
\newacronym[\glslongpluralkey={Primary Datasets}]{pd}{PD}{Primary Dataset}
\newacronym{daq}{DAQ}{data acquisition system}
\newacronym{isr}{ISR}{Initial State Radiation}
\newacronym{fsr}{FSR}{Final State Radiation}

\newacronym{mva}{MVA}{Multivariate Analysis}
\newacronym{bdt}{BDT}{Boosted Decision Tree}
\newacronym{pog}{POG}{Physics Object Group}
\newacronym{wp}{WP}{Working Point}

\newacronym{sf}{SF}{scale factor}

\newacronym{fom}{FOM}{fakeable object method}

\newacronym{blue}{BLUE}{Best Linear Unbiased Estimator}

\newacronym{atgc}{aTGC}{anomalous triple-gauge boson couplings}

\microtypesetup{protrusion=false}
\tableofcontents
\listoffigures
\listoftables
\cleardoublepage
\phantomsection
\addcontentsline{toc}{chapter}{Notation and conventions}
\chapter*{Notation and Conventions}
\markboth{\MakeUppercase{Notation and Conventions}}{}

\begin{table*}[ht]
	\begin{center}
		{\large
			\begin{tabular}{lp{1.3cm}l}
			\multicolumn{3}{c}{\textbf{List of selected symbols}}\\
			&&\\
			\ensuremath{\mathbf{\mathcal{A}}}    && acceptance \\
			\ensuremath{\boldsymbol{\sigma}}     && cross section \\
			\ensuremath{\boldsymbol{\varepsilon}}&& efficiency \\
			\ensuremath{\mathbf{N_X}}            && event yields (of process X)  \\
			\ensuremath{\mathbf{\fr}}            && fake lepton \\
			\ensuremath{\mathbf{f}}              && fake rate \\
			\ensuremath{\boldsymbol{M_X,m_X}}    && mass (of particle X)        \\
			\ensuremath{\boldsymbol{\MET,\ETslash}}&& missing transverse energy \\
			\ensuremath{\mathbf{\mathcal{L}}}    && luminosity   \\
			\ensuremath{\mathbf{\pr}}            && prompt lepton \\
			\ensuremath{\mathbf{p}}              && prompt rate \\
			\ensuremath{\boldsymbol{\eta}}       && pseudorapidity \\
			\ensuremath{\mathbf{sf}}             && scale factor \\
			\ensuremath{\boldsymbol{\ET}}        && transverse energy \\
			\ensuremath{\boldsymbol{m_T}}        && transverse mass \\
			\ensuremath{\boldsymbol{\pt}}        && transverse momentum \\
			\ensuremath{\mathbf{y}}              && rapidity  \\
		\end{tabular}
		}
	\end{center}
\end{table*}

The \emph{natural units} are used along this dissertation except where it is explicitly indicated.
The natural units in high energy particle physics are defined through
\begin{equation*}
  \hslash=c=1
\end{equation*}
being $\hslash$ the Planck's constant and $c$ the speed of light in the vacuum. Therefore
in this system
\begin{equation*}
 [\text{length}]=[\text{time}]=[\text{energy}]^{-1}=[\text{mass}]^{-1}
\end{equation*}

\vspace*{0.5cm}

\cleardoublepage
\phantomsection
\addcontentsline{toc}{chapter}{Preface}
\chapter*{Preface}
\markboth{\MakeUppercase{PREFACE}}{}

This dissertation is submitted for the degree of Doctor of Philosophy at the University of
Cantabria. The research described herein was conducted under the supervision of Doctor Iv\'an
Vila \'Alvarez and Professor Teresa Rodrigo Anoro in the \emph{Instituto de F\'isica de 
Cantabria} (IFCA) between October 2011 and November 2013. 

This work is ultimately based on the experimental apparatus and data of the CMS experiment
from proton-proton collision provided by the LHC at CERN. Although none of the text of the 
dissertation is taken directly from previously published or collaborative articles, except where 
acknowledgements and references are made, the collaborative efforts with other members of the 
the CMS collaboration have made possible this research. 

\microtypesetup{protrusion=true}

\mainmatter 
\chapter{Theoretical Framework}\label{ch1}
The \gls{sm}~\cite{Halzen:1984mc} is the current theory accepted nowadays in particle physics. 
The \gls{sm} is written down in the language of quantum mechanics where a physical system is described by 
its \emph{state} and a physical process is understood as the \emph{transition} from one state to
different state. Moreover, the quantum mechanics language does not allow to ask for any 
predictions other than probabilities. In this context, we can calculate the probability for a given 
transition to occur, i.e. cross sections and decay\glsadd{ind:decay} rates. Those are the only observables we can
expect to predict using the quantum mechanics formalism~\cite{Sakurai1993Modern}.

The \gls{sm}\glsadd{ind:sm} of particle physics is a relativistic \gls{qft}\glsadd{ind:qft}, a Lagrangian
($\mathcal{L}$) fully describes the theory~\cite{Peskin:1995qft} and the system described by
the Lagrangian is composed by quantum fields and interpreted as particles. The fields are split
in the interaction (boson) fields and fermion fields which receive those interactions. In this 
picture, the interaction fields are bosons carrying the quanta of the forces and the fermion 
fields are interpreted as excited states of the medium (vacuum). The Lagrangian describes the 
coupling between forces and fermionic fields and how they modify their internal and/or 
dynamical state as a consequence of the interactions. Therefore, the \gls{sm}\glsadd{ind:sm} of particle physics
describes matter as a collection of particles, called fermions (\emph{quarks} and \emph{leptons}) 
interacting between them by interchanging quanta of force (particles called \emph{bosons}). Note 
that in order to keep the theory inside the special relativity scope, the Lagrangian has to be 
built assuring its invariance under Lorentz transformations.

The \gls{sm}\glsadd{ind:sm} describes three of the four known fundamental forces in nature: electromagnetism,
weak force and strong force\footnote{The fourth fundamental force, the gravity, 
is still not described using a gauge theory as the \gls{sm} is.} by using the concept of 
\emph{local gauge invariance}. The Lagrangian of the system has to be invariant under some 
internal or local gauge transformations that in fact determine the interaction: 
\emph{gauge theories}.

This chapter reviews the main features of the \gls{sm}\glsadd{ind:sm} of particle physics without 
entering in full details. The subject is widely covered by many books, in 
particular~\cite{Peskin:1995qft},~\cite{Halzen:1984mc} and~\cite{Sakurai1993Modern} have been used
as references for this chapter. The chapter starts up with a panoramic view of
gauge theories, then provides a qualitative description of the \gls{sm} of particle physics and
its component constituents. A brief summary of the Quantum Cromodynamics Lagrangian and its
properties is described. Also, a brief description of the Electroweak sector is shown introducing
the Higgs mechanism to give mass to the gauge bosons and to the fermionic fields. After this summary
of the \gls{sm}, some shortcomings of the theory are exposed giving theoretical support to believe 
that the \gls{sm} is not the ultimate theory, although it works remarkably well at the energies in which 
have been tested. Finally, a qualitative introduction to physics at hadron colliders is described
in order to contextualise the theory with our particular experiment.

\section{Gauge Theories and the Standard Model}
The gauge theories rely essentially on the assertion that every physical system should be invariant 
under any artifice used to describe it. In special or general relativity, that statement is applied
to the coordinates used to describe nature, which should play no role in the formulation of the
physical laws~\cite{Utiyama:1956sy}. This is closely related to the fact that the position of any
system must be given with respect, \emph{relative} to, any other system (observer, reference 
frame, ...). This is what we understand as \emph{gauge variables}; once you fixed the arbitrariness
(choose a gauge), you can use this variables to describe a physical system through a 
Lagrangian~\cite{Rovelli:arXiv1308.5599}. 
But if you choose another gauge, the Lagrangian, i.e. the physical laws, has to keep exactly the 
same structure (using the new gauge variables, of course). We say, then, that the Lagrangian is 
\emph{gauge-invariant}.

Every local transformation of the gauge variables that keeps the Lagrangian invariant is associated to
a conserved current, called \emph{Noether current}, carrying a conserved charge, called 
\emph{Noether charge} (Noether's theorem~\cite{Noether:1918zz}). That group of local transformations
can be described using group theory through a symmetry group\footnote{In particular, a continuous 
local transformation is described by Lie groups. Lie groups contain an infinite number of 
elements, but the elements can be written in terms of a finite number of 
parameters~\cite{Chevalley199912}.}, being the generators of that group the Noether charges. In 
turn, Noether current describes the interaction resulting from the considered 
symmetry. This idea was used to formulate the \gls{sm} of particle physics 
using some internal degrees of freedom for the quantized fields which can be represented by
$SU(3)_C\times SU(2)_L\times U(1)_Y$. 

To elucidate these ideas, let's assume a Lagrangian description of a quantum system built from
free quantum fields, $\Psi(x^{\mu})\in\mathbb{C}$, which are interpreted as probability density fields. 
The terms of the Lagrangian should use only $|\Psi(x^{\mu})|^{2n}$ $(n=1,\dots)$ combinations in 
order to assure the invariance of the Lagrangian, therefore whether we use $\Psi(x^{\mu})$ or 
$\Psi'(x^{\mu})=e^{i\xi^a(x^{\mu})T_a}\Psi(x^{\mu})$, the Lagrangian is going to describe the same
physical system\footnote{Note that $T_a$ are the generators of the local transformation.}. This is 
the natural way in which gauge theories\footnote{A better name for these theories would be \emph{phase}
theories, but the historical misconception of the term remains.} emerge. Since the Lagrangian depends 
also on the space-time derivatives of the field, $\Lagrangian(\Psi,\partial_{\mu}\Psi)$, once the phase 
transformation is decided, we should check how the space-time derivatives behave in 
order to assure the Lagrangian invariance. The covariant requirement generalises the derivative
introducing the \emph{covariant derivative} as~\cite{Peskin:1995qft},
\begin{equation}
	D_{\mu}=\partial_{\mu}-igA_{\mu}^aT_a
\end{equation}
which is the derivative operator to use in the Lagrangian instead of $\partial_{\mu}$ only. Notice
that the inclusion of the transformation group introduces one field for each generator of the group, 
$A_{\mu}^a\equiv\frac{1}{g}\partial_{\mu}\xi^a$, for the sake of Lagrangian invariance. Hence, the 
starting Lagrangian describing a free particle system becomes a Lagrangian describing a particle 
interacting with the fields $A_{\mu}^a$ as soon as a local gauge transformation has been introduced.
Postulating that a quantum field obeys a local symmetry, a vector field must appear and interact 
with the quantum field. Furthermore, there will be conserved currents, as consequence of Noether's
theorem, and its associated conserved charge which characterises the system and so a 
quantum number will emerge. 

\section{Quantum fields: matter and interactions}
Matter and interactions are described as fields, spinors $\psi\in\mathbb{C}^4$, using relativistic
\gls{qft}\glsadd{ind:qft}. Those fields live in a Minkowskian space-time\footnote{Using a Geometric Algebra
approach, it is possible to describe the Dirac equation in terms of a space-time algebra
(STA) where the usual Dirac matrices are an explicit basis for the 
STA~\cite{Hestenes03spacetimephysics}.} defined 
by the Dirac matrices, $\gamma^{\mu}$, and are 
interpreted as probability density fields\footnote{The fields can be interpreted as excitation states
of the vacuum~\cite{Peskin:1995qft}.}, such that the observables associated with those 
fields depend of $\apsi\psi$, where the \emph{adjoint spinor} $\apsi=\dpsi\gamma^0$ is defined in 
order to obtain Lorentz scalars. Therefore, invariant requirement on the Lagrangian implies that the
terms in the Lagrangian should be formed by functions of $\apsi\psi$ or even combinations with/of
other fields. 

The fermionic fields describing matter are represented by 4-component Dirac spinors. The Lagrangian
associated to a free fermion of mass $m$ and the derived Dirac equation of motion are:
\begin{equation}
	\Lagrangian=i\apsi\gamma^{\mu}\partial_{\mu}\psi-m\apsi\psi
	   \qquad\qquad\left(i\gamma^{\mu}\partial_{\mu}-m\right)\psi=0
	   \label{ch1:eq:dirac}
\end{equation}
The solutions of the Dirac equations forming a eigenvalue problem are of the form: 
$\psi(x)=u(x)e^{\pm ip\cdot x}$, which are eigenstates of the \emph{helicity} operator
\begin{equation}
	\hat{h}\equiv\frac{\hat{\mathbf{p}}\cdot\hat{\mathbf{S}}}{|\hat{\mathbf{p}}|}
\end{equation}
where $\mathbf{\hat{p}}$ is the momentum operator and $\hat{\mathbf{S}}$ the spin operator. 
Spin-$1/2$ particles with its spin direction parallel to the linear momentum have $+1/2$ helicity
and they are called \emph{right-handed} states, and spin-$1/2$ particles with its spin direction 
anti-parallel to the linear momentum have $-1/2$ helicity, \emph{left-handed} states. As we will
see in next section, the right- and left-handed states do not behave in the same way when 
weakly interacting. Obviously, the two states are not disconnected since it is always possible to
perform a Lorentz transformation to a reference system where the particle momentum has the same
value but opposite sign, keeping the spin invariant. However, this is not possible for massless 
particles as the neutrinos. Thus, it is possible to project each fermion, except for the neutrinos
which only exist as left-handed states, into its left-handed ($\psi_{L}$) and right-handed 
($\psi_{R}$) components, the so called Weyl representation, using the projection operators 
$P_L=\frac{1}{2}(1-\gamma^5)$ and $P_R=\frac{1}{2}(1+\gamma^5)$, 
where $\gamma^5=i\gamma^0\gamma^1\gamma^2\gamma^3$
\begin{equation}
	\psi_{L}=P_L\psi=\frac{1}{2}(1-\gamma^5)\psi\quad\psi_{R}=P_R\psi=\frac{1}{2}(1+\gamma^5)\psi
\end{equation}
and $\psi=\psi_R+\psi_L$.

The fermionic elementary particles have been classified according their properties and quantum
numbers in leptons and quarks and organised in families of identical structure differing only
for their mass. Table~\ref{ch1:table:fermions} shows the fermion classification in the 
\gls{sm} and the relevant quantum numbers for each particle~\footnote{The anti-particle related
to each particle is described by the same mass and quantum numbers but the charge reversed.}:
the electric charge ($Q_{EM}$), the third component of the weak isospin $T_3$ and 
the weak hypercharge $Y$, defined by the Gell-Mann and Nishijma relation $Q_{EM}=T_3+Y/2$. Note 
that all the leptons and quarks have spin 1/2.
\begin{table}[!htpb]
	\centering
	\begin{tabular}{c c c c c c c}\hline\hline
		\multicolumn{3}{c}{} & $Q_{EM}$ & $Y$ & $T_3$ & Interactions\\ \hline
		\multicolumn{3}{c}{LEPTONS}     &          &     &       &\\ \hline
		\multirow{2}{*}{$\Ldoublet{\nu_{e_L}}{e_L}$} & \multirow{2}{*}{$\Ldoublet{\nu_{\mu_L}}{\mu_L}$} &
			\multirow{2}{*}{$\Ldoublet{\nu_{\tau_L}}{\tau_L}$} & 0 & -1 & +1/2 & weak \\
		  & & & -1 & -1 & -1/2 & weak,EM \\\hline
		\multicolumn{3}{c}{QUARKS}     &          &     &       &\\ \hline
		\multirow{2}{*}{$\Ldoublet{u_L}{{d'}_L}$} & \multirow{2}{*}{$\Ldoublet{c_L}{{s'}_L}$} &
			\multirow{2}{*}{$\Ldoublet{t_L}{{b'}_L}$} & +2/3 & +1/3 & +1/2 & weak,EM,strong \\
		  & & & -1/3 & +1/3 & -1/2 & weak,EM,strong \\
		$u_R$ & $c_R$ & $t_R$ & +2/3 & +4/3 & 0 & weak,EM,strong\\
		$d_R$ & $s_R$ & $b_R$ & -1/3 & -2/3 & 0 & weak,EM,strong\\\hline
	\end{tabular}
	\caption[Fermion properties]{Fermionic fields in the SM. The main quantum numbers and 
	        the interaction that are sensitive to feel are shown, but the colour charge. 
		All the fermions have spin $1/2$ and for each listed particle, there is a 
		corresponding antiparticle with same mass but with opposite value of $Q$.
		The left-handed states are grouped into weak isospin doublets represented by
		one-array columns.}\label{ch1:table:fermions}
\end{table}


The interactions between fermions are described by bosonic vector fields, which in turn are related to
a local symmetry in the system. A free scalar boson with mass $m$ is represented by a scalar 
field $\phi\in\mathbb{C}$, dynamically described by the Klein-Gordon Lagrangian, from which its
equation of motion is derived,
\begin{equation}
	\Lagrangian=(\partial_{\mu}\phi)^{\dagger}(\partial^{\mu}\phi)-m\phi^{\dagger}\phi
	   \qquad\qquad\left(\Box-m\right)\phi=0
	\label{ch1:eq:KleinGordon}
\end{equation}
In the case of a vector boson $B_{\mu}$ of mass $m$, the dynamics for the free boson is obtained with
the Lagrangian,
\begin{equation}
	\Lagrangian=-\frac{1}{4}F^{\mu\nu}F_{\mu\nu}+\frac{1}{2}B_{\mu}B^{\mu}
\end{equation}
where $F_{\mu\nu}$ is the dynamic term of the field,
\begin{equation}
	F_{\mu\nu}=\partial_{\mu}B_{\nu}-\partial_{\nu}B_{\mu}
\end{equation}

The \gls{sm} describes three of the four elementary interactions between particles. The 
interactions are modelled with the $U(1)_Y\times SU(2)_L\times SU(3)_C$ symmetry group describing 
the fermion fields transformation in the different internal spaces. The $SU(3)_C$ group of \gls{qcd}\glsadd{ind:qcd}
deals with the colour charge content of the quarks mediated by eight massless coloured bosons called
gluons. Every quark can be represented in the colour charge space as a $SU(3)_C$ triplet and the 
leptons are colourless $SU(3)_C$ singlets, therefore quarks are sensible to the strong force whereas
leptons are not. The \gls{qcd} Lagrangian characterising the quark coupling with the colour charge
current is,
\begin{equation}
	\mathcal{L}_\mathrm{QCD} = \bar{\psi}\left(i\gamma^\mu D_\mu - m\right) \psi -
	          \frac{1}{4}G^a_{\mu \nu} G^{\mu \nu}_a 
		  \label{ch1:eq:qcdL}
\end{equation}
where $G^a_{\mu\nu}$ are the gluon field strength,
\begin{equation}
	G^a_{\mu\nu}=\partial_{\mu}A_{\nu}^a-\partial_{\nu}A_{\mu}^a
	    +g_cf^{abc}A_{\mu}^b A_{\nu}^c\,,\qquad a=1,\dots,8
\end{equation}
built from the \emph{structure constants} $f_{abc}$ of the $SU(3)_C$ group; the 8 gauge field 
\emph{gluons} $A_{\mu}^a$ and the dimensionless coupling strength $g_c$.
The covariant derivative to keep gauge invariance is defined as
\begin{equation}
	D_{\mu}=\partial_{\mu}-ig_cA_{\mu}^aT_a
\end{equation}
where $T_a$ are the generators of the non-Abelian $SU(3)_C$ group. The essential properties and features of 
\gls{qcd}\glsadd{ind:qcd} are summarised as:
\begin{itemize}
	\item Quarks carry colour and electric charge
	\item Colour charge is exchanged by eight bicoloured, spin-1 and massless \emph{gluons}
	\item Gluons themselves carry colour charge, so they can interact with other gluons
	\item The fundamental \gls{qcd} vertices are:
		\begin{itemize}
		\item quark-gluon interactions $\begin{array}{c}
			\resizebox{3cm}{!}{\begin{fmffile}{qcdVertexQQ} 
  \fmfframe(1,2)(2,2){
	\begin{fmfgraph*}(110,62) 
		\fmfleft{i1,i2}		\fmfright{o1}
		\fmflabel{$q$}{i1}
		\fmflabel{$\bar{q}$}{i2}
		\fmf{fermion}{v1,i2}
		\fmf{fermion}{i1,v1}
		\fmf{gluon}{v1,o1}
		\fmflabel{$g$}{o1}
	\end{fmfgraph*}
}
\end{fmffile}
}\end{array}$ $\propto g_c$
		\item 3-gluon self-interactions $\begin{array}{c}
			\resizebox{3cm}{!}{\begin{fmffile}{qcdVertex3G} 
  \fmfframe(1,2)(2,2){
	\begin{fmfgraph*}(110,62) 
		\fmfleft{i1,i2}		\fmfright{o1}
		\fmflabel{$g$}{i1}
		\fmflabel{$g$}{i2}
		\fmf{gluon}{v1,i2}
		\fmf{gluon}{i1,v1}
		\fmf{gluon}{v1,o1}
		\fmflabel{$g$}{o1}
	\end{fmfgraph*}
}
\end{fmffile}
}\end{array}$ $\propto g_c$
		\item 4-gluon self-interactions $\begin{array}{c}
			\resizebox{3cm}{!}{\begin{fmffile}{qcdVertex4G} 
  \fmfframe(1,2)(2,2){
	\begin{fmfgraph*}(110,62) 
		\fmfleft{i1,i2}		\fmfright{o1,o2}
		\fmflabel{$g$}{i1}
		\fmflabel{$g$}{i2}
		\fmf{gluon}{v1,i2}
		\fmf{gluon}{i1,v1}
		\fmf{gluon}{v1,o1}
		\fmf{gluon}{v1,o2}
		\fmflabel{$g$}{o1}
		\fmflabel{$g$}{o2}
	\end{fmfgraph*}
}
\end{fmffile}
}\end{array}$ $\propto g_c^2$
		\end{itemize}
	\item Colour confinement: the coupling strength\footnote{The coupling strength can be 
		renormalised as	$\alpha_s=\frac{g_c^2}{4\pi}$} $\alpha_s$ increases with decreasing 
		energies (or large distances), as a consequence only colour neutral hadrons can be
		observed in nature~\cite{Halzen:1984mc}
	\item Asymptotic freedom: the coupling strength $\alpha_s$ decreases with increasing energies 
		(or short distances) becoming small enough to treat coloured particles as 
		free~\cite{Halzen:1984mc}. The coupling can be approximated as 
		$\alpha_s\simeq1/(\beta_0ln(k^2/\Lambda_{QCD}))$, being $\beta_0$ a constant, $k$ the 
		energy of the process and $\Lambda_{QCD}$ the \gls{qcd} scale. Therefore in this regime 
		the QCD calculations can be performed as expansions of $\alpha_s$, perturbatively.
\end{itemize}

The role played by \gls{qcd} in high energy proton colliders, such as the \gls{lhc}\glsadd{ind:lhc},
is fundamental. 
Section~\ref{ch1:sec:physicsathadroncolliders} offers a brief summary of the underlying physics
used to describe high energy collisions at hadron colliders.

\subsection{The Electro-Weak Interaction}
The electroweak interaction was postulated\footnote{The electroweak theory as unified theory of
\gls{qed}\glsadd{ind:qed} and weak force was first described by Glashow~\cite{Glashow:1961tr}, 
Weinberg~\cite{PhysRevLett.19.1264} and Salam~\cite{Salam1964168}.}
through the introduction of the symmetry group $SU(2)_L\times U(1)_Y$ for the fermionic fields, 
emerging a set of four vector fields $W_{\mu}^i$ ($i=1,2,3$) related to the $SU(2)_L$ group and
$B_{\mu}$ corresponding to the $U(1)_Y$ symmetry. The fermionic fields are split into isospin
doublets for the left-handed states and isospin singlets for right-handed states, meaning that
the doublets couple to the $SU(2)_L$ gauge bosons while the singlet states do not. As a 
consequence of requiring gauge invariance the covariant derivative is defined as:
\begin{equation}
	D_{\mu}=\partial_{\mu}-igT_iW_{\mu}^i-ig'\frac{Y}{2}B_{\mu}
\label{ch1:eq:d_ew}
\end{equation}
where $T_i$,$g$ and $Y$,$g'$ are the generators and coupling strengths of the $SU(2)_L$ and $U(1)_Y$
groups, respectively. The kinematic terms for the gauge fields included in the Lagrangian make use 
of the field strength tensors,
\begin{align}
	W^i_{\mu\nu} &= \partial_{\mu}W^i_{\nu}-\partial_{\nu}W^i_{\mu}+g\epsilon^{ijk}W_{\mu}^jW_{\nu}^k\\
	B_{\mu\nu}  &= \partial_{\mu}B_{\mu}-\partial_{\nu}B_{\mu}
\end{align}
where $\epsilon^{ijk}$ is the total antisymmetric tensor. The \gls{ym}\glsadd{ind:ym} 
\gls{ew}\glsadd{ind:ew} Lagrangian can be expressed in the following form:
\begin{align}	
	\Lagrangian_{YM}=\Lagrangian_{kin}+\Lagrangian_{charged}+ & \Lagrangian_{neutral} = \notag\\
	        & - \frac{1}{4}W_{\mu\nu}^iW_i^{\mu\nu} - \frac{1}{4}B_{\mu\nu}B^{\mu\nu}\notag\\ 
		& - \apsi_Li\gamma^{\mu}\left(\partial_{\mu}-igT_iW_{\mu}^i-ig'\frac{Y}{2}B_{\mu}\right)\psi_L
	              \label{ch1:eq:Lagrangian_ew_YM}\\
		& - \apsi_Ri\gamma^{\mu}\left(\partial_{\mu}-ig'\frac{Y}{2}B_{\mu}\right)\psi_R\notag
\end{align}

The experimental constraint of having two neutral currents~\cite{Peskin:1995qft}, only one of them with 
parity conserved, was resolved by transforming the original basis. Defining 
the weak mixing angle as,
\begin{align}
	sin\theta_W=\frac{g'}{\sqrt{g^2+g'^2}}\label{ch1:eq:Wangle1}\\
	cos\theta_W=\frac{g}{\sqrt{g^2+g'^2}\label{ch1:eq:Wangle2}}
\end{align}
it is possible to perform a rotation by an angle $\theta_W$ in the neutral sector revealing new
physic vector fields (with mass=0),
\begin{align}	
	&W_{\mu}^{\pm}=\frac{1}{\sqrt{2}}(W_{\mu}^1\pm W_{\mu}^2)\label{ch1:eq:Wpm}\\
	&Z_{\mu}=cos\theta_W\,W_{\mu}^3-sin\theta_W\,B_{\mu}\label{ch1:eq:Z}\\
	&A_{\mu}=cos\theta_W\,B_{\mu}+sin\theta_W\,W_{\mu}^3\label{ch1:eq:A}
\end{align}
Rearranging the Lagrangian using those new fields after the rotation angle the neutral sector is expressed as,
\begin{align}
	\Lagrangian_{neutral}&=  -A_{\mu}\left\{\apsi_L\gamma^{\mu}\left(gT^3sin\theta_W+g'\frac{Y}{2}cos\theta_W\right)\psi_L+
  		\apsi_R\gamma^{\mu}\left(g'\frac{Y}{2}cos\theta_W\right)\psi_R\right\}\notag\\
		-& Z_{\mu}\left\{\apsi\gamma^{\mu}\left(gT^3cos\theta_W-g'\frac{Y}{2}sin\theta_W\right)\psi-
  		\apsi_R\gamma^{\mu}\left(g'\frac{Y}{2}sin\theta_W\right)\psi_R\right\}\label{ch1:eq:EWneutral}
\end{align}
The relations~\eqref{ch1:eq:Wangle1} and~\eqref{ch1:eq:Wangle2} materialise the
electroweak unification with the link between the elementary electromagnetic charge $e$ and the
weak mixing angle $\theta_W$ and the coupling strengths of the weak isospin and the hypercharge,
\begin{equation}	
	e=gsin\theta_W=g'cos\theta_W
	\label{ch1:eq:ewithW}
\end{equation}
Making use of it and the Gell-Mann and Nishijma relation $Q=T^3+Y/2$ in the equation~\eqref{ch1:eq:EWneutral},
\begin{align}
	\Lagrangian_{neutral}= A_{\mu}\left\{e\apsi\gamma^{\mu}Q\psi\right\} &+\notag\\
	+& Z_{\mu}\left\{\frac{g}{cos\theta_W}\apsi\gamma^{\mu}\left(T^3\frac{1}{2}(1-\gamma^5)-Qsin^2\theta_W\right)\psi\right\}
	\label{ch1:eq:EWneutral_rearranged}
\end{align}
where it is assumed that $T^3$ is zero when applied over a right-hand fermion, so using only 
the left-hand components $\frac{1}{2}(1-\gamma^5)\psi$ of the fermion field.
The electromagnetic and neutral weak currents are,
\begin{align}
	&J^{em}_{\mu}=e\apsi\gamma_{\mu}Q\psi\\
	&J^{NC}_{\mu}=\frac{1}{cos\theta_W}\apsi\gamma_{\mu}\left(T^3\frac{1}{2}(1-\gamma_5)-Qsin^2\theta_W\right)\psi
\end{align}
where the third component of the weak isospin current is
\begin{equation}
	J^3_{\mu}=\apsi\gamma_{\mu}\left(T^3\frac{1}{2}(1-\gamma^5)\right)\psi
\end{equation}
Therefore it is straightforward to find the relation between electromagnetic and neutral weak currents,
\begin{equation}
	J^{NC}_{\mu}=J^{3}_{\mu}-sin^2\theta_W J^{em}_{\mu}
\end{equation}
 
We have successfully achieved the objective of expressing the observed neutral currents in terms of 
the currents $J_{\mu}^3$ and $J_{\mu}^{Y}$ belonging to symmetry groups $SU(2)_L$ and $U(1)_Y$. 

To complete the picture, the charged currents, 
\begin{align}
	J_{\mu}^+=\frac{1}{\sqrt{2}}\apsi\gamma_{\mu}\left(T^1-T^2\right)\frac{1}{2}(1-\gamma^5)\psi\\
	J_{\mu}^-=\frac{1}{\sqrt{2}}\apsi\gamma_{\mu}\left(T^1+T^2\right)\frac{1}{2}(1-\gamma^5)\psi
\end{align}
are also expressed in terms of the $J_{\mu}^i$,
\begin{align}
	J_{\mu}^+=\frac{1}{\sqrt{2}}(J_{\mu}^1-J_{\mu}^2)\\
	J_{\mu}^-=\frac{1}{\sqrt{2}}(J_{\mu}^1+J_{\mu}^2)
\end{align}

Up to know the theory sketched here has been constructed requiring massless fermions and gauge bosons, since 
the presence of mass terms for the gauge fields destroys the gauge invariance of the Lagrangian. But 
experimental results prove the fermions and the $W^{\pm}$ and $Z$ bosons to be massive. The inclusion
of the mass terms in a gauge invariant way is provided by the spontaneous symmetry breaking 
mechanism.

\subsection{The spontaneous symmetry breaking mechanism}\label{ch1:sec:goldstone}
\begin{figure}[!htpb]
	\centering
	\includegraphics[scale=0.6]{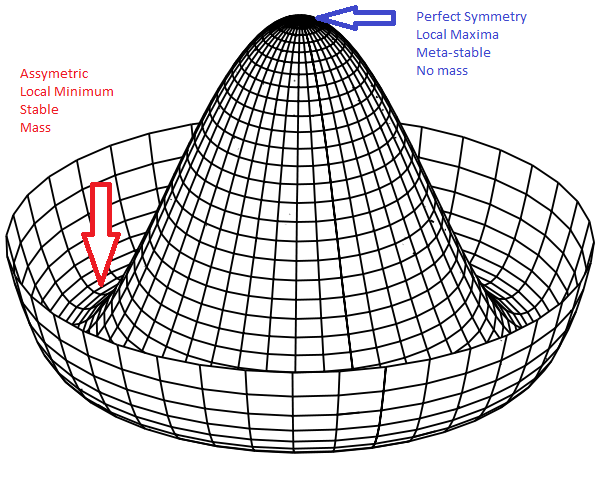}
	\caption[Goldstone potential example]{The potential $V(\phi)$ for a complex scalar field 
		with $\mu^2<0$ and $\lambda > 0$.}\label{ch1:fig:higgsPotential}
\end{figure}

The \emph{Goldstone's theorem}~\cite{Goldstone:1961eq} states that for every spontaneously broken 
continuous symmetry, the theory must contain a massless particle. The massless fields arising 
through spontaneous symmetry breaking are called \emph{Goldstone bosons}. The Goldstone theorem 
deals with global continuous symmetries of a system and reveals a ground state which does not possess
the global continuous symmetry of the Lagrangian. Choosing an explicit ground state the symmetries
is \emph{spontaneously broken}. We can visualise this effect using a theory with a complex scalar
self-interacting field $\phi$ described by the Lagrangian,
\begin{equation}	
	\Lagrangian=T-V(\phi)=(\partial_{\mu}\phi)^*(\partial^{\mu}\phi)-\left(\mu^2\phi^*\phi+\lambda(\phi^*\phi)^2\right),
	\qquad (\lambda>0)
	\label{ch1:eq:LexampleHiggs}
\end{equation}
which is invariant under the transformation $\phi\rightarrow e^{i\alpha}\phi$, i.e., the Lagrangian
possesses a $U(1)$ global symmetry. The $\phi^4$ term shows that the four-particle vertex exists 
with coupling $\lambda$, so $\phi$ is a self-interacting field. There are two possible forms of the 
potential depending the sign of $\mu$. The case $\mu^2 >0$ is simply describing a scalar field with
mass $\mu^2$. The ground state of the system (the vacuum) corresponds to $\phi=0$. 
However, the case $\mu^2<0$ introduces a wrong sign for the mass term ($\phi^2$) of the field, since the 
relative sign of the mass term and the kinetic energy $T$ is always negative. We can express 
$\phi$ in its real and complex components $\phi=\frac{1}{\sqrt{2}}\left(\phi_1+i\phi_2\right)$ to show that 
the potential $V(\phi)$ has a circle of minima in the $\phi_1$, $\phi_2$ plane with radius $v$,
\begin{equation}\label{ch1:eq:minimumHiggsPotential}
	\phi_1^2+\phi_2^2=v^2\,,\quad v^2=-\frac{\mu^2}{\lambda}
\end{equation}
as shown in figure~\ref{ch1:fig:higgsPotential}. 
Translating the field $\phi$ to a minimum energy
position, which without loss of generality we may take as the point $\phi_1=v$, $\phi_2=0$, and
perturbing the field around the stable minima,
\begin{equation}
	\phi(x)=\sqrt{\frac{1}{2}}\left[v+\eta(x)+i\xi(x)\right]
\end{equation}
we can expand the original Lagrangian~\eqref{ch1:eq:LexampleHiggs} in terms of $\eta$ and $\xi$ and
obtain,
\begin{equation}
	\Lagrangian = \frac{1}{2}(\partial_{\mu}\xi)^2+\frac{1}{2}(\partial_{\mu}\eta)^2+\mu^2\eta^2
	   +\mathcal{O}(\eta^3,\xi^3,\dots)
	   \label{ch1:eq:LexampleHiggs2}
\end{equation}
Therefore, the third term of the Lagrangian~\eqref{ch1:eq:LexampleHiggs2} has the form of a mass term
with the right sign, $-1/2m_{\eta}\eta^2$, so the $\eta$-mass is $m_{\eta}=\sqrt{-2\mu^2}$. Note 
that the two first terms represent the kinetic energy of the $\xi$ and $\eta$ fields respectively, but
there is no mass term for $\xi$, the massless Goldstone boson. The theory contained a symmetry,
i.e. the field $\phi$ was invariant under rotation represented by the circle of minima ground 
states, but that symmetry of the Lagrangian has apparently been broken by our choice of the 
ground state ($\phi_1=v$, $\phi_2=0$) around which to do our perturbation calculations. Thus,
the Lagrangian using the field around the ground state do not possess that symmetry, we say 
that the symmetry has been \emph{spontaneously broken}\footnote{Or more accurately, the symmetry is not
apparent in the ground state.}. And due to the spontaneous symmetry breaking the mass of the
$\eta$ field has been \emph{revealed}. So far, the theory still contains a massless gauge boson, but
demanding local invariance we shall see that the Goldstone boson is \emph{absorbed}. This is known 
as the Higgs mechanism~\cite{PhysRevLett.13.508}.

\subsubsection{The Higgs Mechanism}
The spontaneous symmetry breaking mechanism of a \emph{local} gauge symmetry, in particular 
$SU(2)_L\times U(1)_Y$ forces to include the covariant derivative~\eqref{ch1:eq:d_ew}. Also we need
to introduce a scalar field in the multiplet spinor representation belonging to $SU(2)_L\times U(1)_Y$,
\begin{equation}
	\phi = \left(\begin{array}{c} \phi^{+}\\\phi^0\end{array}\right)\;, \quad\text{with}
		\begin{cases} \phi^+=\frac{1}{\sqrt{2}}(\phi_1+i\phi_2)\\
			      \phi^0=\frac{1}{\sqrt{2}}(\phi_3+i\phi_4)
 	        \end{cases}
	\label{ch1:eq:higgs_fields}
\end{equation}
and add a term to the Yang-Mills electroweak Lagrangian~\eqref{ch1:eq:Lagrangian_ew_YM},
\begin{equation}
	\Lagrangian_{h}=(D_{\mu}\phi)^{\dagger}(D^{\mu}\phi)-V(|\phi|)\,,
	\label{ch1:eq:higgsLagrangian}
\end{equation}
being the Higgs potential $V(|\phi|)=\mu^2\phi^{\dagger}\phi+\lambda(\phi^{\dagger}\phi)^2$, and
$\mu^2<0$, $\lambda>0$ as we have seen in the previous section. Analogously to 
section~\ref{ch1:sec:goldstone} the potential $V(\phi)$ has its minimum at a finite value of
$\phi^{\dagger}\phi$ where
\begin{equation*}
	\phi^{\dagger}\phi=\frac{1}{2}(\phi^2_1+\phi_2^2+\phi^2_3+\phi^2_4)=-\;\frac{\mu^2}{2\lambda}\,,
\end{equation*}
We can choose a particular minimum,
\begin{equation*}
	\phi_1=\phi_2=\phi_4=0\,,\quad\phi_3^2=-\;\frac{\mu^2}{\lambda}\equiv v^2\,
\end{equation*}
and substituting in equation~\eqref{ch1:eq:higgs_fields}, the value of the field $\phi$ in the minimum is
\begin{equation}
	\phi_0=\sqrt{\frac{1}{2}}\left(\begin{array}{c} 0 \\ v\end{array}\right)
		\label{ch1:eq:vacuumchoice}
\end{equation}
It is possible to expand $\phi(x)$ about this particular vacuum
\begin{equation}
	\phi(x)=\sqrt{\frac{1}{2}}\left(\begin{array}{c} 0 \\ v+H(x)\end{array}\right)\,
		\label{ch1:eq:phiaboutvacuum}
\end{equation}
where the only remaining scalar field is $H(x)$, the Higgs field. The expanded $\phi(x)$ field
is substituted into the Lagrangian~\eqref{ch1:eq:higgsLagrangian}. The broken Lagrangian
obtained turns out to be invariant under $U(1)_{EM}$, i.e. the vacuum is still invariant under
electromagnetic interaction and therefore the associated gauge boson, the photon, remains massless.
But the choice of the vacuum described in equation~\eqref{ch1:eq:vacuumchoice} breaks both 
$SU(2)_L\times U(1)_Y$ and therefore mass terms appear for the associated gauge bosons. 
Re-expressing the Higgs Lagrangian (eq.~\eqref{ch1:eq:higgsLagrangian}) around the minimum,
\begin{equation}
   \begin{split}
	\Lagrangian_{h}&=(D_{\mu}\phi)^{\dagger}(D^{\mu}\phi)-V(|\phi|)= \\
          & {\textstyle = \frac{1}{2}\left(\left(\partial_{\mu}-igT_iW^i_{\mu}-ig'\frac{B_{\mu}}{2}\right)
	              \left(\begin{array}{c} 0 \\ v+ H\end{array}\right)\right)^{\dagger}
	  \left(\left(\partial_{\mu}-igT_iW^i_{\mu}-ig'\frac{B_{\mu}}{2}\right)
		      \left(\begin{array}{c} 0 \\ v+ H\end{array}\right)\right) }\\
 	     &\qquad \qquad  +\frac{\mu^2}{2}\left(v+H\right)^2-\frac{\lambda}{16}\left(v+H\right)^4
   \end{split}
   \label{ch1:eq:HLexplicit}
\end{equation}
The first term in equation~\eqref{ch1:eq:HLexplicit} is expressing the kinetic energy associated to
the Higgs boson, its interaction with the gauge bosons and, we will see, it is also expressing the
gauge boson masses. The second term is the quadratic term of the Higgs boson,~\ie the mass term,
\begin{equation}
	\frac{\mu^2}{2}\left(v+H\right)^2=\mu^2\frac{1}{\sqrt{2}}
               \left(\begin{array}{cc}0 & v+H\end{array}\right)
	\frac{1}{\sqrt{2}}\left(\begin{array}{c}0 \\ v+H\end{array}\right) = {\mu^2}|\phi|^2=-\frac{1}{2}m_H^2|\phi|^2
\end{equation}
Thus, recalling the vacuum expectation value, $v=-\mu^2/\lambda$, the mass for the Higgs field is 
$m_{H}=v\sqrt{2\lambda}$. The vacuum expectation value can be re-expressed as a function 
of the Fermi constant $v=\frac{1}{G_F\sqrt{2}}\simeq 246.22\;GeV$. The coupling $\lambda$ is a 
free parameter of the theory, so the mass is not predicted by the theory and has to be measured
experimentally. Note that the last term is expressing the Higgs self-coupling . 

Once we have obtained the mass term for the Higgs field, we develop 
equation~\eqref{ch1:eq:HLexplicit} focusing on the Higgs interactions with the gauge bosons,
\begin{equation}
 \begin{split}
   \Lagrangian_{h_i}=\frac{\left(v+H\right)^2}{8}g^2 & \left(W_{\mu}^1(W^1)^{\mu}+W_{\mu}^2(W^2)^{\mu}\right)+\\
          +&\frac{\left(v+H\right)^2}{8}g^2\left(\begin{array}{cc}B_{\mu} & W_{\mu}^3\end{array}\right)
	     \left(\begin{array}{cc} g^2 & g'g\\ g'g & g'^2\end{array}\right)
	      \left(\begin{array}{c}B_{\mu}\\W_{\mu}^3\end{array}\right)
 \end{split}
 \label{ch1:eq:Hi}
\end{equation}
Now, equation~\eqref{ch1:eq:Hi} reveals why the basis transformation using the weak angle defined at
equations~\eqref{ch1:eq:Wangle1} and~\eqref{ch1:eq:Wangle2} and therefore explains the definitions of the 
physical gauge bosons $W^{\pm}_{\mu}$, $Z_{\mu}$ and $A_{\mu}$ 
(equations~\eqref{ch1:eq:Wpm},~\eqref{ch1:eq:Z} and~\eqref{ch1:eq:Z}). Thus, we can express the interaction 
Lagrangian~\eqref{ch1:eq:Hi} in terms of these bosons,
\begin{eqnarray}
    \Lagrangian_{h_W}&=&\frac{v^2g^2}{4}W_{\mu}^+(W^-)^{\mu}+\frac{vg^2}{2}HW_{\mu}^+(W^-)^{\mu}+
              \frac{g^2}{4}H^2W_{\mu}^+(W^-)^{\mu}\label{ch1:eq:HiW}\qquad\qquad\\
    \Lagrangian_{h_{AZ}}&=&\frac{v^2}{8}(g^2+g'^2)Z_{\mu}Z^{\mu}+\frac{v}{4}(g^2+g'^2)HZ_{\mu}Z^{\mu}+
              \frac{(g^2+g'^2)}{8}H^2Z_{\mu}Z^{\mu}\label{ch1:eq:HiZA}
\end{eqnarray}
where we have split the interaction Lagrangian in two terms, each one referring to the involved gauge
boson, $\Lagrangian_{h_i}=\Lagrangian_{h_W}+\Lagrangian_{h_{AZ}}$. 
The $\Lagrangian_{h_W}$ Lagrangian contains a quadratic term in the gauge field plus the interaction
with the Higgs field. The quadratic term is identified as the mass,
\begin{equation}
	m^2_{W^{\pm}}W_{\mu}^+(W^-)^{\mu}=\frac{v^2g^2}{4}W_{\mu}^+(W^-)^{\mu}\implies 
	 m_{W^{\pm}}=\frac{1}{2}vg
\end{equation}
and analogously for the $Z$ boson, $m_{Z}=\frac{1}{2}v\sqrt{g^2+g'^2}$. Notice that the Lagrangian 
does not contain mass terms for the $A_{\mu}$ field neither any coupling term between the Higgs and 
$A_{\mu}$, meaning that the Higgs mechanism in the \gls{sm} predicts a massless photon and no direct
couplings between the Higgs and the photon. Summarising the gauge boson masses:
\begin{align}
	&m_{W^{\pm}}=\frac{1}{2}vg\\
	&m_{Z^0}=\frac{1}{2}v\sqrt{g^2+g'^2}\\
	&m_{A} = 0
\end{align}
On the other hand, both Lagrangians contain coupling terms between the Higgs and the gauge bosons, 
where it can be observed that the coupling strength is proportional to the gauge boson mass,
\begin{eqnarray}
      \frac{vg^2}{2}HW_{\mu}^+(W^-)^{\mu} \implies y_W=\frac{vg^2}{2}=gm_{W}\\
      \frac{vg^2}{2}HZ_{\mu}Z^{\mu} \implies y_Z=\frac{v}{4}=\frac{g m_{Z}}{2cos\theta_W}\\
\end{eqnarray}

Summarising, the mass acquired by the gauge bosons in the \gls{sm} is because of their interaction
with the Higgs boson,~\ie $y_W,~y_Z\neq0$, and because of its non-vanishing vacuum expectation 
value, $v\neq0$.

\subsection{The Fermionic sector}
So far, the Higgs mechanism has been useful to generate massive gauge bosons in a gauge invariant
way. But fermions mass terms have not been put into the Lagrangian because the left- and right-handed
components of the various fermion fields have different gauge quantum numbers and so simple mass 
terms violate gauge invariance. But the great feature of the Higgs mechanism is that the same Higgs
multiplet which generates $W^{\pm}$ and $Z^0$ masses is also sufficient to give masses to the 
leptons and quarks.  It is possible to link the left-handed fermions weak isospin doublets
$\psi_L\equiv\chi_L$\footnote{Note $\psi_L=\left(\begin{array}{c}a_L \\b_L\end{array}\right)$, 
where $a_L$, $b_L$ are the the left-handed states of lepton and quarks. The explicit doublets 
can be found in Table~\ref{ch1:table:fermions}.}, the right-handed fermions weak isospin singlet 
$\psi_R\equiv\ell_R\equiv q_R$ and the spinor under $SU(2)$ scalar field $\phi$,~\ie the Higgs field,
\begin{align}
	\Lagrangian_{Yukawa}&=-\sum_{\substack{\chi_L=1,2,3\\\ell=e,\mu,\tau}}Y_{\ell}\left[(\bar{\chi}_{L}\cdot\phi)
                        \ell_R+h.c.\right]\notag\\
			&-\sum_{\substack{\chi_L=1,2,3\\q=d,s,b}}Y_{q}\left[(\bar{\chi}_{L}\cdot\phi)q_R+h.c.\right]\\
		 &-\sum_{\substack{\chi_L=1,2,3\\q=u,c,t}}Y_{q}\left[\epsilon^{ab}\bar{\chi}_{L_a}\phi_b^{\dagger}q_R+h.c.\right]
			\notag
\end{align}
where the $\epsilon^{ab}$ is the total antisymmetric tensor, the indices 1,2,3 refer to 
the lepton and quark generations and $q_R$, $\ell_R$ indices refer to the quark and lepton
flavour, respectively (see Table~\ref{ch1:table:fermions}).The $Y_{\ell}$, $Y_{q}$ are dimensionless
couplings called \emph{Yukawa couplings}, they are the coupling terms of the fermions with the scalar
field $\phi$. Replacing the field $\phi$ with this expression about its vacuum expectation 
value~\eqref{ch1:eq:phiaboutvacuum},
\begin{equation}
	\Lagrangian_{Yukawa}=
	-\sum_{\substack{f=e,\mu,\tau\\d,s,b\\u,c,t}}\frac{1}{\sqrt{2}}Y_f
		\left[v\bar{f}f+\bar{f}f H\right]\\
\end{equation}
and the mass terms of the fermions are 
\begin{equation}
	m_f=\frac{1}{\sqrt{2}}Y_f v\,,\quad f=e,\mu,\tau,d,s,b,u,c,t
\end{equation}

The derivation used above does not make use of the fact that additional coupling terms 
mixing generations can exist. In fact we did an implicit diagonalisation of the Higgs
couplings by choosing a new basis for the quark fields, so in this basis, we choose the weak states
(u,c,t) to have definite masses, but then (d,s,b) will not. Nevertheless the flavour or weak 
states (d,s,b) can be expressed as a linear combination of the mass eigenstates (d',s',b') through
a $3\times3$ \gls{ckm}\glsadd{ind:ckm} unitary matrix in the form,
\begin{equation}
	\left(\begin{array}{c}d'\\s'\\b'\end{array}\right)=V_{CKM}\left(\begin{array}{c}d\\s\\b\end{array}\right)\,,
		\text{where }\quad
		V_{CKM}=\left(\begin{array}{ccc}
				 V_{ud} & V_{us} & V_{ub}\\
				 V_{cd} & V_{cs} & V_{cb}\\
				 V_{td} & V_{ts} & V_{tb}
			      \end{array}\right)
\end{equation}
The off-diagonal terms in \gls{ckm} allow weak-interaction transitions between quark generations. 
The picture for the leptons is different due to the fact that right-handed neutrinos are not charged
under any \gls{sm} group, but including a Higgs coupling with the right-handed neutrinos would give
a neutrino mass comparable to that of the electron. But we know from experiment that neutrino masses 
are extremely small, $\mathcal{O}(eV)$. This extreme suppression of the neutrino mass would be naturally
explained if the right-handed neutrino states do not exist\footnote{In generalisation of the \gls{sm},
neutrinos can acquire Majorana mass terms that are naturally very small respecting the constrains of
lepton flavour mixing. In this case the weak states chosen to have definite masses are 
($e$,$\mu$,$\tau$) and, as the quarks case, it is possible to express the flavour states 
($\nu_e$,$\nu_{\mu}$,$\nu_{\tau}$) using a linear combination of the mass eigenstates 
($\nu_1$,$\nu_2$,$\nu_3$) using a $3\times3$ matrix called \emph{Pontecorvo-Maki-Nakagawa-Sakata} (PMNS) 
matrix.}. Therefore, there is no right-handed states for neutrinos in the \gls{sm} and there is no
equivalent to the \gls{ckm}\glsadd{ind:ckm} matrix in the lepton sector. Thus, there are no transition between leptons
of different generations.

\begin{figure}[!htpb]
	\centering
	\includegraphics[scale=0.45]{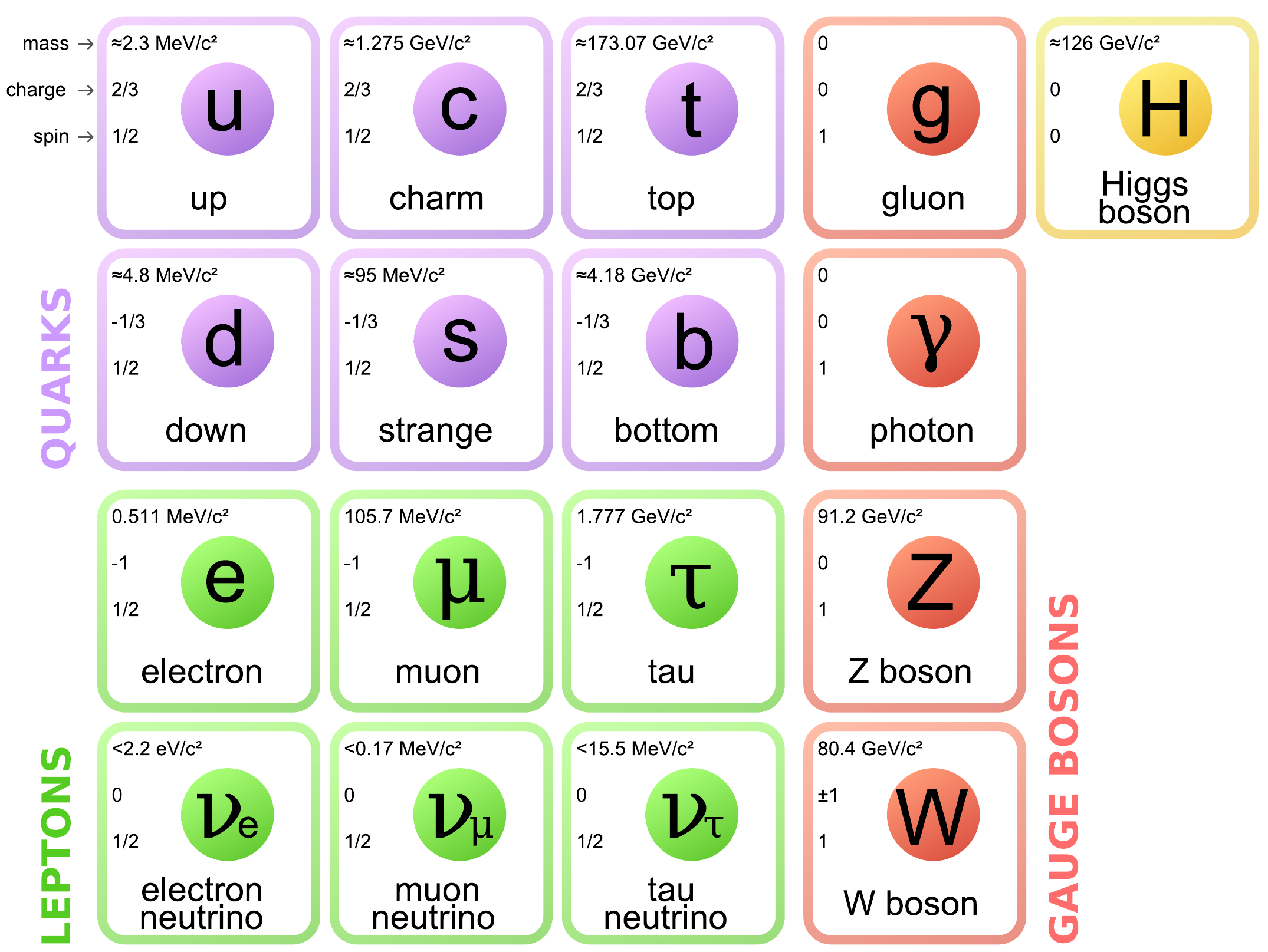}
	\caption[Standard Model constituents summary]{Constituents of the \gls{sm} summarised.
	}\label{ch1:fig:constituentsSM}
\end{figure}

\paragraph{}
The $SU(3)_C\times SU(2)_L\times U(1)_Y$ gauge theory of quarks and leptons plus the Higgs
mechanism to provide mass to the fermions and $W^{\pm}$ and $Z^0$ gauge bosons do an excellent
job of accounting for the symmetries and conservation laws that are observed in elementary particle
phenomena. By imposing gauge invariance and renormalizability the Lagrangian built is able to 
predict which symmetries should be exact, as lepton flavour for instance, and which should be approximate, 
as P or C violation in weak currents. Furthermore, the \gls{sm} has predicted the existence of $W$, 
$Z$ boson, gluons and top and charm quarks before these particles were observed. The predicted 
properties of the \gls{sm} have resisted many experimental precision tests with high accuracy.
In July 2012, CMS~\cite{Chatrchyan:2012ufa} and ATLAS~\cite{Aad:2012tfa} reported the observation 
of a new particle compatible with the \gls{sm} Higgs, the last missing ingredient predicted by the 
\gls{sm}. Therefore, the \gls{sm} of particle physics has turned out to be an impressive theory
which uses relatively few constituents to explain our current known universe. 
Figure~\ref{ch1:fig:constituentsSM} summarises the building blocks of the \gls{sm} showing the
matter content (spin-1/2 fermions) and the gauge interaction mediators (spin-1 bosons) of the
forces with some of its main features. 

However, in the theory remain a wide range of unresolved issues~\cite{ho1991invitation}:
\begin{itemize}
	\item The \gls{sm} does not include the remaining force of the Nature: gravity
	\item Hierarchy problem. Attempting the extension of \gls{sm} as a broken symmetry
		of a larger symmetry, a \gls{gut}\glsadd{ind:gut} at some energy scale, one would expect a
		Higgs mass comparable to the underlying mass scale of the fundamental 
		interactions. If not it seems to require dramatic and even bizarre cancellations
		in the renormalised value of the Higgs boson mass.
	\item Strong CP problem. Experimentally, the CP symmetry is an exact symmetry in the 
		strong interaction sector. Theoretically, however, the strong interaction naturally
		contains a term that breaks CP symmetry, relating matter to antimatter. There is no 
		explanation in the \gls{sm} for the absence of this term.	
	\item High number of unrelated and arbitrary parameters. The \gls{sm} is described by nineteen
		numerical constants which have to be taken from experiments
	\item Matter-antimatter asymmetry. There is an asymmetry in our universe favouring matter 
		against the antimatter which it is not present in the \gls{sm}. The CP-violation
		of the \gls{sm} provides an effect far too small to account for the original
		imbalance in the early Universe.
	\item Dark matter and dark energy are not described neither predicted in the \gls{sm}
\end{itemize}
Although the above sketched shortcomings support the belief that the \gls{sm} is not the ultimate
theory, it is a good model at relatively low energies (at least far away from the Planck scale).
So far, all the precision tests performed have been passed with high accuracy. 

\section{Physics at Hadron Colliders}\label{ch1:sec:physicsathadroncolliders}
Hadron collisions are governed by the strong interaction force, given that hadrons are described
as bound states of strongly interacting quarks and gluons (\emph{partons}). There, \gls{qcd} plays a
fundamental role in the attempt to describe the physics produced by a proton-proton collision. The 
theoretical calculations of the low-energy regime of the interacting partons inside the proton are
subject to non-perturbative (or \emph{soft}) \gls{qcd}. However, when considering a high-energy
collision, the interactions take place at small distances between the partons of the two incoming 
protons giving rise to collisions with a high-momentum transference, called \emph{hard scatterings}.
The partons involved in the hard scattering can be considered as free 
partons\footnote{See \emph{asymptotic freedom} in previous section.} and a perturbative expansions
in $\alpha_s$ can be used in order to calculate the observables of the hard scattering. Both the 
soft and hard aspects of hadron collider collisions can be split by virtue of the 
\emph{factorisation theorems}~\cite{Collins:1989gx}~\cite{Drell:1970yt}. The perturbative and 
process-dependent scattering can be separated from the non-perturbative but universal 
(\ie process-independent) structure inside the proton. Qualitatively, the hard scattering taking 
place with a momentum transfer $Q^2$ has typical interaction timescale of $\sim 1/Q$. The soft 
interactions inside the proton have an energy typically close to the \gls{qcd} scale, \ie 
$\Lambda_{QCD}$, therefore with
a typical interaction timescale of $1/\Lambda_{QCD}$. Hard scattering processes, as vector boson
production, usually have $Q\gg\Lambda_{QCD}$, the time scale of the hard scattering is much shorter
than the soft physics inside the proton. Therefore, during hard scattering interaction, the internal
structure of the proton is not going to change significantly and thus can be determined
before the interaction took place. The internal structure of a hadron is summarised with a set
of probability distributions of quarks and gluons, called \gls{pdf}\glsadd{ind:pdf}. These \gls{pdf}s encapsulate
all the non-perturbative \gls{qcd} that determines the probability of finding a parton of a given
flavour and momentum inside a hadron. The calculation of an interaction cross section of a hard
scattering involves the amplitudes for hard scattering between partons which must be convoluted with
the \gls{pdf} in order to incorporate the probability of finding the necessary partons and their
energies for the hard scattering.
\begin{figure}[!htpb]
	\begin{center}
	 \input{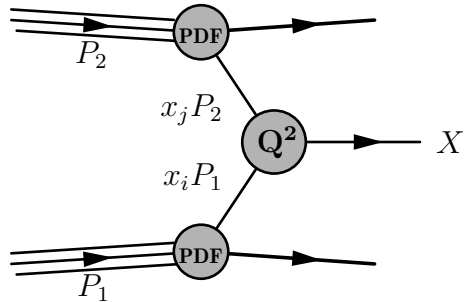}
	\end{center}
	\caption[Hard scattering process]
	        {Hard scattering process representation. Two partons interact exchanging $Q$ 
			momentum from the incoming protons and as a result of the hard scattering
			the particle $X$ is produced. The PDFs are encapsulating the soft QCD 
		        content, a parton is selected out of each hadron carrying a fraction of
		the hadron momentum, $xP$ described by the PDFs.}\label{ch1:fig:hardscattering}
\end{figure}

\begin{figure}[!htpb]
	\centering
	\includegraphics[scale=0.65]{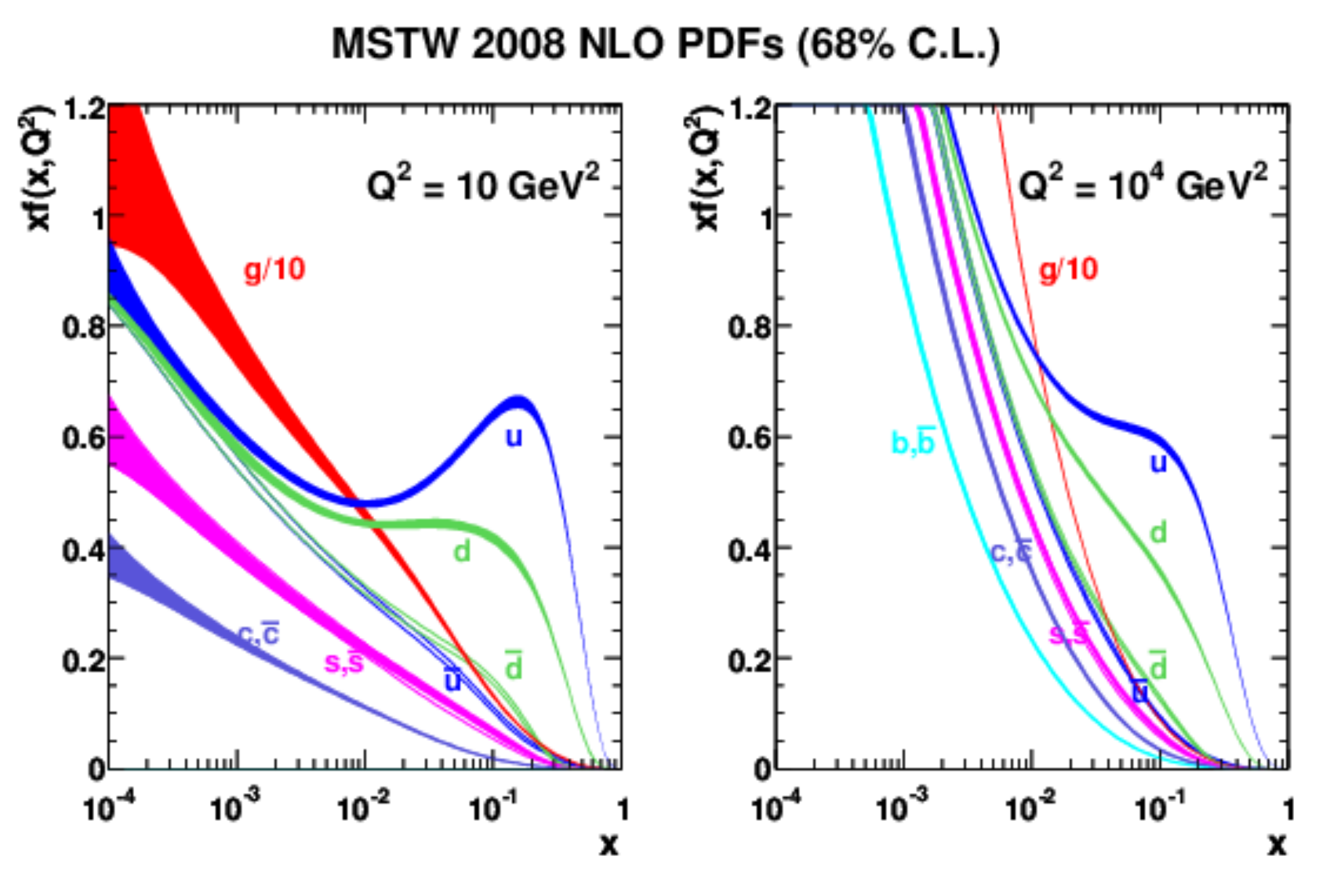}
	\caption[MSTW2008 PDFs example]{Parton distribution functions calculated by the MSTW 
	group~\cite{Martin:2009iq}. The \gls{pdf}s are shown for two different energy 
	scales.}\label{ch1:fig:mstwPDF}
\end{figure}

Any hadron is composed by \emph{valence} quarks, which in the case of the proton are two $u$ and one $d$
quarks, confined within the hadron. The total sum of valence quarks charge yields the overall 
charge of the hadron. Those valence quarks are continuously interacting by exchange of gluons and
those gluons can also self-interact to produce more gluons or produce additional quark-antiquark 
pairs, called \emph{sea quarks}. The momentum of the hadron is distributed amongst the valence quarks,
the sea quarks and the gluons and the \gls{pdf}s describe the probability to find a parton with a
given fraction of the hadron momentum, \ie $p_i=xP$, being $p_i$ the momentum of the parton $i$ 
carrying a fraction $x$ of the hadron momentum $P$. At low energies, the three valence quarks
essentially carry all of the hadron's momentum. But when the energy transfer is large, 
$Q^2\gtrsim 1\GeV$, the other substructure components of the hadron, sea quarks and gluons, can
be resolved. The hadron substructure depends on $Q$ because partons at high $x$ tend to radiate
and drop down to lower values of $x$, while at the same time additional new partons at low $x$
arise from radiation. Therefore, when increasing the energy of the hadron, the hadron momentum is 
shared amongst a larger number of constituents, although the valence quarks tend to carry a significant
fraction of the hadron momentum. The \gls{pdf}s take the form $f_i(x,Q^2)$, where $i$ is the
parton type, $x$ is the momentum fraction and $Q^2$ is the scale. The \gls{pdf} cannot be calculated
from first principles due to the presence of non-perturbative effects, but the evolution of the 
\gls{pdf}s with $Q^2$ can be calculated. The \gls{pdf}s are obtained by a fit to experimental data 
at one scale and then evolved to different scales. An example of \gls{pdf}s at $Q=\sqrt{10}\GeV$ and 
$Q=100\GeV$ from the MSTW group can be found in Figure~\ref{ch1:fig:mstwPDF}.

\chapter{\WZ production at hadron colliders}\label{ch2}

This chapter is a phenomenological and experimental review of the \wpmz production mechanisms at 
hadron colliders.  The main production Feynman diagrams are revisited and the up-to-date \wpmz 
cross section and \wzp/\wzm ratio theoretical predictions for \wpmz are surveyed. Some explicit
calculations for the analysis phase space are done using standard tools. Then, the motivations for 
the study and measurement of \wpmz and ratio are pinpointed before finalising the chapter with the 
status of all \wpmz measurements performed so far.

\section{The \WZ diboson production}\label{ch2:sec:wzproduction}
The \wpmz pairs are produced at hadron colliders mainly at \gls{lo}\glsadd{ind:lo} in 
$\alpha_s$, by quark-antiquark annihilation
which proceeds via t- and u-channel quark exchange and s-channel \W-boson exchange as seen in
the Feynman diagrams of Figure~\ref{ch2:fig:wzatlo}. Charge conservation requires a quark up-type
and an antiquark down-type for the \wzp production. Conversely, a down-type quark and an
up-type antiquark are needed for the \wzm. Note that the quark and antiquark are not required
to be from the same generation, but the small off-diagonal elements of the \gls{ckm}\glsadd{ind:ckm} matrix 
highly suppress cross-generation \WZ production. In a proton-proton machine as the 
\gls{lhc}\glsadd{ind:lhc}, the predominance of $u$ quarks enhances the $\wzp$ production\footnote{Recall
the quark valence content of the proton: $uud$.} via $u\bar{d}$ type interactions taking
the antiquark among the sea-quarks. Therefore it is expected more \wzp production than
\wzm, \ie $\sigma_{\wzm}/\sigma_{\wzp} < 1$.
\begin{figure}[!htpb]
	\begin{center}
	 \input{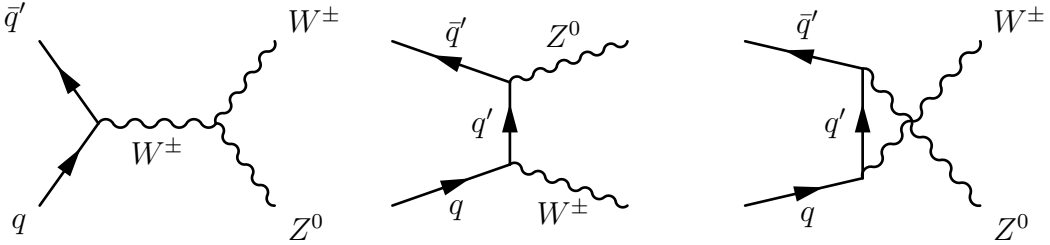}
	\end{center}
	\caption[Feynman diagrams for \WZ production at Born-level]
	        {Feynman diagrams for the \gls{lo}\glsadd{ind:lo} quark-antiquark annihilation for \WZ production. 
		It is shown from left to right the s-channel, t-channel and u-channel.
		Diagrams interchanging the \W and \Z are not shown.}\label{ch2:fig:wzatlo}
\end{figure}
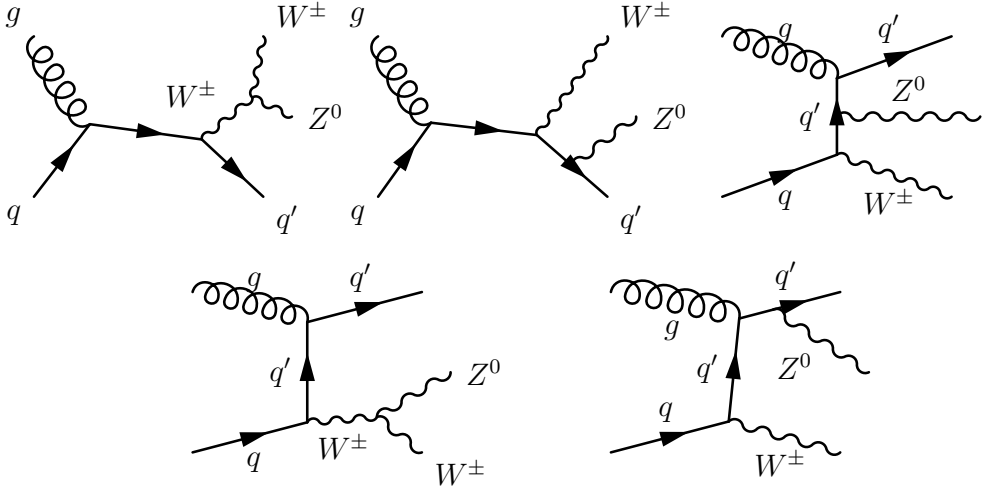
\begin{figure}[!htpb]
	\begin{center}
	 \begin{fmffile}{wzNLOGQ} 
	\resizebox{\textwidth}{!}
	{
	 \begin{tabular}{c}
	  \fmfframe(5,17)(20,1)
   	  {
		\begin{fmfgraph*}(110,62) 
			\fmfleft{i1,i2}		\fmfright{o1,o2,o3}
			\fmflabel{$g$}{i2}
			\fmflabel{$q$}{i1}
			\fmflabel{$q'$}{o1}
			\fmf{gluon}{i2,v1}
			\fmf{fermion}{i1,v1,v2}
			\fmf{fermion}{v2,o1}
			\fmf{boson,label=$W^{\pm}$}{v2,v3}
			\fmflabel{$Z^0$}{o2}
			\fmflabel{$W^{\pm}$}{o3}
			\fmf{boson}{v3,o2}
			\fmf{boson}{v3,o3}
		\end{fmfgraph*}
		\hspace*{0.5cm}
		\begin{fmfgraph*}(110,62) 
			\fmfleft{i1,i2}		\fmfright{o1,o2,o3}
			\fmf{gluon}{i2,v1}
			\fmf{fermion}{i1,v1,v2}
			\fmflabel{$g$}{i2}
			\fmflabel{$q$}{i1}
			\fmf{fermion}{v2,o1}
			\fmflabel{$q'$}{o1}
			\fmf{boson}{v2,v3}
			\fmflabel{$Z^0$}{o2}
			\fmflabel{$W^{\pm}$}{o3}
			\fmf{boson}{v3,o3}
			\fmffreeze
			\fmf{phantom}{v2,I1,o1}
			\fmf{boson,tension=0}{I1,o2}
		\end{fmfgraph*}
		\hspace*{0.5cm}	\begin{fmfgraph*}(110,62) 
			\fmfleft{i1,i2,i3}  \fmfright{o1,o2,o3}
			\fmf{fermion,label=$q$}{i1,v1}
			\fmf{fermion,tension=0.6,label=$q'$,l.side=left}{v1,v2}
			\fmf{phantom}{v1,V1,v2}
			\fmf{gluon,label=$g$}{v2,i3}
			\fmf{boson,label=$W^{\pm}$}{v1,o1}
			\fmf{boson,tension=0,label=$Z^0$,l.side=left}{V1,o2}
			\fmf{fermion,label=$q'$}{v2,o3}
		\end{fmfgraph*}
	} \\ \\
	\fmfframe(5,17)(20,1)
   	{
		\hspace*{0.5cm} \begin{fmfgraph*}(110,62) 
			\fmfleft{i1,i2,i3}  \fmfright{o1,o2,o3}
			\fmf{fermion,label=$q$}{i1,v1}
			\fmf{fermion,tension=0.6,label=$q'$,l.side=left}{v1,v2}
			\fmf{gluon,label=$g$}{v2,i3}
			\fmf{phantom}{v1,o1}
			\fmf{fermion,label=$q'$}{v2,o3}
			\fmffreeze
			\fmf{phantom}{v1,I2,I3,o1}
			\fmf{boson,tension=0.7,label=$W^{\pm}$,l.side=right}{v1,I2}
			\fmf{boson}{I2,o2} \fmflabel{$Z^0$}{o2}
			\fmf{boson,tension=0.5}{I2,o1} \fmflabel{$W^{\pm}$}{o1}
		\end{fmfgraph*}
		\hspace*{1.5cm}	\begin{fmfgraph*}(110,62) 
			\fmfleft{i1,i2,i3}  \fmfright{o1,o2,o3}
			\fmf{fermion,label=$q$}{i1,v1}
			\fmf{fermion,tension=0.6,label=$q'$,l.side=left}{v1,v2}
			\fmf{gluon,label=$g$}{v2,i3}
			\fmf{fermion,label=$q'$,l.side=left}{v2,o3}
			\fmf{phantom}{v2,I1,I2,o3}
			\fmf{boson,tension=0,label=$Z^0$}{I1,o2}
			\fmf{boson,label=$W^{\pm}$}{v1,o1}
		\end{fmfgraph*}
       	   } 
	 \end{tabular}
	} 
\end{fmffile}

	\end{center}
	\caption[Feynman diagrams for \WZ production at NLO: gluon-(anti)quark interactions]
	        {Feynman diagrams with NLO corrections: gluon-(anti)quark interactions
		for the \WZ production process. Diagrams obtained by interchanging
		the \W and \Z and equivalent diagrams using antiquarks are not shown.
	}\label{ch2:fig:wzatnlogq}
\end{figure}

The inclusion of \gls{nlo}\glsadd{ind:nlo} corrections from \gls{qcd}\glsadd{ind:qcd} allows \WZ production to be induced by
gluon-quark or gluon-antiquark interactions~\cite{PhysRevD.44.3477} where a quark is present
in the final state. Additional \gls{nlo}\glsadd{ind:nlo} corrections include quark-antiquark interactions with 
virtual corrections and with gluon bremsstrahl\"ung~\cite{Campanario:2010hp} in the final state.
The Feynman diagrams for these \gls{nlo}\glsadd{ind:nlo} corrections are shown in Figure~\ref{ch2:fig:wzatnlogq},
Figure~\ref{ch2:fig:wzatnloqqbgbremss} and Figure~\ref{ch2:fig:wzatnloqqbvirtual}. These diagrams must
be convoluted with the proton \gls{pdf}\glsadd{ind:pdf}s in order to calculate the full 
cross section for \wpmz production at a $pp$ collider.
\begin{figure}[!htpb]
	\begin{center}
	 \input{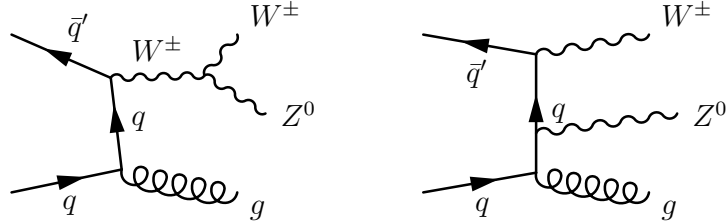}
	\end{center}
	\caption[Feynman diagrams for \WZ production at NLO: gluon-antiquark contributions]
		{Feynman diagrams with NLO corrections: quark-antiquark contributions
		with gluon bremssstrahl\"ung in the final state. Diagrams obtained
		by interchanging the \W and \Z and equivalent diagrams using antiquarks are
		not shown.}\label{ch2:fig:wzatnloqqbgbremss}
\end{figure}
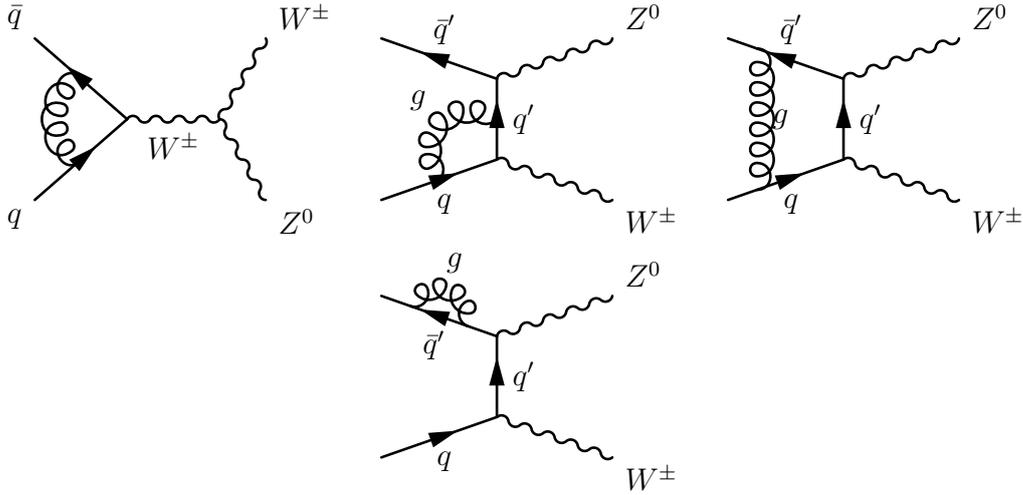
\begin{figure}[!htpb]
	\begin{center}
	 \begin{fmffile}{wzNLOQQBVIRTUAL} 
	\resizebox{\textwidth}{!}
	{
	 \begin{tabular}{c}
	  \fmfframe(5,17)(17,1)
   	  {
		\begin{fmfgraph*}(110,62) 
			\fmfleft{i1,i2}		\fmfright{o1,o2}
			\fmflabel{$q$}{i1}
			\fmflabel{$\bar{q}$}{i2}
			\fmf{plain}{v1,V1,i2}
			\fmf{plain}{i1,V2,v1}
			\fmf{boson,label=$W^{\pm}$}{v1,v2}
			\fmflabel{$Z^0$}{o1}
			\fmflabel{$W^{\pm}$}{o2}
			\fmf{boson}{v2,o1}
			\fmf{boson}{v2,o2}
			\fmffreeze
			\fmf{phantom_arrow}{v1,i2}
			\fmf{phantom_arrow}{i1,v1}
			\fmf{gluon,right=0.5}{V1,V2}
		\end{fmfgraph*}
		\hspace*{0.5cm}
		\begin{fmfgraph*}(110,62) 
			\fmfleft{i1,i2}  \fmfright{o1,o2}
			\fmf{fermion,label=$q$}{i1,v1} 
			\fmf{fermion,label=$q'$,l.side=right}{v1,v2}
			\fmf{fermion,label=$\bar{q}'$}{v2,i2}  
			\fmf{boson}{v2,o2} \fmflabel{$Z^0$}{o2}
			\fmf{boson}{v1,o1} \fmflabel{$W^{\pm}$}{o1}
			\fmffreeze
			\fmf{phantom}{i1,I0,I1,I2,v1}
			\fmf{phantom}{v1,V0,V1,V2,v2}
			\fmf{gluon,tension=0,left,label=$g$}{I1,V1}
		\end{fmfgraph*}
		\hspace*{0.5cm}	\begin{fmfgraph*}(110,62) 
			\fmfleft{i1,i2}  \fmfright{o1,o2}
			\fmf{fermion,label=$q$}{i1,v1} 
			\fmf{fermion,label=$q'$,l.side=right}{v1,v2}
			\fmf{fermion,label=$\bar{q}'$}{v2,i2}  
			\fmf{boson}{v2,o2} \fmflabel{$Z^0$}{o2}
			\fmf{boson}{v1,o1} \fmflabel{$W^{\pm}$}{o1}
			\fmffreeze
			\fmf{phantom}{i1,I0,I1,I2,v1}
			\fmf{phantom}{i2,V0,V1,V2,v2}
			\fmf{gluon,tension=0,label=$g$}{I0,V0}
		\end{fmfgraph*}
	  } \\ \\
	  \fmfframe(5,17)(17,1)
   	  {
		\begin{fmfgraph*}(110,62) 
			\fmfleft{i1,i2}  \fmfright{o1,o2}
			\fmf{fermion,label=$q$}{i1,v1} 
			\fmf{fermion,label=$q'$,l.side=right}{v1,v2}
			\fmf{fermion,label=$\bar{q}'$,l.side=left}{v2,i2}  
			\fmf{boson}{v2,o2} \fmflabel{$Z^0$}{o2}
			\fmf{boson}{v1,o1} \fmflabel{$W^{\pm}$}{o1}
			\fmffreeze
			\fmf{phantom}{i2,V0,V1,V2,v2}
			\fmf{gluon,tension=0,left,label=$g$}{V0,V2}
		\end{fmfgraph*}
       	   } 
         \end{tabular}
	} 
\end{fmffile}

	\end{center}
	\caption[Feynman diagrams for \WZ production at NLO: virtual contributions]
	        {Feynman diagrams with \gls{nlo} corrections: virtual contributions
		with internal gluon loops to the \WZ production through quark-antiquark 
		annihilation. Diagrams obtained by interchanging the \W and \Z
		and equivalent diagrams with the loop affecting other legs are not shown.
	}\label{ch2:fig:wzatnloqqbvirtual}
\end{figure}

The most updated cross section predictions for the $W^+Z$, $W^-Z$, \wpmz and the ratio 
$\sigma_{\wzm}/\sigma_{\wzp}$ at 
\gls{nlo}~\cite{Campbell:2011bn} for 7\TeV and 8\TeV of centre of mass proton-proton collisions
are reported in Table~\ref{ch2:tab:xspredicted}. The cross section values were obtained including full
spin correlations, the \Z and \W widths and keeping the vector bosons on-shell. Including the \drellyan
interference introduces divergences to the \wpmz cross section at very low $m_{\drellyan}$, however
it is possible to define a mass window around the \Z mass-pole as the phase space to be used. In
Table~\ref{ch2:tab:xspredmassrange} the \wpmz cross section is obtained using the \MCFM software~\cite{MCFM}
and \gls{pdf}\glsadd{ind:pdf} \gls{mstw8}\glsadd{ind:mstw8}~\cite{Martin:2009iq} doing the same 
procedure as reference~\cite{Campbell:2011bn}
but including the \drellyan interference. In this case, the bosons are allowed to
be off-shell and therefore singly resonant boson diagrams are included in the calculations. This
prediction has been calculated because the analysis phase space is exactly defined in this \Z mass
window.
\begin{table}[!hbtp]
	\centering
	\begin{tabular}{l r r}\hline\hline
		              &  7\TeV   & 8\TeV \\\hline
	$\sigma_{\wzp}$ (pb) & $11.88(1)^{+0.65}_{-0.50}$              &  $14.28(1)^{+0.75}_{-0.57}$       \\
	$\sigma_{\wzm}$ (pb) & $6.69(0)^{+0.37}_{-0.29}$               &  $8.40(0)^{+0.45}_{-0.34\%}$        \\
	$\sigma_{\wpmz}$ (pb) & $18.57(1)^{+0.75}_{-0.58}$             &  $22.88(1)^{+0.88}_{-0.66}$  \\
	$\sigma_{\wzm}/\sigma_{\wzp}$ & $0.563^{+0.002}_{-0.001}$      &  $0.588^{+0.001}_{-0.001}$  \\	\hline
	\end{tabular}
	\caption[NLO cross sections for \WZ production at LHC]{\gls{nlo} cross section predictions 
		for \WZ	production and ratio between \wzm and \wzp at $pp$ collisions with
		a centre of mass energy of 7\TeV and 8\TeV extracted from~\cite{Campbell:2011bn}. 
		Renormalisation and factorisation scales are set equal to the average mass of the W and Z.
	 	The vector bosons are kept on-shell, with no decays included.}\label{ch2:tab:xspredicted}
\end{table}

\begin{table}[!hbtp]
	\centering
	\begin{tabular}{l r r}\hline\hline
		              &  7\TeV   & 8\TeV \\\hline
	$\sigma_{\wzp}$ (pb) & $11.37(1)^{+0.55}_{-0.47}$  &  $13.86(1)^{+0.73}_{-0.40}$       \\
	$\sigma_{\wzm}$ (pb) & $6.40(1)^{+0.36}_{-0.27}$   &  $8.05(1)^{+0.43}_{-0.33}$        \\
	$\sigma_{\wpmz}$ (pb) & $17.77(2)^{+0.66}_{-0.54}$  &  $21.91(2)^{+0.85}_{-0.52}$  \\
	$\sigma_{\wzm}/\sigma_{\wzp}$ & $0.563^{+0.002}_{-0.001}$   &  $0.581^{+0.001}_{-0.001}$  \\	\hline
	\end{tabular}
	\caption[\WZ cross section production with restricted phase space]{\gls{nlo} cross-section for \WZ
	production and ratio between \wzm and \wzp at $pp$ collisions with a centre of mass energy
	of 7\TeV and 8\TeV calculated
	using \MCFM and \gls{pdf}\glsadd{ind:pdf} \gls{mstw8}\glsadd{ind:mstw8} sets in the Z mass window $91.1876\pm20\GeVcc$ 
	phase space. The upper and lower deviations are obtained by varying the renormalisation 
        and factorisation scales around the central value, $(M_{\Z}+M_{W})/2$, by a factor
        two.}\label{ch2:tab:xspredmassrange}
\end{table}

These calculations show that the \gls{qcd} \gls{nlo} contributions to the \wpmz are important 
and cannot be avoided. The \MCFM program estimates about a 85\% of the cross section comes from
the quark-antiquark annihilation and the remainder from quark-gluon interactions. Note that
antiquark-gluon contributions are found negligible which is expected due to the larger availability
of quarks than anti-quarks in the proton \gls{pdf}\glsadd{ind:pdf}s.

Due to the centre of mass energy and the charged initial state needed, and the low probability
of occurrence, diboson production measurements are relatively new. In particular, 
the \wpmz production was observed for the first time at the Tevatron in a $p\bar{p}$ collisions at a 
centre of mass energy of
1.96~\TeV. These \wpmz cross section measurements are valuable to test the \gls{sm}\glsadd{ind:sm} through the 
constraining of its parameters and at the same time, high precision measurements could yield 
information of New Physics processes. Furthermore, the \wpmz production is found to be an important background
to several Higgs production and other channels in Beyond \gls{sm} theories. An accurate
knowledge of the \wpmz production is therefore needed to control the search analyses. The derived 
observable \wzp/\wzm allows to reach more precision because some experimental uncertainties 
are cancelled or highly suppressed. Also the ratio is more sensitive to the \gls{pdf}\glsadd{ind:pdf} used,
therefore it is a potential observable to constrain the \gls{pdf}\glsadd{ind:pdf}
fits.

\begin{table}[!hbtp]
	\centering
	 \begin{tabular}{l c c} \hline\hline
		&  channel & BR [\%] \\ \hline
	  \multirow{4}{*}{ \W }  
	      &	$e\nu_{e}$    & $10.75\pm0.13$ \\
	      & $\mu\nu_{\mu}$& $10.57\pm0.15$ \\
	      & $\tau\nu_{e}$ & $11.25\pm0.20$ \\
	      & hadrons       & $67.60\pm0.27$ \\\hline
	  \multirow{5}{*}{ \Z }  
	      &	$e^+e^-$       & $3.363\pm0.004$ \\
	      & $\mu^+\mu^-$   & $3.366\pm0.007$ \\
	      & $\tau^+\tau^-$ & $3.370\pm0.008$ \\
	      & hadrons        & $69.91\pm0.06$ \\
	      & $\nu_{\ell}\bar{\nu_{\ell}}$ & $20.00\pm0.06$ \\\hline
	 \end{tabular}
	\caption[Branching ratios for \W and \Z]{Decay probabilities, \ie branching ratios, for \W 
		and \Z. Hadrons tag is including all decays involving at least one 
		(anti)quark. The value are obtained from ref.~\cite{PhysRevD.86.010001}}\label{ch2:tab:wandzbr}
\end{table}
The aim is this thesis is to measure the cross section and ratio observable for the \WZ
production using the fully leptonic decay\glsadd{ind:decay} of the \W and \Z gauge bosons. Despite the fact that leptonic 
decays only represent a small faction of possible \W and \Z decays, as can be seen in 
Table~\ref{ch2:tab:wzbr}, they produce a clean 3-lepton signature\glsadd{ind:signature} in detectors,
having relatively small backgrounds. Almost 90\% of the time \W and/or \Z decays to hadrons but 
this signature has very large experimental backgrounds
as QCD multi-jet production and a fuzzy experimental signature full of jets, missing energy and 
no higher momentum leptons make the event reconstruction an intricate and challenging process.
\begin{table}[!hbtp]
	\centering
	 \begin{tabular}{l c c} \hline\hline
		&  channel & BR [\%] \\ \hline
	  \multirow{3}{*}{ \WZ }  
	      &	$\ell'\nu_{\ell'}\ell^+\ell^-$ & $3.29\pm0.08$ \\
	      & $\ell^{\pm}+\nu_{\ell'}\bar{\nu_{\ell'}}$         & $6.51\pm0.06$ \\ 
	      & hadrons 	               & $90.34\pm0.07$ \\ \hline
	 \end{tabular}
	\caption[Branching ratios for \WZ]{Decay probabilities for \WZ. Hadrons tag is 
	including all decays involving at least one (anti)quark. $\ell$ indicate sum over
	all type of lepton ($e,\mu,\tau$).}\label{ch2:tab:wzbr}
\end{table}

\section{Previous Measurements}
The \WZ cross section production was measured for the first time at the $p\bar{p}$ Tevatron 
collider experiments \DZERO and CDF with a centre of mass energy of \comene=1.96~\TeV.
After the \gls{lhc} successfully started, the \gls{atlas} and \gls{cms}\glsadd{ind:cms} experiments have also measured
the \WZ production cross section at $pp$ collisions with \comene=7~\TeV and \comene=8~\TeV.
Only the analyses considering a fully leptonic signature are reported here.

The latest measurement of the \WZ cross section production at Tevatron experiment's
\DZERO~\cite{Abazov:2012cj} reported 75 candidate events in the fully leptonic 
decay channel\glsadd{ind:decaychannel} $\WZ\rightarrow\ell'\nu\ell^+\ell^-$ 
(where $\ell',\ell\in{e,\mu}$) using 8.6\fbinv 
integrated luminosity. The \WZ cross section measured in the phase space\footnote{Due 
to \drellyan interference, the \wpmz cross section diverges at very low $m_{\drellyan}$, 
therefore an appropriate phase space must be defined to avoid the divergences at low mass when 
the cross section is calculated.} $M_{\ell^+\ell^-}\in[66,116]$ was
$\sigma_{WZ}=4.50\pm0.61(\text{stat})^{+0.16}_{-0.25}(\text{sys})$~pb, in agreement with the 
\gls{sm} \gls{nlo} prediction from \MCFM~\cite{Campbell:1999ah} $\sigma_{WZ}^{theo}=3.21\pm0.19$~pb. 

The CDF experiment has also measured the \WZ production cross section at \comene=1.96\TeV using
in that case \lumi=7.1\fbinv. The production cross section in the phase space 
$M_{\ell^+\ell^-}\in[60,120]$ was measured to be 
$\sigma_{\WZ}=3.9^{+0.6}_{-0.5}(\text{stat})^{+0.6}_{-0.4}(\text{sys})$~pb, also
in agreement with the SM \gls{nlo} prediction from \MCFM, $\sigma_{WZ}^{theo}=3.46\pm0.21$~pb.

New centre of mass energies have become available with the \gls{lhc} $pp$ collider and its experiments 
have presented results at the reached energies. \gls{atlas} collaboration has published \WZ cross section 
measurements at \comene=7\TeV~\cite{Aad2012341},~\cite{Aad:2012twa} using almost the full available 
integrated luminosity \lumi=4.6\fbinv, obtaining a measurement in the phase space 
$M_{\ell^+\ell^-}\in[66,116]$ of 
$\sigma_{WZ}=19.0^{+1.4}_{-1.3}(\text{stat})\pm0.9(\text{sys})\pm0.4(\text{lumi})$ pb compatible
within uncertainty errors with the \gls{sm} prediction $\sigma_{WZ}^{theo}=17.6^{+1.1}_{+1.0}$~pb. \gls{atlas}
has also presented results~\cite{ATLAS:2013fma} at \comene=8\TeV in the same phase space 
using an integrated luminosity of 13\fbinv. Its result 
$\sigma_{WZ}=20.3^{+0.8}_{-0.7}(\text{stat})^{+1.2}_{-1.1}(\text{sys})^{+0.7}_{-0.6}(\text{lumi})$~pb
is also compatible with \gls{sm} prediction of $\sigma_{WZ}^{theo}=20.3\pm0.8$~pb. 

The \gls{cms} collaboration has published a public note~\cite{CMS:2011dqa} based on \lumi=1.1\fbinv.
The phase space was defined to be $M_{\ell^+\ell^-}~\in~[60,120]$, the \WZ cross section was 
measured to be $\sigma_{WZ}=17.0\pm2.4(\text{stat})\pm1.1(\text{sys})\pm1.0(\text{lumi})$~pb.
Notice that the theoretical prediction was not reported in the reference. This thesis is in fact 
the core of the \gls{cms} public measurement using the full 2011 and 2012 
luminosity~\cite{CMS-PAS-SMP-12-006}, which at the time of this writing the analysis was in
preparation process.

\paragraph{}
\begin{table}[!hbtp]\footnotesize
	\centering
	\resizebox{\textwidth}{!}
	{
		\begin{tabular}{l c c c c c }\hline\hline
		& \comene (\TeV) &  \lumi (\fbinv) & $M_{\ell^+\ell^-}(\GeVcc)$ & Measured $\sigma_{\WZ}$ (pb) 
			                                       & Theoretical $\sigma_{\WZ}$ (pb) \\\hline
\DZERO  &  1.96 &   8.6  &   [66,116]  & $4.50\pm0.61(\text{stat})^{+0.16}_{-0.25}(\text{sys})$ & $3.21\pm0.19$\\
CDF-II  &  1.96 &   7.1  &   [60,120]  & $3.9^{+0.6}_{-0.5}(\text{stat})^{+0.6}_{-0.4}(\text{sys})$ & $3.46\pm0.21$\\
ATLAS   &  7    &   4.6  &   [66,116]  & $19.0^{+1.4}_{-1.3}(\text{stat})\pm0.9(\text{sys})\pm0.4(\text{lumi})$ & $17.6^{+1.1}_{+1.0}$\\
CMS     &  7    &   1.09 &   [60,120]  & $17.0\pm2.4(\text{stat})\pm1.1(\text{sys})\pm1.0(\text{lumi})$ & $--$\\
ATLAS   &  8    &   13   &   [66,116]  & $20.3^{+0.8}_{-0.7}(\text{stat})^{+1.2}_{-1.1}(\text{sys})^{+0.7}_{-0.6}(\text{lumi})$ & $20.3\pm0.8$\\\hline
		\end{tabular}
	}
	\caption[Latest available \WZ cross section measurements]
	        {Latest available \WZ cross section production measurements through the fully
		leptonic decay. Each row shows the cross section measured in each experiment quoting
	        the centre of mass energy, the integrated luminosity used and the theoretical
		\gls{sm} cross section predicted at \gls{nlo} in \gls{qcd} using mainly the \MCFM 
 		generator}\label{ch2:tab:wzxsuptonow}
\end{table}

Table~\ref{ch2:tab:wzxsuptonow} summarises all the latest available \WZ cross section measurements 
up to now and the corresponding theoretical \gls{sm} predictions.
In summary, all the measured values up to now are in agreement with the \gls{sm} predictions 
within uncertainty errors. However, the central values are slightly shifted with respect to the predicted
ones, although always within 2$\sigma$ in the worst of the cases.

Notice that there are no previous measurements of the ratio $\sigma_{\wzp}/\sigma_{\wzm}$ to be 
reported; this thesis is the first measurement of this ratio observable.

\chapter{The Experiment}\label{ch3}
The experiment used for the \WZ cross section measurement is introduced in this chapter.
The chapter is split in two sections, one section describing the main features of the \gls{lhc} machine
and the experiments located therein, and the other section focused on the \gls{cms} experiment.
The aim of the \gls{lhc} section is to show some of the machine parameters and the performance of 
the last three years of running relevant to the analysis performed in this thesis. The section
is mainly based on the technical references~\cite{Pettersson:291782} and~\cite{Benedikt:823808},
so the reader is invited to look at the references for more technical and interesting details. 
The \gls{cms} section describes the main characteristics of the detector, introducing some 
physical quantities used in the analysis. The section enumerates the 
different subdetector systems and their main features relevant to the analysis. As the \gls{lhc}
section, the level of detail is kept in a mere description of the main components, nevertheless
the interested reader can find more details in the technical references~\cite{CmsTDRI:2006} 
and~\cite{CmsTDRII:2007}.

\section{The Large Hadron Collider}
The \gls{lhc}\glsadd{ind:lhc} is a proton-proton collider with a nominal design centre of mass energy of 
14 \TeV~\cite{Pettersson:291782}~\cite{Benedikt:823808} expected be reached in 2014-2015 when 
restarting the machine after the upgrade stop of 2013. Its primary motivation is to elucidate the
nature of electroweak symmetry breaking, for which the Higgs mechanism is presumed to be 
responsible. However, there are alternatives that invoke more symmetry, such as supersymmetry,
or invoke new forces or constituents, such as strongly-broken electroweak symmetry, technicolor,
etc or even a yet unknown mechanism. 
Furthermore, there are high hopes for discoveries that could pave the way toward a unified
theory. These discoveries could take the form of supersymmetry or extra dimensions, the latter often
requiring modification of gravity at the \TeV scale. Hence, there are many compelling reasons to 
investigate the \TeV energy scale. Hadron colliders are well suited to the task of exploring new 
energy domains, and the region of 1 \TeV constituent \comene can be explored if the proton energy
 and the luminosity are high enough. The beam energy (7 \TeV) and the design luminosity 
(\Lumi = $10^{34}~\percms$) of the \gls{lhc} have been chosen in order to study physics 
at the \TeV energy scale.

The \gls{lhc} machine is installed in the tunnel where the \gls{lep}\glsadd{ind:lep} was previously installed. The 
underground infrastructure of \gls{lep}\glsadd{ind:lep}~\cite{Lep:1984} basically consisted of a 26.7 km long ring
tunnel lined with concrete. It included experimental areas at four points (2, 4, 6 and 8), each 
incorporating experimental and service caverns. For the \gls{lhc} project, the existing LEP tunnel 
has been re-used after the complete dismantling of the LEP machine. In addition, new structures have
been added, including experimental and service caverns destined to accommodate two new experiments 
at points 1 and 5.
\begin{figure}[!htpb]
\centering
\includegraphics[scale=0.5]{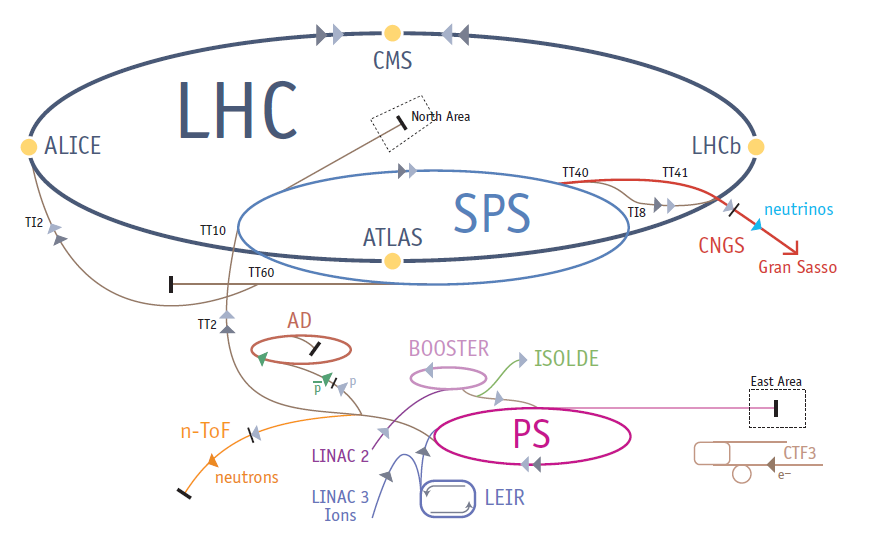}
\caption[LHC ring schema]{Schematic overview of the CERN accelerator complex.}\label{ch3:fig:lhcScheme}
\end{figure}

The \gls{lhc} produces proton-proton collisions at high centre of mass energy. For this, protons must 
be accelerated to velocities close to the speed of light and be focused at some points (where detectors
 are placed). In order to achieve this, a large variety of magnets are set in the \gls{lhc} ring: dipoles, 
quadrupoles, sextupoles, decapoles, etc. The trajectory of the beam is curved using superconducting 
magnets. The \gls{lhc} has more than 1200 superconducting magnetic dipoles of 8.3 Tesla operating at a 
temperature of 1.9 Kelvin, so the accelerator has a large cryogenic system, with superfluid Helium 
pumped into the magnet systems. The machine parameters relevant for the operation of the \gls{cms} 
detector are listed in Table~\ref{ch3:tab:lhcparam}. The nominal parameters, corresponding to the 
nominal design to reach a centre of mass energy of 14~\TeV, are compared with the 2010, 2011 and 
2012 running periods.
\begin{table}[!htpb]
	\centering
	\begin{tabular}{rllll}\hline\hline
		Parameter & 2010 & 2011 & 2012 & Nominal\\ \hline
		Energy per proton, $E$ [\TeV]         & 3.5    & 3.5  & 4    & 7    \\	
		Peak Luminosity   $[10^{33}~\percms]$ & 0.2    & 3.6  & 7.7  & 10   \\
		Bunch separation  $[ns]$              & 150    & 75/50& 50   & 25   \\
		Maximum number of bunches, $k_B$      & 368    & 1380 & 1380 & 2808 \\ 
		Particles per bunch, $N_p$ $[10^{11}]$& 1.2    & 1.5  & 1.7  & 1.15 \\
		Beta value at IP, $\beta^*$ [$m$]     & 3.5    & 1.0  & 0.6  & 0.55 \\ \hline
	\end{tabular}
	\caption[LHC parameters for 2010, 2011 and 2012]{LHC parameters for 2010, 2011 
		and 2011 compared with the nominal values. The point dependent parameter values 
		($\beta^*$, peak luminosity) are taken at point 5,~\ie at 
		CMS.}\label{ch3:tab:lhcparam}
\end{table}

Each of the 1232 dipole magnets has radio frequency cavities 
providing a kick that results in an increase in the proton energy of 0.5 \MeV/turn.  The nominal 
energy of each proton beam is 7 \TeV and the design luminosity of \Lumi~=~$10^{34}~\percms$ 
leads to around 1 billion proton-proton interactions per second. The luminosity is given by:
\begin{equation}
  L = \dfrac{\gamma f k_B N^2_p}{4 \pi \epsilon_n \beta^*} F 
  \label{ch3:eq:lhcparameters}
\end{equation}
which gives the main parameters at the \gls{lhc}:
\begin{itemize}
 \item $\gamma$: Lorentz factor.
 \item $f$: revolution frequency.
 \item $k_B$: number of proton bunches per beam.
 \item $N_p$: number of protons per bunch.
 \item $\epsilon_n$: normalised transverse emittance (with a design value of 3.75~\micron).
 \item $\beta^*$: betatron function at the \gls{ip}\glsadd{ind:ip}.
 \item $F$: reduction factor due to the beam crossing angle at the \gls{ip}
\end{itemize}
Particles to be injected to the \gls{lhc} are first prepared in a set of previous accelerators, to rise 
the energy gradually. LINAC-2 is a linear particle accelerator that increases the proton energy up
to 50 \MeV. Protons are accelerated up to 1.4~\GeV at the next system, the \gls{psb}. They reach an
energy of 26 \GeV at the \gls{ps} before they enter the last accelerator previous to the \gls{lhc},
the SPS, where the final energy is 450 \GeV. This operation is repeated 12 times for each 
counter-rotating beam. Finally, protons are transferred to the \gls{lhc} ring. A schema of the \gls{lhc} 
and the previous accelerators can be seen in Figure~\ref{ch3:fig:lhcScheme}.

Along the \gls{lhc} circumference there are six detectors installed: \gls{alice}~\cite{AliceTD:2008}, 
\gls{atlas}~\cite{AtlasTD:2008}, \gls{cms}~\cite{CmsTDRII:2007}~\cite{CmsTDRI:2006}, 
\gls{lhcb}~\cite{Lhcb:2008}, \gls{lhcf}~\cite{Lhcf:2008} and \gls{totem}~\cite{Totem:2008}. The first four 
detectors are installed in huge underground caverns, built in points 2,1,5 and 8 respectively, 
around the four collision points. 
\begin{figure}[!htpb]
	\centering
	\includegraphics[scale=0.5]{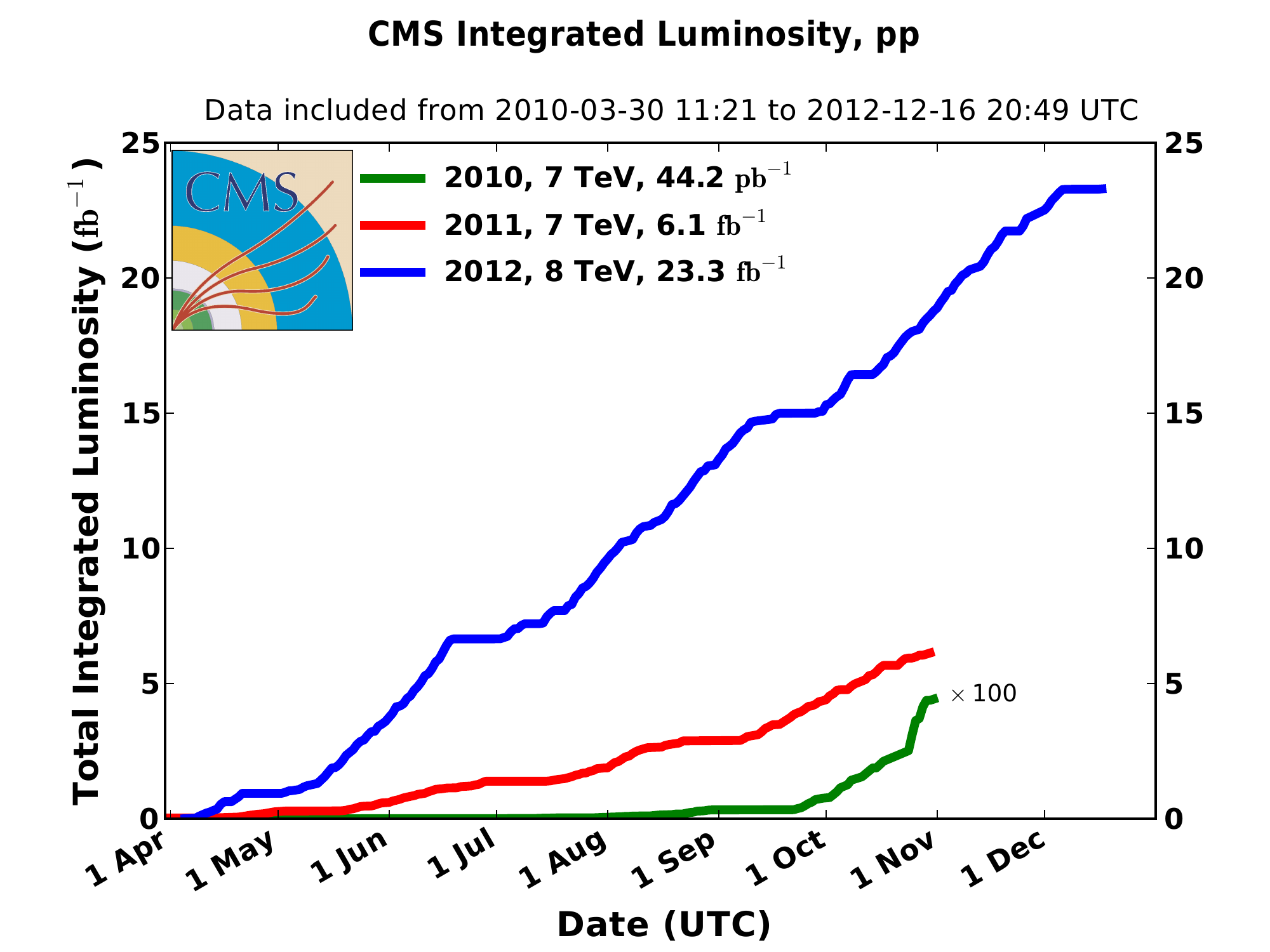}
	\caption[Delivered luminosity to CMS]{Cumulative luminosity delivered to CMS versus day
	during stable beams and for proton-proton collision. This is shown for 2010 (green), 2011 (red) 
	and 2012 (blue) data-taking.}\label{ch3:fig:lumicumulative}
\end{figure}

\gls{cms} and \gls{atlas} are general purpose detectors that share the same physics goals, so a 
cross-check between their results can be made, although they have different designs. The other 
experiments are specialised in different topics, such as heavy flavor physics and precise
measurements in the case of the \gls{lhcb} or heavy ion studies for \gls{alice}.

\subsection{Performance}
\begin{figure}[!htpb]
	\centering
	\includegraphics[scale=0.6]{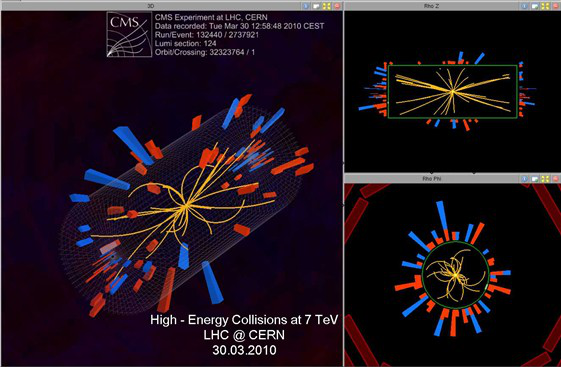}
	\caption{First CMS Event recorded at 7~\TeV \comene.}\label{ch3:fig:firstCMSevent}
\end{figure}

In September 10th 2008, the \gls{lhc} started up with proton beams circulating successfully in the
main ring for the first time. Nevertheless, nine days later a faulty electrical connection caused
a chain of damages which delayed further operations for fourteen months. On November 2009 the machine
came back to work and the first proton-proton collision at 450~\GeV per beam were recorded.
In March 2010, the LHC provided the world's highest energy collisions of 2.36~\TeV and
quickly increased the collision energy to 7~\TeV.  One of the first events recorded by the CMS 
detector on 30th March can be seen in Figure~\ref{ch3:fig:firstCMSevent}. The intensity of the 
beams was progressively increased up to reach a peak luminosity of 
2 x $10^{32}$~${\rm cm}^{-2}$~${\rm s}^{-1}$, around 2\% of the design luminosity, 
at the end of the run. 
The total integrated luminosity delivered to the CMS experiment for proton-proton collisions was 
47.03~\pbinv, from which 43.17~\pbinv were recorded. The maximum instantaneous luminosity 
recorded was $\sim$ 205~\mubinv per second. More details for the 2010 Run are summarised in 
Table~\ref{ch3:tab:lhcrun2010}. For Heavy Ion collisions the total integrated luminosity 
delivered was 9.56~\mubinv and 8.7~\mubinv recorded.
\begin{table}[!htpb]
	\centering
	\begin{tabular}{rl}\hline\hline
		Parameter & Value \\ \hline
		Peak Instantaneous Stable Luminosity	& $198.8$ x $10^{30}~\percms$ \\	
		Max Luminosity Delivered in one Fill        & $6.2~\pbinv$\\
		Maximum Luminosity Delivered in one Day     & $5.4~\pbinv$\\
		Maximum Luminosity Delivered in one Week    & $15.2~\pbinv$\\ 
		Maximum Luminosity Delivered in one Month   & $30.0~\pbinv$ \\ 
		Maximum Colliding Bunches                   & 348 \\ \hline
	\end{tabular}
	\caption[LHC performance summary for 2010 run]{LHC and CMS report for 2010 Run}\label{ch3:tab:lhcrun2010}
\end{table}

During 2011 the \gls{lhc} delivered 5.74~\fbinv and \gls{cms} recorded 5.21~\fbinv of data, reaching
a detector recording efficiency of about 91\%. The 2011 proton-proton Run started in mid March and ended
at the end of October, when the Heavy Ion run started. Information about the 2011 year records of the
LHC and CMS for p\-p collisions are reported in Table~\ref{ch3:tab:lhcrun2011}. 
\begin{table}[!hbtp]
	\centering
	\begin{tabular}{rl}\hline\hline
		Parameter & Value \\ \hline
		Peak Instantaneous Stable Luminosity	& $3547.6$ x $10^{30}~\percms$ \\	
		Max Luminosity Delivered in one Fill        & $123.1~\pbinv$\\ 
		Maximum Luminosity Delivered in one Day     & $135.6~\pbinv$\\
		Maximum Luminosity Delivered in one Week    & $537.9~\pbinv$ \\ 
		Maximum Luminosity Delivered in one Month   & $1614.9~\pbinv$ \\ 
		Maximum Colliding Bunches                   & 1331 \\
		Maximum Interactions per Crossing (pileup)      & 31 \\ \hline
	\end{tabular}
	\caption[LHC performance summary for 2011 run]{LHC and CMS report for 2011 Run.\label{ch3:tab:lhcrun2011}}
\end{table}
The Technical Stop (TS) for accelerator maintenance done in September 2011 divides the data in two
different periods: before the TS (Run2011A) and after (Run2011B). After the TS the accelerator 
conditions changed (\ie more colliding bunches and smaller $\beta^*$), giving as a result an 
increase of the instantaneous luminosity, which raised the \gls{pileup} conditions in the detector. 
A maximum value of 31 \gls{pileup} events per bunch crossing was reached. This increase of the instantaneous luminosity 
also gave a gain in the total luminosity delivered by \gls{lhc} and recorded by \gls{cms}, as 
can be seen in Figure~\ref{ch3:fig:lumicumulative} as a function of time.

\begin{table}[!hbtp]
	\centering
	\begin{tabular}{rl}\hline\hline
		Parameter & Value \\ \hline
		Peak Instantaneous Stable Luminosity	& $7670.2$ x $10^{30}~\percms$ \\	
		Max Luminosity Delivered in one Fill        & $246.3~\pbinv$\\ 
		Maximum Luminosity Delivered in one Day     & $286.1~\pbinv$ \\
		Maximum Luminosity Delivered in one Week    & $1300.6~\pbinv$ \\ 
		Maximum Luminosity Delivered in one Month   & $3693.1~\pbinv$ \\ 
		Maximum Colliding Bunches                   & 1380 \\
		Maximum Interactions per Crossing (pileup)  & 35 \\ \hline
	\end{tabular}
	\caption[LHC performance summary for 2012 run]{LHC and CMS report for 2012 Run.\label{ch3:tab:lhcrun2012}}
\end{table}
The 2012 run started with an increased centre of mass energy of 8~\TeV reaching a new record
in centre of mass energy. The data taking started in April 2012 and 23.30~\fbinv were delivered by
the \gls{lhc} whilst \gls{cms} recorded 21.79~\fbinv. Three TS in April, June and September 2012 
split the data in four different periods: Run2012A-B-C-D. After every TS, as the 2011 run, the 
accelerator conditions changed, increasing its performance. As example, the peak instantaneous 
luminosity was doubled with respect to the previous year. The impressive performance obtained in the 
2012 run is shown in Table~\ref{ch3:tab:lhcrun2012}

Several of the parameters have already reached their nominal value during the 2010-2012 
running period. Some of the machine parameters are shown in Table~\ref{ch3:tab:lhcparam}, where
the nominal design values are compared with the values of the 2010, 2011 and 2012 running. In 
particular, the number of protons per bunch was overcome already in 2010. The 2011 and 2012 have
been a high \gls{pileup} environment because the higher separation between bunches with respect 
to the nominal one was compensated with a low value of $\beta^*$ and a higher number of protons per
bunch, achieving a high instantaneous luminosity (almost the nominal one) but increasing the \gls{pileup}
environment.

\section{The Detector}\label{ch3:sec:cms}
The \acrfull{cms} is a general purpose detector designed to see a wide range of particles 
and phenomena produced in high-energy proton-proton collisions in the \gls{lhc}. 
The main goal of the experiment is the discovery of new particles,
such as the Higgs boson or supersymmetric partners of the Standard Model 
particles~\cite{CmsTDRII:2007}. The scale of the experiment would make impossible to deal with it
unless the efforts for building, managing and studying the results would be shared along a wide
world collaboration. The \gls{cms}\glsadd{ind:cms} Collaboration is composed of about 181 institutions from 38 
countries, a total of more than 3000 people. 

The experiment name shows the importance given to the muon system and to the relative compactness of the 
detector as a result of a novel design fitting both the electromagnetic and hadronic calorimeters
inside of its solenoidal magnet. With a length of 21.6~m and a diameter of 14.6~m, these dimensions make
\gls{cms} smaller than the other main purpose detector at the \gls{lhc}, \gls{atlas}, but much more 
dense, with a total weight of 12500 tons; \gls{cms} follows the standard cylindrical symmetry 
topology where the detector's subsystems are disposed in concentric layers, like a cylindrical onion. 
The detector is divided mainly in two different regions: barrel and endcaps.

\begin{figure}[!htpb]
	\centering
	\includegraphics[scale=0.5]{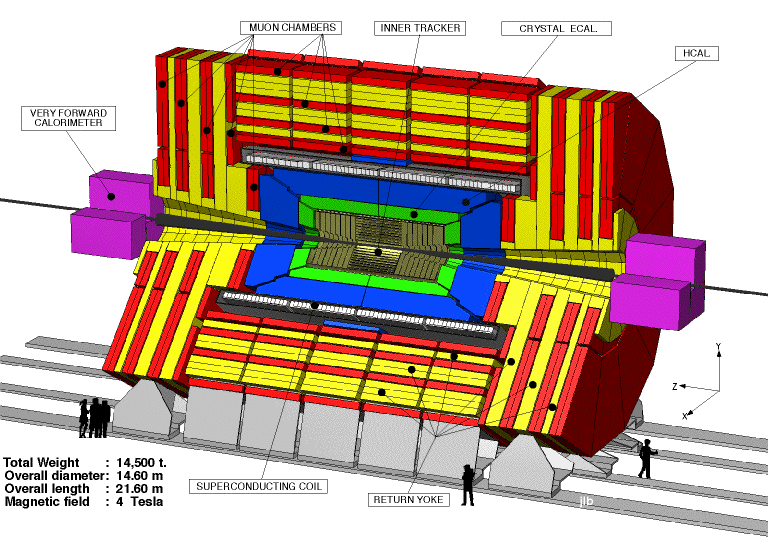}
	\caption[CMS Overview]{A cut-away view of the \gls{cms} detector, showing the main 
	components and subdetector systems.}\label{ch3:fig:cmsOverview}
\end{figure}

\gls{cms} is composed, from the beam line outward, of a vertex detector, a silicon tracker, 
Electromagnetic and Hadronic Calorimeters (ECAL and HCAL) and muon spectrometer.
Figure~\ref{ch3:fig:cmsOverview} shows a schematic view of the detector.
\gls{cms} adopted a coordinate system where the origin is centred at the nominal collision point 
inside the experiment, the y-axis points vertically upward, and the x-axis points radially
inward toward the centre of the \gls{lhc}. The azimuthal angle $\phi$ is measured from
the x-axis in the XY plane, $\phi=arctan(y/x)$ and the polar angle $\theta$ is measured from the z-axis,
$\theta=arctan(\sqrt{x^2+y^2}/z)$. The Figure~\ref{ch3:fig:cmsOverview} contains the reference frame
along with the detector.

The largest group of collisions produced at \gls{lhc}, two protons interacting via strong force, 
tend to carry almost all the momenta along the z-axis, so particle physicists have traditionally
defined a more convenient quantity to describe the particle deflection from the beam pipe, the
\emph{rapidity},
\begin{equation}
	y\equiv\frac{1}{2}ln\left(\frac{E+p_z}{E-p_z}\right)
\end{equation}
which in the relativistic limit, $E\simeq|\vec{p}|$, reduces to a simple function of the polar angle,
\begin{equation}
	\eta\equiv-ln\left(tan\frac{\theta}{2}\right)
\end{equation}
This quantity is called \emph{pseudorapidity} and it is an interesting observable because the occupancy
of the detector is approximately equal in $\eta$ intervals. 

The momentum and energy measured transverse to the beam direction, denoted by 
\pt and \ET respectively, are computed from the x and y components. The imbalance of energy measured 
in the transverse plane is denoted by \MET (missing transverse energy)~\cite{CmsTDRI:2006}. These are
convenient quantities since in hadron colliders the energy balance of an event is known only in the 
transverse plane.

The detector requirements for \gls{cms} to meet the goals of the \gls{lhc} physics program can be
summarised as follows:
\begin{itemize}
	\item Good muon identification and momentum resolution over a wide range of momenta in the
		region $|\eta|<2.5$, good dimuon mass resolution ($\sim 1$\% at 100 \GeV), and the
		ability to determine unambiguously the charge of muons with $p<1 \TeV$.
	\item Good charged particle momentum resolution and reconstruction efficiency in the inner 
		tracker. Efficient triggering and offline tagging of $\tau$ and b-jets, requiring 
		pixel detectors close to the interaction region.
	\item Good electromagnetic energy resolution, good diphoton and dielectron mass resolution 
		($\sim 1$\% at 100 \GeV), wide geometric coverage  ($|\eta|<$~2.5), measurement 
		of the direction of photons and correct localisation of the primary interaction 
		vertex, $\pi_0$ rejection and efficient photon and lepton isolation at high 
		luminosities.
	\item Good missing transverse energy and dijet mass resolution, requiring hadron 
		calorimeters with a large hermetic geometric coverage ($|\eta| <$ 5) and with fine 
		lateral segmentation ($\Delta \eta $ x $\Delta \phi < $ 0.1 x 0.1).
\end{itemize}

\subsection{Magnetic Field}\label{ch3:sec:solenoid}
The magnetic field at \gls{cms} is designed to achieve a momentum resolution of 
$\Delta{\rm p/p}$ $\sim$ 10 \% at $p =  1 \TeV$, as part of the physics requires an unambiguous 
determination of the muon charge for muons with a momentum of  $\sim$ 1~\TeV and the measurement 
of narrow states decaying into muons. 

For a particle of charge $q$ moving  inside a magnetic field $B$, the transverse momentum can be 
inferred from the radius of curvature of its trajectory, $r$,
\begin{equation*}
	\pt=qrB
\end{equation*}
Therefore, in order to achieve the design requirements, a powerful magnet was installed.
A magnetic field of 4 Tesla is given by a superconducting solenoid about 14 meters long with
an inner diameter of 5.9 meters. The return field is large enough to saturate 1.5 m of iron, 
providing a consistent 2 T field throughout the outer muon system and allowing 4 muon stations 
to be integrated to ensure robustness and full geometric coverage~\cite{CMSMagnet:1997}. 
Table~\ref{ch3:tab:SolenoidParameters} shows the design parameters for the \gls{cms} Solenoid. 
\begin{table}[!hbtp]
  \centering
  \begin{tabular}{cc}\hline\hline
    Parameter & Nominal value \\ \hline
    Field                   & 4 T        \\
    Length               & 12.9 m         \\
    Inner diameter & 5.9 m \\  
    Number of turns & 2168 \\
    Current                 & 19.14 kA \\ 
    Stored energy     & 2.6 GJ \\    \hline
  \end{tabular}
  \caption[CMS superconducting solenoid parameters]{Main parameters of the \gls{cms}
   superconducting solenoid.\label{ch3:tab:SolenoidParameters}}
\end{table}
The \gls{cms} collaboration decided to start to operate the magnet at a central magnetic field
intensity of 3.8 T. After the first years of operation, once the ageing of the coil is better 
understood, the collaboration may decide to increase the operation mode to its nominal value.

\subsection{Inner Tracker System}\label{lab:cmsTracker}
A robust tracking and detailed vertex reconstruction are able to identify and precisely measure
vertices, muons, electrons, and the charged tracks within the jets over a large energy range. 
The inner tracker system~\cite{CMSTracker:1997} is based on silicon semiconductor technology.
The actual technological implementation is driven by the detector occupancy; we can differentiate
three radial regions:
\begin{itemize}
	\item Closest to the interaction vertex where the particle flux is the highest 
		($\sim~10^7$~/\cm/s at r~$\sim$~10~cm), pixel detectors are placed. The size
		of a pixel is $\sim$100 x 150~$\mu$m$^2$,  giving an occupancy of about
		$10^{-4}$ per pixel per \gls{lhc} crossing.
	\item In the intermediate region (20~$<$~r~$<$~55 cm), the particle flux is low enough to
		enable use of silicon microstrip detectors with a minimum cell size of 
		10~cm~x~80~$\mu$m., leading to an occupancy of $\sim$ 2-3 \% per \gls{lhc} crossing.
	\item In the outermost region (r~$>$~55 cm) of the inner tracker, the particle flux drops 
		sufficiently to allow use of larger-pitch silicon microstrips with a maximum cell
		size of 25~cm~x~180~$\mu$m, whilst keeping the occupancy at $\sim$~1\%.

\end{itemize}
Providing precise measurements of the trajectories of all charged particles is the main goal of
the inner tracker. Nevertheless, its resolution is also sufficient to distinguish a secondary
vertex in a single collision event corresponding to displaced tracks. This displaced tracks are
the distinctive characteristic of long-lived hadrons containing $b$ or $c$ quarks. Therefore, this
allows to discriminate between prompt\glsadd{ind:prompt} leptons produced from the decay of vector bosons and secondary
leptons coming from the semileptonic decay of heavy flavour hadrons. 

The closest region to the interaction vertex is the barrel region, which consists of 3 layers of 
hybrid pixel detectors, situated at radii of 4.4, 7.3 and 10.2 \cm from the beam line. The size 
of these pixels is 100~x~150~$\micron^2$. The pixel detector provides tracking 
points in both the $r-\phi$ (with resolution of 10\micron) and $r-z$ (resolution 20\micron) planes.
This design, providing a z resolution on par with the $r-\phi$ resolution allows a successful 
secondary vertex reconstruction in three dimensions. 

\begin{figure}[!htpb]
	\centering
	\includegraphics[scale=0.7]{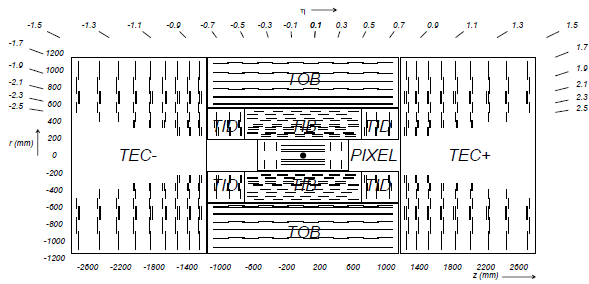}
	\caption[Schematic Tracker Layout]{Schematic vision of the \gls{cms} Tracker.
	Single lines represent detector modules, double lines indicate back-to-back modules which
        deliver stereo hits.}\label{ch3:fig:TrackerLayout}
\end{figure}

Ten layers of silicon microstrip detectors are placed at radii between 20 and 110 \cm. Each strip
has a length between 10~\cm and 25~\cm and with a width of 180~\micron. The forward region is formed by 2 endcaps 
(Tracker Endcap, \ie TEC), each having 2 pixel and 9 microstrip layers.
In total, the inner tracker comprises 66 million pixels and 9.6 million silicon strips. The
barrel part is separated into an Inner (TIB) and an Outer Barrel (TOB). In order to avoid excessively 
shallow track crossing angles, the TIB is shorter than the TOB, and there are an additional 3 Inner 
Disks (TIC) in the transition region between the barrel and endcap parts, on each side of the TIB.
The total area of the pixel detector is $\sim$~1~m$^2$, whilst that of the silicon strip detectors
is 200~m$^2$, providing coverage up to $|\eta|~<$~2.4. This is the largest silicon microstrip 
detector ever built. The material inside the active volume of the tracker increases from
$\sim$ 0.4 $X_0$ at $\eta$ = 0 to around 1 $X_0$ at $|\eta| \sim$ 1.6, before decreasing to 
$\sim$ 0.6 $X_0$ at $|\eta|$ = 2.5. 
The layout of the \gls{cms} tracker is shown in Figure~\ref{ch3:fig:TrackerLayout}.

Hits in the silicon pixel and strips are used as input to reconstruction algorithms which connect
them together into tracks and calculate the associated momenta. The momentum resolution of the 
tracker for $|\eta|<1.6$ is,
\begin{equation}
	\frac{\sigma(\pt)}{\pt}=\left(\frac{\pt}{\GeVc}\right)\cdot0.015\%\oplus0.5\%
\end{equation}
and with the relative error increasing in the forward region to a maximum in $|\eta|=2.5$ of,
\begin{equation}
	\frac{\sigma(\pt)}{\pt}=\left(\frac{\pt}{\GeVc}\right)\cdot0.060\%\oplus0.5\%
\end{equation}
The first term accounts for the curvature measurement which becomes less precise for high-momentum
tracks that bend only slightly in the magnetic field. The second term, the interactions with the 
tracker material such as multiple scattering.

\begin{figure}[!htpb]
	\centering
	\begin{subfigure}[t]{0.45\textwidth}
		\includegraphics[width=0.94\textwidth]{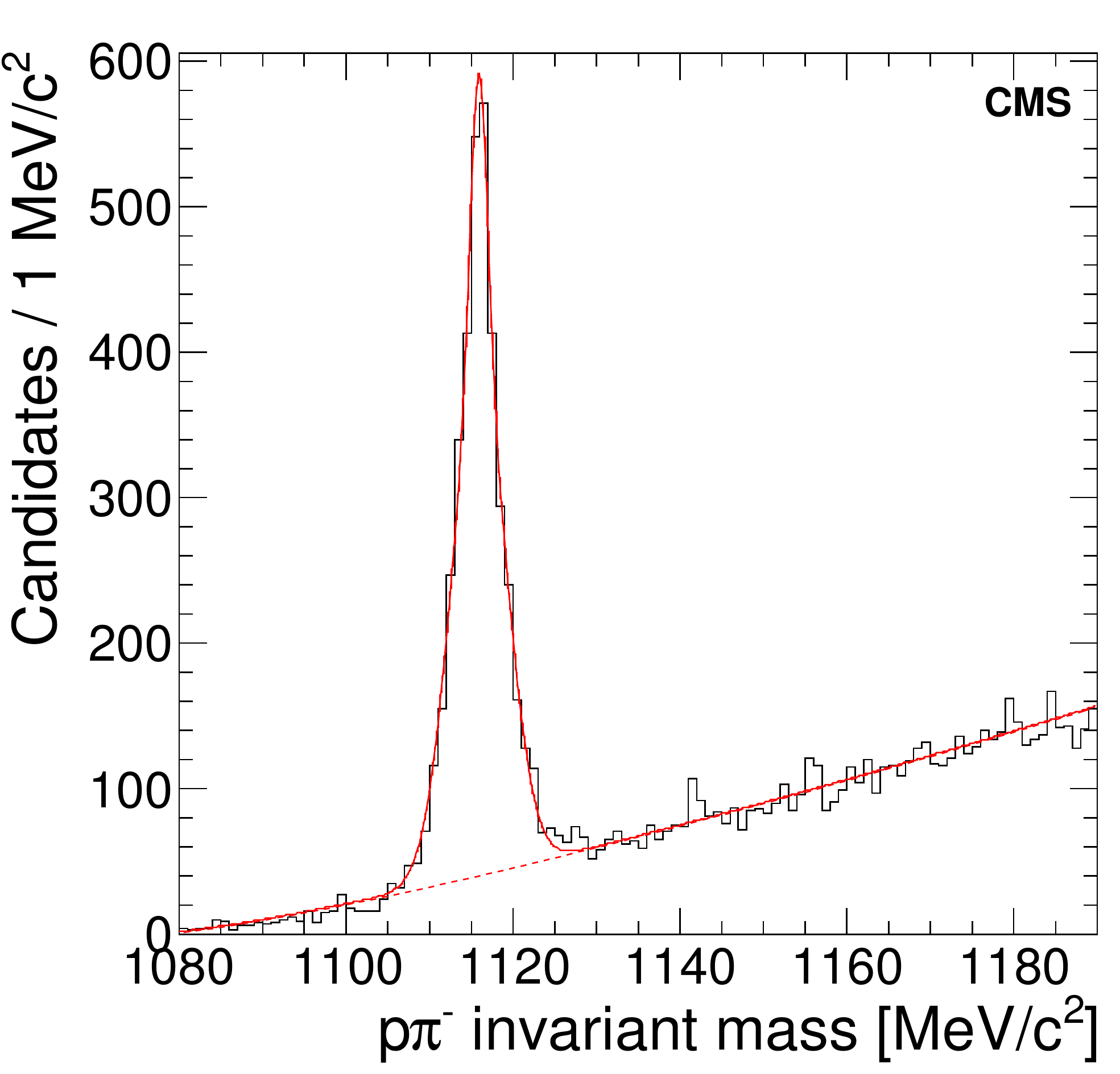}
		\caption{Invariant mass distribution of	$p\pi^-$, and charge conjugate, with a 
			fit to $\Lambda^0$.}\label{ch3:fig:lambda0}
	\end{subfigure}\quad
	\begin{subfigure}[t]{0.45\textwidth}
		\includegraphics[width=\textwidth]{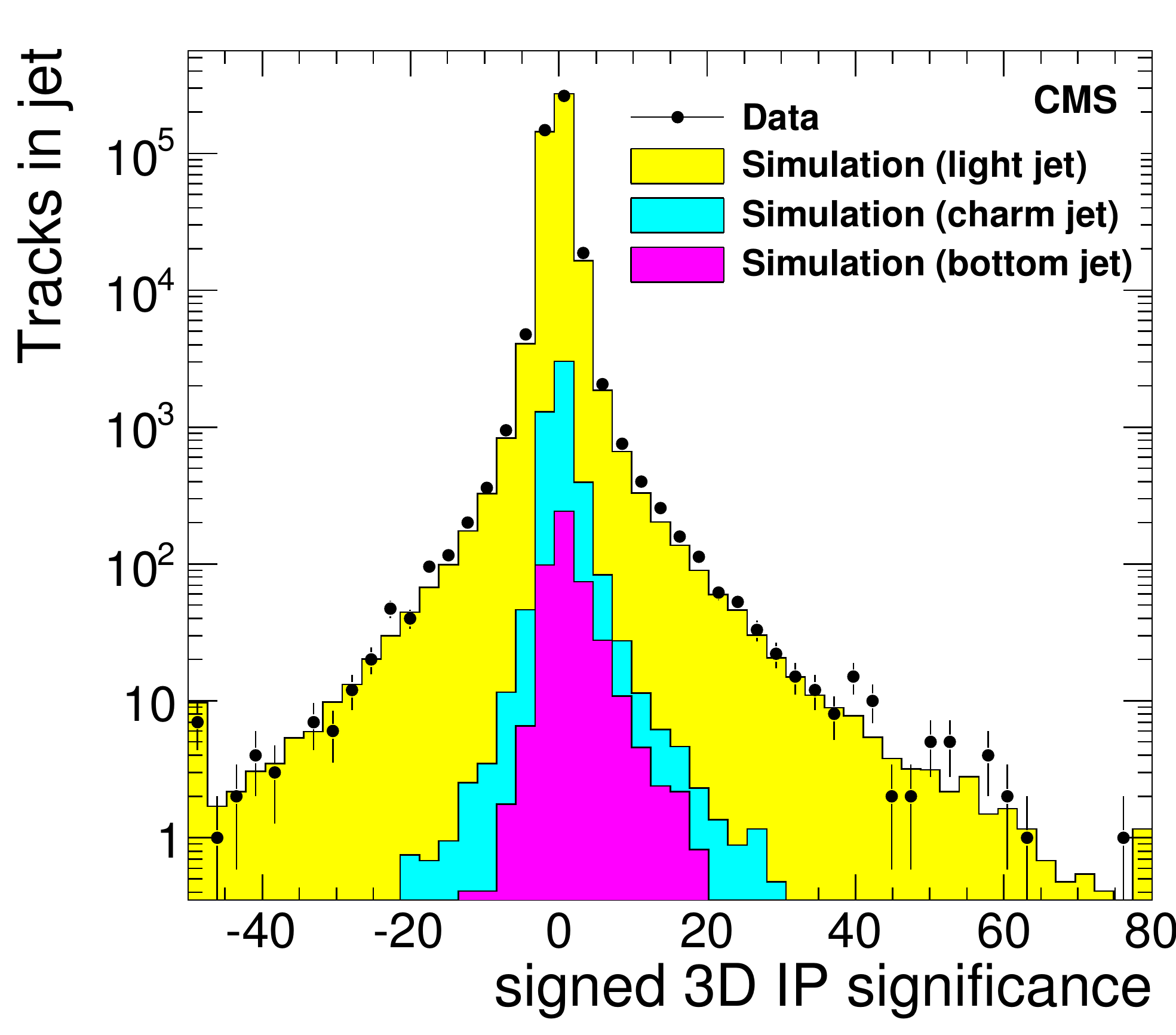}
		\caption{Distribution of the significance of the three-dimensional impact
		parameter for all tracks in a jet}\label{ch3:fig:ip3d}
	\end{subfigure}
	\caption[Tracker performance]{Tracker performance plots from early LHC operations (2010) with
	centre-of-mass energies of 0.9 and 2.36~\TeV. Figures extracted from 
	Reference~\cite{2010EPJC70.1165K}}\label{ch3:fig:trkperformance}
\end{figure}

The impressive performance of the tracker detector is illustrated in 
Figure~\ref{ch3:fig:trkperformance}.

\subsection{Electromagnetic Calorimeter}\label{lab:cmsECAL}
\begin{figure}[!hbtp]
	\centering
	\includegraphics[scale=0.25]{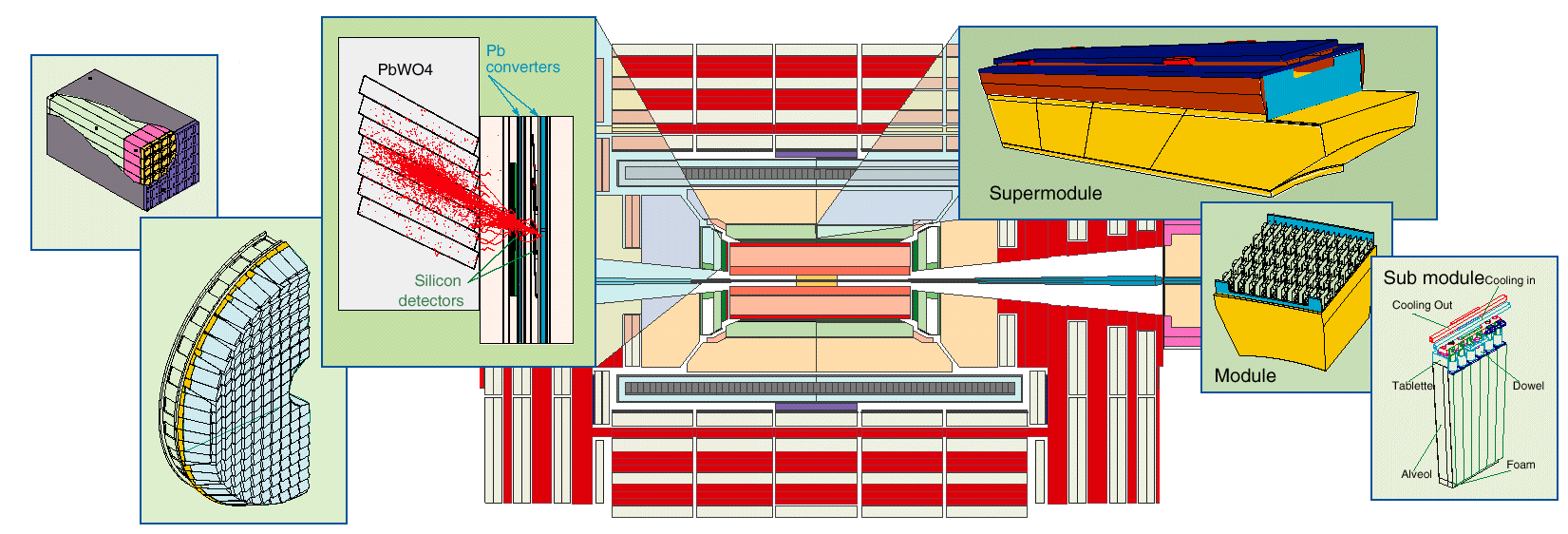}
	\caption[Electromagnetic calorimeter layout]{Layout of the \gls{cms} Electromagnetic Calorimeter 
	showing several features.}
	\label{ch3:fig:ECAL}
\end{figure}

The \gls{cms} \gls{ecal}\glsadd{ind:ecal} is designed to measure precisely the energy of photons and 
electrons and measure the energy of jets as a result of electromagnetic showers. Also, it provides
hermetic coverage for measuring missing transverse energy~\cite{CMSECAL:1997}. The other important
task for the \gls{ecal} is to achieve good efficiency for electron and photon identification as well
as excellent background rejection against hadrons and jets.  The large solenoid radius allows the 
calorimetry to be located inside the solenoid. A scintillating crystal calorimeter offers the best
performance for energy resolution since most of the energy from electrons or photons is deposited
within the homogeneous crystal volume of the calorimeter. High density crystals with a small 
Moli\`ere radius allow a very compact electromagnetic calorimeter system. Lead tungstate (PbWO$_4$)
crystals provide a high density (8.28~g/\cm$^3$), short radiation length (0.89~\cm) and small
Moli\`ere radius (2.2~\cm), leading to rapidly progressing, tightly contained showers for 
high-energy electrons and photons. Also, lead tungstate is a fast scintillator (80\% of the light
is emitted within 25~ns) and it is relatively easy to produce from readily available raw materials.
Furthermore, substantial experience and production capacity were available. These characteristics
drove the choice for the technology in the \gls{ecal} detector. The \gls{ecal} is a hermetic, 
homogeneous calorimeter comprising 61200 PbWO$_4$ crystals mounted in the central barrel part and
closed by 7324 crystals in each of the two endcaps. In Figure~\ref{ch3:fig:ECAL} a layout of the 
ECAL is shown.

The crystals emit a blue-green scintillation light peaking near 425~nm, which is collected by 
silicon avalanche photodiodes (APD) in the barrel and vacuum phototriodes (VPTs) in the endcaps. The
APDs and VPTs produce electrical signals which correlate with the multiplicity of detected photons,
allowing the calculation of \emph{energy deposits} left in each crystal. %

The barrel section (EB) has an inner radius of 129~\cm. It is structured as 36 identical 
supermodules, each covering half the barrel length and corresponding to a pseudorapidity
interval of 0 $< |\eta| <$ 1.479. The crystals are quasi-projective (the axes are tilted
at $3^o$ with respect to the line from the nominal vertex position) and cover 0.0174 (\ie  1$^o$)
in $\Delta\phi$ and $\Delta\eta$. The crystals have a front face cross section of 
$\sim$ 22 x 22~mm$^2$ and a length of 230~mm, corresponding to 25.8 $X_0$. 

The endcaps (EE), are situated at a distance of 314~cm from the vertex and cover a pseudorapidity
range of 1.479 $< |\eta| <$ 3.0. They consist of semi-circular aluminium plates from 
which are cantilevered structural units of 5x5 crystals, known as supercrystals. The endcap 
crystals, like the barrel crystals, off-point from the nominal vertex position, but are 
arranged in an x-y grid (\ie not an $\eta$-$\phi$ grid). They are all identical and have a front 
face cross section of 28.6 x 28.6~mm$^2$ and a length of 220~mm (24.7 $X_0$). Neutral pions 
($\pi^0$) can inadvertently mimic high-energy photons when they decay into two closely-spaced lower
energy photons that \gls{ecal} picks up together. This problem is localised mostly in the endcap 
regions, thus a preshower detector, with finer granularity than \gls{ecal}, has been installed in front 
of the \gls{ecal} to prevent such false signals. The preshower device is placed in front of the 
crystal calorimeter over much of the endcap pseudorapidity range. The active elements of this 
device are 2 planes of silicon strip detectors, with a pitch of 1.9~mm, which lie behind disks of 
lead absorber at depths of 2 $X_0$ and 3 $X_0$.

Energy deposits in individual crystals are combined into clusters of energy, which are further 
grouped into superclusters in the reconstruction algorithms, serving as starting point for 
identification of photons and electrons in the detector. The \gls{ecal} achieves an energy 
resolution given as
\begin{equation}
	\frac{\sigma(E)}{E}=\frac{1}{\sqrt{E/\GeV}}\cdot2.8\%\oplus\frac{1}{E/\GeV}\cdot12\%\oplus0.3\%
\end{equation}
being the first term the stochastic term, the second one the electronic noise and the last one due 
to detector non-uniformity and calibration uncertainties. 

\begin{figure}[!htpb]
	\centering
	\begin{subfigure}[b]{0.45\textwidth}
		\includegraphics[width=\textwidth]{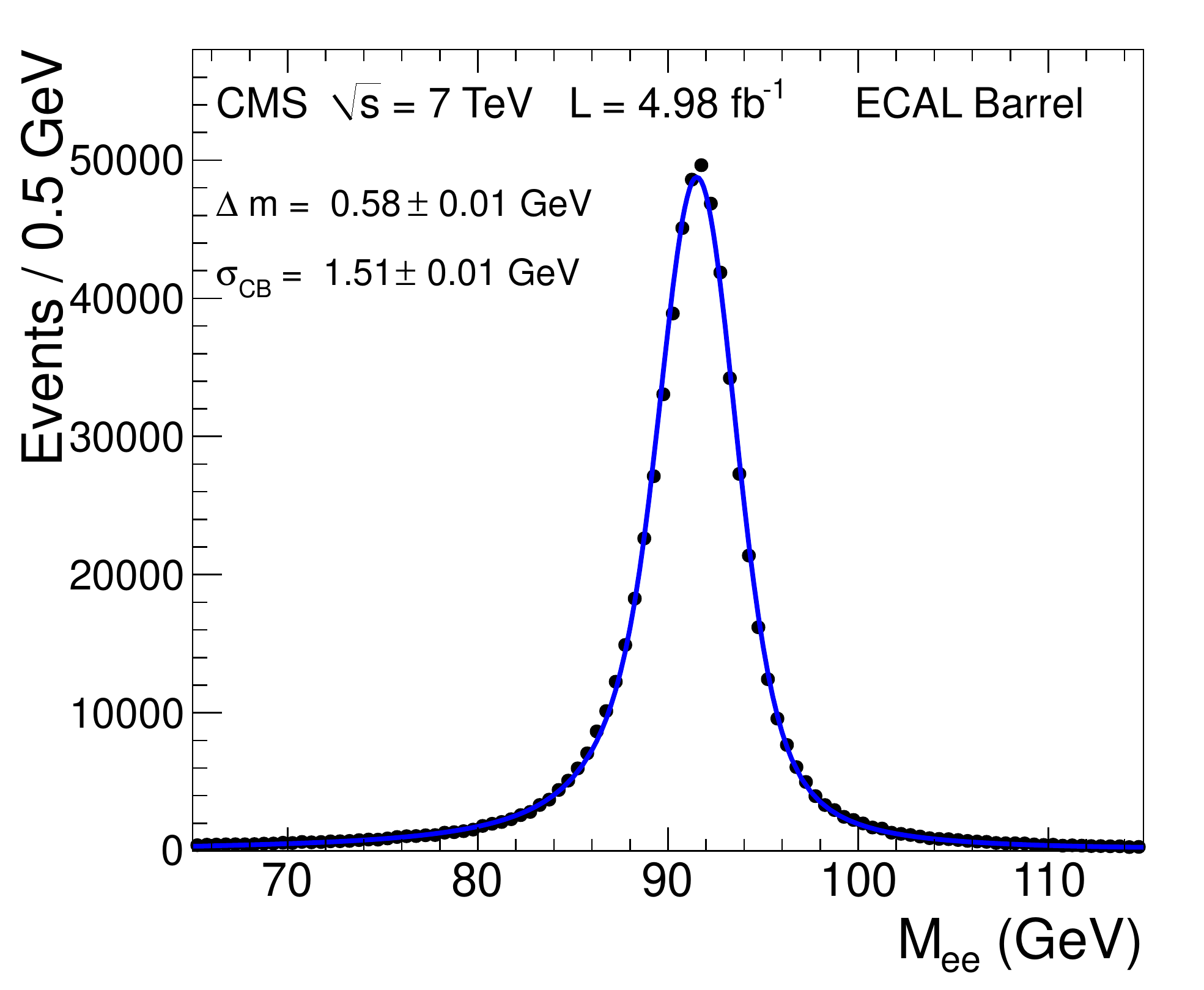}
		\caption{Electrons from the barrel}\label{ch3:fig:zeeEB}
	\end{subfigure}\quad
	\begin{subfigure}[b]{0.45\textwidth}
		\includegraphics[width=\textwidth]{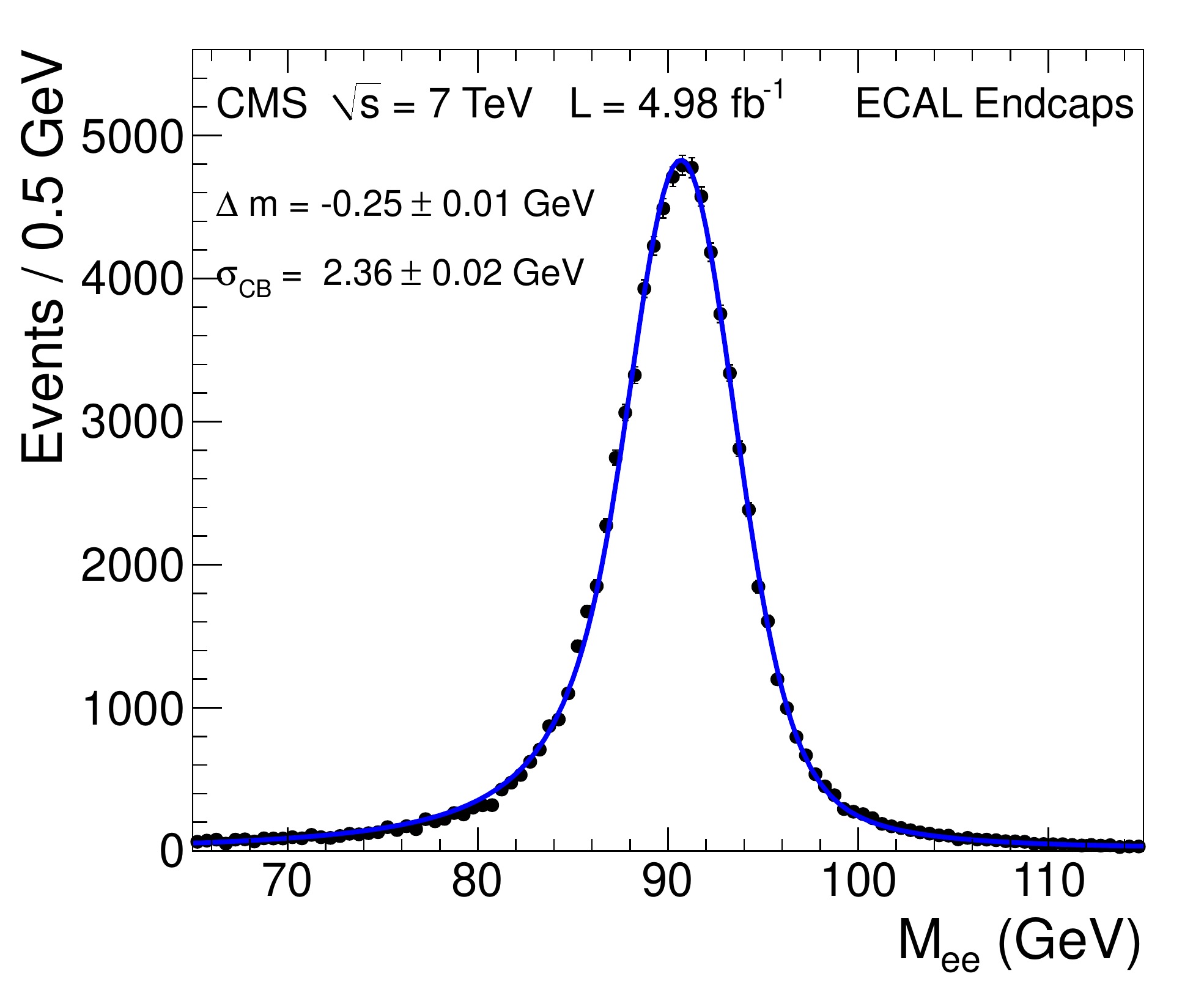}
		\caption{Electrons from the endcap}\label{ch3:fig:zeeEE}
	\end{subfigure}
	\caption[ECAL performance]{Dielectron invariant mass distributions extracted from 2011
	data taking period.}\label{ch3:fig:zee}
\end{figure}

Figure~\ref{ch3:fig:zee} shows the performance of the \gls{ecal} through the dielectron invariant
mass distributions for \Z boson decays with both electrons in the barrel 
(Figure~\ref{ch3:fig:zeeEB}) and both in the endcap (Figure~\ref{ch3:fig:zeeEE}).

\subsection{Hadron Calorimeter}\label{lab:cmsHCAL}
The \gls{hcal}\glsadd{ind:hcal}~\cite{CMSHCAL:1997} surrounds the electromagnetic calorimeter and acts in conjunction 
with it to measure the energies and directions of jet particles. Also, it provides hermetic coverage
for measuring energy. It is strongly influenced by the choice of magnet 
parameters since most of the \gls{cms} calorimetry is located inside the magnet coil. The \gls{hcal}
is designed to detect particles which primarily interact with atomic nuclei via the strong force. 
An important requirement of the \gls{hcal} is to minimise the non-Gaussian tails in the energy 
resolution and to provide good containment and hermeticity for the \MET measurement. Hence, the 
\gls{hcal} design maximises material inside the magnet coil in terms of interaction lengths. This
is complemented by an additional layer of scintillators, referred to as the hadron outer (HO) 
detector, lining the outside of the coil. Brass has been chosen as absorber material, as it has a 
reasonably short interaction length, is easy to produce and is non-magnetic. 
\begin{figure}[!htpb]
	\centering
	\includegraphics[scale=0.8]{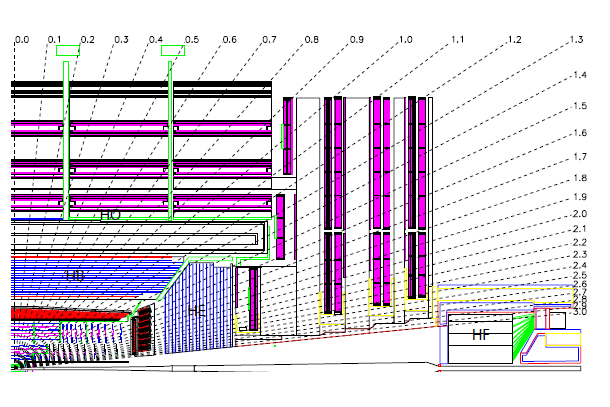}
	\caption[HCAL layout]{Layout of the \gls{cms} Hadron Calorimeter showing the different
	modules which are composed.}\label{ch3:fig:HCAL}
\end{figure}
The \gls{hcal} is a sampling calorimeter with alternating layers of brass and scintillator. The
brass acts as a non-ferromagnetic absorber, capable of withstanding the intense magnetic field and 
providing 5.82 interaction lengths of material in the barrel to facilitate the development of 
hadronic showers. The scintillator consists of tiles along with wavelength-shifting fibre. Hadrons
leaving the \gls{ecal} interact with the scintillating material to produce a broad spectrum of
photons which are then absorbed in the fibre and re-emitted in a more narrow range to which the 
photodetectors are sensitive. Brass is replaced with steel and scintillating fibre with quartz in the endcap, 
which are both better able to withstand the higher radiation dose in that region.

The \gls{hcal} is formed by the hadron barrel (HB), the hadron outer (HO), the hadron endcap (HE)
and the hadron forward (HF) calorimeters, see Figure~\ref{ch3:fig:HCAL}. The full system covers
a pseudorapidity range of $|\eta|<5$.

The granularity of the sampling in the 3 parts of the HCAL (HB, HO and HF) has been chosen such 
that the jet energy resolution, as a function of \ET, is similar in all 3 parts. The resolution for
the barrel and endcap has been measured to be,
\begin{equation}
	\frac{\sigma(E)}{E}=\frac{1}{\sqrt{E/\GeV}}\cdot85\%\oplus7.4\%
\end{equation}
with statistical fluctuations and constant terms in analogy to those discussed in the \gls{ecal} section.
The inferior performance relative to the \gls{ecal} is due both to its operating principle of 
sampling the shower rather than absorbing all produced energy in high-resolution crystals and 
also to the intrinsically lower particle multiplicity in hadronic showers with respect the
electromagnetic showers, leading to wider statistical fluctuations.

\subsection{Muon System}\label{lab:cmsMuonSystem}
The main purposes of the muon system are to identify muons, to be able to do it quickly in order to
select the event with muon content for recording (\emph{muon trigger}) and to measure muon momentum.
Performance requirements follow the physics goals, including the maximum reach for unexpected 
discoveries, and the background environment of \gls{lhc} at its highest luminosity. A robust 4 T 
solenoid-based system is the key to the \gls{cms} design~\cite{CMSMuonSystem:1997}. The advanced
muon spectrometer has the following functionality and performance:
\begin{itemize}
	\item Muon identification: at least 16 layers of material is present up to $|\eta|$ = 2.4 
		with no acceptance losses.
	\item Muon Trigger: the combination of precise muon chambers and dedicated fast trigger 
		detectors provide unambiguous beam crossing identification and trigger 
		on single and multi-muon events with well defined \pt thresholds from a
		few \GeV to 100~\GeV up to $|\eta|$=2.1.
	\item Standalone momentum resolution from 8 to 15\% $\Delta$\pt/\pt at 10~\GeV and 20
		to 40\% at 1~\TeV.
	\item Global momentum resolution after matching with the Central Tracker: from 1.0\% to
		1.5\% at 10~\GeV, and from 6\% to 17\% at 1~\TeV. Momentum-dependent spatial
		position matching at 1~\TeV less than 1~mm in the bending plane and less than 
		10~mm in the non-bending plane.
	\item Charge assignment correct to 99\% confidence up to the kinematic limit of 7~\TeV.
	\item Capability of withstanding the high radiation and interaction background expected
		at the \gls{lhc}.
\end{itemize}

The muon system uses three different technologies~\cite{Blum:2008zza}~\cite{Santonico:1981sc}
to detect and measure the muons; drift tubes (DT) in the barrel region, cathode strip chambers
(CSC) in the endcap region, and resistive plate chambers (RPC) in both the barrel and endcap. 
\begin{figure}[!hbtp]
	\centering
	\includegraphics[scale=0.42]{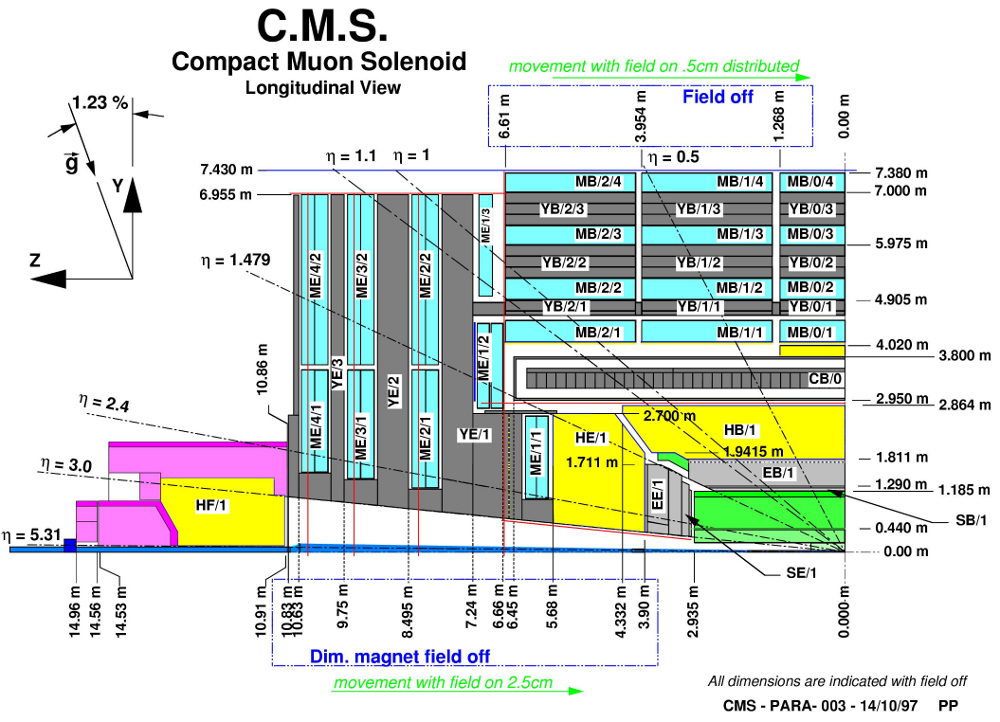}
	\caption[Layout of the CMS Muon System]{Layout of the CMS Muon system showing the 
	disposed stations for barrel and endcap}\label{ch3:fig:cmsMuonSystem}
\end{figure}
To select events with muon content for recording, the trigger (see 
Section~\ref{ch3:subsec:trigger}) in the barrel region is generated using a mean-timer to 
identify patterns. In the endcap the trigger is generated from the cathode readout patterns and the
wire timing. For both barrel and endcap the RPCs provide an additional trigger signal which has a
different sensitivity to backgrounds. All the muon chambers are aligned roughly perpendicular to 
the muon trajectories and distributed to provide hermetic coverage over the $\eta$ range from 0 to 
2.4. The barrel DTs cover roughly from $|\eta|$~=~0 to $|\eta|$~=~1.3, where the neutron-induced 
background is small, the muon rate is low and the residual magnetic field in the chambers is low. 
The endcap CSCs cover from $|\eta|$~=~0.9 to $|\eta|$~=~2.4. In this region, the muon rate, as well
as the neutron-induced background rate, is high, and the magnetic field is also high. The RPCs cover
the region from $|\eta|$~=~0 to $|\eta|$~=~2.1, used both in the barrel and endcap regions. RPCs 
provide a fast response with good time resolution, but with a coarser position resolution than the 
DTs or CSCs. RPCs can therefore identify unambiguously the correct bunch crossing.

The DTs or CSCs and the RPCs operate within the first level trigger system, providing two independent
and complementary sources of information. The complete system results in a robust, precise and 
flexible trigger device. Four stations of detectors are arranged in cylinders interleaved with the 
iron yoke in the muon barrel (MB) region. The segmentation along the beam direction follows the five
wheels of the yoke (labelled YB-2 for the farthest wheel in -z, and YB+2 for the farthest in +z).
In each of the endcaps, the CSCs and RPCs are arranged in four disks perpendicular to the beam, and
in concentric rings; three rings in the innermost station, and two in the others. In total, the 
muon system contains of order 25000~m$^2$ of active detection planes, and nearly 1 million 
electronic channels. A longitudinal view of the muon system with the labelled stations is shown
in Figure~\ref{ch3:fig:cmsMuonSystem}. 
\begin{figure}[!hbtp]
	\centering
	\includegraphics[scale=0.35]{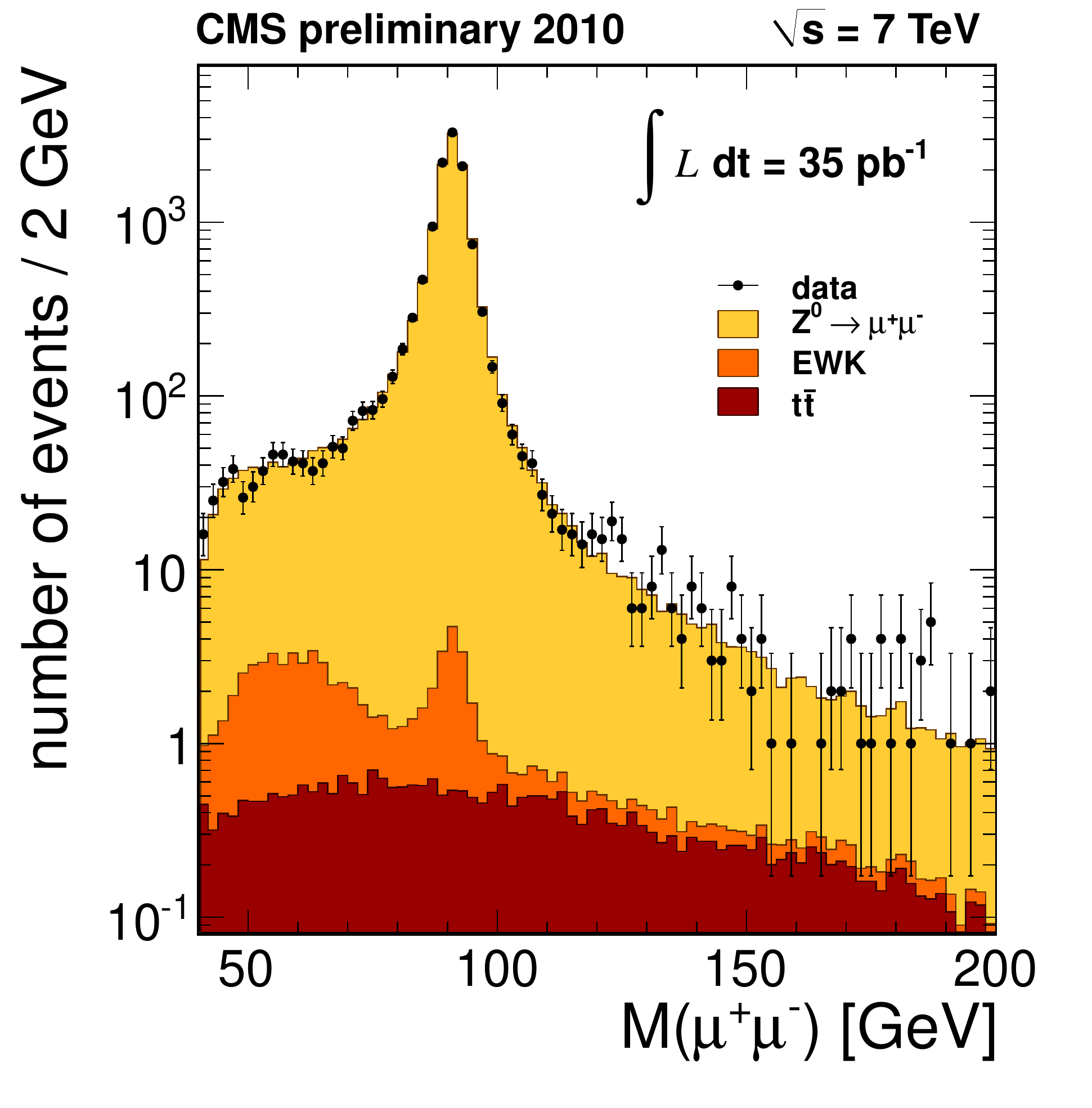}
	\caption[Muon spectrometer performance]{Dimuon invariant mass distribution for 
	\Z boson decays. Data extracted from 2011 run at \comene=7~\TeV using 35~\pbinv of
	data. The experimental data are compared with the expected contributions using simulated 
        data.}\label{ch3:fig:zmumu}
\end{figure}

The great performance of the muon spectrometer combined with the inner tracker detector is
illustrated with the dimuon invariant mass distribution for \Z boson decays of 
Figure~\ref{ch3:fig:zmumu}.

\subsection{Trigger System and Data Acquisition}\label{ch3:subsec:trigger}
\begin{figure}[!htpb]
	\centering
	\includegraphics[scale=0.5]{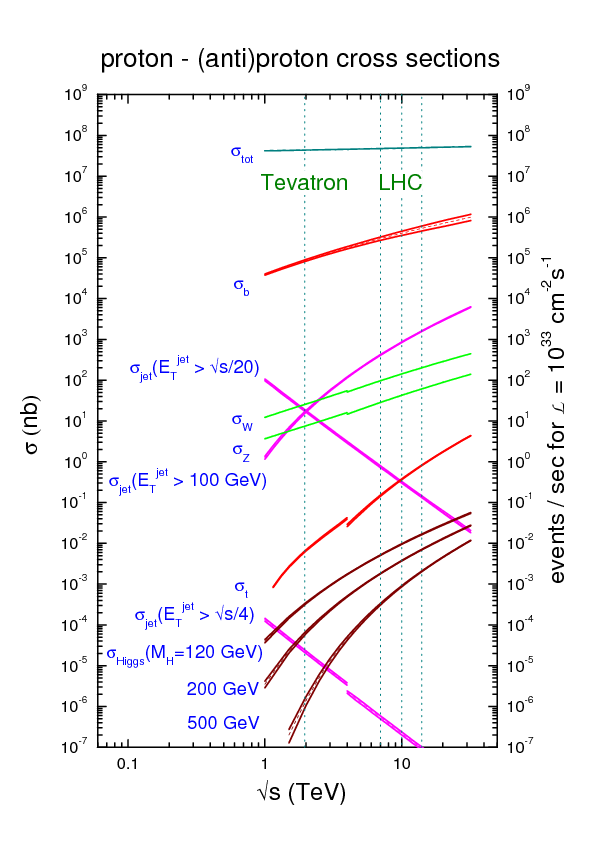}
	\caption[Cross section at hadron colliders]{Production rates of some representative processes
	at hadron colliders as a function of its centre of mass energy. The discontinuity is due to 
	the Tevatron being a proton-antiproton collider while LHC is a proton-proton machine. It
	can be appreciated the tiny production rate for high energy process with respect to the 
	inelastic or low energy processes.}\label{ch3:fig:xspp}
\end{figure}

The huge amount of data delivered by \gls{lhc}\footnote{At design luminosity of \gls{lhc}, 
beams cross at a frequency of 40~MHz leading to collisions on the order of $10^8$ per 
second delivered to \gls{cms}.} makes it impossible for the acquisition system to 
record all of them. Indeed, there is no need to store every collision just due to the fact that 
most of them have "no physic interest": a lot of bunch crossings are going to interact 
inelastically or to produce low energy interactions. Depending on what is going to be measured in
a particular analysis a large number of events should be recorded before an interesting event is 
produced. Figure~\ref{ch3:fig:xspp} shows the cross section or production rate of some 
representative processes at hadron colliders as a function of its centre of mass energy. It can be
observed the huge differences along high energy processes and inelastic or low energy processes.
A flexible and highly configurable system has been deployed, able to make quick decisions on which 
event is worth keeping and which will not be as interesting for analysis. This system is called
the \emph{\gls{gltrigger}\glsadd{ind:gltrigger} system}~\cite{Cittolin:578006}. The \gls{gltrigger}\glsadd{ind:gltrigger} has to provide a huge event reduction
factor and at the same time must maintain high efficiency for the few interesting events among 
millions of background ones. Furthermore, it is required to be flexible enough to easily adapt to 
the different running conditions and physics targets. This flexibility is accomplished by a combined
hardware system made of largely programmable electronics, called \emph{Level-1} (L1) Trigger, and a
software based system for the online event filter, the so called \gls{hlt}\glsadd{ind:hlt}. Using the \gls{gltrigger}\glsadd{ind:gltrigger} system,
the output rate is reduced about a factor $10^7$, delivering to the data processing centres a rate down to a few
hundred Hz.

The L1 \gls{gltrigger}\glsadd{ind:gltrigger} is designed with an output rate of 100~kHz, using information from calorimeters and
muon system and some correlation of information between these systems. The L1 system is able to 
build some coarse high level objects, such as muons, electrons, jets, \MET, photons, etc; which
in turn are used to make decisions about whether to keep or discard a particular event. The total 
time allocated for the transit and for reaching a decision from a particular beam crossing is 
3.2~$\mu s$. This time is in part determined by the size of the \gls{lhc} detectors and the 
underground caverns, as signals from the front-end electronics have to reach the services cavern 
housing the L1 \gls{gltrigger}\glsadd{ind:gltrigger} logic and return back to the detector front-end electronics. Given that each
25~ns a beam crossing is produced, roughly 128 beam crossings are produced while the L1 \gls{gltrigger}\glsadd{ind:gltrigger} is
deciding about the initial event. During this latency time, the detector data must be held in 
buffers, while \gls{gltrigger}\glsadd{ind:gltrigger} data are collected from the front-end electronics and decisions reached that
discard a large fraction of events while retaining the small fraction of interactions of interest 
(nearly 1 crossing in 1000). Of the total latency, the time allocated to Level-1 \gls{gltrigger}\glsadd{ind:gltrigger} 
calculations is less than 1~$\mu s$. 
\begin{figure}[!htpb]
	\centering
	\includegraphics[scale=0.15]{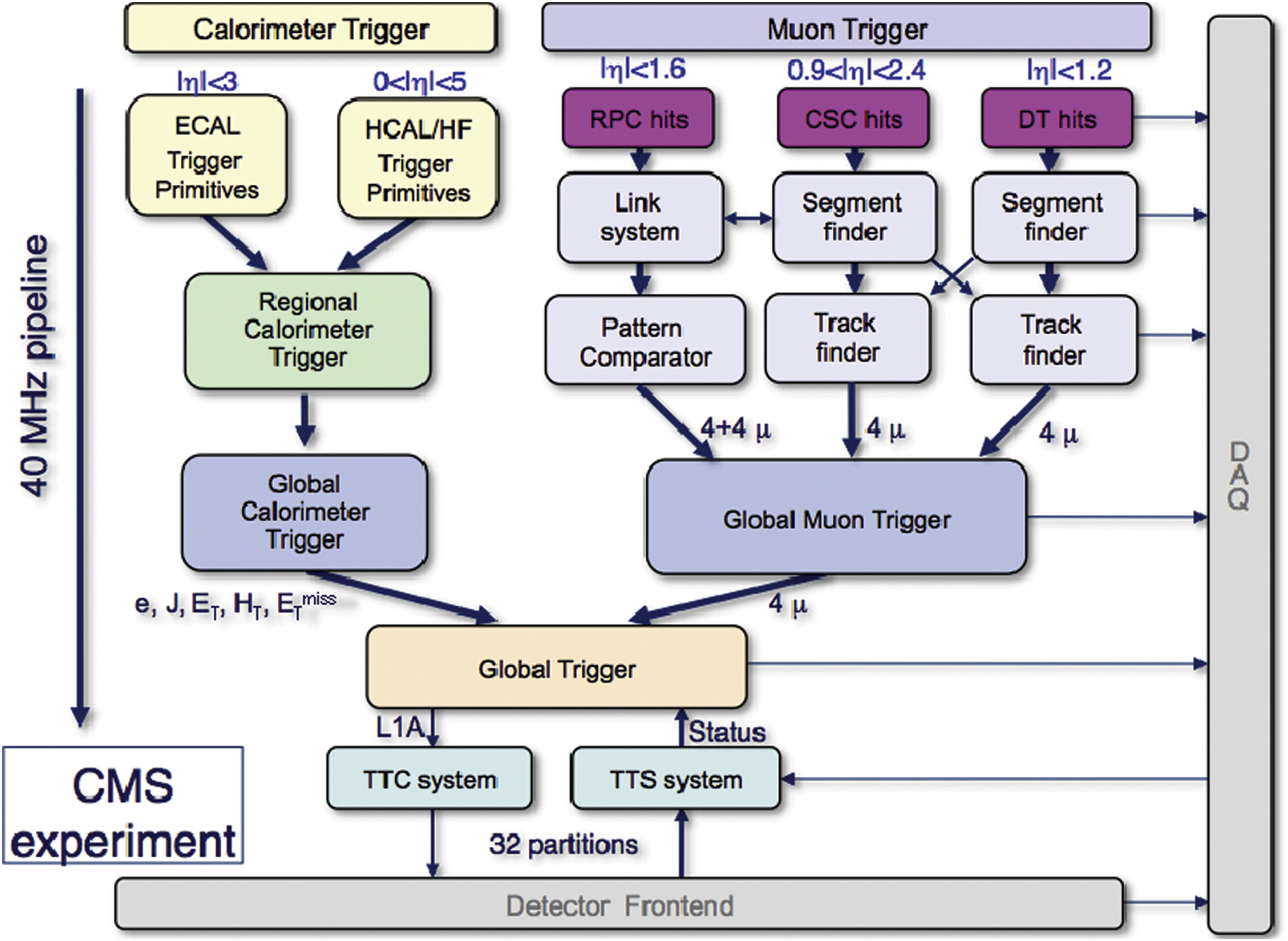}
	\caption[L1 \gls{gltrigger}\glsadd{ind:gltrigger} scheme]{Scheme for the Level-1 trigger system showing the main 
	subsystems used to make decisions.}\label{ch3:fig:L1Trigger}
\end{figure}

The data from the pipelines are transferred to front-end readout buffers. After further signal 
processing, zero-suppression and data-compression, the data are placed in dual-port memories for
access by the \gls{daq}\glsadd{ind:daq}. Each event, with a size of about 1.5 MB (in proton-proton interactions), 
is contained in several hundred front-end readout buffers. Data from a given event 
are transferred to processors where each processor runs the same \gls{hlt}\glsadd{ind:hlt} software code to reduce 
the L1 output rate of 100~kHz to 300 or 400~Hz for massive and permanent storage. The \gls{hlt}\glsadd{ind:hlt}
algorithms, unlike the L1, have access to the complete read-out data, with the possibility to perform complex 
calculations similar to those made in the offline analysis software. The strategies of \gls{hlt}\glsadd{ind:hlt}
development are focused to discard events as soon as possible, therefore the \gls{hlt}\glsadd{ind:hlt} is 
implemented through sequential levels (Level-2 and Level-3) where stricter conditions are checked 
as the level is increased. The \gls{hlt}\glsadd{ind:hlt} system is split in \emph{\gls{gltrigger}\glsadd{ind:gltrigger} paths}: each \gls{gltrigger}\glsadd{ind:gltrigger} path
is targeting different physics processes, therefore each \gls{gltrigger}\glsadd{ind:gltrigger} path may deal with different 
physics objects. There are \gls{gltrigger}\glsadd{ind:gltrigger} paths requiring to have events with at least one single 
high-\pt electron with a \pt higher than a preselected threshold, others requiring a pair of 
electrons, other paths are devoted to muons or \MET, etc. Since events firing each type of \gls{gltrigger}\glsadd{ind:gltrigger}
path are generally independent, the data is naturally sorted into \glspl{pd} based on \gls{gltrigger}\glsadd{ind:gltrigger} path.
Of course, the \glspl{pd} are biased in favour of events with certain properties, but this is 
exactly what was intended when developing the \gls{gltrigger}\glsadd{ind:gltrigger} system. This effect has to be considered
or corrected to ensure there is no bias propagation on the final result of any analysis.

\chapter{Analysis Framework}\label{ch4}

\gls{cms} was described in chapter~\ref{ch3} through its main subdetector systems which 
generate an electrical signal response when particle pass through them. As discussed
in the previous chapter, the individual subdetectors can only offer a readout of hits in the
tracking and muon detectors, energy deposits in the calorimeters and other basic electronic signals.
Therefore, the identities and trajectories of the particles which induced that detector response 
should be inferred using reconstruction algorithms in order to get the ingredients, \ie the \emph{physic
objects}, to recover and reconstruct the collision event. This chapter introduces the different 
algorithms used in the \gls{cms} collaboration to obtain the physics objects 
used in the analysis out of the raw data collected from the subdetectors. The chapter focuses on
the physics objects used in this thesis work; in particular the \emph{muons}, \emph{electrons} and 
the reconstruction of the \emph{transverse energy} is covered in detail. 
Previously, a brief overview of the software framework and the event data model used in the 
collaboration is introduced to contextualise the software physics objects and algorithms applied to 
the raw data. 

\section{Software Framework}
The \gls{cms}\glsadd{ind:cms} collaboration designed and deployed a collection of software, called \emph{CMSSW}, 
built to facilitate the development and deployment of reconstruction and analysis software. The 
CMSSW is based on a framework, an \gls{edm}\glsadd{ind:edm} and the services needed by the simulation, calibration
and alignment, and reconstruction modules that process the raw event data coming from the \gls{cms} 
detector systems~\cite{CMS:2005aa}. Furthermore, the main reconstruction processes from raw 
data to final high level physics objects, as well as simulation, calibration and alignment are done in a 
centralised way using standard and collaboration-agreed algorithms and methods. This allows to use 
the same reconstructed physics objects for all analyses although the framework also allows to tune 
any step of the reconstruction's chain if an analysis requires so. 

\begin{figure}[!htpb]
	\centering
	\includegraphics[scale=0.45]{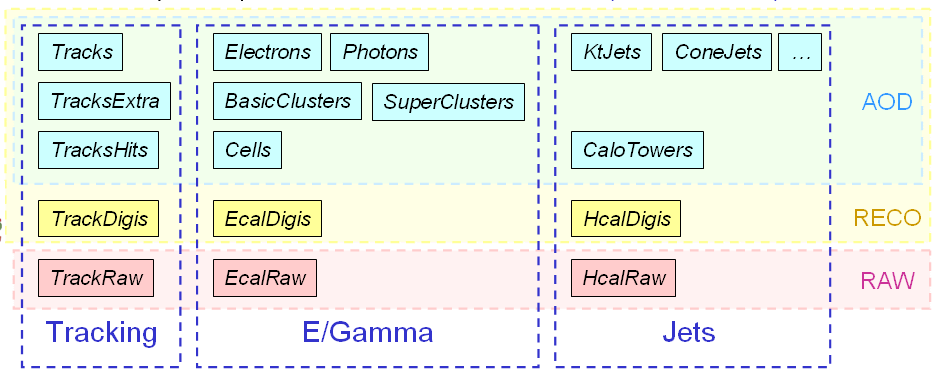}
	\caption[Event content example]{Event content example showing some physics objects in the 
	C++ \emph{Event} container. The information payload carried by each data tier is delimited
	by dotted lines. The "Raw" collections are the readout provided by the subdetector systems 
	of \gls{cms}, these collections are processed to provide digitised grouped signals 
        ("Digi" collections), which are the input for the reconstruction algorithms. After the
        reconstruction step, the \emph{Event} contains intermediate level objects, as 
	\emph{TracksHits}, \emph{SuperCluster} or \emph{CaloTowers} which in turn are used to build
        the high level physics objects, \emph{Tracks}, \emph{Electrons} or 
        \emph{CaloJets}.}\label{ch4:fig:eventcontent}
\end{figure}

The readout of the detector electronics and signals, \ie the raw data, for each proton-proton 
bunch crossing are stored in a C++ container called \emph{Event}. The \emph{Event} is the main 
concept in the \gls{cms} \gls{edm}\glsadd{ind:edm}. From a software point of view, an \emph{Event} starts as a 
collection of the raw data from \gls{cms} or simulated data for a given collision. As the event data
are processed, products are stored in the \emph{Event} as reconstructed data objects. The 
\emph{Event} thus holds all data that was taken from the detectors as well as all data derived from
them. The \emph{Event} also contains metadata describing the configuration of the software used for 
the reconstruction of each contained data object and the conditions and calibration data used for 
such reconstruction~\cite{CMSWorkbook:2012}. This allows to keep track of the different steps as 
well as to introduce any variation in the reconstruction chain. The \emph{Event} data are output to
files in ROOT~\cite{root} format, storing raw plus reconstructed data. Due to the huge amount of 
information dealt with in the \gls{cms} experiment and the finite available resources, the data 
size must be reduced. This is accomplished by selecting only interesting candidate events or by removing 
irrelevant information depending on the stage of an analysis. The former case is achieved by the 
so called Trigger system which is covered in Chapter~\ref{ch5}, the latter defines what is called
\emph{data tier}: each data tier gathers data of each step. Thus,
the RAW tier is defined by all the signals coming from the detectors, the RECO tier groups all the
relevant collections processed from RAW needed for the physic reconstruction plus the collection 
raised from the reconstruction itself. There is another data tier designated as Analysis Object 
Data (AOD) which is a subset of RECO, sufficient for most kinds of physics analyses, so usually 
the root files delivered to physicists only contain the AOD information. 
Figure~\ref{ch4:fig:eventcontent} shows an example of \emph{Event} container with schematic 
information of its contents and the data tiers involved.

\section{Physics Objects}\label{ch4:sec:physicsobjects}
All the reconstructed information is based on interpreting the readout electronics and signals from
each subdetector and associating them to decay vertices, trajectories, energy or particle identities.
These quantities have already physical meaning and can be grouped to form high level physics objects
as electrons, muons, photons, jets and missing transverse energy. Therefore, the high level physics 
objects are built from intermediate level objects like tracks, superclusters of energy, \etc 

\subsection{Particle flow reconstruction}\label{ch4:subsec:pf}
The particle flow reconstruction paradigm is becoming a new standard in the event reconstruction. 
The particle-flow event reconstruction~\cite{CMS-PAS-PFT-09-001} combines the information from all 
subdetectors in \gls{cms} to identify and reconstruct individually all particles produced in a 
collision event. The particles reconstructed (particle flow candidates), namely charged hadrons, 
neutral hadrons, photons, muons and electrons, are used to construct a wide variety of higher-level 
particle-based objects and observables such as jets, missing transverse energy, lepton and photon
isolation, tau identification, b-jet tagging, \etc 
\begin{figure}[!htpb]
	\centering
	\includegraphics[width=\textwidth]{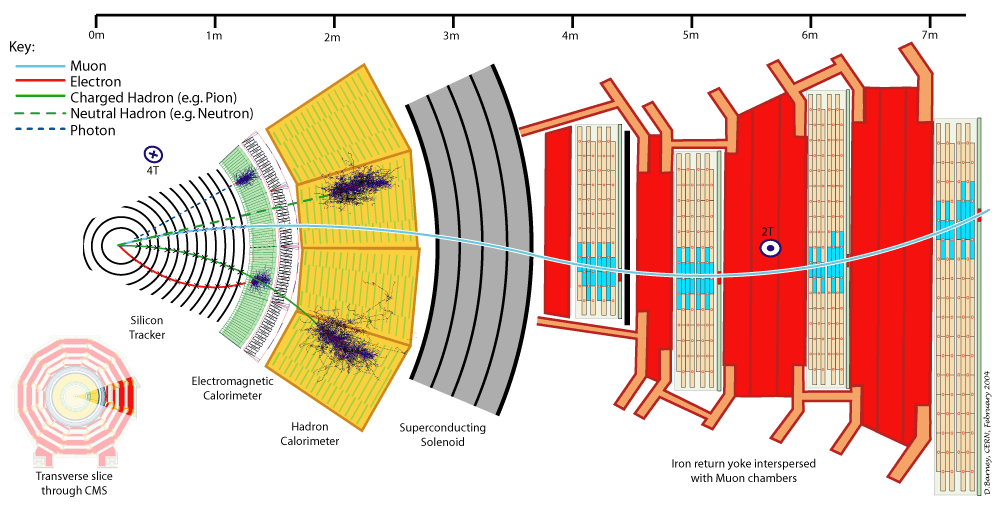}
	\caption[Detector signature of several particles]{Detector signature of several particles 
	showing their main interaction with the detector subsystems. The particle flow algorithm
	uses all the available information from all the subsystems and combine it to obtain the
	particle candidates.}\label{ch4:fig:cmsslice}
\end{figure}
The muons are reconstructed first, accounting for all segments in the muon chambers while removing 
related tracks in the tracker and energy deposits in the calorimeter before moving on the electrons
and jets. The electron reconstruction and identification follows using the remaining energy deposits 
of the \gls{ecal}\glsadd{ind:ecal} and the electron tracks of the tracker. The tracks are refitted with this 
information and the particle-flow electron is built, removing the corresponding tracks and 
\gls{ecal} clusters used. Tighter quality criteria are applied to the remaining tracks in order to 
reject fake tracks. The surviving tracks may rise to charged hadrons, photons or neutral hadrons, and 
more rarely to additional muons. Using several procedures to connect tracks with \gls{ecal} and 
\gls{hcal}\glsadd{ind:hcal} energy gives rise to particle-flow charged hadrons. In the process, energy compatibility 
between tracks and energy clusters in the \gls{ecal} and \gls{hcal}\glsadd{ind:hcal} together with ordering algorithms
allow to obtain neutral hadrons and photon candidates.

\subsection{Muon Reconstruction}\label{ch4:subsec:muons}
Although the muon is not a stable particle, from the point of view of \gls{cms} it behaves as such. 
The muon's interaction is very similar to the electron, but because of its 200 times heavier mass,
and the fact that the radiated power ($P$) of an accelerated particle in presence of electromagnetic fields is 
inversely proportional to the mass of that particle (in fact, $P\propto m^{-4}$), muons do not emit
as much bremsstrahl\"ung radiation as electrons. This feature allows muons to penetrate further than
electrons, barely interacting with the electromagnetic calorimeter and passing through the
flux-return yoke. Thus, the main muon signature in \gls{cms} is a track in the inner tracker 
matched with hits in some of the muon system's detectors and a slight energy deposits in the 
\gls{ecal}.

Muon reconstruction starts from local pattern recognition in the muon systems and tracker, followed
by the so called \emph{stand-alone}, \emph{tracker muon} or \emph{global} reconstruction 
algorithms~\cite{1748-0221-7-10-P10002}. These three stages of reconstruction lead to objects of 
different levels providing by their combination a robust and efficient final muon candidate.

\begin{figure}[!htpb]
	\centering
	\includegraphics[scale=0.6]{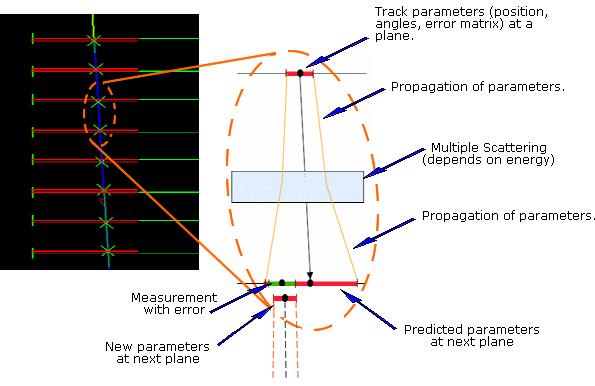}
	\caption[Kalman filter fitting track overview]{Description of the Kalman Filter
	technique used to fit the track parameters. The figure shows the extrapolation of the vector
	state from a surface to another in a schematic way.}\label{ch4:fig:Kalmanfit}
\end{figure}

\paragraph{\indent Stand-alone muon track (STA)}\mbox{}\\
The \emph{stand-alone} algorithm integrates information from all the muon subsystems.
It starts reconstructing hit positions in the DT, CSC and RPC subsystems and building
segments of the trajectory in the surface of the chambers with the hits using a 
pattern recognition. A vector (track position, momentum and direction) is associated to 
each segment. The innermost vector is used as the seed to fit the muon trajectory using a
Kalman Filter technique~\cite{Kalman:1960}. The innermost vectors are propagated to the next 
chamber layer surface (prediction) and a compatible segment (measurement) is searched; once there, the
predicted vector is compared with the measurement in the surface and the trajectory parameters 
are updated accordingly~\cite{Fruhwirth:178627} (see Figure~\ref{ch4:fig:Kalmanfit}). 
The operation is performed until the outermost 
chamber is reached. Material effects like multiple scattering and energy losses due to ionisation
and bremsstrahl\"ung in the muon chambers and return yoke are considered in the vector state
prediction. At each segment a cut is applied in the quality of the track fit, a $\chi^2$ cut,
in order to evaluate the incremental $\chi^2$ of the track fit due the new state and reject 
possible bad hits, mostly due to showering, delta rays and pair production. After the outward 
trajectory fitting, the same Kalman Filter technique
is applied inward to define track parameters in the innermost muon station. From there, the 
trajectory is extrapolated to the point of closest approach to the beam line and optionally 
constrained with a vertex condition. There are additional constrains to accept a trajectory as a
muon track: at least two measurements must be present in the fit, moreover, at least one of them must 
come from the DT or CSC chambers. This allows rejection of fake DT/CSC segments due to combinatorics.

\paragraph{\indent Global muon track}\mbox{}\\
A global muon\glsadd{ind:global} track is obtained by combining the stand-alone muon tracks with 
independently-reconstructed tracks from the inner tracker. Those inner tracks are propagated to the
inside surface of the muon detector and matched with the stand-alone tracks compatible in terms of
momentum, position and direction. Once the match is accomplished, the hits from both collections 
are used as input for a new, global fit. The final trajectory is extrapolated to the interaction
region to obtain the track parameters there. Arbitration and quality algorithms are applied in
order to minimise possible ambiguities and poor matches between the inner and stand-alone tracks.
The global muon track takes advantage from both the tracker detector and muon spectrometer to obtain
a more accurate description of muon's properties. The tracker can in general provide a much higher 
momentum resolution than the muon system due to its high precision and the greater multiplicity of 
hits available for the track fit, but at high energies, $p\sim100\GeV$, the reduced bending of the 
particle limits the resolution of the inner tracker fit. At low momentum, the best momentum 
resolution for muons is obtained from the silicon inner tracker. At higher momentum, however, 
adding hits at large radii from the muon spectrometer can significantly improve the curvature
measurement and thus provide a better momentum resolution. This analysis uses global muons as the
primary algorithm to reconstruct muons although cross-checks with the tracker muon algorithm is
also done in order to improve the quality of the object.

\paragraph{\indent Tracker muon}\mbox{}\\
Besides \emph{stand-alone} and \emph{global muons}, the third algorithm for muon reconstruction is the so called
\emph{tracker muon} which considers all tracks reconstructed from the inner tracker and looks for 
compatible signatures in the calorimeters and muon system. For each track 
in the silicon tracker, the algorithm searches for compatible segments in the muon detectors. In
particular each track with $\pt>0.5~\GeV$ and $p>2.5~\GeV$ acts as seed and it is considered as a muon candidate
if it can be matched to at least one muon segment. Energy depositions compatible with a muon 
hypothesis can also be used for muon identification. The tracker muon approach is particularly 
useful for low-\pt analyses where the global algorithm degrades.

\begin{figure}[!htpb]
	\centering
	\includegraphics[scale=0.3]{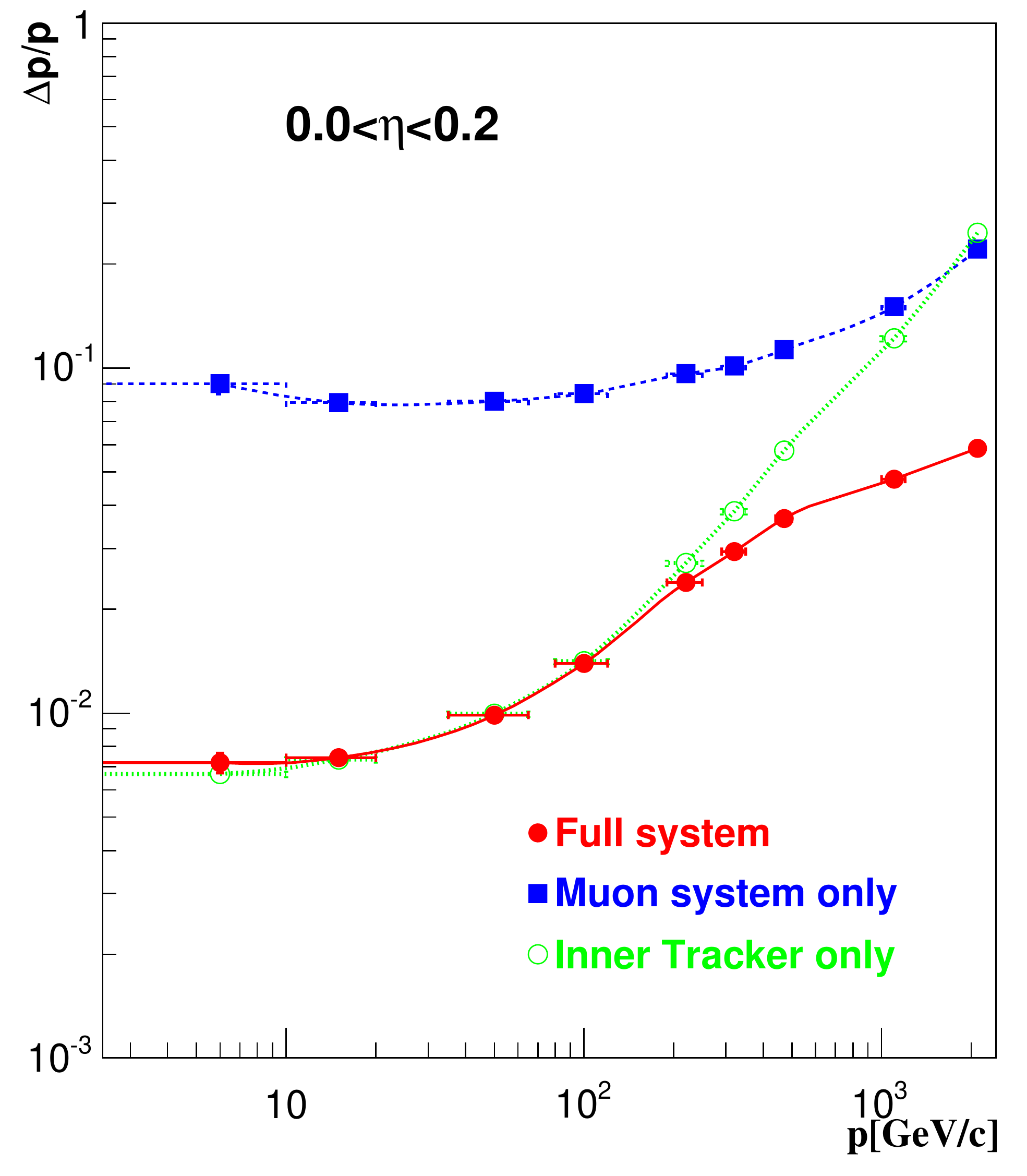}
	\includegraphics[scale=0.3]{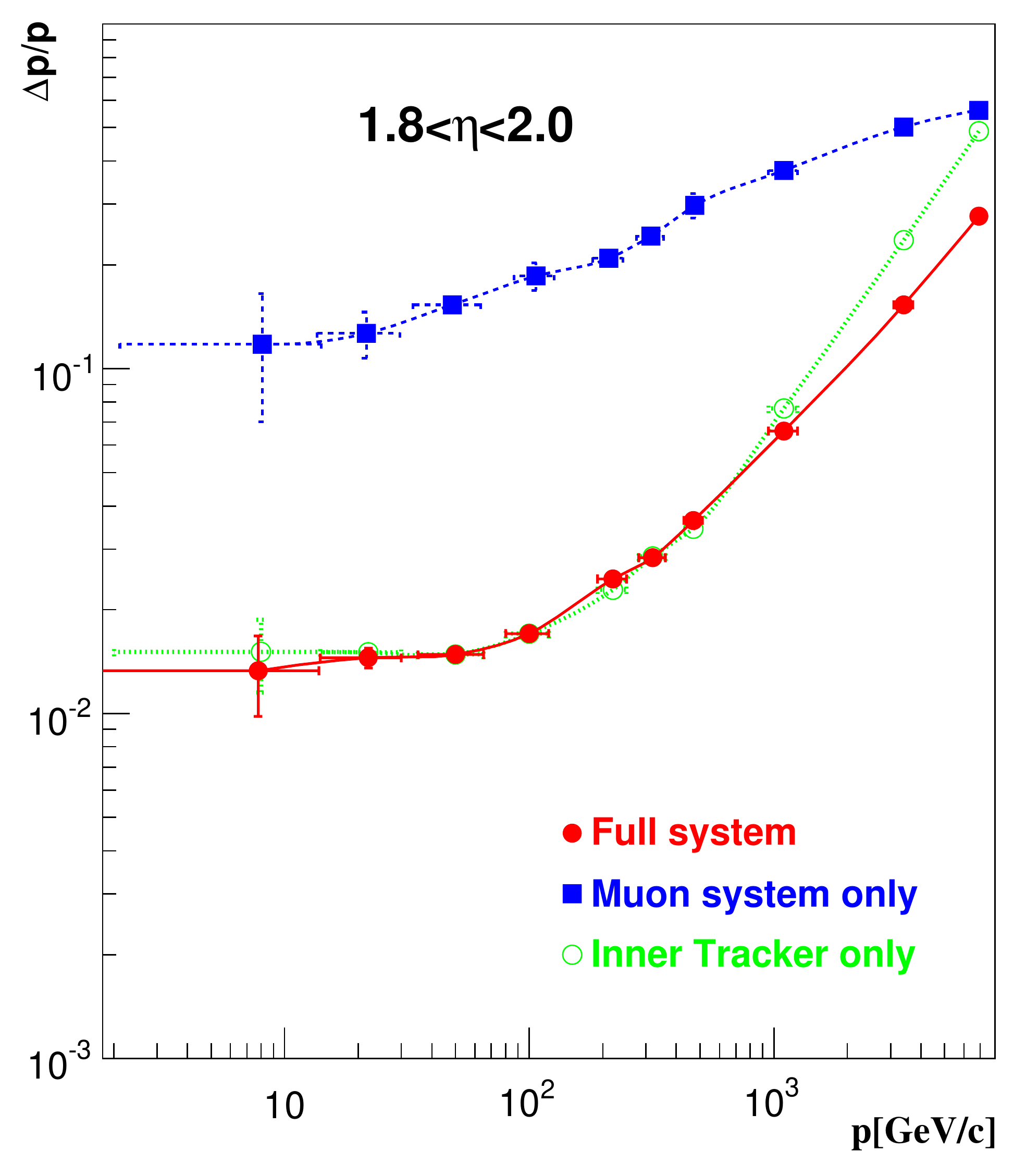}
	\caption[Muon momentum resolutions]{Momentum resolution for the tracker (green), stand-alone
		(blue) and global (red) algorithms for barrel (left) and endcap (right). The plots are
		made using simulated data.}\label{ch4:fig:muonRes}
\end{figure}

\paragraph{}
The three algorithms cover a wide range in momentum keeping the momentum measurement and 
resolution within the challenging design requirements. Figure~\ref{ch4:fig:muonRes} shows 
the momentum resolution for the three algorithms described. For values 
below 200~\GeV the measurement of the momentum is dominated by the tracker resolution whilst for
higher values, the muon spectrometer becomes essential to achieve good resolution.

\begin{figure}[!htpb]
	\centering
	\includegraphics[scale=0.6]{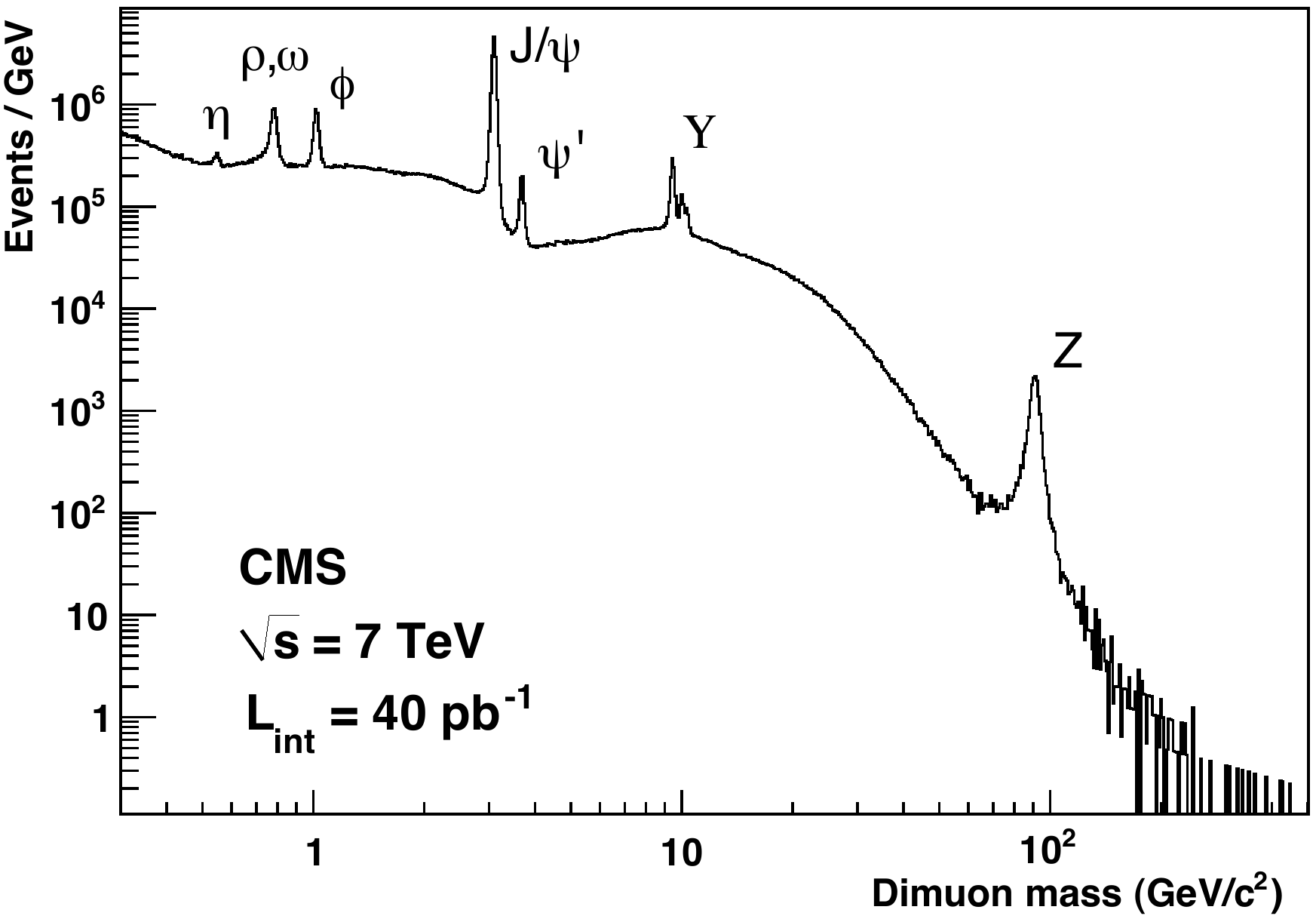}
	\caption[Dimuon mass spectra]{Invariant mass spectra of opposite-signed muon pairs using
	40~\pbinv from 2010 run. Several mass peaks for low and high mass resonances can be 
        appreciated.}\label{ch4:fig:dimuSpectra}
\end{figure}
As an example of the versatility of the muon reconstruction algorithms, Figure~\ref{ch4:fig:dimuSpectra}
shows the reconstruction of the dimuon spectrum for the first 40~\pbinv of data collected in 2010
at 7~\TeV centre of mass energy. This plot shows the high muon momentum resolution in the 
detector, which is able to resolve the invariant mass of a wide range of resonances, covering a 
large kinematic region from $\pt\sim500\MeV$ to 1 \TeV.

\begin{figure}[!htpb]
	\centering
	\includegraphics[scale=0.6]{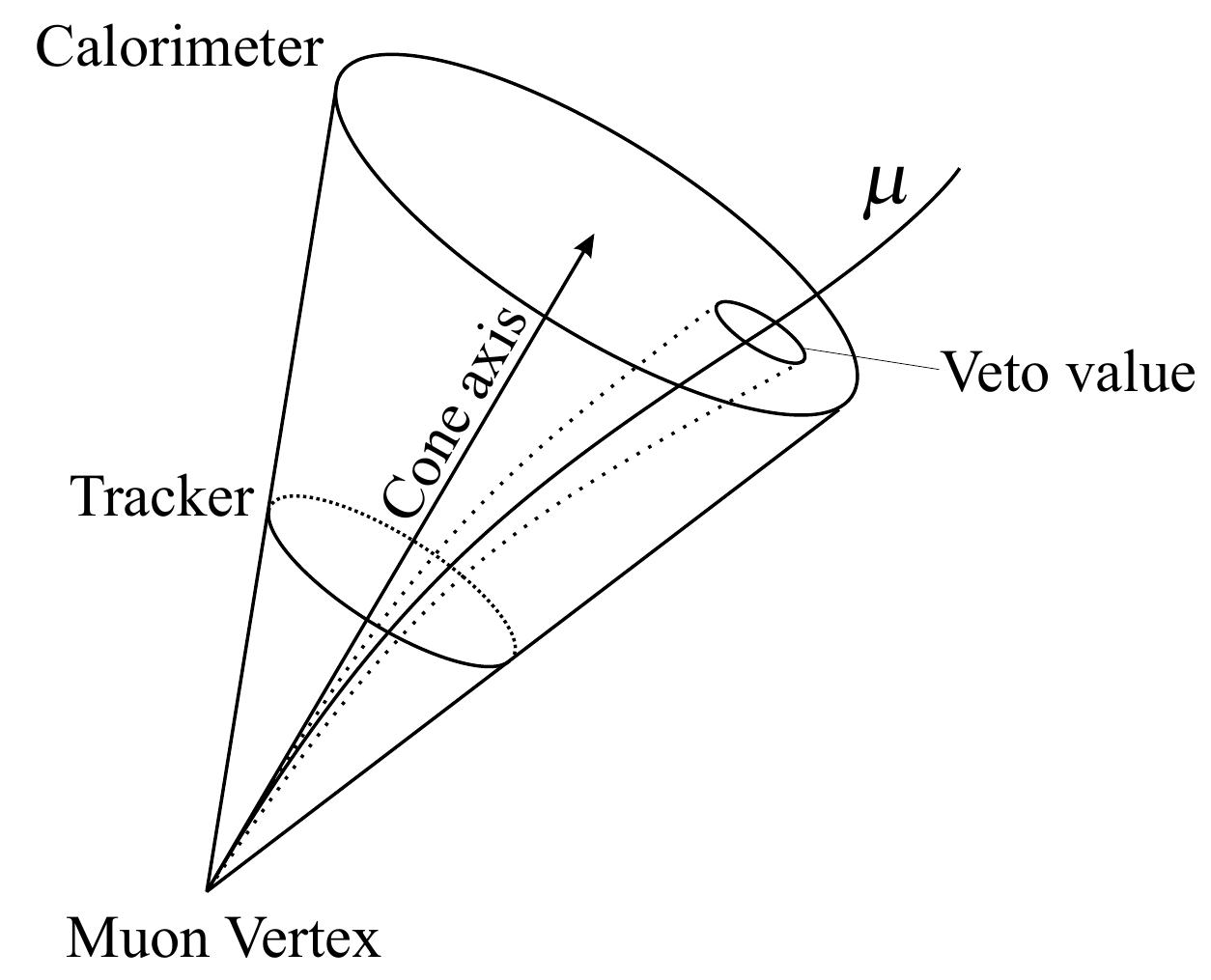}
	\caption[Isolation cone for muon]{Schematic representation of an isolation cone around a muon. 
	The veto value corresponds to the energy contribution of the muon itself which is going to be
	subtracted from the energy deposit of the cone.}\label{ch4:fig:isoCone}
\end{figure}

\paragraph{}
Another important variable in order to discriminate between prompt\glsadd{ind:prompt} leptons coming from gauge bosons
and leptons from heavy quark decays is the isolation of the muon. Leptons coming from jets are 
expected to be surrounded by hadronic activity whilst the prompt leptons from \W or \Z are not. The 
isolation of a muon is evaluated using an algorithm which checks the total energy deposited in a cone
around the muon. The deposit can be the transverse energy in a calorimeter or the sum of transverse
momenta of reconstructed charged-particle tracks. The cone axis is chosen according to the muon 
direction with a procedure that is tailored to the specific properties of each isolation algorithm.
The geometrical definition of the cone is given by the condition $\Delta R\leq \Delta R_{max}$, 
where $\Delta R=\sqrt{\Delta\eta^2+\Delta\phi^2}$, being $\Delta\eta$ and $\Delta\phi$ the 
distances in pseudorapidity and azimuthal angle, respectively, between the deposit and the cone 
axis. As the muon itself contributes to the energy measurement inside the cone, it is subtracted
to improve the discriminating power of the isolation algorithm. Figure~\ref{ch4:fig:isoCone}
illustrates schematically the isolation cone defined around a muon. 
The algorithm used for the \WZ analysis performed in this thesis work is a particle flow 
candidate-based approach (see Section~\ref{ch6:sec:muonselection}). The intense luminosity provided
by the \gls{lhc}\glsadd{ind:lhc} creates and environment where each bunch crossing can lead to dozens of individual 
pp collisions. Most of the recorded events contain only one hard scattering interaction but in the 
same event various other proton collisions are present due to finite resolution of 
the data acquisition system. These other proton collisions, typically soft, are known as 
\emph{pileup} (see Section~\ref{ch5:subsec:pileup} for details) and lead to a degradation of the 
event reconstruction due to the electronic signal contamination produced by its product particles. 
In particular, the hadronic activity is increased, especially in the forward region of the detector. 
The high resolution of the silicon tracker allows a proper reconstruction of the hard scattering 
interaction point\glsadd{ind:ip}, called \emph{primary vertex}, and other points (vertices) making possible the 
association of charged particles to distinct vertices. This mitigates the effect of the pileup given
that the soft collisions are displaced from the primary vertex but there is no possibility to 
associate a neutral particle to a vertex because neutral particles leave no signature in the 
tracker. In this case, finding the vertex associated to a neutral particle relies on the 
calorimeters. Therefore, observables highly dependent on the calorimeters such as jets, missing 
transverse energy, \etc, are very sensitive to the pileup environment. In the case of the particle flow, 
those effects are propagated along all the particles reconstructed. Building an isolation variable
which uses particle-flow candidates allows to cope with and control the intrinsically high pileup 
environment of the \gls{lhc}.

\subsection{Electron Reconstruction}\label{ch4:subsec:Electrons}
The primary electron signature in \gls{cms} is composed of a single track matched to an energy 
deposit in the \gls{ecal}. As electrons highly interact with matter, when traversing the silicon 
layers of the inner tracker they radiate bremsstrahl\"ung photons and, since
the electron direction can change significantly in presence of the 4 T solenoidal magnetic field,
the energy reaches the \gls{ecal} with a spread in $\phi$ which is \pt dependent. The amount of
bremsstrahl\"ung emitted when integrating along the electron trajectory can be very large. 
Furthermore, the conversion of secondary photons in the tracker material might lead to showering
patterns and entail energy loss. The CMSSW software provides two complementary algorithms at the 
track seeding stage. The \emph{\gls{ecal} driven} seeding and the \emph{tracker driven} seeding,
more suitable for low \pt electrons and electrons inside jets~\cite{CMS-PAS-EGM-10-004}. 

The \emph{\gls{ecal} driven} algorithm is optimised for isolated electrons with transverse momentum
$\gtrsim10~\GeV$,
\ie the relevant region of this work. It starts in the electromagnetic calorimeter by grouping one or
more associated clusters of energy deposits into the so called \emph{superclusters}. Due to the 
electron's trajectory bending in the magnetic field and radiating as it passes through the tracker
material, the superclusters usually are spread in $\phi$ although narrow in $\eta$. Using the found
superclusters, the algorithm matches them to pairs or triplets of hits in the innermost layers of 
the tracker. The \emph{tracker driven} seeding algorithm starts from standard tracks reconstructed
from the inner tracker which are extrapolated to the \gls{ecal} searching for bremsstrahl\"ung clusters. 
The energy of the cluster, $E$, and the momentum from the track, $p$, are compared using the $E/p$ 
ratio. The seed of the track is promoted to electron seed if $E/p$ is close to unity. The 
\emph{tracker driven} seeding has been primarily developed and optimised for low \pt, non-isolated
electrons, nevertheless additional isolated electrons can be recovered using this approach combined
with the \emph{\gls{ecal} driven}, in particular in the \gls{ecal} crack regions 
($\eta\simeq0$ and $|\eta|\simeq1.5$). 

Both seeding algorithms are merged into a single collection, keeping track of its provenance. The
seeds are used to initiate a dedicated electron tracking algorithm to fit the electron trajectories,
the \gls{gsf}\glsadd{ind:gsf}~\cite{0954-3899-31-9-N01}, which takes into account a model
of the typical electron energy loss when moving through the tracker. The \gls{gsf}\glsadd{ind:gsf} algorithm 
describes the bremmstrahl\"ung energy loss probability distributions by a superposition of several
gaussians which model the Bethe-Heitler functions~\cite{PhysRevA.34.5126}. The momentum, energy and 
point of origin of the electron trajectory are assigned based on the track parameters at the distance
of closest approach to the nominal beam spot, while energy is determined from a combination of tracker
and \gls{ecal} information. 

After the track building and fitting a preselection is applied to the electron candidates in order 
to reduce the rate of jets faking electrons. The preselection is made very loose so as to maximise the
reconstruction efficiency and satisfy a large number of possible analyses. In the case of electrons
with \emph{\gls{ecal} driven} seeds, some cuts have been already applied at the seeding level, 
requiring the transverse energy of the electron to be greater than 4~\GeV and also the ratio 
of energy deposited in the \gls{hcal} \vs the \gls{ecal} in the supercluster region must fall below
0.15, as significant deposits in the \gls{hcal} would indicate hadronic activity from a jet. In 
addition, the displacement between the supercluster centroid and its associated track must 
satisfy $\Delta\eta<0.02$ and $\Delta\phi<0.15$.

\begin{figure}[!htpb]
	\centering
	\includegraphics[scale=0.35]{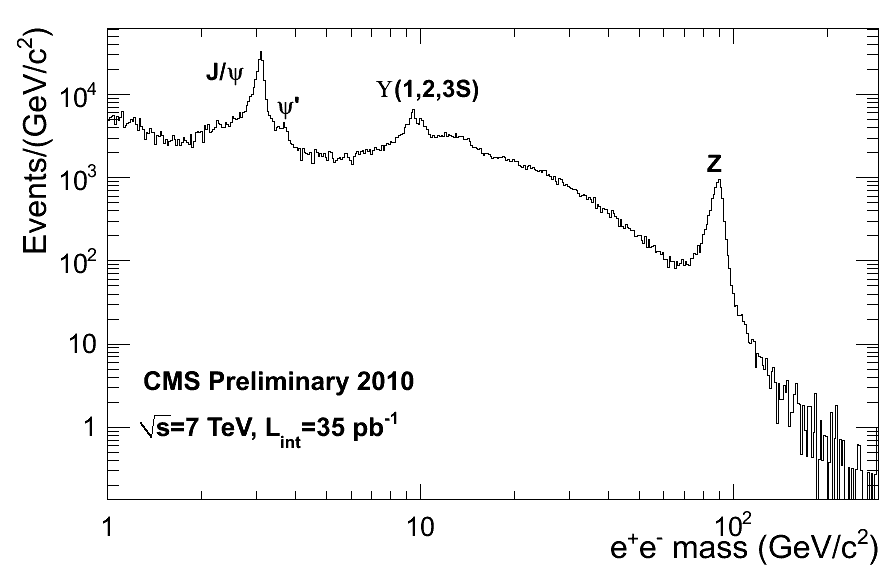}
	\caption[Dielectron mass spectra]{Invariant mass spectra of opposite-sign electron pairs using
	35~\pbinv from 2010 run. Several mass peaks for low and high resonances can be 
        appreciated.}\label{ch4:fig:diElectronSpectra}
\end{figure}

The electron momentum is best estimated if the energy measured in the \gls{ecal} is combined with
the momentum provided by the tracker. In accordance to the respective sensitivity to bremsstral\"ung
induced effects, $E$ (calorimeter measurement) and $p$ (tracker measurement) are either combined or 
only one measurement is used. The tracker measurement is more effective at low energies as well as 
those regions where the precision of the \gls{ecal} is poor, but in general, the \gls{ecal} 
dominates the measurement and resolution. Figure~\ref{ch4:fig:diElectronSpectra} shows the 
dielectron mass spectrum for 2010 data. The mass resolution is worse compared to 
the muon case (Fig.~\ref{ch4:fig:dimuSpectra}) which is able to resolve many more resonances in the
low mass region than electrons.

The isolation variable used in the electron case is analogous to the muons. More details about
the concrete isolation used in the analysis are found in Section~\ref{ch6:sec:electronselection}.

\subsection{Jet Reconstruction}\label{ch4:subsec:jets}
Jets are not used in this analysis, except indirectly in missing transverse energy 
reconstruction (sec.~\ref{ch4:subsec:MET}). The hadrons coming from the fragmentation of 
quarks and gluons produce signals in the \gls{ecal} and \gls{hcal} calorimeters, and if they
are charged, also in the inner tracker. The calorimeter signals are clustered into collimated objects
composed of stable particles, called jets. The neutral particles that partially compose a jet
do not leave tracks in the inner tracker, therefore the jet reconstruction significantly relies
on the calorimeters which introduces ambiguity in jet reconstruction. The \gls{cms} collaboration
makes use of a wide range of jet reconstruction and clustering algorithms~\cite{Chatrchyan:2011ds}.
In particular, the particle flow jet reconstruction (through the missing transverse energy 
reconstruction) is used in this thesis work. The jet momentum and spatial resolutions are expected
to be improved with respect to calorimeter jets, as the use of the tracking detectors and of the 
excellent granularity of the \gls{ecal} allows to resolve and precisely measure charged hadrons and
photons inside jets, which constitute approximately 90\% of the jet energy. 

The input to the particle-flow jet reconstruction algorithm is a collection of energy deposits 
which have a high likelihood of belonging to a jet. The energy deposits are clustered by means of the
"anti-$k_{T}$" clustering algorithm~\cite{Cacciari:2008gp}, which is based on successive pair-wise 
recombination of particles according to the distances between any two particles and the distance of
any particle to the beam. The algorithm starts with a high-momentum particle as seed to the jet and 
successively adds nearby particles to the jet with weights corresponding to their momenta. The 
"anti-$k_T$" algorithm does not change its results neither by the presence of soft particles which 
results in \gls{qcd} divergences, \ie it is \emph{infrared safe}, nor by collinear
splitting (it automatically recombines collinear partons), \ie \emph{collinear safe}. These 
properties lead to a robust event interpretation in terms of partons allowing the application of
the algorithm in theoretical calculations to arbitrary order for meaningful comparisons with 
experimental data.

\subsection{Missing Transverse Energy Reconstruction}\label{ch4:subsec:MET}
The missing transverse energy, \MET or \ETslash, is the main physics quantity used to indicate
the presence of undetected neutrinos. The interacting partons of a hard interaction at a hadron
collider may carry significant longitudinal boost with respect to the lab frame, but should have
negligible momentum in the transverse plane. This fact can be used to infer undetected particles, 
such neutrinos. Therefore, a significant 
imbalance of the vector sum of the transverse momenta of the decay products would indicate the 
direction and momentum of a particle which escaped the detector without interacting. 
This quantity is particularly sensitive to detector malfunctions and detector resolution effects. 
Many sources such as finite measurement resolution, finite reconstruction efficiency, fake tracks, fake clusters, 
etc., can contribute to a spurious observed \MET. Also, the pileup conditions affect 
the estimation of the real \MET resolution, increasing the distribution tails.

The \gls{cms} collaboration uses several techniques to estimate the transverse missing energy in 
the events~\cite{Chatrchyan:2011tn}. 

\paragraph{\indent \MET Calorimetric based (CaloMET)}\mbox{}\\
The \emph{CaloMET} is determined using measurements relying mostly on calorimetric information; in this
case, the \MET is defined as,
\begin{equation}
	\left(\VEtmiss\right)_{CaloMET} = -\sum_{n}\vec{E}_T(n)-\sum_{\mu}\vec{p}_T(\mu)
		+\sum_{\mu}\vec{E}_T(\mu)
		\label{ch4:eq:calomet}
\end{equation}
where $n$ iterates over all energy deposits in the calorimeters, and $\vec{E}_T(n)$ is the 
transverse projection of a vector with magnitude equal to the selected energy deposit, pointing
from the interaction point to the deposit. Explicitly,
\begin{equation}
	\vec{E}_T(n) = - E_n sin\theta_n cos\phi_n \hat{\mathbf{x}} + 
		E_n sin\theta_n sin\phi_n \hat{\mathbf{y}}  
\end{equation}
being $E_n$ the calorimeter inputs and $\phi$, $\theta$, $\hat{\mathbf{x}}$ and $\hat{\mathbf{y}}$
the coordinates related quantities defined in Section~\ref{ch3:sec:cms}. The index $\mu$ refers
to muons and the terms associated are correcting the muon's energy deposit in the 
\gls{ecal}. The muons are minimum ionising particles that transverse the calorimeters almost 
unaffected (the average energy deposits are a few gigaelectronvolts). Hence, to correct for the muon response the 
actual muon momentum measurement from the central tracker and muon system, $\pt^{\mu}$, is used to 
replace the energy measured along the muon trajectory in the calorimeter. The calorimeter \MET 
calculation can be improved by correcting for several effects. In particular, jets can be corrected
to the particle level using the \gls{jec}\glsadd{ind:jec}~\cite{CMS-DP-2012-006} and the \MET is recalculated using
these corrected jets. The \emph{type-I corrections} for the \MET use these \gls{jec}\glsadd{ind:jec}s for all jets 
whose energies are above a threshold and having less than 90\% of their energy in the \gls{ecal}.
These corrections can be up to a factor of two of the initial uncorrected \MET. The remaining soft 
jets below the threshold and energy deposits not clustered in any jet are considered by applying 
a second correction which is referred to as the \emph{type-II correction}. 

\paragraph{\indent \MET Track Corrected based (tcMET)}\mbox{}\\ 
The \emph{tcMET} algorithm uses the \MET measured in the calorimeters, \ie the CaloMET, with further 
corrections using information of the tracker. For each track measured in the tracker, its transverse
momentum, \pt, is included in the \MET whilst the predicted calorimetric energy deposit is removed. 
This approach takes advantage of the better resolution of the tracker versus the \gls{ecal}, 
allowing an overall resolution improvement and a better description of the \MET distribution in the
tails. For all tracks not identified as electrons or muons, the predicted energy deposition is 
extracted from simulations of single pions and extrapolated to the calorimeters. No correction is 
applied for very high \pt tracks ($\gtrsim100~\GeV$) whose energy is already well measured by the 
calorimeters and there is no gain from the tracker, whereas for the low-\pt tracks ($\lesssim2~\GeV$)
no response from the calorimeter is assumed so the measured momentum from the tracker is taken. 
Therefore, the \emph{tcMET} is defined as follows,
\begin{equation}
	\left(\VEtmiss\right)_{tcMET} = \left(\VEtmiss\right)_{CaloMET} -\sum_{tracks}\vec{p}_T(track)
		+\sum_{tracks}\vec{E}_T(track).
		\label{ch4:eq:tcmet}
\end{equation}

\paragraph{\indent Particle Flow \MET (PFMET)}\mbox{}\\
The full extent of the capabilities of the \gls{cms} detector can be reached using a particle 
flow approach to calculate the \MET. All detector information is included to reconstruct the event
decay products. It is simple to use the particle flow particle candidates to estimate the \MET by 
subtracting vectorially all the particle flow transverse momenta.
\begin{equation}
	\left(\VEtmiss\right)_{PF} = - \sum_{\text{PF-cand.}}\vec{p}_T({PF})
	\label{ch4:eq:pfmet}
\end{equation}

\begin{figure}[!htpb]
	\centering
	\includegraphics[scale=0.6]{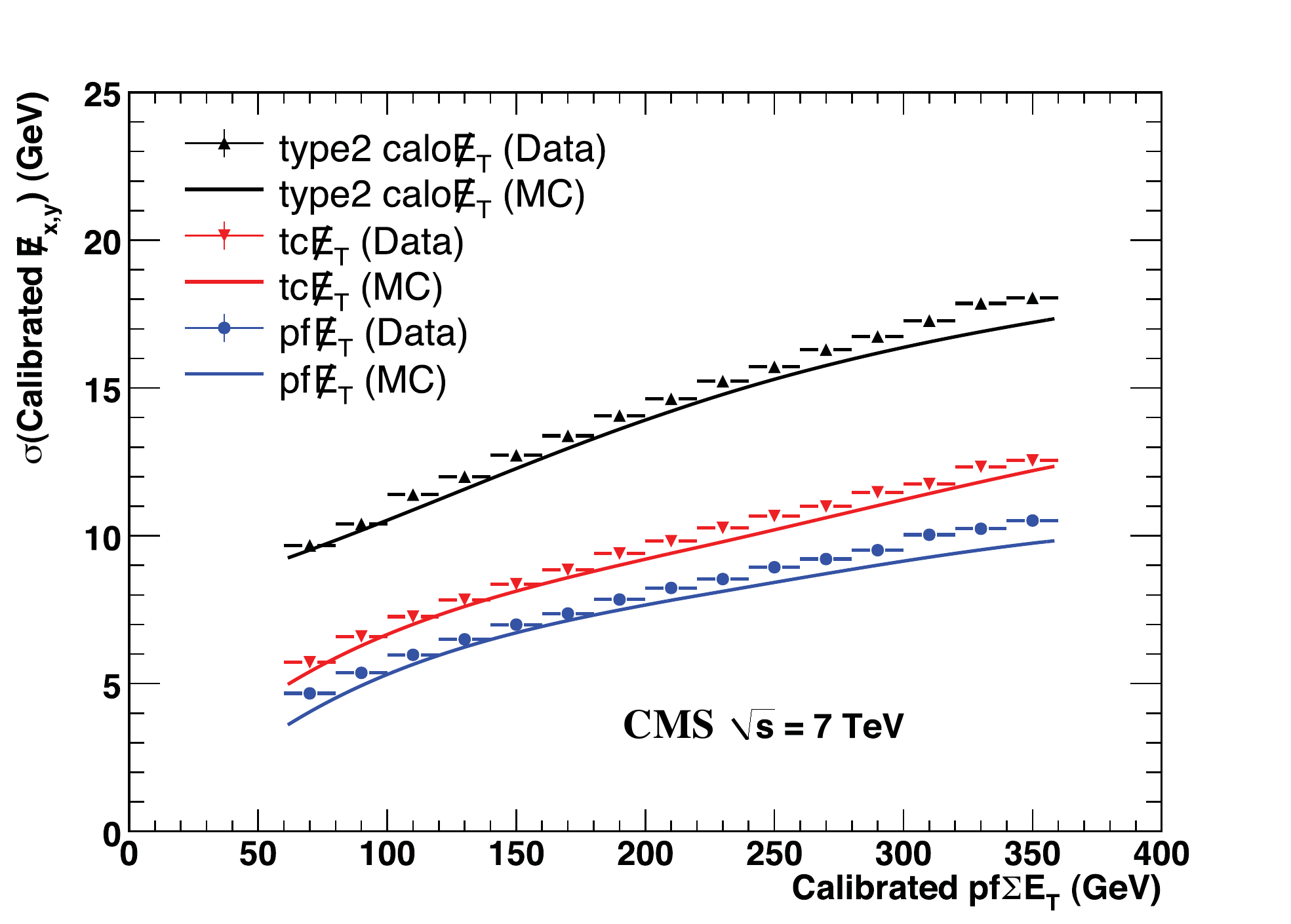}
	\caption[Transverse missing energy resolution]{\MET gaussian core resolution with respect the 
		total transverse energy in the event ($pf\Sigma\ET$) evaluated through the particle 
		flow algorithm. The data are selected to have at least two jets with \pt\textgreater
		25~\GeV. The plot was made using 2010 data.}\label{ch4:fig:metResComp}
\end{figure}
Figure~\ref{ch4:fig:metResComp} shows the comparison for the \MET resolution of the three 
algorithms described \vs the sum of the transverse energy of all the particles in an event using
the particle flow algorithm to reconstruct the particles, \ie $pf\Sigma\ET$. The resolution of the 
\MET is improved by the \emph{tcMET} and \emph{PFMET} by a factor of 2, compared to \emph{CaloMET}.
\MET resolution is affected by several factors: electronic noise, pile-up, underlying events, 
statistical sampling nature of the energy deposits in individual calorimeter towers, systematic 
effects due to non-linearities, cracks and dead material, \etc As a result of this, it is clear 
that the higher pileup conditions of the \gls{lhc} have a direct influence on the \MET measurement.

\chapter{Data Corrections and Monte Carlo tuning}\label{ch5}

The data provided by the detector and reconstructed by the software have to be corrected 
for data-taking and processing effects before they can be analysed. Furthermore, this
data must be compared with theoretical predictions to allow hypothesis falsifiability. The high
energy community uses the so called \gls{mc}\glsadd{ind:mc} techniques to simulate experimental
outcomes to be compared with the data obtained from the experiments. The present chapter is
devoted to enumerate the peculiarities of the data-taking process through the system developed 
to select "interesting events" and the corrections applied to the data:
efficiencies, \emph{\glspl{sf}} and \gls{pileup} effects are described. The last section
of the chapter specifies the \gls{mc}\glsadd{ind:mc} approach and describes how it can be compared with the 
experimental data.

\section{Data corrections}~\label{ch5:sec:datacorr}
Both the \glspl{pd} and the physics objects used in an analysis suffer from a biased realisation
of a physical system, \ie they are not ideal. The detector sub-systems have cracks, dead or
blind zones; they have also latency times, etc. Furthermore, the algorithms used to reconstruct, 
identify or isolate an object, and to make decisions in the \gls{gltrigger}\glsadd{ind:gltrigger} system are inherently
finite and usually do not use all the available information or the information is biased by speed,
storage or other computational-related requirements giving rise to 
inefficiencies. As an example, the trigger paths used for selecting a given \gls{pd}\glsadd{ind:pd} have 
inefficiencies related to the coarse granularity of the decision system and the use of local
reconstruction; the reconstruction algorithms to build the physics objects are also subjugated to 
the blind zones of the detectors and also to the intrinsic nature of reconstructing hits and energy 
deposits to fit trajectories, assign energies or discriminate between high-level objects. 
This loss of information introduces inefficiencies on the trigger selection, identification, 
reconstruction, and to any stage needed to obtain high level objects from raw electronic data. 
These inefficiencies can be split by sources and be evaluated with probabilistic theory using 
simulation or data techniques. 

Besides the inefficiencies, the measures can also be biased due to subdetectors mis-alignments or 
wrong material description used by the reconstruction algorithms to scale some observables such
the transverse momentum or the energy of a reconstructed particle. Also, the multiple interactions
produced at each beam crossing, the so called~\emph{\gls{pileup}}, are going to affect the quality 
of the reconstruction. These effects can be observed and properly corrected. 

The corrections outlined above are described in this section. Nevertheless, there 
are more data corrections to be considered, but they are already included in the standard 
reconstruction of \CMSSW, so the interested reader can see them in more detail in the references
of Chapter~\ref{ch4}.

\subsection{Efficiencies: Tag and Probe method}\label{ch5:subsec:tap}
As we have emphasised, after the full reconstruction process there is no warranty that all the 
final particles passing through the detectors have been successfully detected and/or reconstructed.
Given the huge amount of variables which enter in the detection and reconstruction problems and 
the inherent lack of information, the efficiency problem is attacked from a probabilistic 
perspective in order to quantify the inefficiencies introduced by any procedure used in the data 
chain processing. The efficiency of some procedure can be defined as the probability that such a 
procedure does its job. If a frequentist approach is adopted, the probability can be interpreted as
frequencies and it is possible to evaluate the rate of "work well-done" by counting the output of 
the procedure with respect to the initial number of elements in which the procedure is applied. The
raising problem when dealing with data is the lack of knowledge of the initial objects, as this
is exactly what was lost so there is no information about it. Using simulation data, the objects
are always controlled, but then introduces dependence in the simulation model, inserting in turn
uncertainties in the measures. There is a tacitly generic \gls{data-driven}\glsadd{ind:data-driven} technique to evaluate
these efficiencies in order to be incorporated to the analysis: the \gls{tap}\glsadd{ind:tap} method.

The \gls{tap}\glsadd{ind:tap} method allows to select an unbiased sample of physics objects in data by exploiting
di-object resonances like \Z or \JPsi which can be used to measure some object efficiencies from data.
In brief, the resonance is reconstructed as pairs of objects (typically leptons) with one leg passing a 
tight identification 
(\emph{tag}) and the other one passing a loose identification (\emph{probe}). The tight leg "tags"
the event as an event containing two same flavour leptons just because it was possible to reconstruct
an invariant mass with the dilepton system around the resonance, and because the tight requirements
are assuring the identity of the tag lepton. Therefore, it is possible to check any property of the
probe lepton given that the event has to contain two leptons. The passing probes 
are defined in such a way according to whatever is the efficiency to measure. The efficiency is
then evaluated considering the ratio between passing probes over the total probes. 
\begin{equation}
	\varepsilon=\frac{N_{\text{passing probes}}}{N_{\text{probes}}}
	\label{ch5:eq:tap}
\end{equation}
Equation~\eqref{ch5:eq:tap} is barely used because of the background contamination
of the experimental data; instead a complex version of it is used taking into account the 
data contamination through a fitting procedure. Usually, the selected probe sample is contaminated 
by events which do not contain a resonance, these events should be subtracted in order not to bias 
the measurement including non-resonant events. This is accomplished by fitting a signal-background 
model to the dilepton invariant mass, separately, for the passing probes and tags, and the failing
probes and tags in order to calculate the efficiency as the ratio of the signal yields extracted 
from the fitting. The procedure is repeated in bins of some probe variables to compute efficiency 
histograms as function of those variables. In summary, the signal is modelled by a probability 
density function describing the invariant resonance ($S(x)$, $x$ is designing whatever dependency)
and two probability density 
functions are used to describe the background because of the different expected origin for the 
passing probes ($PB(x)$) and the failing probes, ($FB(x)$). The passing distribution is
modelled using the addition of signal and passing-probe background distributions, where at this
point the efficiency is introduced,
\begin{equation*}
	N_S\cdot\varepsilon\cdot S(x)+N_{PB}\cdot PB(x)
\end{equation*}
and the failing probes, using again the efficiency, is related with the failing probe background
distribution.
\begin{equation*}
	N_S\cdot(1-\varepsilon)\cdot S(x)+N_{FB}\cdot FB(x)
\end{equation*}
\begin{figure}[!htpb]
	\centering
	\includegraphics[width=0.8\textwidth]{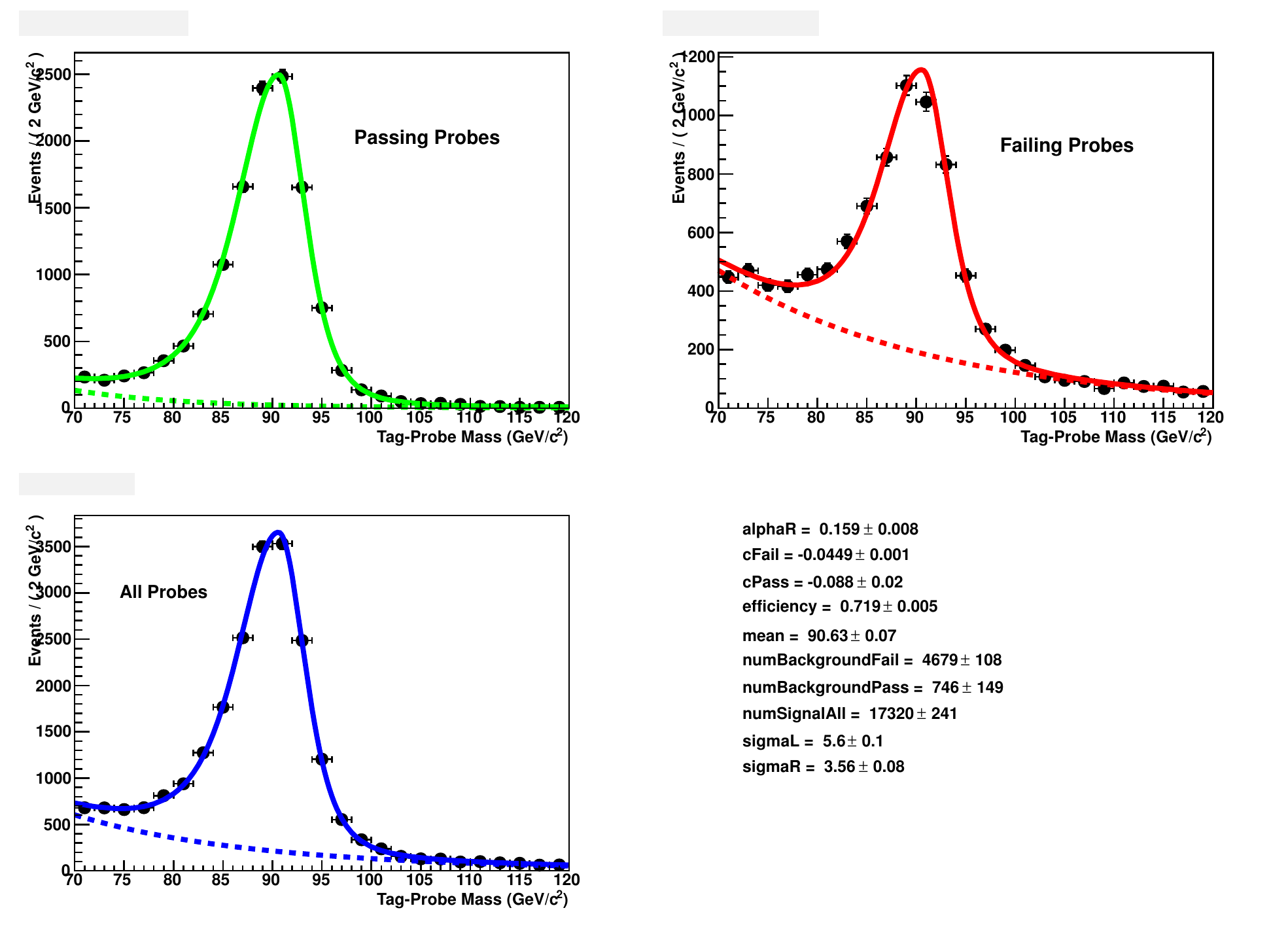}
	\caption[Fitting procedure for the tag and probe method]{Passing probe and failing probe
	simultaneously fit for the calculation of the reconstruction efficiency of electrons.
	The \Z mass peak and the background contribution is extracted from the fit. 
        The fit output is shown in the table, in particular the estimated efficiency.}\label{ch5:fig:tnp}
\end{figure}
A simultaneous fit is performed between both distributions getting as output of the fit the
efficiency $\varepsilon$ and other parameters such as the number of signal events $N_S$, the
number of passing background events $N_{PB}$, etc. Figure~\ref{ch5:fig:tnp} illustrates the 
fitting procedure. Electrons from the 2011 data period have been used to extract the reconstruction
efficiency. The fit is performed for a given pseudorapidity and \pt bin.

The efficiencies are not only calculated in data, the tag and probe method is also used in the 
simulated data and the same efficiencies than in the experimental data are obtained. As the 
simulation does not reflect exactly the material description of the detector, nor failures in the 
subdetector systems nor whatever inherent instrumental effects in the real data taking, there may be differences between
the efficiencies obtained with the experimental and the simulated data. These differences are taken
into account with the \emph{\acrlong{sf}[s]}\glsadd{ind:sf}, the ratio between the efficiency obtained with 
simulated versus experimental data. As will be seen in Chapter~\ref{ch8}, the analysis 
actually uses these scale factors instead of the efficiencies.

\subsection{Momentum and energy scales}\label{ch5:subsec:energyscale}
The momentum and energy measurement can be biased due the limited knowledge of the physical 
configuration of the detector\footnote{Mis-alignments of the 
subsystems and between them, mis-calibration of the calorimeters, \etc} and the limited 
capability of the
reconstruction algorithms. Uncertainties in the magnetic field should also be included
when the tracks are used to measure the momentum. The effects of these sources in the momentum
or energy measurements can be observed and therefore corrected, and the remaining systematic 
uncertainties are estimated after these corrections. There are several methods in \gls{cms}\glsadd{ind:cms} to 
accomplish this goal~\cite{CMS-PAS-EGM-10-004}~\cite{1748-0221-7-10-P10002} but all of them 
rely on the same strategy: use well-known resonances (\JPsi, $\Upsilon$, \Z) to correct the 
momentum (for muons) or energy (for electrons) scale. A bias in the (transverse) momentum of a
muon or in the energy of a electron can be detected building dimuon (dielectron) systems with
its invariant mass around a well-known resonance because the resonance mass peak will be
displaced from its nominal value. 
\begin{figure}[!hbtp]
	\centering
	\begin{subfigure}[b]{0.45\textwidth}
		\centering
		\includegraphics[width=\textwidth]{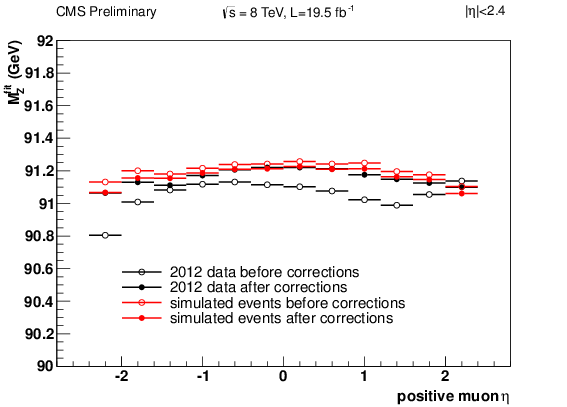}
	\subcaption{Positive legs}
	\end{subfigure}
	\begin{subfigure}[b]{0.45\textwidth}
		\centering
		\includegraphics[width=\textwidth]{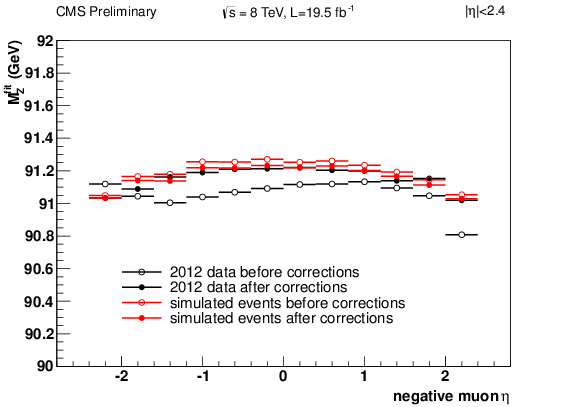}
	\subcaption{Negative legs}
	\end{subfigure}
	\caption[Muon momentum scale]{The dimuon mass distribution is fitted to extract the 
		\Z mass, $M_{Z}^{\ell\ell}$. The plots show the different $M_{Z}^{\ell\ell}$
		obtained when fitted in different pseudorapidity bins, before (empty dots) 
		and after (filled dots) applying the momentum scale correction. The black dots
		refer to data obtained during the 2012 run period whilst the red ones to 
	simulation data.}\label{ch5:fig:momentumscale}
\end{figure}
Using this fact it is possible to correct the momentum or energy 
and discover all the possible biases and understand their sources by plotting the invariant mass 
as a function of some sensitive kinematic variables. The correction is applied to the transverse
momentum of a muon or the energy of an electron by scaling its value, $\pt^c=(1-a(x))\pt$, where
$a(x)$ is referring to any possible variable dependence of the correction. 
Figure~\ref{ch5:fig:momentumscale} shows the fitted mean value for the invariant mass distribution
of a dimuon mass system with and without the momentum scale correction applied, with respect the 
pseudorapidity of the muons.

\subsection{Pileup effects}\label{ch5:subsec:pileup}
\begin{figure}[!hbtp]
	\centering
	\includegraphics[width=0.7\textwidth]{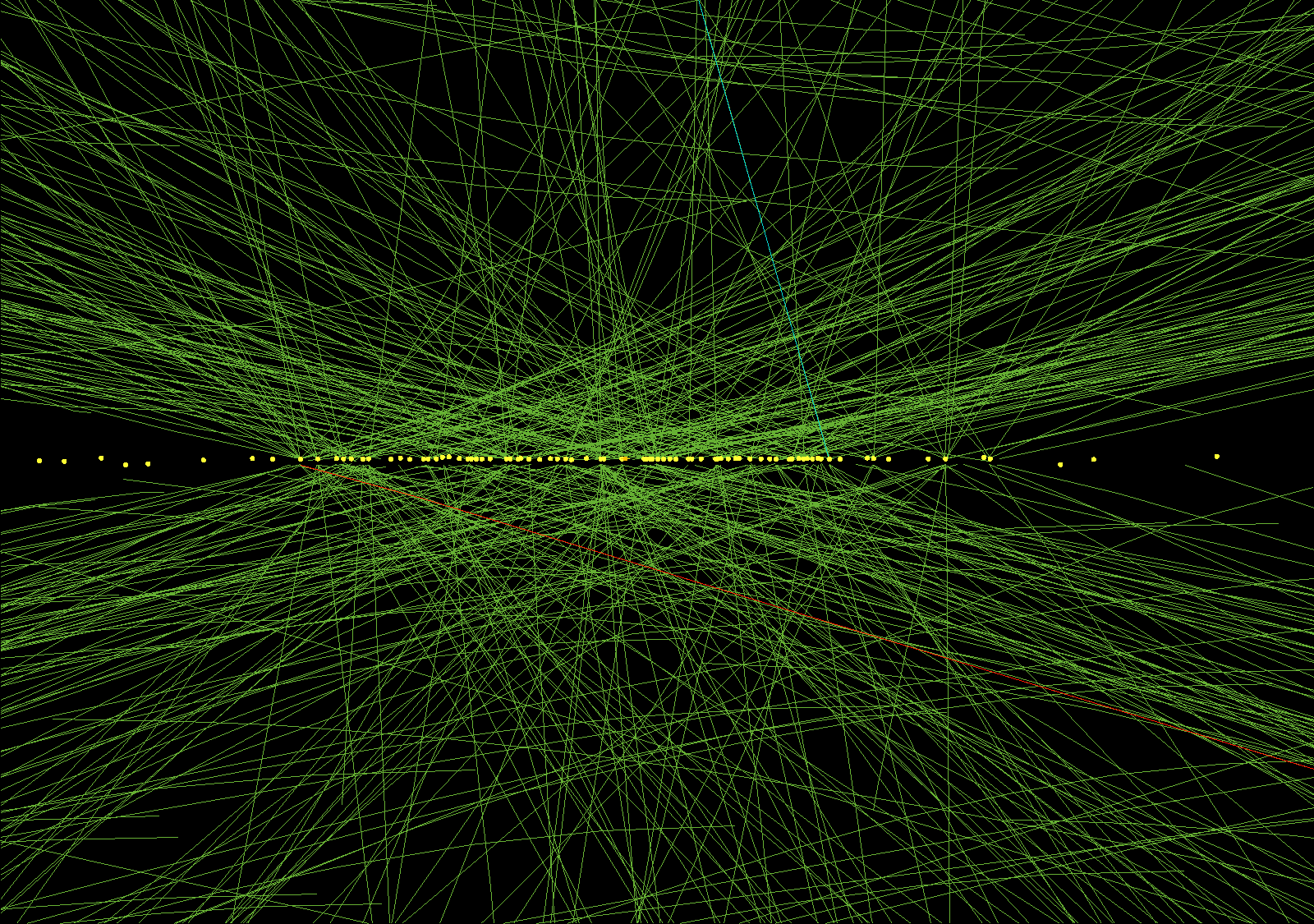}
	\caption[High pileup event]{Reconstructed event from the 2012 run period (extracted from a
	high-pileup run 198609) showing 78 reconstructed vertices (yellow dots) in one beam
	crossing. The event is 
	displayed in the $Rho-Z$ view, so the vertices can be seen along the beam pipe. 
	Following the track reconstruction (green, red and blue lines), the tracks are grouped into
	the vertices, each one representing a proton-proton collision.}\label{ch5:fig:pileupevent}
\end{figure}

The instantaneous luminosity reached in 2011 and 2012 along with the machine parameters used by the 
\gls{lhc} (see Chapter~\ref{ch3}) to accomplish this high rate of collisions make each proton bunch 
crossing highly likely to give more than one interaction. Indeed, dozens of collisions can occur
in the same bunch crossing, which the \gls{cms} detector records them 
as the same event; they are "piled up" together with the hard scattering. This secondary collisions 
are known as \emph{\gls{pileup}\glsadd{ind:pileup}} events.
\begin{figure}[!htpb]
	\centering
	\includegraphics[width=\textwidth]{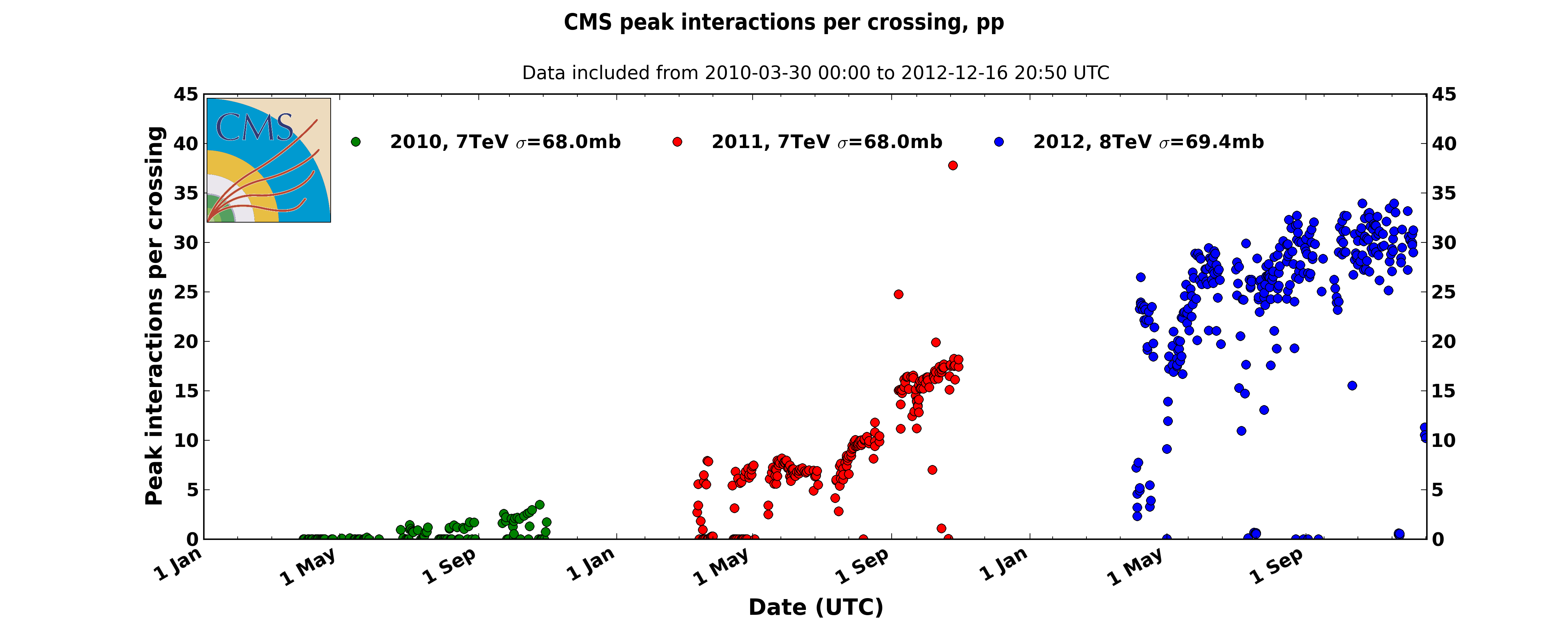}
	\caption[Peak number of interactions per beam crossing, 2011 and 2012]{Peak number of 
		interactions per beam crossing for the 2010 (green), 2011 (red) and 2012 (blue) 
		data. The mean number of interactions integrated along the full data-period for 
		2011 is 9.1, and for 2012 is 20.7}\label{ch5:fig:peakPU}
\end{figure}
Figure~\ref{ch5:fig:peakPU} shows the peak number of 
interactions per beam crossing during the 2010, 2011 and 2012 run periods. An important 
increment in the number of interactions can be observed along the two last periods, giving a mean 
number of interactions per beam crossing about 9 for the 2011 run period, and around 21 for 2012.
Therefore, besides the hard interaction, the event is populated with particles coming from 
secondary interactions, mostly soft, which contaminate the primary hard scattering and complicate 
the reconstruction process. A typical event with pileup is characterised by many primary 
vertices along the beam line, as can be seen in Figure~\ref{ch5:fig:pileupevent}.
Multiple overlapping interactions lead to an enhanced detector occupancy
which are almost saturated by the particles produced from different vertices. A direct consequence 
is that the number of jets is much increased as well as the density of the reconstructed tracks and
the mean energy deposited in the detector. Calorimetric measurements result particularly sensitive
to such conditions, biasing the isolation and identification of the objects. 

\begin{figure}[!htpb]
	\centering
	\includegraphics[width=0.8\textwidth]{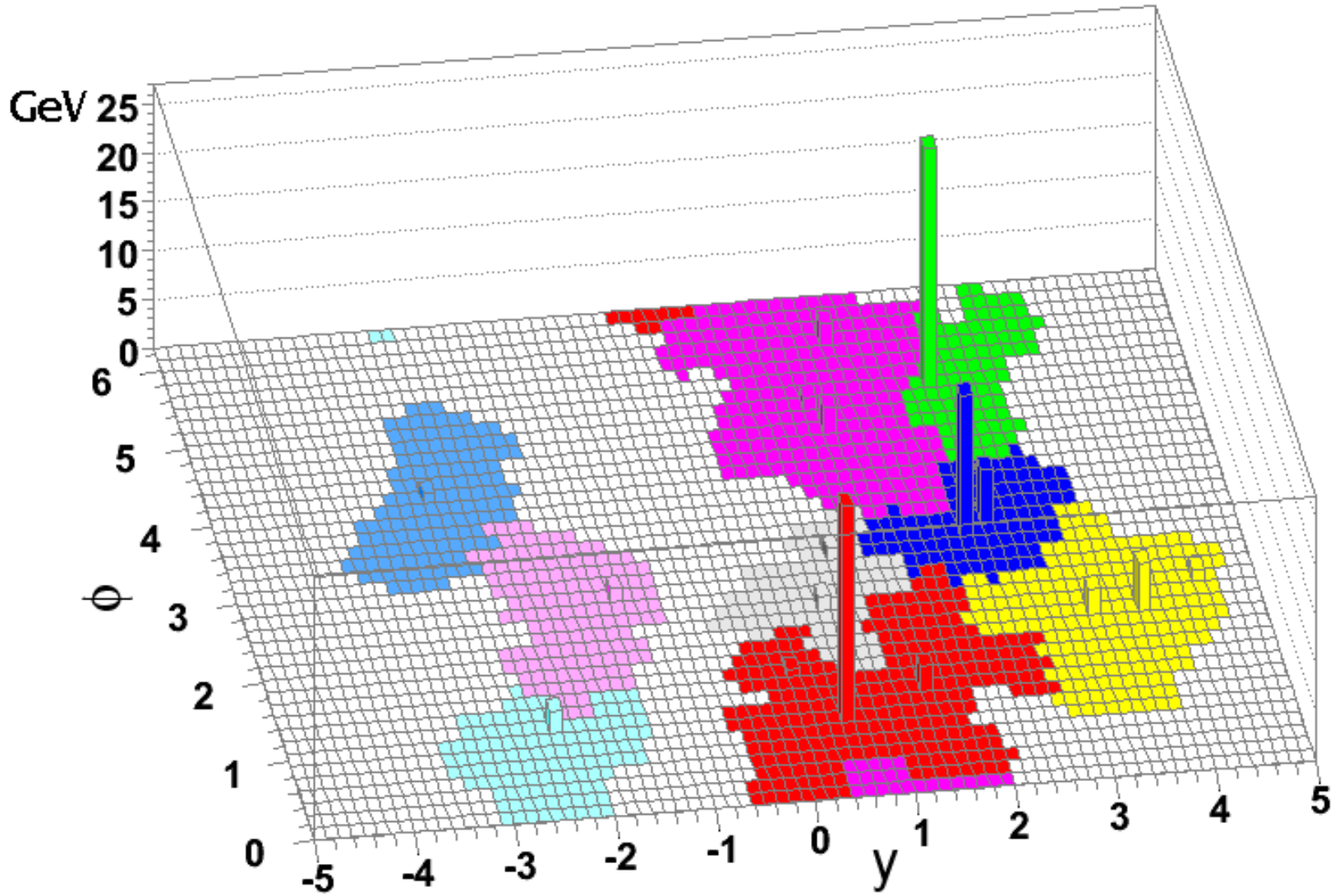}
	\caption[Jet area defined by \FASTJET algorithm]{Calorimeter's energy deposits in the $y-\phi$ plane.
	Each colour is associated to a different clustered jet. The $k_t$ algorithm was used to 
	cluster the energy deposit cells and the \FASTJET algorithm to associate to each jet
        an area, represented with the same colour of the jet.}\label{ch5:fig:jetarea}
\end{figure}

The net effect of the pileup events presence in the main event is that the measured energy of the 
jets is overestimated. This is because particles coming from vertices different from the hard 
scattering primary vertex contribute with tracks or energy deposition in the calorimeters, leading 
to an increase of the total energy measured within the jet cone. Therefore, to correct these effects 
an event-by-event and jet-by-jet treatment is applied to the event. The main algorithm used in 
\gls{cms} to correct for the measured energy is the \FASTJET~\cite{Cacciari:2007fd} algorithm, that
works correctly for any infrared safe jet reconstruction algorithm, and estimates the energy 
contribution due to \gls{pileup} for each reconstructed jet which can then be subtracted from the 
jet's energy to yield a result which more closely represents the energy of the initiating parton. 
The algorithm introduces an "abstract" area for each jet in order to account for the energy density of
the jet. However, a jet consists of point-like particles which themselves have no intrinsic area, 
therefore it defines a sensible area by adding additional, infinitely soft particles (called 
ghosts) and identifies the region in rapidity and $\phi$ where the ghosts are clustered with a 
given jet. The extent of this region gives a measure of the dimensionless jet area, $A$.
Figure~\ref{ch5:fig:jetarea} shows several jets clustered using the $k_t$ algorithm; the 
reconstructed jets are associated to the high energy deposits in the calorimeter's cells
and the shaded areas surrounding them were constructed in the way described above. 
This area is then a measure of the jet's susceptibility to pileup contamination. The algorithm
provides also a parameter measuring the overall level of diffuse noise\footnote{The \gls{pileup},
although it could be some extent of the underlying event (see section~\ref{ch5:subsec:ue} for 
definition).} $\rho$ (in \GeV/Area) as the median value of $\pt^{jet}/A^{jet}$ taken over all jets.
The jet energy correction is done by subtracting from the measured energy of the jet the corresponding
$\rho\cdot A^{jet}$.

As the \FASTJET algorithm extracts the energy density per unit area $\rho$ due to the 
contribution of pileup and underlying events, it can be used to correct the sensitive quantities 
related with the reconstructed objects, in particular isolation and identification of leptons, and
the particle flow \MET object which is built from the event's jets and other particles. The 
\MET then uses directly the jet energy correction when it is calculated using jets corrected for 
\gls{pileup} events. Moreover, the isolation of the muon and electron can also take advantage of 
the $\rho$ estimation to subtract it in the isolation cones defined for the leptons. The specific
pileup correction applied to the lepton objects used in this analysis is detailed in 
Chapter~\ref{ch6}.

\section{The Monte Carlo approach for event simulation}\label{ch5:sec:simulation}
Any physics experiment is conceived to increment the knowledge on the particular field in which
the experiment is designed. To that end, most of the time the outcome of the experiment is compared
with theoretical predictions based on physical models which try to describe the
studied phenomenon. In high energy particle collisions, the physics involved is currently described
by the \gls{sm} of particle physics, as has been established in Chapter~\ref{ch1}. The 
theoretical predictions offered by the \gls{sm} are mainly particle production rates (or cross
sections), \ie probabilities for a particular process to occur. The comparison of that kind of 
theoretical predictions with the experimental data is, then, intrinsically statistical. The task to map 
the cross section for various processes onto the discrete event structure of the 
experimental data is extremely challenging. Typically hundreds of particles are produced in every
high energy collision and all the species of the \gls{sm} and maybe some beyond, are involved with
momentum ranges that spread along several orders of magnitude. Furthermore, theoretical calculations in
\gls{qcd} processes involve the intrinsically non-perturbative and unsolved problem of confinement
(see Section~\ref{ch1:sec:physicsathadroncolliders}). These particular, a priori, intractable 
problems have been overcome with a wide range of techniques based on \gls{mc}\glsadd{ind:mc} 
simulation~\cite{Metropolis1949Monte}. Roughly speaking, a random number generator is interfaced 
with the equations governing a certain process in order to produce a large number of simulated
collision events. This is done through highly specialised software called \emph{event generators}.
The event generators produce as well the decays of unstable particles that do not escape from the 
detector. These particles are the outcomes of the event generator, apart from the history
of all the decay chain, and their passage through the detector has to be simulated. The simulation 
step mimics the energy deposits, hits and any other material response due to
the passage of the particles through the subdetectors, and again \gls{mc}\glsadd{ind:mc} methodology is applied 
to deal with it. The standard package used by almost the entire community in particle physics
and other areas is the \GEANTfour~\cite{Allison:2006ve} framework\footnote{In this work, we 
are not going to describe the physics and methodology involving the simulation stage. We point
the interested reader to the cited reference and in particular the web page of the package where 
additional and very complete information can be found.}. In order to obtain the same 
electronic signals produced by the read-out of the real detector, the emulation of the data 
acquisition system and read-out is also performed. 

The simulation of collision events is divided in several steps by which event generators, usually
different programs for each step, build up the hadron-hadron collision involving a hard
scattering process of interest. The basic phases of the process that needs to be simulated are
the primary hard subprocess (\emph{hard scattering}), parton showers associated with the incoming
and outgoing coloured participants in the subprocess (\emph{parton showering}), 
non-perturbative interactions that convert the showers into outgoing hadrons and connect them to 
the incoming beam hadrons (\emph{hadronisation}), secondary interactions that give rise to the 
\emph{\gls{ue}\glsadd{ind:ue}}, and the decays of unstable particles which are fed to the detector 
simulator. Not all stages are relevant in all processes. In particular, 
the soft \gls{qcd} type, which are the most produced in hadron-hadron collision, rely on 
phenomenological models and make use of the \gls{pdf}\glsadd{ind:pdf}\footnote{As was seen in 
Section~\ref{ch1:sec:physicsathadroncolliders}}. The main stages of the generation of a 
hadron-hadron collision event are briefly described below without entering in full details. The 
interested reader can found a deeper development of the subject in~\cite{Buckley:2011ms}.

\subsection{Hard scattering process}
The hard scattering process involves large momentum transfers between the implicated particles.
These are the processes in which most of the analyses carried out at the detectors of the \gls{lhc}
are interested, and these are the processes able to produce heavy particles or jets with high
transverse momenta. The cross section for a scattering subprocess $ab\rightarrow n$, where
$n$ denotes the number of final state particles, at hadron colliders can be computed in collinear 
factorisation through~\cite{Ellis:1991qj} 
\begin{eqnarray}
\sigma_{h_1h_2\rightarrow n}
&=&
\sum\limits_{a,b}\,\int\limits_{0}^{1}\text{d}x_a\text{d} x_b\,\int\,
f^{h_1}_{a}(x_a,\mu_F)f^{h_2}_{b}(x_b,\mu_F)\,
\text{d}\hat\sigma_{ab\to n}(\mu_F,\mu_R)\notag\\
&=&
\sum\limits_{a,b}\,\int\limits_{0}^{1}\text{d} x_a\text{d} x_b\,\int\text{d}\Phi_n\,
f^{h_1}_{a}(x_a,\mu_F)f^{h_2}_{b}(x_b,\mu_F)\label{ch5:eq:master4xsec}\\
&&\qquad\times\frac{1}{2\hat s}|\mathcal{M}_{ab\to n}|^2(\Phi_n;\mu_F,\mu_R)\,,
\nonumber
\end{eqnarray}
where
\begin{itemize}
      \item $f^{h}_{a}(x,\mu)$ are the \glspl{pdf}, which depend on the momentum fraction 
      $x$ of parton $a$ with respect to its parent hadron $h$, and on the factorisation 
      scale $\mu_F$;
\item $\hat\sigma_{ab\to n}$ denotes the parton-level cross section for the 
      production of the final state $n$ through the initial partons $a$ and $b$.
      It depends on the momenta given by the final-state phase space $\Phi_n$, 
      on $\mu_F$ and on the renormalisation scale $\mu_R$.
      The fully differential parton-level cross section is given by the product 
      of the corresponding matrix element squared, averaged over initial-state 
      spin and colour degrees of freedom, $|\mathcal{M}_{ab\to n}|^2$, and the 
      parton flux $1/(2\hat s) = 1/(2x_ax_bs)$, where $s$ is the hadronic 
      centre-of-mass energy squared.
\item $\text{d}\Phi_n$ denotes the differential phase space element over the $n$ 
      final-state particles.
\end{itemize}
The \glspl{pdf} carry the non-perturbative \gls{qcd} part, which describe the probability of
a parton to have some fraction of the total hadronic momentum; these functions are extracted 
experimentally (see Section~\ref{ch1:sec:physicsathadroncolliders}). The parton-level probabilities
are contained in the matrix element squared $|\mathcal{M}_{ab\to n}|^2(\Phi_n;\mu_F,\mu_R)$ and can
be evaluated using Feynman diagrams. All multi-purpose event generators provide a comprehensive 
list of \gls{lo} matrix elements and the corresponding phase-space parametrisations for $2\to1$,
$2\to2$ and some $2\to3$ production channels in the \gls{sm} and some of its new physics 
extensions. Some event generators provide \gls{nlo}\glsadd{ind:nlo} or even \gls{nnlo}\glsadd{ind:nnlo}. Moreover there are a
wide variety of event generators specialised in particular processes or family of processes. A 
couple of examples of event generators used in this work are \PYTHIA~\cite{Sjostrand:2006za} and
\MADGRAPH~\cite{Alwall:2011uj}.

\subsection{Parton showering}
The hard scattering interaction is well described using perturbative \gls{qcd} due to the notion of 
asymptotic freedom in strong interactions. Nevertheless, to give an inclusive picture of the process,
including the internal structure of the jets and the distributions of accompanying particles, any 
fixed order, as is used in the matrix elements, is not sufficient. The effect of all higher orders
can be simulated through a parton shower algorithm, which is typically formulated as an evolution in 
momentum transfer down from the high scales associated with the hard process to the low scales, of 
order 1 GeV, associated with confinement of the partons it describes into hadrons. 
In summary, scattered, annihilated and created partons radiate gluons, and as gluons themselves 
are coloured, this radiation give rise to further gluon radiation and parton multiplication. The
radiation can be produced before the parton scattering or annihilating process, the so 
called \gls{isr}\glsadd{ind:isr}, or after, called 
\gls{fsr}\glsadd{ind:fsr}. The parton shower algorithms will provide
the partonic final state to be added to the final state products of the hard scattering. At that 
point, the interaction scale of the partons has fallen during the parton showering, 
eventually initiating the process of \emph{hadronisation}, \ie the partons are bound into colourless
hadrons (see Subsection~\ref{ch5:subsec:had}).

\subsection{Underlying event}\label{ch5:subsec:ue}
Besides the gluon radiation, several parton-parton interactions can occur within a single 
hadron-hadron collision and can be modelled by \emph{multiple parton interactions} which can 
produce additional observable jets. Most multiple parton interactions are relatively soft, however,
and do not lead to easily identifiable additional jets. Instead, they contribute to building up the
total amount of scattered energy and cause colour exchanges between the remnants, thereby 
increasing the number of particles produced in the hadronisation stage. This additional activity
is known as the \emph{underlying event}.

\subsection{Hadronisation}\label{ch5:subsec:had}
The outcome of the different algorithms described above to populate and describe a 
hadron-hadron collision is obtained as coloured partons carrying some momentum and energy, \ie
the partonic final state. Nevertheless, the partons are coloured so they are going to bind into
colourless hadrons. The description of this process is done with phenomenological models due to the 
non-perturbative nature of the problem\footnote{The only available rigorous approach, 
lattice-\gls{qcd}, is formulated in Euclidean space-time, whereas time evolution of partons into
hadrons is inherently Minkowskian.}. The event generators usually implement the Lund string 
model~\cite{Andersson:1983ia}, in which quarks are bound together with a gluon string. For quarks
travelling away from one another, this string becomes stretched and stores energy, eventually 
breaking to produce new $q\bar{q}$ pairs when an energy threshold is reached. The process
is repeated until the energy available is below a threshold. The initial partons have been 
progressing to a collection of colourless bound states. These resulting hadrons are typically 
collimated along the direction of the initial hard parton, forming a coherent jet of particles.

\subsection{Pileup}
The \gls{pileup}\glsadd{ind:pileup} contribution (see Section~\ref{ch5:subsec:pileup}) to a hard scattering event
is copied in simulation by superimposing some number of simulated soft interaction events
on top of each nominal event, following the interaction multiplicity distribution observed in the
experimental data. This distribution is used to re-weight the simulated data to produce the exact 
data distribution.

\begin{figure}[!htpb]
	\centering
	\includegraphics[width=\textwidth]{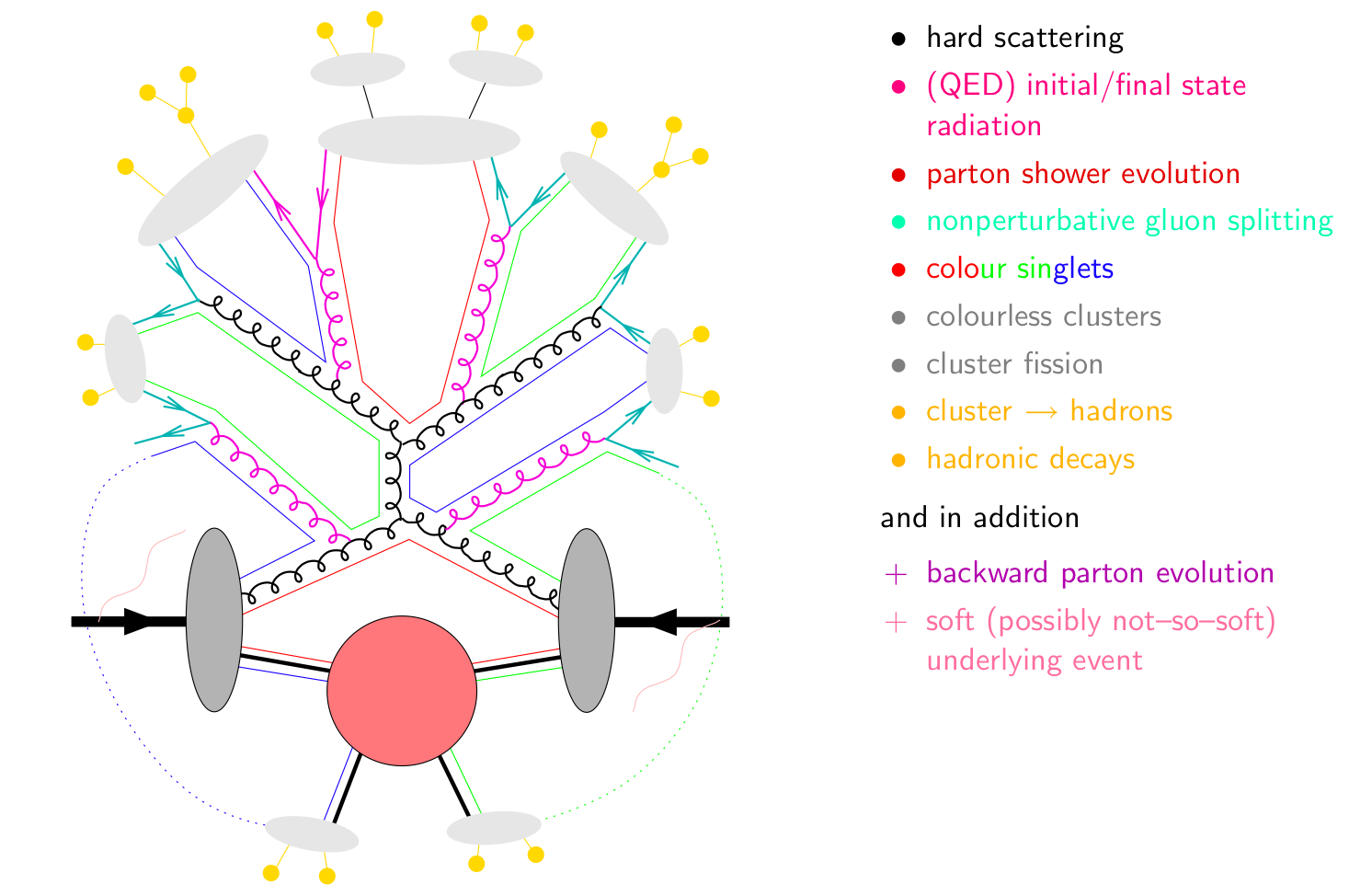}
	\caption[Proton-proton event at LHC]{Schematic proton-proton collision showing a 
	gluon-gluon hard scattering. The grey blobs are the incoming protons, the red blob
	represents the multiple interactions giving the underlying event. The figure was
	extracted from 
        \url{http://www.gk-eichtheorien.physik.uni-mainz.de/Dateien/Zeppenfeld-3.pdf}.
        }\label{ch5:fig:eventLHC}
\end{figure}

\paragraph{}
The full process to simulate a hard scattering event at \gls{lhc} starts by defining the process 
we want to generate defining a particular equation~\eqref{ch5:eq:master4xsec}. With a random number 
generator, the chosen \gls{pdf} is sampled in order to determine the initial momentum of the 
partons. The probability for a proton to radiate a photon before the collision is also 
considered (\gls{qed}\glsadd{ind:qed}-\gls{isr}\glsadd{ind:isr}\glsadd{ind:isr}) modifying the initial available proton energy. Then, the random 
generator is used again to sample the differential cross section of
equation~\eqref{ch5:eq:master4xsec}, defining momenta for the final state particles. Once the final 
state particles from the hard scattering are obtained, the \gls{isr}\glsadd{ind:isr}\glsadd{ind:isr} and \gls{fsr}\glsadd{ind:fsr}\glsadd{ind:fsr} are modelled by 
a parton shower algorithm which increases the gluon and quark multiplicity of the event,
adding new partons to the final state. In parallel, a number generator is used again to select the
number of interactions which occurred in the same bunch crossing, adding more particles to the event. The
full partonic content is matched to be able to hadronise. Finally, all the final particles are 
evolved into stable or unstable particles ready to be detected by the detector subsystems.
Figure~\ref{ch5:fig:eventLHC} shows a full proton-proton event with the main stages described here.
After the event generation, the final particles outcoming from the \gls{mc}\glsadd{ind:mc} programs are introduced
to the \GEANTfour simulator whence the particles passing through the whole \gls{cms} are simulated. 
The simulator returns a set of energy deposits and hits in the sensitive detectors which in turn 
are sent to a read-out emulator to obtain the final output as it is obtained in a real \gls{cms} 
collision.

\chapter{Selection of \WZ events}\label{ch6}
The chapter describes the event topology and expected signature of the \WZ production decaying 
leptonically and establishes our signal definition. The possible sources of noise, the backgrounds, 
are determined and it is examined how they can mimic our signal in the \gls{cms} detector. The 
selection strategy of muons and electrons is analysed thoroughly, detailing the several approaches
of lepton identification and isolation, and also the online data selection. For each of the 
selection processes, their efficiencies are calculated and presented. The final step is focused 
in the analysis strategy which is based in sequential cuts to suppress the remnant background 
achieving high signal purity. In this context, the observables used and their role in the analysis
are described. Several distributions are shown to control the event selection and illustrate the 
logic of the analysis strategy.

\section{The signal signature and backgrounds}\label{ch6:sec:topology}
The $\WZ\rightarrow\ell\nu\ell^{'+}\ell^{'-}$ decay\glsadd{ind:decay} 
\gls{signature}\glsadd{ind:signature} is defined by two opposite-charged,
same-flavour, high-\pt, isolated leptons, whose invariant mass is compatible with the one of the \Z 
boson, together with a third high-\pt, isolated lepton and a significant deficit of transverse
energy, \MET, associated with the escaping neutrino.
\begin{figure}[htpb]
	\begin{center}
	 \input{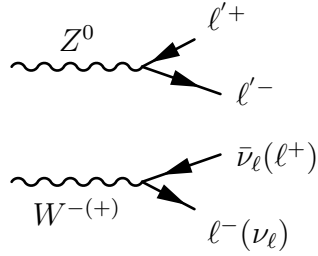}
	\end{center}
	\caption[\WZ leptonic decay]{Schematic diagram for the leptonic decay of the \WZ process. The
		event is characterised by a couple of high-\pt, opposite-charged, same-flavour and
		isolated leptons decaying from the \Z boson besides a third high-\pt, isolated 
		lepton from the \W. A significant amount of \MET is expected due to the undetected
	        neutrino from the \W.}\label{ch6:fig:decayWZ}
\end{figure}
The same signature can be obtained from other high-energy processes introducing noise to the signal
\WZ process we want to measure; these noise processes are called \emph{backgrounds}. All 
the production processes leading to three high-\pt leptons in the final state should be categorised and
studied in order to suppress them or, if it is not possible, to control them. Such processes are called
the \emph{physical backgrounds} in opposition with the \emph{instrumental} background. This later 
background appears due to inefficiencies in the detection and/or reconstruction and to the limited 
coverage of the detector (\emph{detector acceptance}); spurious leptons are reconstructed
from jets or other mis-identified particles transforming an event which originally did not contain
three leptons into one that does. The main physical background is the $ZZ$ production process in 
its full leptonic decay, when one lepton is either no-reconstructed or falling outside
the detector acceptance. The trilinear boson couplings are also physical backgrounds but due to
their low production cross section they should only be considered when the accumulated data are large
enough. These background have been considered in the 8~\TeV analysis, however even with the large
amount of data (19.6~\fbinv) collected they barely contribute. The instrumental sources are 
summarised below:
\begin{description}
	\item[QCD] production processes related with \gls{qcd}, where final state
		particles are hadrons, therefore jets. This background is 
		generically called \gls{qcd}. Three high-\pt, isolated spurious 
		(\emph{fakes}) leptons should be reconstructed making it unlikely. Despite of
		the huge production rate of \gls{qcd} processes, their contribution in this
		analysis is negligible.
	\item[W+jets] the \W production in association with jets has a very high production rate
		(around 31.3~nb). The \W hadronic decay can be ignored as it needs three fake
		leptons (and the production rate is much lower than \gls{qcd} backgrounds). The
		leptonic decay provides a true high-\pt, isolated lepton, therefore two fakes
		leptons are required to mimic the signal topology. This is, again, very unlikely
		and the contribution is negligible.
	\item[Z+jets] the production of $Z/\gamma^*$ in association with jets is going to be relevant in
		the leptonic decay of the \Z. Moreover, unlike the other two backgrounds described
		above, this process contains a genuine \Z in the final state making more difficult
		to reject it through an invariant mass requirement. Thus, the process only needs one
		fake lepton and some \MET to present the same signature as \WZ.
	\item[Single top] a top quark can be produced in association with a \W boson. When the 
		top quark decays through the weak force, it decays almost exclusively to a \W boson
		and a bottom quark. The bottom quark can eventually decay leptonically, thus it is 
		possible to find a third lepton, although not isolated, together with the two 
		leptons	from the $W$s. The lepton coming from the quark is considered as a fake 
		lepton because it should not pass the isolation criteria. The \W and top quark 
		hadronic decays can contribute also when the jets are mis-identified as leptons, 
		however the low probability of reconstructing more than one fake lepton (as we will
		see in Chapter~\ref{ch7}) together with the topology of the single top production
		makes these contributions negligible.
	\item[$\boldsymbol{\ttbar}$] As the single top case, the weak decay of the top (anti-top) quark 
		produces $W^-$ ($W^+$) boson and anti-bottom (bottom) quark. Therefore, two 
		high-\pt, opposite-charged, isolated leptons and \MET from the $W$s are going
		to be present in the final state signature, together with the non-isolated 
		lepton from the quarks or any properly fake lepton. This background source 
		together with the $Z+jets$ are the dominant instrumental backgrounds, as it 
		will see in chapters~\ref{ch7} and~\ref{ch8}.
	\item[$\mathbf{WW}$] the \W diboson production in its leptonic decay mimics the topology of the
		signal events by acquiring a third fake lepton from the underlying or pileup events.
		Given that the third lepton is not reconstructed from high energy jets and the
		absence of a genuine \Z, this background is easily contained.
	\item[$\mathbf{Z}\boldsymbol{\gamma}$] The \Z production can be accompanied by a initial 
		state radiation, where the photon is produced by the incoming partons, or a final
		state radiation, where the photon is radiated by one of the charged leptons from 
		the \Z decay. A photon conversion into leptons (mostly electrons) can be produced 
		when the photon interacts with the detector, giving the third missing lepton to 
		complete the signal signature.
\end{description}

The leptonic decay of the \WZ process consider four possible final states. The \W muonic decay
and \Z muonic and electronic decays, and \W electronic decay and also \Z muonic and electronic 
decays. This four possibilities with final states defined by its lepton presence ($3\mu$, $2\mu1e$,
$1\mu2e$ and $3e$) allow us to split the \WZ final state into four independent and exhaustive
channels and use them to analyse independently the \WZ process.

\section{Online selection of \WZ candidate events}\label{ch6:sec:onlineselection}
The \gls{lhc} delivered proton-proton collisions at 7~\TeV of centre of mass energy during the year
2011 splitting the data taking into two major \emph{run periods}, A and B, separated by a short 
technical stop. The next year, 2012, the centre of mass energy of the collisions reached the 8~\TeV
and there were four major run periods (A, B, C and D) again separated by technical stops. During
this period, \gls{cms} recorded 5.56~\fbinv of $pp$ collision data for 2011 and 21.79~\fbinv for
2012. During the data taking, each subdetector of \gls{cms} experiences some amount of downtime
due to equipment failures, meaning that some fraction of the recorded luminosity cannot be used for
general analyses which rely on the integration of the full detector. Consequently, the 
collaboration certifies a list of runs suitable for physics publication, which for 2011 was 
an integrated luminosity of 4.9~\fbinv and for 2012, 19.6~\fbinv.

As it was explained in Chapter~\ref{ch5}, the events selected by the \gls{gltrigger}\glsadd{ind:gltrigger} system
are sorted in \glspl{pd} based on trigger paths. The topology of the \WZ process in its leptonic 
decay makes suitable the use of trigger paths looking for at least two high-\pt leptons, in this
analysis we consider the \verb/DoubleMu/ and \verb/DoubleElectron/ \glspl{pd} where events must fire
a trigger looking for a pair of muons or a pair of electrons, respectively. The 2012 period includes
also the dataset selected with events with at least one muon and one electron, the \verb/MuEG/ 
\gls{pd}\glsadd{ind:pd}. To control the recorded event rate, each of these triggers imposes momentum thresholds on 
the candidate objects (amongst some loose object identification and isolation requirements), with 
these thresholds increasing as the instantaneous luminosity increases. These \gls{hlt}\glsadd{ind:hlt} paths 
are each seeded by a L1 trigger path requiring one or two low-level detector 
objects with thresholds lower than those imposed at higher levels. Some loose quality cuts in 
identification and/or isolation are also imposed on the candidates, cuts which must become stricter 
in the analysis level to avoid a biased analysis. Tables~\ref{ch6:tab:hlt2011} 
and~\ref{ch6:tab:hlt2012} give the value of these thresholds corresponding to various run ranges 
for 2011 and 2012, respectively.

\begin{table}[htpb]	
	\centering
	\begin{subtable}[b]{0.6\textwidth}
		\centering
		      \resizebox{\textwidth}{!}
		      {
			\begin{tabular}{c cc c cc}\hline\hline
				Run Range     & \multicolumn{5}{c}{DoubleElectron (\ET)} \\\hline
				  & \multicolumn{2}{c}{L1} &  & \multicolumn{2}{c}{HLT}\\\cline{2-3}\cline{5-6}
			160329-170901          & 12 & 0 &  & 17 & 8 \\
			171050-EndYear         & 12 & 5  &  & 17 & 8 \\\hline
			\end{tabular}
		      }
		      \caption{Double Electron trigger paths}\label{ch6:tab:hlt2011DE}
	      \end{subtable}\vskip 1em
	\begin{subtable}[b]{0.6\textwidth}
		\centering
		      \resizebox{\textwidth}{!}
		      {
			\begin{tabular}{c cc c cc}\hline\hline
				Run Range     & \multicolumn{5}{c}{DoubleMuon (\pt)} \\\hline
				  & \multicolumn{2}{c}{L1} &  & \multicolumn{2}{c}{HLT}\\\cline{2-3}\cline{5-6}
			160329-165208          & 3 & 3     &  & 7 & 7 \\
			165364-178419          & 3 & 3  &     & 13 & 8 \\
			178420-EndYear         & 3 & 3  &     & 17 & 8 \\\hline
			\end{tabular}
		      }
		      \caption{Double Muon trigger paths}\label{ch6:tab:hlt2011DM}
	\end{subtable}
	\caption[Online selection trigger for 2011]{L1 and HLT energy and momentum thresholds for the 
	PDs of 2011 data period. The \emph{EndYear} label is referring to the last run 
	available before the end of the	data taking period. Besides the energy thresholds, the trigger
	paths are also requiring some loose quality cuts.}
	\label{ch6:tab:hlt2011}
\end{table}

\begin{table}[htpb]	
	\centering
	\begin{subtable}[b]{0.6\textwidth}
		\centering
		      \resizebox{\textwidth}{!}
		      {
			\begin{tabular}{c cc c cc}\hline\hline
				Run Range     & \multicolumn{5}{c}{DoubleElectron (\ET)} \\\hline
				  & \multicolumn{2}{c}{L1} &  & \multicolumn{2}{c}{HLT}\\\cline{2-3}\cline{5-6}
			190456-EndYear          & 13 & 7 &  & 17 & 8 \\\hline
			\end{tabular}
		      }
		      \caption{Double Electron trigger paths}\label{ch6:tab:hlt2012DE}
	\end{subtable}\vskip 1em
	\begin{subtable}[b]{0.6\textwidth}
		\centering
		      \resizebox{\textwidth}{!}
		      {
			\begin{tabular}{c cc c cc}\hline\hline
				Run Range     & \multicolumn{5}{c}{DoubleMuon (\pt)} \\\hline
				  & \multicolumn{2}{c}{L1} &  & \multicolumn{2}{c}{HLT}\\\cline{2-3}\cline{5-6}
			190546-196027          & 10 & 0    &  & 17 & 8 \\
			196046-EndYear         & 10 & 0 OR 3.5  &     & 17 & 8 \\\hline
			\end{tabular}
		      }
		      \caption{Double Muon trigger paths}\label{ch6:tab:hlt2012DM}
	\end{subtable}\vskip 1em
	\begin{subtable}[b]{0.6\textwidth}
		      \centering
		      \resizebox{\textwidth}{!}
		      {
			\begin{tabular}{c cc c cc}\hline\hline
				Run Range     & \multicolumn{5}{c}{MuEG (\pt,\ET)} \\\hline
				  & \multicolumn{2}{c}{L1} &  & \multicolumn{2}{c}{HLT}\\\cline{2-3}\cline{5-6}
			190456-EndYear        &  12 &   7     &   & 17 & 8 \\
			190456-196027         &  0  &   12    &   & 8 & 12 \\
			196046-EndYear        & 0 OR 3.5 & 12  & & 8 & 12 \\\hline
			\end{tabular}
		      }
		      \caption{Muon-Electron trigger paths thresholds. The first column for L1 and HLT is
		      referring to the muon object whilst the second column to the electron.}\label{ch6:tab:hlt2012MuEG}
	\end{subtable}
	\caption[Online selection trigger for 2012]{L1 and HLT energy and momentum thresholds 
	for the PDs of 2012 data period. Besides the energy and momentum thresholds, the trigger paths are 
	also requiring some quality cuts. In particular, the 0 value of some muon triggers implies that 
	there is no momentum requirement for the muon, but just the quality requirements.}
	\label{ch6:tab:hlt2012}
\end{table}

Besides the signal trigger paths to select events suitable for the \WZ analysis, we have selected 
other trigger paths, called \emph{utility trigger}, used to perform tag and probe efficiencies. 
These trigger paths use further requirements for one of the leptons but very loose on the other, 
which is going to be used to calculate the efficiencies. 

\subsection{Trigger efficiencies}\label{ch6:subsec:triggerEff}
As it was explained in Section~\ref{ch5:sec:datacorr}, the data selected by the
trigger system are inherently biased to favour certain types of physical processes and, therefore, 
it is mandatory to take into account this bias. The usual way to proceed is by calculating the 
trigger efficiencies,~\ie the probability that given an object which should have been fired the 
trigger path, this trigger path was actually fired. There are several approaches, some involving 
simulation data and others just involving data from the experiment. The simulation data approach 
uses data samples simulated with the \gls{mc} techniques\footnote{Henceforth throughout this work, 
the data samples generated using \gls{mc} techniques may just be called \emph{\gls{mc} samples}, 
despite of the abuse of language.} described at ~\ref{ch5:sec:simulation}; 
also including the trigger logic and the trigger paths. Then, the simulated data will also lost
some events due to trigger inefficiencies and can be compared directly with the experimental data.
The drawback of this method is the high dependence on the detector simulation which in turn has
to be taken into account too. Besides, the trigger paths are evolving rapidly (defining new paths,
prescaling others,~\etc) because of the continuously changing conditions of data taking (increasing 
of instantaneous luminosity, increasing pileup, detector equipment failures,~\etc); as meaning 
that the \gls{mc} samples and the experimental data samples do not share exactly the same trigger
paths. The calculation of the trigger efficiencies using the \emph{tag and probe} method 
(see~\ref{ch5:subsec:tap}) address this problem using exclusively experimental data.
Then, the extracted efficiencies can be used on the \gls{mc} samples to weight the simulated
event accordingly to the trigger efficiency applicable to that event. In this way, the simulated
data incorporate the trigger inefficiencies, mimicking the effect of having been "selected" through
a real trigger system and becoming suitable to be trigger-comparable with the experimental data.
\begin{table}[htb]
	\centering
	\begin{subtable}[b]{0.8\textwidth}
		\centering
	 \resizebox{\textwidth}{!}
	 {
	  \begin{tabular}{lcccc}\hline\hline
           &$0.0< |\eta| \leq 0.8$ & $0.8< |\eta| \leq 1.2$ & $1.2< |\eta| \leq 2.1$ & $2.1< |\eta| \leq 2.4$ \\ \hline 
$10.0< p_t \leq 13.0$ & $0.0700\pm0.0114$ & $0.0715\pm0.0107$ & $0.1042\pm0.0069$ & $0.1307\pm0.0141$ \\ 
$13.0< p_t \leq 15.0$ & $0.9084\pm0.0103$ & $0.9050\pm0.0108$ & $0.8996\pm0.0069$ & $0.8073\pm0.0163$ \\ 
$15.0< p_t \leq 17.0$ & $0.9393\pm0.0064$ & $0.9126\pm0.0088$ & $0.9161\pm0.0055$ & $0.8478\pm0.0128$ \\ 
$17.0< p_t \leq 20.0$ & $0.9656\pm0.0029$ & $0.9504\pm0.0044$ & $0.9448\pm0.0030$ & $0.8876\pm0.0075$ \\ 
$20.0< p_t \leq 30.0$ & $0.9648\pm0.0007$ & $0.9516\pm0.0013$ & $0.9480\pm0.0009$ & $0.8757\pm0.0026$ \\ 
$30.0< p_t \leq \infty$ & $0.9666\pm0.0003$ & $0.9521\pm0.0005$ & $0.9485\pm0.0004$ & $0.8772\pm0.0012$ \\ \hline
	 \end{tabular}
         }
	\caption{Leading leg}\label{ch6:tab:twleadingmu}
       \end{subtable}\vskip 1em
	\begin{subtable}[b]{0.8\textwidth}
		\centering
	\resizebox{\textwidth}{!}
	 {
	  \begin{tabular}{lcccc}\hline\hline
           &$0.0< |\eta| \leq 0.8$ & $0.8< |\eta| \leq 1.2$ & $1.2< |\eta| \leq 2.1$ & $2.1< |\eta| \leq 2.4$ \\ \hline 
$10.0< p_t \leq 13.0$ & $0.9604\pm0.0092$ & $0.9417\pm0.0098$ & $0.9472\pm0.0052$ & $0.8951\pm0.0129$ \\ 
$13.0< p_t \leq 15.0$ & $0.9589\pm0.0075$ & $0.9464\pm0.0086$ & $0.9519\pm0.0050$ & $0.8964\pm0.0130$ \\ 
$15.0< p_t \leq 17.0$ & $0.9711\pm0.0047$ & $0.9401\pm0.0075$ & $0.9518\pm0.0043$ & $0.9000\pm0.0109$ \\ 
$17.0< p_t \leq 20.0$ & $0.9669\pm0.0029$ & $0.9535\pm0.0043$ & $0.9576\pm0.0027$ & $0.9166\pm0.0066$ \\ 
$20.0< p_t \leq 30.0$ & $0.9655\pm0.0007$ & $0.9535\pm0.0013$ & $0.9558\pm0.0009$ & $0.9031\pm0.0023$ \\ 
$30.0< p_t \leq \infty$ & $0.9670\pm0.0003$ & $0.9537\pm0.0005$ & $0.9530\pm0.0004$ & $0.8992\pm0.0011$ \\ \hline
	  \end{tabular}
         }
	 \caption{Trailing leg}\label{ch6:tab:twtrailingmu}
       \end{subtable}
       \caption[Muon trigger efficiencies for 2011]{Muon trigger efficiencies extracted with a tag and probe 
       method in \pt and $\eta$ bins, for 2011 data. The errors shown are statistical.}\label{ch6:tab:triggermu}
\end{table}

\begin{table}[htb]
	\centering
	\begin{subtable}[b]{0.45\textwidth}
	\resizebox{1.1\textwidth}{!}
	 {
	  \begin{tabular}{lcccc}\hline\hline
           &$0.0< |\eta| \leq 1.5$ & $1.5< |\eta| \leq 2.5$ \\ \hline 
$10.0< p_t \leq 20.0$ & $0.5061\pm0.0037$ & $0.3176\pm0.0059$ \\ 
$20.0< p_t \leq 30.0$ & $0.9849\pm0.0003$ & $0.9774\pm0.0007$ \\ 
$30.0< p_t \leq \infty$ & $0.9928\pm0.0001$ & $0.9938\pm0.0001$ \\ \hline
	  \end{tabular}
         }
	\caption{Leading leg}\label{ch6:tab:twleadinelec}
       \end{subtable}
       \quad\quad
	\begin{subtable}[b]{0.45\textwidth}
	\resizebox{1.1\textwidth}{!}
	 {
	  \begin{tabular}{lcccc}\hline\hline
           &$0.0< |\eta| \leq 1.5$ & $1.5< |\eta| \leq 2.5$ \\ \hline 
$10.0< p_t \leq 20.0$ & $0.9854\pm0.0009$ & $0.9938\pm0.0011$ \\ 
$20.0< p_t \leq 30.0$ & $0.9923\pm0.0002$ & $0.9953\pm0.0003$ \\ 
$30.0< p_t \leq \infty$ & $0.9948\pm0.0001$ & $0.9956\pm0.0001$ \\\hline 
	  \end{tabular}
         }
	 \caption{Trailing leg}\label{ch6:tab:twtrailingelec}
       \end{subtable}
       \caption[Electron trigger efficiencies for 2011]{Electron trigger efficiencies extracted 
       with a tag and probe method in bins of \pt and $\eta$ for 2011 data. The errors shown
       are statistical.}\label{ch6:tab:triggerelec}
\end{table}

The trigger efficiencies can be obtained from independent \glspl{pd} selected with trigger paths
with looser criteria than the ones used to select the analysis sample. A tag and probe method is 
used, where the tag has been matched with the lepton which fired the trigger to avoid the bias on
the probe lepton. The efficiencies obtained with the tag and probe are trigger efficiencies per 
lepton. The analysis requires two leptons to select the event (double lepton trigger paths), one 
lepton is the high-\pt object and it is called the \emph{leading leg} and the other one is the 
low-\pt object and it is called the \emph{trailing leg}. Therefore, the trigger efficiencies are 
calculated for the leading and the trailing lepton independently and are interpreted as the 
probability of a lepton passing one leg trigger requirement\footnote{The trailing efficiency 
is actually evaluated requiring a leading lepton in the event, therefore is a conditional 
probability.}. Tables~\ref{ch6:tab:triggermu} and~\ref{ch6:tab:triggerelec} tabulate the 
efficiency per leg for the 2011 analysis muon and electron object, respectively, calculated with 
the tag and probe method.

\begin{figure}[htb]
	\centering
	\begin{subfigure}[b]{0.45\textwidth}
		\includegraphics[width=0.8\textwidth]{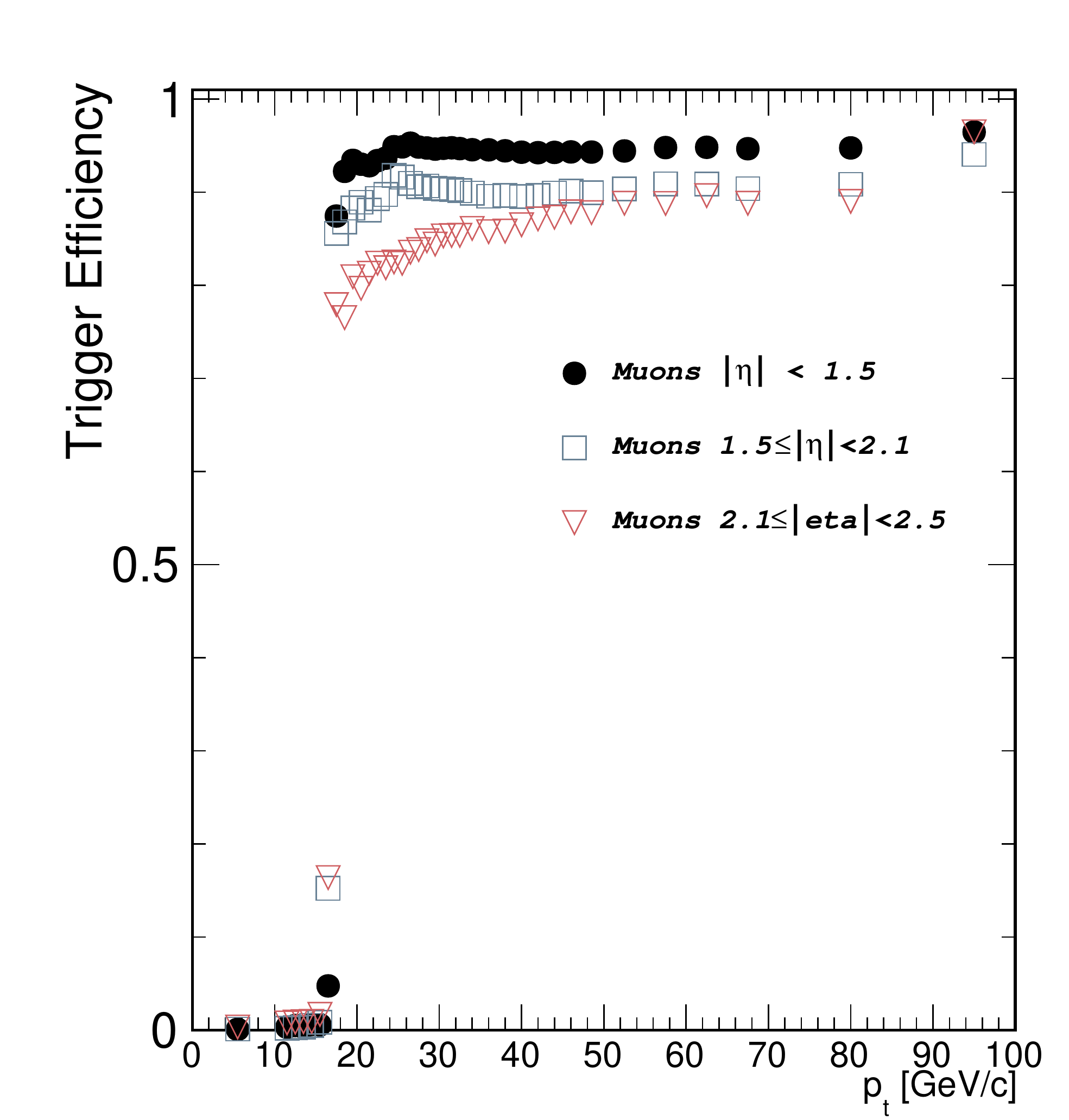}
		\caption{Double\_Mu17\_Mu8 trigger leading leg}
	\end{subfigure}\quad
	\centering
	\begin{subfigure}[b]{0.45\textwidth}
		\includegraphics[width=0.8\textwidth]{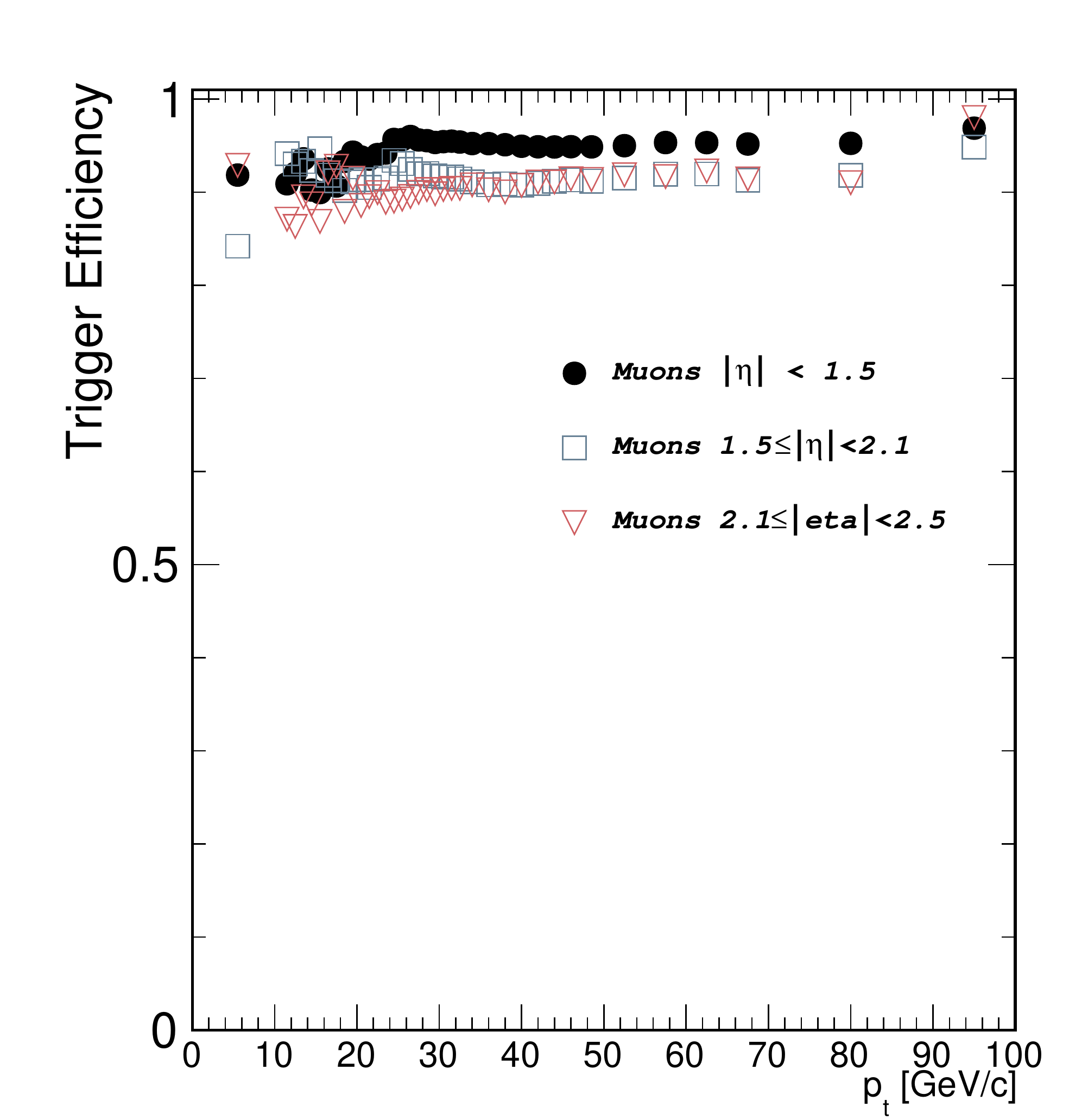}
		\caption{Double\_Mu17\_Mu8 trigger trailing leg}
	\end{subfigure}
	\begin{subfigure}[b]{0.45\textwidth}
		\includegraphics[width=0.8\textwidth]{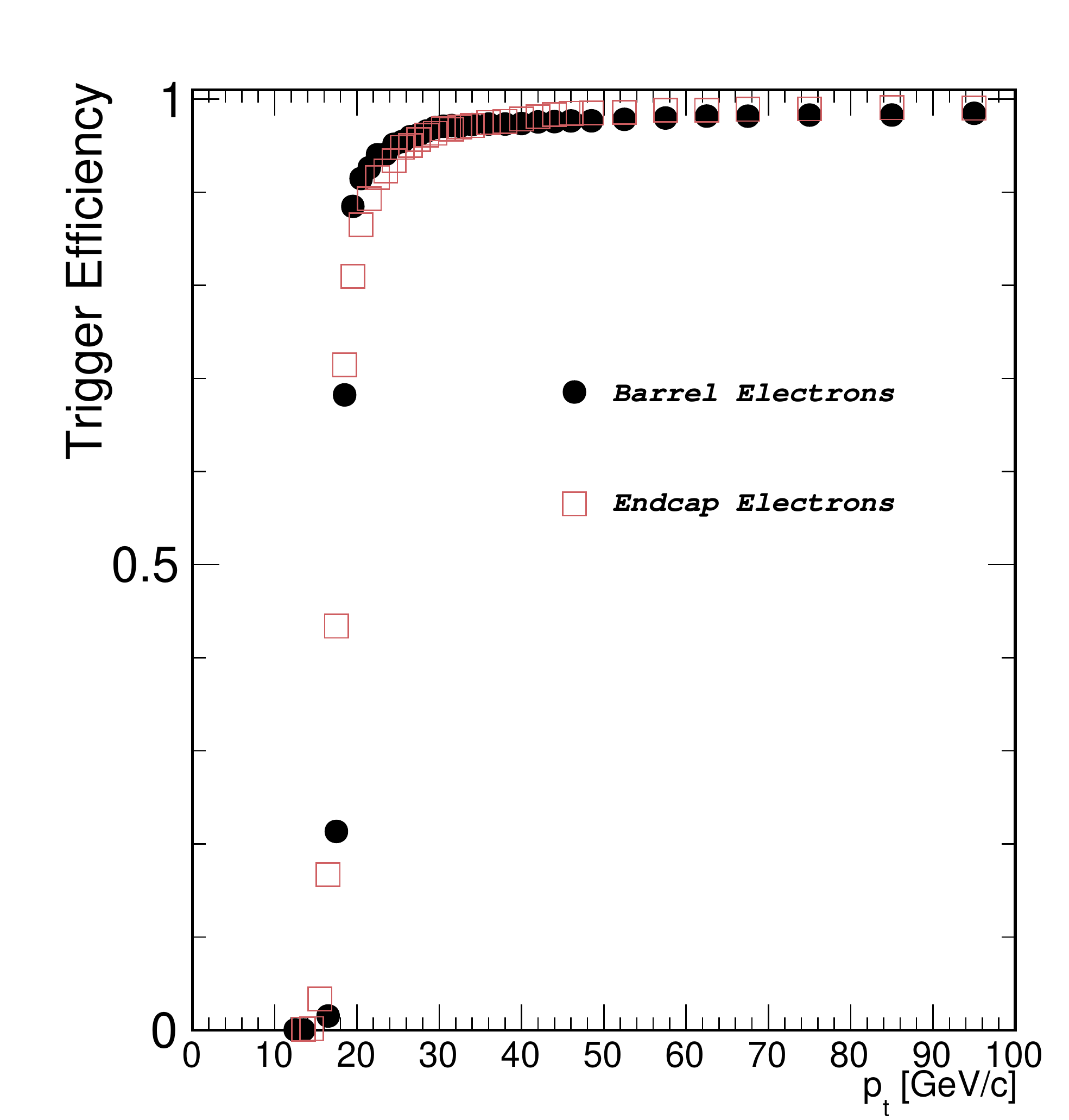}
		\caption{Double\_Ele17\_Ele8 trigger leading leg}
	\end{subfigure}\quad
	\centering
	\begin{subfigure}[b]{0.45\textwidth}
		\includegraphics[width=0.8\textwidth]{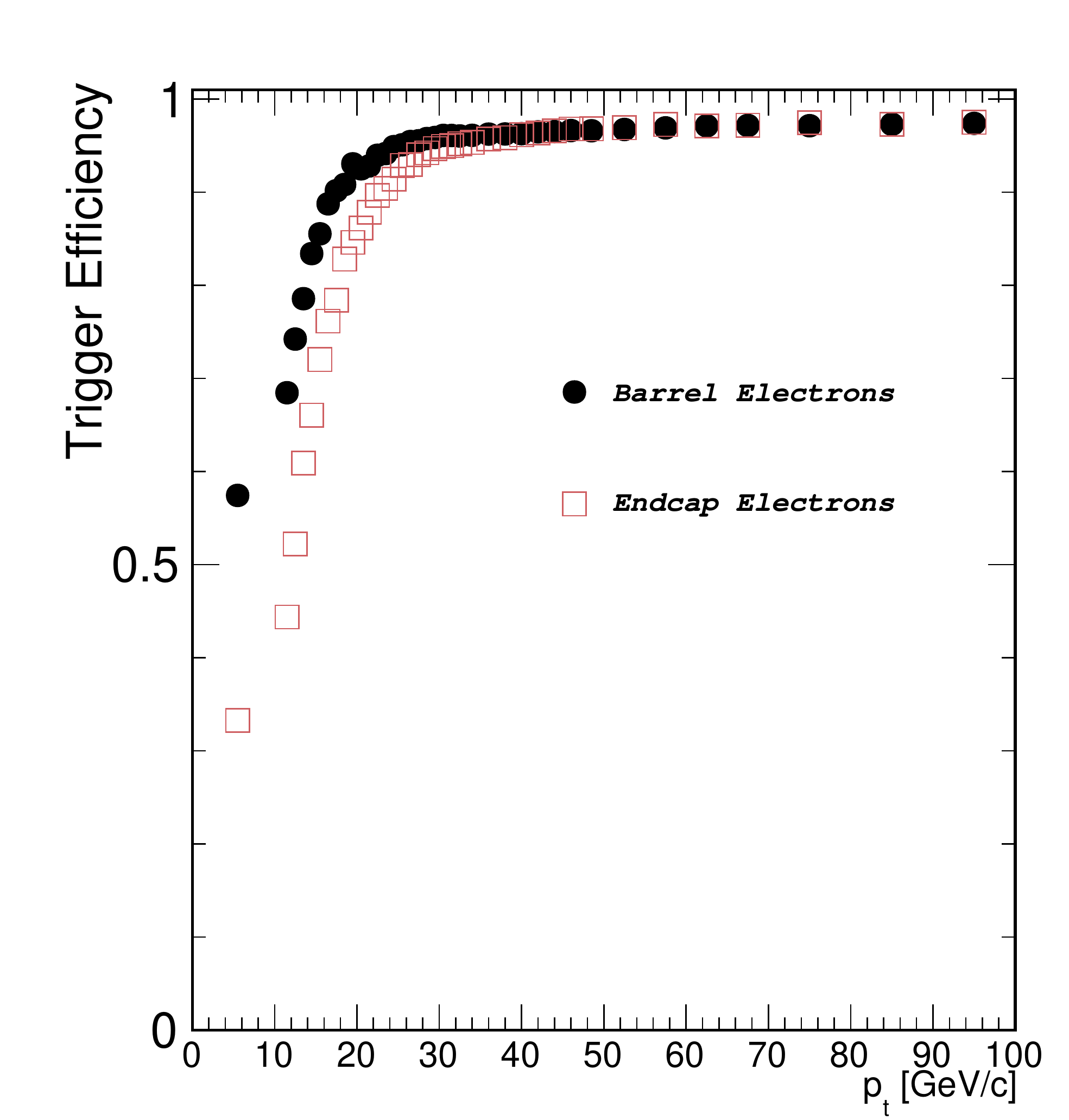}
		\caption{Double\_Ele17\_Ele8 trigger trailing leg}
	\end{subfigure}
	\caption[Turn-on curves for trigger efficiencies]{Trigger efficiencies with 
	respect to the offline selection for 2012 data. Each curve is plotted in function of \pt
	of the lepton for the different $\eta$ regions considered. The turn-on curves guide the 
	analysis transverse momentum requirements: in order to avoid biased samples the leptons 
	should be cut in a transverse momentum placed in the plateau of the turn-on
	curve.}\label{ch6:fig:turn-on}
\end{figure}

Figures~\ref{ch6:fig:turn-on} show the trigger efficiencies per leg for muons and electrons
of the 2012 analysis in function of the transverse momentum of the lepton. Each curve is plotted
for the different pseudorapidity regions considered. The figures illustrate the transverse momentum
cut used in the leptons of the analysis, which are localised in the plateau of the turn-on curves. 

The efficiencies per leg have been interpreted as the probability of a lepton passing one leg 
trigger requirement and can be used to build a probability function per event (not per leg) 
which takes into account the possibilities for each combination of the final state leptons to 
fire the double trigger. Therefore, as the experimental data is selected through the triggers,
in order to be able of compare the simulated data with the experimental one, this effect is 
introduced in the simulated data using weights. Each \gls{mc} event is weighted with 
equation~\eqref{ch6:eq:trweightpure} in the case of same flavour channel $eee$ and $\mu\mu\mu$ 
because it is possible to use all the electrons (muons) in the event to fire the 
\verb/DoubleElectron/ (\verb/DoubleMu/) trigger. And the equation~\eqref{ch6:eq:trweightmixed} is 
used for $ee\mu$ and $\mu\mu e$ channels where only the same flavour leptons may be used to fire the 
Double trigger path, since the third different-flavour lepton is not entering in the possible 
combinations. 
\begin{subequations}
	\begin{equation}
		\mathcal{P}(\text{Pass}|3\ell) = 1-\Big[(1-\varepsilon^L_1)(1-\varepsilon^L_2)(1-\varepsilon^L_3)+
		\sum_{\substack{i,j,k=1\\i\neq j\neq k}}^{3}\varepsilon_i^L(1-\varepsilon_j^T)(1-\varepsilon_k^T)\Big]
		\label{ch6:eq:trweightpure}
	\end{equation}
	\begin{equation}
	\mathcal{P}(\text{Pass}|2\ell) = 1-\Big[(1-\varepsilon^L_1)(1-\varepsilon^L_2)+
		\sum_{\substack{i,j=1\\i\neq j}}^{2}\varepsilon_i^L(1-\varepsilon_j^T)\Big]
		\label{ch6:eq:trweightmixed}
	\end{equation}
	\label{ch6:eq:triggerweight}
\end{subequations}
where $\varepsilon^L_i=\varepsilon^L_i(\pt^i,\eta^i)$ is the efficiency per leg of the trailing 
leg, evaluated on the i-lepton of the event and $\varepsilon^T_i=\varepsilon^T_i(\pt^i,\eta^i)$ is 
the efficiency per leg of the leading leg, evaluated on the i-lepton of the event. 

After the event weighting, the simulated data incorporate the trigger inefficiencies and, thus,
the probability to "store" the event by a trigger system.

\section{Muon selection}\label{ch6:sec:muonselection}
Muon selection is designed to achieve high efficiency for muons coming from \W or \Z decays, 
hereafter called \emph{prompt\glsadd{ind:prompt} muons}, keeping a high rejection in those which are not. Muons
are restricted to be within the pseudorapidity acceptance ($|\eta|<2.4$) of the muon and tracking
system, and have to fulfil various track quality requirements. To avoid any bias introduced by the
trigger thresholds, the transverse momentum is chosen to be higher than trigger thresholds used and in 
the plateau of the trigger efficiency turn-on curve. The muons are required to be 
reconstructed (see Section~\ref{ch4:subsec:muons}) using the tracker and the muon 
spectrometer,~\ie a \emph{Global} muon, and the
relative $\chi^2$ over the degrees of freedom (normalised $\chi^2$) of the global fit has a
quality cut; this is useful to reject muons from hadrons decaying in flight and kaons punching
through the calorimeter. In addition, for a track, more than 10 hits in the inner tracker are 
required to guarantee a good fit of the track, at least another hit is required in the pixel 
detector and at least one hit in the muon spectrometer. Moreover, the muon must be matched to 
track segments in two different muon stations which suppress accidental track-to-segment matches 
and punch-through. The transverse impact parameter of the track is also restricted around the 
beam spot in order to reject cosmic ray muons; the longitudinal impact parameter 
is also restricted to be around of the beam spot to further suppress cosmic muons, muons from decays
in flight and tracks from \gls{pileup}. Finally, the requirement of the relative error fit 
$\Delta\pt/\pt$ better than 10\% is applied to reject poorly measured muons. 
\begin{table}[htpb]
	\centering
	\begin{tabular}{lrrr}\hline\hline
		                                                                   &  & 2011 & 2012 \\\hline
		\multicolumn{2}{l}{Max. relative \pt resolution, $\Delta\pt/\pt$}     & 0.1  & 0.1 \\
		\multicolumn{2}{l}{Max. normalised $\chi^2$ for global fit}           & 10   & 10  \\
		\multicolumn{2}{l}{Min. number of pixel hits}                         & 1    & 1  \\
		\multicolumn{2}{l}{Min. number of inner tracker hits}                 & 11   & -- \\  
		\multicolumn{2}{l}{Min. number of layers in the inner tracker with hits} & --& 6 \\
		\multicolumn{2}{l}{Min. number of matched muon segments}              & 2    & 2 \\
		\multicolumn{2}{l}{Max. transverse impact parameter, $|d_0|$ [cm]}   & 0.2   & 0.01 (0.02)\\
		\multicolumn{2}{l}{Max. longitudinal impact parameter, $|d_z|$ [cm]} & 0.1   & 0.1 \\\hline
     		\multicolumn{2}{l}{Max. relative particle flow isolation $Iso_{PF}/p_{T}$}  & 0.12 & -- \\
     		\multirow{4}{*}{Min. MVA isolation output}  &      ($\pt\leq20\GeV$, MB)   & -- & 0.86 \\
     							    &      ($\pt\leq20\GeV$, ME)   & -- & 0.82\\
     							    &      ($\pt>20\GeV$, MB)   & -- & 0.82 \\
     							    &      ($\pt>20\GeV$, ME)   & -- & 0.86 \\\hline
	\end{tabular}
	\caption[Muon selection requirements]{Selection requirements imposed on muons in 
	the 2011 and 2012 analyses. Besides these requirements, in the case of 2012 analysis,
	the muon object must be reconstructed with a particle flow algorithm and it can
	be a global or a tracker muon, not just a global as in the 2011 case. The $d_0$ value in the
	2012 column is shown for muons with $\pt<20\GeV$ and for muons with $\pt>20\GeV$ in 
        parentheses. The isolation requirements are also included. The \emph{MB} and \emph{ME} 
	labels are referring to barrel and endcap muons,~\ie muons with $|\eta|$ lower or higher
	than 1.479.}\label{ch6:tab:muonreq}
\end{table}
The requirements described above have been used
for the 2011 analysis, the 2012 muon objects use the same requirements and add new ones. The muon
is also required to be reconstructed using a particle flow algorithm and, therefore, the global 
requirement is relaxed and it also admitted a tracker muon. Furthermore, the minimum number of
inner tracker hits is substituted by the equivalent cut in the number of tracker layers with hits.
Also the impact parameter cuts have been tight. Table~\ref{ch6:tab:muonreq} shows the quality
cuts for the 2011 and 2012 muon objects.

Besides the aforementioned quality requirements, the muon is required to be isolated. For this 
purpose, a particle flow isolation is used. The isolation variable is built using the particle 
flow candidates found in the event through,
\begin{equation}
	Iso_{PF} = \sum_{PF} \ET^{ChHad}+max\left(0,\sum_{PF}\pt^{NeutHad}+\sum_{PF}\ET^{\gamma}-d\beta\right),
	\label{ch6:eq:muoniso}
\end{equation}
where $\sum_{PF}\pt^{ChHad}$ is the \pt sum of the charged hadron candidates with\linebreak
\mbox{$|d_z^{candidate}-d_z^{\mu}|< 0.1\;cm$}, $\sum_{PF}\ET^{NeutHad}$ is the scalar sum of 
transverse energy of the neutral hadron candidates and $\sum_{PF}\ET^{\gamma}$ is the scalar sum of 
transverse energy of the photon candidates. The charged, neutral and photon candidates have been 
defined within a $\Delta R <0.3$ cone around the muon candidate. The $d\beta$ correction is defined 
as $d\beta=0.5\sum \pt^{PU}$ being $\sum\pt^{PU}$ the \pt sum of the charged particles in the 
$\Delta R < 0.3$ cone around the muon, but with particles not originating from the primary 
vertex\footnote{Note that this is a \gls{pileup} correction similar to that provided by the 
\FASTJET algorithm of Section~\ref{ch5:subsec:pileup}.}. The 0.5 factor corresponds to a naive 
average of neutral to charged particles (which has been measured in Reference~\cite{CMS:2010eua} 
and Higgs searches~\cite{Chatrchyan:2012vp} and~\cite{Chatrchyan:2011tz}). In summary, the muon 
isolation is calculated by estimating all the charged hadrons around the isolation cone and adding
the neutral candidates which have been corrected by the \gls{pileup}, avoiding possible \gls{pileup}
over corrections (when the correction is negative). 
\begin{figure}[htb]
	\centering
	\begin{subfigure}[b]{0.45\textwidth}
		\includegraphics[width=\textwidth]{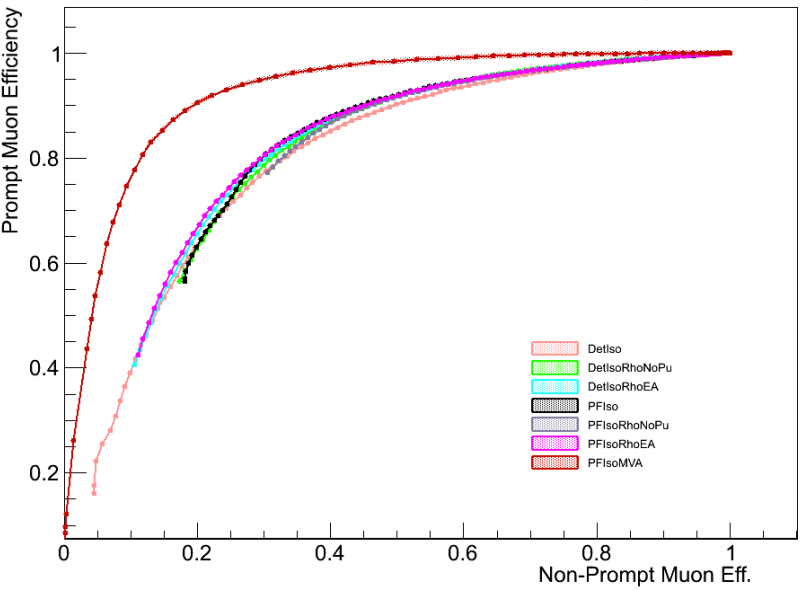}
		\caption{Muons with \pt lower than 20~\GeV and $|\eta|< 1.48$ (barrel).}
	\end{subfigure}\quad
	\begin{subfigure}[b]{0.45\textwidth}
		\includegraphics[width=\textwidth]{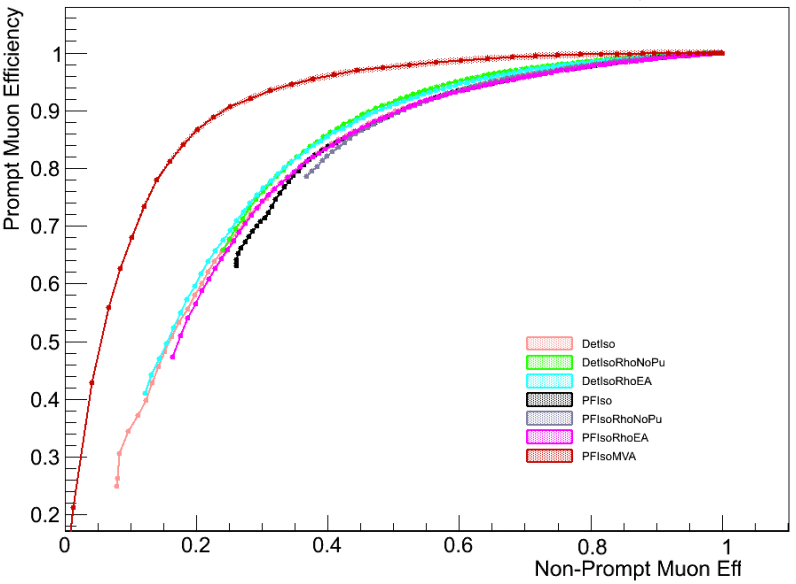}
		\caption{Muons with \pt lower than 20~\GeV and $|\eta|\geq 1.48$ (endcap).}
	\end{subfigure}
	\centering
	\begin{subfigure}[b]{0.45\textwidth}
		\includegraphics[width=\textwidth]{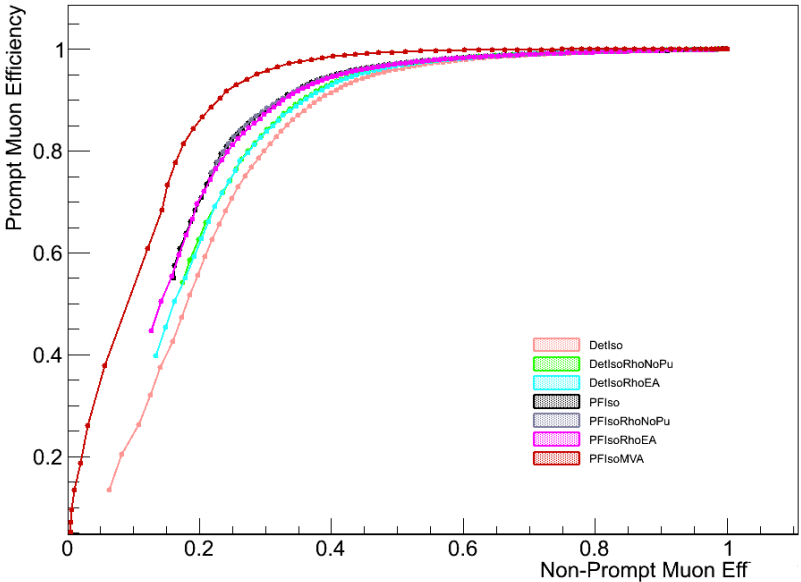}
		\caption{Barrel muons with \pt higher than 20~\GeV}
	\end{subfigure}\quad
	\begin{subfigure}[b]{0.45\textwidth}
		\includegraphics[width=\textwidth]{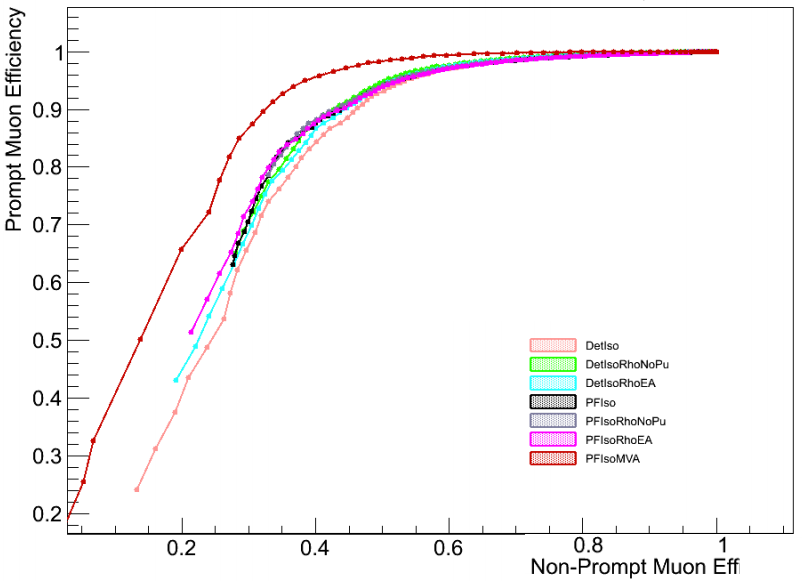}
		\caption{Endcap muons with \pt higher than 20~\GeV}
	\end{subfigure}
	\caption[ROC curves for several muon isolation strategies]{The prompt muon efficiency with 
	respect to the fake muon efficiency is shown in the plots. The curve is build by
	varying the cuts thresholds of the MVA output and evaluating the signal and background
	efficiency in each point (the so called ROC curves, from Receiver Operating Characteristic 
	in signal detection theory). The MVA approach used in this analysis is labelled as 
	"PFIsoMVA" and represented by red lines and dots. There are several muon isolation 
	strategies compared, in particular the one used in 2011 analysis is represented by black
 	lines and dots and labelled "PFIso". It may be observed the improvement reached with the
	MVA-isolation approach with respect the sequential 2011 one, the purity of the 2012 muons
	is substantially higher.}\label{ch6:fig:mvaiso}
\end{figure}

The increasing \gls{pileup} conditions in the 2012 running period have motivated a change in the 
isolation strategy in order to reduce the dependence of this variable with the number of 
\gls{pileup} vertices and increase both prompt efficiency and fake muon rejection. It is 
accomplished with a particle flow isolation variable implemented using a \gls{mva}\glsadd{ind:mva}\glsadd{ind:mva}: 
the discrimination power of the $\Delta R$ between the muon and other particles are used to discriminate
between isolated and non-isolated muons. Some of the input variables are $\sum_{PF} \ET^{ChHad}$,
$max\left(0,\sum_{PF}\pt^{NeutHad}+\sum_{PF}\ET^{\gamma}-d\beta\right)$, 
$\sum \Delta R(\mu,PF^{ChHad})$,~\etc Once the \gls{bdt}\glsadd{ind:bdt} is trained in a enriched sample of 
isolated muons (sample defined with a \Z resonance similar to the tag and probe method) in 
order to discriminate between prompt and non-isolated muons, it can be used to classify the 
muons between prompt and non-isolated. The improvement accomplished using this \gls{mva}\glsadd{ind:mva}\glsadd{ind:mva} 
approach for isolation versus the sequential cut isolation approaches are shown in 
Figures~\ref{ch6:fig:mvaiso}. The \gls{bdt}\glsadd{ind:bdt} output is dependent of the kinematics of the muon, 
thus it has been split the muons into barrel and endcap and between lower and higher than 
\pt~=20~\GeV as it can be seen in Table~\ref{ch6:tab:muonreq}. 

The values of the requirements used to identify and isolate the muon objects have been optimised
by the \gls{cms} collaboration providing a few different baseline selection and isolation working
points suitable for different target analyses~\cite{MuonIDWeb}. The main advantage of this 
approach is that efficiencies related with the muon objects are provided in a centralised way
for the whole collaboration. A dedicated working group\footnote{Muon \gls{pog}\glsadd{ind:pog}, there are analogous
working groups for each of the physics objects used in the \gls{cms} analysis framework.} is in 
charge to define, optimise and support the recommended muon selections. 

\subsection{Muon efficiencies and scale factors}\label{ch6:subsec:muoneff}
From the hits in the tracker and muon spectrometer systems up to the high level physic object used 
for analysis, the final muon has been built in three main stages: the reconstruction, explained in 
Section~\ref{ch4:subsec:muons}, the identification and the isolation (both described in the 
previous sections). Each process introduces inefficiencies which may be calculated using the tag 
and probe method. The efficiency of the final selected muon object can be factorised as the 
efficiencies of each stage, given that the output of one process is the input of the next, and they
are applied sequentially. Thus,
\begin{equation*}
	\varepsilon_{\mu} = \varepsilon_{hlt|iso}\cdot\varepsilon_{iso|id}\cdot\varepsilon_{id|reco}\cdot\varepsilon_{reco}
\end{equation*}
where $\varepsilon_{reco}$ stands for the efficiency of the reconstruction process, using as probes
the available inner and stand-alone tracks and evaluating how many global muon\glsadd{ind:global} tracks are obtained.
Note that an extra term should be included before the reconstruction efficiency, it is the track
reconstruction efficiency, but it has been measured to be compatible with the unity, thus it is not
shown explicitly. The $\varepsilon_{id|reco}$ is the identification efficiency which can be calculated 
using as probes the previous global tracks and checking how many pass the identification criteria 
described in previous section. The $\varepsilon_{iso|id}$ stands for the efficiency of isolation, and
analogously the probes are defined as the identified leptons and the passing probes, those passing 
the isolation criteria. And finally $\varepsilon_{hlt|iso}$ is the trigger efficiency (both 
L1 and \gls{hlt}\glsadd{ind:hlt}) and it is calculated from a tag and probe using as probe sample the identified 
and isolated muons, then it is checked if they pass the trigger leg. 

\begin{table}[htbp]
	\centering
	\begin{subtable}[b]{0.5\textwidth}
		\centering
		\resizebox{\textwidth}{!}
		{
		\begin{tabular}{ccc}\hline\hline
			&$0< |\eta| \leq 1.2$ & $1.2< |\eta| \leq 2.5$ \\ \hline 
			$10< p_t \leq 20$ & $0.933\pm0.009$ & $0.964\pm0.007$ \\ 
			$20< p_t \leq 30$ & $0.951\pm0.002$ & $0.944\pm0.002$ \\ 
			$30< p_t \leq 40$ & $0.971\pm0.001$ & $0.954\pm0.001$ \\ 
			$40< p_t \leq 50$ & $0.980\pm0.008$ & $0.965\pm0.001$ \\ 
			$50< p_t \leq 60$ & $0.985\pm0.001$ & $0.968\pm0.002$ \\ 
			$60< p_t \leq 80$ & $0.978\pm0.002$ & $0.967\pm0.004$ \\ 
			$80< p_t \leq \infty$ & $0.979\pm0.006$ & $0.984\pm0.016$ \\ \hline
		\end{tabular}
	        }
		\caption{Muon scale factors for 2011 data.}
	\end{subtable}\vskip 1em
	\begin{subtable}[b]{0.7\textwidth}
		\centering
		\resizebox{\textwidth}{!}
		{
		 \begin{tabular}{cccc}\hline\hline
		            &$0.0< |\eta| \leq 0.9$ & $0.9< |\eta| \leq 1.2$ & $1.2< |\eta| \leq 2.4$ \\ \hline 
		 $10< p_t \leq 15$ & $0.9923\pm0.012$ & $0.971\pm0.358$ & $1.002\pm0.005$ \\ 
		 $15< p_t \leq 20$ & $0.9611\pm0.006$ & $0.951\pm0.005$ & $0.995\pm0.003$ \\ 
		 $20< p_t \leq 25$ & $0.9821\pm0.002$ & $0.982\pm0.001$ & $1.020\pm0.002$ \\ 
		 $25< p_t \leq 30$ & $1.0000\pm0.001$ & $0.993\pm0.002$ & $1.019\pm0.001$ \\ 
		 $30< p_t \leq 50$ & $0.9928\pm0.0002$ & $0.9911\pm0.0004$ & $1.0018\pm0.0003$ \\ 
		 $50< p_t \leq \infty$ & $0.994\pm0.001$ & $0.991\pm0.001$ & $1.005\pm0.001$ \\\hline 
		 \end{tabular}
	 	}
		 \caption{Muon scale factors for 2012 data}
	\end{subtable}
	\caption[Muon scale factors]{Muon reconstruction, isolation and identification scale factors
		applied to the Monte Carlo sample events in bins of pseudorapidity
		an transverse momentum. Scale factors calculated for the 2011 analysis.}\label{ch6:tab:sfmuon}
\end{table}

The efficiencies are calculated both in the simulated samples and in the experimental data in the 
way described above using a tag and probe method as it was explained in Section~\ref{ch5:subsec:tap}.
The cross section measurement 
is performed by keeping the experimental observed data uncorrected, as it will be explained in 
Chapter~\ref{ch8}, whereas the simulated data is assuming the efficiency corrections through the 
application of the \glspl{sf}\footnote{\emph{Id est}, the ratio between the efficiencies calculated in 
\gls{mc} versus experimental data} in the simulated samples. Table~\ref{ch6:tab:sfmuon} shows the 
total \gls{sf}\glsadd{ind:sf},~\ie reconstruction, identification and isolation, for the selected muons to be 
applied to the \gls{mc} samples. Notice that the trigger efficiency is not introduced in the
\glspl{sf} due to the weight method technique described at Section~\ref{ch6:subsec:triggerEff}. The
effect of the trigger in the \gls{mc} samples is simulated by the 
equations~\eqref{ch6:eq:triggerweight} which make use of the trigger efficiencies per leg evaluated
in experimental data.

\section{Electron selection}\label{ch6:sec:electronselection}
\begin{figure}[htpb]
	\centering
	\begin{subfigure}[b]{0.45\textwidth}
		\includegraphics[width=\textwidth]{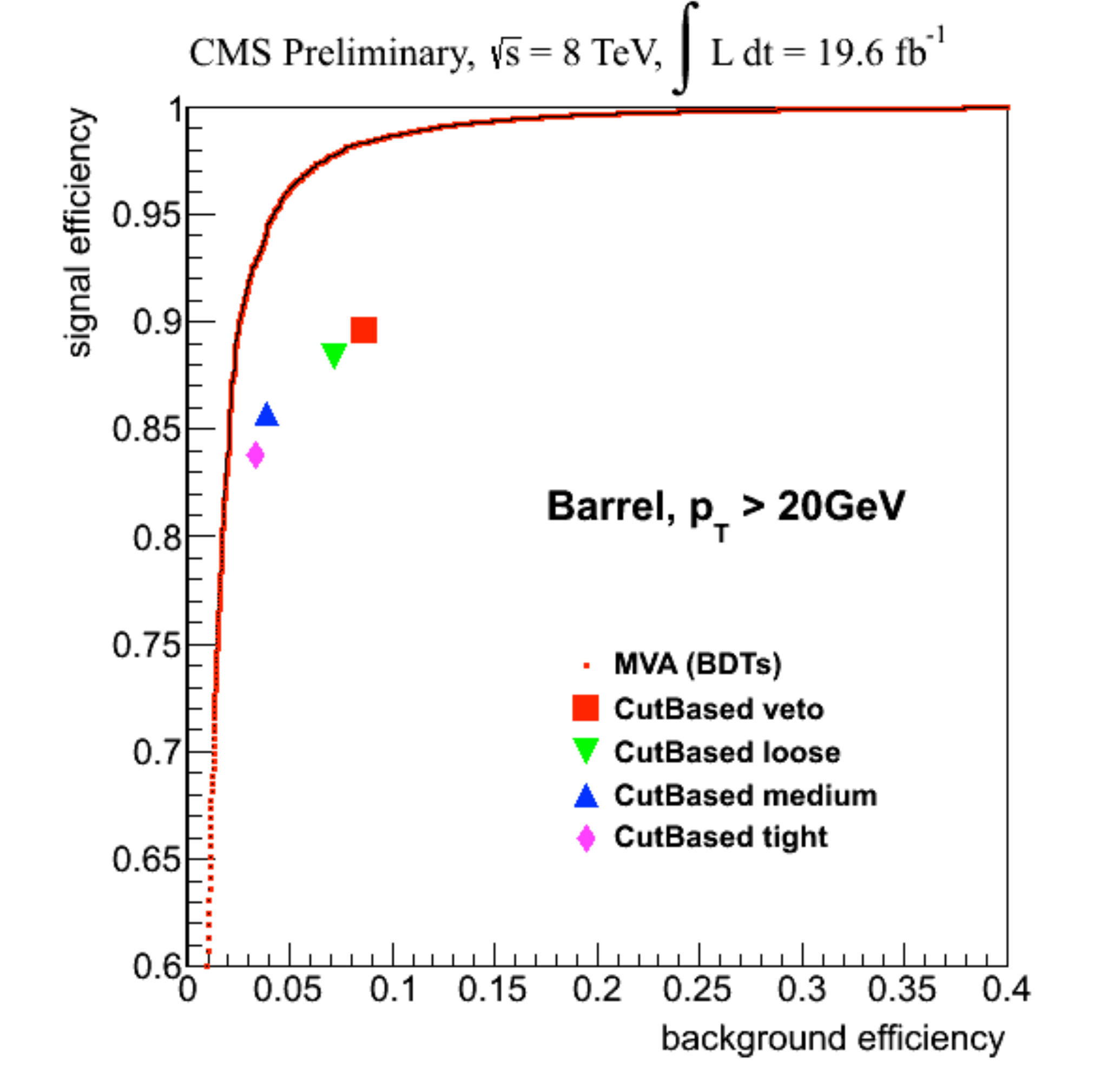}
		\caption{Barrel electrons}
	\end{subfigure}\quad
	\begin{subfigure}[b]{0.45\textwidth}
		\includegraphics[width=\textwidth]{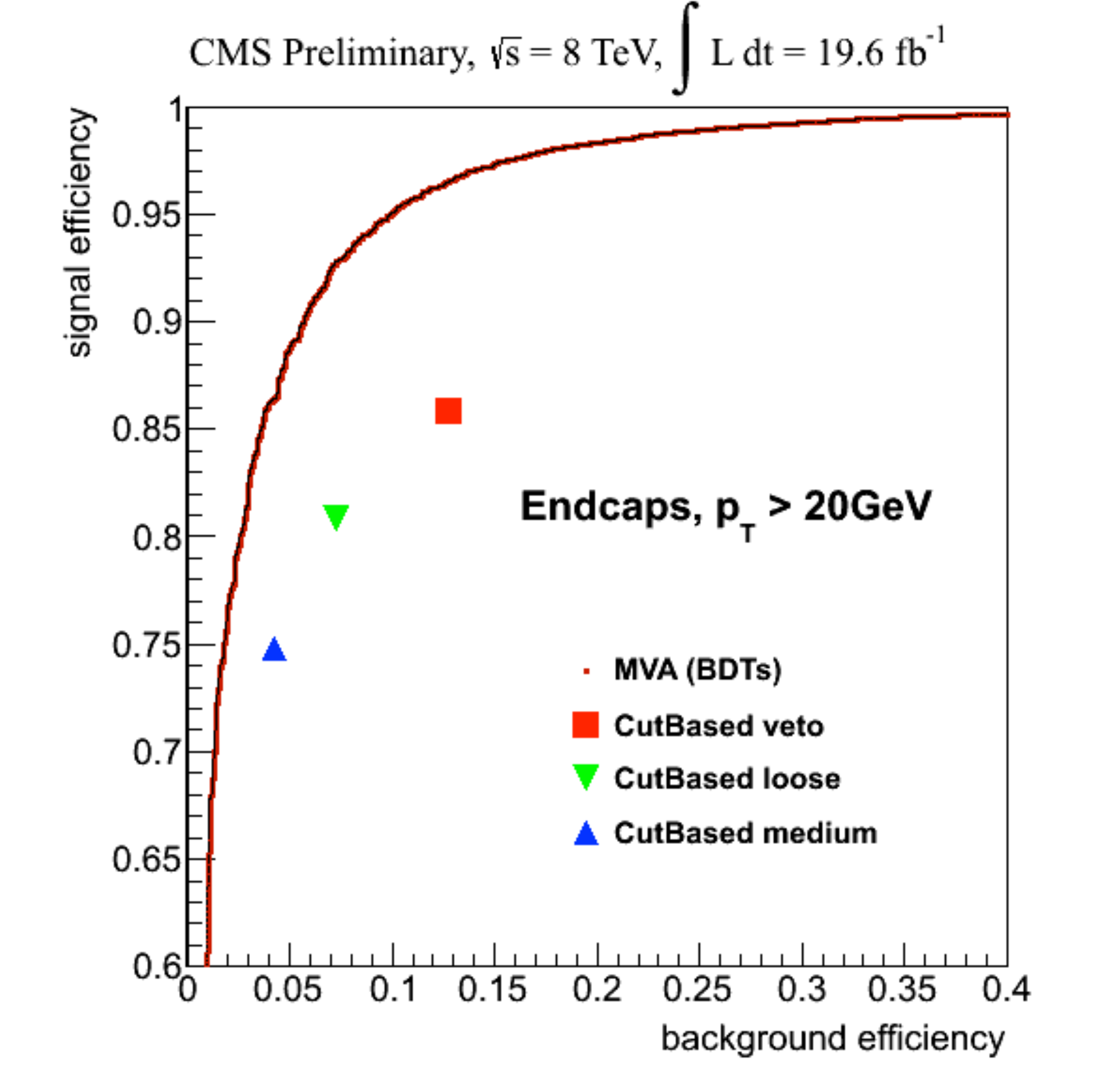}
		\caption{Endcap electrons}
	\end{subfigure}
	\caption[ROC curves for electron identification]{The prompt electron efficiency with 
	respect to the fake lepton efficiency is shown in the plots. The curve is build by
	varying the cuts thresholds of the MVA output and evaluating the signal and background
	efficiency in each point. The cut based points are referring to different 
	WP. The MVA-based identification improves significantly the purity of the 
	electrons.}\label{ch6:fig:roccurves}
\end{figure}

The electron objects, after the reconstruction process (see Section~\ref{ch4:subsec:Electrons}), 
are selected on a basis of few discriminating variables in terms of identification, selection and
photon conversion rejection. The main discriminating variables used are described in 
Table~\ref{ch6:tab:elecvar}. 
\begin{table}[htpb]
	\centering
	\begin{tabular}{lp{8cm}}\hline\hline
		$\sigma_{i\eta i\eta}$ & supercluster $\eta$ width \\
                $\sigma_{i\phi i\phi}$ & supercluster $\phi$ width \\
		$\Delta\eta_{in}$ & distance in the $\eta$-plane between the track and the 
		  supercluster \\
                $\Delta\phi_{in}$ & distance in the $\phi$-plane between the track and the 
		  supercluster\\
		$f_{brem}$  & the fraction of the total momentum carried away by bremsstrahl\"ung\\
	        $E_{HCAL}/E_{ECAL}\equiv H/E$ & Ratio between energy deposited in the 
		   \acrshort{hcal}~\vs \acrshort{ecal}\\
		$N_{SC}^a$        & Number of additional clusters from supercluster\\
		$1/E_{SC}-1/p_{TRK}$  &	The difference between the inverse of the Energy in the supercluster and the
			momentum measured with the tracker \\\hline
	\end{tabular}
	\caption[Discriminating variables for electron identification]{Main discriminating variables
	for electron identification, used in both cut based and MVA approach. Note the $i$ label in
        the calorimeter's supercluster related variables is indicating than any measurement ç
	(distance or width) is taken as a number of crystals rather than 
	distance.}\label{ch6:tab:elecvar}
\end{table}
The \gls{cms} collaboration has developed two main approaches to identify electrons: a cut based 
approach~\cite{CMS-PAS-EGM-10-004} and a \gls{mva}\glsadd{ind:mva}\glsadd{ind:mva} discriminator approach~\cite{CMS-PAS-HIG-12-016}.
Both of them have been optimised to select electrons from \Z and \W (\emph{prompt\glsadd{ind:prompt}} electrons) and 
reject fakes from jets or photon conversion. Threshold cuts in the discriminant variables can be 
tuned, defining \gls{wp}\glsadd{ind:wp} of different efficiency on the prompt and rejection on the fakes. This 
analysis uses the \gls{mva}\glsadd{ind:mva}\glsadd{ind:mva} approach using \gls{bdt}\glsadd{ind:bdt} to identify prompt electrons given the optimal 
efficiency in selecting signal keeping a low rate of selected fakes, with respect the cut based 
approach as it can see in Figures~\ref{ch6:fig:roccurves}. 

The training of the multivariate algorithm is performed with the combination of those variables
in a \gls{bdt}\glsadd{ind:bdt}. Since the calorimeter response is different between endcap and barrel, and also the 
track reconstruction differs for low-\pt and high-\pt electrons, the training is done in different
regions of $\eta$ and \pt. Further details can be obtained from Reference~\cite{CMS-PAS-HIG-12-016}.
Besides the electron identification, all electrons are required to be within detector acceptance 
($|\eta|<2.5$) and further requirements are applied in order to avoid a possible bias due to the 
trigger selection; the trigger paths for electrons use some loose cuts in some of the variables of
Table~\ref{ch6:tab:elecvar} which have to be taken into account in the analysis. These extra 
requirements are shown in Table~\ref{ch6:tab:elecreq} and it summarises as follows. 
The electromagnetic shower shape (see Section~\ref{ch4:subsec:Electrons}) is checked to test the 
compatibility with the isolated electron hypothesis. Therefore, the shower is evaluated for
the width of the electromagnetic cluster in terms of pseudorapidity ($\sigma_{i\eta i\eta}$), the
difference between the measured position of the \gls{ecal}\glsadd{ind:ecal} supercluster and the associated track 
($\Delta\phi_{in}$ and $\Delta\eta_{in}$) and the ratio of the energy deposited in the \gls{hcal}\glsadd{ind:hcal} 
over the \gls{ecal}\glsadd{ind:ecal} ($E_{HCAL}/E_{ECAL}$). Furthermore, it is required that electron tracks have no 
missing hits in the innermost regions in order to reject $e^+e^-$ pair created from photons of the 
hard interaction. These photon have a high probability to convert to an $e^+e^-$ pair in the 
tracker, thus those tracks are likely to miss hits in the innermost region of the tracker.
\begin{table}[htpb]
	\centering
	\resizebox{\textwidth}{!}
	{
	\begin{tabular}{lrrrcrr}\hline\hline
		&  & \multicolumn{2}{c}{2011} & & \multicolumn{2}{c}{2012} \\\hline
				     &  & EB & EE            &&        EB     & EE       \\\cline{3-4}\cline{6-7}
		\multicolumn{2}{l}{Max. $\sigma_{i\eta i\eta}$}      & 0.01   & 0.03  && 0.01 & 0.03   \\
		\multicolumn{2}{l}{Max. $|\Delta\phi_{in}|$}         & 0.15  & 0.1   && 0.15 & 0.1    \\
		\multicolumn{2}{l}{Max. $|\Delta\eta_{in}|$}                 & 0.007 & 0.009  && 0.007 & 0.009  \\
		\multicolumn{2}{l}{Max. $E_{HCAL}/E_{ECAL}$}                 & 0.12  & 0.1   && 0.12 & 0.1 \\  
	\multicolumn{2}{l}{Max. transverse impact parameter, $|d_0|$ [cm]}   &\multicolumn{2}{c}{0.02} &&\multicolumn{2}{c}{0.02} \\
	\multicolumn{2}{l}{Max. longitudinal impact parameter, $|d_z|$ [cm]} &\multicolumn{2}{c}{0.1}  &&\multicolumn{2}{c}{0.1} \\
	\multicolumn{2}{l}{Max. $\sum_{HCAL}\ET\pt$}&\multicolumn{2}{c}{0.2} 
		&&\multicolumn{2}{c}{0.2} \\
	\multicolumn{2}{l}{Max. $\sum_{ECAL}\ET/\pt$}&\multicolumn{2}{c}{0.2} 
		& &\multicolumn{2}{c}{0.2} \\
	\multicolumn{2}{l}{Max. $\sum_{tracker}\ET/\pt$}& \multicolumn{2}{c}{0.2} 
		& &\multicolumn{2}{c}{0.2} \\\hline
	\multirow{4}{*}{Min. MVA identification output}  
			 &      ($\pt\leq20\GeV$, $|\eta|\leq R1$)&\multicolumn{2}{c}{0.14}   &&\multicolumn{2}{c}{0.0}  \\
			 & ($\pt\leq20\GeV$, $\eta\in(R1,1.479]$)&\multicolumn{2}{c}{0.53}    &&\multicolumn{2}{c}{0.1} \\
                         &   ($\pt\leq20\GeV$, $\eta\in(1.479,2.5]$)&\multicolumn{2}{c}{0.54} &&\multicolumn{2}{c}{0.62} \\
                         &         ($\pt>20\GeV$, $|\eta|\leq R1$)&\multicolumn{2}{c}{0.95}   &&\multicolumn{2}{c}{0.94}  \\
 			 &    ($\pt>20\GeV$, $\eta\in(R1,1.479]$)&\multicolumn{2}{c}{0.95}    &&\multicolumn{2}{c}{0.85} \\
                         &      ($\pt>20\GeV$, $\eta\in(1.479,2.5]$)&\multicolumn{2}{c}{0.88} &&\multicolumn{2}{c}{0.92} \\\hline
	\multicolumn{2}{l}{Max. relative particle flow isolation $Iso_{PF}/\pt$}& 0.13  & 0.09&&\multicolumn{2}{c}{0.15} \\\hline
	\end{tabular}
	}
	\caption[Electron selection requirements]{Selection requirements imposed on electrons in 
	the 2011 and 2012 analyses. The electrons are split in barrel electrons (EB), for those 
	having a $|\eta|<1.479$ and endcap electrons (EE). The detector based isolation variables 
	are the scalar sum of the energy deposits in the electromagnetic ($\sum_{ECAL}\ET$) and
	hadronic ($\sum_{HCAL}\ET$) calorimeters and also the sum of the transverse momentum in the
	tracker ($\sum_{tracker}\ET$). Those sum are performed around the electron candidate by
	defining a $\Delta R=0.3$ cone around it.}\label{ch6:tab:elecreq}
\end{table}

The electron isolation is computed using a particle flow approach~\cite{CMS-PAS-HIG-12-016} similar 
to the muon case (Equation~\eqref{ch6:eq:muoniso}) but using the density noise estimation $\rho$ and 
the jet area $A_{eff}$ approach from the \FASTJET method described at Section~\ref{ch5:subsec:pileup}
to control the \gls{pileup} environment. This \gls{pileup} correction in the isolation is especially
important in the 2012 analysis.

\begin{equation}
	Iso_{PF} = \sum_{PF} \ET^{ChHad}+max\left(0,\sum_{PF}\pt^{NeutHad}+
	          \sum_{PF}\ET^{\gamma}-\rho\cdot A_{eff}\right)
	\label{ch6:eq:eleciso}
\end{equation}

\subsection{Electron efficiencies and scale factors}\label{ch6:subsec:eleceff}
Analogously to the muon case (Section~\ref{ch6:subsec:muoneff}), the efficiencies for the electron
selection were calculated with the tag and probe method. At each stage of the selection, it has
been evaluated its efficiency in experimental and simulated data, and compared both values to obtain 
the \glspl{sf}. Table~\ref{ch6:tab:elecsf} shows the electron \glspl{sf} for the total selection 
process.
\begin{table}[htpb]
	\centering
	\begin{subtable}[b]{\textwidth}
		\centering
		\resizebox{0.8\textwidth}{!}
		{
		\begin{tabular}{cccc}\hline\hline
			&$0< |\eta| \leq 1.44$ & $1.44< |\eta| \leq 1.56$ & $1.56< |\eta| \leq 2.5$ \\ \hline 
			$10< p_t \leq 15$ & $0.96\pm0.03$ & $0.92\pm0.15$ & $1.0\pm0.7$ \\
			$15< p_t \leq 20$ & $0.96\pm0.02$ & $1.01\pm0.12$ & $1.0\pm0.6$ \\ 
			$20< p_t \leq 25$ & $0.95\pm0.01$ & $0.99\pm0.04$ & $1.01\pm0.02$ \\ 
			$25< p_t \leq 50$ & $0.99\pm0.01$ & $1.01\pm0.02$ & $1.01\pm0.02$ \\ 
			$50< p_t \leq \infty$ & $0.98\pm0.01$ & $1.00\pm0.11$ & $1.01\pm0.02$ \\ \hline
		\end{tabular}
		}
		\caption{Electron scale factors for 2011 data.}
	\end{subtable}\vskip 1em
	\begin{subtable}[b]{\textwidth}
		\centering
		\resizebox{\textwidth}{!}
		{
		\begin{tabular}{cccccc}\hline\hline
			&$0< |\eta| \leq 0.8$ & $0.8< |\eta| \leq 1.44$ & $1.44< |\eta| \leq 1.56$
			    & $1.56< |\eta| \leq 2.0$ & $2.0< |\eta| \leq 2.5$ \\ \hline 
			$10< p_t \leq 15$ & $0.66\pm0.02$ 
			        & $0.73\pm0.03$ & $0.81\pm0.09$ & $0.61\pm0.04$ & $0.64\pm0.03$ \\ 
			$15< p_t \leq 20$ & $0.901\pm0.009$ 
			        & $0.942\pm0.014$ & $0.86\pm0.07$ & $0.83\pm0.02$ & $0.76\pm0.02$ \\ 
			$20< p_t \leq 30$ & $0.943\pm0.003$ 
			        & $0.950\pm0.004$ & $0.92\pm0.01$ & $0.923\pm0.005$ & $0.972\pm0.006$ \\ 
			$30< p_t \leq 40$ & $0.9614\pm0.0009$ 
			        & $0.94\pm0.15$ & $0.965\pm0.005$ & $0.9249\pm0.0015$ & $1.0\pm0.6$ \\ 
			$40< p_t \leq 50$ & $0.9763\pm0.0006$ 
			        & $0.97\pm0.09$ & $0.954\pm0.004$ & $0.9608\pm0.0011$ & $0.982\pm0.002$ \\ 
			$50< p_t \leq \infty$ & $0.9742\pm0.0012$ 
			          & $0.9698\pm0.0013$ & $0.986\pm0.009$ & $1.0\pm0.2$ & $0.970\pm0.003$ \\ \hline
		\end{tabular}
		}
		\caption{Electron scale factors for 2012 data}
	\end{subtable}
	\caption[Electron scale factors]{Electron reconstruction, isolation and identification 
		scale factors applied to the Monte Carlo sample events in bins of pseudorapidity
		an transverse momentum. Errors are statistical.}\label{ch6:tab:elecsf}
\end{table}

\section{Neutrino selection}\label{ch6:sec:neutrinosel}
The neutrino leaves no signal in the detector systems, nevertheless is possible to infer its 
presence by the \MET observable (introduced and described in Section~\ref{ch4:subsec:MET}) which
is constructed employing the energy-momentum conservation in the transverse plane of the proton 
beams. The signal events are expected to have a significant amount of \MET because of the \W 
leptonic decay $\W\to\ell\nu_{\ell}$ whilst the main backgrounds (as Drell-Yan) are restricted to 
measure spurious \MET, coming from detector resolution and/or from \MET not properly reconstructed.

The analysis makes use of the particle flow algorithm to estimate the \MET which takes advantage 
of the full detector information to reconstruct the event decay products. The \MET is estimate by 
subtracting vectorially all the particle flow candidates in the transverse plane,
\begin{equation}
	\left(\VEtmiss\right)_{PF} = - \sum_{\text{PF-cand.}}\vec{p}_T({PF})
	\label{ch6:eq:pfmet}
\end{equation}
Notice that this observable is considerably sensitive to the \gls{pileup} environment and some 
corrections are needed to avoid fake missing transverse energy reconstruction. Further details
about such corrections and performance behaviour description have been explained at 
Section~\ref{ch4:subsec:MET}.

\section{Event selection}\label{ch6:sec:eventselection}
The \WZ cross section analysis is essentially based in counting events. Counting how many events
we have recorded that have the expected signature of the \WZ production when the gauge bosons 
decay leptonically. The signal was classified in four different categories depending the final state 
leptons considered (see Section~\ref{ch6:sec:topology}) allowing us to perform four independent 
counting analysis using four orthogonal samples. The leptonic signature 
$\ell^{\pm}\ell'^{+}\ell'^-$ is going to be defined by the gauge bosons decays 
$\W\rightarrow\ell^{\pm}\nu_{\ell}$ and $\Z\rightarrow\ell'^+\ell'^-$, being these final state 
leptons high-\pt isolated muons and electrons. In the previous section we have focused precisely on 
optimising the selection requirements in order to select such kind of leptons. Notice that we are 
also considering as signal the $\tau$ decay of the gauge bosons as long as they decay 
leptonically, getting a muon or electron in the final state. In Chapter~\ref{ch7} we will see how
to deal with this kind of signal.

The three high-\pt lepton final state is in fact a remarkably restrictive requirement. We have 
detailed the potential sources of noise in Section~\ref{ch6:sec:topology} and we have anticipated that
the only (important) irreducible background is the $\Z\Z$ production which, however, has a 
production rate almost an order of magnitude lower than the \WZ process. Therefore, the analysis has
such a clean signature than the signal-to-noise ratio is considerably high. Furthermore, the 
presence of the \Z resonance and the \W allows to reduce the remnant instrumental background. The
\Z resonance is tested using the invariant mass of the two same flavour, opposite charged leptons,
being the invariant mass defined as,
\begin{equation}
	M_{\ell_1\ell_2}=\sqrt{E_{1}^2+E_{2}^2-|\mathbf{p_1}+\mathbf{p_2}|^2},
	\label{ch6:eq:invariantmass}
\end{equation}
where $E_i$ is the energy and $\mathbf{p}_i$ the 3-momentum of the i-lepton. In the highly
relativistic regime of the collider experiments, the energy of the lepton is much higher than its
rest mass ($E\gg m$), therefore the previous expression becomes
\begin{equation}
	M_{\ell_1\ell_2}=2p_{t_1}p_{t_2}\left(cosh\left(\eta_1-\eta_2\right)-cos\left(\phi_1-\phi_2\right)\right),
	\label{ch6:eq:invariantmassHEP}
\end{equation}
being $\eta_i$ and $\phi_i$ the pseudorapidity and the azimuthal angle of the i-lepton, 
respectively.

Since it is not possible to reconstruct the invariant mass of the \W gauge boson because of the 
neutrino's presence, which it is not leaving any electronic signal in the detector, the transverse
mass is defined through the missing transverse energy and a lepton to build the \W mass in the 
transverse plane~\cite{PhysRevD.86.010001},
\begin{equation}
	M_T(W)=\sqrt{2\pt^{\ell}\MET\left(1-cos\phi_{\ptvec^\ell,\VEtmiss}\right)},
\end{equation}
where $\phi_{\ptvec^\ell,\VEtmiss}$ is the angle between the transverse momentum vector of the 
lepton with the transverse missing energy vector. Note that this observable is invariant under 
Lorentz boosts in the $z$ direction.

\paragraph*{}
The analysis is performed in three main sequential steps filtering events that have to fulfil
some requirements. The main stages of the analysis are:
\begin{enumerate}
	\item Lepton preselection
		\begin{itemize}
			\item The event contains exactly three leptons fulfilling the requirements 
				of Sections~\ref{ch6:sec:muonselection}.
			\item Two of the event leptons must have at least a transverse momentum 
				higher than 20~\GeV and the third higher than 10~\GeV. 
			\item The leptons must be inside the detector acceptance which implies 
				$|\eta|<2.4$~($2.5$) for muons (electrons).
		\end{itemize}
		As it will show along this chapter, these requirements filters most of the 
		instrumental background events with more than one fake lepton, meaning $QCD$ and 
		$W+Jets$ mostly, leaving a sample mainly composed by instrumental background with 
		one fake lepton,~\ie $Z/\gamma^*+jets$ and $\ttbar$, besides of the irreducible 
		background.
	\item \Z candidate selection
		\begin{itemize} 
			\item The event must contain two opposite-charged, same flavour leptons.
				The invariant mass of these leptons must be compatible with the
				\Z resonance, meaning that the invariant mass of the dilepton
				system must be inside $M_{Z}\pm20~\GeV$, where the \Z nominal
				mass~\cite{PhysRevD.86.010001} is $M_{Z}=~91.1876~\GeV$. In case that
				more than one lepton pair satisfies these requirements, the pair 
				with invariant mass closer to the nominal one is chosen.
		\end{itemize}
		These stage of the analysis rejects the remnant background without a real \Z 
		resonance: $\ttbar$ and single top and $WW$.
	\item \W candidate selection
		\begin{itemize}
			\item The available third lepton not associated with the \Z is required to
				have a transverse momentum higher than 20~\GeV. This implies that
				whenever a lepton has a $\pt<20~\GeV$ it must come from the \Z 
				decay.
			\item This lepton is required to be outside a cone of $\Delta R= 0.1$ 
				around either of the \Z candidate leptons. The requirement is 
				rejecting asymmetric internal photon
				conversions\footnote{Asymmetric 
				conversions~\cite{Dalitz:1951aj}
				is a process whereby one lepton takes most of the photon energy
				and the second lepton is very soft and not measured. There are 
				two types of photon conversion important for this analysis. The 
				first type is an external conversion in which a photon radiated 
				by the collision interacts with the detector material and produces
				a lepton pair, primarily electrons but very rarely muons. The second
				type of	conversion is the internal photon conversion, where the photon
				is virtual and does not interact with the detector. Internal
				photon conversion can produce muons almost as often than 
				electrons.} mainly from $\Z/\gamma^*+Jets$. The \Z lepton candidate,
				which radiates the virtual photon, is going to be close to the 
				lepton created from the asymmetric conversion~\cite{PhysRev.98.1355}.
			\item The missing transverse energy of the event must be higher 
				than $30~\GeV$ in order to take into account the undetected 
				neutrino. 
		\end{itemize}
		This final selection tries to reject events with the presence of spurious \MET
		due to its resolution as the the Drell-Yan process and the $ZZ$ background. 
\end{enumerate}
Any event that does not fulfil these requirements is rejected. These cuts are applied to the 
experimental data previously obtained with the double lepton trigger paths (see 
Section~\ref{ch6:sec:onlineselection}) and to the simulated data\footnote{The simulated data used 
in this analysis are explained in detail in Chapter~\ref{ch7}} to which the 
corrections described before (\acrlong{sf}[s], trigger weights, \gls{pileup} reweighting, etc.) have
been applied. The simulated predictions are compared with the experimental data at each stage
of the analysis mainly by comparing the event distributions of sensitive observables. 
\begin{figure}[htbp]
	\centering
	\begin{subfigure}[b]{0.35\textwidth}
		\includegraphics[width=\textwidth]{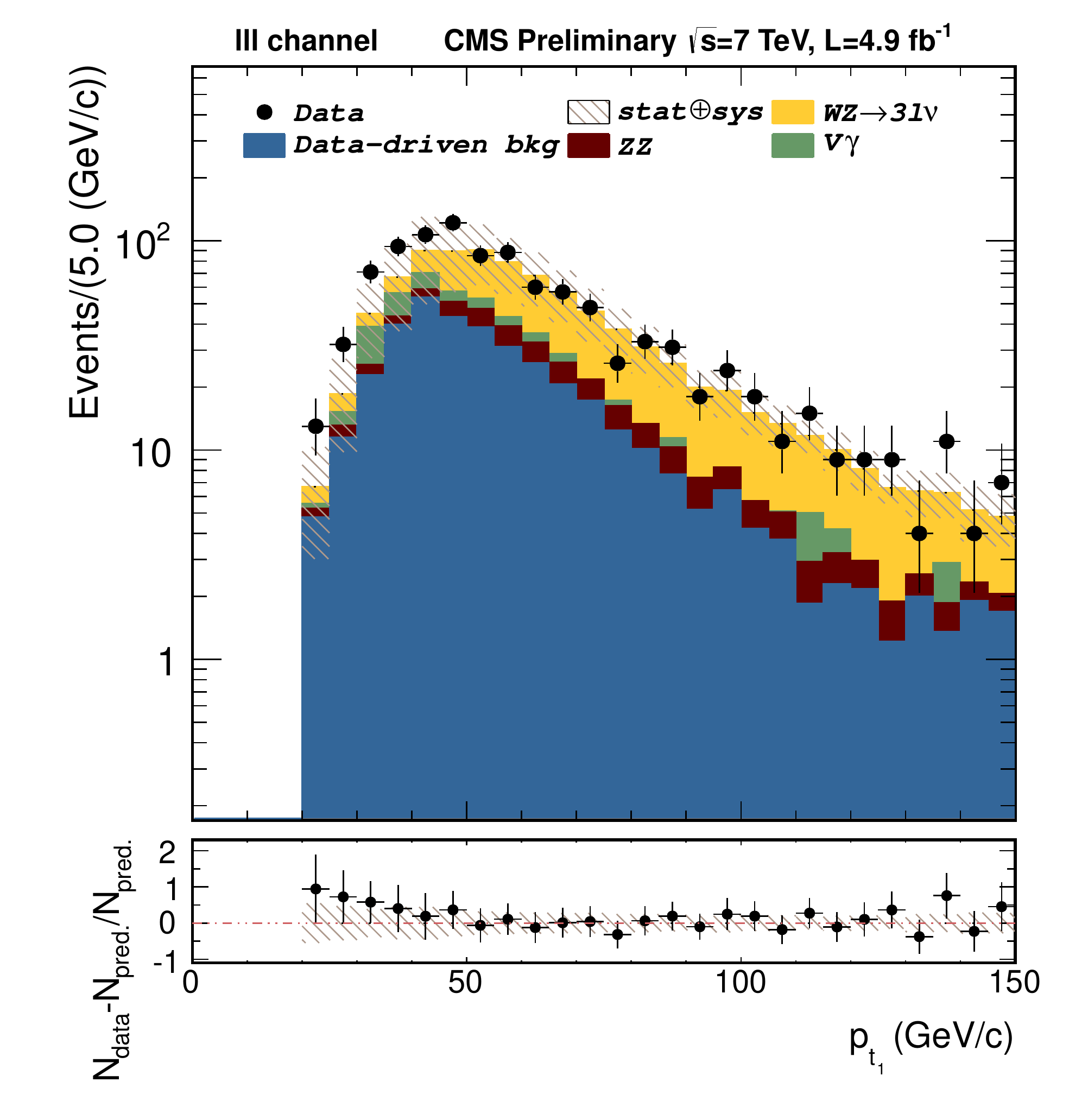}
		\caption{Higher-\pt lepton transverse momentum}
	\end{subfigure}\quad
	\begin{subfigure}[b]{0.35\textwidth}
		\includegraphics[width=\textwidth]{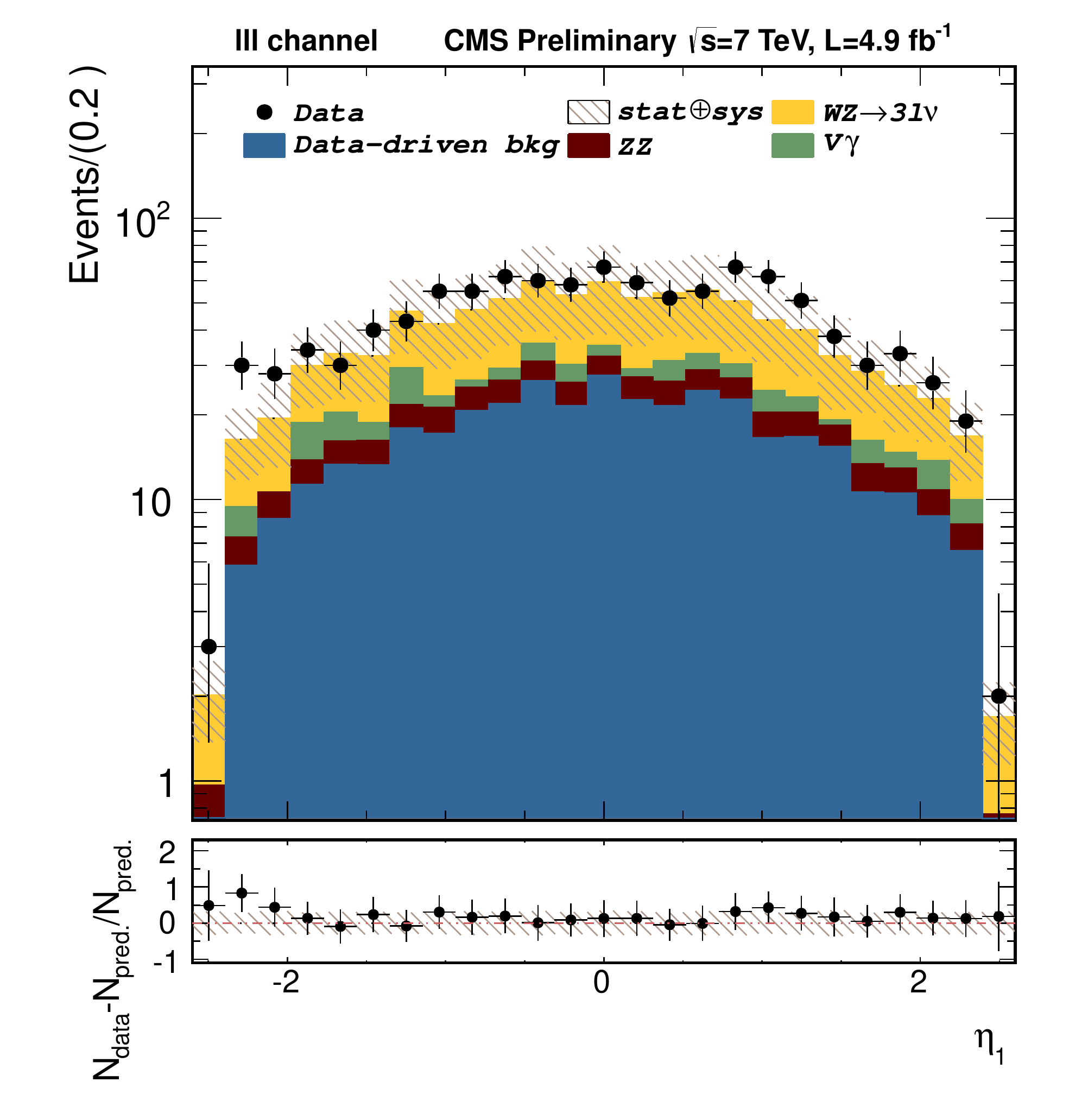}
		\caption{Higher-\pt lepton pseudorapidity}
	\end{subfigure}
	\begin{subfigure}[b]{0.35\textwidth}
		\includegraphics[width=\textwidth]{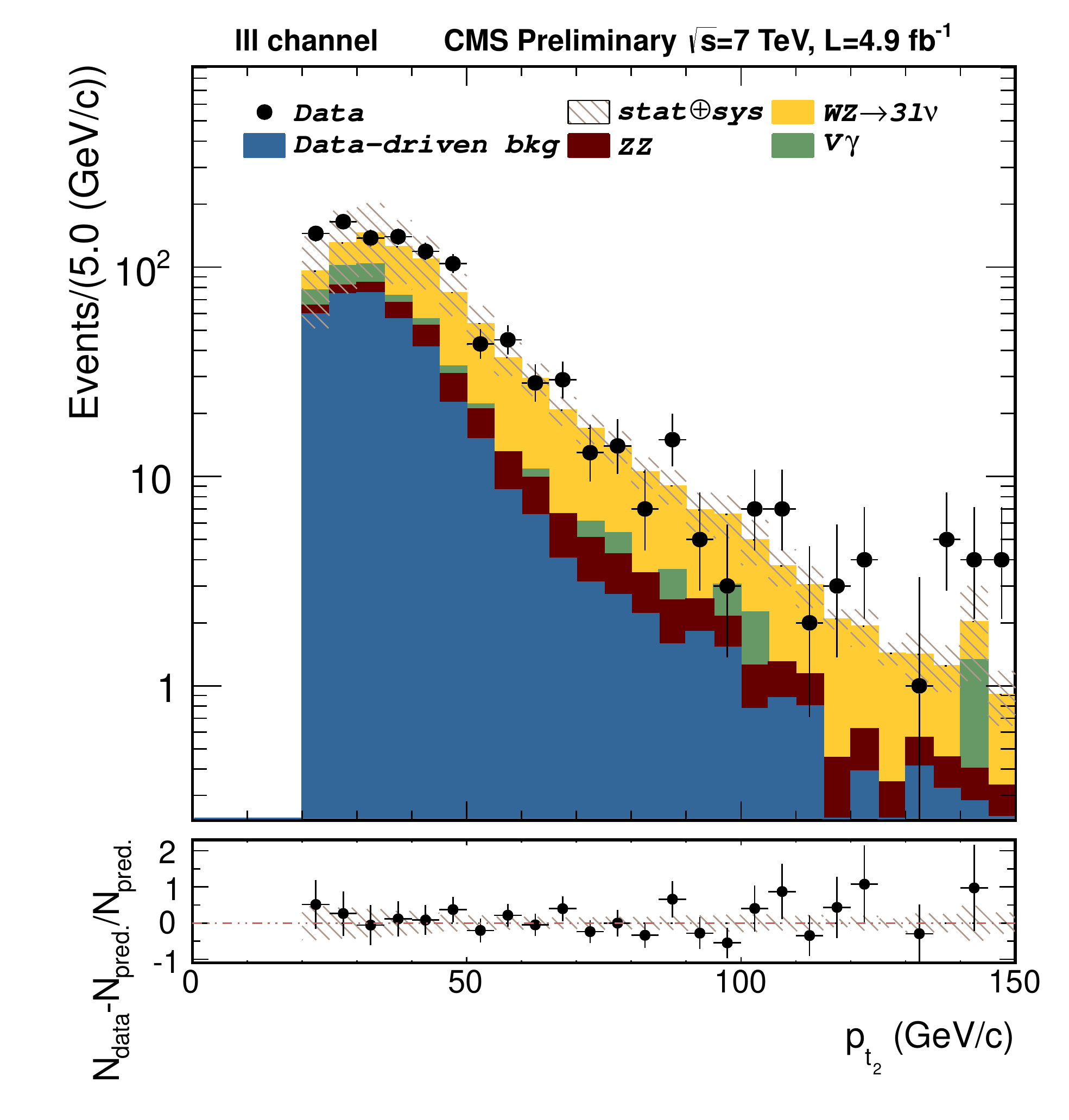}
		\caption{Middle-\pt lepton transverse momentum}
	\end{subfigure}\quad
	\begin{subfigure}[b]{0.35\textwidth}
		\includegraphics[width=\textwidth]{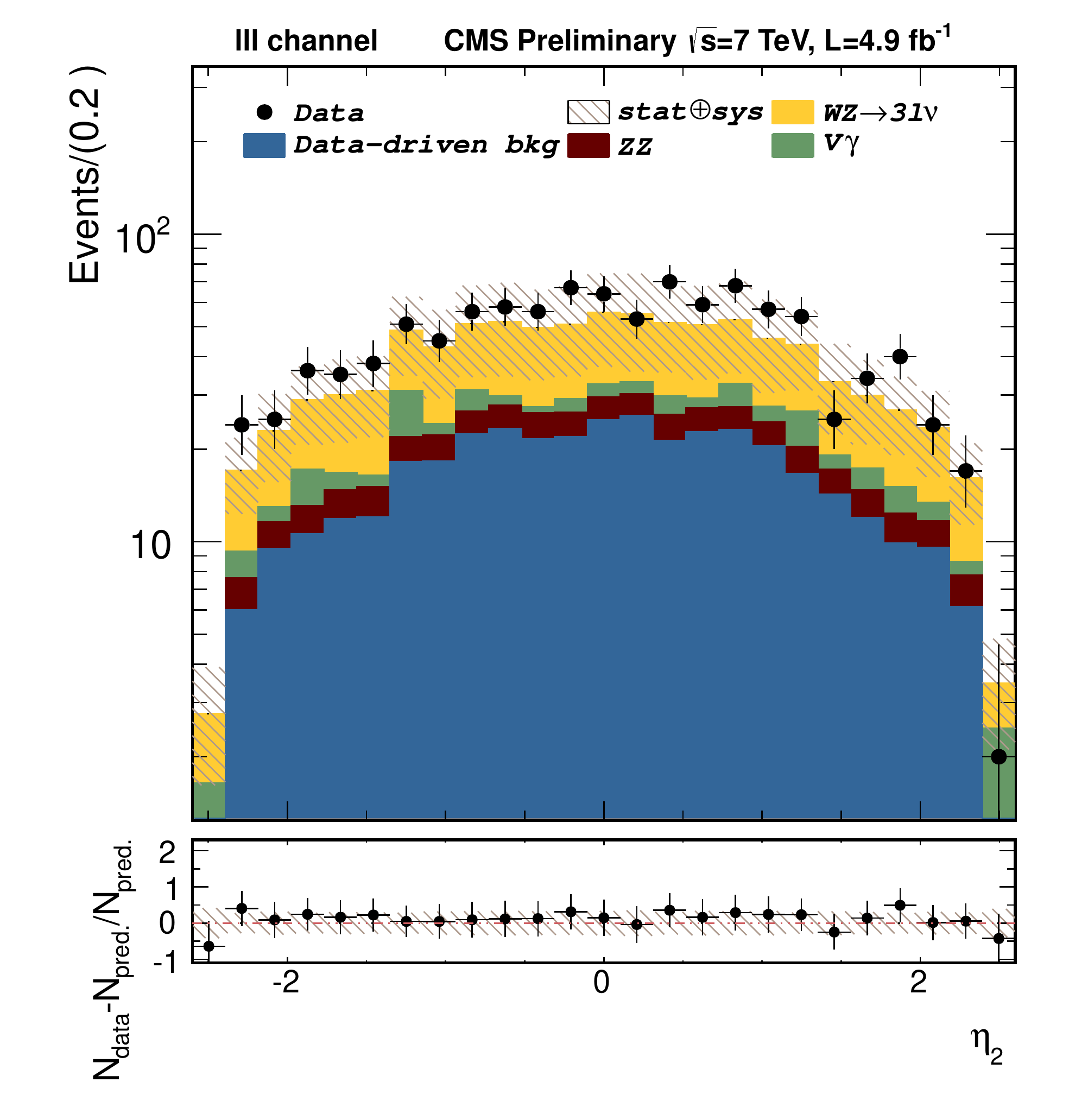}
		\caption{Middle-\pt lepton pseudorapidity}
	\end{subfigure}
	\begin{subfigure}[b]{0.35\textwidth}
		\includegraphics[width=\textwidth]{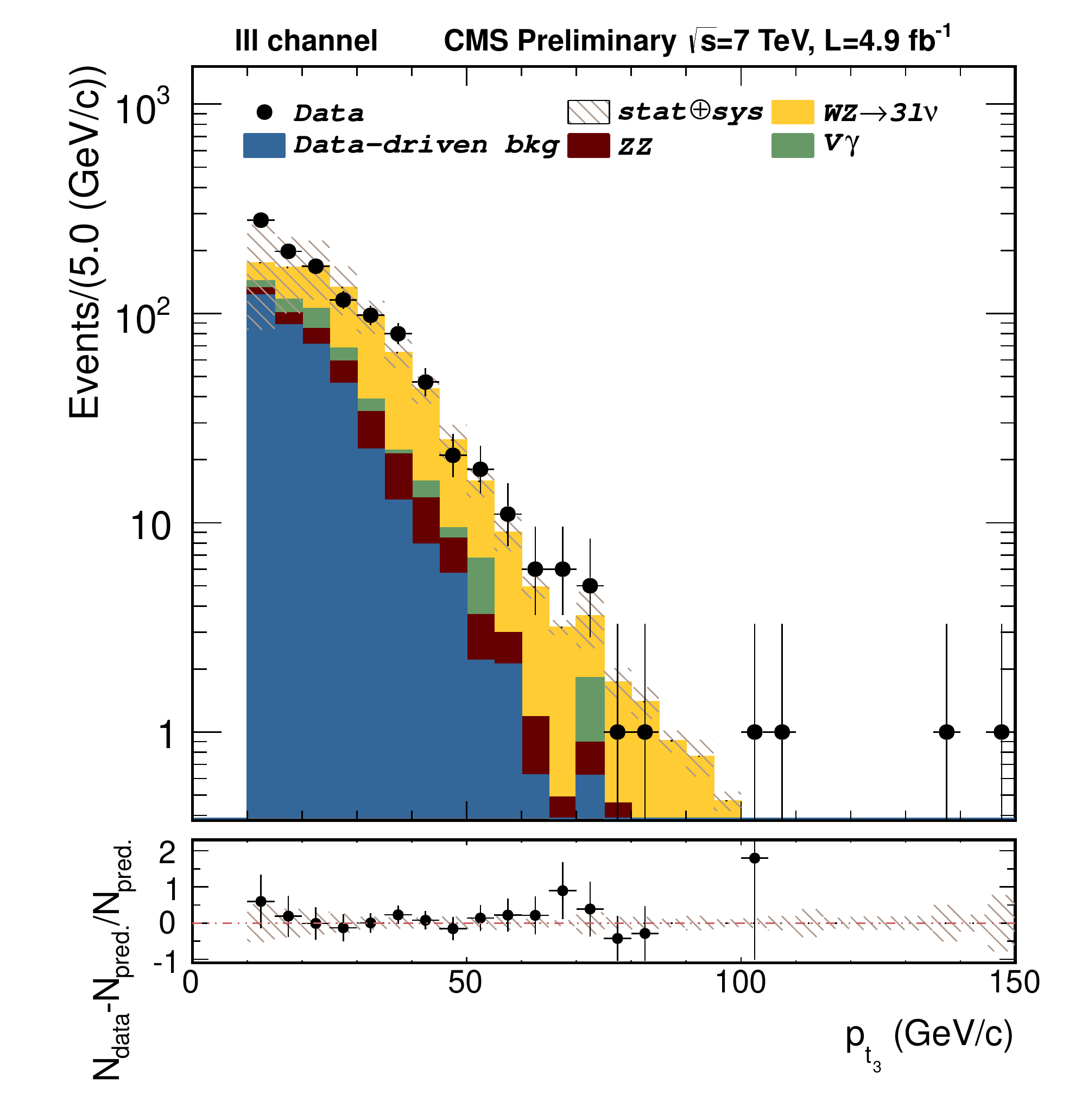}
		\caption{Trailing-\pt lepton transverse momentum}
	\end{subfigure}\quad
	\begin{subfigure}[b]{0.35\textwidth}
		\includegraphics[width=\textwidth]{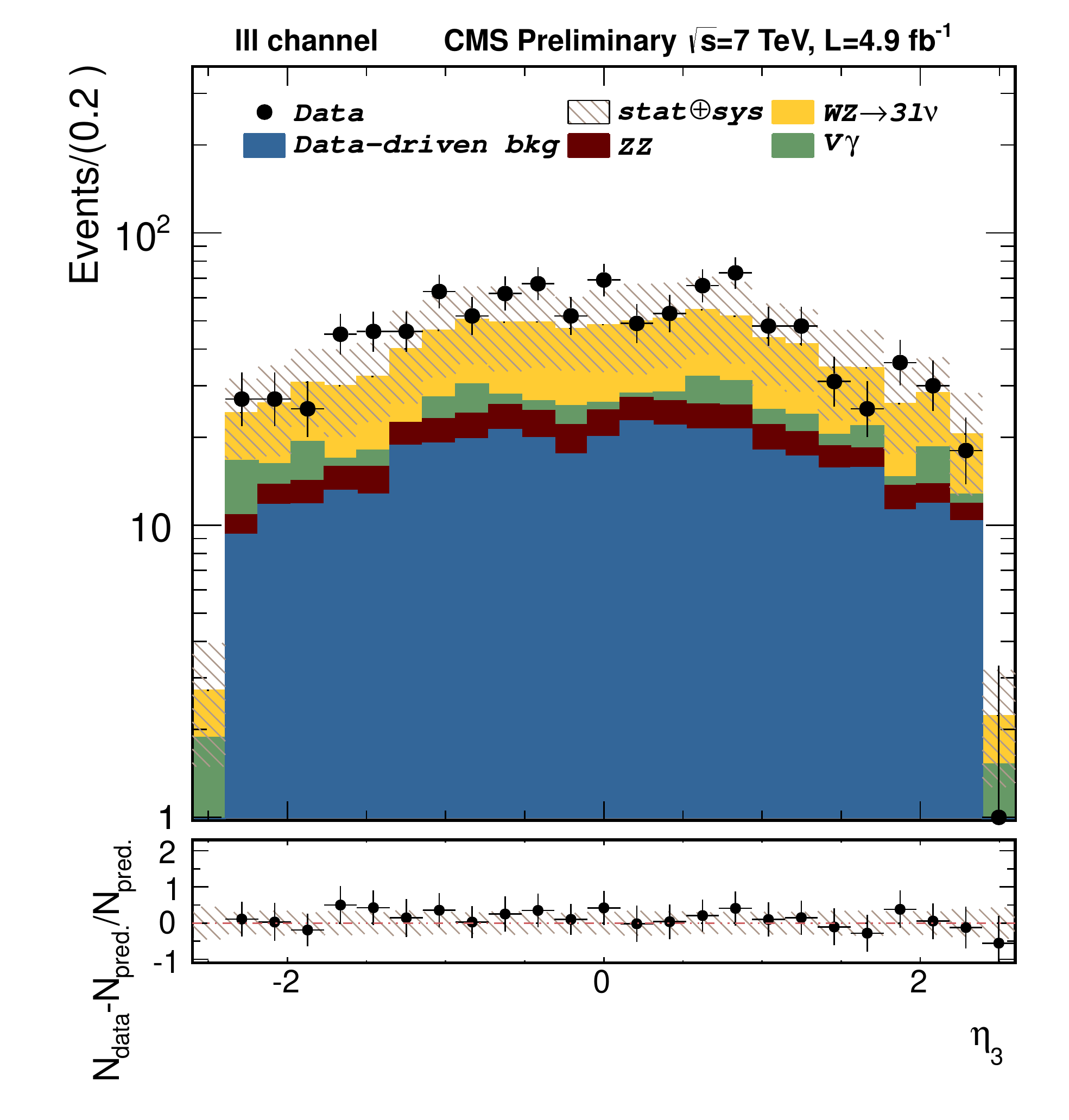}
		\caption{Trailing-\pt lepton pseudorapidity}
	\end{subfigure}
	\caption[Lepton kinematic distributions at preselection level]{Lepton kinematic 
		distributions at preselection level. The distributions are built adding up
		the four final state channels. Statistical and systematic uncertainties are 
		included. The data versus \gls{mc} prediction is shown in the lower plot. Data 
		corresponding to 2011 analysis.}
		\label{ch6:fig:preselection}
\end{figure}
Figures~\ref{ch6:fig:preselection} show the transverse momentum and pseudorapidity of the three 
leptons selected imposing the preselection requirements. It can be observed that the dominant 
background consists of the so called \emph{Data-driven background} which is primarily composed by
$Z/\gamma^*+Jets$ and \ttbar (see Chapter~\ref{ch7} where the background treatment is detailed). 
The instrumental background coming from more than one fake lepton is already mitigate at the 
preselection level. The number of signal events is essentially of the same order than the 
background.
\begin{figure}[htb]
	\centering
	\begin{subfigure}[b]{0.45\textwidth}
		\includegraphics[width=\textwidth]{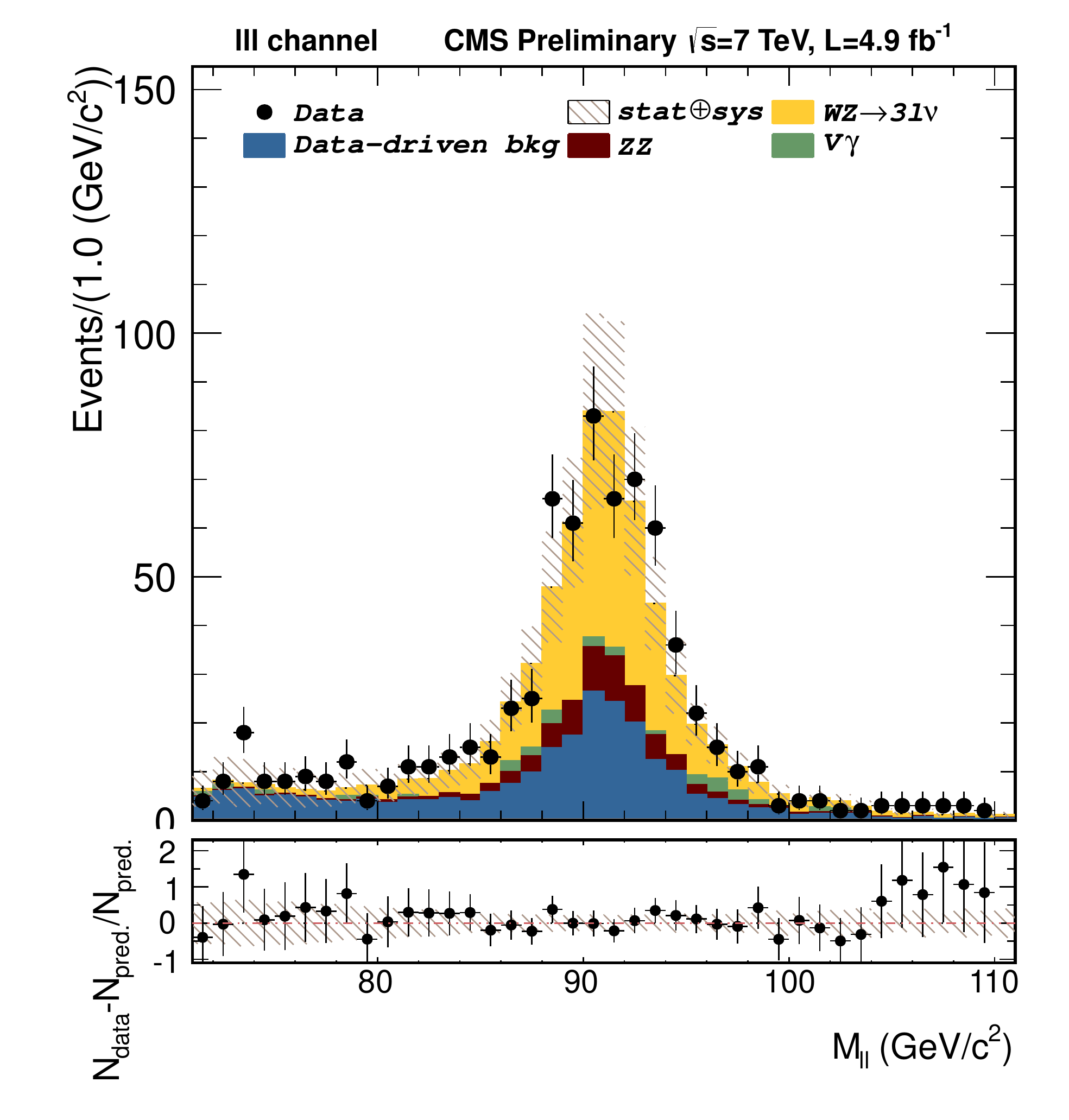}
		\caption{Invariant mass of the \Z-candidate dilepton system}
	\end{subfigure}\quad
	\begin{subfigure}[b]{0.45\textwidth}
		\includegraphics[width=\textwidth]{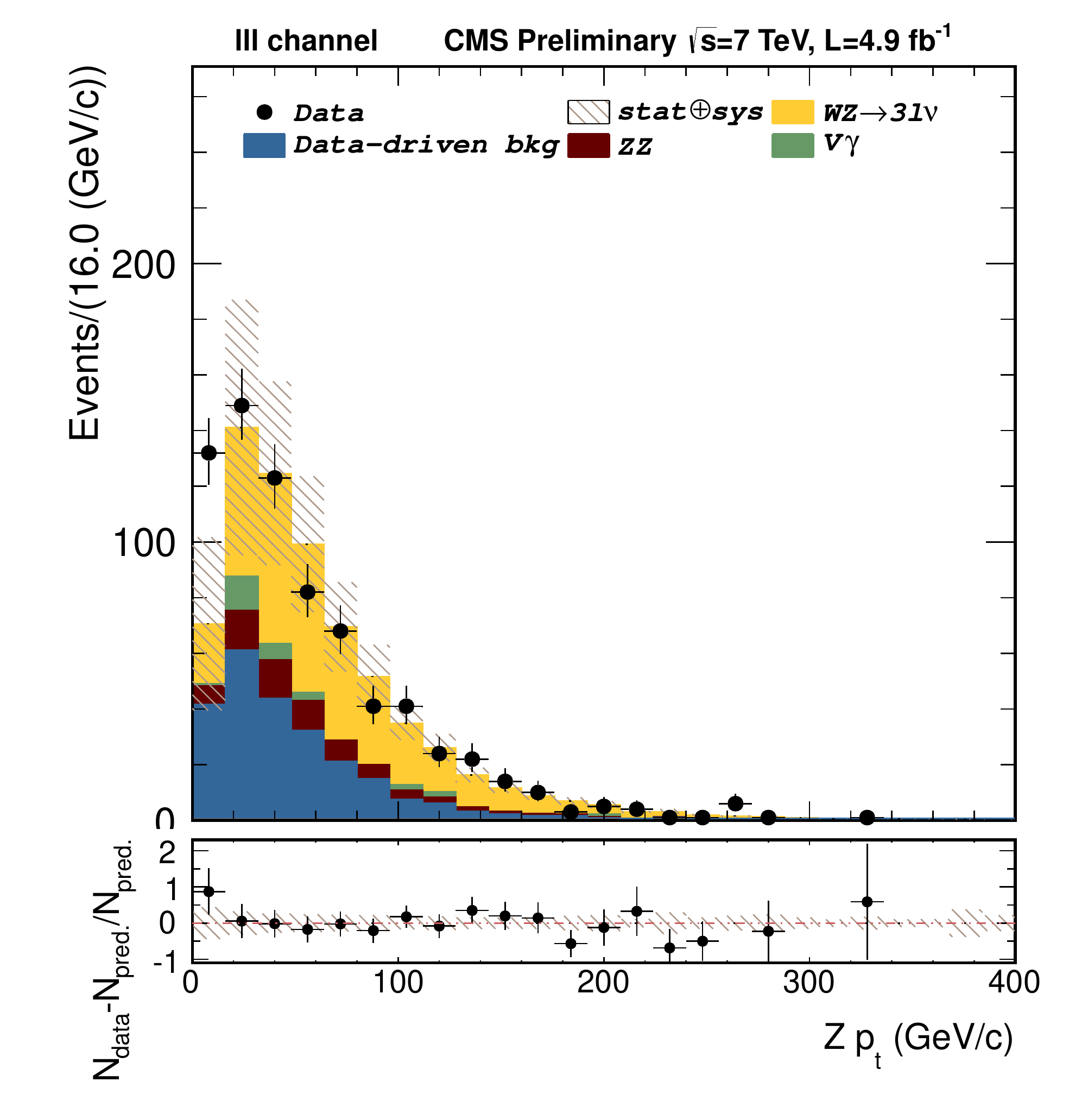}
		\caption{Transverse momentum of the \Z-candidate dilepton system}
	\end{subfigure}
	\caption[\Z-candidate dilepton system distribution at \Z selection stage]{\Z-candidate
		dilepton system related distributions after applied the \Z selection step. The
		distributions are built adding up the four final state channels. Statistical and 
		systematic uncertainties are included. The experimental data versus the \gls{mc}
		prediction is shown in the lower plot. Data corresponding to 2011 analysis.}
		\label{ch6:fig:zcand}
\end{figure}

\begin{figure}[htb]
	\centering
	\begin{subfigure}[b]{0.45\textwidth}
		\includegraphics[width=0.9\textwidth]{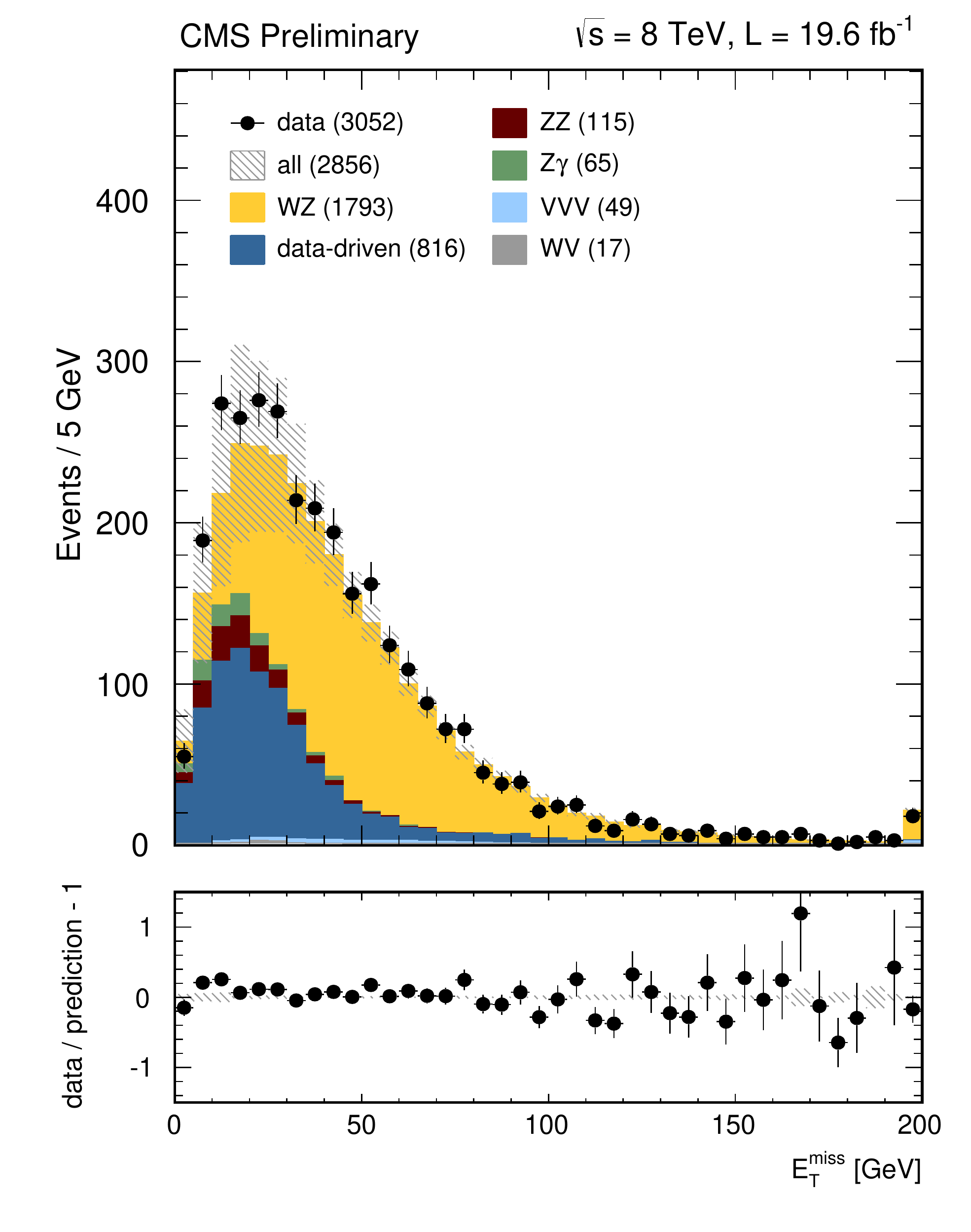}
		\caption{\MET distribution}\label{ch6:subfig:zcand::MET}
	\end{subfigure}
	\begin{subfigure}[b]{0.45\textwidth}
		\includegraphics[width=0.9\textwidth]{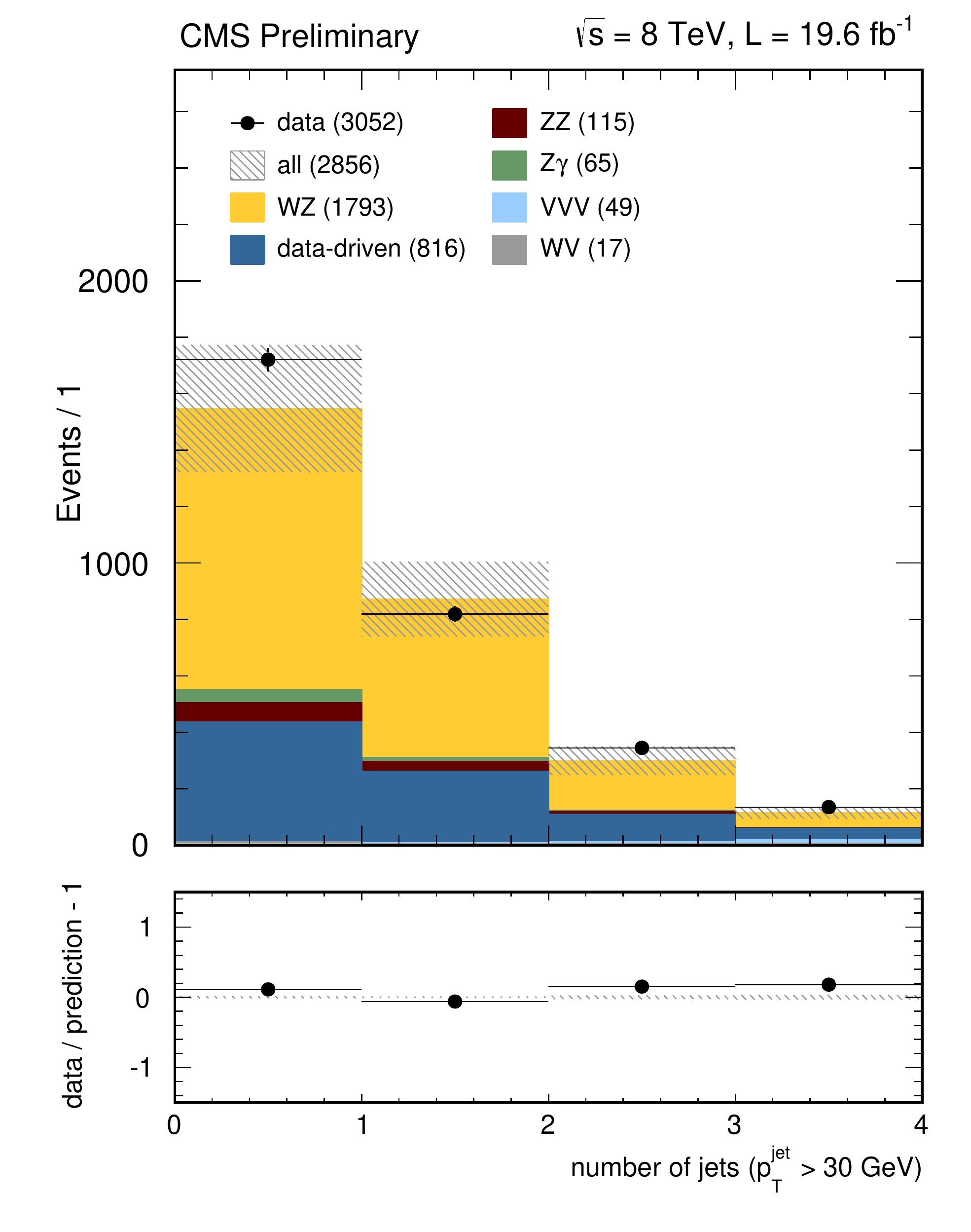}
		\caption{Number of jets distribution}
	\end{subfigure}
	\caption[\MET and number of jets distribution at \Z selection stage]{Distributions at the \Z-candidate
		selection stage. The distributions are built adding up the four final state channels. 
		Statistical and systematic uncertainties are included. The experimental data versus
		the \gls{mc} prediction is shown in the lower plots. Data corresponding to 2012 
		analysis.}\label{ch6:fig:zcand_2012}
\end{figure}

After the \Z candidate selection, it is possible to build the invariant mass of the dilepton system
selected and check some other interesting observables. At this point, the non-peaking backgrounds 
should have diminished and it can be observed that the predicted signal is primarily composing the 
experimental data. See for instance, Figures~\ref{ch6:fig:zcand} where the invariant mass 
and the transverse momentum of the dilepton system is plotted.
The Figure~\ref{ch6:subfig:zcand::MET} displays the \MET distribution at \Z-candidate selection. This 
figure illustrates the \MET cut which is applied when requiring the \W candidate; the remnant 
background populates the low \MET region, likely due to the non-presence of real \MET, whilst the 
signal trends to be in higher \MET regions.

\begin{figure}[htb]
	\centering
	\begin{subfigure}[b]{0.45\textwidth}
		\includegraphics[width=\textwidth]{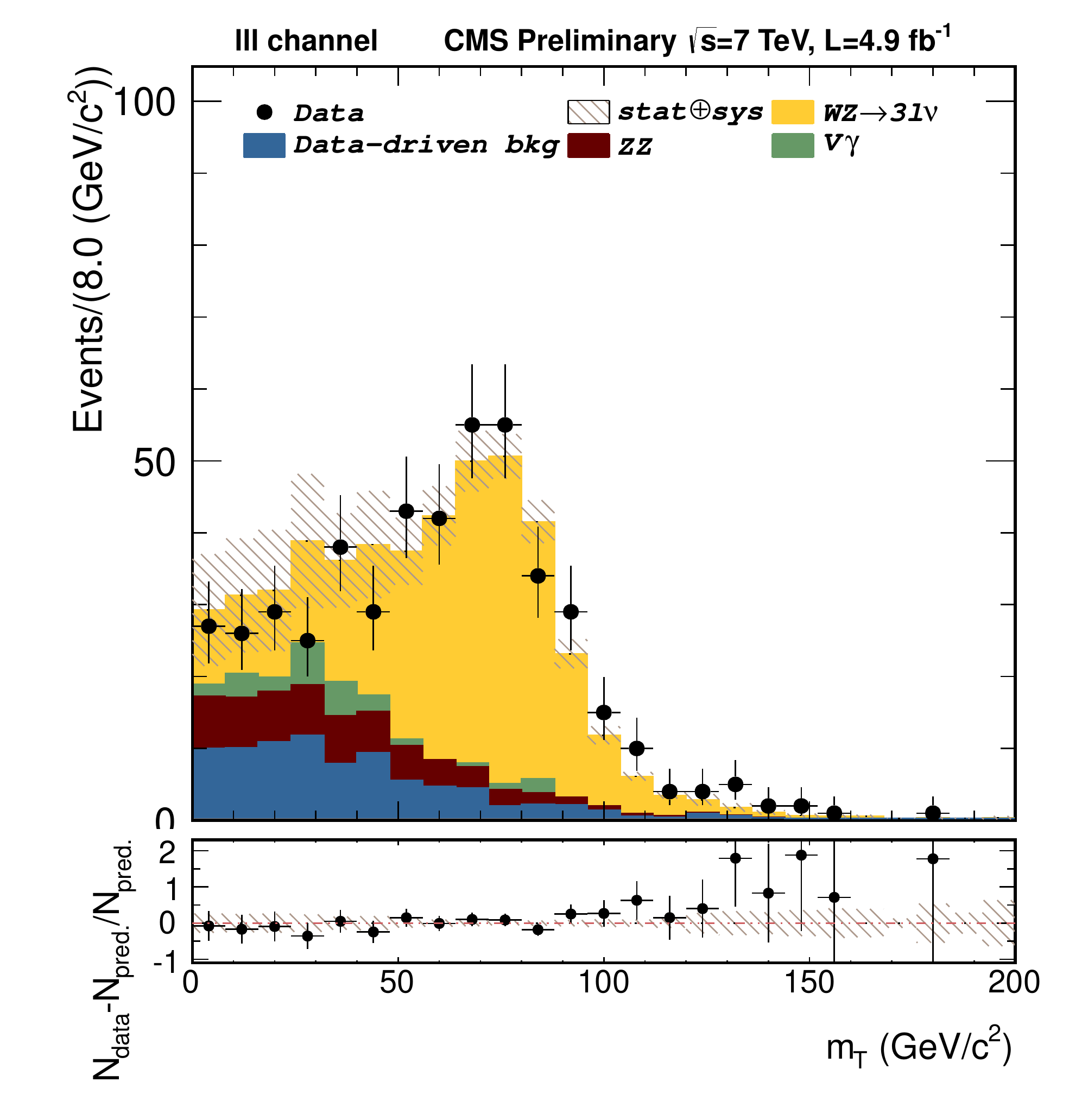}
		\caption{All the \W-selection requirements are applied but the \MET 
		cut.}\label{ch6:fig:wcand::MT}
	\end{subfigure}\quad
	\begin{subfigure}[b]{0.45\textwidth}
		\includegraphics[width=\textwidth]{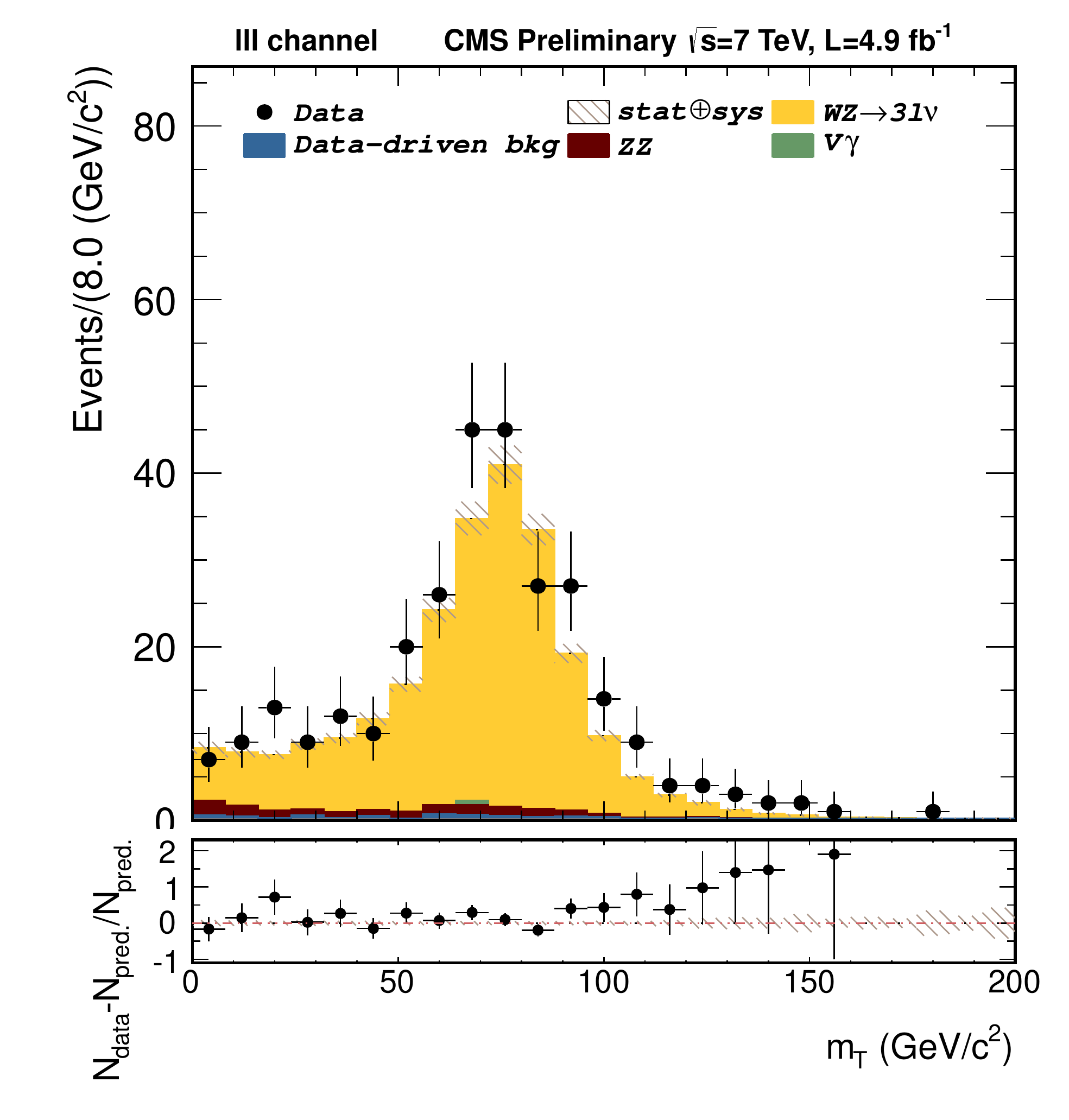}
		\caption{All the \W-selection requirements are applied}
		\label{ch6:fig:wcandfinal::MT}
	\end{subfigure}
	\caption[Transverse mass at \W selection stage]{Transverse mass of the \W lepton candidate
		and the \MET at \W selection stage. The distribution is built adding up the four 
		final state channels. Statistical and systematic uncertainties are included. The 
		experimental data versus the \gls{mc} prediction is shown in the lower plot. Data 
		corresponding to 2011 analysis.}\label{ch6:fig:wcand::TransverseMass}
\end{figure}

\begin{figure}[!htbp]
	\centering
	\begin{subfigure}[b]{0.45\textwidth}
		\includegraphics[width=0.9\textwidth]{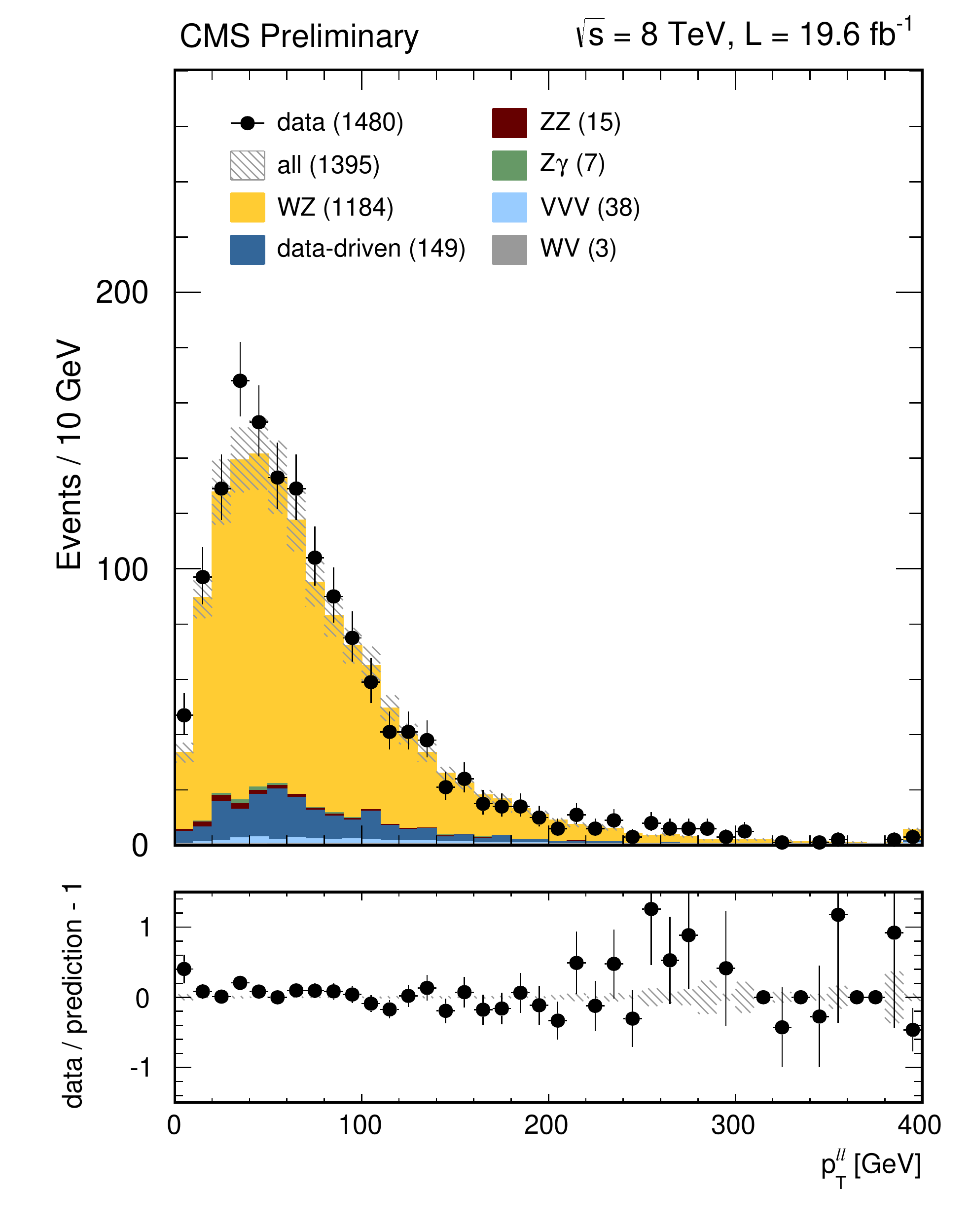}
		\caption{Transverse momentum of the \Z system}
	\end{subfigure}\quad
	\begin{subfigure}[b]{0.45\textwidth}
		\includegraphics[width=0.9\textwidth]{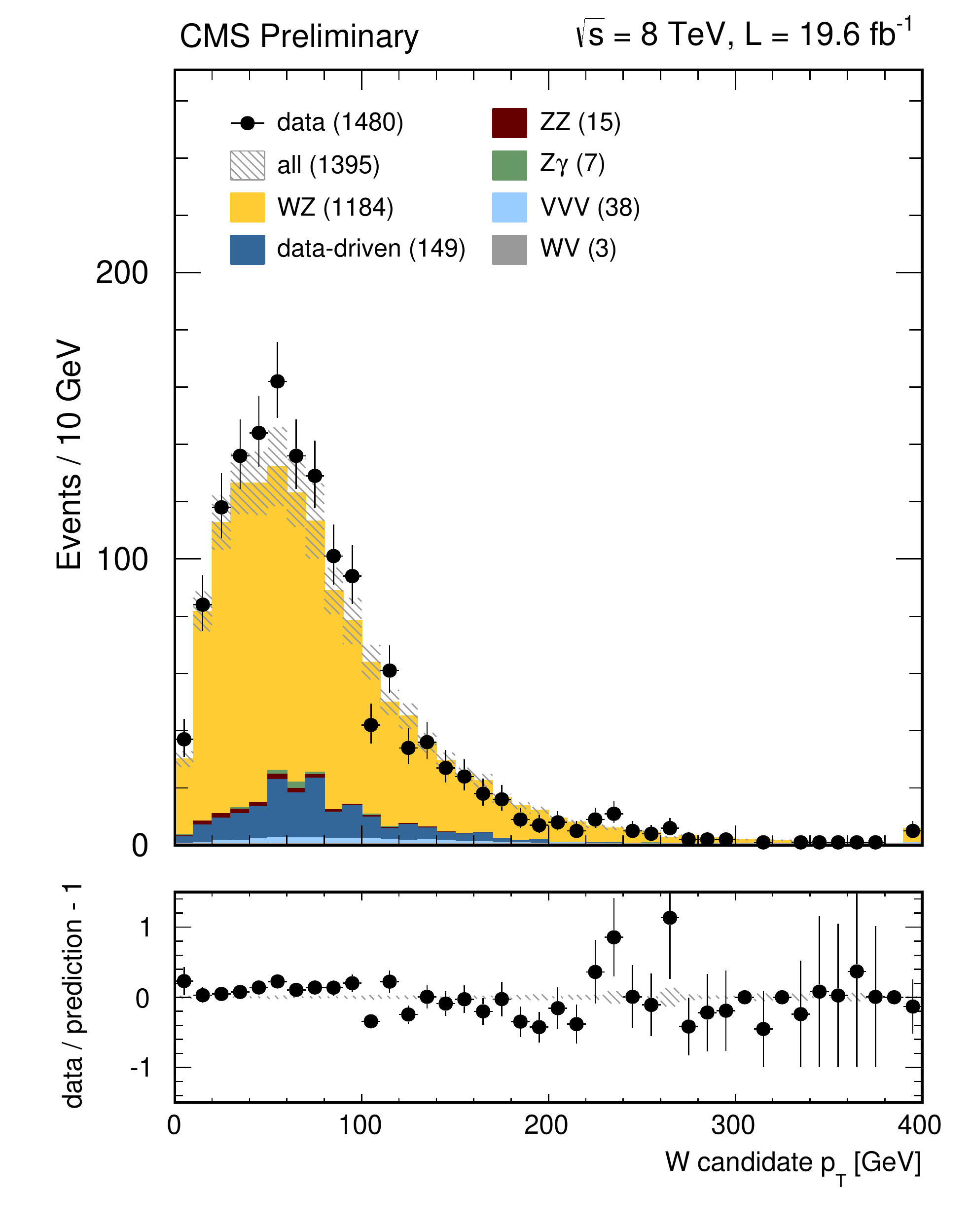}
		\caption{Transverse momentum of the \W system}
	\end{subfigure}
	\caption[Transverse momentum of the \W and \Z candidates]{Distributions after all the
		selection stages are applied. The distributions are built adding up the four final 
		state channels. Statistical and systematic uncertainties are included. The 
		experimental data versus the \gls{mc} prediction is shown in the lower plots. 
		Data corresponding to 2012 analysis.}\label{ch6:fig:wcand::Pt_WandZ}
\end{figure}

Using the remaining lepton which is going to be assigned as \W candidate, it is possible to build 
the transverse mass of the \MET and this \W-candidate lepton. The Figure~\ref{ch6:fig:wcand::MT}
shows this transverse mass just before requiring the \MET cut. The $m_T(\ell_W,\MET)$ distribution
exhibits the Jacobian peak in the \WZ \gls{mc} prediction whilst the background are mainly focused
to low regions because of the spurious \MET, due to resolution, used to build the observable. In the
Figure~\ref{ch6:fig:wcandfinal::MT} when all the requirements of the \W selection stage are applied,
and in particular the \MET cut, almost all the background has disappeared, clearing the transverse 
mass around the \W invariant mass. Figures~\ref{ch6:fig:wcand::Pt_WandZ} show the transverse 
momentum of the \Z and \W system after all the selection is applied.

\begin{figure}[!htbp]
	\centering
	\begin{subfigure}[b]{0.45\textwidth}
		\includegraphics[width=\textwidth]{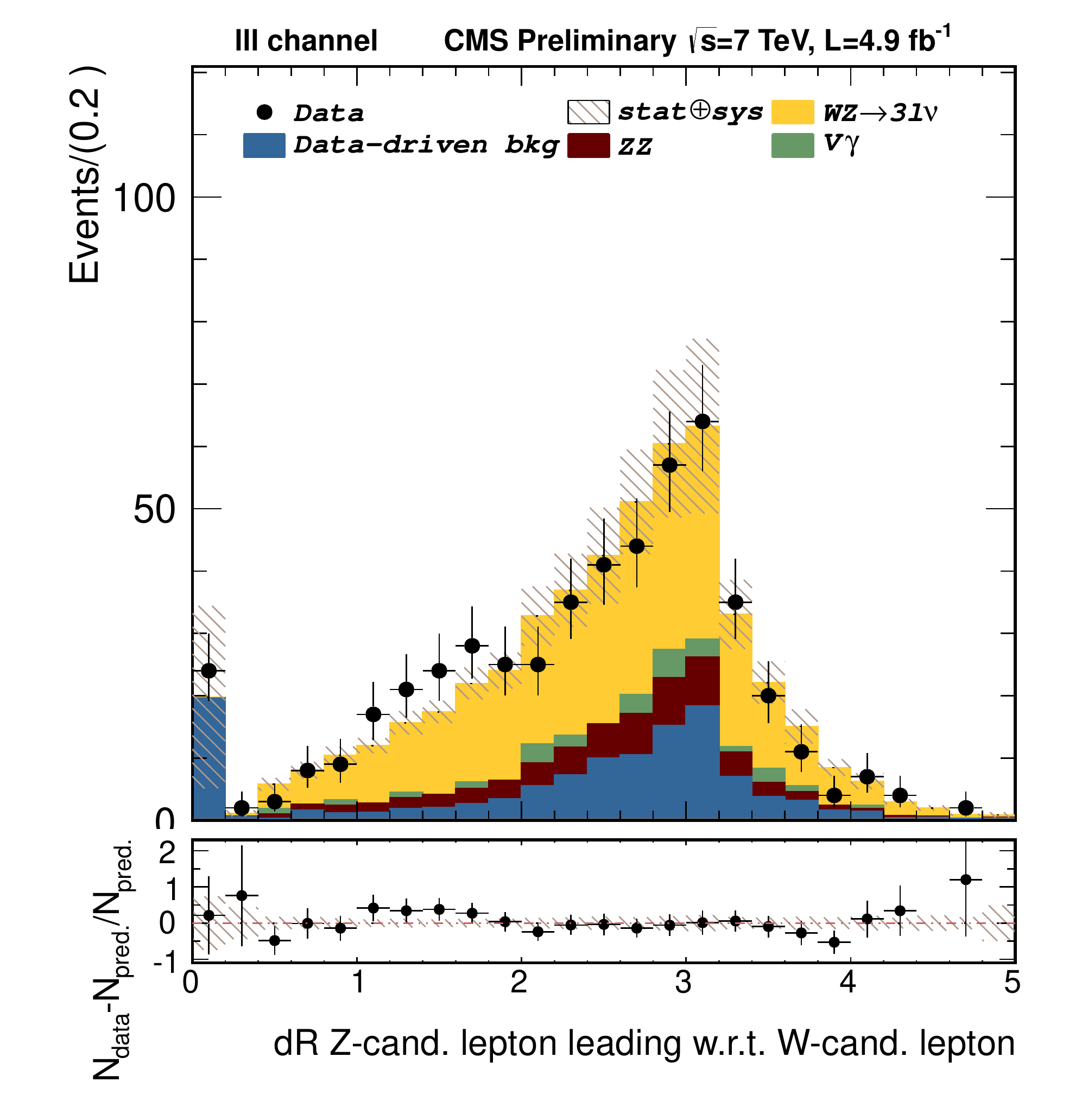}
		\caption{Leading \Z lepton candidate checked with the \W lepton}
	\end{subfigure}\quad
	\begin{subfigure}[b]{0.45\textwidth}
		\includegraphics[width=\textwidth]{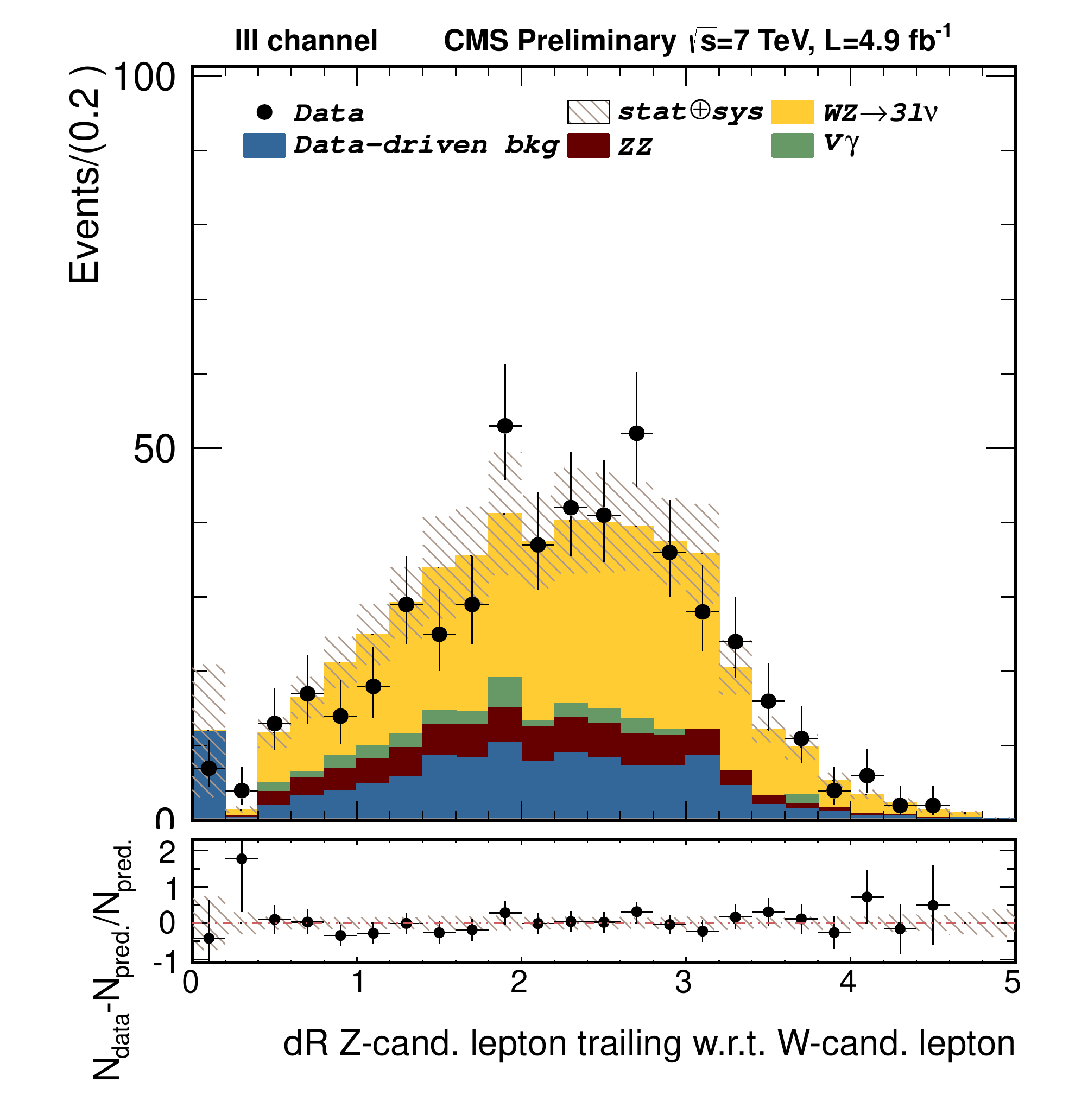}
		\caption{Trailing \Z lepton candidate checked with the \W lepton}
	\end{subfigure}
	\caption[$\Delta R$ distribution between \Z and \W leptons]{$\Delta R$ distribution between
		the \Z and \W candidates. The distribution is built adding up the four 
		final state channels. Statistical and systematic uncertainties are included. The 
		experimental data versus the \gls{mc} prediction is shown in the lower plot. Data 
		corresponding to 2011 analysis.}\label{ch6:fig:wcand::dR}
\end{figure}

Furthermore, Figures~\ref{ch6:fig:wcand::dR} shows the angular distances between the two 
\Z-candidate leptons with the \W-candidate lepton in order to check the internal photon conversion. 
The angular distance is estimated with,
\begin{equation}
	\Delta R=\sqrt{\Delta\eta^2+\Delta\phi^2}
\end{equation}
where $\Delta\eta=\eta_1-\eta_2$ is the difference in pseudorapidity and $\Delta\phi=\phi_1-\phi_2$
the azimuthal angle difference between the two leptons. This $\Delta R$ defines a cone generated by
the revolution in the three-dimensional space of the surface created by the two linear momentum 
vectors of the leptons. Hence, the figures are showing the angular separation between the 
\Z-candidates and the \W-candidate leptons: the lowest region of $\Delta R$ is likely notifying an
internal photon conversion. As expected this region is populated only by Drell-Yan events.

The control distributions plots reveal a good agreement between the \gls{mc} prediction 
for the \gls{sm} processes and the experimental data. All the distributions are consistent within
the statistical and systematic\footnote{A detailed treatment of the systematic uncertainties 
considered in this analysis is elaborated in Chapter~\ref{ch8}.} uncertainties. The 
Appendix~\ref{app:distributions} is filled with more detailed distributions, both for the 7 and 
8~\TeV analyses, at each stage of the selection and for each final lepton state channel separately.

Finally, a quantitative view of the analysis is given in Tables~\ref{ch6:tab:events11} 
and~\ref{ch6:tab:events12}, showing the experimental and \gls{mc} prediction data events obtained 
for the four lepton final state channels.
\begin{table}[!htbp]
	\centering
	\begin{subtable}[b]{0.45\textwidth}
		\resizebox{\textwidth}{!}
		{
\begin{tabular}{ r  l  l  l }\hline\hline
 {              }                    & {\bf Z cand.}      & {\bf W cand.} \\ \hline
 {Data-driven bkg.}                  & 32     $\pm$ 2     & 2.1 $\pm$ 0.4 \\ 
 $ZZ$                                & 14.07  $\pm$ 0.06  & 1.95 $\pm$ 0.02 \\ 
 $V\gamma$                           & 12     $\pm$ 4     & 0.00 $\pm$ 0.00 \\ 
 {$WZ\rightarrow3\ell\nu$ }          & 66.2   $\pm$ 0.6   & 44.8 $\pm$ 0.5 \\ \hline
 {\bf Total expect. }                & 124    $\pm$ 4     & 48.8 $\pm$ 0.6 \\ 
 {\bf Data       }                   & 117                & 64              \\ \hline
\end{tabular}

	        }
		\caption{Three electron final state}
	\end{subtable}\quad
	\begin{subtable}[b]{0.45\textwidth}
		\resizebox{\textwidth}{!}
		{
\begin{tabular}{ r  l  l  l }\hline\hline
 {              }                & {\bf Z cand.}  & {\bf W cand.} \\ \hline
 {Data-driven bkg}               & 46    $\pm$ 3      & 1.44 $\pm$ 0.3 \\ 
 $ZZ$                            & 13.87 $\pm$ 0.07   & 3.46 $\pm$ 0.04 \\ 
 $V\gamma$                       & $(9 \pm 9)\cdot10^{5}$  & 0.00 $\pm$ 0.00 \\ 
 {$WZ\rightarrow3\ell\nu$ }      & 78.7 $\pm$ 0.6     & 50.4 $\pm$ 0.5 \\ \hline
 {\bf Total expect.}             & 139 $\pm$ 3        & 55.3 $\pm$ 0.5 \\ 
 {Data }                         & 162                & 62              \\ \hline
\end{tabular}

		}
		\caption{Two electron and one muon final state}
	\end{subtable}\vskip 1em
	\centering
	\begin{subtable}[b]{0.45\textwidth}
		\resizebox{\textwidth}{!}
		{
\begin{tabular}{ r   l  l  l }\hline\hline
 {              }                & {\bf Z cand.}  & {\bf W cand.} \\ \hline
 {Data-driven bkg.}              & 97    $\pm$ 3    & 2.5 $\pm$ 0.4 \\ 
 $ZZ$                            & 21.66 $\pm$ 0.09 & 2.68 $\pm$ 0.03 \\ 
 $V\gamma$                       & 15    $\pm$ 4    & 0.5 $\pm$ 0.5 \\ 
 {$WZ\rightarrow3\ell\nu$ }      & 83.3  $\pm$ 0.6  & 55.7 $\pm$ 0.5 \\ \hline
 {\bf Total expect. }            & 217   $\pm$ 5    & 61.4 $\pm$ 0.8 \\ 
 {\bf Data }                     & 178              & 70               \\ \hline
\end{tabular}

		}
		\caption{Two muons and one electron final state}
	\end{subtable}\quad
	\begin{subtable}[b]{0.45\textwidth}
		\resizebox{\textwidth}{!}
		{
\begin{tabular}{ r l  l  l }\hline\hline
 {              }                 & {\bf Z cand.}     & {\bf W cand.} \\ \hline
 { Data-driven bkg.}              & 68 $\pm$ 3        & 1.70 $\pm$ 0.2 \\ 
 $ZZ$                             & 18.85 $\pm$ 0.07  & 4.83 $\pm$ 0.03 \\ 
 $V\gamma$                        & 0.00 $\pm$ 0.00   & 0.00 $\pm$ 0.00 \\
 {$WZ\rightarrow3\ell\nu$ }       & 117.2 $\pm$ 0.7   & 75.07 $\pm$ 0.6 \\ \hline
 { \bf Total expect. }            & 204  $\pm$ 3      &  81.6 $\pm$ 0.6 \\ 
 { \bf Data }                     & 272               & 97               \\\hline
\end{tabular}

		}
		\caption{Three muons final state}
	\end{subtable}
	\caption[Number of total events at several stages of the selection for 2011]{Number of events at the
		different stages of the signal selection in the four leptonic channels investigated
		for the 2011 analysis.
		The data driven background (mainly \ttbar plus Drell-Yan) estimation and the \gls{mc}
		background samples used are described in Chapter~\ref{ch7}. The \WZ signal event 
		yields are obtained from applying the signal selection to \gls{mc} simulated events.
		The errors shown are statistical only.}\label{ch6:tab:events11}
\end{table}

\begin{table}[!htbp]
	\centering
	\begin{subtable}[b]{0.45\textwidth}
		\resizebox{\textwidth}{!}
		{
		\begin{tabular}{ r  l  l  l }\hline\hline
 {              }                 & {\bf Z cand.}   & {\bf W cand.} \\ \hline
 {Data-driven bkg.}    		  & 75  $ \pm$ 3    & 14.8 $\pm$ 1.4\\ 
 $ZZ$                             & 19.0 $\pm$ 0.1  & 2.43 $\pm$ 0.04\\ 
 $V\gamma$                        & 22   $\pm$ 3    & 2.4 $\pm$ 0.9\\ 
 $WV$                             & 1.7 $\pm$ 0.8   & 0.1 $\pm$ 0.1\\
 $VVV$                            & 7.6 $\pm$ 0.3   & 6.1 $\pm$ 0.3\\
 {$WZ\rightarrow3\ell\nu$ }       & 281.0 $\pm$ 1.7 & 193.9 $\pm$ 1.4\\ \hline
 {\bf Total expect. }             & 406 $\pm$ 5     & 220 $\pm$ 3\\ 
 {\bf Data       }                & 442             & 235             \\ \hline
\end{tabular}

	        }
		\caption{Three electron final state}
	\end{subtable}\quad
	\begin{subtable}[b]{0.45\textwidth}
		\resizebox{\textwidth}{!}
		{
		\begin{tabular}{ r  l  l  l }\hline\hline
 {              }                & {\bf Z cand.}  & {\bf W cand.} \\ \hline
 {Data-driven bkg}               & 214 $\pm$ 7    & 27  $\pm$ 3\\
 $ZZ$                            & 22.7 $\pm$ 0.2 & 3.11 $\pm$ 0.04\\ 
 $V\gamma$                       & 1.2 $\pm$ 0.7  & 0.4 $\pm$ 0.4\\
 $WV$                            & 0.5 $\pm$ 0.1  & 0.1 $\pm$ 0.1\\
 $VVV$                           & 10.7 $\pm$ 0.4 & 7.9 $\pm$ 0.3\\
 {$WZ\rightarrow3\ell\nu$ }      & 380  $\pm$ 2   & 245.8 $\pm$ 1.6 \\ \hline
 {\bf Total expect.}             & 629  $\pm$7    & 284 $\pm$ 3 \\ 
 {Data }                         & 613            & 288            \\ \hline
\end{tabular}

		}
		\caption{Two electron and one muon final state}
	\end{subtable}\vskip 1em
	\centering
	\begin{subtable}[b]{0.45\textwidth}
		\resizebox{\textwidth}{!}
		{
		\begin{tabular}{ r  l  l  l }\hline\hline
 {              }                & {\bf Z cand.}  & {\bf W cand.} \\ \hline
 {Data-driven bkg}               & 150   $\pm$ 5   & 48 $\pm$ 3   \\
 $ZZ$                            & 31.5 $\pm$ 0.2  & 3.9 $\pm$ 0.1  \\ 
 $V\gamma$                       & 41   $\pm$ 4    & 3.8 $\pm$ 1.2  \\
 $WV$                            & 0.7 $\pm$ 0.2   & 0.2 $\pm$ 0.1\\
 $VVV$                           & 13.1 $\pm$ 0.4  & 10.4 $\pm$ 0.4 \\
 {$WZ\rightarrow3\ell\nu$ }      & 466  $\pm$ 2    & 316 $\pm$ 2\\ \hline
 {\bf Total expect.}             & 703  $\pm$ 7    & 382   $\pm$ 4\\ 
 {Data }                         & 790             & 400             \\ \hline
\end{tabular}

		}
		\caption{Two muons and one electron final state}
	\end{subtable}\quad
	\begin{subtable}[b]{0.45\textwidth}
		\resizebox{\textwidth}{!}
		{
		\begin{tabular}{ r  l  l  l }\hline\hline
 {              }                & {\bf Z cand.}  & {\bf W cand.} \\ \hline
 {Data-driven bkg}               & 377   $\pm$ 10    & 59  $\pm$ 5 \\
 $ZZ$                            & 42.1 $\pm$ 0.2    & 5.8 $\pm$ 0.1  \\ 
 $V\gamma$                       & 0.8 $\pm$ 0.5     & 0.0 $\pm$ 0.0  \\
 $WV$                            & 14     $\pm$ 2    & 2.2 $\pm$ 0.7  \\
 $VVV$                           & 18.0 $\pm$ 0.5    & 13.4 $\pm$ 0.4 \\
 {$WZ\rightarrow3\ell\nu$ }      & 666   $\pm$ 3     & 428 $\pm$ 2\\ \hline
 {\bf Total expect.}             & 1118 $\pm$ 10     &  508 $\pm$ 5\\ 
 {Data }                         & 1207              & 557            \\ \hline
\end{tabular}

		}
		\caption{Three muons final state}
	\end{subtable}
	\caption[Number of total events at several stages of the selection for 2012]{Number of events at the
		different stages of the signal selection in the four leptonic channels investigated for the
		2012 analysis.
		The data driven background (mainly \ttbar plus Drell-Yan) estimation and the \gls{mc}
		background samples used are described in Chapter~\ref{ch7}. The \WZ signal event 
		yields are obtained from applying the signal selection to \gls{mc} simulated events.
		The errors shown are statistical only.}\label{ch6:tab:events12}
\end{table}

\chapter{Background Studies}\label{ch7}
The main backgrounds processes which contaminate the \WZ signal are classified in two categories 
depending on their origin: instrumental and irreducible physics backgrounds. The instrumental background,
mostly jet-induced, is estimated using a method based on the experimental data which exploits the 
\WZ lepton isolation and identification criteria. In this chapter, the method is explained in detail and the main 
formulae obtained, then the results of the method are validated and subsequently applied, obtaining
the jet-induced background contribution to the \WZ signal. The second part of this Chapter is 
devoted to explain the irreducible physics background components, which are estimated using a 
\gls{mc} simulation.

\section{Contamination of the \WZ signal}
In the previous Chapter, Section~\ref{ch6:sec:topology}, the main sources of noise were identified 
for the \WZ leptonic final state signature. In summary, the background processes contributing to the
three lepton final states were categorised into two groups.
\begin{itemize}
	\item Backgrounds where some of the identified final three lepton states are originated by 
		leptons	no-promptly produced by \W or \Z decays, meaning particles misidentified 
		as leptons or leptons in jets, mostly coming from heavy flavour decays. The main 
		processes contributing to this jet-induced source of background are:
		\begin{itemize}
			\item QCD, with three non-prompt leptons or fakes;
			\item W+Jets, with two fake leptons;
			\item Drell-Yan plus jets, WW+Jets, \ttbar and single top, providing 
				one fake lepton
		\end{itemize}
	\item Backgrounds with one or more prompt leptons no-reconstructed, or other backgrounds,
		\begin{itemize}
			\item ZZ production, where one lepton is lost or outside acceptance
			\item VVV ($V=\gamma,\W,\Z$) production, triple gauge boson decays can 
				mimic exactly the
				signal selection, as WWW; or one decay leptons is outside the 
				fiducial volume, as in WWZ. Nevertheless, the production rate 
				of this processes is very small compared with the \WZ production 
				rate; the 2012 data taking period 
				achieved enough integrated luminosity to observe some of this processes.
				Thus, this background is just considered for 2012 data.
			\item $V\gamma$ ($V=\Z,\W$), mostly a third electron\footnote{Although it can
				be a muon, but with much lesser probability.} appears because of 
				the photon interaction with the detector material (external photon
				conversion) and one of the leptons of the created pair is lost.
		\end{itemize}
\end{itemize}
The jet-induced background is not well modelled in the simulations, therefore it has been estimated 
using a method based upon experimental data,~\ie a \emph{data-driven} method. The method is called 
\gls{fom}\glsadd{ind:fom}. The contribution of the second category,~\ie signal-like events where the 
three lepton are prompts, is estimated with a \gls{mc} simulation.

\section{The fakeable object method}
The fakeable object method is a \glsadd{ind:data-driven} method used to estimate the background contribution
caused by the so called \emph{fake} leptons. The method is based in the \emph{matrix} 
method~\cite[pp.~334--337]{Behnke:1517556}, widely used in high energy physics to estimate the
composition of collected data. In brief, the matrix method starts from a set 
of cuts $S$ which have been applied to a dataset in order to enhance the number of signal events. After 
applying these selection cuts to a dataset $N(S)$ events are selected. The number of signal 
events $\nu_S$ can be estimated by considering the efficiency of the selection cuts,
\begin{equation}
      \nu(S)=\varepsilon_S\nu_S
      \label{ch7:eq:nest}
\end{equation}
being $\nu(S)$ the number of events after the selection, and it may be estimated by
$\hat{\nu}(S)=N(S)$, being $N(S)$ the number of events measured after applying the selection cuts.

The selected sample is not only composed by signal events, as Equation~\eqref{ch7:eq:nest} is 
assuming. Indeed, a realistic scenario would consider that before the selection, $n$ different
sources of background processes contribute to the expected number of events, and the selection 
criteria $S$ do not totally reject all background contributions. Therefore, the number of selected
events after the selection cuts is
\begin{equation}
      \nu(S)=\varepsilon_S\nu_S+\varepsilon_{S|B_1}\nu_{B_{1}}+\dots+\varepsilon_{S|B_n}\nu_{B_n}
      \label{ch7:eq:nestandB}
\end{equation}
where $\nu_{B_i}$ is the number of events of the i-background and $\varepsilon_{S|B_i}$ is the 
efficiency of the selection cuts $S$ to select the i-background source. Another set of selection
criteria $B_{i}$ may be introduced in order to select enriched regions of the different 
i-background sources. Using these new selection cuts, the dataset will be split in n-different
samples, 
\begin{equation}
     \nu(B_i)=\varepsilon_{B_i|S}\nu_S+\dots+\varepsilon_{B_i}\nu_{B_i}+\dots+
          \varepsilon_{B_i|B_n}\nu_{B_n}
\end{equation}
being $\nu(B_i)$ the number of selected events using the selection cuts $B_i$, 
$\varepsilon_{B_i|S}$ is the efficiency of the selection cuts $B_i$ to select the signal events,
$\varepsilon_{B_i}$ is the efficiency of the selection cuts $B_i$ to select the i-background source
and $\varepsilon_{B_i|B_n}$ the efficiency of the selection cuts $B_i$ to select the n-background 
source. Therefore, we end up with a total of $n+1$ selection cuts each of them aiming to select
a given component of the primary dataset, Equations~\eqref{ch7:eq:nest} and~\eqref{ch7:eq:nestandB}
can be expressed with a linear system of equations in matrix form
\begin{equation}
   \begin{pmatrix}
      \nu(S) \\ 
      \nu(B_i) \\
      \vdots \\
      \nu(B_n)
   \end{pmatrix} = 
   \begin{pmatrix}
             \varepsilon_S       & \varepsilon_{S|B_1} & \dots & \varepsilon_{S|B_n} \\
             \varepsilon_{B_1|S} & \varepsilon_{B_1}   & \dots & \varepsilon_{B_1|B_n} \\
	     \vdots              &  \vdots             & \ddots & \vdots \\
             \varepsilon_{B_n|S} & \varepsilon_{B_n|B_1}   & \dots & \varepsilon_{|B_n}
   \end{pmatrix}
   \begin{pmatrix}
      \nu_S \\ 
      \nu_{B_i} \\
      \vdots \\
      \nu_{B_n}
   \end{pmatrix}
   \label{ch7:eq:mm}
\end{equation}
or in the equivalent vector form 
$\boldsymbol{\nu}_{\mathbf{sel}}=\boldsymbol{\varepsilon}\boldsymbol{\nu}$.
Thus, we may use the number of measured events $N_{sel}$ to estimate $\nu$,
\begin{equation}
	\boldsymbol{N}_{\text{\bf sel}}=\boldsymbol{\varepsilon}\boldsymbol{\hat{\nu}}
   \label{ch7:eq:mmvect}
\end{equation}
being $\boldsymbol{N}_{\text{\bf sel}}$ the vector of events measured using the different
criteria $S$ and $B_i$, $\boldsymbol{\varepsilon}$ is the efficiency matrix, and 
$\boldsymbol{\hat{\nu}}$ is the vector of the actual number of events for the different processes. 
Therefore, inverting the 
efficiency matrix, it is possible to estimate the original signal and background contributions
\begin{equation}
	\boldsymbol{\hat{\nu}}=\boldsymbol{\varepsilon}^{-1}\mathbf{N}_{\text{\bf sel}}
	\label{ch7:eq:mminv}
\end{equation}
Estimating the efficiency matrix, by \gls{mc} techniques or data-driven methods, and 
performing the $n+1$ sample selection, it is possible to estimate the background contribution to a 
given datasample.

\paragraph*{}

The \gls{fom}\glsadd{ind:fom} shares the strategy of the matrix method outlined above where the source of 
background to be estimated is caused by \emph{fake leptons}. The \emph{fake} meaning is dependent 
of each analysis; in the \WZ context, a fake lepton could be:
\begin{itemize}
	\item a true ``fake`` lepton, for example, a jet misidentified and reconstructed as an electron,
	\item a real lepton from a heavy hadron decay
\end{itemize}
A prompt lepton from a \W or \Z gauge boson is expected not to have hadronic activity 
surrounding it and to fulfil the identification requirements presented in the previous Chapter 
(see Section~\ref{ch6:sec:muonselection} and~\ref{ch6:sec:electronselection}). Conversely,
a jet-induced lepton is expected to be poorly isolated, and also not coming from the primary 
vertex. Therefore, it is possible to relax the isolation and identification criteria of the 
leptons, building a sample of \emph{loose}\glsadd{ind:loose} or also called 
\emph{fakeable}\glsadd{ind:fakeable} leptons, in order to 
study these isolation and identification properties. 

In addition, the loose definition is used to build a experimental data sample of three final state
fakeables in the signal region. The sample is split in four exhaustive subsamples\footnote{In the 
mathematical group theory sense, thus the four subsamples fully cover and complete the sample} by 
evaluating the category of the fakeables,~\ie whether they pass or fail the analysis cuts. 
Therefore this sample built with three loose leptons $N_{3L}$ is split in a subsample which all of 
the three loose leptons do not pass the tight analysis cuts $N_{t0}$, another subsample which only 
one of the fakeable leptons pass the tight analysis cuts $N_{t1}$, a further subsample which two 
of the fakeable leptons pass the tight analysis cuts $N_{t2}$, and finally the total sample is 
completed with the subsample which all of the three loose leptons pass the tight analysis cuts 
$N_{t3}$. This can be expressed in a succinct way by
\begin{equation}
	{N_{3L}}=N_{t0}\cup N_{t1}\cup N_{t2} \cup N_{t3}
\end{equation}
Notice that the identification of this notation with the one used to introduced the matrix method
is straightforward. The number of selected events using the signal selection criteria 
$\nu(S)$, or more precisely its estimator $N_{sel}$, is recognised as $N_{t3}$, and the number of 
selected events using a enriched region of the i-source of background $\nu(B_i)$ may be identified 
with $N_{ti}$ being $i=0,1,2$. 

Thus, in order to apply the Equation~\eqref{ch7:eq:mminv} the efficiency matrix has to be estimated. 
The probability to detect in the analysis a fake\glsadd{ind:fake} lepton is assumed to be a \emph{universal} 
property only dependent on the detector acceptance and resolution by means of the transverse 
momentum and pseudorapidity of the lepton\footnote{Throughout this Chapter is possible to find the
expression ``lepton kinematics`` referring to the lepton transverse momentum and pseudorapidity 
which determine the quality of the detector measurement.}, and it is fundamentally determined
by the lepton isolation and identification criteria. Therefore, it is possible to extract this 
probability using a data sample which only contains such fake leptons. 
\begin{figure}[htpb]
	\centering
	\includegraphics[width=0.45\textwidth]{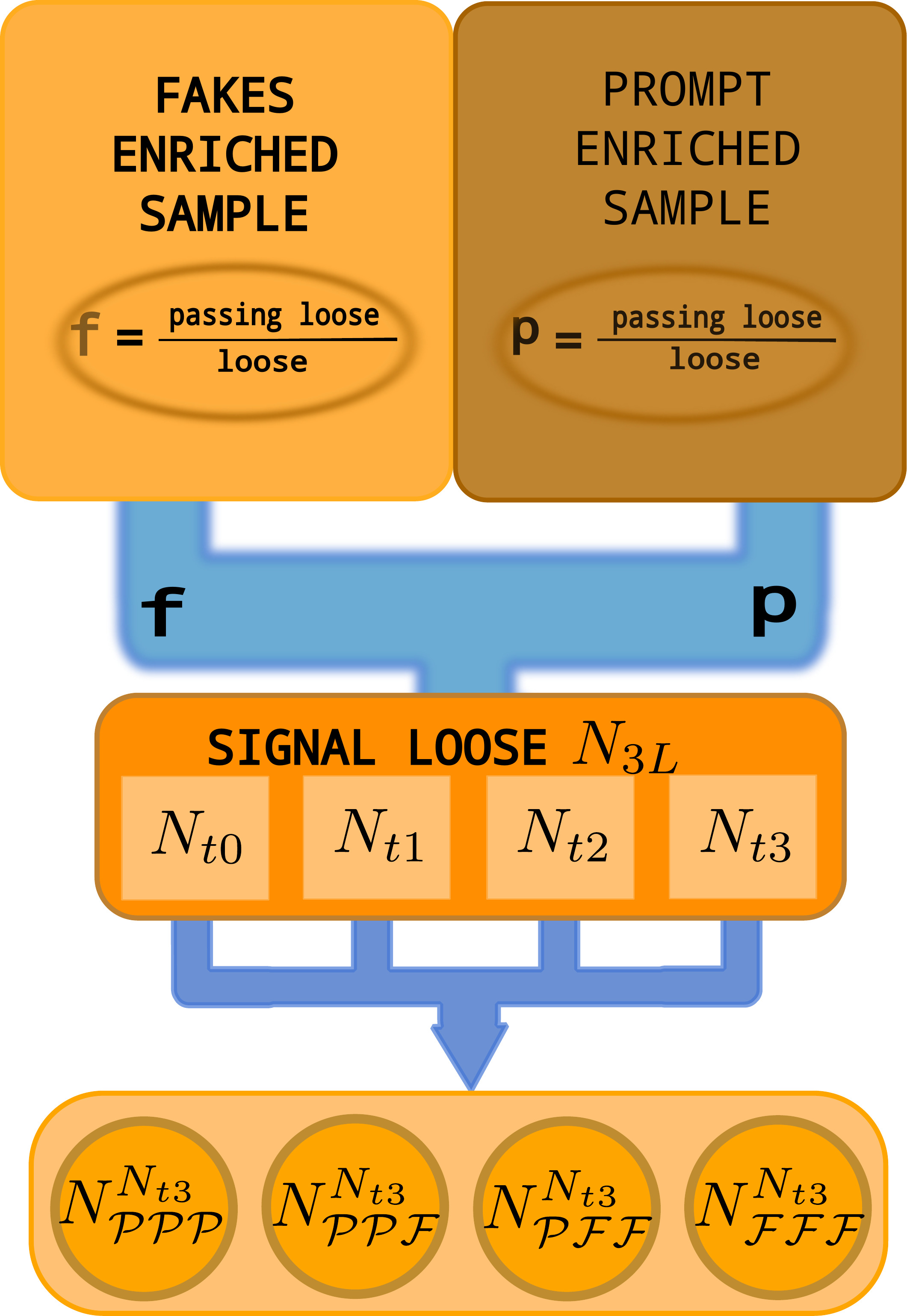}
	\caption[Schematic illustration of the fakeable object method]{Schematic illustration of
	the fakeable object method. The Fake and Prompt enriched samples are used to obtain
	the fake and prompt rates, respectively. The Signal Loose sample is built by applying to
	the analysis the loose selection to the lepton objects; each $N_{ti}$ subsample is defined
	by their content of i-tight leptons. Afterwards, each event of the four exhaustive loose
	subsamples is weighted with functions (Equations~\eqref{ch7:eq:fomEstimation}) of $p$, $f$ 
	and $N_{ti}$ and combined to obtain the several estimations of the jet-induced 
	background and prompt content represented as $N_{\pr\pr\pr}^{N_{t3}}$, 
	$N_{\pr\pr\fr}^{N_{t3}}$, $N_{\pr\fr\fr}^{N_{t3}}$ and 
	$N_{\fr\fr\fr}^{N_{t3}}$.}\label{ch7:fig:fomdiagram}
\end{figure}
This may be accomplished by
generating a sample of fakeable leptons enriched of fake leptons and counting how those 
fakeables would pass or fail the full isolation and identification analysis' criteria. The ratio of
the passing tight criteria leptons (\emph{tight} leptons) over all the loose leptons is called the 
\emph{fake ratio}. This fake ratio along with the \emph{prompt rate},~\ie the probability to a 
prompt lepton to pass the tight analysis cuts (which can be extracted using tag and probe methods),
are the efficiency factors measured from experimental data needed to build the efficiency matrix of 
Equation~\eqref{ch7:eq:mminv}. Therefore, the method have been built to estimate the jet-induced 
background contribution to the three final state \WZ analysis,~\ie the $\nu_S$ and the $\nu_{B_i}$ 
which in the \gls{fom} notation are $N_{\pr\pr\pr}^{N_{t3}}$ for the signal,
and $N_{\pr\pr\fr}^{N_{t3}}$, $N_{\pr\fr\fr}^{N_{t3}}$ and $N_{\fr\fr\fr}^{N_{t3}}$ for the 
backgrounds. The full workflow of the method outlined in the last paragraphs is schematised in 
Figure~\ref{ch7:fig:fomdiagram}.

A formal derivation of the method is now conducted. Some definitions are introduced here to fix the 
notation and nomenclature and to delimit the scope of this probabilistic problem,
\begin{itemize}
	\item Definitions related with true information
		\begin{definition}[Prompt Lepton, \pr]
			Lepton from \W or \Z decay\glsadd{ind:prompt}
			\label{ch7:def:prompt}
		\end{definition}
		\begin{definition}[Fake lepton, \fr] 
			Jet-induced or misidentified lepton\glsadd{ind:fake}
			\label{ch7:def:fake}
		\end{definition}
		\begin{definition}[$N_{XYX}$] 
			Number of events with just X,Y,Z leptons, being X,Y,Z fakes (\fr) or prompt (\pr)
			\label{ch7:def:Ntrue}
		\end{definition}
	\item Definition related with measured information
		\begin{definition}[Loose Lepton, L]
			Lepton passing the relaxed cuts (see Tables~\ref{ch7:tab:loose2011}
			and~\ref{ch7:tab:loose2012}) mainly related with isolation\glsadd{ind:loose}
			\label{ch7:def:loose}
		\end{definition}
		\begin{definition}[Tight Lepton, T]
			Loose lepton passing the more restrictive isolation and identification analysis cuts 
			\label{ch7:def:tight}
		\end{definition}
		\begin{definition}[Fail Lepton, F]
			Loose lepton failing the isolation and identification analysis cuts,~\ie
			a no-tight lepton
			\label{ch7:def:fail}
		\end{definition}
		\begin{definition}[Prompt rate, p]
			Probability that a prompt lepton pass the tight cuts, $P(T|\pr)$
			\label{ch7:def:pr}
		\end{definition}
		\begin{definition}[Fake rate, f]
			Probability that a fake lepton pass the tight cuts, $P(T|\fr)$ 
			\label{ch7:def:fr}
		\end{definition}
		\begin{definition}[$N_{ti}$]
			Number of events containing just three loose leptons of which i leptons are
			tight
			\label{ch7:def:Nmeas}
		\end{definition}
\end{itemize}

The loose or fakeable lepton definition is accomplished by relaxing isolation and some of the 
identification requirements of the \WZ analysis leptons whilst the other requirements are kept as 
in the main analysis. Tables~\ref{ch7:tab:loose2011} and~\ref{ch7:tab:loose2012} define the 
loose lepton for 2011 and 2012 analyses, respectively. The tables are just showing the modified 
requirements of the main lepton criteria of Chapter~\ref{ch6} (Tables~\ref{ch6:tab:muonreq} 
and~\ref{ch6:tab:elecreq}) where it is understood that the other requirements are also applied. 
\begin{table}[!hpbt]
	\centering
	\begin{subtable}[b]{0.8\textwidth}
		\centering
		\begin{tabular}{l c c c }\hline\hline
			                  &  Loose  & Tight     & Fail \\\hline
			$ISO_{PF}/\pt$    &  $<1.0$ & $<0.12$   & $[0.12,1.0)$\\\hline
		\end{tabular}
		\caption{Muon loose definition. The loose definition relaxes the isolation 
		requirement, while the other requirements defined in the 2011 column of the
		Table~\ref{ch6:tab:muonreq} are also applied.\\ \\ \\}\label{ch7:subtab:loosemuon}
	\end{subtable}
	\begin{subtable}[b]{0.8\textwidth}
	        \centering
		\begin{tabular}{l  c c c }\hline\hline
			                  &  Loose  & Tight   & Fail \\\hline
			$ISO_{PF}/\pt$    &   NOT REQUIRED    & $<0.13\; (0.09)$ & $\ge 0.13 (0.09)$\\
					  &                   &  AND    &  OR   \\
			MVA ID            &   NOT REQUIRED    & $> cut(p_t,\eta)$ & $\le cut(p_t,eta)$\\ \hline
		\end{tabular}
	\caption{Electron loose definition. The loose definition removes the multivariate 
		identification (MVA ID) and the isolation requirement, while the other requirements
		defined in the 2011 column of the Table~\ref{ch6:tab:elecreq} are also applied. The
		MVA ID values, \pt and $\eta$ dependent, for the tight category are detailed in the
		aforementioned table. The relative isolation contains two cuts, one cut is applied 
		to barrel electrons whereas the other one (in parenthesis) is applied to endcap electrons.
	        }\label{ch7:subtab:looseelectron}
	\end{subtable}
	\caption[Loose lepton definition for 2011 analysis]{Loose lepton definitions for 2011 
	analysis.}\label{ch7:tab:loose2011}
\end{table}

\begin{table}[!hpbt]
	\centering
	\begin{subtable}[b]{0.8\textwidth}
		\centering
		\begin{tabular}{l c c c }\hline\hline
			                  &  Loose        & Tight           & Fail \\\hline
			$|d_0|$ [cm]      &  $<0.2$       & $<0.01(0.02)$   & $[0.01(0.02),0.2)$ \\
			                  &               &   AND           &   OR   \\
			MVA ISO           &  $\ge -0.6$   & $\ge cut(\pt,\eta)$& $(-0.6,cut(\pt,\eta)]$\\\hline
		\end{tabular}
		\caption{Muon loose definition. Loose definition relaxes the transverse impact 
			parameter and the multivariate isolation (MVA ISO), the other requirements
			specified in the 2012 column of Table~\ref{ch6:tab:muonreq} are also
			applied. The MVA ISO values for the tight category are detailed in 
			the aforementioned table. The $d_0$ parameter contains two cuts, one cut is
			applied to muons with $p_t\leq20~\GeV$ whereas the other one (in parenthesis) 
			is applied to muons with $p_t>20~\GeV$.\\ \\}\label{ch7:subtab:loosemuon12}
	\end{subtable}
	\begin{subtable}[b]{0.8\textwidth}
		\centering
		\begin{tabular}{l  c c c }\hline\hline
			                  &  Loose        & Tight   & Fail \\\hline
			$|d_0|$ [cm]      &  NOT REQUIRED & $<0.02$ & $\ge 0.02$ \\
			                  &               &   AND   &   OR      \\
			$|d_z|$ [cm]      &  NOT REQUIRED & $<0.1$  & $\ge 0.1$ \\
			                  &               &   AND   &   OR   \\
			$ISO_{PF}/\pt$    &   NOT REQUIRED& $<0.15$ & $\ge 0.15$\\
					  &               &  AND    &  OR   \\
			MVA ID            &   NOT REQUIRED& $> cut(p_t,\eta)$ & $\le cut(p_t,eta)$\\ \hline
		\end{tabular}
	\caption{Electron loose definition. The loose definition removes the transverse and 
		longitudinal impact parameter, the relative isolation and the multivariate 
		identification (MVA ID), whilst the other requirements specified in the 2012
		column of Table~\ref{ch6:tab:elecreq} are also applied. The MVA ID values for
		the tight category are detailed in the aforementioned 
		table.}\label{ch7:subtab:looseelectron12}
	\end{subtable}
	\caption[Loose lepton definition for 2012 analysis]{Loose lepton definitions for 2012
	analysis.}\label{ch7:tab:loose2012}
\end{table}
\pagebreak
\subsection{Fakeable object method derivation}
The definitions~\ref{ch7:def:prompt},~\ref{ch7:def:fake} and~\ref{ch7:def:Ntrue} specify
the true composition of the sample, whereas definitions from~\ref{ch7:def:loose} to~\ref{ch7:def:Nmeas}
refer to measurable quantities. The first equality relating the measurable 
observables with the true ones is straightforward obtained,
\begin{equation}
	N_{3L}=N_{\pr\pr\pr}+N_{\pr\pr\fr}+N_{\pr\fr\fr}+N_{\fr\fr\fr}=N_{t0}+N_{t1}+N_{t2}+N_{t3}
\end{equation}
\emph{Id est}, the sample selected using three loose-criteria leptons ($N_{3L}$) consists of 
four exhaustive subsamples. Regarding the origin of the true leptons, the leptons only can be
prompt-prompt-prompt ($N_{\pr\pr\fr}$), prompt-prompt-fake ($N_{\pr\pr\fr}$), prompt-fake-fake
($N_{\pr\fr\fr}$) or fake-fake-fake ($N_{\fr\fr\fr}$). Furthermore, regarding the measured category 
of the three selected leptons, the leptons only can be classified as tight-tight-tight ($N_{t3}$),
tight-tight-fail ($N_{t2}$), tight-fail-fail ($N_{t1}$) and fail-fail-fail ($N_{t0}$).

The contributions of the several prompt-fake combinations which pass the full analysis cuts can
be obtained using probabilistic relations,
\begin{subequations}
	\begin{align}
		N_{\fr\fr\fr}^{N_{t3}} &= \sum_{events}^{N_{t3}}P(\fr\fr\fr|TTT)P(TTT) \\
		N_{\pr\fr\fr}^{N_{t3}} &= \sum_{events}^{N_{t3}}P(\pr\fr\fr|TTT)P(TTT) \\
		N_{\pr\pr\fr}^{N_{t3}} &= \sum_{events}^{N_{t3}}P(\pr\pr\fr|TTT)P(TTT) \\
		N_{\pr\pr\pr}^{N_{t3}} &= \sum_{events}^{N_{t3}}P(\pr\pr\pr|TTT)P(TTT)
	\end{align}
	\label{ch7:eq:bkgcontribution}
\end{subequations}
Here, $P(TTT)$ is the probability that, given an event with three loose leptons, these 
leptons pass the analysis cuts, therefore $N_{t3}=\sum_{events}^{N_{t3}}P(TTT)$. The conditional
probabilities $P(\fr\fr\fr|TTT)$, $P(\pr\fr\fr|TTT)$, $P(\pr\pr\fr|TTT)$ and $P(\pr\pr\pr|TTT)$
are quantifying the probability of, given three tight leptons, that these leptons are all fakes,
one prompt and two fakes, two prompt and one fake and all prompt, respectively. Thus, the left
side of Equations~\eqref{ch7:eq:bkgcontribution} designates the estimation of each prompt and fake
contribution to the \WZ three lepton final state analysis. Therefore, each event passing the 
\WZ analysis cuts is weighted by the conditional probability (lepton kinematic dependent) for each
lepton to be fake or prompt given that each lepton pass the tight analysis criteria. The problem is
focused in finding these conditional probabilities. 

Nevertheless, the probabilities available are the fake and prompt rates,~\ie the probabilities which
have a fake or prompt lepton to be tight (as definitions~\ref{ch7:def:fr} and~\ref{ch7:def:pr} 
established),
\begin{subequations}
	\begin{equation}
		P(T|\fr)\equiv f\qquad\qquad \text{(Fake rate)}
		\label{ch7:eq:fr}
	\end{equation}
	\begin{equation}
		P(T|\pr)\equiv p\qquad\quad \text{(Prompt rate)}
		\label{ch7:eq:pr}
	\end{equation}
\end{subequations}
In an analysis with one final state lepton, the probability that a lepton passes the tight cuts, is 
evaluated through the full space of lepton types,~\ie prompt and fake,
\begin{gather}
	N_{T} = \sum_{events}^{N_L}\left[P(T|\pr)P(\pr)+P(T|\fr)P(\fr)\right] 
		= pN_{\pr}+fN_{\fr}
	\intertext{and analogously for the fail case,}
	N_{F} = \sum_{events}^{N_L}\left[P(F|\pr)P(\pr)+P(F|\fr)P(\fr)\right] 
		= (1-p)N_{\pr}+(1-f)N_{\fr}
\end{gather}

Assuming that the fake and prompt rates are independent of the number of leptons in the event,
Equations~\eqref{ch7:eq:bkgcontribution} can be equivalently 
expressed through their true contribution content by using the prompt and fake efficiencies,
\begin{subequations}
	\begin{align}
   	 N_{\fr\fr\fr}^{N_{t3}} &= \sum_{events}^{N_{3L}}P(TTT|\fr\fr\fr)P(\fr\fr\fr) = f^3N_{\fr\fr\fr} \\
         N_{\pr\fr\fr}^{N_{t3}} &= \sum_{events}^{N_{3L}}P(TTT|\pr\fr\fr)P(\pr\fr\fr) = pf^2N_{\pr\fr\fr} \\
     N_{\pr\pr\fr}^{N_{t3}} &= \sum_{events}^{N_{3L}}P(TTT|\pr\pr\fr)P(\pr\pr\fr) = p^2fN_{\pr\pr\fr}\label{ch7:subeq:ppfnt3}\\
         N_{\pr\pr\pr}^{N_{t3}} &= \sum_{events}^{N_{3L}}P(TTT|\pr\pr\pr)P(\pr\pr\pr) = p^3N_{\pr\pr\pr}		
	\end{align}                                                                    		
	\label{ch7:eq:fomEstimation}
\end{subequations}
Therefore, the problem has been restated to find the estimated number of background 
contributions, which it is resolved within the matrix method in the Equation~\eqref{ch7:eq:mminv}. 
Thus, the equivalent equations but for the three lepton final state cases, which are, in fact, the 
linear system of equations of the matrix method (Equation~\eqref{ch7:eq:mm}) particularised to the 
fakes backgrounds, are
\begin{subequations}
  \begin{align}
	        N_{t0} &=(1-p)^3N_{\pr\pr\pr}+(1-p)^2(1-f)N_{\pr\pr\fr}+&\notag\\
		       &\qquad\quad\quad +(1-p)(1-f)^2N_{\pr\fr\fr}+(1-f)^3N_{\fr\fr\fr}\\
		N_{t1} &= 3p(1-p)^2N_{\pr\pr\pr}+\left[2p(1-p)(1-f)+f(1-p)^2\right]N_{\pr\pr\fr} +&\notag\\
		       &\qquad\qquad +\left[2f(1-f)(1-p)+p(1-f)^2\right]N_{\pr\fr\fr} +&\\
	               &\qquad\qquad\qquad\qquad\qquad\qquad\qquad\qquad\qquad+3f(1-f)^2N_{\fr\fr\fr}&\notag \\
		N_{t2} &= 3p^2(1-p)N_{\pr\pr\pr}+\left[2pf(1-p)+p^2(1-f)\right]N_{\pr\pr\fr} +\notag\\
		       &\qquad\qquad  +\left[2pf(1-f)+(1-p)f^2\right]N_{\pr\fr\fr} +3f^2(1-f)N_{\fr\fr\fr} \\
	        N_{t3} &=p^3N_{\pr\pr\pr}+p^2fN_{\pr\pr\fr}+pf^2N_{\pr\fr\fr}+f^3N_{\fr\fr\fr}
	        \label{ch7:subeq:fomNt3}
  \end{align}
  \label{ch7:eq:fomNti}
\end{subequations}
where the sum over the total events has been explicitly made. The equations are derived for same 
lepton pseudorapidity, transverse momentum and flavour, for the sake of clarity. The generalisation 
considering different $\eta$, \pt and lepton flavour is straightforward but complicates the notation
and does not introduced any new insight, it will be derived later.

The linear Equations~\eqref{ch7:eq:fomNti} are inverted to obtain the estimation of each source,
equivalent to Equation~\eqref{ch7:eq:mminv},
\begin{subequations}
	\begin{align}
		N_{\pr\pr\pr}&=\frac{1}{(p-f)^3}\left\{(1-f)^3N_{t3}-f(1-f)^2N_{t2}-\right.\notag\\
			   &\qquad\qquad\qquad\qquad\qquad\qquad\left.-f^2(1-f)N_{t1}+f^3N_{t0}\right\}\\ 
 		N_{\pr\pr\fr}&=\frac{1}{(p-f)^3}\left\{-3(1-p)(1-f)^2N_{t3}+\right.\notag\\
			     &\qquad\qquad\left. +\left[2f(1-p)(1-f)+p(1-f)^2\right]N_{t2} - \right.\label{ch7:subeq:ppf}\\ 
			 &\qquad\qquad\qquad\qquad \left.-\left[f^2(1-p)+2pf(1-f)\right]N_{t1}+3pf^2N_{t0}\right\}\notag\\
 		N_{\pr\fr\fr}&=\frac{1}{(p-f)^3}\left\{3(1-p)^2(1-f)N_{t3} -\right.\notag\\
			 &\qquad\qquad\left. -\left[f(1-p)^2+2p(1-p)(1-f)\right]N_{t2} + \right.\\
			 &\qquad\qquad\qquad \left.+\left[2pf(1-p)+p^2(1-f)\right]N_{t1}-3p^2fN_{t0}\right\}\notag\\ 
	       N_{\fr\fr\fr}&=\frac{1}{(p-f)^3}\left\{-(1-p)^3N_{t3}+p(1-p)^2N_{t2}-\right.\notag\\
			 &\qquad\qquad\qquad\qquad\qquad\qquad\left.-p^2(1-p)N_{t1}+p^3N_{t0}\right\}
        \end{align}
	\label{ch7:eq:fomNPF}
\end{subequations}

The above equations give the estimated number of events composed by the available combinations 
of prompt and fakes leptons to make up three lepton final states, from the isolation and 
identification lepton categories (tight and fail). Now, Equations~\eqref{ch7:eq:fomEstimation} 
(through the inclusion of Equations~\eqref{ch7:eq:fomNPF} on them)
have all the ingredients to estimate the contribution of each type of fake lepton background to the
final analysis. In particular, the $N_{\pr\pr\pr}^{N_{t3}}$ is the estimated number of signal events
(because of the three prompts in the final state) before subtracting any irreducible process ($ZZ$,
$VVV$), which is part of the three prompt component.

The derived Equations~\eqref{ch7:eq:fomEstimation} and~\eqref{ch7:eq:fomNPF} may be expressed in form of 
\emph{weighting rules} which takes into account the flavour dependence and the lepton kinematic 
not explicitly considered and being able to estimate not only the normalised factors of each background
source but also the distribution of these processes. Each event is weighted following the reported 
rules of Table~\ref{ch7:tab:weightingrules} depending of the category of the lepton (fail or tight)
and the contribution which is being estimated (fake or prompt).
\begin{table}[!htpb]
	\centering
	\begin{tabular}{l r r}\hline\hline
		&  Tight & Fail \\\hline\\[-2ex] 
		w(\pr)  &  $\frac{1}{p-f}p(1-f)$ & $\frac{1}{p-f}pf$ \\[1ex]
		w(\fr)  &  $\frac{1}{p-f}f(1-p)$ & $\frac{1}{p-f}pf$ \\ \\[-2ex]\hline
	\end{tabular}
	\caption[Weighting rules of the fakeable object method estimation]{Each loose event is 
	weighted by the combination of prompt and fake rates reported in the table depending the
	category of the lepton. The factors are applied lepton by lepton in each event. The table 
	is showing the weight that should be applied to each lepton which is going to be estimated as 
	\emph{prompt} (first row) or \emph{fake} (second row) given that the lepton passed the 
	tight analysis cuts (column \emph{Tight}) or failed these cuts (column \emph{Fail}).
	The weighted event contributes to the current estimation depending of the estimation
	and the number of tight and fail leptons $N_{ti}$ have, following the 
	prescription of Equations~\eqref{ch7:eq:fomNPF}}\label{ch7:tab:weightingrules}
\end{table}

\subsection{Lepton fake and prompt rates determination}\label{ch7:subsec:prfr}
Adopting the loose lepton definition, a jet enriched sample is selected in the experimental data
from a combination of single-lepton trigger paths. The trigger paths have the lowest \pt threshold
in order to select the sample mainly with $QCD$ events. The loose lepton identification requirements
shown in Tables~\ref{ch7:tab:loose2011} and~\ref{ch7:tab:loose2012} are tighter than the trigger
requirements in order to suppress a possible trigger-induced bias to the fake rate. 
The jet enriched sample may still contain prompt leptons from real \W and \Z decays. In order to
suppress contamination due to signal leptons from the decay of \W and \Z bosons it is also required 
that the missing transverse energy of the event to be less than 20 \GeV and the \W transverse mass 
less than 20~\GeV. The muons from Drell-Yan and \Z decays are removed with the
$m_{\mu\mu} > 20$~\GeV and the $m_{\mu\mu} \notin [76,106]$~\GeV constraints. For electrons the W 
transverse mass cut is not applied, and the Z-peak veto is enlarged to $m_{\mathrm{ee}} \notin 
[60,120]$~\GeV. The remaining electroweak (W/Z+jets) contribution, which clearly biases the fake rate
at high \pt, is removed using the background estimations as provided by the corresponding 
\gls{mc} simulations.

The jet-enriched sample is composed of fakeables, leptons which pass the loose isolation and 
identification cuts. Thus the sample space $\Omega$ is defined as
\begin{equation}
   \Omega\equiv\bigcup_{events}L
\end{equation}
After the application of the tight cuts to the fakeables, the sample space $\Omega$ is divided with 
fakeables passing the tight criteria cuts $T$ and fakeables not passing (or failing) that cuts
$\neg T\equiv F$,
\begin{equation}
   \Omega\equiv L =T \oplus \neg T=T\oplus F ,\quad
   \text{ thus } P(T)+P(F) = 1
\end{equation}
where it is implicitly included the sum over all events. 

Given that the sample was biased selecting mostly jet events, all the leptons 
of the sample should be fakes (at least ideally). We can count how many leptons pass the tight cuts and 
assign this ratio,~\ie the fake rate, as the probability (using a frequentist approach to 
probabilities):
\begin{equation}
	f(\pt,\eta)=P(T|\fr)_{\pt,\eta}=\frac{M_{t1}(\pt,\eta)}{M_L(\pt,\eta)}\;,
\end{equation}
where $M_{t1}$ is the number of leptons passing the tight cuts and $M_L$ is the total number of loose
leptons for a given lepton transverse momentum and pseudorapidity.

\begin{figure}[!hbtp]
	\centering
	\begin{subfigure}[b]{0.45\textwidth}
		\centering
		\includegraphics[width=\textwidth]{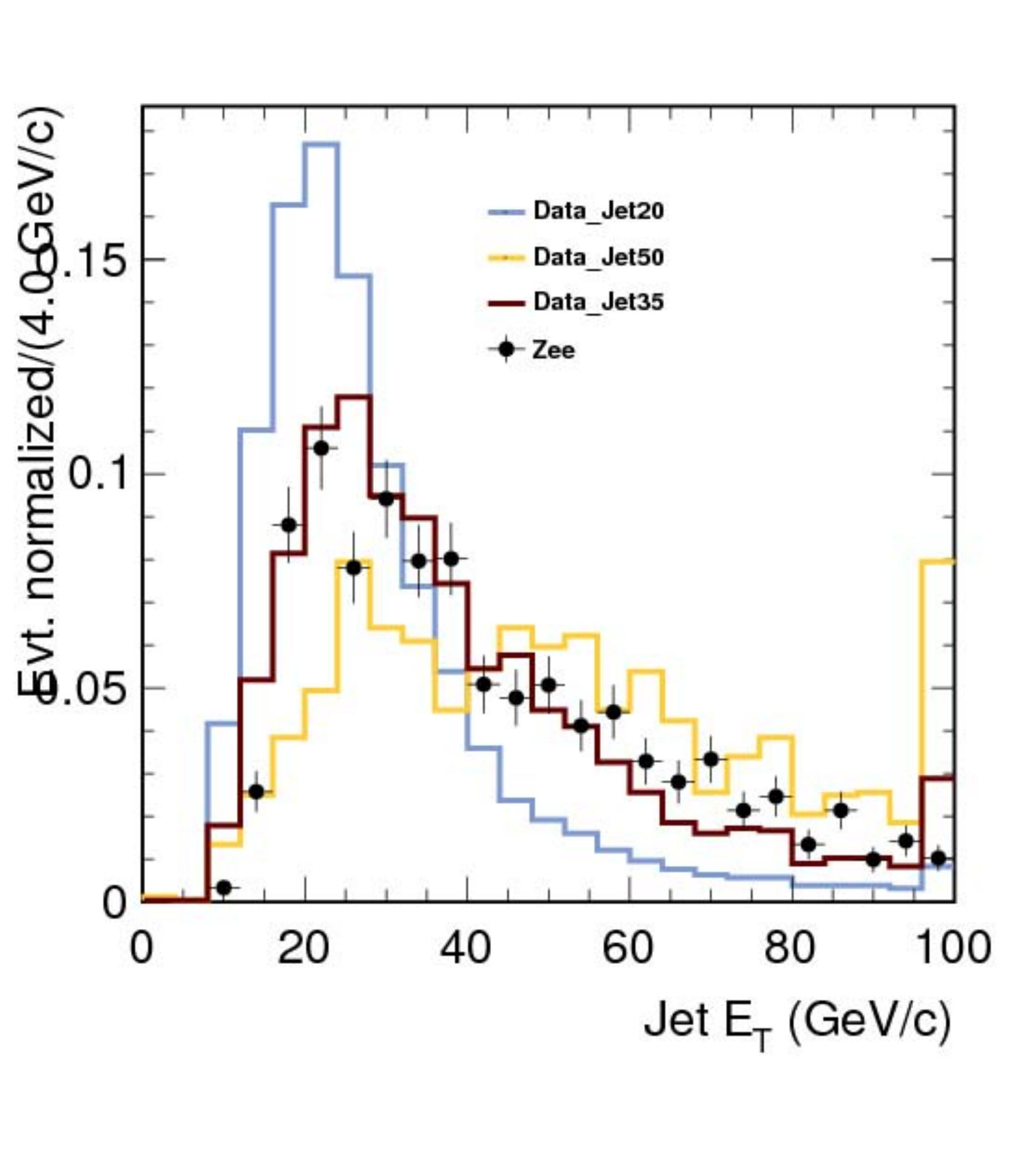}
		\vspace*{-1.2cm}
		\caption{Drell-Yan MC sample matching.\\}\label{ch7:fig:ET35}
	\end{subfigure}\quad
	\begin{subfigure}[b]{0.45\textwidth}
		\centering
		\includegraphics[width=\textwidth]{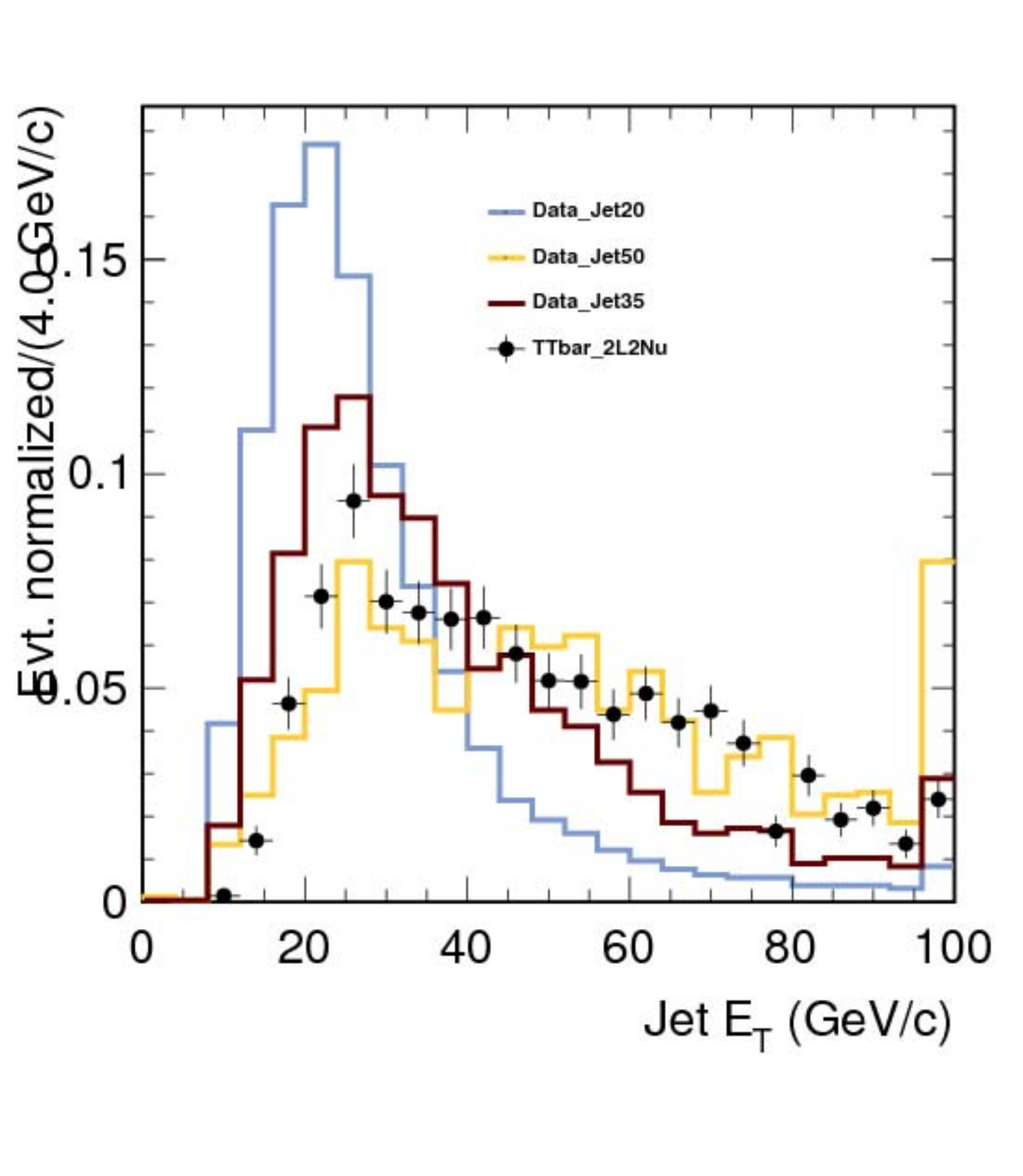}
		\vspace*{-1.2cm}
		\caption{\ttbar MC sample matching.}\label{ch7:fig:ET50}
	\end{subfigure}		
	\caption[Transverse energy spectra of fakeable jets]{Transverse energy distribution of the
	jets which contain a fakeable lepton for 2011 electrons in the fake-enriched sample used to 
	calculate the fake rate. A cut in the transverse energy of the leading jet of the 
	fake-enriched sample is applied giving as a result a variation in the \ET jet distribution
	of the fakeable jets. The \ET jet spectra for the \gls{mc} are also plotted. The curve
	built with leading jet with transverse energy higher than 35~\GeV matches with the Z+Jets 
	\gls{mc} distribution, whereas the curve with $E_{th}^{LeadJet}>50~\GeV$ matches with the 
	\gls{mc} \ttbar distribution.}\label{ch7:fig:ETSPECTRA}
\end{figure}

The fakeable sample should be as similar as possible to the jet content and lepton isolation 
distribution of the instrumental background contributions we want to estimate for our analysis. 
The contribution of the one fake component to the analysis,~\ie \PPF, is expected to be composed
mainly by Drell-Yan, \ttbar, single top and WW. Moreover, we assume (and it will be verified along 
this Section) the hypothesis that before the \W candidate requirement, the \PPF contribution is
essentially Drell-Yan due to the real \Z boson presence. But 
after requiring the \W candidate and therefore requiring high \MET we would expect a strong reduction
of Drell-Yan allowing the $\ttbar$ events emerge up. Accepting that hypothesis, the fake-enriched sample
defined to extract the fake rates should follow the jet or lepton isolation distribution of Drell-Yan
and \ttbar events\footnote{Strictly speaking, it must follow every background present but as it is 
shown in Table~\ref{ch7:tab:fomResults2011} the \PFF and \FFF distributions are almost 
negligible}. Using the \ET of the leading jet to be above a given threshold, $E_{th}^{LeadJet}$,
it is possible to modify the hardness of the \ET jet spectra of the sample and also the isolation
of the leptons due to the high correlation between both quantities. Therefore, the $E_{th}^{LeadJet}$
is cut in different values and the \ET of the jets with a lepton inside\footnote{In a $\Delta R$=0.3}
are plotted and compared with the Drell-Yan and \ttbar distributions simulated with \gls{mc}. Using
the same $E_{ŧh}^{LeadJet}$ cut, the relative isolation distribution of the leptons is also plotted
\begin{figure}[!hbtp]
	\centering
	\begin{subfigure}[b]{0.45\textwidth}
		\centering
		\includegraphics[width=\textwidth]{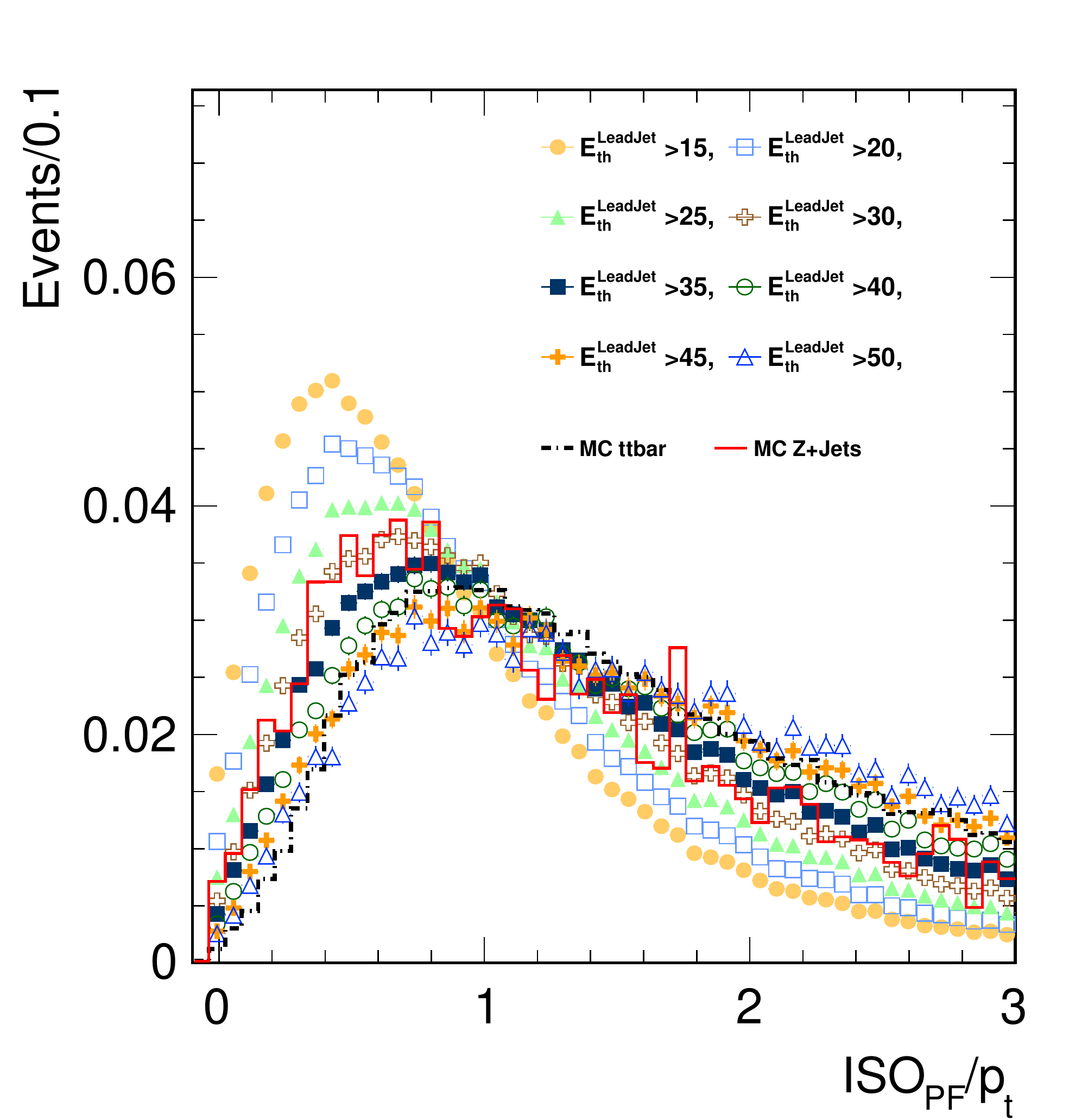}
		\caption{Relative isolation using a full Monte Carlo sample for Z+Jets and
		\ttbar processes.\\}\label{ch7:fig:ETISOMC}
	\end{subfigure}\quad
	\begin{subfigure}[b]{0.45\textwidth}
		\centering
		\includegraphics[width=\textwidth]{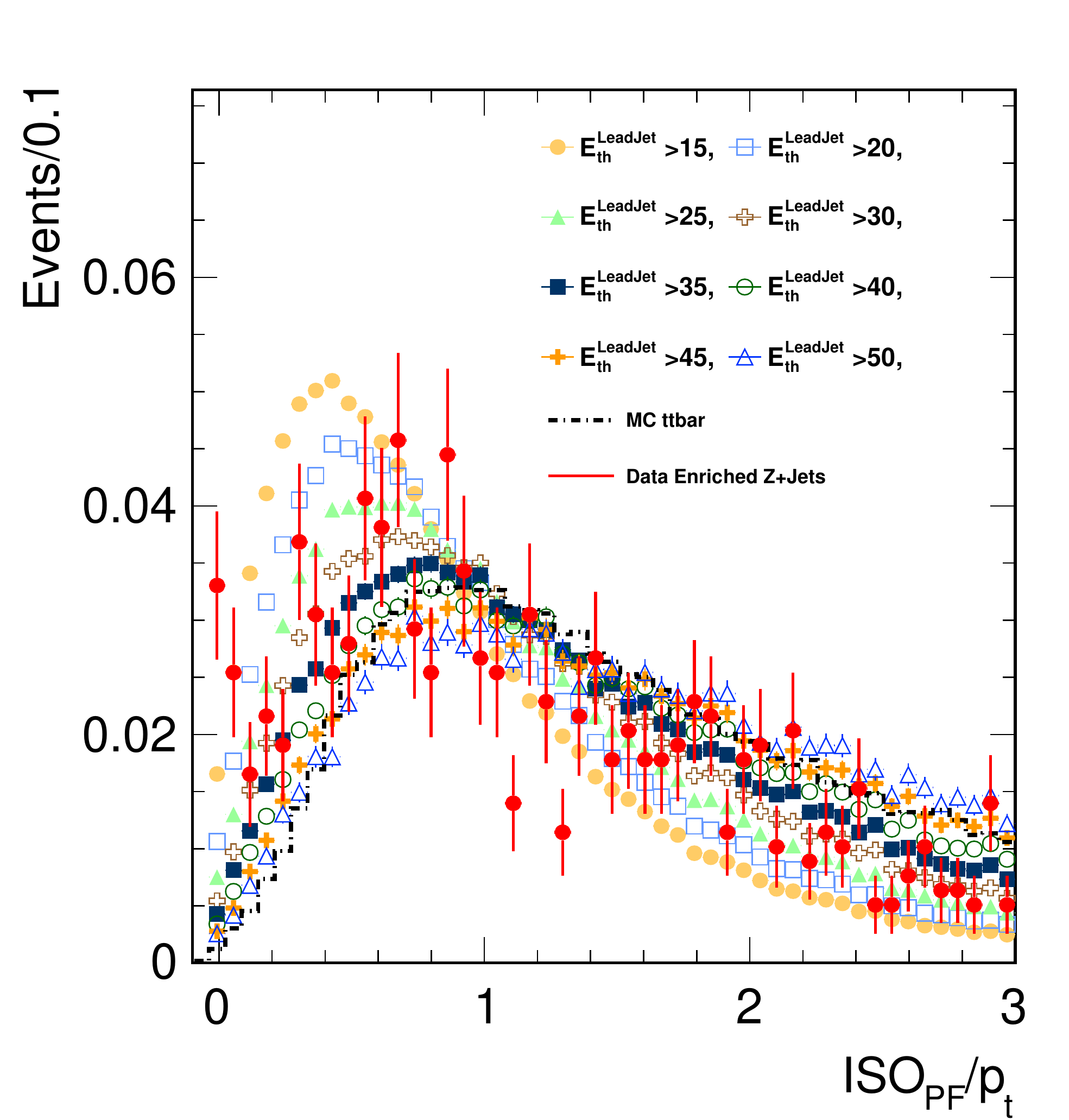}
		\caption{Relative isolation using a Z+Jets enriched experimental data and a MC
		sample for the \ttbar processes.}\label{ch7:fig:ETISODATA}
	\end{subfigure}		
	\caption[Relative isolation of the loose leptons in the fake-enriched sample]{Relative
	isolation distribution of the loose 2011 muons in the fake-enriched sample used to calculate
	the fake rate. A cut in the transverse energy of the leading jet of the fake-enriched 
	sample is applied giving as a result a variation in the relative isolation of the loose 
	leptons. The relative isolation of the fakes for the \gls{mc} are also plotted. The curve
	built with leading jet with transverse energy higher than 30~\GeV matches with the Z+Jets 
	\gls{mc} distribution, whereas the curve with $E_{th}^{LeadJet}>45~\GeV$ matches with the 
	\gls{mc} \ttbar distribution.}\label{ch7:fig:ETISO}
\end{figure}
and compared again with the \gls{mc} Drell-Yan and \ttbar distributions. Figure~\ref{ch7:fig:ETISOMC}
shows the relative isolation distribution of the 2011 muons, while Figure~\ref{ch7:fig:ETSPECTRA}
is showing the transverse energy spectra of the jets associated to an electron. Although the \ET of the 
jets with a lepton inside can also be used to obtain the matching with the \gls{mc} distributions, the 
isolation distribution is preferred because is a direct observable giving information of the 
hadronic activity of the lepton; in contrast, looking at the \ET of the jet with a lepton inside,
the information of the lepton, which is our primary goal, is obtained through the energy of the 
surrounding jet, introducing the jet reconstruction, the jet-lepton matching, and other 
convoluted effects. 

Applying a $\Psi$-test~\cite[pp.~294--305]{Jaynes2003} to evaluate the amount of plausibility has the \gls{mc} 
predicted distribution when it is found a particular set of observed data, the matching
data distribution with a Z+Jets is accomplished with a transverse energy cut in the Leading 
jet of the fake-enriched sample higher than 30 (35)~\GeV for muons (electrons), whilst for the 
\ttbar sample is better to use a transverse energy higher than 45-50~\GeV. These values, shown 
for leptons on the 2011 analysis, were found to be consistent along the 2012 too. Notice that the
data-driven method has been ``contaminated'' with the inclusion of the \gls{mc} prediction for the
relative isolation distributions. Nevertheless, although there are other 
possibilities\footnote{A Z+Jets and \ttbar enriched regions could be selected using only 
experimental data and use them to match the isolation distributions. This approach was tried in 
this analysis, but for the \ttbar case there was not enough events in 2011 to reach enough accuracy 
for the distribution shapes.} to determine the fakeable leptons kinematics, the \gls{mc} approach is 
simple enough and, in the case of the \ttbar, is a process well described theoretically. The $\Z+Jets$ 
is better estimated using a control region in data, defined with events fulfilling the \Z mass 
resonance window ($M_{Z}\pm10~\GeV$) and rejecting the presence of real \MET through a cut in its 
resolution $\MET<20~\GeV$. A very pure $Z+Jets$ sample is collected with these cuts where it is 
used the lepton not associate to the \Z resonance as the fake lepton in order to be compared its 
isolation with the isolation of the loose leptons in the jet-enriched sample. 
Figure~\ref{ch7:fig:ETISODATA} shows the relative isolation distribution of the fakeables of the
$Z+Jets$ sample; the limited number of events avoids us to extract a solid conclusion concerning
the optimal \ET cut on the leading jet of the jet-enriched sample, relying on \gls{mc} for selecting 
this cut (Figure~\ref{ch7:fig:ETISOMC}).

Therefore, the fake rates have been extracted, in $\eta$ and \pt of the leptons, by counting the 
number of passing loose leptons over the total loose leptons selected in a jet-enriched sample 
biased by requiring a transverse energy of the leading jet higher than 30 (35)~\GeV for 
muons (electrons) and 50~\GeV for both leptons, to account for the Z+Jets and \ttbar 
composition, respectively. Tables~\ref{ch7:tab:FRJet502011} and~\ref{ch7:tab:FRJet502012} 
present the fake rate matrices for muons and electrons used in the 2011 and 2012 analyses.
\begin{table}[!htbp]
	\centering
	\begin{subtable}[b]{0.9\textwidth}
		\centering
		\begin{tabular}{lcccc}\hline\hline
		     &$|\eta|\in[0,1]$ & $\eta\in(1,1.48]$ & $\eta\in(1.48,2]$ & $\eta\in(2,2.5]$ \\ \hline 
		        $10< p_t \leq 15$ & $3.4\pm0.7$ & $5.0\pm1.2$ & $5.7\pm1.5$ & $5\pm2$ \\ 
			$15< p_t \leq 20$ & $0.8\pm0.4$ & $2.9\pm1.3$ & $6.2\pm2.2$ & $0.04\pm0.03$ \\ 
			$20< p_t \leq 25$ & $1.1\pm0.1$ & $1.5\pm0.2$ & $2.5\pm0.3$ & $3.1\pm0.6$ \\ 
			$25< p_t \leq 30$ & $1.0\pm0.1$ & $0.9\pm0.2$ & $1.6\pm0.3$ & $2.3\pm0.6$ \\ 
			$30< p_t \leq \infty$ & $1.6\pm0.1$ & $1.7\pm0.2$ & $2.6\pm0.3$ & $2.2\pm0.4$ \\ \hline
		\end{tabular}
	\caption{Measured muon fake rates in function of transverse momentum and $\eta$ of the muon.
	Note that values in $10^{-2}$. Errors are statistical only}\label{ch7:subtab:muFR11}
	\end{subtable}
	\vskip 1em
	\begin{subtable}[b]{0.9\textwidth}
		\centering
		\begin{tabular}{lcccc}\hline\hline
		     &$|\eta|\in[0,1]$ & $\eta\in(1,1.48]$ & $\eta\in(1.48,2]$ & $\eta\in(2,2.5]$ \\ \hline 
			$10< p_t \leq 18$  & $6\pm4$     & $4\pm3$     & $1\pm1$ & $3\pm3$ \\ 
			$18< p_t \leq 26$  & $2.6\pm1.2$ & $3.4\pm1.6$ & $1\pm1$ & $4\pm2$ \\ 
			$26< p_t \leq 34$  & $6.1\pm1.5$ & $6.0\pm1.8$ & $5.4\pm1.8$ & $5.1\pm1.8$ \\ 
			$34< p_t \leq \infty$& $5.2\pm1.4$ & $2.4\pm1.2$ & $3.3\pm1.5$ & $4.4\pm1.6$ \\\hline
		\end{tabular}
	\caption{Measured electron fake rates in function of transverse momentum and $\eta$ of the electron. 
		Values in $10^{-2}$. Errors are statistical only}\label{ch7:subtab:elecFR11}
	\end{subtable}
	\caption[Fake rates for 2011 analysis, extracted with $E_{th}^{LeadJet}>50~\GeV$]{Measured 
		lepton fakes rates from the jet-enriched sample described in the text, using a 
		transverse energy of the leading jet higher than 50~\GeV for 2011 analysis. The 
		values in the tables are in $10^{-2}$.}\label{ch7:tab:FRJet502011}
\end{table}
\begin{table}[!htbp]
	\centering
	\begin{subtable}[b]{0.9\textwidth}
		\centering
		\begin{tabular}{ccccc}\hline\hline
		     &$|\eta|\in[0,1]$ & $\eta\in(1,1.48]$ & $\eta\in(1.48,2]$ & $\eta\in(2,2.5]$ \\ \hline 
			$10< p_t \leq 15$ & $7.6\pm1.1$& $7.6\pm1.7$& $9\pm3$ & $18\pm6$ \\ 
			$15< p_t \leq 20$ & $6.3\pm1.9$& $6\pm3$    & $9\pm5$ & $38\pm14$ \\ 
			$20< p_t \leq 25$ & $7.7\pm1.0$& $9.2\pm1.8$& $9\pm3$ & $10\pm4$ \\ 
			$25< p_t \leq 30$ & $7.0\pm1.5$& $11\pm3$   & $9\pm3$ & $13\pm6$ \\ 
                    $30< p_t \leq \infty$ & $7.4\pm1.9$& $8\pm3$    & $7\pm3$ & $21\pm11$ \\ \hline
		\end{tabular}

	\caption{Measured muon fake rates in function of transverse momentum and $\eta$ of the muon.
	Note that values in $10^{-2}$. Errors are statistical only}\label{ch7:subtab:muFR12}
	\end{subtable}
	\vskip 1em
	\begin{subtable}[b]{0.9\textwidth}
		\centering
		\begin{tabular}{ccccc}\hline\hline
		     &$|\eta|\in[0,1]$ & $\eta\in(1,1.48]$ & $\eta\in(1.48,2]$ & $\eta\in(2,2.5]$ \\ \hline 
			$10< p_t \leq 15$ & $4.3\pm0.5$ & $3.4\pm0.4$ & $0.8\pm0.2$ & $1.6\pm0.5$ \\ 
			$15< p_t \leq 20$ & $4.8\pm0.3$ & $6.5\pm0.3$ & $2.7\pm0.1$ & $3.0\pm0.2$ \\ 
			$20< p_t \leq 25$ & $4.1\pm0.2$ & $6.0\pm0.3$ & $3.1\pm0.2$ & $2.4\pm0.2$ \\ 
			$25< p_t \leq 30$ & $3.8\pm0.3$ & $5.5\pm0.5$ & $2.2\pm0.3$ & $2.1\pm0.3$ \\ 
		    $30< p_t \leq \infty$ & $2.8\pm0.6$ & $4.8\pm0.9$ & $2.5\pm0.6$ & $2.4\pm0.5$ \\ \hline
		\end{tabular}
	\caption{Measured electron fake rates in function of transverse momentum and $\eta$ of the electron. 
		Values in $10^{-2}$. Errors are statistical only}\label{ch7:subtab:elecFR12}
	\end{subtable}
	\caption[Fake rates for 2012 analysis, extracted with $E_{th}^{LeadJet}>50~\GeV$]{Measured 
		lepton fakes rates from the jet-enriched sample described in the text, using a 
		transverse energy of the leading jet higher than 50~\GeV for 2012 analysis. 
		The values in the tables are in $10^{-2}$.}\label{ch7:tab:FRJet502012}
\end{table}

\begin{figure}[!htbp]
	\centering
	\begin{subfigure}[b]{0.45\textwidth}
		\includegraphics[width=\textwidth]{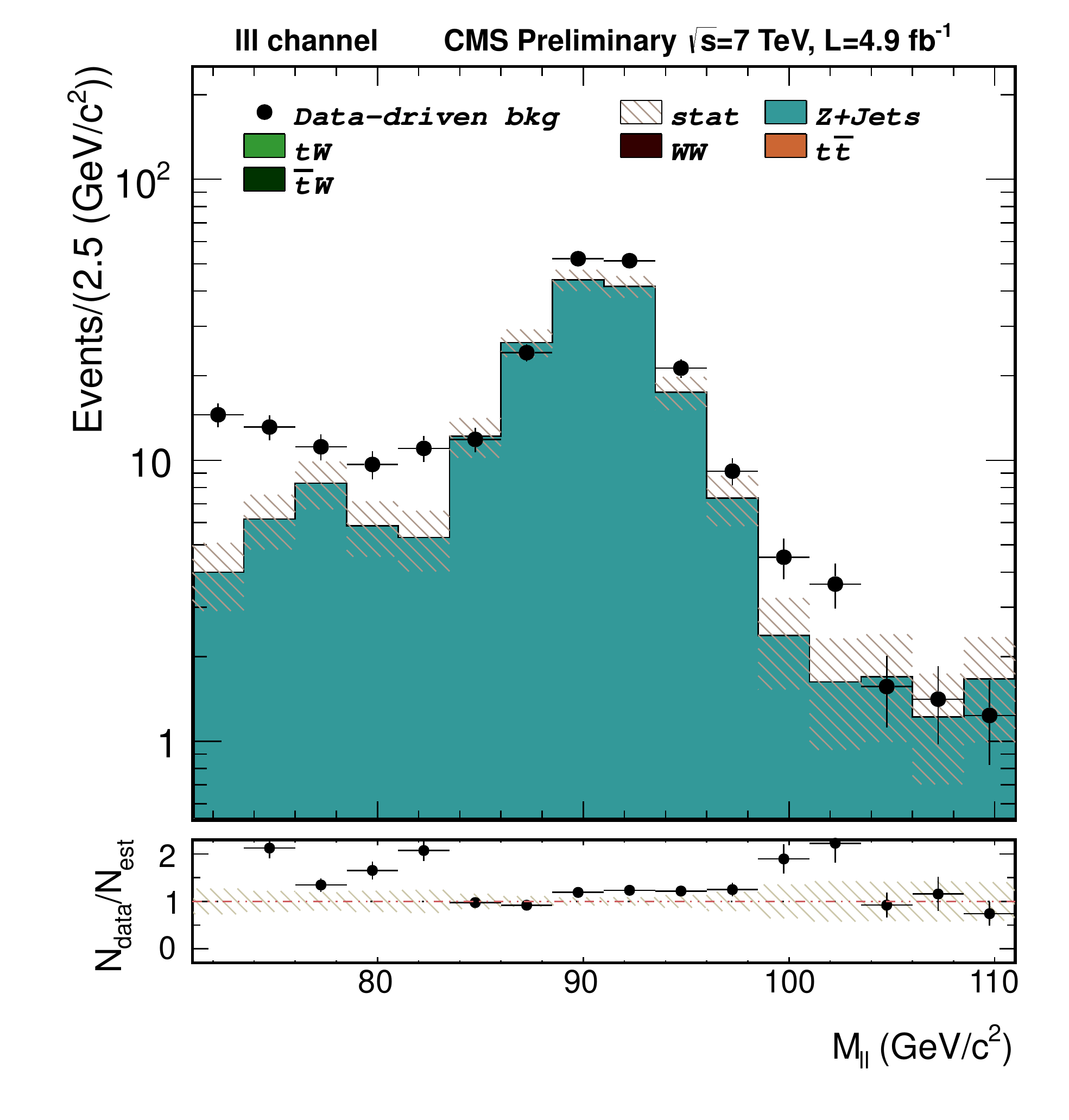}
		\caption{Invariant mass distribution of the same flavour, opposite-signed lepton 
		system}
	\end{subfigure}\quad
	\begin{subfigure}[b]{0.45\textwidth}
		\includegraphics[width=\textwidth]{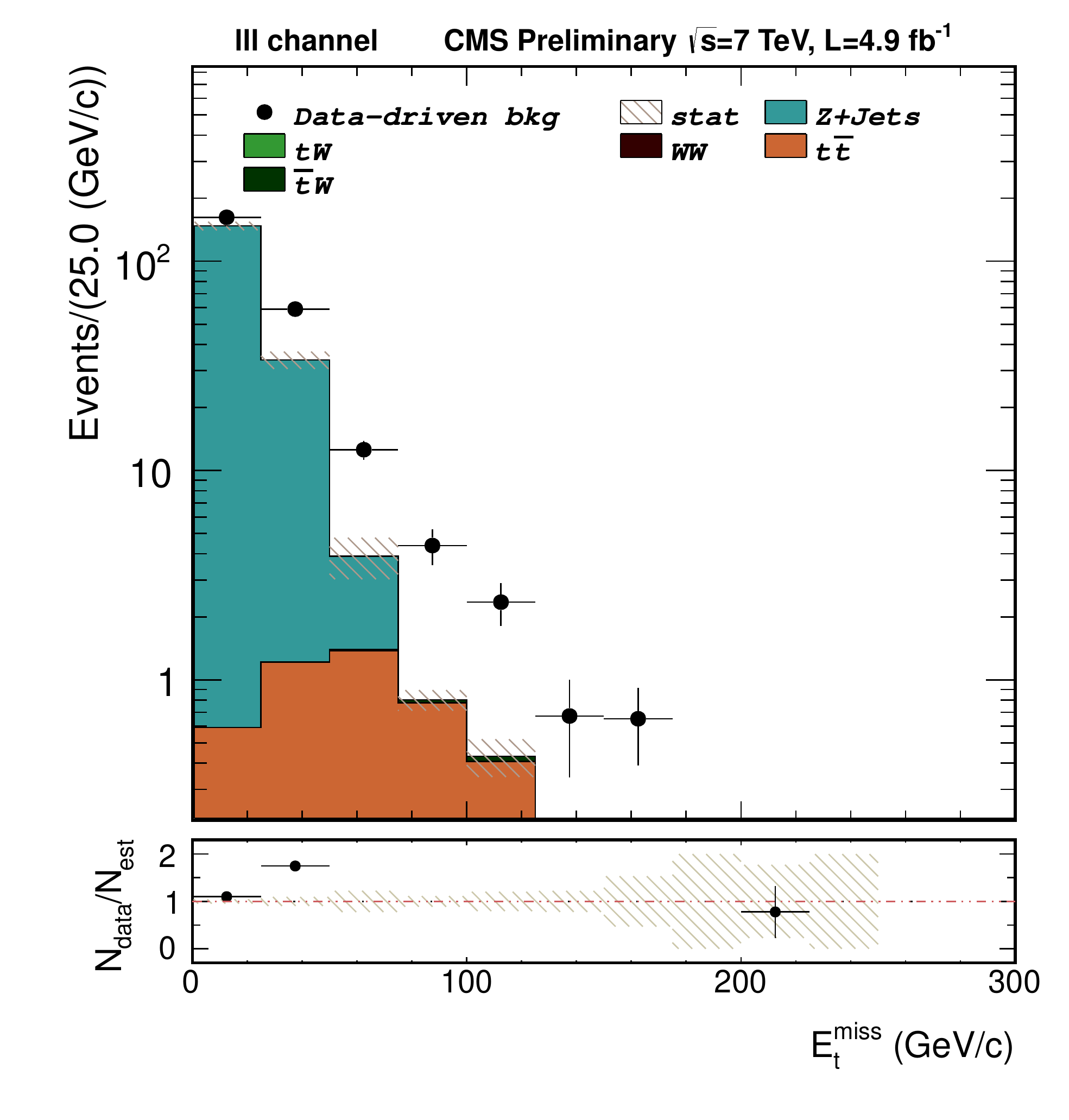}
		\caption{Missing transverse energy event distribution}
	\end{subfigure}
	\caption[Jet-induced background composition plots, requiring a \Z-candidate]{Jet-induced 
		background composition after
		the \Z-candidate analysis requirement is applied. We may appreciate that the dominant
		contribution is coming from the Z+Jets process. The figures are showing the four 
		channels added up in logarithmic scale. The dot markers are the \PPF contribution 
		estimated with the data-driven method, the other samples are Monte Carlo 
		based. Data from 2011 run.}\label{ch7:fig:ddcomposition_afterZ}
\end{figure}

\begin{figure}[!htpp]
	\centering
	\begin{subfigure}[b]{0.45\textwidth}
		\includegraphics[width=\textwidth]{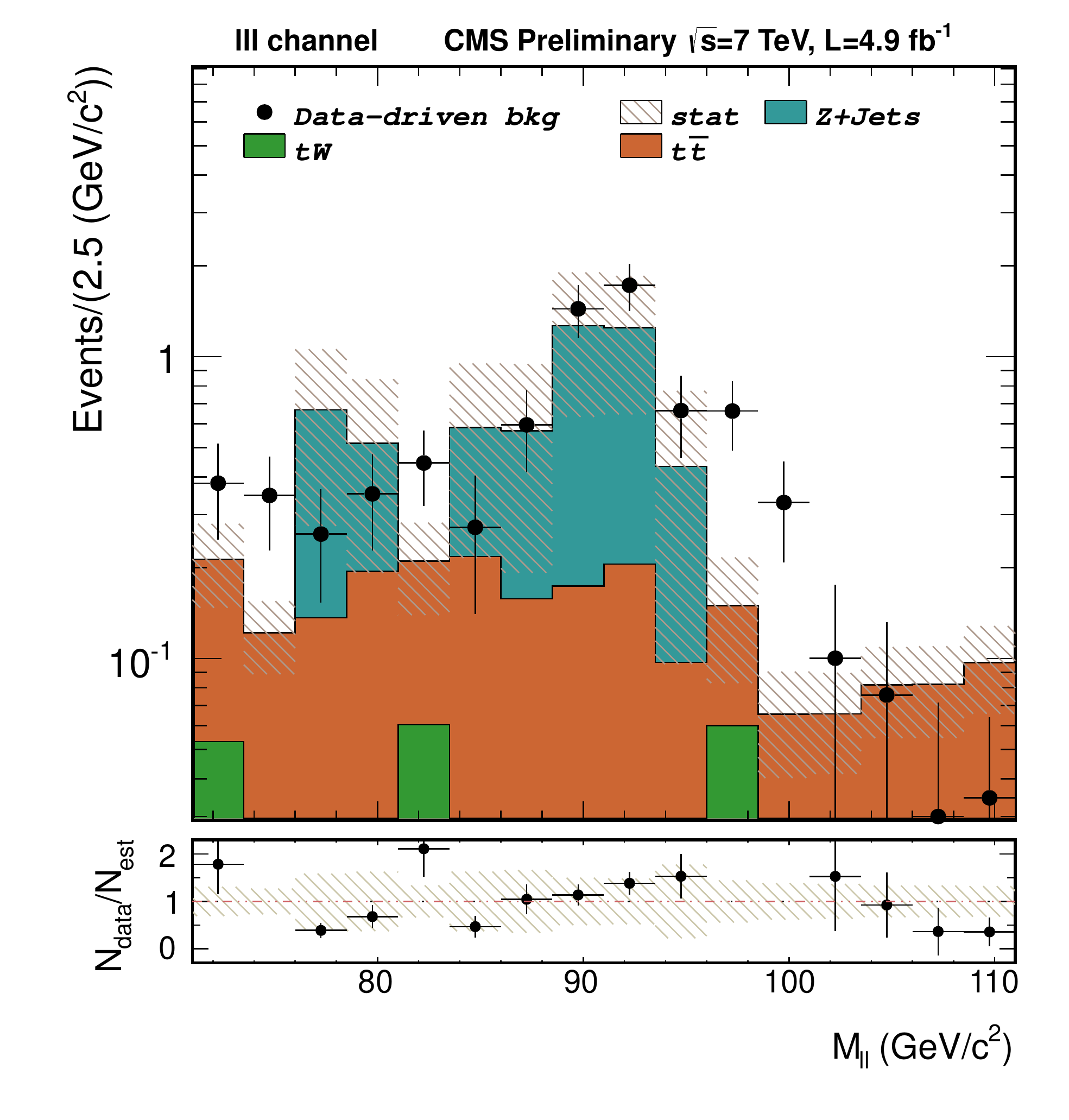}
		\caption{Invariant mass distribution of the same flavour, opposite-signed lepton 
		system}
	\end{subfigure}\quad
	\begin{subfigure}[b]{0.45\textwidth}
		\includegraphics[width=\textwidth]{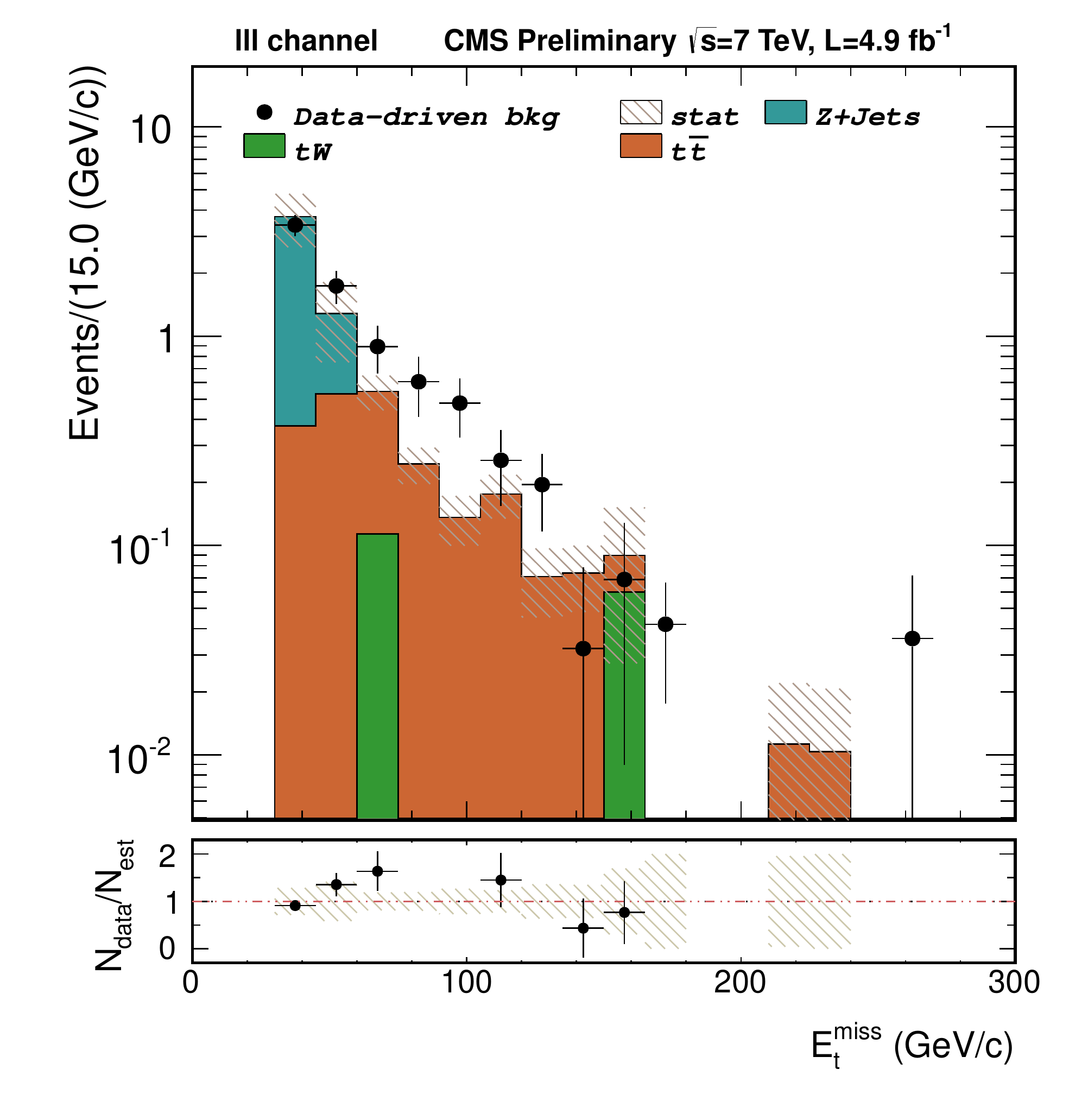}
		\caption{Missing transverse energy event distribution}
	\end{subfigure}
	\caption[Jet-induced background composition plots]{Jet-induced background composition after
		the \MET analysis requirement is applied. The composition is almost even mixture
		of \ttbar and Drell-Yan events. The figures are showing the four channels added up. 
		The dot markers are the \PPF contribution estimated with the data-driven 
		method,~\ie $N_{\pr\pr\fr}^{N_{t3}}$, the other samples are Monte Carlo 
		based. Data from 2011 run.}\label{ch7:fig:ddcomposition}
\end{figure}

The selection of the leading jet transverse energy cut depends on the jet-induced background 
composition in the \WZ analysis. It was already argued that the major contribution of this 
background is going to come from the processes with one fake lepton and, thus, two prompt (\PPF
in the notation defined at the beginning of the Chapter). This hypothesis will be probed in the 
Section~\ref{ch7:subsec:fomResults}. The other assumption made is that the \PPF contribution is
mainly composed by Drell-Yan and \ttbar events. This assumption has naively checked by comparing the 
data-driven estimation with Z+Jets and \ttbar \gls{mc} \PPF samples. The data-driven estimation has
been obtained using several fake rate matrices extracted with different $E_{th}^{LeadJet}$ cuts 
trying to avoid possible bias due to the fake matrix choice. 
Figures~\ref{ch7:fig:ddcomposition_afterZ} show that the fakes lepton background in 
the \WZ analysis is mainly dominated by the Drell-Yan before the W lepton candidate requirement.
After the \MET requirement, Figures~\ref{ch7:fig:ddcomposition} show the suppression of the Drell-Yan 
component, allowing the $\ttbar$ sample to contribute at the same level of events. 
Therefore, the $\pr\pr\fr$ contribution is going to be composed mainly by Drell-Yan and $\ttbar$
events as we have already assumed.

\paragraph{}\mbox{}\\
For the prompt rate extraction, a tag and probe method is used with the fakeable leptons defining 
the probes. The measured prompt rates are shown in Tables~\ref{ch7:tab:PR2011} 
and~\ref{ch7:tab:PR2012}.
\begin{table}[!htbp]
	\centering
	\begin{subtable}[b]{0.8\textwidth}
		\centering
		\resizebox{\textwidth}{!}
		{
		\begin{tabular}{ccccc}\hline\hline
	           &$0.0< |\eta| \leq 1.0$ & $1.0< |\eta| \leq 1.48$ & $1.489< |\eta| \leq 2.0$ & $2.0< |\eta| \leq 2.4$ \\ \hline 
		   $10< p_t \leq 15$ & $0.7780\pm0.0024$ & $0.7780\pm0.0024$ & $0.7780\pm0.0024$ & $0.7780\pm0.0024$ \\ 
		   $15< p_t \leq 20$ & $0.7780\pm0.0024$ & $0.7780\pm0.0024$ & $0.7780\pm0.0024$ & $0.7780\pm0.0024$ \\ 
		   $20< p_t \leq 25$ & $0.8670\pm0.0015$ & $0.8670\pm0.0015$ & $0.8670\pm0.0015$ & $0.8670\pm0.0015$ \\ 
		   $25< p_t \leq 30$ & $0.9059\pm0.0009$ & $0.9059\pm0.0009$ & $0.9059\pm0.0009$ & $0.9059\pm0.0009$ \\ 
		   $30< p_t \leq \infty$ & $0.9489\pm0.0002$ & $0.9489\pm0.0002$ & $0.9489\pm0.0002$ & $0.9489\pm0.0002$ \\\hline
		\end{tabular}
		}
		\caption{Measured muon prompt rates in bins of $p_t$ and $\eta$ of the muon. Errors are statistical only}
		\label{ch7:subtab:muonPR11}
	\end{subtable}
	\vskip 1em
	\begin{subtable}[b]{0.8\textwidth}
		\centering
		\resizebox{\textwidth}{!}
		{
		\begin{tabular}{ccccc}\hline\hline
		   &$0.0< |\eta| \leq 1.0$ & $1.0< |\eta| \leq 1.48$ & $1.48< |\eta| \leq 2.0$ & $2.0< |\eta| \leq 2.5$ \\ \hline 
		   $10< p_t \leq 15$ & $0.8145\pm0.0033$ & $0.8145\pm0.0033$ & $0.6048\pm0.0053$ & $0.6048\pm0.0053$ \\ 
		   $15< p_t \leq 20$ & $0.8145\pm0.0033$ & $0.8145\pm0.0033$ & $0.6048\pm0.0053$ & $0.6048\pm0.0053$ \\ 
		   $20< p_t \leq 25$ & $0.9031\pm0.0014$ & $0.9031\pm0.0014$ & $0.8058\pm0.0026$ & $0.8058\pm0.0026$ \\ 
		   $25< p_t \leq 30$ & $0.9244\pm0.0009$ & $0.9244\pm0.0009$ & $0.8406\pm0.0020$ & $0.8406\pm0.0020$ \\ 
		   $30< p_t \leq \infty$ & $0.9572\pm0.0002$ & $0.9572\pm0.0002$ & $0.8980\pm0.0005$ & $0.8980\pm0.0005$ \\\hline
		\end{tabular}
		}
		\caption{Measured electron prompt rates in bins of $p_t$ and $\eta$ of the electron.
		Errors are statistical only}
		\label{ch7:subtab:elecPR11}
	\end{subtable}
	\caption[Prompt rates for 2011 analysis]{Measured lepton prompt rates for 2011 analysis,
		using a tag and probe method.}\label{ch7:tab:PR2011}
\end{table}
\begin{table}[!htbp]
	\centering
	\begin{subtable}[b]{0.8\textwidth}
		\centering
		\resizebox{!}{!}
		{
		\begin{tabular}{ccccc}\hline\hline
	          &$0.0< |\eta| \leq 1.48$ & $1.48< |\eta| \leq 2.5$ \\ \hline 
			$10< p_t \leq 15$ & $0.7119\pm0.0003$ & $0.7582\pm0.0006$ \\ 
			$15< p_t \leq 20$ & $0.8049\pm0.0018$ & $0.8495\pm0.0001$ \\ 
			$20< p_t \leq 25$ & $0.9027\pm0.0008$ & $0.8948\pm0.0012$ \\ 
			$25< p_t \leq 50$ & $0.9741\pm0.0001$ & $0.9627\pm0.0002$ \\ 
			$50< p_t \leq \infty$ & $0.9900\pm0.0001$ & $0.9875\pm0.0003$ \\\hline 
		\end{tabular}
		}
	\caption{Measured muon prompt rates in bins of $p_t$ and $\eta$ of the muon. Errors are statistical only}
	\label{ch7:subtab:muonPR12}
	\end{subtable}
	\vskip 1em
	\begin{subtable}[b]{0.8\textwidth}
		\centering
		\resizebox{\textwidth}{!}
		{
		\begin{tabular}{ccccc}\hline\hline
		   &$0.0< |\eta| \leq 1.0$ & $1.0< |\eta| \leq 1.479$ & $1.479< |\eta| \leq 2.0$ & $2.0< |\eta| \leq 2.5$ \\ \hline 
		   $10.0< p_t \leq 15.0$ & $0.8145\pm0.0033$ & $0.8145\pm0.0033$ & $0.6048\pm0.0053$ & $0.6048\pm0.0053$ \\ 
		   $15.0< p_t \leq 20.0$ & $0.8145\pm0.0033$ & $0.8145\pm0.0033$ & $0.6048\pm0.0053$ & $0.6048\pm0.0053$ \\ 
		   $20.0< p_t \leq 25.0$ & $0.9031\pm0.0014$ & $0.9031\pm0.0014$ & $0.8058\pm0.0026$ & $0.8058\pm0.0026$ \\ 
		   $25.0< p_t \leq 30.0$ & $0.9244\pm0.0009$ & $0.9244\pm0.0009$ & $0.8406\pm0.0020$ & $0.8406\pm0.0020$ \\ 
		   $30.0< p_t \leq \infty$ & $0.9572\pm0.0002$ & $0.9572\pm0.0002$ & $0.8980\pm0.0005$ & $0.8980\pm0.0005$ \\ \hline
		\end{tabular}
		}
		\caption{Measured electron prompt rates in bins of $p_t$ and $\eta$ of the electron. Errors are statistical only}
		\label{ch7:subtab:elecPR12}
	\end{subtable}
	\caption[Prompt rates for 2012 analysis]{Measured lepton prompt rates for 2012 analysis, 
		using a tag and probe method.}\label{ch7:tab:PR2012}
\end{table}

\subsection{Estimation of data-driven contribution}\label{ch7:subsec:fomResults}
The jet-induced background contribution to the \WZ signal is estimated through the 
Equations~\eqref{ch7:eq:fomEstimation}. In particular, $N_{\pr\pr\pr}^{N_{t_3}}$ is in fact 
the data-driven expectation signal given that the equation is estimating the contribution of three
prompt leptons to the final state. It may be noticed, however, that any process with three prompt
leptons,~\ie an irreducible background as, for instance, the $\ensuremath{ZZ}$ full leptonic decay where one 
lepton is lost, will contribute to $N_{\pr\pr\pr}^{N_{t3}}$. Therefore, such processes have to 
be subtracted, using the corresponding \gls{mc}. Equivalently, the \WZ signal expectation may be 
obtained using Equation~\ref{ch7:subeq:fomNt3} by,
\begin{equation}
	N_{\pr\pr\pr}^{N_{t3}}=N_{t3}-\left(N_{\pr\pr\fr}^{N_{t3}}
	    +N_{\pr\fr\fr}^{N_{t3}}+N_{\fr\fr\fr}^{N_{t3}}\right)
\end{equation}
The \WZ signal expectation (plus the irreducible backgrounds) is estimated subtracting to 
the total events passing all the selection cuts, the estimation of the \PPF, \PFF and \FFF 
\begin{table}[!hpbt]
	\centering
	\begin{subtable}[b]{0.45\textwidth}
		\centering
		\begin{tabular}{l  cccc}  \hline\hline
		& $3e$  & $2e1\mu$ &  $2\mu1e$  & $3\mu$ \\ \hline 
		$N_{t0}$&   1   &   8      &   3        &   6    \\
		$N_{t1}$&  20   &  16      &  21        &  29    \\
		$N_{t2}$&  63   &  80      &  98        & 119    \\
		$N_{t3}$&  64   &  62      &  70        &  97    \\ \hline 
		\end{tabular}
		\caption{2011 analysis table}
	\end{subtable}\quad
	\begin{subtable}[b]{0.45\textwidth}
		\centering
		\begin{tabular}{l  cccc}  \hline\hline
		& $3e$  & $2e1\mu$ &  $2\mu1e$  & $3\mu$ \\ \hline 
		$N_{t0}$&   7   &   1      &   3        &   5    \\
		$N_{t1}$&  81   &  33      &  71        &  53    \\
		$N_{t2}$&  270  &  207     &  497       & 327    \\
		$N_{t3}$&  235  &  288     &  400       &  557   \\ \hline 
		\end{tabular}
		\caption{2012 analysis table}
	\end{subtable}
	\caption[Yields for the fakeable signal samples for 2011]{Number of selected events with 
	fakeable objects for each decay channel and selection category, where $N_{ti}$ is the actual 
	selected number of events with $i$-tight and $(3-i)$ fail leptons. Data from 2011
	analysis.}\label{ch7:tab:fomNtiYields}
\end{table}
jet-induced background contributions. The estimated number of events obtained in the \WZ analysis 
using the Equations~\eqref{ch7:eq:fomEstimation} and the fake and prompt rates obtained in the last 
Section (Tables~\ref{ch7:tab:FRJet502011} and~\ref{ch7:tab:PR2011}, respectively) are reported in 
Table~\ref{ch7:tab:fomResults2011} split by measured channels. The actual measured $N_{ti}$ yields
are given in Table~\ref{ch7:tab:fomNtiYields}.  
\begin{table}[!hpbt]
	\centering
	\begin{subtable}[b]{\textwidth}
	  \centering
          \begin{tabular}{l c c c c}\hline \hline
          	&      $3e$         &     $2e1\mu$     &  $2\mu1e$         &  $3\mu$ \\ \hline
          $N_{\pr\pr\pr}^{N_{t3}}$  & $61.92\pm8.07$    &$ 60.56\pm7.93$   & $67.67\pm8.42$    & $95.30\pm9.89$ \\
          $N_{\pr\pr\fr}^{N_{t3}}$  & $ 2.06\pm0.45$    &$ 1.44\pm 0.28$   & $ 2.32\pm0.42$    & $ 1.70\pm0.24$ \\
          $N_{\pr\fr\fr}^{N_{t3}}$  & $ 0.022\pm0.011$  &$ 0.0028\pm0.0041$& $ 0.0088\pm0.0050$& $ 0.0060\pm0.0028$ \\
          $N_{\fr\fr\fr}^{N_{t3}}$  & $ 0.0001\pm0.0002$&$ 0.0002\pm0.0001$& $ 0\pm0$          & $ 0\pm0$ \\ \hline
          $\sum N$         & $ 64$        &$  62         $   & $ 70             $& $97        $\\\hline
          \end{tabular}
	  \caption{2011 analysis table}
        \end{subtable}
	\vskip 1em
	\begin{subtable}[b]{\textwidth}
	  \centering
          \begin{tabular}{l c c c c}\hline \hline
          	&      $3e$         &     $2e1\mu$     &  $2\mu1e$         &  $3\mu$ \\ \hline
          $N_{\pr\pr\pr}^{N_{t3}}$  & $220.16\pm15.74$ &$ 260.89\pm17.62$ & $352.08\pm20.82$& $498.00\pm24.70$ \\
          $N_{\pr\pr\fr}^{N_{t3}}$  & $14.67\pm1.44$   &$ 27.02\pm 2.93$  & $ 47.32\pm3.44$ & $57.11\pm4.76$ \\
          $N_{\pr\fr\fr}^{N_{t3}}$  & $0.169\pm0.058$  &$ 0.09\pm0.13$    & $ 0.59\pm0.24$  & $1.86\pm0.52$ \\
          $N_{\fr\fr\fr}^{N_{t3}}$  & $0.0008\pm0.0010$&$ 0.0014\pm0.0012$& $ 0.011\pm0.013$& $0.030\pm0.027$ \\ \hline
          $\sum N$         & $ 235$        &$  288         $   & $ 400             $& $557        $\\\hline
          \end{tabular}
	  \caption{2012 analysis table}
        \end{subtable}
	\caption[Fakeable object method estimation]{Number of events estimated in the using the 
	fakeable object method. The last row is the analytic sum of previous
	rows and should be exactly $N_{t3}$. It can be appreciated the negligible contribution of 
	the \PFF and \FFF processes, meaning the jet-induced background is mostly due to the 
	\PPF processes.}\label{ch7:tab:fomResults2011}
\end{table}
From table~\ref{ch7:tab:fomResults2011} one may observe that the contribution to the signal 
selection $N_{t3}$ is clearly dominated by the \PPF component, being negligible in the analysis 
the \PFF and \FFF contributions.

\subsection{Validation of the method}
Several closure tests have been carried out to assess the validity of the data-driven
approach to estimate the fake-lepton background contribution. A closure test may be performed
by obtaining a pure \PPF sample, thus in that case one may eliminate all the non \PPF terms from
Equation~\ref{ch7:subeq:fomNt3},
\begin{equation}
	N_{t3}=p^2fN_{\pr\pr\fr}\equiv N_{\pr\pr\fr}^{N_{t3}}
	\label{ch7:eq:closuretest}
\end{equation}
We can check the method reliability and its assumptions by checking the validity of the above 
equation in a pure \PPF sample. 

The \PPF background in the \WZ analysis is expected to be dominated by \ttbar and Drell-Yan 
processes (see Section~\ref{ch7:subsec:prfr}). Therefore, it has been applied the \gls{fom}
method to a \ttbar and Drell-Yan \gls{mc} simulated samples and to a \ttbar and Drell-Yan enriched 
data samples. Both approaches have caveats: the data-driven applied in the \gls{mc} samples uses a 
fake rate matrices extracted from a jet-enriched data sample described in 
Section~\ref{ch7:subsec:prfr} whilst it should be use fake rates matrices extracted from a simulated 
\gls{mc} data. Conversely, the data-driven applied to the \PPF enriched samples suffers contamination
from other background contributions, due to the impossibility to select a pure \PPF sample using 
experimental data, thus it does not fulfil the requirements needed to apply the 
Equation~\eqref{ch7:eq:closuretest}. 

The \gls{mc}-based closure tests have been performed in the 2011 analysis applying the \gls{fom} to
the $Z/\gamma\rightarrow\ell^+\ell^-+Jets$ \gls{mc} sample and to the 
$\ttbar\rightarrow 2\ell 2\nu 2b$, both described in Table~\ref{ch7:tab:mcsamples11}. The 
Table~\ref{ch7:tab:ctMCttbar} shows the closure test results for the \ttbar dileptonic sample and 
the results for the Drell-Yan sample are in Table~\ref{ch7:tab:ctMCZJets}.
\begin{table}[!htpb]
	\centering
	\begin{tabular}{l  c c c c}\hline\hline
		&            $3e$   &           $2e1\mu$    &           $2\mu1e$  &  $3\mu$   \\ \hline
	$N_{\pr\pr\fr}^{N_{t3}}$    & $0.220\pm0.011$ & $0.396\pm0.012$ & $0.607\pm0.014$ & $0.843\pm0.015$ \\ 
	$N_{t3}$                    & $0.29\pm0.06$   & $0.44\pm0.07$   & $0.65\pm0.08$   & $0.64\pm0.08$ \\ \hline
	$\delta\varepsilon\cdot 100$& $         24\%$ & $         10\%$ & $         7\%$  & $         32\%$\\
	$N_{\sigma}$                & $1.2$           & $0.6$           & $0.5 $          & $ 2.5$ \\\hline
	\end{tabular}
	\caption[Closure test for MC \ttbar for 2011 analysis]{Data-driven estimation
	($N_{\pr\pr\fr}^{N_{t3}}$) and yields from the nominal analysis selection 
	($N_{t3}$), obtained with the 
	dileptonic \ttbar simulated sample. Errors are statistical only. The $\delta\varepsilon$
	is the relative difference between both central 
	measurements and $N_{\sigma}$ the number of standard deviations between both measures.
	The fake rates used for this estimation were obtained from a jet-enriched sample with a
	$E_{th}^{LeadJet}>50~\GeV$ cut. The yields are weighted to a luminosity 
	of 4.9~\fbinv}\label{ch7:tab:ctMCttbar}
\end{table}
\begin{table}[!htbp]
	\centering
	\begin{tabular}{l c c c c}\hline\hline
		&            $3e$   &           $2e1\mu$    &           $2\mu1e$   &  $3\mu$   \\ \hline
		$N_{\pr\pr\fr}^{N_{t3}}$     & $27.3\pm1.1$  & $30.27\pm1.1$ &  $149\pm2$   & $42.5\pm1.4$\\ 
		$N_{t3}$                     & $19\pm3$      & $53\pm4$      &  $106\pm6$   & $91\pm6$\\ \hline
		$\delta\varepsilon\cdot 100$ & $33\%$        & $   43\%$     & $         29\%$   & $53\%$\\
		$N_{\sigma}$                 & $2.6$         & $5.1$          & $6.9$             & $8.31$ \\\hline
	\end{tabular}
	\caption[Closure test for Z+Jets MC sample for 2011]{Data-driven estimation 
	($N_{\pr\pr\fr}^{N_{t3}}$) and yields from the nominal analysis selection
	($N_{t3}$) obtained with the $Z+Jets$ MC simulated sample. Errors shown are statistical only.
	The $\delta\varepsilon$ is the relative difference 
	between both central measurements and $N_{\sigma}$ the number of standard deviations 
	between both measures. The fake rates used for this estimation were obtained from a 
	jet-enriched sample with a $E_{th}^{LeadJet}>30(35)~\GeV$ cut. Yields weighted to an 
	arbitrary luminosity.}\label{ch7:tab:ctMCZJets}
\end{table}
In both cases, the \PPF contribution has been obtained by applying the \gls{fom} to the simulated 
samples, while concurrently, the selection cuts of the main \WZ analysis were applied to the same
simulated sample. The closure test (Equation~\eqref{ch7:eq:closuretest}) predicts the same values 
the \PPF contribution and for the $N_{t3}$ yields within uncertainties, which can be tested using
the \emph{number of standard deviation} observable, $N_{\sigma}$, defined as
\begin{equation}
	N_{\sigma}=\frac{|\bar{a}-\bar{b}|}{\sqrt{\sigma^2_a+\sigma^2_b}},
	\label{ch7:eq:numberofsigma}
\end{equation}
being $\bar{a}$ and $\bar{b}$ the central values of two measurements, and $\sigma_a$, $\sigma_b$
their respective associated errors. Thus, the $\sigma_{n}$ observable is quantifying the 
compatibility of two measurements, and as a rule of thumb a $N_{\sigma} < 3$ is implying a 
compatible measurements\footnote{Assuming independent Gaussian errors} (with probability of 99.7\%).

The results obtained in the aforementioned tables show compatible values for the \ttbar closure 
test whereas is not the case for the $Z+Jets$ \gls{mc} sample. As it was argued in previous
sections, the \gls{qcd} sector of the Z+Jets \gls{mc} simulation is expected to be poorly
modelled, unlike the \ttbar process. This is illustrated by the incompatible \PPF estimation with 
the main analysis results of the Z+Jets closure test, because of using a fake rate matrix 
extracted from a experimental dataset which was defined through the $E_{th}^{LeadJet}>30 (35)~\GeV$
cut matching the experimental Z+Jet enriched region. Therefore, to be conclusive it is mandatory
to check the closure test in experimental data, in particular the Z+Jets region.

A \ttbar enriched region have been extracted from the experimental data using the selection cuts
given in Chapter~\ref{ch6} and modifying some of the selection cuts,
\begin{itemize}
	\item same flavour, opposite-charged leptons should fulfil 
		$M_{\ell\ell} \notin M_{Z}\pm25~\GeV/c$
	\item number of jets in the event $>$ 2, at least 1 b-tagged
	\item $\MET>40~\GeV$
\end{itemize}
Vetoing the \Z candidate and requiring a high amount of \MET will reduce significantly the 
Drell-Yan process, and the b-tagging~\footnote{Method to identify jets originating from bottom quarks.
See details in Reference~\cite{Chatrchyan:2012bra}} requirement reduces the remnant WW. In addition,
the \WZ, ZZ and $V\gamma$ \gls{mc} samples have been processed, denoted as $N_{\pr\pr\pr}^{MC}$, 
to account the possible contamination to the signal region. Table~\ref{ch7:tab:ctDATAttbar} 
shows the obtained yields once the \gls{mc} subtraction of $N_{\pr\pr\pr}^{MC}$ have been done, 
and the data-driven estimation for the \PPF contribution. Notice the small number of events 
available after the selection cuts, which although a remarkably different central values are 
obtained between the estimation of the method and the analysis, the measurements are completely
compatible as it may see in the $N_{\sigma}$ row.
\begin{table}[!htpb]
	\centering
	\begin{tabular}{l  c c c c}\hline\hline
		&            $3e$   &            $2e$ &         $2\mu$    &        $3\mu$  \\ \hline
		$N_{\pr\pr\fr}^{N_{t3}}$       & $0.4\pm0.2$ & $0.56\pm0.15$ & $0.8\pm0.2$ & $0.66\pm0.15$\\ 
		$|N_{t3}-N_{\pr\pr\pr}^{MC}|$  & $3\pm3$     & $0.4\pm1.0$   & $4\pm2$     & $3\pm3$\\ \hline
		$\delta\varepsilon\cdot100$    & $  85\%$    & $35\%$        & $      82\%$& $      78\%$ \\
		$N_{\sigma}$                   & $0.8$       & $0.15$        & $1.5$       & $0.69$ \\\hline     
	\end{tabular}
	\caption[Closure test for \ttbar region for 2011]{Data-driven estimation
	$N_{\pr\pr\fr}^{N_{t3}}$ and yields from nominal analysis $N_{t3}$ obtained with the 
	\ttbar region sample. Errors shown are statistical only.
	The \PPF contribution has been obtained applying the fakeable object method to the 
	enriched-\ttbar region sample and subtracted the \PPP contribution with simulated data.
	The selection cuts of the \WZ analysis are applied to the same region to obtain the 
	$N_{t3}$ yields. The $\delta\varepsilon$ is the relative difference between both central
	measurements and $N_{\sigma}$ the number of standard deviations between both measures. 
	Fake rates used extracted with the $E_{th}^{LeadJet}>50~\GeV$ cut in the jet-enriched 
	sample (see Section~\ref{ch7:subsec:prfr}). Yields correspond to the available luminosity 
	of the 2011 run period: 4.9~\fbinv}\label{ch7:tab:ctDATAttbar}
\end{table}

The last closure test performed is in the $Z+Jets$ region. The experimental data have been enriched
with Z+Jets processes using the nominal selection cuts of the \WZ analysis but modifying some of
them,
\begin{itemize}
	\item same flavour, opposite-charged leptons should fulfil 
		$M_{\ell\ell} \in M_{Z}\pm15~\GeV$,
	\item third lepton $p_t>10\;GeV/c$,
	\item $\MET<20\;GeV/c$
\end{itemize}
The \Z mass window have been reduced in order to assure a better quality of the \Z-candidates and 
the transverse momentum cut of the third lepton has been lowered to increase the number of events.
The Z+Jets process does not contain real \MET, therefore the associated cut has been reverted and
reduced. As in the case of the \ttbar region, the \WZ, ZZ and $V\gamma$ processes have been 
incorporated with a \gls{mc} simulated samples and subtracted to the $N_{t3}$ yields. The results 
of the test are reported in Table~\ref{ch7:tab:ctDATAZJets}. All the channels close the test, 
although it is worth to mention that whence the mixed channels have an impressive accuracy, the
pure electronic and muonic channel present more differences between the \gls{fom} estimation and
the standard analysis. Nevertheless, the results are consistent in the experimental data case of
the Z+Jets sample in contrast with the \gls{mc} simulation (Table~\ref{ch7:tab:ctMCZJets})
leading to the not-well-modelled Z+Jets' initial assumption discussed previously further 
plausibility.
\begin{table}[!htpb]
	\centering
	\begin{tabular}{l c c c c}\hline\hline
	&            $3e$ &            $2e$ &         $2\mu$  &        $3\mu$  \\ \hline
	$N_{\pr\pr\fr}^{N_{t3}}$       & $18\pm2$      & $23\pm2$ & $23\pm2$  & $35\pm2$\\ 
        $|N_{t3}-N_{\pr\pr\pr}^{MC}|$  & $6\pm5$       & $24\pm6$ & $22\pm7$  & $63\pm9$ \\ \hline
          $\delta\varepsilon\cdot100$  & $189\%$       & $3\%$    & $ 2\%$    & $45\%$ \\
	$N_{\sigma}$		       & $2.4$         & $0.1$    & $0.1$     & $3.0$ \\ \hline
	\end{tabular}
	\caption[Closure test for enriched Z+Jets region for 2011]{Data-driven estimation
	$N_{\pr\pr\fr}^{N_{t3}}$ and yields from nominal analysis $N_{t3}$ obtained with the 
	enriched Z+Jets region sample. Errors shown are statistical only.
	The \PPF contribution has been obtained applying the fakeable object method to the 
	enriched Z+Jets region sample and subtracted the \PPP contribution with simulated data.
	The selection cuts of the \WZ analysis are applied to the same region to obtain the 
	$N_{t3}$ yields. The $\delta\varepsilon$ is the relative difference between both central
	measurements and $N_{\sigma}$ the number of standard deviations between both measures. 
	Fake rates used extracted with the $E_{th}^{LeadJet}>30(35)~\GeV$ cut in the jet-enriched 
	sample (see Section~\ref{ch7:subsec:prfr}). Yields correspond to the available luminosity 
	of the 2011 run period: 4.9~\fbinv}\label{ch7:tab:ctDATAZJets}
\end{table}

\subsection{Systematic uncertainties}\label{ch7:subsec:sysuncertainties}
The jet-enriched sample used to built the fake rate has been selected in such a way that the
relative isolation spectra of the fakeable of this sample and the background samples (mainly 
$t\bar{t}$ and $Z+Jets$) are as similar as possible by introducing a $E_T$ cut in the leading jet
as it was discussed in Section~\ref{ch7:subsec:prfr}. The optimal cut value for a \ttbar sample has 
found to be $E_T^{LeadJet} > 50~\GeV$ whilst for the case of $Z+jet$ has found a 
$E_T^{LeadJet} > 30(35)~\GeV$ cut, for muons (electrons).

Those results have been obtained by matching the $E_T$ spectra of the fakeable jet in the 
jet-enriched, $Z+Jets$ and \ttbar samples (see Figures~\ref{ch7:fig:ETISO}) and confirmed by the
several experimental and \gls{mc} simulated data closure tests performed in the last section. 
However, the background contribution is not expected to be homogeneous but a mixture of \ttbar and
$Z+Jets$ process (see Section~\ref{ch7:subsec:prfr}). The $E_T^{LeadJet} > 50\GeVc$ choice as the 
nominal cut to extract the fake rates is taken into account by recalculating the data-driven using 
the $E_T^{LeadJet} > 30(35)~\GeV$ which mimics $Z+Jets$ and assigning a systematic using the 
differences between both estimations. The fake rates extracted using the 
$E_{th}^{LeadJet}>30(35)~\GeV$ cut are shown in Tables~\ref{ch7:tab:FRJet302011} 
and~\ref{ch7:tab:FRJet302012}. The obtained differences are shown in Table~\ref{ch7:tab:fomsys}. 
\begin{table}[!htbp]
	\centering
	\begin{tabular}{l cccc}\hline\hline
		& $3e$ & $2e1\mu$ & $2\mu1e$ & $3\mu$ \\\hline
		$E_T^{LeadJet} > 15~\GeV$           & $60\pm8$ & $57\pm8$ & $64\pm9$ & $91\pm10$ \\
		$E_T^{LeadJet} > 50\GeVc$ (nominal) & $62\pm8$ & $61\pm8$ & $68\pm8$ & $95\pm10$ \\\hline
		$\delta_{sys}\cdot 100$       &  2.4\%         &   5.5\%        &    5.3\%       &  5.0\%  \\\hline
	\end{tabular}
	\caption[Fakeable object method systematic for 2011]{Signal estimation $N_{\pr\pr\pr}^{N_{t3}}$ 
	for the data-driven using different fake rate sets. The relative differences between them are 
	used as systematic for the method, $\delta_{sys}$}\label{ch7:tab:fomsys}
\end{table}
Although it was propagated the errors associated to the Equations~\eqref{ch7:eq:fomEstimation}
of the fakeable object method, the systematic errors obtained are negligible with respect to the 
systematic uncertainty associated to the transverse energy of the leading jet cut, and 
consequently they are not considered in this analysis.

\begin{table}[!htbp]
	\centering
	\begin{subtable}[b]{0.8\textwidth}
		\centering
	\resizebox{\textwidth}{!}
	{
		\begin{tabular}{lcccc}\hline\hline
		     &$|\eta|\in[0,1]$ & $\eta\in(1,1.48]$ & $\eta\in(1.48,2]$ & $\eta\in(2,2.5]$ \\ \hline 
		        $10< p_t \leq 15$ & $0.041\pm0.003$ & $0.052\pm0.004$ & $0.061\pm0.005$ & $0.057\pm0.008$ \\ 
			$15< p_t \leq 20$ & $0.018\pm0.002$ & $0.024\pm0.004$ & $0.028\pm0.005$ & $0.019\pm0.007$ \\ 
			$20< p_t \leq 25$ & $0.021\pm0.001$ & $0.032\pm0.002$& $0.037\pm0.002$& $0.049\pm0.004$ \\ 
			$25< p_t \leq 30$ & $0.045\pm0.002$ & $0.071\pm0.004$ & $0.090\pm0.005$ & $0.112\pm0.010$ \\ 
 		     $30< p_t \leq\infty$ & $0.075\pm0.002$ & $0.087\pm0.003$ & $0.109\pm0.005$ & $0.151\pm0.010$ \\ \hline
		\end{tabular}
	}
	\caption{Measured muon fake rates in function of transverse momentum and $\eta$ of the muon.
	Errors are statistical only}\label{ch7:subtab:muJET30FR11}
	\end{subtable}
	\vskip 1em
	\begin{subtable}[b]{0.8\textwidth}
		\centering
	\resizebox{\textwidth}{!}
	{
		\begin{tabular}{lcccc}\hline\hline
		     &$|\eta|\in[0,1]$ & $\eta\in(1,1.48]$ & $\eta\in(1.48,2]$ & $\eta\in(2,2.5]$ \\ \hline 
			$10< p_t \leq 15$ & $0.066\pm0.015$ & $0.040\pm0.012$ & $0.016\pm0.009$ & $0.023\pm0.013$ \\ 
			$15< p_t \leq 20$ & $0.057\pm0.009$ & $0.054\pm0.010$ & $0.018\pm0.007$ & $0.041\pm0.013$ \\ 
			$20< p_t \leq 25$ & $0.085\pm0.009$ & $0.064\pm0.011$ & $0.057\pm0.010$ & $0.046\pm0.009$ \\ 
			$25< p_t \leq 30$ & $0.085\pm0.012$& $0.067\pm0.015$ & $0.052\pm0.012$ & $0.070\pm0.015$ \\ 
		     $30< p_t \leq\infty$ & $0.09\pm0.02$   & $0.06\pm0.02$   & $0.07\pm0.02$   & $0.07\pm0.02$ \\ \hline
		\end{tabular}
	}
	\caption{Measured electron fake rates in function of transverse momentum and $\eta$ of the electron. 
	Errors are statistical only}\label{ch7:subtab:elecJET30FR11}
	\end{subtable}
	\caption[Fake rates for 2011 analysis, extracted with $E_{th}^{LeadJet}>30~\GeV$]{Measured 
		lepton fakes rates from the jet-enriched sample described in the text, using a 
		transverse energy of the leading jet higher than 30(35)~\GeV for muons (electrons) 
		for 2011 analysis. }\label{ch7:tab:FRJet302011}
\end{table}
\begin{table}[!htbp]
	\centering
	\begin{subtable}[b]{0.8\textwidth}
		\centering
	\resizebox{\textwidth}{!}
	{
		\begin{tabular}{ccccc}\hline\hline
		     &$|\eta|\in[0,1]$ & $\eta\in(1,1.48]$ & $\eta\in(1.48,2]$ & $\eta\in(2,2.5]$ \\ \hline 
			$10< p_t \leq 15$ & $0.084\pm0.005$ & $0.100\pm0.008$ & $0.128\pm0.012$ & $0.20\pm0.02$ \\ 
			$15< p_t \leq 20$ & $0.073\pm0.009$ & $0.080\pm0.014$ & $0.10\pm0.02$ & $0.21\pm0.04$ \\ 
			$20< p_t \leq 25$ & $0.098\pm0.005$ & $0.134\pm0.009$ & $0.11\pm0.01$ & $0.12\pm0.02$ \\ 
			$25< p_t \leq 30$ & $0.136\pm0.009$ & $0.191\pm0.016$ & $0.16\pm0.02$ & $0.23\pm0.04$ \\ 
			$30< p_t \leq\infty$ & $0.19\pm0.02$ & $0.25\pm0.03$ & $0.23\pm0.04$ & $0.31\pm0.10$ \\ \hline
		\end{tabular}
	}
	\caption{Measured muon fake rates in function of transverse momentum and $\eta$ of the muon.
	Errors are statistical only}\label{ch7:subtab:muJET30FR12}
	\end{subtable}
	\vskip 1em
	\begin{subtable}[b]{0.8\textwidth}
		\centering
	\resizebox{\textwidth}{!}
	{
		\begin{tabular}{ccccc}\hline\hline
		     &$|\eta|\in[0,1]$ & $\eta\in(1,1.48]$ & $\eta\in(1.48,2]$ & $\eta\in(2,2.5]$ \\ \hline 
			$10< p_t \leq 15$ & $0.045\pm0.005$ & $0.033\pm0.004$ & $0.008\pm0.002$ & $0.021\pm0.005$ \\ 
			$15< p_t \leq 20$ & $0.044\pm0.003$ & $0.049\pm0.003$ & $0.017\pm0.001$ & $0.017\pm0.002$ \\ 
			$20< p_t \leq 25$ & $0.041\pm0.002$ & $0.064\pm0.003$ & $0.025\pm0.002$ & $0.025\pm0.002$ \\ 
			$25< p_t \leq 30$ & $0.059\pm0.003$ & $0.101\pm0.005$ & $0.041\pm0.003$ & $0.043\pm0.003$ \\ 
		    $30< p_t \leq \infty$ & $0.084\pm0.006$ & $0.111\pm0.009$ & $0.058\pm0.006$ & $0.066\pm0.005$ \\ \hline
		\end{tabular}

	}
	\caption{Measured electron fake rates in function of transverse momentum and $\eta$ of the electron. 
	Errors are statistical only}\label{ch7:subtab:elecJET30FR12}
	\end{subtable}
	\caption[Fake rates for 2012 analysis, extracted with $E_{th}^{LeadJet}>30~\GeV$]{Measured 
		lepton fakes rates from the jet-enriched sample described in the text, using a 
		transverse energy of the leading jet higher than 30(35)~\GeV for muons (electrons)
		for 2012 analysis.}\label{ch7:tab:FRJet302012}
\end{table}

\section{Irreducible backgrounds}\label{ch7:sec:mcbkg}
The contribution of other prompt components to the signal (ZZ, $V\gamma$ and VVV for the 2012 
analysis) has been subtracted using \gls{mc} simulated samples, taking as background estimation
the number of events passing the signal selection. The samples have been simulated in a centralised
way for the whole \gls{cms}\glsadd{ind:cms} collaboration for the sake of coherence and consistency. Thus, any
analysis performed within the collaboration makes use of the same simulation input. The massive
simulation of \gls{mc} samples are structured in production campaigns, whence the detector 
conditions (alignment, magnetic field, dead detector regions, \etc) are kept frozen along the 
production. The generation of the physics processes is accomplished by the use of different
generator programs depending the process to simulate. As is explained in detail in 
Section~\ref{ch5:sec:simulation}, once the process is generated, the outcome
is sent to the detector simulator based in \GEANTfour, to simulate the particles passing through
the detector. The list of the samples used in this analysis, with some relevant information as the
production campaign, the \gls{mc} program used to generate the sample, the cross section considered 
and the internal name in \gls{cms}\glsadd{ind:cms} are detailed in Tables~\ref{ch7:tab:mcsamples11} 
and~\ref{ch7:tab:mcsamples12}.

In order to take into account the reconstruction, identification and isolation lepton efficiencies
(studied in Chapter~\ref{ch6}) that are present in the experimental data, these efficiencies have 
been measured in data, $\varepsilon$, and in the simulation, $\varepsilon_{sim}$, using tag and 
probe methods. Each \gls{mc} simulated event has been weighted by the efficiencies scale 
factors $SF=\varepsilon/\varepsilon_{sim}$, mimicking the simulated events with the data 
inefficiencies behaviour. 

As the data is selected using trigger requirements the \gls{mc} events should contain 
only events with the same trigger paths accepted. This approach is difficult to accomplish due to 
the asynchrony between \gls{mc} sample massive production and the continuous development of the 
trigger paths. Instead, as it was discussed in Section~\ref{ch6:subsec:triggerEff} of the previous
Chapter, it has been measured, in $\eta$ and \pt bins, the lepton trigger efficiencies of any 
double trigger used in the analysis and interpreted as the probability of a lepton passing one 
leg trigger requirement. Moreover, it has been built a probability function which is used to weight
each \gls{mc} event (Equations~\ref{ch6:eq:triggerweight}).

In addition to the object corrections, the \gls{mc} samples need to be reweighted in order to match
the \gls{pileup} interactions present in the data. This is needed because the simulated data is 
produced before the acquisition of the experimental data and, therefore, there is no clue about the
number of additional interactions per bunch crossing. The approach used in the \gls{cms}\glsadd{ind:cms} 
collaboration to simulate the \gls{pileup} is based in the last data-taking period. A distribution
representing the mean number of interactions seen during the last data-taking is built and used
\begin{figure}[!htpb]
	\centering
	\includegraphics[width=0.6\textwidth]{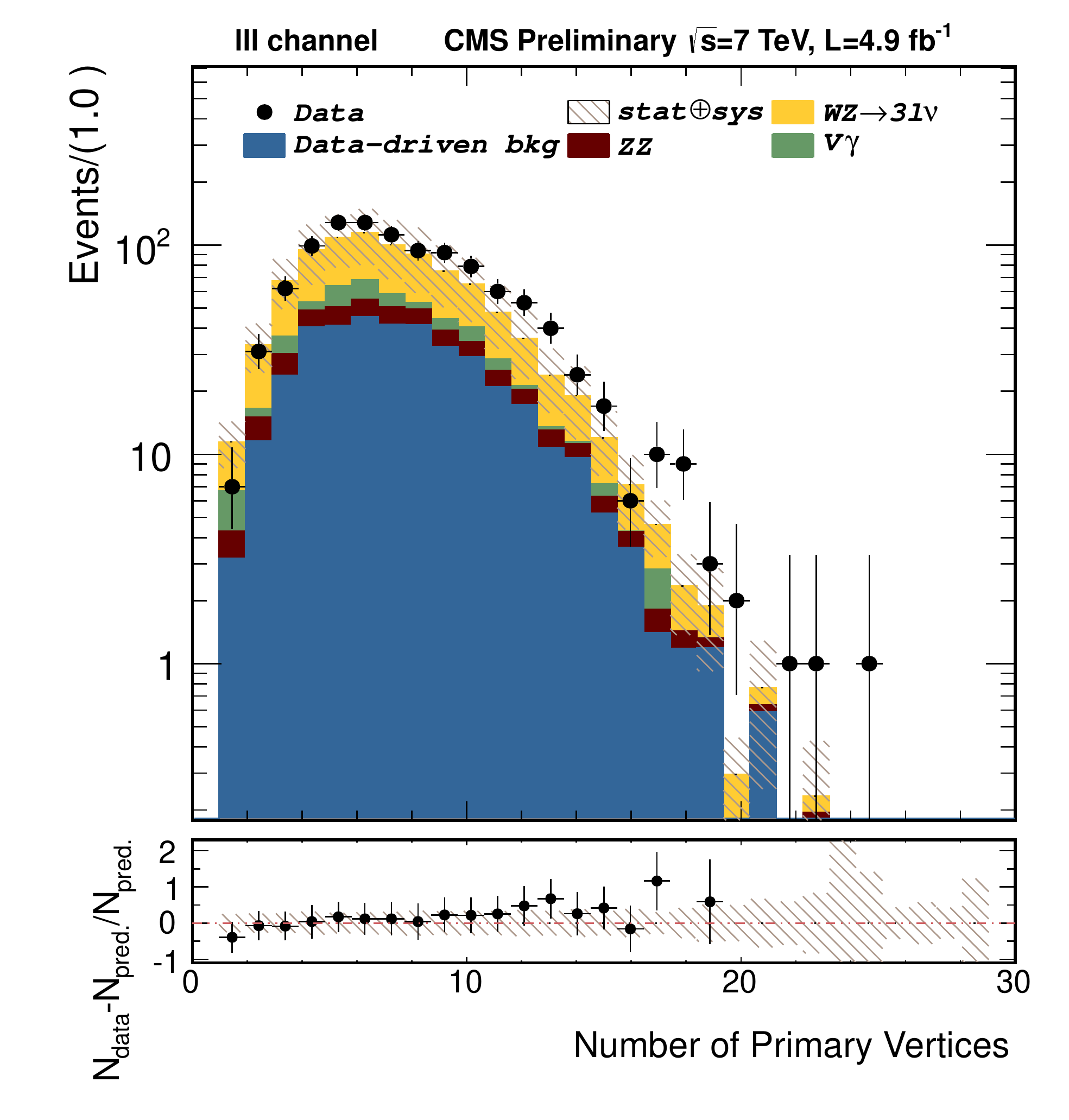}
	\caption[Number of primary vertices distribution for 2011]{Number of primary vertices 
	reconstructed per event distribution in the \WZ 2011 analysis. The lepton preselection
	stage have been applied}\label{ch7:fig:pileup}
\end{figure}
as input for a given production campaign. For each event, the mean number of interactions per bunch
crossing is chosen from the input distribution. This sets the instantaneous luminosity to 
be simulated for all of the bunch crossings in that event. For each bunch crossing, the number of 
interactions is randomly sampled from a Poisson distribution with a mean equal to the value chosen
before. When compare the simulated with the experimental data, the number of interactions observed
in the simulated data should match the experimental data, therefore a event reweighting 
($N_{int.}/N_{int.}^{MC}$) is performed in the simulation to obtain the same number of interaction
distribution. Figure~\ref{ch7:fig:pileup} shows the experimental data compared with the \gls{mc}
simulation for the number of primary vertices distribution at each event, which is a sensible 
observable for the \gls{pileup}.

\begin{table}[!hbtp]\centering
	\resizebox{\textwidth}{!}
	{
		\begin{tabular}{lllr}	\hline\hline
			\bf Signal  & Nickname & {\bf  MC dataset name} & $\sigma\cdot BR$ $[pb]$  \\ \hline
			$WZ\rightarrow3\ell1\nu$     & WZJet3L1Nu 
			          & /WZJetsTo3LNu\_TuneZ2\_7TeV-madgraph-tauola/*/AODSIM & $0.879$ (NLO) \\\hline
			\\\\\hline\hline
			\bf Background & Nickname & {\bf  MC dataset name} & $\sigma\cdot BR$ $[pb]$  \\ \hline
			\multicolumn{1}{l}{
		  	    \multirow{6}{*}{ $Z/\gamma^*\rightarrow\ell\ell+Jets$ } } 
			       & DYee     & /DYToEE\_M-10To20\_CT10\_TuneZ2\_7TeV-powheg-pythia/*/AODSIM   & $3319.61$ (NNLO)  \\
			\multicolumn{1}{l}{}
			        & Zee      & /DYToEE\_M-20\_CT10\_TuneZ2\_7TeV-powheg-pythia/*/AODSIM       & $1666$ (NNLO)     \\
			\multicolumn{1}{l}{} 
			        & DYtautau & /DYToTauTau\_M-10To20\_TuneZ2\_7TeV-pythia6-tauola/*/AODSIM    & $31319.61$ (NNLO) \\
			\multicolumn{1}{l}{}      
			        & Ztautau  & /DYToTauTau\_M-20\_CT10\_TuneZ2\_7TeV-powheg-pythia/*/AODSIM   & $1666$ (NNLO)  \\
			\multicolumn{1}{l}{}      
			        & DYmumu   & /DYToMuMu\_M-10To20\_CT10\_TuneZ2\_7TeV-powheg-pythia/*/AODSIM & $31319.61$ (NNLO) \\
			\multicolumn{1}{l}{}      
			        & Zmumu    & /DYToMuMu\_M-20\_CT10\_TuneZ2\_7TeV-powheg-pythia/*/AODSIM     & $1666$ (NNLO) \\
			$W\rightarrow\ell\nu+Jets$   
			        & WJetsToLNu & /WJetsToLNu\_TuneZ2\_7TeV-madgraph-tauola/*/AODSIM           & $31314$ (NNLO)\\
			$V\gamma+jets$   & PhotonVJets& /GVJets\_7TeV-madgraph	      & $165$  (LO)  \\
			$WW\rightarrow X$  & WW & /WW\_TuneZ2\_7TeV\_pythia6\_tauola/*/AODSIM              & $47$ (NLO)\\
			$tW\rightarrow X$ & tW  & /T\_TuneZ2\_tW-channel-DR\_7TeV-powheg-tauola/*/AODSIM         & $7.87$ (NLO) \\
			$\bar{t}W\rightarrow X$ & tbarW & /Tbar\_TuneZ2\_t-channel\_7TeV-powheg-tauola/*/AODSIM & $7.87$ (NLO) \\ 
			$t\bar{t}\rightarrow X$ & TTbar & /TTJets\_TuneZ2\_7TeV-madgraph-tauola/*/AODSIM     & $163$ (NLO)  \\
			$t\bar{t}\rightarrow 2\ell 2\nu 2b$ 
			                  & TTTo2L2Nu2B & /TTTo2L2Nu2B\_7TeV-powheg-pythia6/*/AODSIM & $17.10$ (NLO) \\ 
			\multirow{3}{*}{ $ZZ\rightarrow4\ell$ } 
			            & ZZ4e     & /ZZTo4e\_7TeV\_powheg\_pythia6/*/AODSIM        & $0.0154$ (NLO)\\ 
			\multicolumn{1}{l}{}   
			                  & ZZ4mu    & /ZZTo4m\_7TeV\_powheg\_pythia6/*/AODSIM  & $0.0154$ (NLO)\\ 
			\multicolumn{1}{l}{}  & ZZ4tau   & /ZZTo4tau\_7TeV\_powheg\_pythia6/*/AODSIM    & $0.0154$ (NLO)\\ 
			\multirow{3}{*}{ $ZZ\rightarrow2\ell2\ell'$ }  
			                  & ZZ2e2mu  & /ZZTo2e2mu\_7TeV\_powheg\_pythia6/*/AODSIM & $0.0308$ (NLO)\\ 
			\multicolumn{1}{l}{}  &ZZ2e2tau & /ZZTo2e2tau\_7TeV\_powheg\_pythia6/*/AODSIM  & $0.0308$ (NLO)\\ 
			\multicolumn{1}{l}{}  & ZZ2mu2tau& /ZZTo2mu2tau\_7TeV\_powheg\_pythia6/*/AODSIM & $0.0308$ (NLO)\\ 
			$ZZ\rightarrow X$  & ZZ  & /ZZ\_TuneZ2\_7TeV\_pythia6\_tauola/*/AODSIM & $7.67$ (NLO)\\ \hline 
			\multicolumn{4}{l}{$*$ Fall11-PU\_S6\_START42\_V148-v1}     
		\end{tabular}
	}
	\caption[Monte Carlo samples used in 2011 analysis]{Summary of Standard Model processes, 
	Monte Carlo simulated samples and cross section times 
	branching ratio values used for this analysis in 2011. Mostly of the samples were generated with 
	\PYTHIA, in some cases \MADGRAPH was used. The parton shower was
	included by interfacing \POWHEG to the main generator program, and the tau-lepton decays was dealt
	with \TAUOLA.}\label{ch7:tab:mcsamples11}
\end{table}

\begin{table}[!htpb]
	\centering
	\resizebox{\textwidth}{!}
	{
		\begin{tabular}{lllr}\hline\hline
                {\bf Signal} & {Nickname} & {\bf MC dataset name}  & $\sigma\cdot\rm{BR}$~[pb]\\\hline
		$\WZ\to\ell\nu\ell\ell$   & WZ  & /WZJetsTo3LNu\_TuneZ2\_8TeV-madgraph-tauola/[1] &   1.058\\
                 \\\\\hline\hline 
		 {\bf Nickname} & {Background process} & {\bf MC dataset name}  & $\sigma\cdot\rm{BR}$~[pb]\\\hline
\multirow{3}{*}{top}    & $t\bar{t}\to 2\ell 2\nu 2b$                          & /TTTo2L2Nu2B\_8TeV-powheg-pythia6/[3]                                  &  23.640\\
                        & $t$W                                                 & /T\_tW-channel-DR\_TuneZ2star\_8TeV-powheg-tauola/[1]                  &  11.177\\
                        & $\bar{t}$W                                           & /Tbar\_tW-channel-DR\_TuneZ2star\_8TeV-powheg-tauola/[1]               &  11.177\\
\hline
\multirow{9}{*}{VVV}    & WZZ                                                  & /WZZNoGstarJets\_8TeV-madgraph/[1]                                     &  0.0192\\
                        & ZZZ                                                  & /ZZZNoGstarJets\_8TeV-madgraph/[1]                                     & 0.00459\\
                        & WWZ                                                  & /WWZNoGstarJets\_8TeV-madgraph/[1]                                     &  0.0633\\
                        & WWW                                                  & /WWWJets\_8TeV-madgraph/[1]                                            &  0.0822\\
                        & $t\bar{t}$W                                          & /TTWJets\_8TeV-madgraph/[1]                                            &   0.232\\
                        & $t\bar{t}$Z                                          & /TTZJets\_8TeV-madgraph\_v2/[1]                                        &   0.174\\
                        & $t\bar{t}$WW                                         & /TTWWJets\_8TeV-madgraph/[1]                                           & 0.00204\\
                        & $t\bar{t}\gamma$                                     & /TTGJets\_8TeV-madgraph/[1]                                            &    1.44\\
                        & WW$\gamma$                                           & /WWGJets\_8TeV-madgraph/[1]                                            &   0.528\\
\hline
\multirow{9}{*}{WV}     & W~$\to\ell\nu$                                       & /WJetsToLNu\_TuneZ2Star\_8TeV-madgraph-tarball/[1]                     &   37509\\
                        & W~$\to\ell\nu + b\bar{b}$                            & /WbbJetsToLNu\_Massive\_TuneZ2star\_8TeV-madgraph-pythia6\_tauola/[1]  &    39.9\\
                        & W$\gamma^{*}\to\ell\nu ee$                           & /WGstarToLNu2E\_TuneZ2star\_8TeV-madgraph-tauola/[1]                   &   5.873\\
                        & W$\gamma^{*}\to\ell\nu\mu\mu$                        & /WGstarToLNu2Mu\_TuneZ2star\_7TeV-madgraph-tauola/[1]                  &   1.914\\
                        & W$\gamma^{*}\to\ell\nu\tau\tau$                      & /WGstarToLNu2Tau\_TuneZ2star\_7TeV-madgraph-tauola/[1]                 &   0.336\\
                        & W$\gamma\to\ell\nu\gamma$                            & /WGToLNuG\_TuneZ2star\_8TeV-madgraph-tauola/[1]                        &   553.9\\
                        & WW                                                   & /WW\_TuneZ2star\_8TeV\_pythia6\_tauola/[1]                             &   57.07\\
                        & WZ~$\to q\bar{q}'\ell\ell$                           & /WZJetsTo2L2Q\_TuneZ2star\_8TeV-madgraph-tauola/[1]                    &   2.206\\
                        & WZ~$\to\ell\nu q\bar{q}$                             & /WZJetsTo2QLNu\_8TeV-madgraph/[2]                                      &   1.584\\
\hline
Z$\gamma$               & Z$\gamma\to\ell\ell\gamma$                           & /ZGToLLG\_8TeV-madgraph/[1]                                            &   132.6\\
\hline
\multirow{3}{*}{Z+jets} & Z/$\gamma\to\ell\ell$ ($10 < m_{\ell\ell} < 50$~GeV) & /DYJetsToLL\_M-10To50filter\_8TeV-madgraph/[1]                         &   860.5\\
                        & Z/$\gamma\to\ell\ell$ ($m_{\ell\ell} > 50$~GeV)      & /DYJetsToLL\_M-50\_TuneZ2Star\_8TeV-madgraph-tarball/[1]               &  3532.8\\
                        & Z/$\gamma\to\ell\ell + b\bar{b}$                     & /ZbbToLL\_massive\_M-50\_TuneZ2star\_8TeV-madgraph-pythia6\_tauola/[1] &    94.1\\
\hline
\multirow{11}{*}{ZZ}    & ZZ~$\to 2\mu 2\tau$                                  & /ZZTo2mu2tau\_8TeV-powheg-pythia6/[1]                                  &  0.1767\\
                        & ZZ~$\to 4e$                                          & /ZZTo4e\_8TeV-powheg-pythia6/[1]                                       & 0.07691\\
                        & ZZ~$\to 2e 2\tau$                                    & /ZZTo2e2tau\_8TeV-powheg-pythia6/[1]                                   &  0.1767\\ 
                        & ZZ~$\to 4\mu$                                        & /ZZTo4mu\_8TeV-powheg-pythia6/[1]                                      & 0.07691\\
                        & ZZ~$\to 2e 2\mu$                                     & /ZZTo2e2mu\_8TeV-powheg-pythia6/[1]                                    &  0.1767\\
                        & ZZ~$\to 4\tau$                                       & /ZZTo4tau\_8TeV-powheg-pythia6/[1]                                     & 0.07691\\
                        & ZZ~$\to\ell\ell q\bar{q}$                            & /ZZJetsTo2L2Q\_TuneZ2star\_8TeV-madgraph-tauola/[1]                    &   1.275\\
                        & ZZ~$\to\ell\ell\nu\nu$                               & /ZZJetsTo2L2Nu\_TuneZ2star\_8TeV-madgraph-tauola/[3]                   &   0.365\\
                        & gg~$\to$~ZZ~$\to 2\ell 2\ell$                        & /GluGluToZZTo2L2L\_TuneZ2star\_8TeV-gg2zz-pythia6/[1]                  & 0.00447\\
                        & gg~$\to$~ZZ~$\to 4\ell$                              & /GluGluToZZTo4L\_8TeV-gg2zz-pythia6/[1]                                & 0.00224\\
                        & gg~$\to$~H~$\to$~ZZ~$\to 4\ell$                      & /GluGluToHToZZTo4L\_M-125\_8TeV-powheg-pythia6/[1]                     &  0.0524\\
\hline
\multicolumn{3}{l}{[1] Summer12\_DR53X-PU\_S10\_START53\_V7A-v1/AODSIM}\\
\multicolumn{3}{l}{[2] Summer12\_DR53X-PU\_S10\_START53\_V7C-v1/AODSIM}\\
\multicolumn{3}{l}{[3] Summer12-PU\_S7\_START52\_V9-v1/AODSIM}\\
\end{tabular}
}
\caption[Monte Carlo samples used in 2012 analysis]{Summary of Standard Model processes, Monte Carlo 
	simulated samples and cross section times branching ratio values used for this analysis in 2012.
	Mostly of the samples were generated with \PYTHIA, but in some cases
	\MADGRAPH was used. The parton shower was included by interfacing \POWHEG to the main generator 
	program, and the tau-lepton decays was dealt with \TAUOLA.}\label{ch7:tab:mcsamples12}
\end{table}

\chapter{WZ Cross section measurements}\label{ch8}
The measurement of the cross section at centre of mass energies of 7~\TeV and 8~\TeV is described
along this chapter. The general formula to obtain a cross section from the observed events is
recalled, identifying the necessary elements and describing how they are estimated. Afterwards, 
using the results from previous chapters, the cross section measurements in each channel are 
reported, along with their estimated uncertainties. A detailed description of the sources of 
systematic uncertainties identified are presented and propagated to the measurement. The chapter
concludes with a review of the BLUE method used to combine the four measurements to finally show
the final combined result.

\section{Cross section estimation}\label{ch8:sec:xsestimation}
The probability of a process is given by the \emph{cross section}, $\sigma$. 
Introducing the \emph{instantaneous 
luminosity}, $\lumi$, as the number of incident particles per unit area per unit time 
$[\lumi]=\cm^{-2}\,s^{-1}$, then the event rate of a process $A$ is given by,
\begin{equation}
	\frac{dN_{A}}{dt}=P(A)\lumi\equiv\sigma_A\lumi,
\end{equation}
Integrating along a period of time $T$, and including the effects of measuring in a real detector
by introducing the probability of measure an event $A$ in the detector, $\varepsilon_A$, the number 
of events observed is given by,
\begin{equation}
	N_{A}^{obs}=\sigma_AP(\text{Measured}|A)\left(\int_{T}\lumi dt\right)
	\label{ch8:eq:xsgeneric}
\end{equation}
The probability of measure an event from the process $A$ can be decomposed into the probability 
that an event fall into the geometric acceptance of the detector and once the event 
is inside acceptance, the probability that the event is actually reconstructed and measured.
\begin{equation}
	P(\text{Measured})=P(\text{Measured}|\text{Inside Acc.})
	P(\text{Inside Acc.})\equiv\varepsilon\mathcal{A}
\end{equation}
where it has been renamed,
\begin{align*}
	&P(\text{Inside Acc.})\equiv\mathcal{A},  \,\text{ and }\\
	&P(\text{Measured}|\text{Inside Acc.})\equiv\varepsilon
\end{align*}

The production cross section of a process is measured in a fiducial region constrained
by the geometric acceptance of the detector. The number of 
events observed from the signal selection, correcting for the efficiency that an event inside
acceptance is reconstructed, are then extrapolated to the full phase space of events which includes
events outside the detector and selection acceptance. Thus, the production cross section in the
full phase space is given by
\begin{equation}
	\sigma = \frac{N_S}{\mathcal{A}\cdot\varepsilon\cdot\lumi_{int}}
	\label{ch8:eq:xs}
\end{equation}
being $N_S$ the number of signal observed in the fiducial region, $\mathcal{A}$ and $\varepsilon$ 
represents the kinematic and geometric acceptance and the selection efficiency for the fiducial 
events, respectively, as it has been described above, and $\lumi_{int}$ is the integrated luminosity. 

The $\mathcal{A}$ is determined using \gls{mc} simulation. A sample of \WZ process is simulated in 
the full phase space, then a fiducial volume is defined to count how many events enters in 
this volume. The frequentist approach of probabilities allows to measure the $\mathcal{A}$ as 
the ratio between the events in the fiducial region over the total generated events.
The Table~\ref{ch8:tab:fiducialdef} defines the fiducial volume in the \gls{mc} simulated 
\WZ process by cutting the generated objects. 
\begin{table}[!htpb]
	\centering
	\begin{tabular}{rl}\hline\hline
		$\pt^{\ell}>20,20,10~\GeV$ & for the three leptons decaying from \Z and \W \\
		$\pt^{\nu}>30~\GeV$        & for the neutrino decaying from \W \\
		$|\eta^{\ell}|< 2.4\,(2.5)$  & for the W, Z-muons (electrons) \\\hline
	\end{tabular}
	\caption[Fiducial region definition]{Fiducial and kinematic region definition in the MC simulated
	sample for the \WZ process. The cuts are applied to the generated 
	objects.}\label{ch8:tab:fiducialdef}
\end{table}

Therefore, 
\begin{equation}
	\mathcal{A}=\frac{N_{\text{Generated }\WZ\to3\ell\nu}^{\text{MC Fiducial Volume}}}
	{N_{\text{Generated }\WZ\to3\ell\nu}^{\text{MC Phase Space}}} 
	\label{ch8:eq:accextr}
\end{equation}
where the $N_{\text{Generated }\WZ\to3\ell\nu}^{\text{MC Fiducial Volume}}$ is the number of events generated 
in the \WZ \gls{mc} sample fulfilling Table~\ref{ch8:tab:fiducialdef}, and 
$N_{\text{Generated }\WZ\to3\ell\nu}^{\text{MC Phase Space}}$ is the total number of events in the \WZ 
\gls{mc} sample with $|m_{\ell\ell}-m_{z}|<20~\GeV$ defining the analysis measurement phase 
space\footnote{The generation of the \WZ process includes the interference term of $Z/\gamma^*$
which introduces divergences in the \wpmz cross section theoretical calculation and, consequently,
in the event generation, at very low $m_{\drellyan}$. See details in 
Section~\ref{ch2:sec:wzproduction} of Chapter~\ref{ch2}.}. Notice that each generated event shall
be corrected or reweighted by the \gls{pileup} correction described in Section~\ref{ch7:sec:mcbkg}
from Chapter~\ref{ch7}.

Analogously, the $\varepsilon$ term in Equation~\eqref{ch8:eq:xs} takes account of the probability 
that a \WZ event is actually measured due to detector effects. In particular, the reconstruction,
isolation, identification and trigger efficiencies are included in this term, as well as the 
efficiency of the selection cuts described in Chapter~\ref{ch6}. This correction factor essentially
gives the probability of reconstructing an event, given that all objects in the event would have 
been in the detector. The efficiency can be factorised as,
\begin{equation}
	\varepsilon\equiv P(Measured|\mathcal{A}) = P(PAC|LS)P(LS|TF)P(TF|\mathcal{A})
\end{equation}
where the abbreviation used in the above formula stands for,
\begin{itemize}
	\item PAC: \emph{Pass analysis event cuts} described in
		Section~\ref{ch6:sec:eventselection} of Chapter~\ref{ch6} to select the events.
	\item LS: \emph{Lepton object selection}, which are the quality cuts in reconstruction, 
		identification and isolation specified in Sections~\ref{ch6:sec:muonselection} 
		and~\ref{ch6:sec:electronselection} from Chapter~\ref{ch6} to select the leptons
		objects to be used in the analysis.
	\item TF: \emph{Trigger fired},~\ie an event was stored by the trigger paths described
		in Section~\ref{ch6:sec:onlineselection} of Chapter~\ref{ch6}.
\end{itemize}
Therefore, the probability that an event inside acceptance has actually been measured is split
in the probability that an event inside acceptance has been stored because the trigger decision,
and once the event is stored, the probability that the lepton objects in the event fulfil the
quality criteria of the analysis, and once there are three good-quality leptons, the probability
that the event pass all the analysis cuts. Again, the probabilities are renamed,
\begin{align*}
	&P(PAC|LS)\equiv \varepsilon_{event}\\
	&P(LS|TF)\equiv \varepsilon_{leptons}\\
	&P(TF|\mathcal{A})\equiv\varepsilon_{trigger}
\end{align*}
Notice that $\varepsilon_{leptons}\cdot\varepsilon_{trigger}$ may be factorised into individual 
lepton efficiencies, 
${\varepsilon_{leptons}\cdot\varepsilon_{trigger}=\varepsilon_{\ell_1}\varepsilon_{\ell_2}
\varepsilon_{\ell_3}}$. In turn, each individual lepton efficiency is decomposed in the product 
of the efficiencies of the full reconstruction chain of each object:
\begin{equation}
	\varepsilon_{\ell_i}= \varepsilon_{trigger|iso}\cdot\varepsilon_{iso|id}\cdot
	      \varepsilon_{id|reco}\cdot\varepsilon_{reco}
\end{equation}
The lepton object efficiencies have already been introduced and calculated in 
Sections~\ref{ch6:subsec:muoneff} and~\ref{ch6:subsec:eleceff} using tag and
probe methods. 

The probability of any event to pass the analysis cuts, $\varepsilon_{event}$, may be extracted 
again using \gls{mc} simulated data by re-expressing the corrections $\mathcal{A}\cdot\varepsilon$
as,
\begin{equation}
	\mathcal{A}\cdot\varepsilon=\left(\mathcal{A}\cdot\varepsilon^{sim}\right)
		\left(\frac{\varepsilon}{\varepsilon^{sim}}\right)\equiv\mathcal{C}\cdot\rho,
	\label{ch8:eq:effextr}
\end{equation}
where $\varepsilon^{sim}$ is the efficiency to measure an event inside acceptance in simulation.
The factor $\rho$ corrects the differences in between efficiencies evaluated in experimental with 
respect to the simulation data. This factor involves the lepton \gls{sf}[s]\glsadd{ind:sf} introduced
in Chapter~\ref{ch6}, and allows to leave the observed experimental data uncorrected whereas it is 
the simulated data which assume the efficiency corrections.
The factor $\mathcal{C}$ deals with \gls{mc} simulation only and allows to use the \WZ \gls{mc} 
simulated sample to count how many generated events pass the analysis cuts, and with a frequentist 
approach of probabilities, to measure the acceptance and efficiency as the ratio between 
passing-analysis events over generated events. Therefore, the calculation of event efficiency
is not made explicitly in order to avoid resolution effects of the detector, but it is embedded 
in the $\mathcal{C}$ factor together with the acceptance. In practice, as the event reconstruction in \gls{mc}
simulation also involves the reconstruction efficiencies of the leptons in the events; the 
acceptance, efficiency and the $\rho$ factor are extracted at once using the \WZ \gls{mc} 
simulated sample as,
\begin{equation}
	\mathcal{A}\cdot\varepsilon^{sim}\cdot\rho=\mathcal{C}\cdot\rho
	=\frac{N_{\text{Reconstructed }WZ\to3\ell\nu}^{\text{MC Pass Analysis Cuts}}
	\cdot SF_{leptons}}
	{N_{\text{Generated }\WZ\to3\ell\nu}^{\text{MC Phase Space}}} 
	\label{ch8:eq:effevent}
\end{equation}
where $SF_{leptons}$ are the scale factors used to account the discrepancies of the
$\varepsilon_{leptons}$ efficiencies between data and simulation\footnote{See detailed 
description and obtained values in Sections~\ref{ch6:subsec:muoneff} and~\ref{ch6:subsec:eleceff}
in Chapter~\ref{ch6}.} and it is understood to be applied on an event-by-event level. As the
\gls{mc} sample used to simulate the signal is not an inclusive sample, but 
$\WZ\to\ell'\nu\ell^+\ell^-$ ($\ell,\ell'=\mu,e,\tau$), in order to obtain the inclusive cross section with the 
acceptance and efficiencies $\mathcal{C}\cdot\rho$ extracted from this sample, the cross section 
calculated shall be corrected explicitly by the branching ratio,
\begin{equation}
	\sigma = \frac{N_S}{\mathcal{C}\cdot\rho\cdot\mathcal{BR}(\WZ\to\ell'\nu\ell^+\ell^-)\cdot\lumi_{int}}
	\label{ch8:eq:xscorrected}
\end{equation}
where $\mathcal{BR}(\WZ\to\ell'\nu\ell^+\ell^-)=\mathcal{BR}(\W\to\ell'\nu)\cdot 
\mathcal{BR}(\Z\to\ell^+\ell^-)=0.0329\pm0.0003$~\cite{PhysRevD.86.010001}.

The uncertainty associated to the cross section measurement is reported regarding the source of 
uncertainty,
\begin{equation}
	\Delta\sigma = \left(\Delta \sigma\right)_{\text{stats}}\oplus
	     \left(\Delta\sigma\right)_{\text{sys}}\oplus\left(\Delta\sigma\right)_{\text{lumi}}
\end{equation}
where $\left(\Delta \sigma\right)_{\text{stats}}$ is referring to the statistical uncertainty,
$\left(\Delta\sigma\right)_{\text{sys}}$ is the systematic uncertainty described in
Section~\ref{ch8:sec:sys}, and $\left(\Delta\sigma\right)_{\text{lumi}}$ the systematic uncertainty
associated with the measurement of the luminosity.

\section{Cross section measurements results}
The \WZ analysis performed in this thesis work is based in the Equation~\eqref{ch8:eq:xs}. The several 
ingredients composing the aforementioned equation, described in previous section, have been measured
along the previous Chapters, therefore the last stage of the analysis is to harvest them to obtain 
the cross section measurement. 

\begin{table}[!htpb]
	\centering
		\begin{tabular}{lcccc}\hline\hline
			  &  $3e$         &    $2e1\mu$    &   $1e2\mu$    &   $3\mu$ \\\hline
$N_{\pr\pr\pr}^{N_{t3}}$  & $62\pm8$      &    $61\pm8$    & $68\pm8$      & $95\pm10$ \\
$N_{ZZ}^{MC}$             & $1.95\pm0.02$ & $3.46\pm0.04$  & $2.68\pm0.03$ & $4.83\pm0.03$ \\
$N_{V\gamma}^{MC}$        &    --         &     --         & $0.51\pm0.51$ &  --   \\\hline
$N_S$                     & $60\pm8$      &   $57\pm8$     &$65\pm8$        & $90\pm10$ \\\hline

		\end{tabular}
	\caption[Number of observed signal for 2011 analysis]{Number of observed signal 
	$N_S$ for each 
	measured channel, given the data-driven estimation of the prompt-prompt-prompt sample and
	the subtracted MC-simulated background ZZ and $V\gamma$. Errors shown are originated from 
	the finite number of events simulated in the MC samples and from the statistical errors of
	the prompt and fake rates used to estimate the \PPP contribution. The MC samples are 
	normalised to the integrated luminosity achieved in 2011 data of
	\lumi=4.9~\fbinv.}\label{ch8:tab:NSignal11}
\end{table}
The number of signal observed after the selection cuts is obtained by the data-driven 
\gls{fom}\footnote{Described in detail in previous Chapter~\ref{ch7}}, and subtracting the 
\gls{mc}-estimated background yields\footnote{See detailed description of corrections applied to 
simulated samples in Section~\ref{ch7:sec:mcbkg} from Chapter~\ref{ch7}.}
\begin{equation}
	N_S=N_{\pr\pr\pr}^{N_{t3}}-N_{bkg}^{MC}
\end{equation}
The $N_{bkg}^{MC}$ is composed by the \gls{mc}-simulated processes ZZ, $Z\gamma$ (and VVV 
[V=W,Z,$\gamma$] in the 2012 analysis). The obtained yields for the 2011 and 2012 analyses and the
four measured channels are summarised in Tables~\ref{ch8:tab:NSignal11} 
and~\ref{ch8:tab:NSignal12}.
\begin{table}[!htpb]
	\centering
		\begin{tabular}{lcccc}\hline\hline
	                  &  $3e$           &    $2e1\mu$    &   $1e2\mu$   &   $3\mu$ \\\hline
$N_{\pr\pr\pr}^{N_{t3}}$  & $220\pm15$      &    $260\pm16$  & $352\pm19$   & $498\pm22$ \\
$N_{VVV}^{MC}$            &    $6.1\pm0.3$  &  $7.9\pm0.3$   & $10.4\pm0.4$ &  $13.4\pm0.4$   \\
$N_{ZZ}^{MC}$             & $2.42\pm0.08$   & $3.10\pm0.09$  & $3.9\pm0.1$  & $5.8\pm0.1$ \\
$N_{V\gamma}^{MC}$        &    $2.5\pm0.9$  & $0.4\pm0.4$    & $4.0\pm1.2$  &  $2.2\pm0.7$  \\\hline
$N_S$                     & $209\pm15$      &   $249\pm16$   & $334\pm19$   & $477\pm22$ \\\hline

		\end{tabular}
	\caption[Number of observed signal for 2012 analysis]{Number of observed signal $N_S$ for each 
	measured channel, given the data-driven estimation of the prompt-prompt-prompt sample and
	the subtracted MC-simulated background VVV, ZZ, $V\gamma$. Errors shown are originated from 
	the finite number of events simulated in the MC samples and from the statistical errors of
	the prompt and fake rates used to estimate the \PPP contribution. The MC samples are 
	normalised to the integrated luminosity achieved in 2012 data of 
	\lumi=19.6~\fbinv.}\label{ch8:tab:NSignal12}
\end{table}

The correction factors have been extracted using the procedure explained at previous section, in 
particular, expressions~\eqref{ch8:eq:accextr} and~\eqref{ch8:eq:effextr}, and they are reported in
Tables~\ref{ch8:tab:xsresults11}.

Therefore, the Equation~\eqref{ch8:eq:xs} is used to perform the cross section measurement
$\sigma(pp\to\W\Z\to\ell'\nu\ell^{+}\ell^{-})$ in four different and exhaustive channels: $eee$,
$\mu ee$, $e\mu\mu$ and $\mu\mu\mu$. The final state channels are defined by the leptonic decay
of the $\W\to\ell'\nu_{\ell}$ and $\Z\to\ell^+\ell^-$ ($\ell,\ell'=e,\mu,\tau$), where
the $\tau$-decay is accounted to a given channel when decays to electron or muon. The results for
each channel are given in Tables~\ref{ch8:tab:xsresults11} and~\ref{ch8:tab:xsresults12}. 
\begin{table}[!htpb]
	\centering
	\resizebox{\textwidth}{!}
	{
	\begin{tabular}{lccccc}\hline\hline
	      & $\mathcal{A}$ [\%] &  $\mathcal{C}$ [\%] & $\rho$ & $N_{S}$ & $\sigma(pp\to\W\Z+X)$ [pb]\\\hline
	$eee$ &  $5.09\pm0.03$  &  $1.595\pm0.014$ & $1.01\pm0.01$ & $60\pm8$ &  
    		     $23.00\pm3.10_{\text{stat}}\pm1.39_{\text{sys}}\pm0.51_{\text{lumi}}$\\
	$\mu ee$ & $5.04\pm0.03$  & $1.868\pm0.016$ & $0.96\pm0.01$ & $57\pm8$  &
	             $19.67\pm2.73_{\text{stat}}\pm1.50_{\text{sys}}\pm0.43_{\text{lumi}}$\\
	$e\mu\mu$& $5.04\pm0.03$ & $2.312\pm0.017$ & $0.87\pm0.01$ & $65\pm8$ &
                    $19.81\pm2.60_{\text{stat}}\pm1.55_{\text{sys}}\pm0.44_{\text{lumi}}$\\
        $\mu\mu\mu$&$4.89\pm0.03$& $2.859\pm0.019$ & $0.93\pm0.01$ & $91\pm10$ &
	            $21.02\pm2.30_{\text{stat}}\pm1.47_{\text{sys}}\pm0.46_{\text{lumi}}$\\\hline
	\end{tabular}
	}
	\caption[Cross section measurements in $3e$, $2e1\mu$, $2\mu1e$ and $3\mu$ channels for
	2011 data]{Acceptance ($\mathcal{A}$), acceptance and event efficiency in simulation 
	($\mathcal{C}$), experimental-simulation efficiency factor ($\rho$), number of signal 
	observed and inclusive \WZ cross section, measured by channel. The cross section errors
	are split in statistical, systematic and luminosity origin (see next section). The errors
	reported for the other factors are statistical only. Notice that the low acceptance is due
	to the wide signal definition, where each channel final state is defined from the whole 
	leptonic \WZ decay. Results for 7~\TeV.}\label{ch8:tab:xsresults11}
\end{table}
\begin{table}[!htpb]
	\centering
	\resizebox{\textwidth}{!}
	{
	\begin{tabular}{lccccc}\hline\hline
	      & $\mathcal{A}$ [\%] &  $\mathcal{C}$ [\%] & $\rho$ & $N_{S}$ & $\sigma(pp\to\W\Z+X)$ [pb]\\\hline
	$eee$ &  $4.57\pm0.02$  &   $1.46 \pm 0.05$ & $0.89\pm0.05$ & $209\pm15$ &   
       		$24.92 \pm 1.83_{\text{stat.}} \pm 1.25_{\text{syst}} \pm 1.10_{\text{lumi}}$\\
	$\mu ee$ & $4.58\pm0.02$  & $1.78\pm0.07$ & $0.93\pm0.05$ & $249\pm16$  &
		$23.42 \pm 1.59_{\text{stat}}\pm 1.11_{\text{syst}}\pm 1.03_{\text{lumi}}$\\
	$e\mu\mu$& $4.53\pm0.02$ & $2.22\pm0.09$ & $0.96\pm0.06$ & $334\pm19$ &
		$24.40 \pm 1.46_{\text{stat}}\pm 1.33_{\text{syst}}\pm 1.07_{\text{lumi}}$\\
	$\mu\mu\mu$&$4.46\pm0.02$& $2.90\pm0.12$ & $0.99\pm0.06$ & $477\pm22$ &
		$25.71 \pm 1.27_{\text{stat}}\pm 1.34_{\text{syst}}\pm 1.13_{\text{lumi}}$\\\hline
	\end{tabular}
	}
	\caption[Cross section measurements in $3e$, $2e1\mu$, $2\mu1e$ and $3\mu$ channels for
	2012 data]{Acceptance ($\mathcal{A}$), acceptance and event efficiency in simulation
	($\mathcal{C}$), experimental-simulation efficiency factor ($\rho$), number of signal 
	observed and inclusive \WZ cross section, measured by channel. The cross section errors 
	are split in statistical, systematic and luminosity origin (see next section). 
	The errors reported for the other factors are statistical only. Note that the low 
	acceptance is due to the wide signal definition, where each channel final state is defined 
	from the whole leptonic \WZ decay. Results for 8~\TeV.}\label{ch8:tab:xsresults12}
\end{table}

\section{Systematic uncertainty}\label{ch8:sec:sys}
Each measured observable used to calculate the cross section per channel have been determined with 
a degree of uncertainty. Besides of the statistical uncertainty, related with the finite number of 
measurements performed, sources of systematic uncertainties have been identified and evaluated for 
each term of Equation~\eqref{ch8:eq:xscorrected} and subsequently propagated to the cross section 
measurement per channel. The factorisation performed to the efficiency term, 
$\mathcal{A}\cdot\varepsilon=\mathcal{C}\cdot\rho$, allows to evaluate separately the theoretical 
and experimental sources of systematic uncertainty. The 
$\mathcal{C}=\mathcal{A}\cdot\varepsilon^{sim}$ factor concerns both sources of uncertainty in the 
theoretical models used to generate the \WZ \gls{mc} simulated sample and in the detector 
performance such the scale factors and resolution of the final-state objects. The 
$\rho=\varepsilon/\varepsilon^{sim}$ factor, which corrects the detector efficiency in the 
simulation $\mathcal{C}$ with the detector efficiency in experimental data, is mainly characterised
by the \gls{sf}[s]\glsadd{ind:sf} of the trigger, reconstruction and identification requirements of the lepton 
objects of the analysis, and consequently, affected by the uncertainties of those \gls{sf}[s]\glsadd{ind:sf}. 
The $N_S$ observable, built as $N_S=N_{\pr\pr\pr}^{N_{t3}}-N_{bkg}^{MC}$, involves the systematic 
uncertainty assigned to the data-driven method for the $N_{\pr\pr\pr}^{N_{t3}}$ measurement, and 
the same sources of systematic considered in the $\mathcal{C}$ factor but applied to the \gls{mc} 
background samples. In addition, the cross section uncertainties of each simulated sample is also 
considered. Finally, the uncertainty of the integrated luminosity measurement is included 
as a source of systematic uncertainty as well and propagated to the cross section measurement along 
with all the aforementioned sources.

\paragraph{Systematic uncertainties affecting $\mathcal{C}$}\mbox{}

The considered theoretical uncertainties on $\mathcal{A}$ arises from the \gls{pdf}\glsadd{ind:pdf} set choice to 
simulate the \WZ \gls{mc} sample and the uncertainties quoted for the \gls{pdf}, in addition to the 
\gls{qcd} renormalisation and factorisation scales used to generate the \WZ \gls{mc} simulation.
The \gls{pdf} are fit to a phenomenological function using several parameters ($i=1,\dots,N$) from 
available experimental data (see Section~\ref{ch1:sec:physicsathadroncolliders} of 
Chapter~\ref{ch1}). The best fit is used as the central value \gls{pdf}, but by modifying 
$\pm1\sigma_i$ each fitted parameter separately is possible to obtain as many \glspl{pdf} functions as
parameters the function have. Therefore, each \gls{pdf} collaboration provides a set of $N+1$ 
\gls{pdf}, one for the best fit, used to the \gls{mc} simulations, and the other $N$ used to 
propagate the uncertainties.  Further details may be obtained from 
Reference~\cite{Bourilkov:2006cj}. The effect of the \gls{pdf} uncertainties in the acceptance is 
studied by recalculating the acceptance using the $N$-subset of functions\footnote{In fact, the 
effect is obtained using the \emph{master equations} of Reference~\cite{Bourilkov:2006cj}.}. The 
maximum and minimum variation with respect the acceptance calculated with the central \gls{pdf} 
is taken as systematic uncertainty. The other theoretical systematic associated to the modelling
of the \gls{mc} samples is the choice of the \gls{qcd} renormalisation $\mu_R$ and factorisation
$\mu_F$ scales~\cite{Lepage:1980fj}. The nominal values used in the \WZ simulated sample are 
$\mu_R^0=\mu_F^0=(M_W+M_Z)/2$. The usual variation by a factor of two up and down with respect the 
nominal values ($2\mu_R^0$, $\mu_R^0/2$, $2\mu_F^0$ and $\mu_F^0/2$) is evaluated independently in
the acceptance, taking the biggest difference between the nominal and the varied scales as the 
systematic uncertainty due to the \gls{qcd} scales choice.

\paragraph*{}
The detector performance uncertainties are studied by the uncertainties of each involved parameter 
used to correct the data (see Chapter~\ref{ch5}, Section~\ref{ch5:sec:datacorr}), namely the muon 
momentum and electron energy scales, the \MET scale and resolution and the \gls{pileup} re-weighting. 
The muon momentum and the electron energy scale have been corrected (see Chapter~\ref{ch5}, 
Section~\ref{ch5:subsec:energyscale}), introducing an associated uncertainty in the measurement:
1\% for muon momentum scale and 2\% (5\%) for barrel (endcap) electron energy scale. These 
variations have been used to shift up and down the measured momentum (energy) in the signal and 
background simulated samples and the recalculated yields have been compared with the nominal ones,
taking as systematic uncertainty the difference between them.
The \MET scale and resolution is evaluated from their components. Since the \MET is inferred from 
the sum of the transverse momenta of all the observed particles in the event, the \MET is broken 
into its components (jets, leptons, unclustered energy) and their scales and resolutions are varied
by their uncertainties. The maximum and minimum values for the recalculated \MET are used to 
perform again the analysis selection with all the \gls{mc} samples obtaining new yields. The 
relative yields difference with respect the nominal ones is accounted as the systematic uncertainty.
Finally, the weighting process done in the simulated samples in order to match the interaction 
multiplicity distribution observed in the experimental data assumes a inelastic proton-proton cross 
section of $\sigma_{pp}=73.5$~mb with a 5\% of uncertainty. This uncertainty is used to shift up 
and down the inelastic p--p cross section, resulting in a $\pm1\sigma$ variation in the Poisson 
distribution of the mean number of interactions (see Chapter~\ref{ch7}, 
Section~\ref{ch7:sec:mcbkg}), and therefore, obtaining a new weights to use with the simulated 
samples. Using these new weights the yields obtained are compared with the yields using the nominal
Poisson distribution, and the relative differences are assigned as systematic of the re-weighting 
process.

\paragraph{Systematic uncertainties affecting $\rho$}\mbox{}

The $\rho$ factor, which takes into account the discrepancies in the experimental versus simulated 
data efficiencies, is described by the correction factors,~\ie the \gls{sf}[s]\glsadd{ind:sf}, of the 
efficiencies of the trigger and final-state objects selection outlined in 
Section~\ref{ch5:subsec:tap}. The \gls{sf}[s]\glsadd{ind:sf} have been estimated by the ratio of experimental over
simulated data efficiencies as it is described in Sections~\ref{ch6:subsec:muoneff} 
and~\ref{ch6:subsec:eleceff}. The uncertainties of the efficiencies determination by the tag and 
probe method are propagated to the ratio. The sources of uncertainties are the limited statistics 
for the different categories used to extract the efficiencies, and the different shapes used
for the fits of the Z resonance and backgrounds. The statistical error is of the order of 1\%. 
The systematic uncertainty from the shapes used to fit the mass peak is about another 1-2\%. The
shape systematic was calculated in two different ways: varying the parameters that defined the 
fitting function and, conversely, using different functions for the shapes. Nevertheless, this
systematic is partially cancelled out when calculating the scale factors because of the fact that
the fitting functions for data and simulated samples were change in the same way. 

The effect of this uncertainty to the analysis is studied 
by varying each \gls{sf}\glsadd{ind:sf} independently, using the varied \gls{sf}\glsadd{ind:sf} to weight the simulated samples. 
The obtained yields with the modified scale factor is compared with the nominal \gls{sf}\glsadd{ind:sf}'s yields 
being the relative difference estimated as systematic uncertainty.

\paragraph{Systematic uncertainties affecting $N_S$}\mbox{}

The estimated number of signal is obtained by subtracting to the \PPP estimation of the data 
driven method the background estimated with \gls{mc} simulation. Therefore, the systematic 
uncertainties introduced by the \gls{fom}\glsadd{ind:fom} and the theoretical or measured uncertainties on the 
cross sections assigned to the \gls{mc} simulated backgrounds are considered as systematic sources
of $N_S$. The systematic uncertainty assigned to the \PPP estimation emerges because of the 
particular choice of the leading jet transverse energy cut used to bias the jet-induced enriched
region as it is described in Section~\ref{ch7:subsec:sysuncertainties}. The theoretical 
uncertainty on the cross section assigned to the background estimated with \gls{mc} simulations 
are 14\% (15\%) for the 2011 (2012) ZZ sample extracted from \gls{cms} measurements results in
experimental data~\cite{Chatrchyan:2012sga},~\cite{Chatrchyan:2013oev}; and 7\% (15\%) for the 
$Z\gamma$ 2011 (2012) cross section\footnote{The 2012 uncertainty is in fact an extrapolation of 
the 2011.}, also extracted from \gls{cms} results~\cite{CMS-PAS-EWK-11-009}.

\paragraph{Systematic uncertainties affecting $\lumi$}\mbox{}

The integrated luminosity $\lumi_{int}$ is measured by a dedicated group in the \gls{cms} 
collaboration, providing a centralised measure of $\lumi_{int}$ to be used by all the analyses 
performed. The last 2011 measure~\cite{CMS-PAS-SMP-12-008} of the total integrated 
luminosity is affected by 2.2\% of uncertainty, while for 2012 the uncertainty 
quoted~\cite{CMS-PAS-LUM-12-001} is 4.4\%.

\paragraph*{}
The systematic uncertainties described above have been propagated to the cross section measurement.
Tables~\ref{ch8:tab:syscontribution11} and~\ref{ch8:tab:syscontribution12} reports the relative 
uncertainty on the cross section measurement introduced by each source of systematic uncertainty. 
The data-driven and \MET scale and resolution are the dominant sources of systematic
uncertainties in each cross section measured channel.
\begin{table}[!hbtp]
     \centering
     \begin{tabular}{l c c c c}\hline\hline
                                 & $eee$ [\%]   & $ee\mu$ [\%]   & $e\mu\mu$ [\%]  & $\mu\mu\mu$ [\%] \\\hline
       Lepton and Trigger efficiency&   2.8         &   2.5         &    1.9          &    1.4\\
       Muon momentum scale          &    --         &   0.6         &    0.4          &    1.1\\
       Electron energy scale        &   1.9         &   0.6         &    1.2          &    -- \\
       \MET scale and resolution    &   3.7         &   3.4         &    4.3          &    3.7\\
       Fakeable object method       &   2.5         &   5.8         &    5.6          &    5.2\\
       pile-up re-weighting         &   0.3         &   0.5         &    1.0          &    0.7\\
       PDFs                         &   1.5         &   1.5         &    1.5          &    1.5\\ 
       $\mu_F$, $\mu_R$ scales      &   1.3         &   1.3         &    1.3          &    1.3\\
       Theoretical MC cross-sections&   0.4         &   0.8         &    0.6          &    0.7\\ 
       Acceptance stat. error       &   1.0         &   0.9         &    0.9          &    1.6\\ \hline
     \end{tabular}                          
     \caption[Sources of systematic uncertainty on the cross section measurement in 2011]{Summary 
     of systematic uncertainties considered on the 2011 cross section measurement shown in each 
     measured channel. The values report the relative uncertainty introduced by each source of 
     systematic on the cross section showing its final impact in the measurement, 
     $(\delta\sigma_{WZ})_{sys}^i/\sigma_{WZ}\cdot100$.
     }\label{ch8:tab:syscontribution11}
\end{table}

\begin{table}[!hbtp]
     \centering
     \begin{tabular}{l c c c c}\hline\hline
                                 & $eee$ [\%]   & $ee\mu$ [\%]   & $e\mu\mu$ [\%]  & $\mu\mu\mu$ [\%] \\\hline
 	Lepton and Trigger efficiency  & 1.8 & 1.8 & 1.8 & 1.8\\
 	Muon momentum scale            &  -- & 0.4 & 0.9 & 1.1\\
 	Electron energy scale          & 0.8 & 0.6 & 0.1 &  --\\
 	\MET scale and resolution      & 2.9 & 2.9 & 3.5 & 3.1\\
 	Fakeable object method         & 2.7 & 1.9 & 2.7 & 2.6\\
 	pileup re-weighting            & 0.8 & 1.1 & 0.6 & 0.9\\
 	PDFs                           & 1.4 & 1.4 & 1.4 & 1.4\\
 	$\mu_F$, $\mu_R$ scales        & 1.6 & 1.6 & 1.6 & 1.6\\
 	Theoretical MC cross sections  & 0.3 & 0.2 & 0.3 & 0.2\\
 	Acceptance stat. error         & 0.8 & 1.2 & 1.1 & 1.0\\\hline
     \end{tabular}                          
     \caption[Sources of systematic uncertainty on the cross section measurement in 2012]{Summary 
     of systematic uncertainties considered on the 2012 cross section measurement shown in each 
     measured channel. The values report the relative uncertainty introduced by each source of 
     systematic on the cross section showing its final impact in the measurement, 
     $(\delta\sigma_{WZ})_{sys}^i/\sigma_{WZ}\cdot100$.
     }\label{ch8:tab:syscontribution12}
\end{table}

\section{Cross section combination}\label{ch8:sec:xscombined}
The final cross section estimation is obtained by combining the cross section measurements in the
different channels, taking into account the correlation in the uncertainties using the \gls{blue}\glsadd{ind:blue} 
method~\cite{Lyons:1988rp}. The method defines the combined cross section as the linear combination
of the measured cross sections in each channel
\begin{equation}
	\sigma(\wpmz\to\ell'^{\pm}\nu_{\ell'}\ell^+\ell^-)=\sum_{i=1}^4\alpha_i\sigma_i
	\label{ch8:eq:xsblue}
\end{equation}
where the i-index is referring to the four measured channels $3e$, $1\mu2e$, $1e2\mu$ and $3\mu$,
$\sigma_i$ are the cross section measurements in each channel and $\alpha_i$ are the weights to be 
obtained by minimising the variance. The unbiased requirement implies
\begin{equation}
	\sum_{i=1}^4\alpha_i=1
\end{equation}
And from Equation~\eqref{ch8:eq:xsblue}, the variance is deduced as,
\begin{equation}
	Var(\sigma)=\boldsymbol{\alpha}^T\mathbf{E}\boldsymbol{\alpha}
\end{equation}
being $\boldsymbol{\alpha}$ the vector of weighting factors $\alpha_i$, $\boldsymbol{\alpha}^T$ its 
transpose and $\mathbf{E}$ the error matrix, where its diagonal elements give the variances of the 
individual measurements, while the off-diagonal elements describe the correlations between pair of 
measurements,
\begin{equation}
	\mathbf{E}=
	\begin{pmatrix}
		Var(\sigma_1) & Covar(\sigma_1,\sigma_2) & Cov(\sigma_1,\sigma_3) & Cov(\sigma_1,\sigma_4) \\
		Cov(\sigma_2,\sigma_1) & Var(\sigma_2) & Cov(\sigma_2,\sigma_3) & Cov(\sigma_2,\sigma_4) \\
		Cov(\sigma_3,\sigma_1) & Cov(\sigma_3,\sigma_2) & Var(\sigma_3) & Cov(\sigma_3,\sigma_4) \\
		Cov(\sigma_4,\sigma_2) & Cov(\sigma_4,\sigma_2) & Cov(\sigma_4\sigma_3) & Var(\sigma_4) \\
	\end{pmatrix}
\end{equation}
Notice that the correlations are only considered as fully correlated or uncorrelated,
\begin{equation}
	Cov(\sigma_i,\sigma_j)=r\cdot\sqrt{Var(\sigma_i)\cdot Var(\sigma_j)},\quad r=0\text{ or }1
\end{equation}

The minimisation of the variance is accomplished by the method of Lagrangian multipliers to give,
\begin{equation}
	\boldsymbol{\alpha}=\frac{\mathbf{E}^{-1}\mathbf{U}}{\mathbf{U}^T\mathbf{E}^{-1}\mathbf{U}}
\end{equation}
where $\mathbf{E}^{-1}$ is the inverse of the error matrix, and $\mathbf{U}$ a vector whose four 
components are all unity.

\begin{figure}
	\centering
	\begin{subfigure}[b]{0.45\textwidth}
		\centering
		\includegraphics[width=\textwidth]{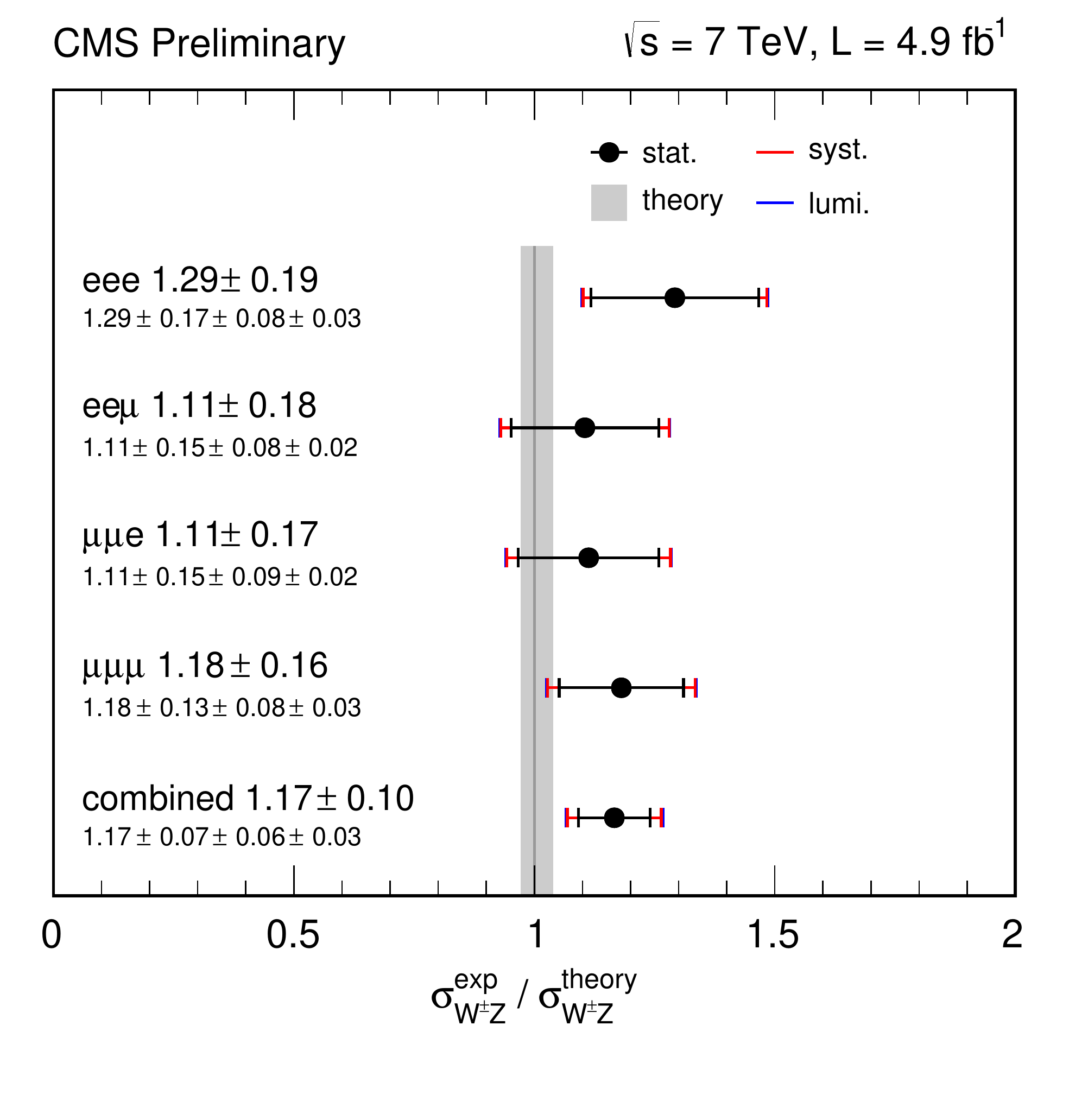}
		\caption{7~\TeV measurement. The NLO prediction is
		$17.8^{+0.7}_{-0.5}$~pb.}\label{ch8:subfig:xs7TeV}
	\end{subfigure}\quad
	\begin{subfigure}[b]{0.45\textwidth}
		\centering
		\includegraphics[width=\textwidth]{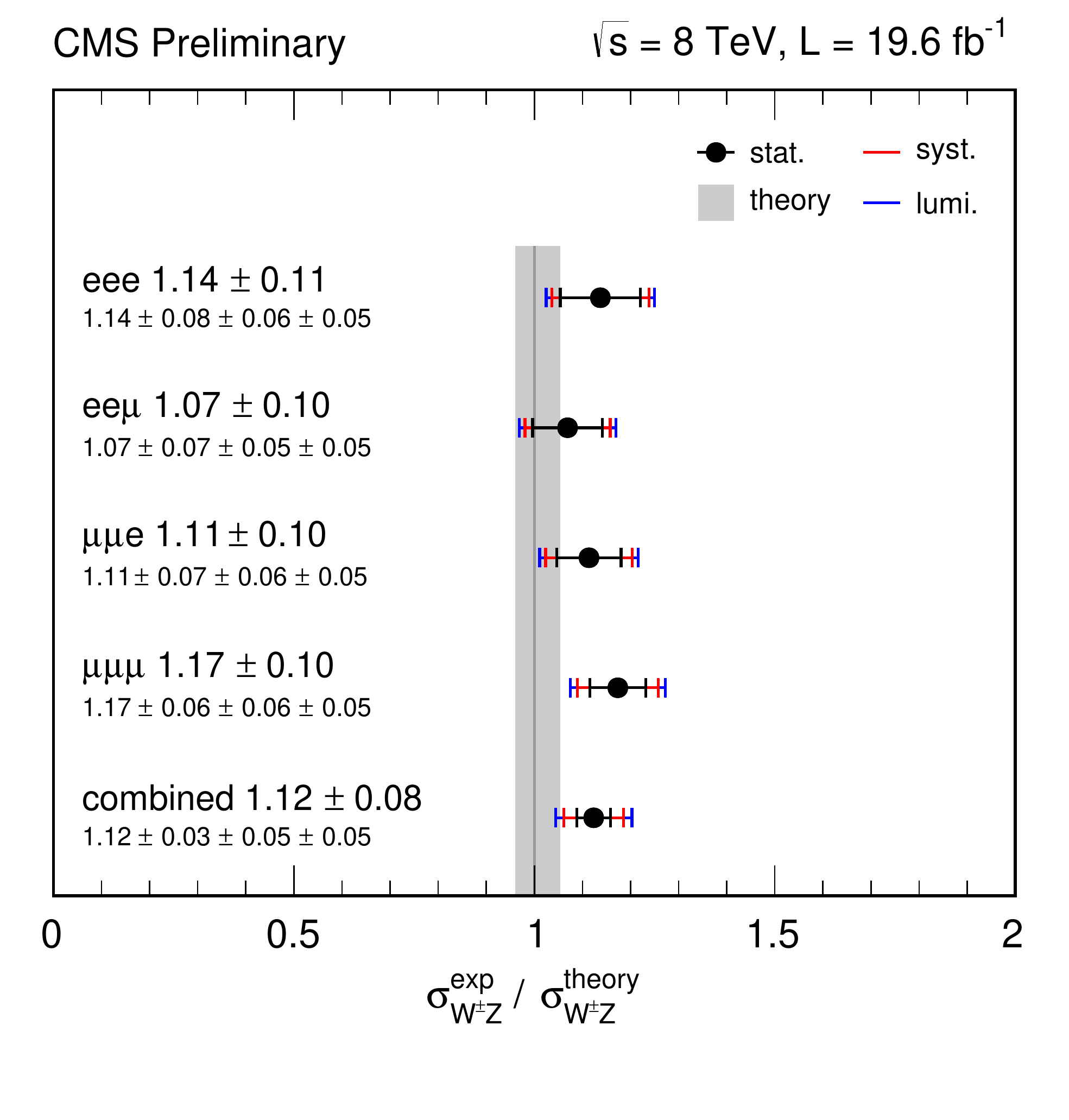}
		\caption{8~\TeV measurement. The NLO prediction is
		$21.9^{+0.9}_{-0.5}$~pb.}\label{ch8:subfig:xs8TeV}
	\end{subfigure}
	\caption[Ratio of measured cross section to the theoretical prediction]{Ratio of
	measured cross section over the theoretical prediction for each measured channel and the 
	BLUE method combined measurement. Each row in the plot shows the measured channel, the	
	ratio value and the uncertainty error (also split in statistical, systematic and 
	luminosity source in the line below each channel).}\label{ch8:fig:xs}
\end{figure}

In order to build the error matrix, the uncertainty errors have been considered
\begin{itemize}
 \item fully correlated between all channels:
 	\begin{itemize}
		\item \MET energy scale and resolution
		\item pileup re-weighting
		\item \gls{pdf} choice
		\item renormalisation and factorisation scales	
		\item cross sections of \gls{mc} estimated backgrounds
	\end{itemize}
 \item fully correlated between electron (muon) channels:
	 \begin{itemize}
		 \item Electron energy (muon momentum) scale
		 \item Electron (muon) reconstruction, identification and isolation scale
			 factors uncertainties from the tag and probe method
	 \end{itemize}
 \item fully correlated between $ee\mu$ and $eee$ ($\mu\mu e$ and $\mu\mu\mu$) channels:
         \begin{itemize}
		 \item Double electron (muon) trigger efficiency
	 \end{itemize}
 \item uncorrelated between channels:
 	\begin{itemize}
		\item Scale factor's statistical errors				
		\item data driven uncertainty
		\item Statistical errors
	\end{itemize}
\end{itemize}
The error matrix obtained for 2011 analysis is
\begin{equation}
	\mathbf{E}=
	\begin{pmatrix}
		11.73 & 1.32 & 1.45 & 1.25 \\
		1.32  & 8.62 & 1.15 & 1.06 \\
		1.45  & 1.15 & 8.09 & 1.16 \\
		1.25  & 1.06 & 1.16 & 6.54 
	\end{pmatrix}
	\;pb^2\,,
\end{equation}
giving a weighting factors of $\alpha_{3e}=0.159$, $\alpha_{1\mu2e}=0.245$, 
$\alpha_{1e2\mu}=0.256$ and $\alpha_{3\mu}=0.340$. The final estimated cross section for the \WZ 
process at 7~\TeV in the phase space $M_{ll}\in 91.1876\pm20 \GeV/c^{2}$ is measured to be:
\begin{equation*}
	\sigma_{7\TeV}(pp\to\WZ+X)= 20.8\pm1.3_{\text{stat}}\pm1.1_{\text{sys}}\pm0.5_{\text{lumi}}\, \text{pb}
\end{equation*}
The measurement is compatible with the \gls{nlo}\glsadd{ind:nlo} prediction obtained in Chapter~\ref{ch2}, 
$17.8^{+0.7}_{-0.5}$~pb, although is noticeable a $1.5$-$\sigma$ deviation from the theoretical 
value. Figure~\ref{ch8:subfig:xs7TeV} shows the ratio between measured and predicted cross 
section in each of the measured channels.

The error matrix obtained for 2012 analysis is
\begin{equation}
	\mathbf{E}=
	\begin{pmatrix}
		4.90  & 1.06 & 1.18 & 1.17 \\
		1.06  & 3.77 & 1.15 & 1.16 \\
		1.18  & 1.15 & 3.93 & 1.35 \\
		1.17  & 1.16 & 1.35 & 3.44 
	\end{pmatrix}
	\;pb^2
\end{equation}
giving a weighting factors of $\alpha_{3e}=0.195$, $\alpha_{1\mu2e}=0.279$, 
$\alpha_{1e2\mu}=0.235$ and $\alpha_{3\mu}=0.290$. The final estimated cross section for the \WZ 
process at 8~\TeV in the phase space $M_{ll}\in 91.1876\pm20 \GeV/c^{2}$ is measured to be:
\begin{equation*}
	\sigma_{8\TeV}(pp\to\WZ+X)= 24.6\pm0.8_{\text{stat}}\pm1.1_{\text{sys}}\pm1.1_{\text{lumi}}\, \text{pb}
\end{equation*}
The measurement is compatible with the \gls{nlo} prediction obtained in Chapter~\ref{ch2}, 
$21.9^{+0.9}_{-0.5}$~pb, and, similar to the 7~\TeV measurement, a $1.3$-$\sigma$ deviation from the
theoretical value can be observed. Figure~\ref{ch8:subfig:xs8TeV} shows the ratio between measured 
and predicted cross section in each of the measured channels.

\chapter{Measurement of \wzm and \wzp cross sections ratio}\label{ch9}
Since the LHC is a proton-proton collider, the \wzp and \wzm cross sections are not equal. As it was
described in chapter~\ref{ch2}, the dominant production mechanism of \wzp bosons involves an
up-type quark and a down-type antiquark while a down-type quark and up-type antiquark is required 
to produce \wzm. The predominance of the valence u-quark in the protons enhances the \wzp production
in front of the \wzm; therefore, an overall excess of $W^+Z$ events over $W^-Z$ is expected. 

This chapter describes the strategy followed to measure the cross section ratio 
$\sigma_{\wzm}/\sigma_{\wzp}$ which is fully based in the \WZ inclusive cross section measurement 
developed in the previous chapters. Therefore, the signal definitions, background description, 
analysis strategy and systematic uncertainties from the previous cross section analysis remain 
valid. Accordingly, in this chapter we will point out the the peculiarities that the charge's split
of the samples introduces to the analysis. Control distributions are also shown along
with the event yields obtained in each measurement. Finally, the cross section ratio is calculated 
in four independent lepton-flavour channels which are combined to obtain the final result. 

\section{Event selection}
The signal definition for the inclusive analysis (see Chapter~\ref{ch6}) is slightly modified to
define two exhaustive regions from the inclusive signal by evaluating the charge of the \W 
candidate lepton. A positive lepton identifies the \wzp signal whilst a negative lepton identifies 
the \wzm. Therefore, the analysis strategy is entirely based in the inclusive cut-based
analysis selection presented in Chapter~\ref{ch6} and performed in these two regions defined by the 
charge of the \W-candidate lepton. Consequently, the \emph{Lepton preselection} and the 
\emph{\Z candidate selection} stages are common to both regions given that the \W-candidate lepton
is still not defined at those levels of the analysis. The number of events for the four 
measured channels are reported in Tables~\ref{ch9:tab:yields11} 
and~\ref{ch9:tab:yields12}. It is worthwhile to mention that, as expected, the yield differences 
between opposite-charged final states are coming from the \WZ production, since every background 
process generates the \W-candidate lepton, whether it is a fake lepton as a lost-by-acceptance 
lepton, with a charge democratically populated.

\begin{table}[!htbp]
	\centering
	\begin{subtable}[b]{0.45\textwidth}
		\resizebox{\textwidth}{!}
		{
		\begin{tabular}{ rcc  }\hline\hline
 {                }              & {\bf $W^+$ }       & {\bf $W^-$ }          \\ \hline
 {Data-driven bkg.}              & 1.0   $\pm$ 0.3  & 1.1  $\pm$ 0.3     \\ 
 $ZZ$                            & 0.99  $\pm$ 0.02 & 0.96 $\pm$ 0.02   \\ 
 $V\gamma$                       & 0     $\pm$ 0    & 0    $\pm$ 0    \\ 
 {$WZ\rightarrow3\ell\nu$ }      & 28.5 $\pm$ 0.4   & 16.3 $\pm$ 0.3   \\ \hline
 {\bf Total expect.}             & 30.5 $\pm$  0.5  & 18.3 $\pm$ 0.4   \\ 
 {\bf Data       }               &      36          &        28        \\ \hline
\end{tabular}

	        }
		\caption{Three electron final state}
	\end{subtable}\quad
	\begin{subtable}[b]{0.45\textwidth}
		\resizebox{\textwidth}{!}
		{
		\begin{tabular}{ r  cc  }\hline\hline
 {                }              & {\bf $W^+$ }       & {\bf $W^-$ }          \\ \hline
 {Data-driven bkg.}              & 0.63 $\pm$ 0.18    & 0.8  $\pm$ 0.2    \\ 
 $ZZ$                            & 1.81 $\pm$ 0.03    & 1.65 $\pm$ 0.02    \\ 
 $V\gamma$                       & 0    $\pm$ 0       & 0    $\pm$ 0    \\ 
 {$WZ\rightarrow3\ell\nu$ }      & 32.3  $\pm$ 0.4    & 18.1 $\pm$ 0.3  \\ \hline
 {\bf Total expect.}             & 34.8  $\pm$ 0.4    & 20.5 $\pm$ 0.4  \\ 
 {\bf Data       }               &        40          & 22                \\ \hline
\end{tabular}

		}
		\caption{Two electron and one muon final state}
	\end{subtable}\vskip 1em
	\centering
	\begin{subtable}[b]{0.45\textwidth}
		\resizebox{\textwidth}{!}
		{
		\begin{tabular}{ r cc }\hline\hline
 {                }              & {\bf $W^+$ }       & {\bf $W^-$ }      \\ \hline
 {Data-driven bkg.}              & 1.1  $\pm$ 0.3  & 1.4   $\pm$ 0.3 \\ 
 $ZZ$                            & 1.37 $\pm$ 0.02 & 1.32  $\pm$ 0.02  \\ 
 $V\gamma$                       & 0.5  $\pm$ 0.5  & 0.003 $\pm$ 0.003\\ 
 {$WZ\rightarrow3\ell\nu$ }      & 35.7 $\pm$ 0.4  & 20.1 $\pm$ 0.3\\ \hline
 {\bf Total expect. }            & 38.7 $\pm$ 0.7  & 22.8  $\pm$ 0.4 \\ 
 {\bf Data       }               & 48              & 22 \\ \hline
\end{tabular}

		}
		\caption{Two muons and one electron final state}
	\end{subtable}\quad
	\begin{subtable}[b]{0.45\textwidth}
		\resizebox{\textwidth}{!}
		{
		\begin{tabular}{ r cc }\hline\hline
 {                }              & {\bf $W^+$ }       & {\bf $W^-$ }          \\ \hline
 {Data-driven bkg.}              & 1.2  $\pm$ 0.2   & 0.56 $\pm$ 0.14    \\ 
 $ZZ$                            & 2.53 $\pm$ 0.02  & 2.30 $\pm$ 0.02    \\ 
 $V\gamma$                       & 0    $\pm$ 0     & 0.00 $\pm$ 0.00   \\ 
 {$WZ\rightarrow3\ell\nu$ }      & 48.2 $\pm$ 0.5   & 26.9 $\pm$ 0.4  \\ \hline
 {\bf Total expect. }            & 51.9 $\pm$ 0.5   & 29.7 $\pm$ 0.4   \\ 
 {\bf Data       }               &         52       &       45        \\ \hline
\end{tabular}

		}
		\caption{Three muons final state}
	\end{subtable}
	\caption[Number of total events in the ratio analysis for 2011]{Number of events selected
		in the four leptonic channels investigated
		for the 2011 analysis. The MC samples are normalised to $\lumi_{int}=4.9~\fbinv$.
		The first column of results is obtained requiring a third positive lepton as \W
		candidate, while the second column is obtained by requiring a negative \W-candidate
		lepton. The errors shown are statistical only.}\label{ch9:tab:yields11}
\end{table}

\begin{table}[!htbp]
	\centering
	\begin{subtable}[b]{0.45\textwidth}
		\resizebox{\textwidth}{!}
		{
		\begin{tabular}{ rcc  }\hline\hline
 {                }              & {\bf $W^+$ }     & {\bf $W^-$ }    \\ \hline
 {Data-driven bkg.}              & 7.4   $\pm$ 1.0  & 7.5  $\pm$ 1.0  \\ 
 $ZZ$                            & 1.22  $\pm$ 0.05 & 1.24 $\pm$ 0.05 \\ 
 $V\gamma$                       & 0.9   $\pm$ 0.5  & 1.5  $\pm$ 0.8  \\ 
 $WV$                            & 0     $\pm$ 0    & 0.1  $\pm$ 0.1  \\ 
 $VVV$                           & 3.3   $\pm$ 0.2  & 2.7  $\pm$ 0.2  \\ 
 {$WZ\rightarrow3\ell\nu$ }      & 118.4 $\pm$ 1.1  & 75.5 $\pm$ 0.9 \\ \hline
 {\bf Total expect.}             & 131.2 $\pm$ 1.6  & 88.5 $\pm$ 1.5  \\ 
 {\bf Data       }               &      138         &        97       \\ \hline
\end{tabular}

	        }
		\caption{Three electron final state}
	\end{subtable}\quad
	\begin{subtable}[b]{0.45\textwidth}
		\resizebox{\textwidth}{!}
		{
		\begin{tabular}{ r  cc  }\hline\hline
 {                }              & {\bf $W^+$ }       & {\bf $W^-$ }    \\ \hline
 {Data-driven bkg.}              & 13   $\pm$ 2       & 14   $\pm$ 2    \\ 
 $ZZ$                            & 1.60 $\pm$ 0.08    & 1.49 $\pm$ 0.07 \\ 
 $V\gamma$                       & 0    $\pm$ 0       & 0.4  $\pm$ 0.4  \\ 
 $WV$                            & 0    $\pm$ 0       & 0    $\pm$ 0    \\ 
 $VVV$                           & 4.3  $\pm$ 0.2     & 3.6  $\pm$ 0.2  \\ 
 {$WZ\rightarrow3\ell\nu$ }      & 150.7$\pm$ 1.3     & 95.2 $\pm$ 1.0  \\ \hline
 {\bf Total expect.}             & 170  $\pm$ 2       & 115  $\pm$ 2    \\ 
 {\bf Data       }               &    179             & 109            \\ \hline
\end{tabular}

		}
		\caption{Two electron and one muon final state}
	\end{subtable}\vskip 1em
	\centering
	\begin{subtable}[b]{0.45\textwidth}
		\resizebox{\textwidth}{!}
		{
		\begin{tabular}{ r  cc  }\hline\hline
 {                }              & {\bf $W^+$ }       & {\bf $W^-$ }    \\ \hline
 {Data-driven bkg.}              & 24   $\pm$ 2       & 24   $\pm$ 2    \\ 
 $ZZ$                            & 2.0  $\pm$ 0.09    & 1.9  $\pm$ 0.09 \\ 
 $V\gamma$                       & 1.8  $\pm$ 0.8     & 2.0  $\pm$ 0.9  \\ 
 $WV$                            & 0.1  $\pm$ 0.1     & 0.1  $\pm$ 0.1  \\ 
 $VVV$                           & 5.7  $\pm$ 0.3     & 4.7  $\pm$ 0.3  \\ 
 {$WZ\rightarrow3\ell\nu$ }      & 192.9$\pm$ 1.5     & 123.0$\pm$ 1.2  \\ \hline
 {\bf Total expect.}             & 227  $\pm$ 3       & 155  $\pm$ 3    \\ 
 {\bf Data       }               &    254             & 146             \\ \hline
\end{tabular}

		}
		\caption{Two muons and one electron final state}
	\end{subtable}\quad
	\begin{subtable}[b]{0.45\textwidth}
		\resizebox{\textwidth}{!}
		{
		\begin{tabular}{ r  cc  }\hline\hline
 {                }              & {\bf $W^+$ }       & {\bf $W^-$ }    \\ \hline
 {Data-driven bkg.}              & 31   $\pm$ 3       & 29   $\pm$ 3    \\ 
 $ZZ$                            & 3.0  $\pm$ 0.1     & 2.8  $\pm$ 0.1 \\ 
 $V\gamma$                       &   0  $\pm$ 0       & 0    $\pm$ 0   \\ 
 $WV$                            & 1.6  $\pm$ 0.7     & 0.6  $\pm$ 0.3  \\ 
 $VVV$                           & 7.4  $\pm$ 0.3     & 6.0  $\pm$ 0.3  \\ 
 {$WZ\rightarrow3\ell\nu$ }      & 263.1$\pm$ 1.7     & 164.9$\pm$ 1.4  \\ \hline
 {\bf Total expect.}             & 306  $\pm$ 4       & 203  $\pm$ 4    \\ 
 {\bf Data       }               &    344             & 213             \\ \hline
\end{tabular}

		}
		\caption{Three muons final state}
	\end{subtable}
	\caption[Number of total events in the ratio analysis for 2012]{Number of events selected
		in the four leptonic channels investigated for the
		2012 analysis. The MC samples are normalised to $\lumi_{int}=19.6~\fbinv$.
		The first column of results is obtained requiring a third positive lepton as \W
		candidate, while the second column is obtained by requiring a negative \W-candidate
		lepton. The errors shown are statistical only.}\label{ch9:tab:yields12}
\end{table}

\begin{figure}[htbp]
	\begin{subfigure}[b]{0.45\textwidth}
		\includegraphics[width=\textwidth]{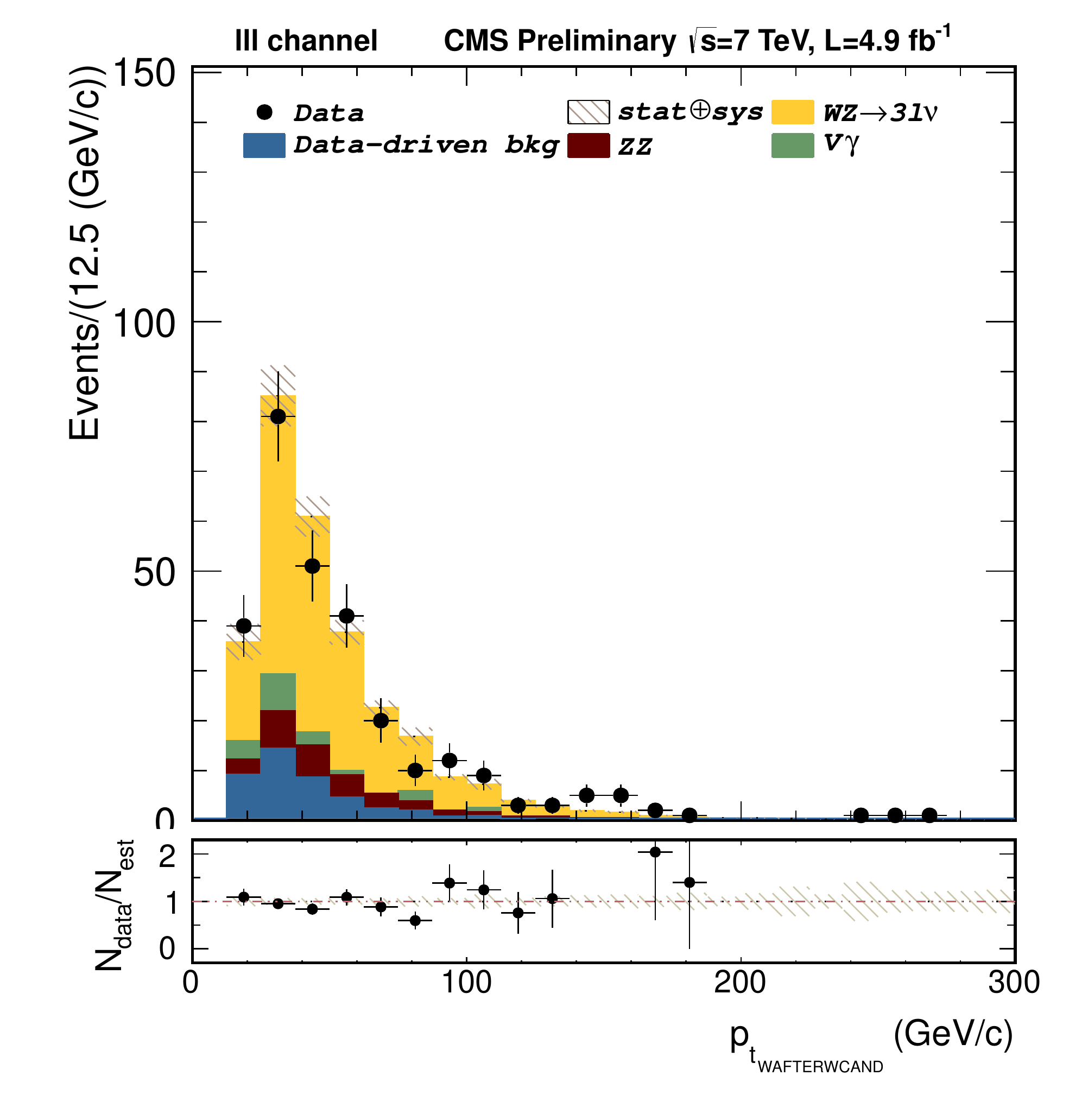}
		\caption{Positively charged \W-candidate lepton}
	\end{subfigure}\quad
	\begin{subfigure}[b]{0.45\textwidth}
		\includegraphics[width=\textwidth]{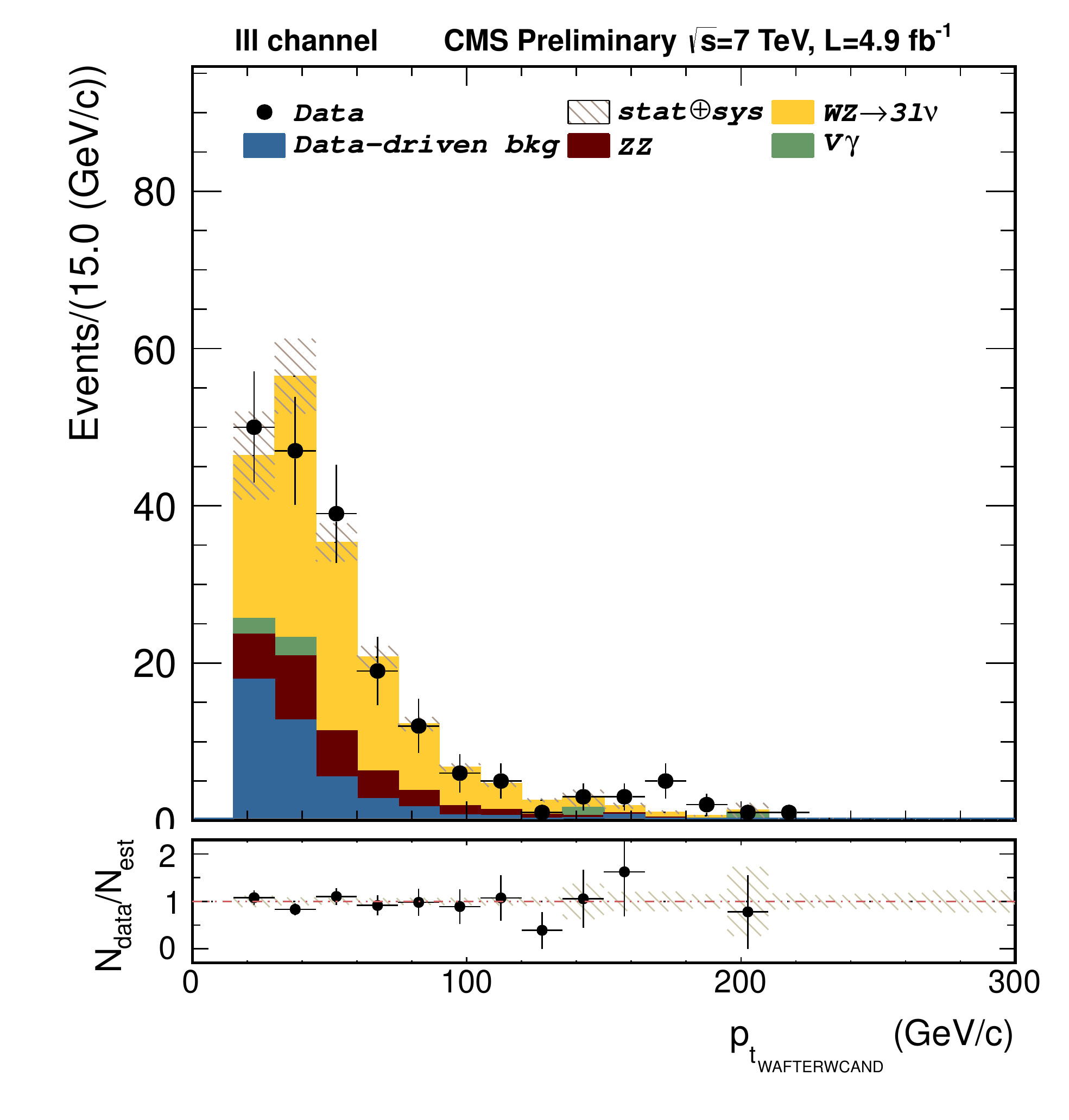}
		\caption{Negatively charged \W-candidate lepton}
	\end{subfigure}
	\caption[Transverse momentum of the \W-lepton candidate for 2011 analysis]{Transverse 
		momentum distributions for the \W-lepton selected candidate once the \W-candidate
		requirement have been applied in the selection. The MC samples are normalised to
		the 2011 integrated luminosity, ${\lumi_{int}=4.9~\fbinv}$.}\label{ch9:fig:wleptonpt}
\end{figure}

\begin{figure}[htbp]
	\begin{subfigure}[b]{0.45\textwidth}
		\includegraphics[width=\textwidth]{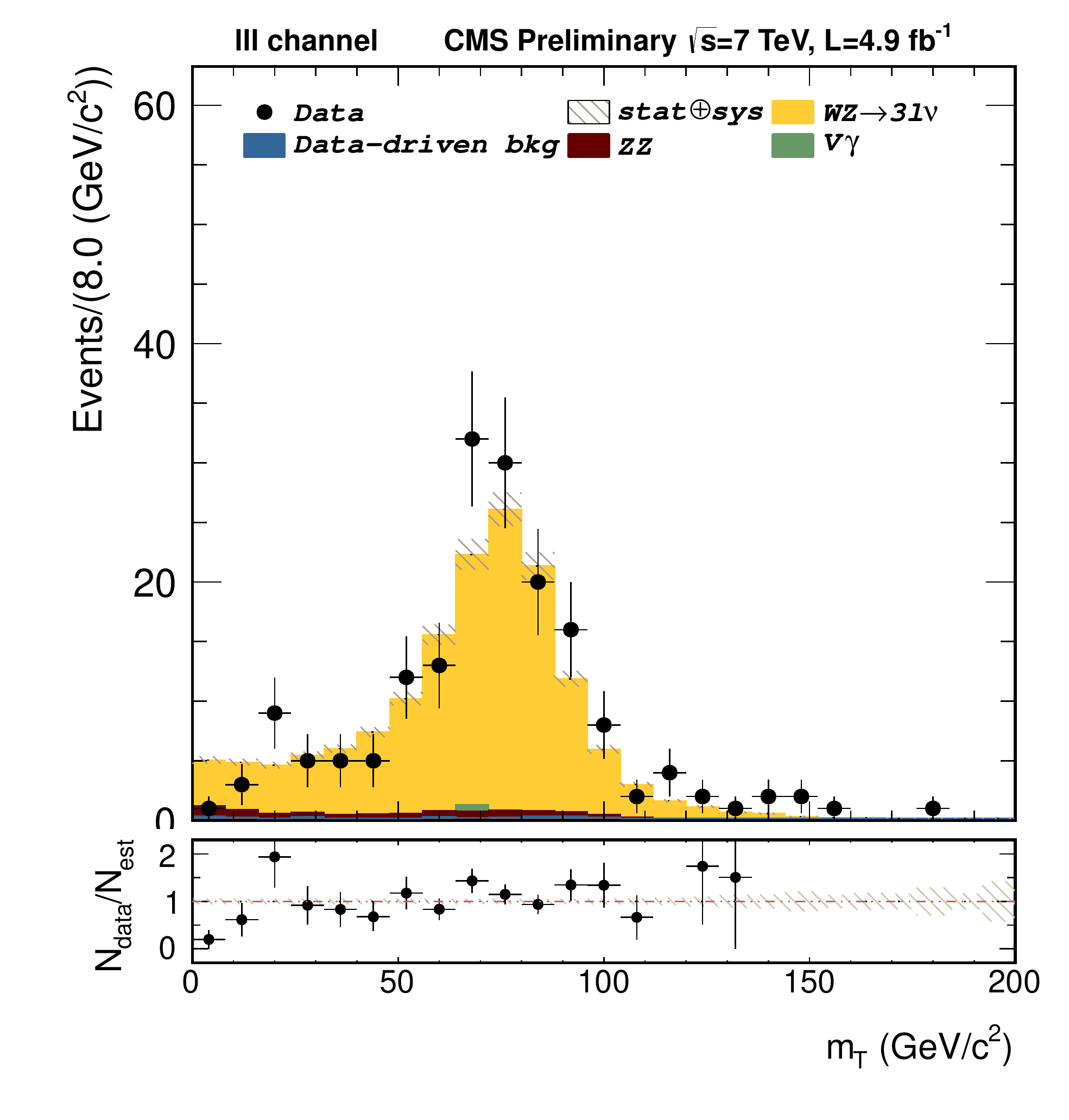}
		\caption{Positively charged \W-candidate lepton}
	\end{subfigure}\quad
	\begin{subfigure}[b]{0.45\textwidth}
		\includegraphics[width=\textwidth]{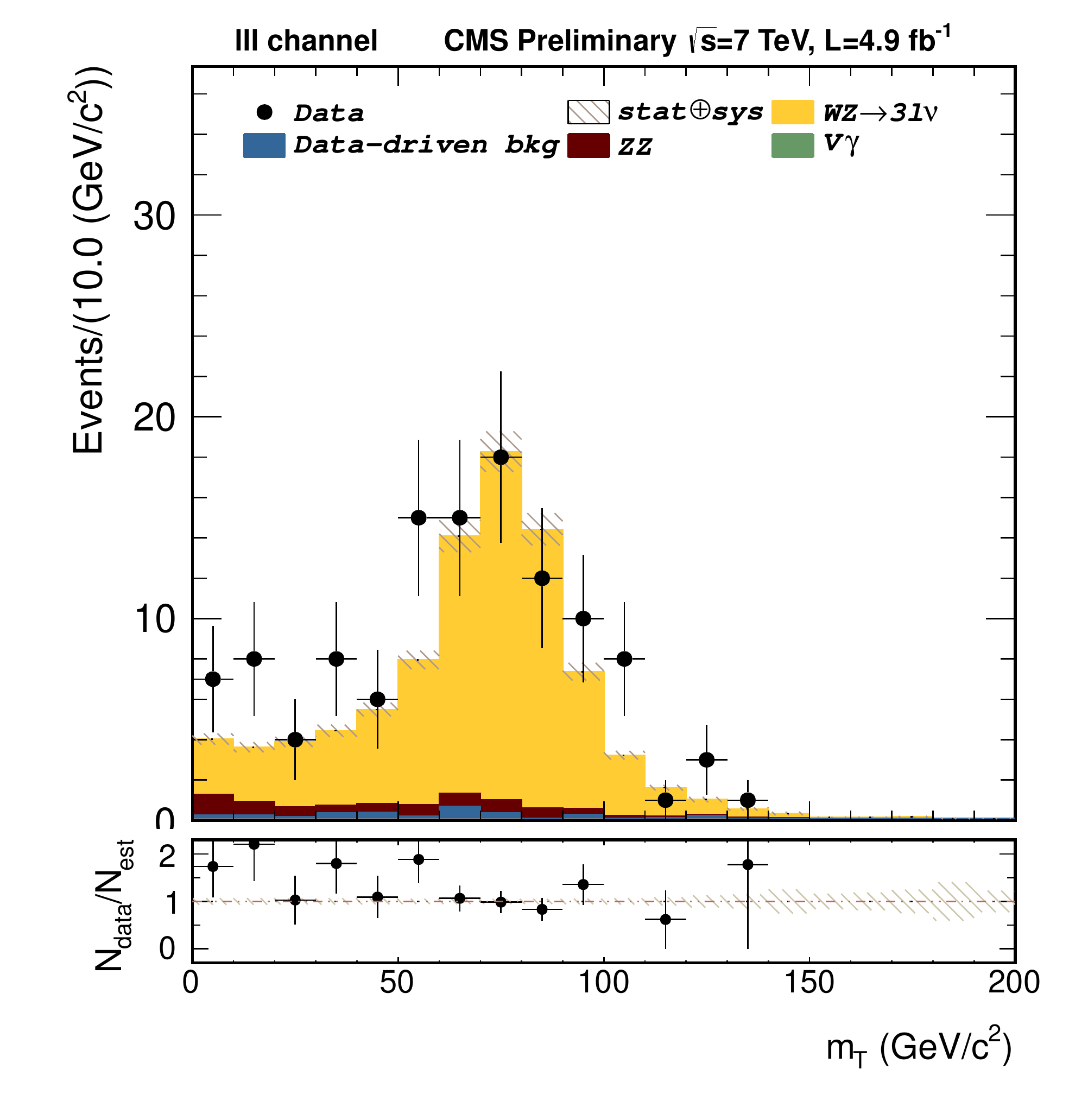}
		\caption{Negatively charged \W-candidate lepton}
	\end{subfigure}
	\caption[Transverse mass of the \W-lepton candidate and \MET for 2011 analysis]{Transverse 
		mass distribution built with the \W-lepton candidate and the \MET
		once the \W-candidate requirement have been applied in the selection. The MC samples
		are normalised to the 2011 integrated 
		luminosity, $\lumi_{int}=4.9~\fbinv$.}\label{ch9:fig:MT}
\end{figure}
\begin{figure}[htbp]
	\begin{subfigure}[b]{0.45\textwidth}
		\includegraphics[width=0.9\textwidth]{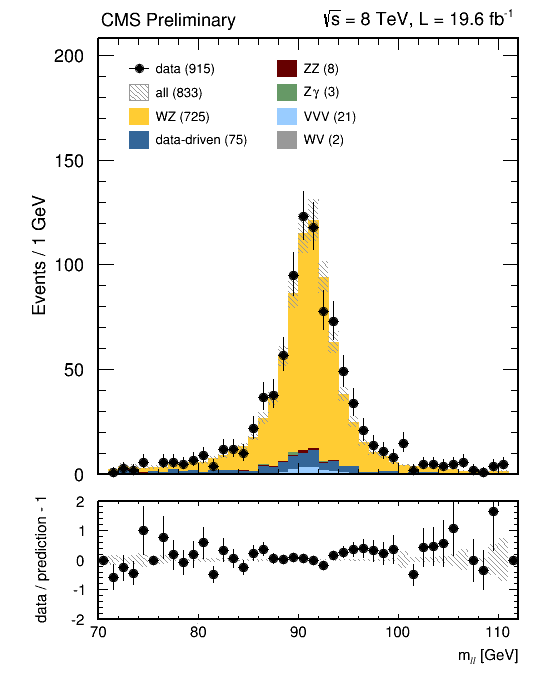}
		\caption{Invariant mass of the \Z-system for the \wzp signal selection}
	\end{subfigure}\quad
	\begin{subfigure}[b]{0.45\textwidth}
		\includegraphics[width=0.9\textwidth]{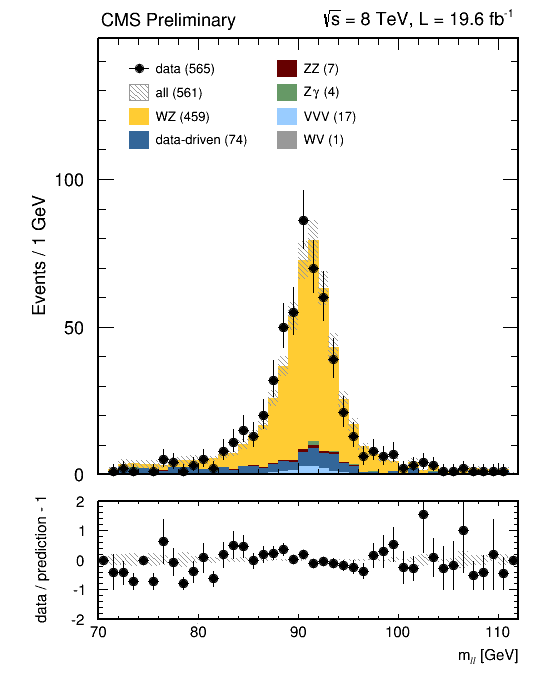}
		\caption{Invariant mass of the \Z-system for the \wzm signal selection}
	\end{subfigure}
	\vskip 1em
	\begin{subfigure}[b]{0.45\textwidth}
		\includegraphics[width=0.9\textwidth]{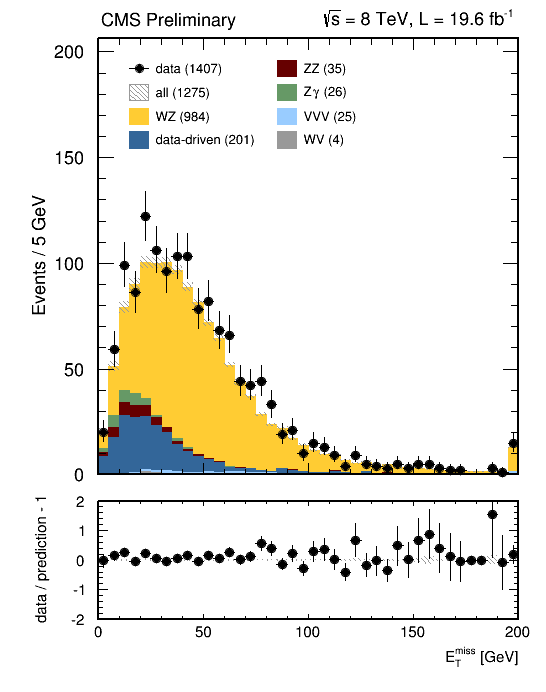}
		\caption{\MET distribution for the \wzp signal selection, before the \MET cut 
		is applied}
	\end{subfigure}\quad
	\begin{subfigure}[b]{0.45\textwidth}
		\includegraphics[width=0.9\textwidth]{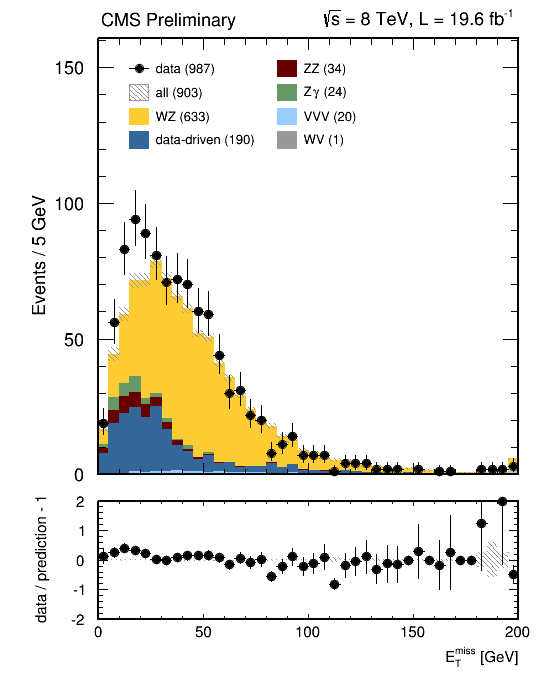}
		\caption{\MET distribution for the \wzm signal selection, before the \MET cut
		is applied}
	\end{subfigure}
	\caption[Several final distributions for 2012 ratio analysis]{Distributions once the full
		steps of the analysis cuts have been performed. The MC samples are normalised to
		the 2012 integrated luminosity, $\lumi_{int}=19.6~\fbinv$.}\label{ch9:fig:severalfigs}
\end{figure}

The observables distributions for both signals are equivalent to the inclusive analysis up to 
\emph{\W-candidate selection} stage when the specific charge for the \W-lepton candidate is
required. Thus, only distributions after the \emph{\W-candidate} stage is required are shown. 
Figures~\ref{ch9:fig:wleptonpt} show the transverse momentum of the \W-lepton candidate
for the \wzp and \wzm once the requirement is applied, just to control the possible differences in
\pt spectra between both opposite-charged lepton. The \pt spectra should be a little bit harder for
the positively charged lepton, as it will be explained in Section~\ref{ch9:sec:ratiomeas}. Moreover,
Figures~\ref{ch9:fig:MT} and Figures~\ref{ch9:fig:severalfigs} summarise some of the main control 
distributions in both signal regions after all the analysis steps have been performed, showing a 
relative good agreement between the experimental data and the data-driven and \gls{mc} predictions.

The Appendix~\ref{app:distributions} contains a bunch of detailed distributions split by measured 
channel, both for 7 and 8~\TeV, at each stage of the selection.

\section{Cross sections ratio measurements}\label{ch9:sec:ratiomeas}
The same technique applied to measured the inclusive cross section 
(Section~\ref{ch8:sec:xsestimation}) is exploited to measure the ratio of the \wzm over \wzp
cross sections. Equation~\eqref{ch8:eq:xscorrected} applied to both \wzp and \wzm is divided to
 obtain,
\begin{equation}
	\frac{\sigma_{\wzm}}{\sigma_{\wzp}}=\frac{N_S^-}{N_S^+}\frac{\mathcal{C}^+}{\mathcal{C}^-}
		\frac{\rho^+}{\rho^-}
	\label{ch9:eq:xsratio}
\end{equation}
Notice that the luminosity term is cancelled but the acceptance and efficiency in simulation, 
$\mathcal{C}$, and efficiency correction, $\rho$, are kept in the equation. 

Keeping the $\mathcal{C}$ term in the above equation is mandatory because of the expected 
topological differences between the \wzp and \wzm production. The \wzp process trends to be
produced more boosted than the \wzm due to the energy distribution of the quarks inside the 
proton. In particular, the up and anti-down quark pairs (\wzp production) have been measured 
to carry the major fraction of the proton energy with higher probability than the down and 
anti-up pairs (\wzm production), as it can be observed, for instance, in the \gls{pdf} fitted by the 
\gls{mstw8}\glsadd{ind:mstw8} group shown in Figure~\ref{ch1:fig:mstwPDF} at Chapter~\ref{ch1}. Therefore, \wzp 
decay products are expected to fall outside the pseudorapidity acceptance of the detector with more
probability than the \wzm decay products and, consequently, the acceptance for \wzp process is 
expected to be slightly lower than the \wzm.

By contrast, the correction efficiency factor $\rho$, applied to the simulated data in order to
compensate the potentially overestimated efficiencies\footnote{Because of the impossibility of 
reproduce with total accuracy a real detector.} of the simulated detector, are not expected to
be excessively charge dependent. In fact, the lepton and trigger efficiencies ratio of negative 
to positive leptons, $\varepsilon^-/\varepsilon^+$, have been measured in several \W charge 
asymmetry analysis at 7~\TeV for both electrons~\cite{CMS-PAS-SMP-12-001} 
and muons~\cite{CMS-PAS-EWK-11-005}, all of them reporting an efficiency ratio compatible with 
unity within the statistical uncertainty. As the same behaviour is reproduced in the simulated
samples, the ratio of positively to negatively charge leptons \glspl{sf} are also compatible with 
unity. Nevertheless, as the efficiency cancellation is valid in small pseudorapidity and \pt regions; 
taking a conservative approach, the $\rho$ terms are kept in equation~\eqref{ch9:eq:xsratio} and
the efficiencies uncertainties estimated in the \W charge asymmetry analyses referenced before,
3\% for electrons and 2.3\% for muons, are propagated as a systematic source of uncertainty due to 
the \glspl{sf} differences between opposite charged leptons. 

The two cross sections \wzp and \wzm measurements were performed, analogously to the inclusive
measurement, in four exhaustive final state regions, to obtain the ingredients for the ratio
Equation~\eqref{ch9:eq:xsratio}. Table~\ref{ch9:tab:ratioresults11} and~\ref{ch9:tab:ratioresults12}
shows the number of observed signal ratio, the acceptance and efficiency ratios and the efficiency
correction ratio between \wzm and \wzp used to obtain the cross section ratios reported. As 
expected, the tables show a slightly higher acceptance for the \wzm process with respect
the \wzp production, between 5-8\% depending of the measured channel. In contrast, the $\rho$ terms
are compatible with unity between the two processes, meaning that this analysis is insensitive to 
the potential differences in the reconstruction, isolation or trigger efficiency between different 
charged lepton objects.
\begin{table}[!htpb]
	\centering
	\begin{tabular}{lcccc}\hline\hline
		&   $\mathcal{C}^+/\mathcal{C}^-$ & $\rho^+/\rho^-$ & $N^-_{S}/N_S^+$ 
			& $\sigma_{\wzm}/\sigma_{\wzp}$\\\hline
	$eee$ &  $0.926\pm0.018$ & $1.00\pm0.02$ & $0.76\pm0.21$ &  
		$0.71\pm0.19_{\text{stat}}\pm0.02_{\text{sys}}$\\
	$\mu ee$ & $0.941\pm0.017$ & $1.01\pm0.03$ & $0.52\pm0.15$ &
		$0.50\pm0.15_{\text{stat}}\pm0.01_{\text{sys}}$\\
	$e\mu\mu$& $0.945\pm0.015$ & $1.00\pm0.02$ & $0.43\pm0.12$ &
		$0.40\pm0.12_{\text{stat}}\pm0.01_{\text{sys}}$\\
        $\mu\mu\mu$& $0.945\pm0.014$ & $1.01\pm0.02$ & $0.87\pm0.19$ & 
		$0.83\pm0.18_{\text{stat}}\pm0.02_{\text{sys}}$\\\hline
	\end{tabular}
	\caption[Cross section ratios by measured channel for 2011 data]{Measured ratios
	between \wzm and \wzp for the acceptance, efficiencies and number of signal at 7~\TeV. 
	The last column contains the cross section ratio measured in the four considered channels. 
	The errors are split in statistical and systematic origin. The errors reported for the other
	acceptance, efficiencies and signal terms are statistical 
	only.}\label{ch9:tab:ratioresults11}
\end{table}

\begin{table}[!htpb]
	\centering
	\begin{tabular}{lcccc}\hline\hline
		&   $\mathcal{C}^+/\mathcal{C}^-$ & $\rho^+/\rho^-$ & $N^-_{S}/N_S^+$ 
			& $\sigma_{\wzm}/\sigma_{\wzp}$\\\hline
	$eee$ &  $0.939\pm0.022$ & $1.01\pm0.04$ & $0.67\pm0.09$ &  
		$0.63\pm0.09_{\text{stat}}\pm0.01_{\text{sys}}$\\
	$\mu ee$ & $0.948\pm0.021$ & $1.00\pm0.03$ & $0.56\pm0.07$ &
		$0.53\pm0.08_{\text{stat}}\pm0.02_{\text{sys}}$\\
	$e\mu\mu$& $0.940\pm0.019$ & $1.00\pm0.03$ & $0.52\pm0.06$ &
		$0.49\pm0.06_{\text{stat}}\pm0.01_{\text{sys}}$\\
        $\mu\mu\mu$& $0.956\pm0.016$ & $1.00\pm0.02$ & $0.58\pm0.06$ & 
		$0.55\pm0.06_{\text{stat}}\pm0.01_{\text{sys}}$\\\hline
	\end{tabular}
	\caption[Cross section ratios by measured channel for 2012 data]{Measured ratios
	between \wzm and \wzp for the acceptance, efficiencies and number of signal at 8~\TeV. 
	The last column contains the cross section ratio measured in the four considered channels.
	The errors are split in statistical and systematic origin. The errors reported for the other
	acceptance, efficiencies and signal terms are statistical 
	only.}\label{ch9:tab:ratioresults12}
\end{table}

The four measurements are combined with the \gls{blue}\glsadd{ind:blue} method described at 
Section~\ref{ch8:sec:xscombined}, where the same error correlation as the inclusive cross section
measurement was used along with the uncertainty originated from possible differences in lepton
efficiencies because of the lepton charge, $\varepsilon^-/\varepsilon^+$, mentioned in the previous
section. The charge ratio efficiency uncertainty is assumed to be fully correlated between the 
channels with the same flavour of the lepton \W-candidate,~\ie $eee$ and $e\mu\mu$ on one hand, and $\mu\mu\mu$ 
and $\mu ee$ on the other. The obtained error matrix for the 7~\TeV analysis is
\begin{equation}
	\mathbf{E}_{7\TeV}=
	\begin{pmatrix}
		0.0372    & <10^{-4} &  0.0003   & <10^{-4} \\
		<10^{-4}  & 0.02125  &  <10^{-4} & 0.0002 \\
		0.0003    & <10^{-4} &  0.01382  & <10^{-4}\\
		<10^{-4}  & 0.0002   &  <10^{-4} & 0.0333  
	\end{pmatrix}
\end{equation}
providing a weighting factors of $\alpha_{3e}=0.151$, $\alpha_{1\mu2e}=0.268$, $\alpha_{1e2\mu}=0.411$ 
and ${\alpha_{3\mu}=0.170}$. It is worthwhile to mention that due to the correlation of the 
systematic uncertainties in both \wzp and \wzm cross sections, the systematic uncertainty of the 
cross section ratio is highly reduced, remaining as main source of systematic uncertainty the ratio 
efficiency introduced. The \wzm, \wzp cross section ratio in the phase space 
defined in $M_{ll}\in 91.1876\pm20 \GeV/c^{2}$ at 7~TeV is measured to be:
\begin{equation}
	\left(\frac{\sigma_{\wzm}}{\sigma_{\wzp}}\right)_{7~\TeV}= 0.547\pm0.075_{\text{stat}}\pm0.011_{\text{sys}}
\end{equation}
in agreement with the \gls{nlo} prediction $0.563^{+0.002}_{-0.001}$, calculated with the \gls{pdf} 
set \gls{mstw8}\glsadd{ind:mstw8}, reported at Chapter~\ref{ch2} (Table~\ref{ch2:tab:xspredmassrange}). The 
arguments rising a differences in acceptance between the \wzp and \wzm production discussed 
before are applicable to argue why the ratio observable is expected to be remarkably more sensitive
to the \gls{pdf}-set choice than the cross section observables. Therefore, it has been used another
\gls{pdf} set, the CT-10~\cite{Lai:2010vv}\glsadd{ind:ct10}, with the \MCFM program to obtain a \gls{nlo} prediction
of the ratio observable. The Table~\ref{ch9:tab:thNLO11} summarises the 7~\TeV \gls{nlo} predictions 
obtained with different \gls{pdf}-sets to be compared with the measured value, and it may be noted
that the prediction obtained with the CT-10 \gls{pdf} set are in excellent agreement with the
measured value within uncertainty errors.
\begin{table}[h]
	\centering
	\begin{tabular}{cclr}\hline\hline
		PDF-set & $\sigma_{\wzm}/\sigma_{\wzp}$ (NLO)&$\sigma_{\wzp}/\sigma_{\wzm}$ (NLO) & $N_{\sigma}$ \\\hline\\[-2ex]
		MSTW08  & $0.563^{+0.002}_{-0.001}$          &$1.776^{+0.006}_{-0.003}$           & 0.21 \\[1ex]
		CT-10   & $0.546^{+0.002}_{-0.001}$          &$1.832^{+0.007}_{-0.003}$           & 0.01 \\\hline
	\end{tabular}
	\caption[Standard model ratio predictions for 7~\TeV]{NLO prediction for the 
	$\sigma_{\wzm}/\sigma_{\wzp}$ and the inverse $\sigma_{\wzp}/\sigma_{\wzm}$ ratios obtained 
	with \MCFM using different PDF sets reported in the first column. The predictions are obtained
	for 7~\TeV centre of mass energy and compared in the last column with the 
	$\sigma_{\wzm}/\sigma_{\wzp}$ measured value $0.547\pm0.075_{\text{stat}}\pm0.011_{\text{sys}}$ 
	or $1.83\pm0.25_{\text{stat}}\pm0.04_{\text{sys}}$ for the inverse ratio, reporting the number 
	of standard deviation (defined at Chapter~\ref{ch7}, 
	Equation~\ref{ch7:eq:numberofsigma}).}\label{ch9:tab:thNLO11}
\end{table}

For completeness, the inverse ratio is also shown and may be compared with the \gls{nlo} prediction
from Table~\ref{ch9:tab:thNLO11}
\begin{equation}
	\left(\frac{\sigma_{\wzp}}{\sigma_{\wzm}}\right)_{7~\TeV}= 1.83\pm0.25_{\text{stat}}\pm0.04_{\text{sys}}
\end{equation}

\paragraph*{}
The 8~\TeV centre of mass energy analysis is performed analogously. The obtained error matrix is
\begin{equation}
	\mathbf{E}_{8TeV}=
	\begin{pmatrix}
		0.0176    & <10^{-4} &  0.0002   & <10^{-4} \\
		<10^{-4}  & 0.0234   &  <10^{-4} & 0.0001   \\
		0.0002    & <10^{-4} &  0.0219   & <10^{-4} \\
		<10^{-4}  & 0.0001   & <10^{-4}  & 0.0113   \\
	\end{pmatrix}
\end{equation}
The ratio in the 8~TeV centre of mass energy analysis is measured to be:
\begin{equation}
	\left(\frac{\sigma_{\wzm}}{\sigma_{\wzp}}\right)_{8~\TeV}= 0.551\pm0.035_{\text{stat}}\pm0.010_{\text{sys}}\,,
\end{equation}
and the inverse ratio
\begin{equation}
	\left(\frac{\sigma_{\wzp}}{\sigma_{\wzm}}\right)_{8~\TeV}= 1.81\pm0.12_{\text{stat}}\pm0.03_{\text{sys}}
\end{equation}

The measured values are in good agreement, within the uncertainty errors, with the \gls{sm} 
predictions reported at Table~\ref{ch9:tab:thNLO12}. As it may be observed in the table, the 
\gls{nlo} prediction calculated with the \MCFM tool using the CT-10 \gls{pdf} set is, as in the
7~\TeV case, in better agreement with the measured value.
\begin{table}[h]
	\centering
	\begin{tabular}{lccr}\hline\hline
		PDF-set & $\sigma_{\wzm}/\sigma_{\wzp}$ (NLO)&$\sigma_{\wzp}/\sigma_{\wzm}$ (NLO) & $N_{\sigma}$ \\\hline\\[-2ex]
		MSTW08  & $0.580\pm0.001$             &$1.724\pm0.003$           & 0.82 \\[1ex]
		CT-10   & $0.563\pm0.001$             &$1.777\pm0.003$           & 0.33 \\\hline
	\end{tabular}
	\caption[Standard model ratio predictions for 8~\TeV]{NLO prediction for the 
	$\sigma_{\wzm}/\sigma_{\wzp}$ and the inverse $\sigma_{\wzp}/\sigma_{\wzm}$ ratios obtained 
	with \MCFM using different PDF sets reported in the first column. The predictions are obtained
	for 8~\TeV centre of mass energy and compared in the last column with the 
	$\sigma_{\wzm}/\sigma_{\wzp}$ measured value $0.551\pm0.035_{\text{stat}}\pm0.010_{\text{sys}}$ 
	or $1.81\pm0.12_{\text{stat}}\pm0.03_{\text{sys}}$ for the inverse ratio, reporting the number 
	of standard deviation (defined at Chapter~\ref{ch7}, 
	Equation~\ref{ch7:eq:numberofsigma}).}\label{ch9:tab:thNLO12}
\end{table}

\chapter{Conclusions}\label{ch10}

The \wzm and \wzp productions from proton-proton collisions have been studied in two centre of mass
energies 7 and 8~\TeV; and, in particular, the inclusive cross section measurement of the \WZ 
production $\sigma(pp\to\WZ+X)$ and the ratio between both processes 
$\sigma(pp\to\wzm+X)/\sigma(pp\to\wzp+X)$ have been performed. The measurements are based in 
data acquired with the \gls{cms} experiment, resulting from proton-proton collisions produced at 
the \gls{lhc}. The total amount of data used for the 7~\TeV analysis is equivalent to 
{\lumi=4.9~\fbinv}, whilst for the 8~\TeV analysis is {\lumi=19.6~\fbinv}. 

The final state particles, used to select the \WZ 
candidates from collision events, are composed of three well-identified, high-\pt and isolated
leptons in addition to substantial \MET. The selected samples of \WZ candidate events are 
compared to the estimation of the background processes, either simulated with \gls{mc}\glsadd{ind:mc} techniques 
or estimated from experimental data. The estimated signal along with the detector acceptance and 
efficiency for identifying the signal events as determined from simulation are included to obtain 
the cross section of the considered process. For the cross section ratio measurement, such
methodology is applied to both signal samples defined by the charge of the \W-candidate lepton and
the obtained cross sections are divided to obtain the ratio. All the measurements are performed
individually for each of the four leptonic final states, $eee$, $\mu ee$, $e\mu\mu$ and 
$\mu\mu\mu$, and the final results are obtained from a best fit linear combination, giving the 
results reported in Table~\ref{ch10:tab:results}. The 7~\TeV analysis is statistically limited, 
however, the 8~\TeV analysis is greatly benefited from the amount of recorded data allowing to 
measure with uncertainty errors dominated by the systematic sources considered, in particular
the luminosity uncertainty error is the same size than the total of the other systematic 
uncertainty errors. The improvement of the luminosity uncertainty error, which was released during
the redaction of this thesis work~\cite{CMS-PAS-LUM-13-001}, will improve the precision of the
measurement.
\begin{table}[!htpb]
	\centering
	\resizebox{\textwidth}{!}
	{
	\begin{tabular}{rll}\hline\hline
	                      &  7~\TeV  (\lumi=4.9~\fbinv)  &  8~\TeV  (\lumi=19.6~\fbinv) \\\hline
	$\sigma(pp\to\WZ+X)$  & $20.8\pm1.3_{\text{stat}}\pm1.1_{\text{sys}}\pm0.5_{\text{lumi}}$ &
		               $24.6\pm0.8_{\text{stat}}\pm1.1_{\text{sys}}\pm1.1_{\text{lumi}}$ \\
$\frac{\sigma(pp\to\wzm+X)}{\sigma(pp\to\wzp+X)}$  & $0.547\pm0.075_{\text{stat}}\pm0.011_{\text{sys}}$ &
			       $0.551\pm0.035_{\text{stat}}\pm0.010_{\text{sys}}$  \\\hline
	\end{tabular}
	}
	\caption[Summary of measurements]{Results obtained for the measurements performed in this
		thesis work.}\label{ch10:tab:results}
\end{table}

The inclusive cross section have been measured to be a slightly higher value than the 
\gls{sm}-\gls{nlo}\glsadd{ind:nlo} predictions, although the number of standard deviations are 1.8 for 7~\TeV and
1.5 for 8~\TeV. In the case of the cross sections ratio, both for 7 and 8~\TeV, are in excellent 
agreement with the \gls{nlo}\glsadd{ind:nlo}. 

\section{Analysis prospects}
The data recorded in \gls{cms} from proton-proton collisions at \comene=7~\TeV have allowed to 
measure the associate \WZ production cross section and, for the first time, the cross sections
ratio between \wzm and \wzp, with results dominated by statistical errors. The amount of collected
data with proton-proton collisions at \comene=8~\TeV, substantially increased with respect to the
2011 data, has allowed to measure with more precision the cross section and the cross sections
ratio, reaching almost the same sensitivity of the \gls{sm} theoretical predictions. Improving the
systematic source of uncertainty's treatment, in particular the luminosity source which is already 
available, will allow to reach a precise measurement with lower errors than the theoretical 
predictions. 

The available data in the 8~\TeV would allow to extend the \WZ electroweak measurements to 
differential cross sections in bins of various kinematic variables, such as $\pt^Z$, which present
a more detailed comparison of theory to measurements. In addition, the presence of 
\gls{atgc}\glsadd{ind:atgc} may also be tested via the $WWZ$ vertex~\cite{Hagiwara:1986vm} and the limits 
significantly improved. Both, the differential cross sections as the \gls{atgc}\glsadd{ind:atgc}
are being calculated, during the writing of this thesis memory, in order to be included in the \gls{cms} \WZ's 
paper in preparation. This paper is based, along with contributions of other \gls{cms} \WZ team 
members, on the contents of this dissertation.

%
\appendix
   \chapter{Extended analysis distributions}\label{app:distributions}
This appendix includes the full set of distributions plots performed for the cross section
and the cross section ratio analyses for both 7~\TeV and 8~\TeV. The distributions were obtained in 
order to study and control the effect of each sequential cut introduced in the signal selection, 
allowing to compare the selected experimental data with the theoretical predictions of the signal
and the irreducible backgrounds, and the data-driven estimation of the instrumental background. 
In addition, each distribution is joined by a bottom plot showing the difference between the
observed experimental data and the estimation, normalised to the estimation. The
Monte Carlo samples used were pileup re-weighted, trigger and scaled factor weighted and normalised 
to the luminosity of the corresponding data set, which is 4.9~\fbinv for the 7~\TeV analysis and 
19.6~\fbinv for the 8~\TeV. The appendix is organised by showing the available distributions 
in each measured channel in addition to the combined channel at each stage of the selection.

\section{Cross section analysis distributions at 7~\TeV}
The distributions shown in this subsection correspond to the 2011 analysis of the inclusive
\WZ cross section using data corresponding to an integrated luminosity of 4.9~\fbinv. The 
ZZ (red in figures) and $V\gamma$ (green) processes were estimated using simulated Monte Carlo 
data, whereas the prompt-prompt-fake background (blue in figures) contribution was estimated 
using the \gls{fom} data-driven. The other instrumental background contributions were found to 
be negligible. The \WZ simulated Monte Carlo sample (yellow in figures) is also shown in order
to compare the theoretical predictions with the experimental data (black dots in figures). 
Systematic and statistical errors are also shown (grey dashed lines in figures). The distributions
are grouped by observable, showing in each figure four columns corresponding to the four measured 
channels, and each row to a stage of the analysis.
\begin{sidewaysfigure}[!htpb]
	\centering
	\begin{subfigure}[b]{0.2\textwidth}
		\includegraphics[width=\textwidth]{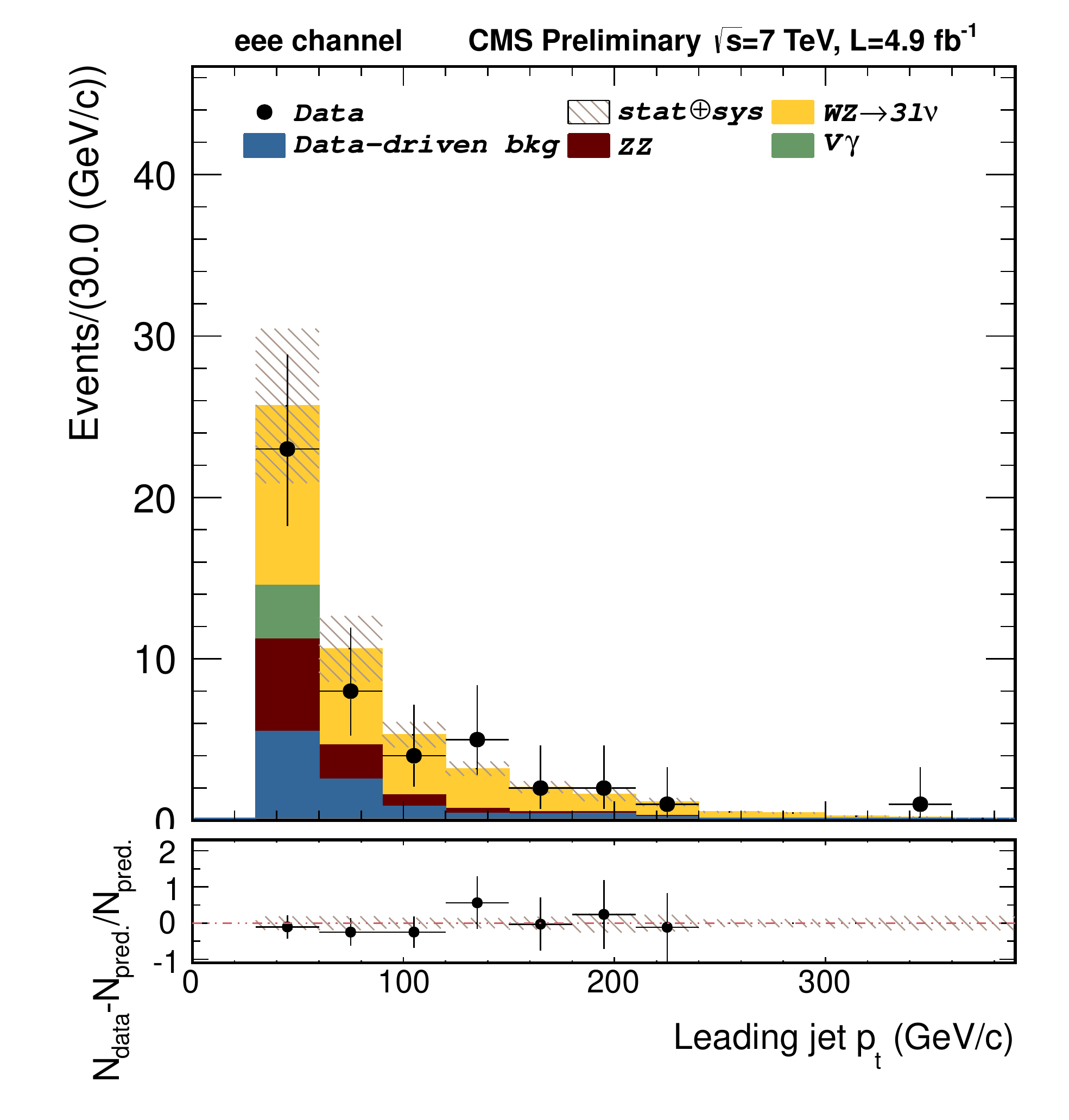}
	\end{subfigure}\quad
	\begin{subfigure}[b]{0.2\textwidth}
		\includegraphics[width=\textwidth]{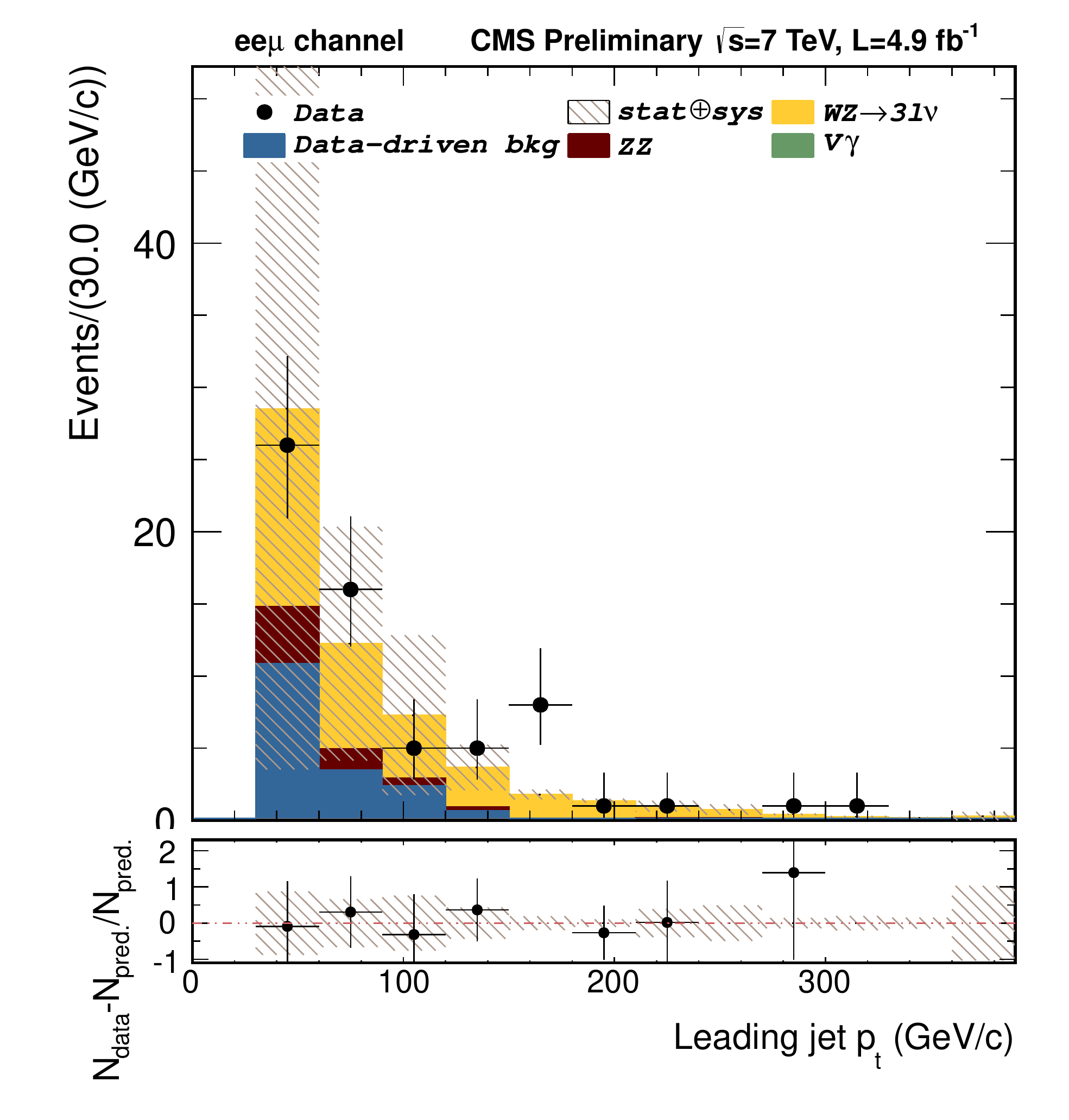}
	\end{subfigure}\quad
	\begin{subfigure}[b]{0.2\textwidth}
		\includegraphics[width=\textwidth]{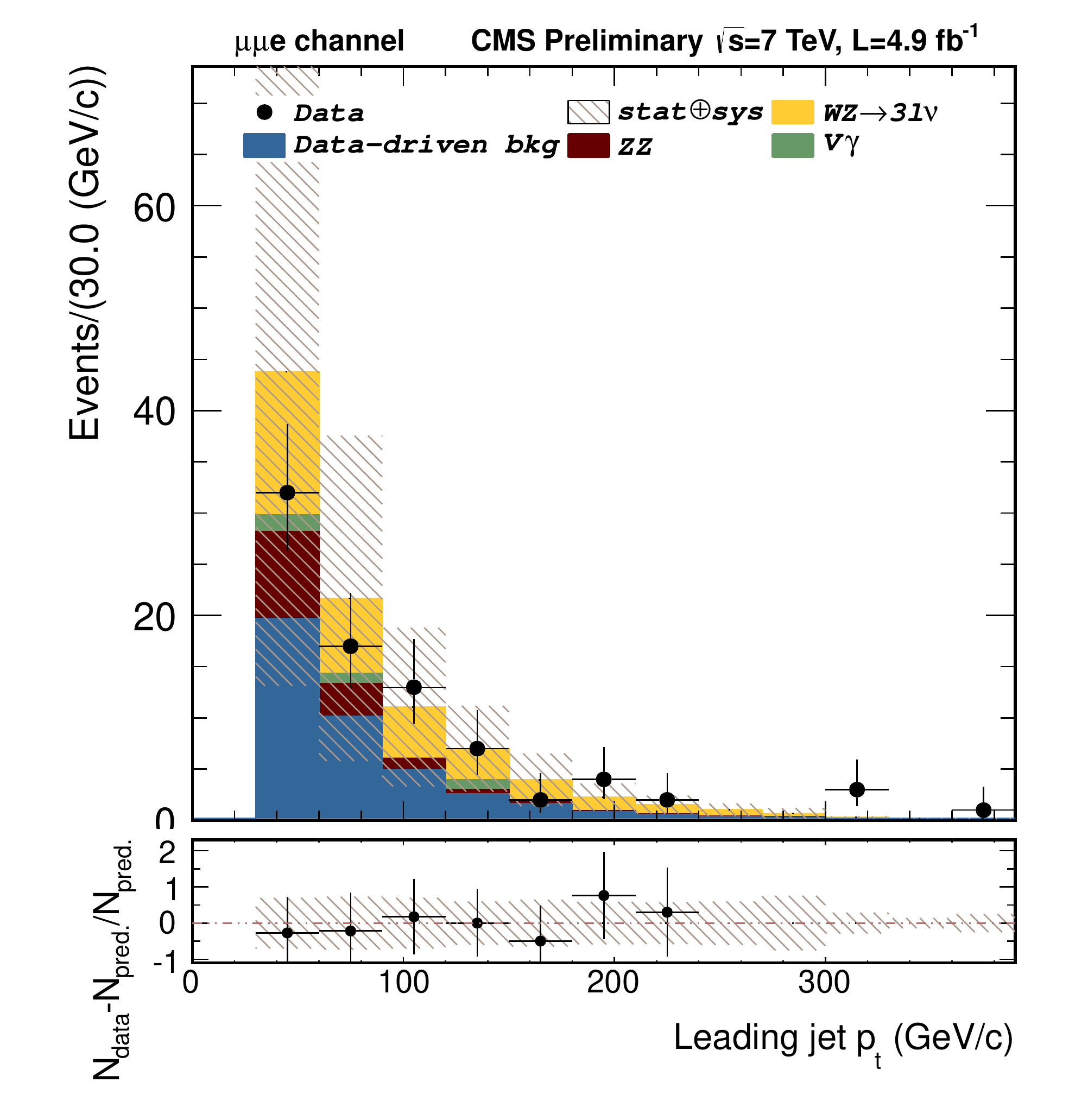}
	\end{subfigure}\quad
	\begin{subfigure}[b]{0.2\textwidth}
		\includegraphics[width=\textwidth]{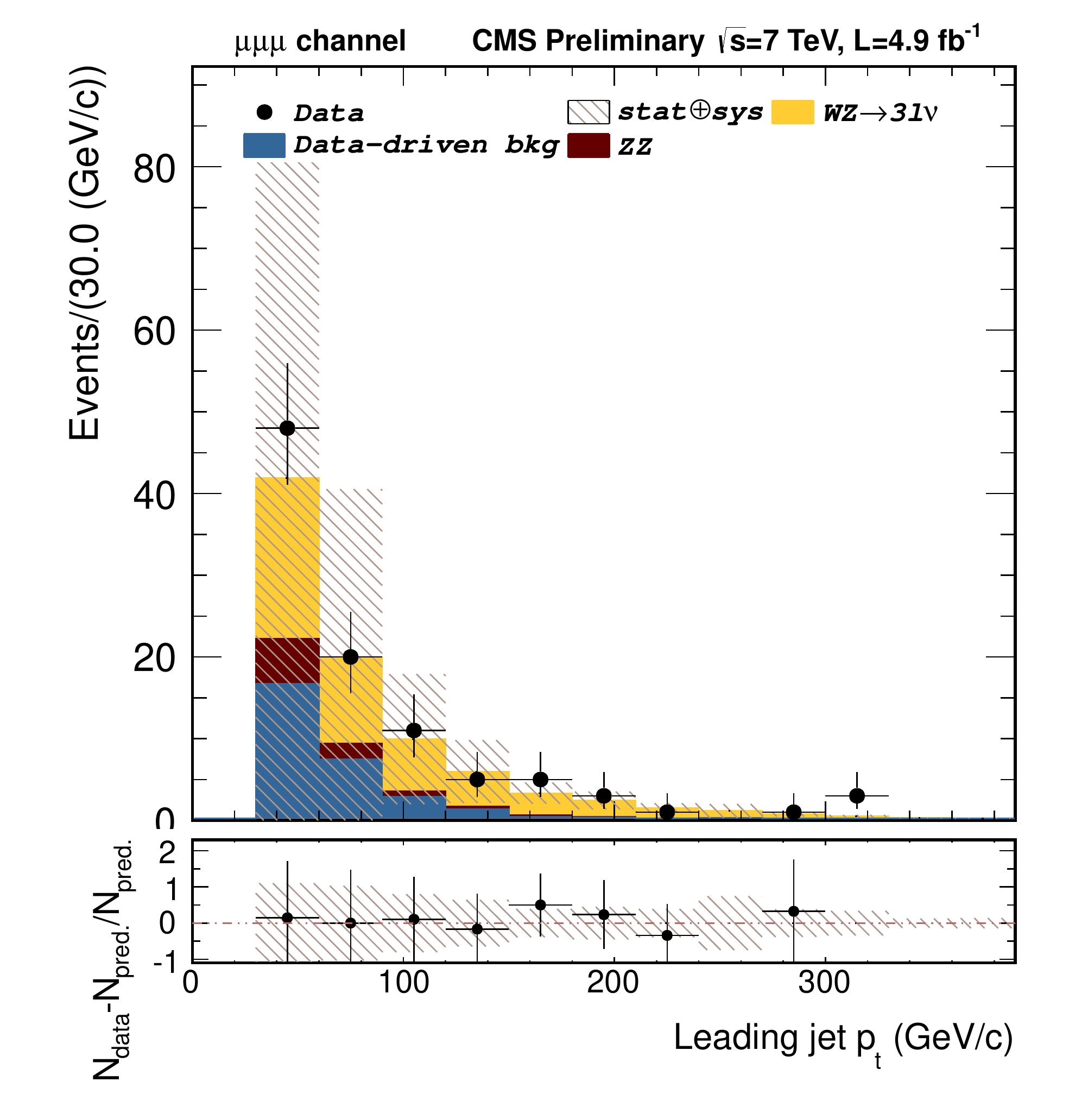}
	\end{subfigure}
	\vskip 1ex
	\centering
	\begin{subfigure}[b]{0.2\textwidth}
		\includegraphics[width=\textwidth]{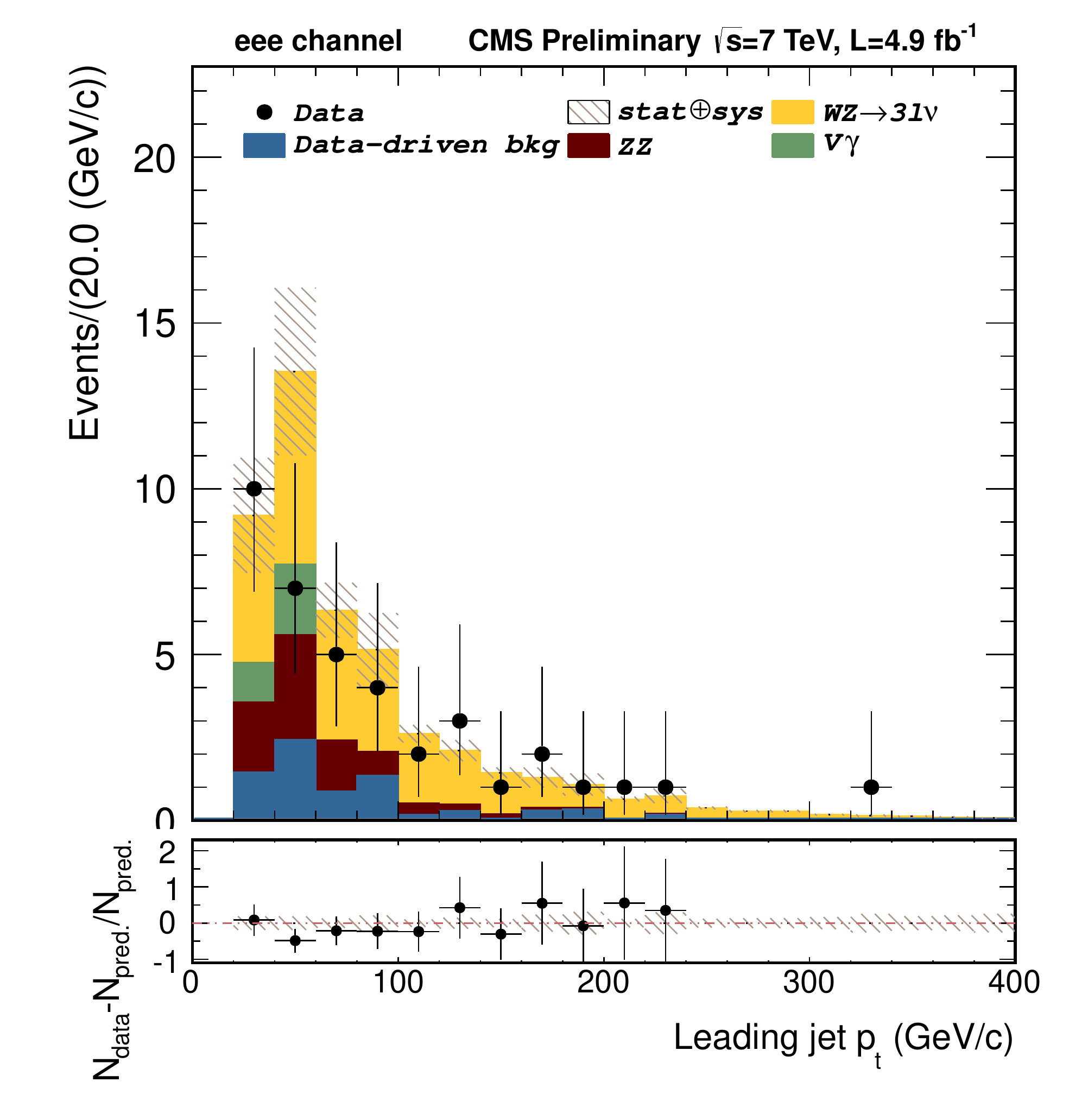}
	\end{subfigure}\quad
	\begin{subfigure}[b]{0.2\textwidth}
		\includegraphics[width=\textwidth]{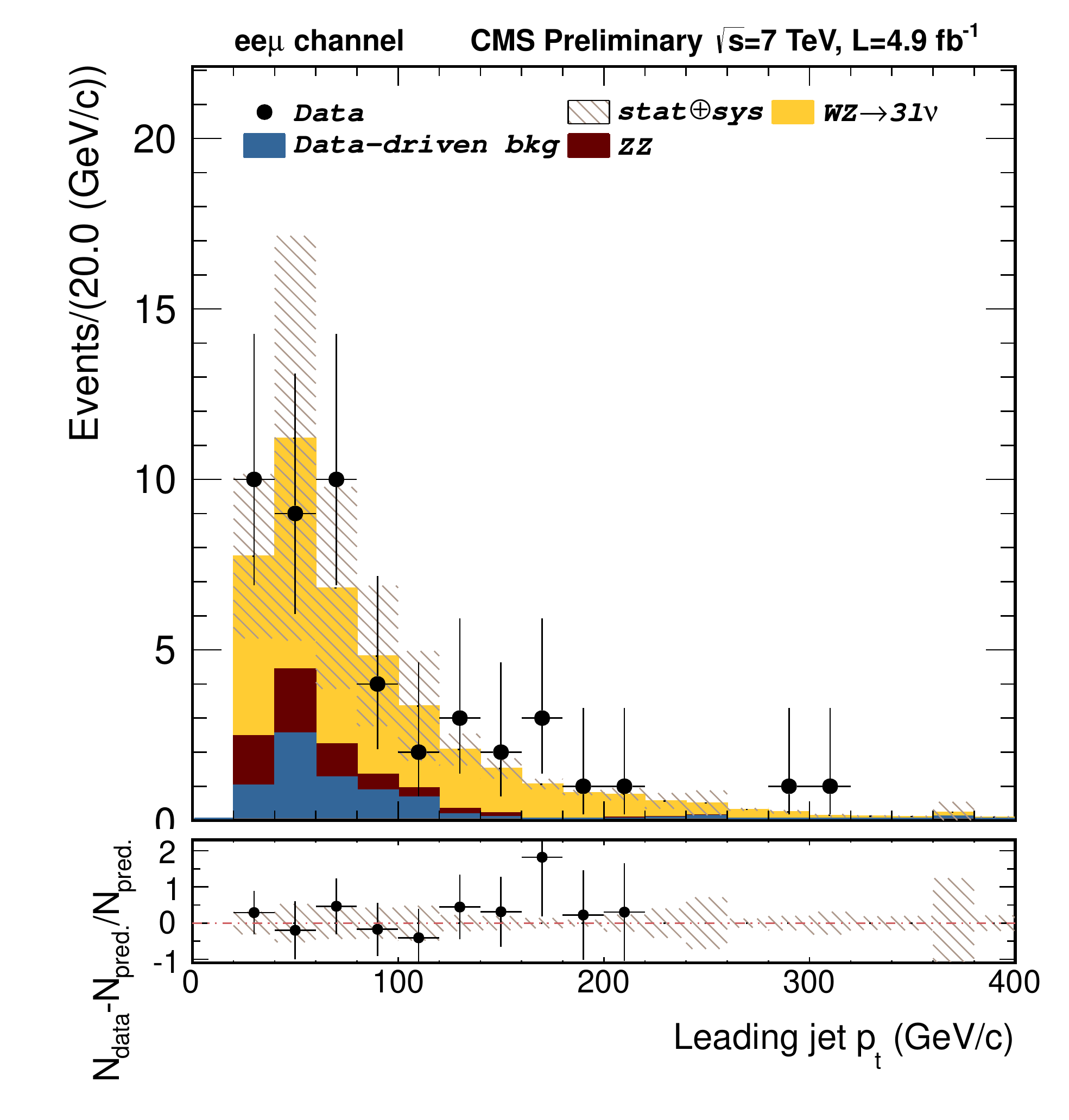}
	\end{subfigure}\quad
	\begin{subfigure}[b]{0.2\textwidth}
		\includegraphics[width=\textwidth]{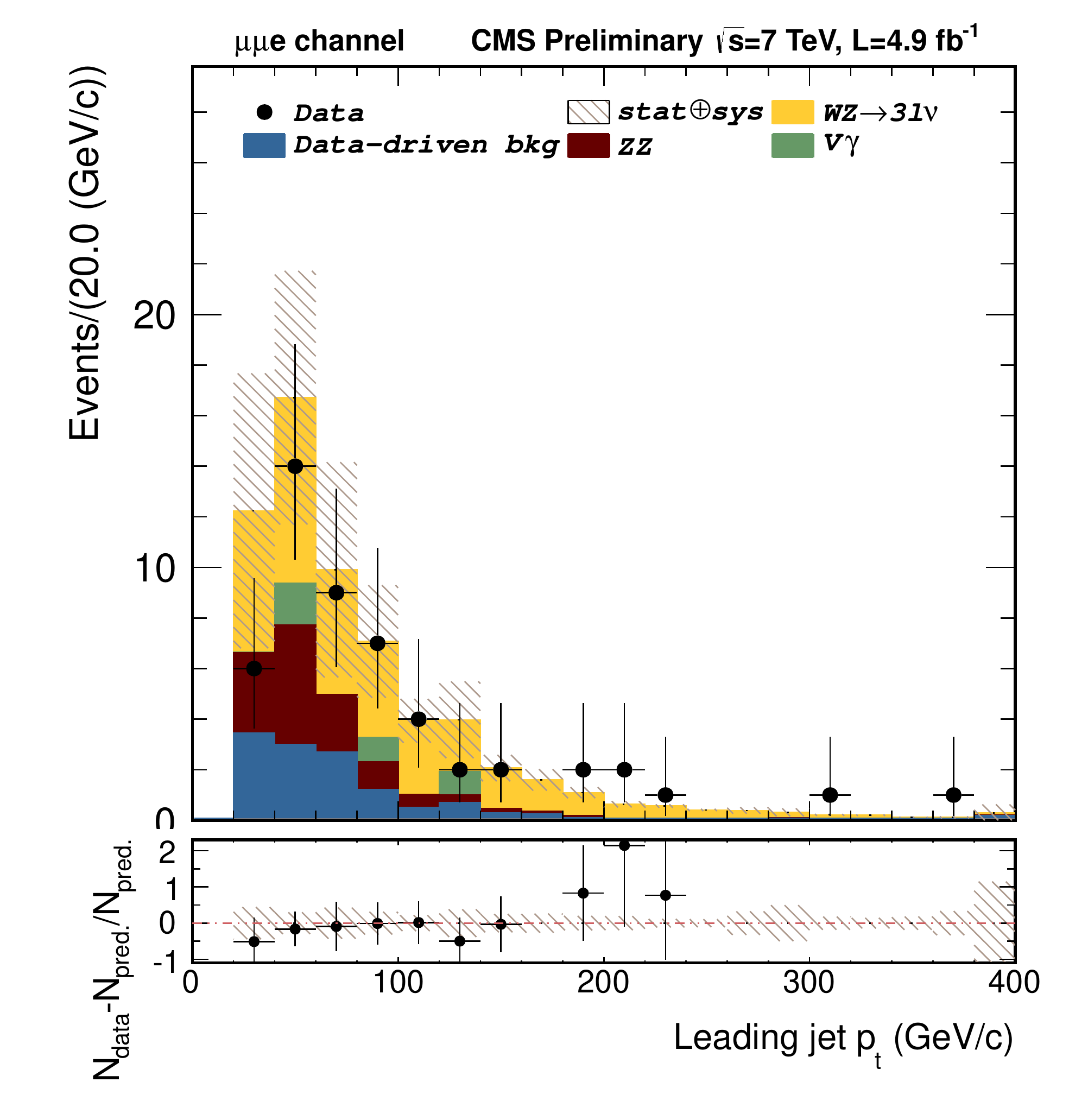}
	\end{subfigure}\quad
	\begin{subfigure}[b]{0.2\textwidth}
		\includegraphics[width=\textwidth]{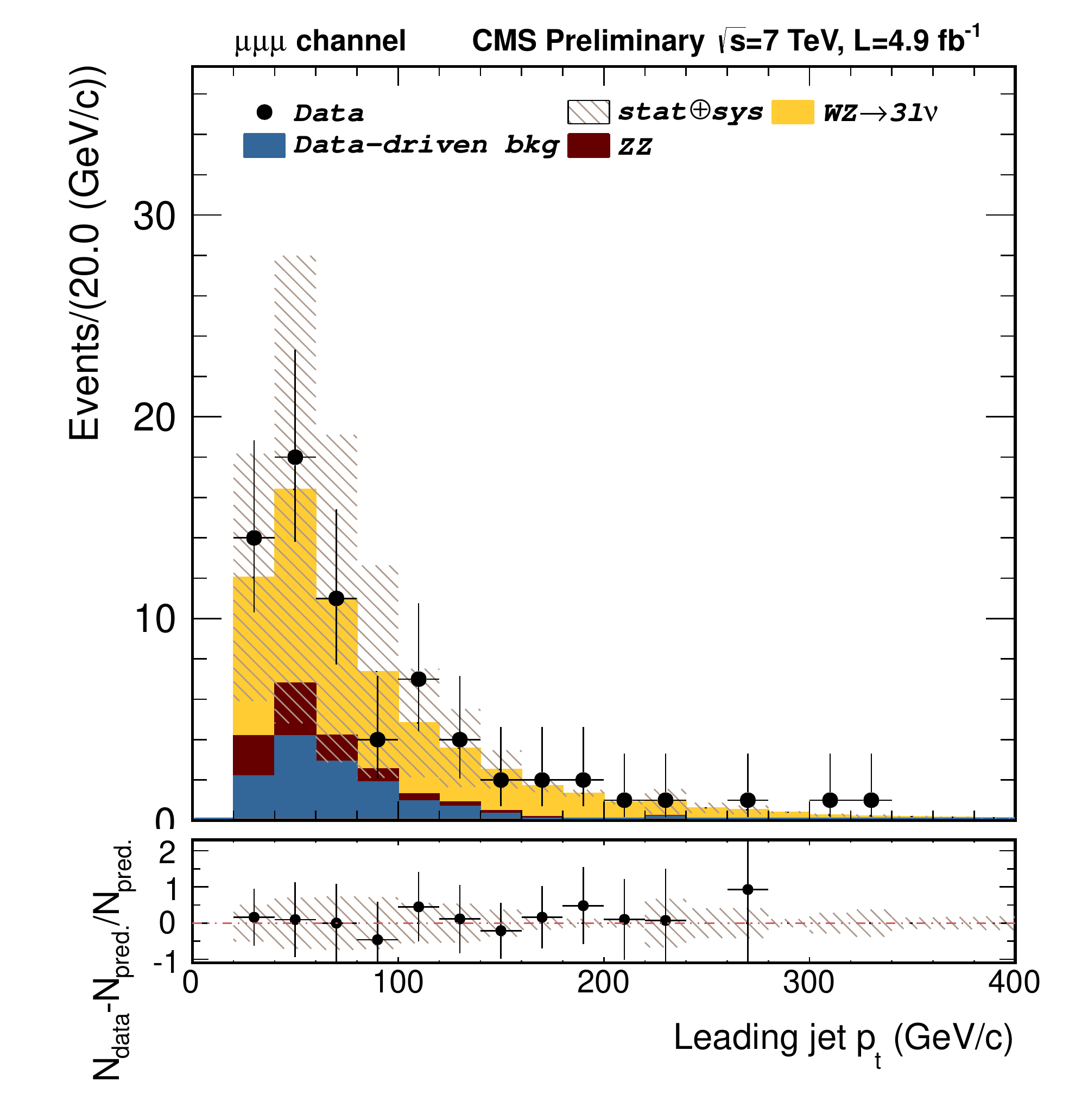}
	\end{subfigure}
	\vskip 1ex
	\begin{subfigure}[b]{0.2\textwidth}
		\includegraphics[width=\textwidth]{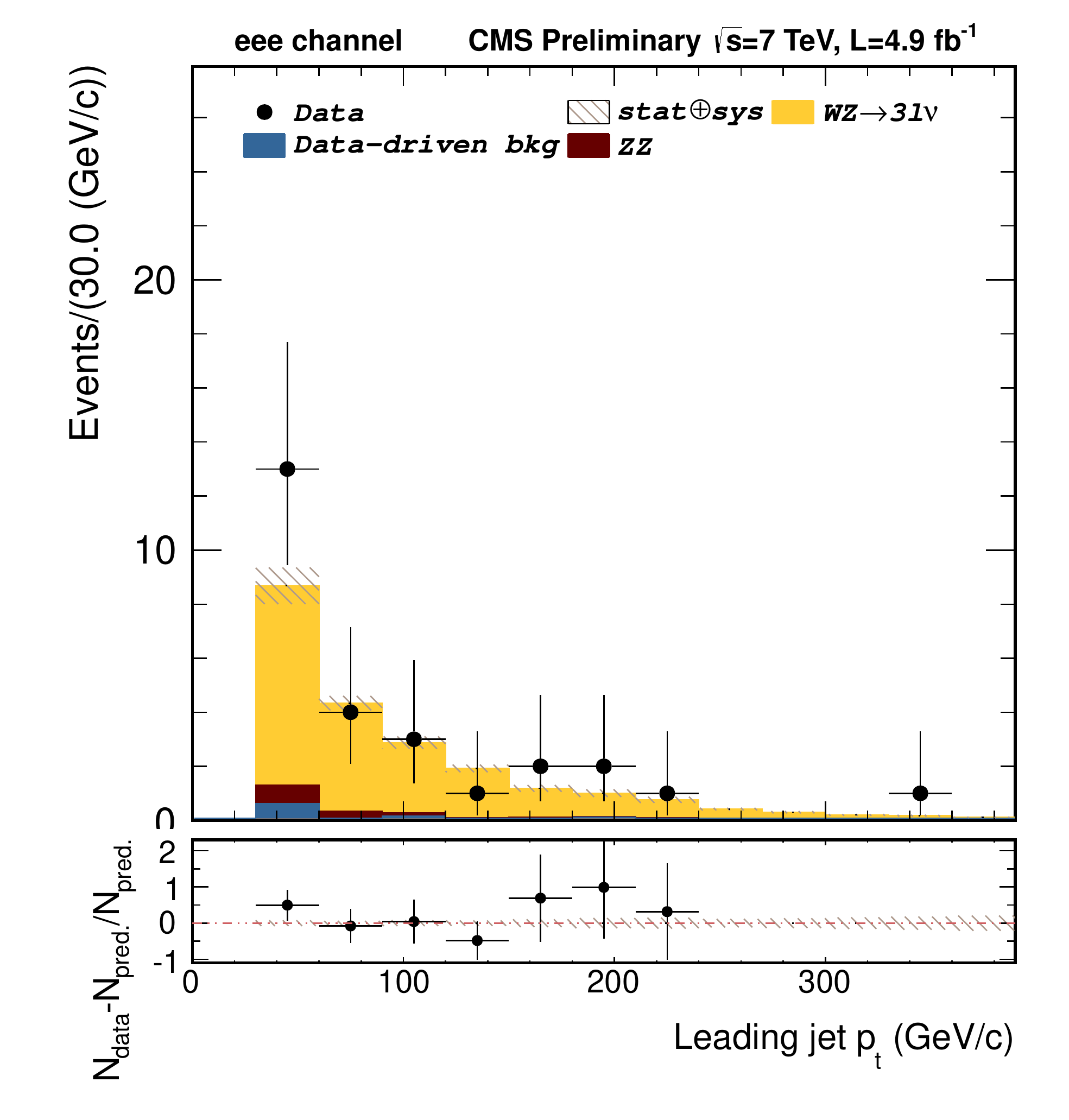}
	\end{subfigure}\quad
	\begin{subfigure}[b]{0.2\textwidth}
		\includegraphics[width=\textwidth]{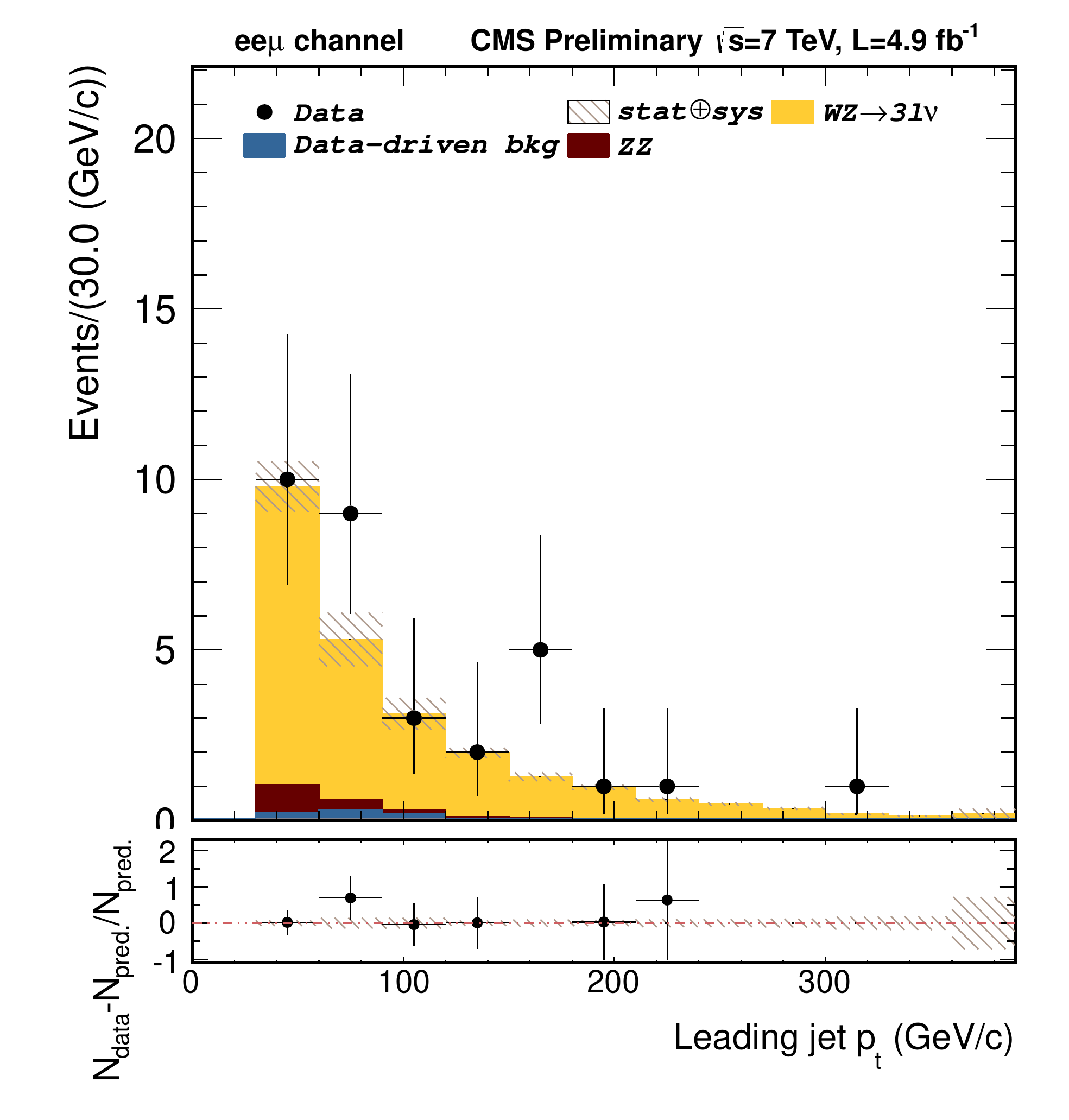}
	\end{subfigure}\quad
	\begin{subfigure}[b]{0.2\textwidth}
		\includegraphics[width=\textwidth]{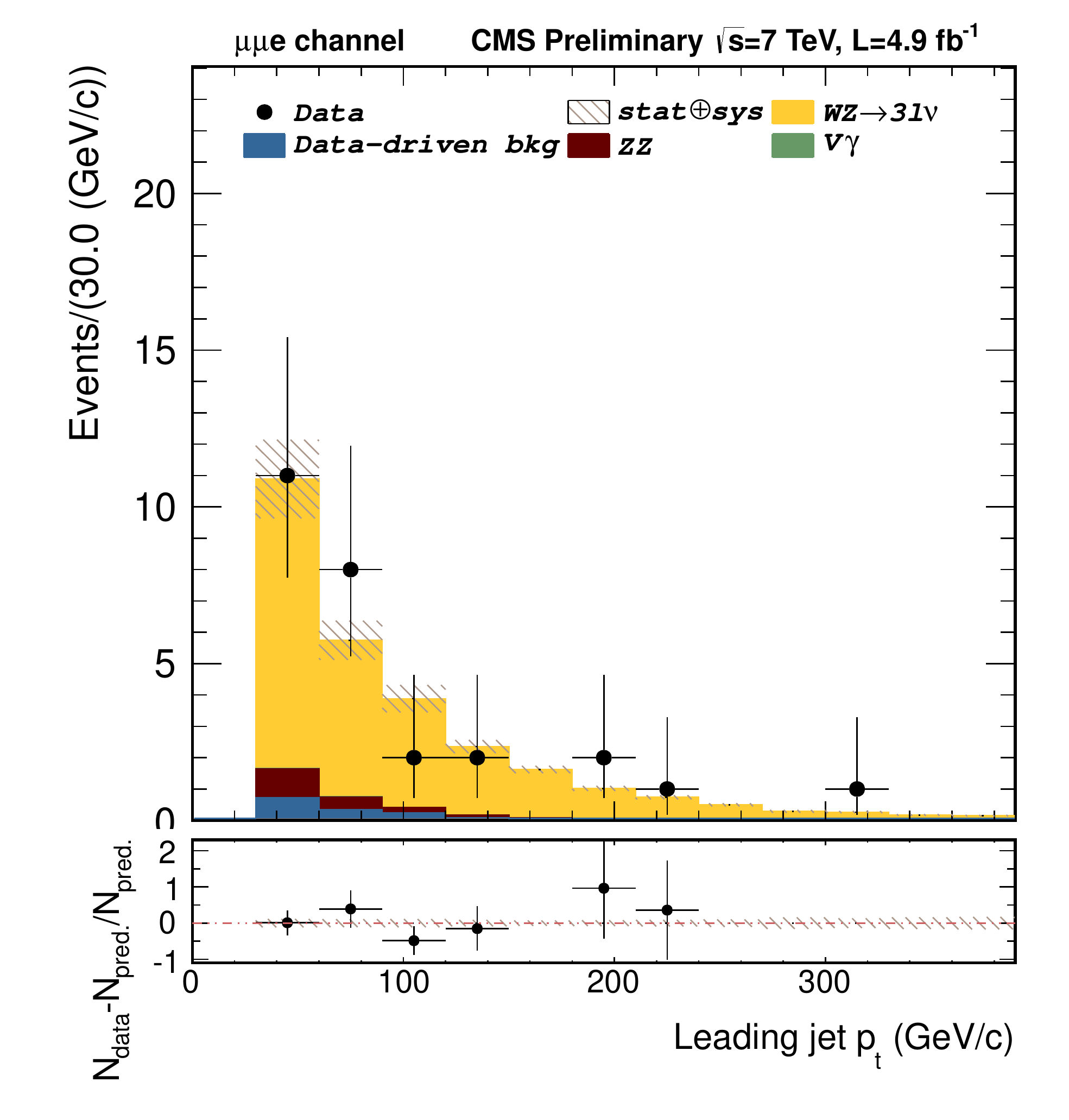}
	\end{subfigure}\quad
	\begin{subfigure}[b]{0.2\textwidth}
		\includegraphics[width=\textwidth]{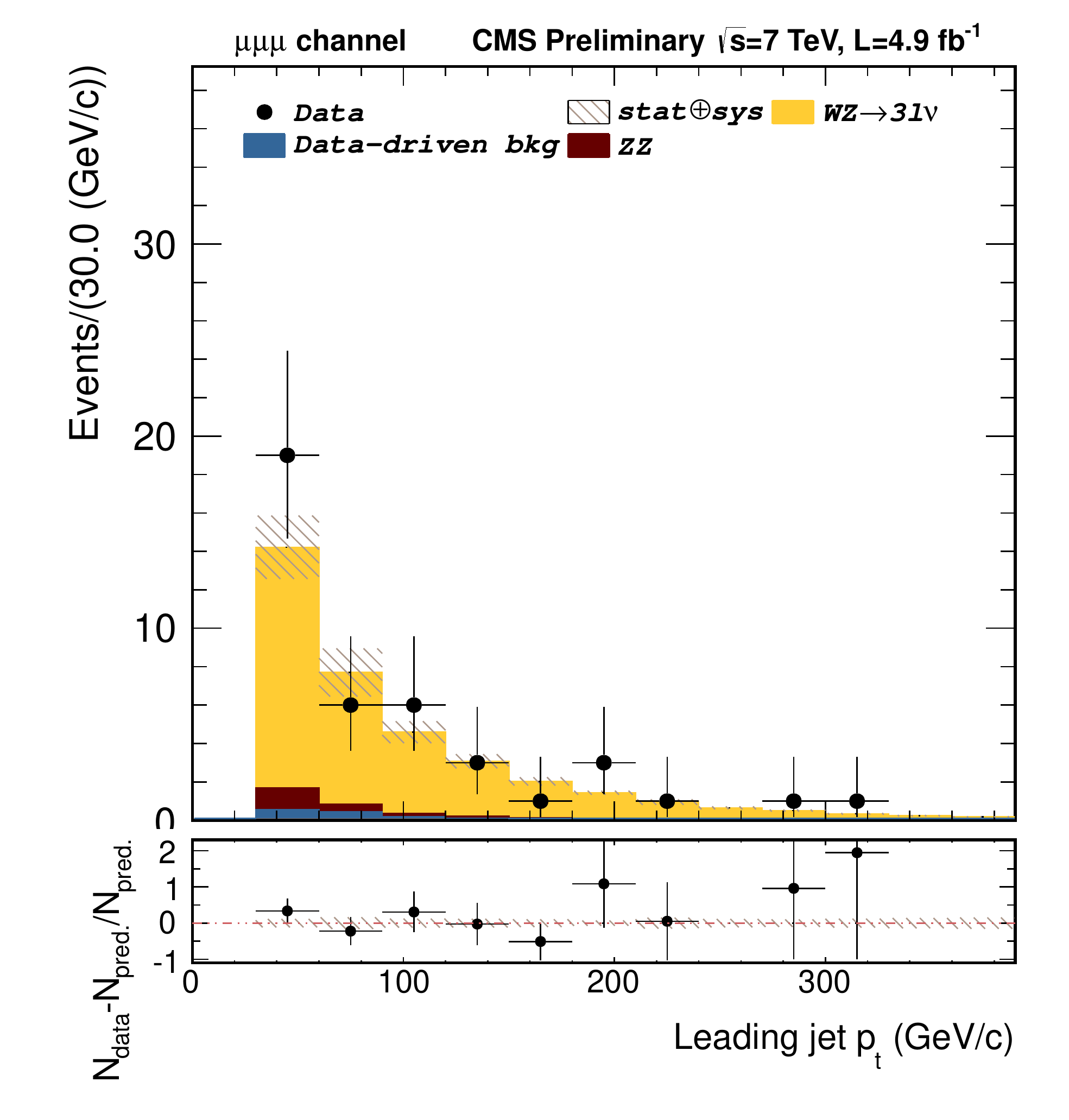}
	\end{subfigure}
	\caption[Transverse momentum of the leading jet at 7 TeV]{Transverse 
	momentum distribution of the leading jet at each event for the
	measured channels $eee$, $\mu ee$, $e\mu\mu$ and $\mu\mu\mu$ (from left to right) and
	after each analysis selection stage: after Z-candidate requirement (up row), after 
	W-candidate, without the \MET cut (middle row) and after W-candidate including \MET
	cut (bottom row).}
\end{sidewaysfigure}

\begin{sidewaysfigure}[!htpb]
	\centering
	\begin{subfigure}[b]{0.2\textwidth}
		\includegraphics[width=\textwidth]{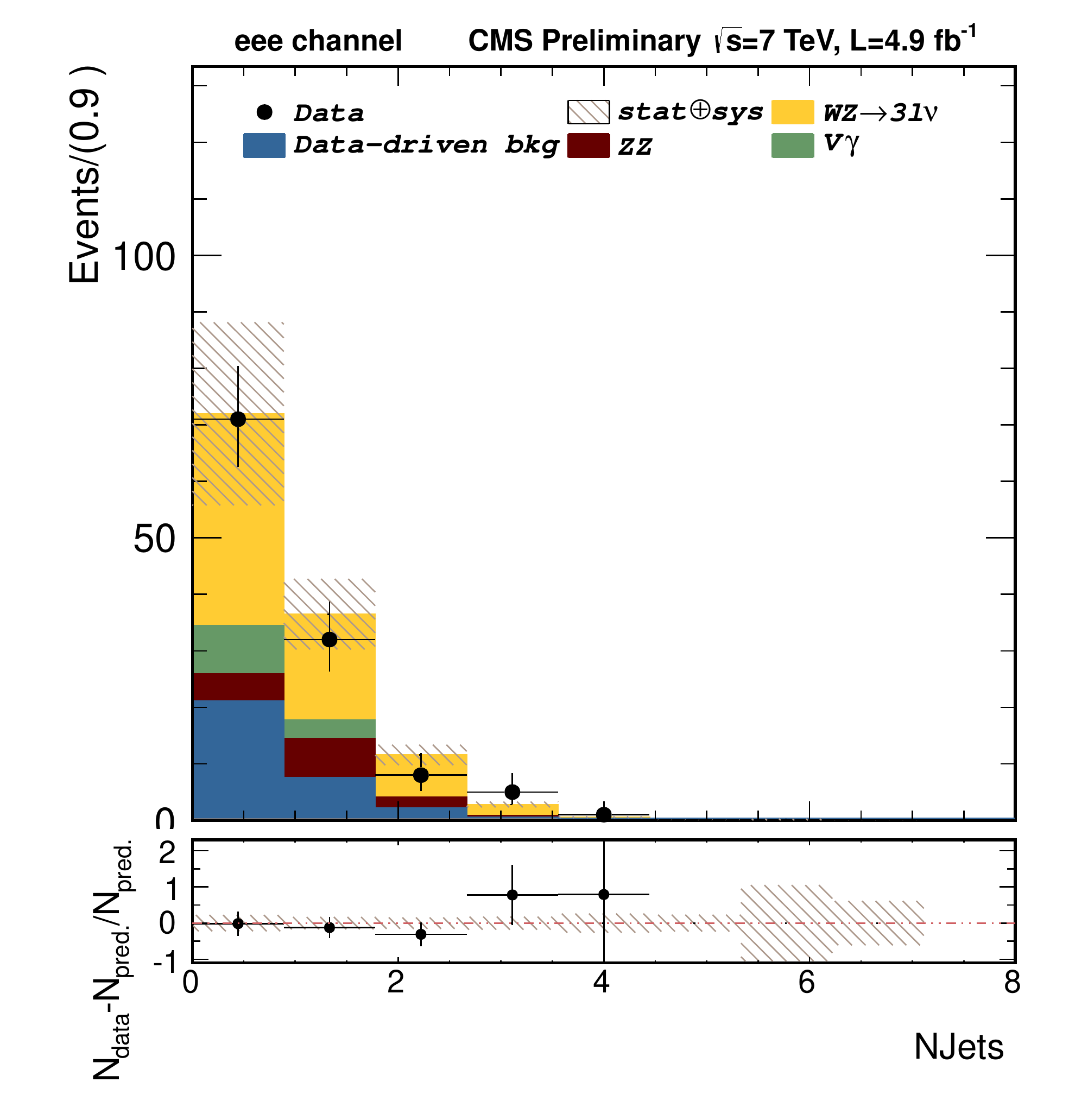}
	\end{subfigure}\quad
	\begin{subfigure}[b]{0.2\textwidth}
		\includegraphics[width=\textwidth]{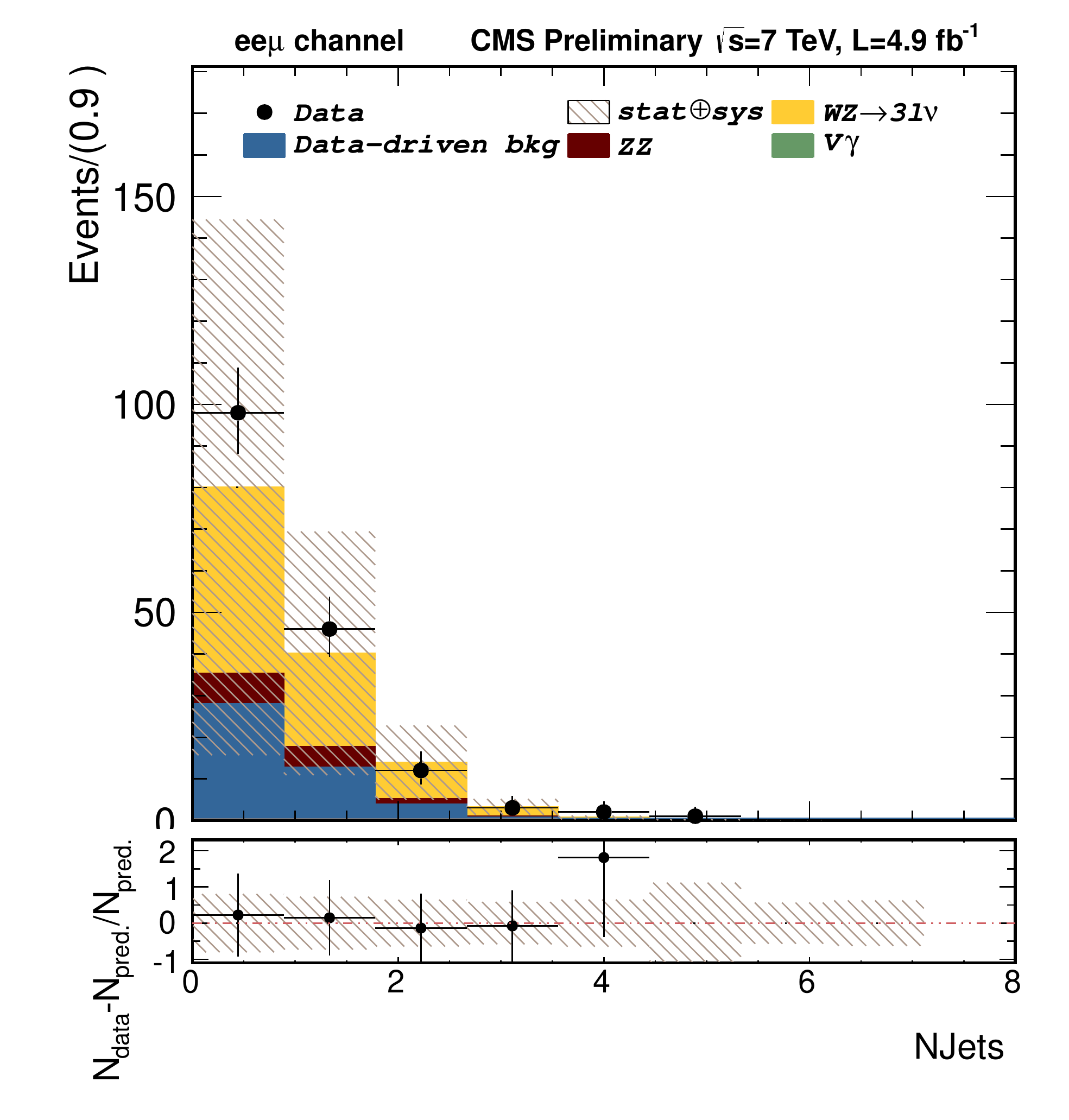}
	\end{subfigure}\quad
	\begin{subfigure}[b]{0.2\textwidth}
		\includegraphics[width=\textwidth]{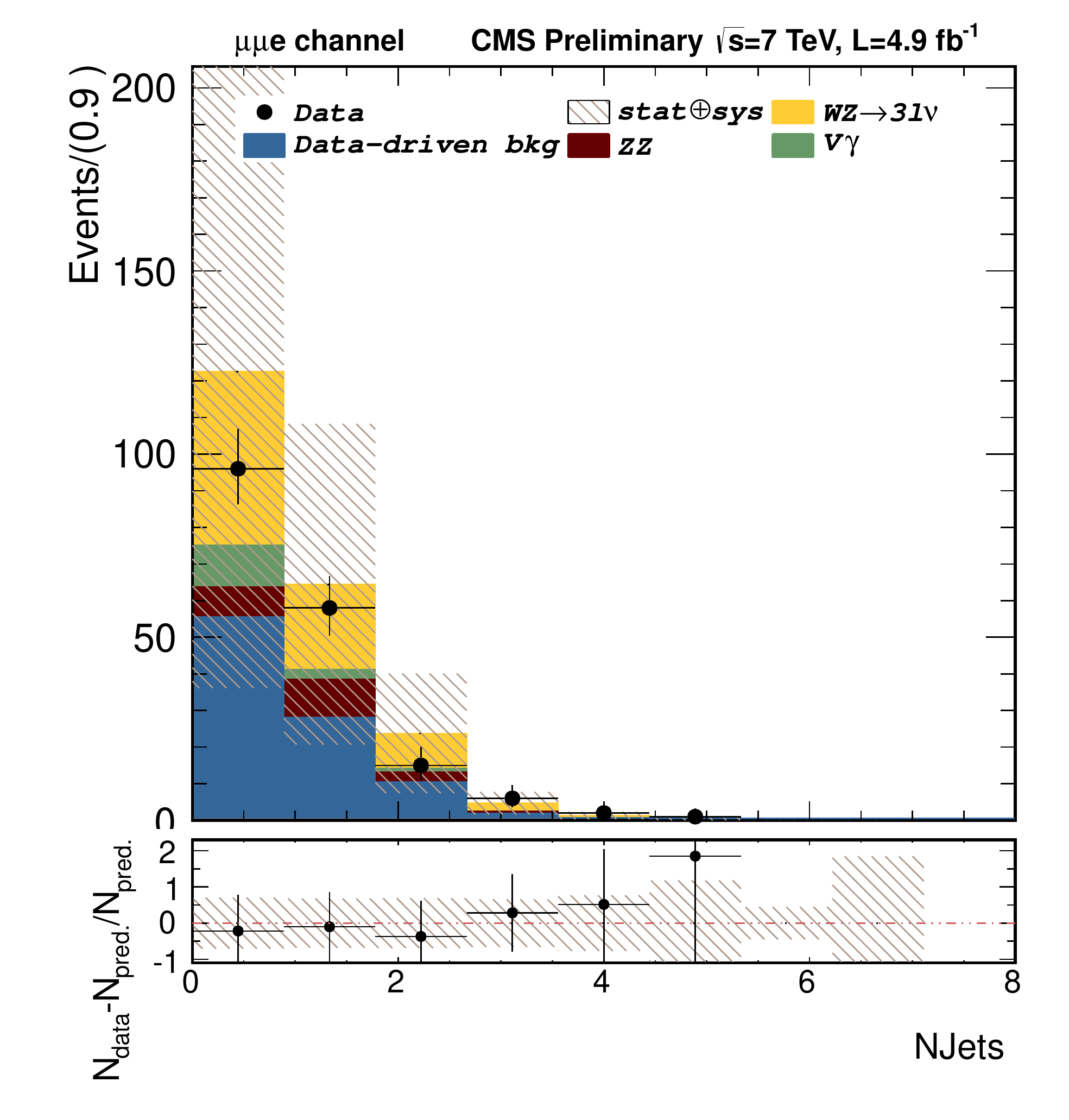}
	\end{subfigure}\quad
	\begin{subfigure}[b]{0.2\textwidth}
		\includegraphics[width=\textwidth]{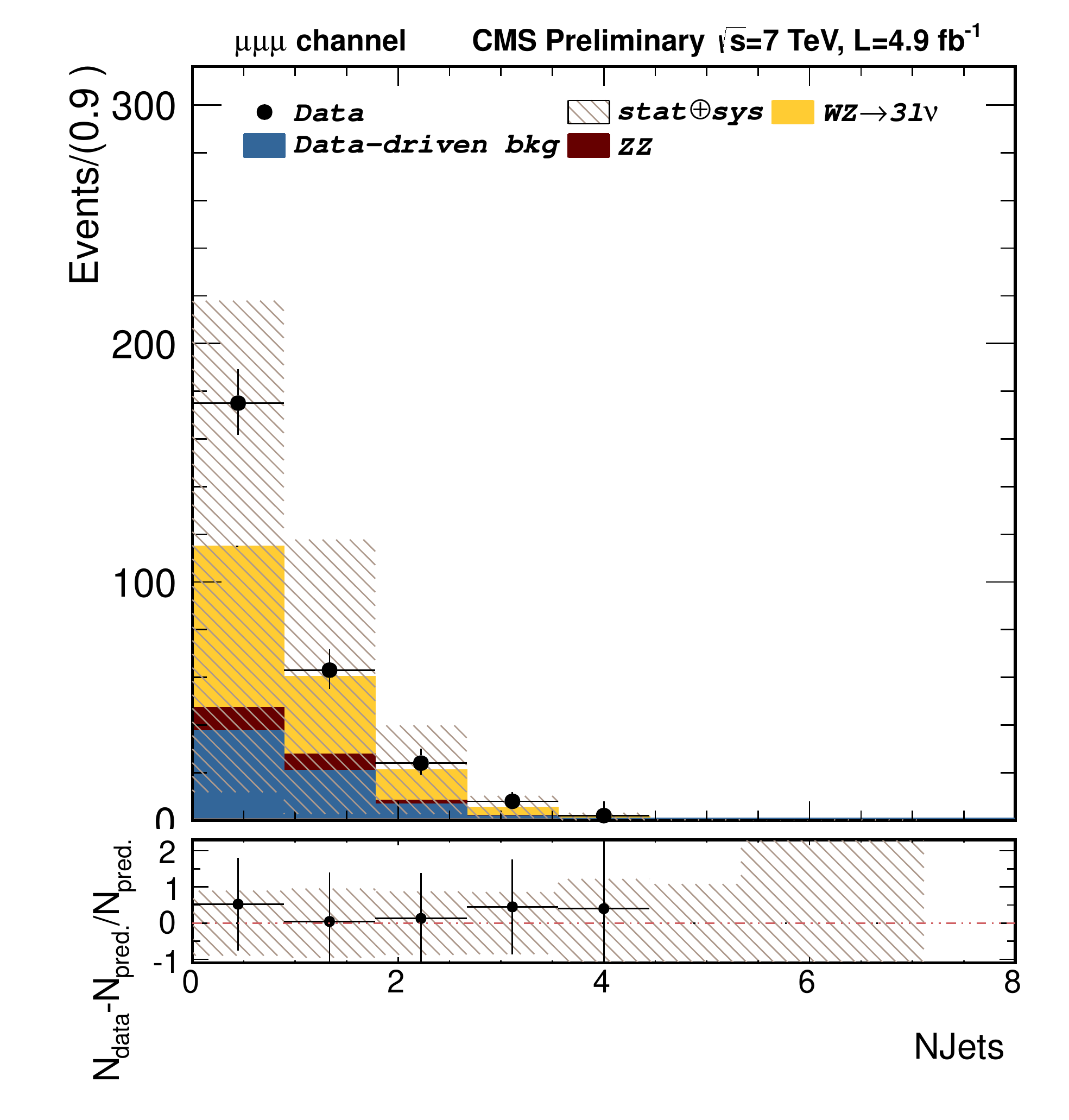}
	\end{subfigure}
	\vskip 1ex
	\centering
	\begin{subfigure}[b]{0.2\textwidth}
		\includegraphics[width=\textwidth]{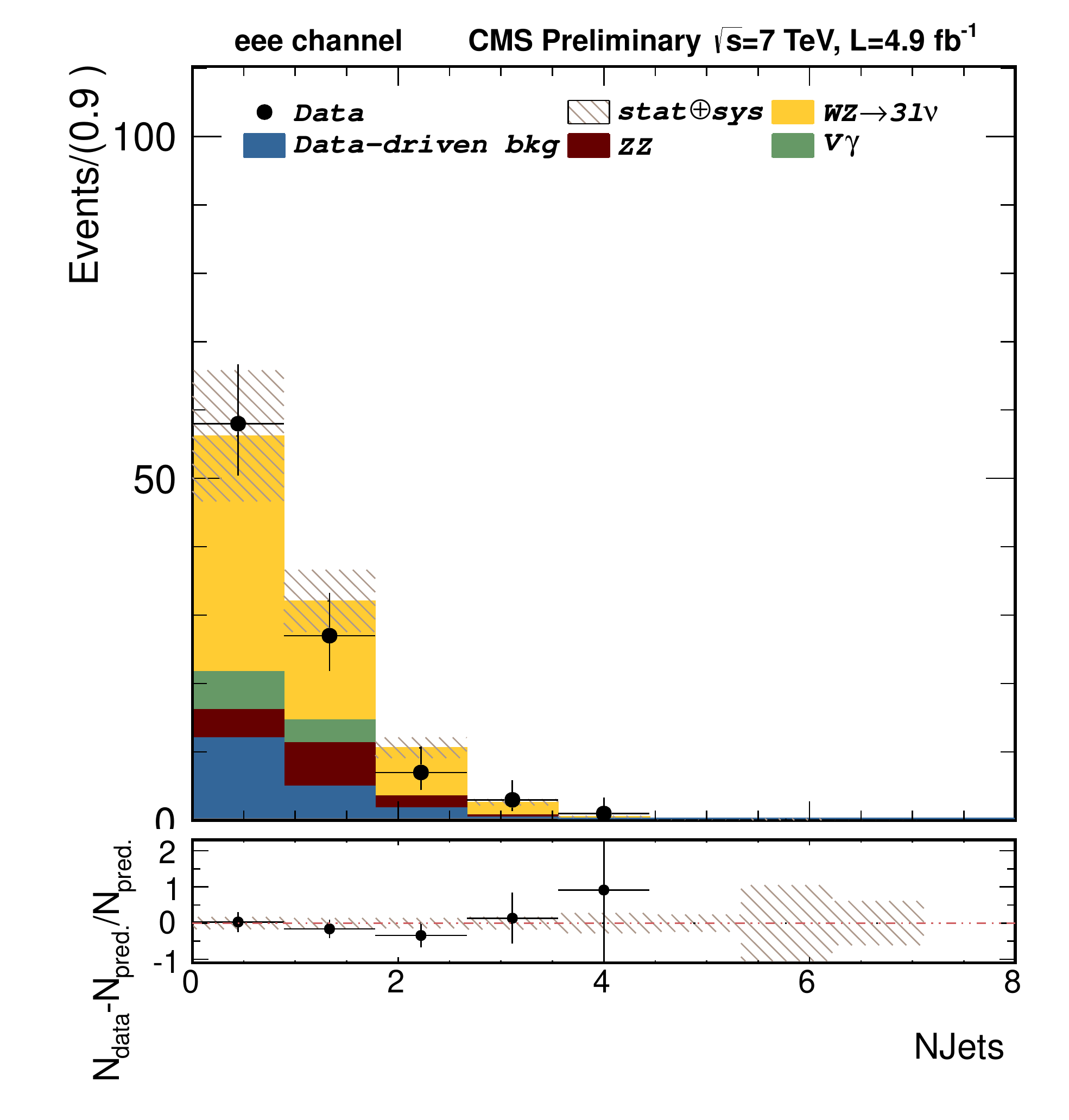}
	\end{subfigure}\quad
	\begin{subfigure}[b]{0.2\textwidth}
		\includegraphics[width=\textwidth]{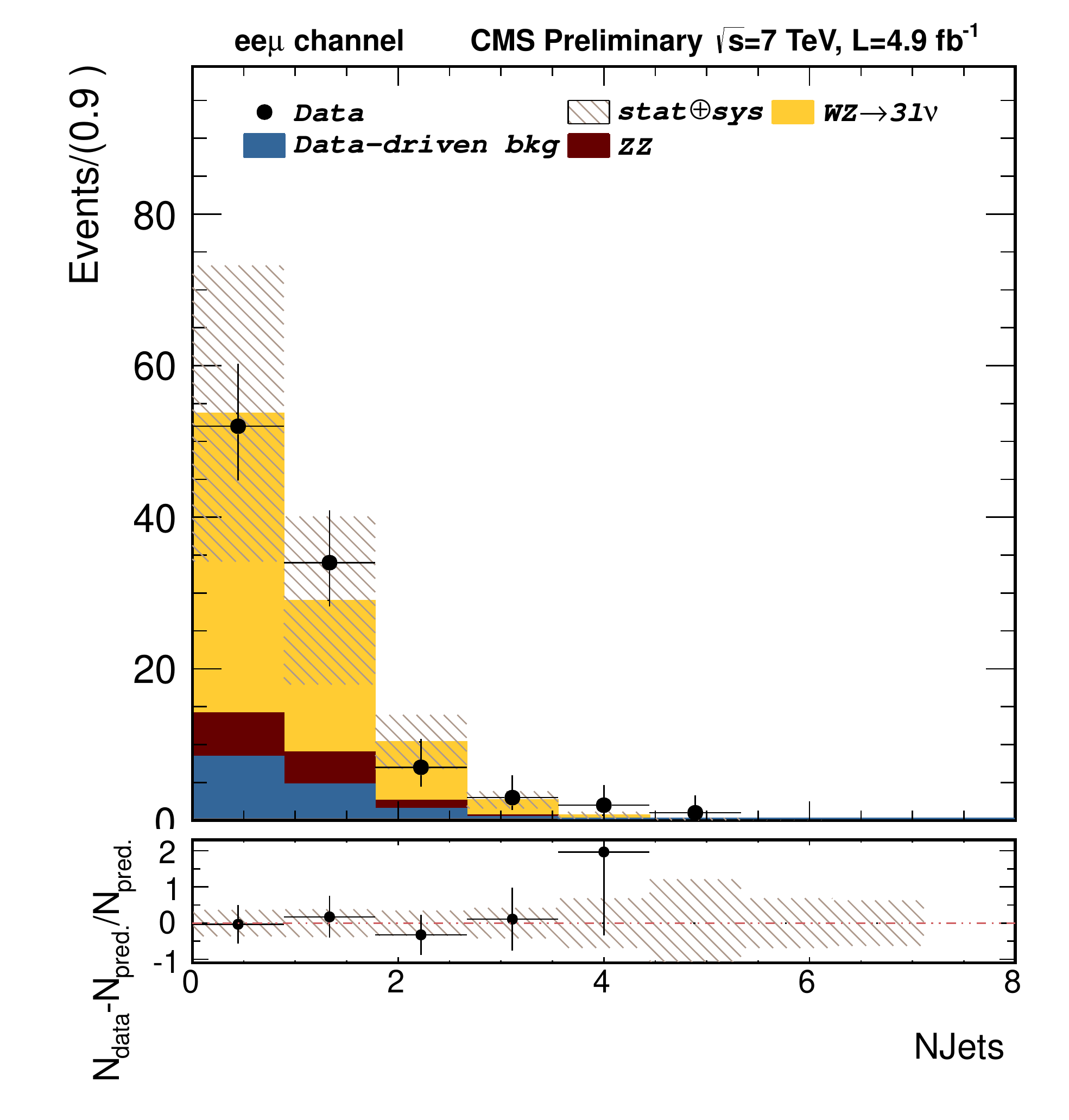}
	\end{subfigure}\quad
	\begin{subfigure}[b]{0.2\textwidth}
		\includegraphics[width=\textwidth]{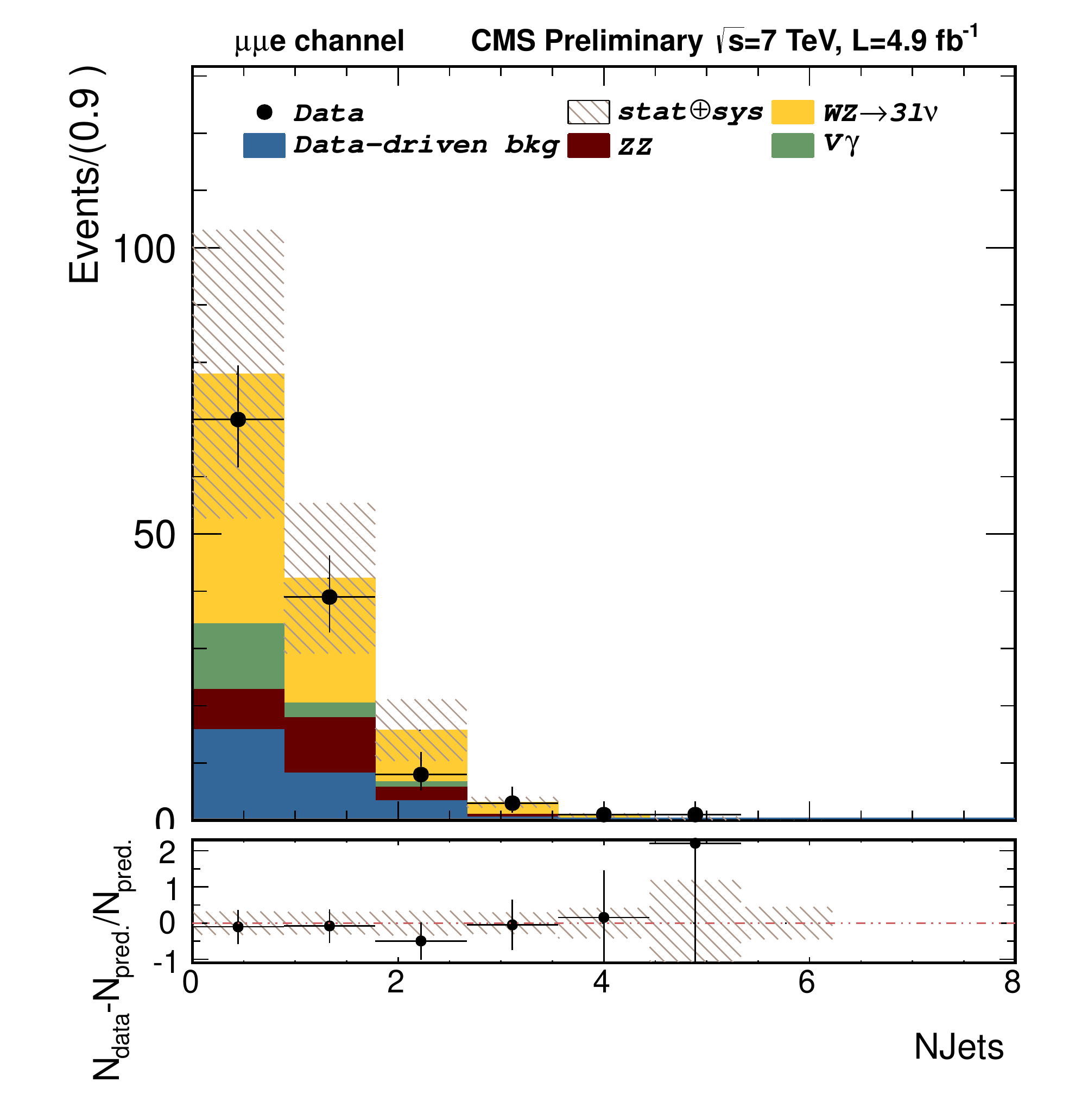}
	\end{subfigure}\quad
	\begin{subfigure}[b]{0.2\textwidth}
		\includegraphics[width=\textwidth]{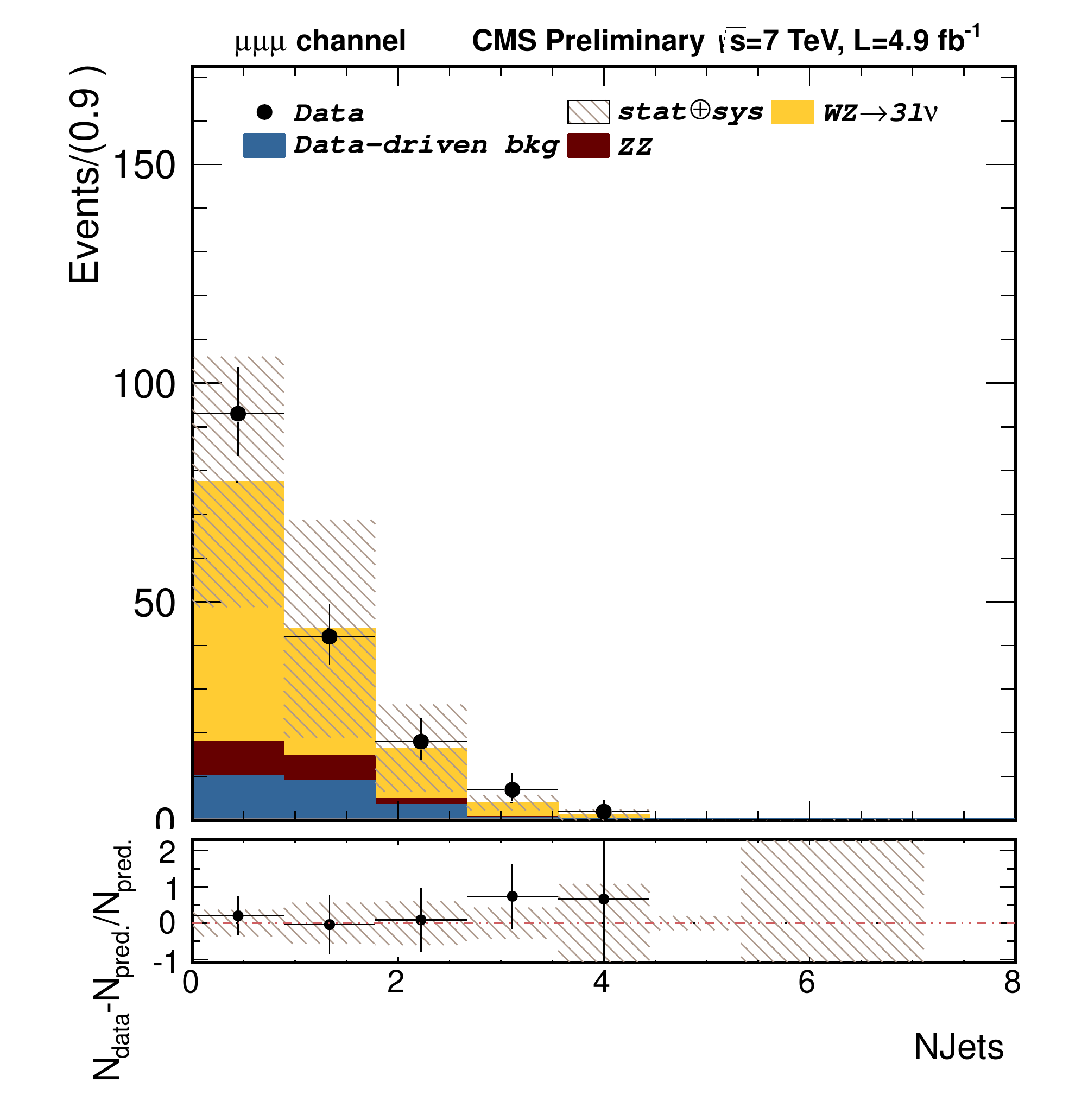}
	\end{subfigure}
	\vskip 1ex
	\begin{subfigure}[b]{0.2\textwidth}
		\includegraphics[width=\textwidth]{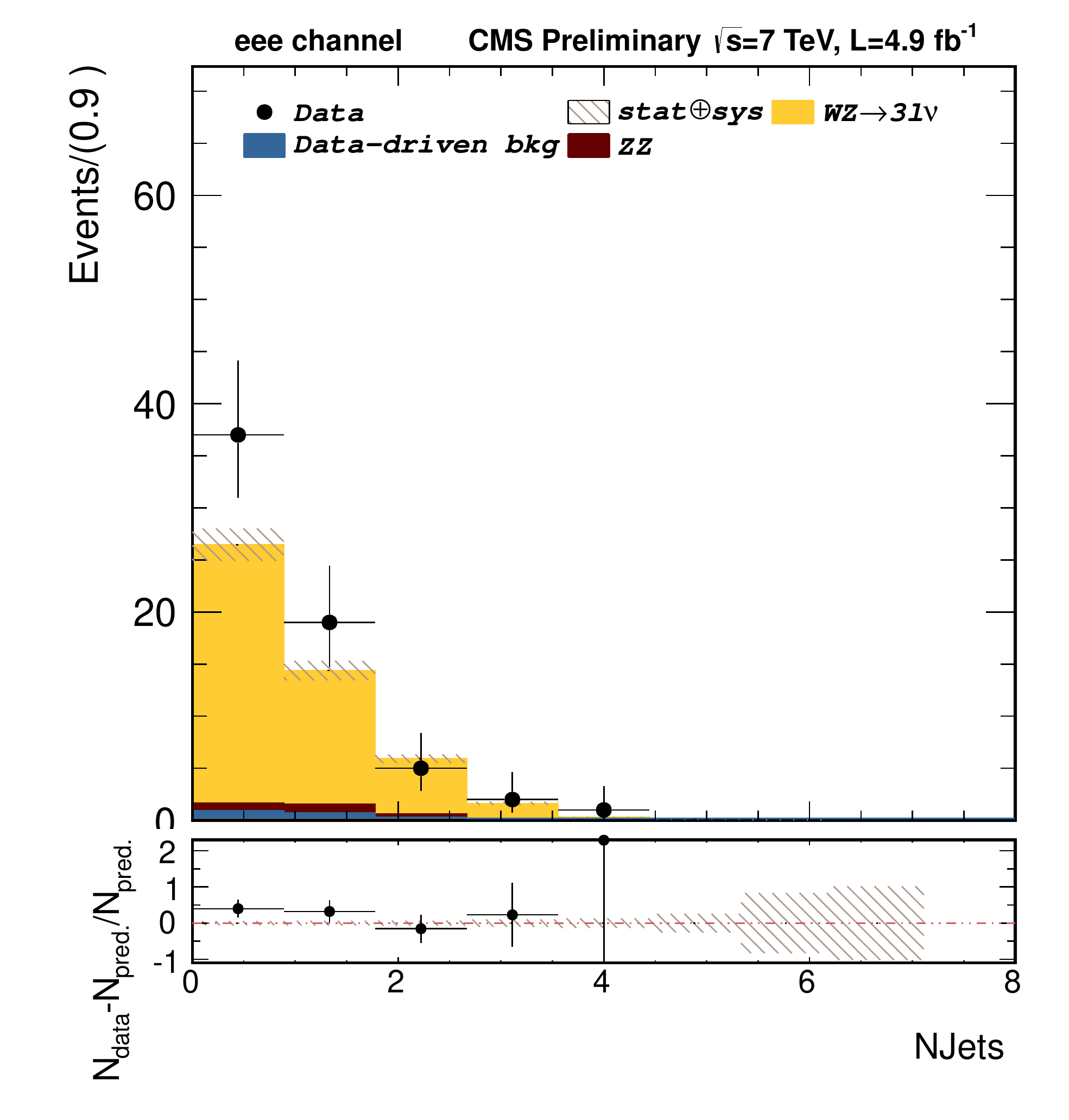}
	\end{subfigure}\quad
	\begin{subfigure}[b]{0.2\textwidth}
		\includegraphics[width=\textwidth]{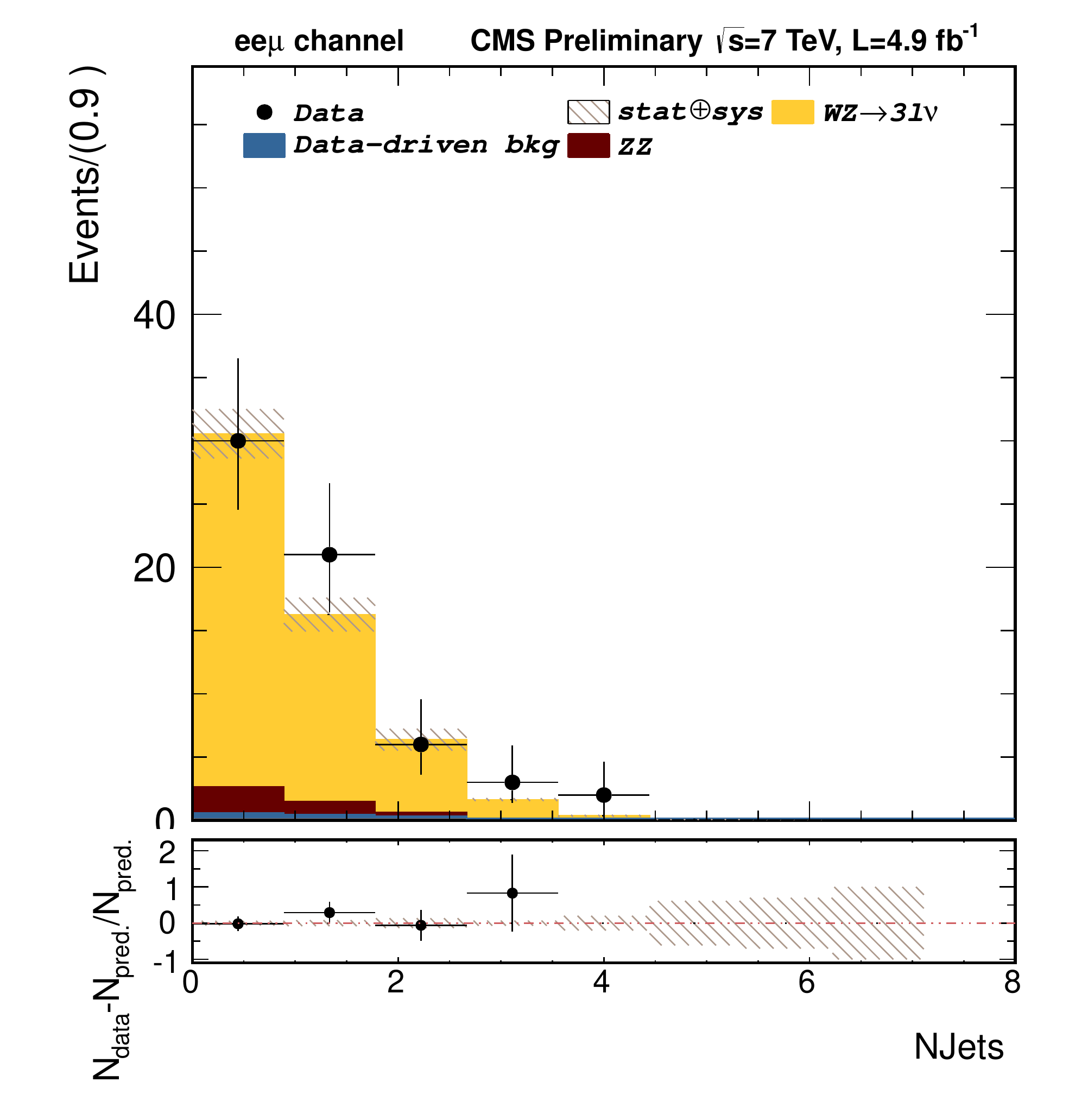}
	\end{subfigure}\quad
	\begin{subfigure}[b]{0.2\textwidth}
		\includegraphics[width=\textwidth]{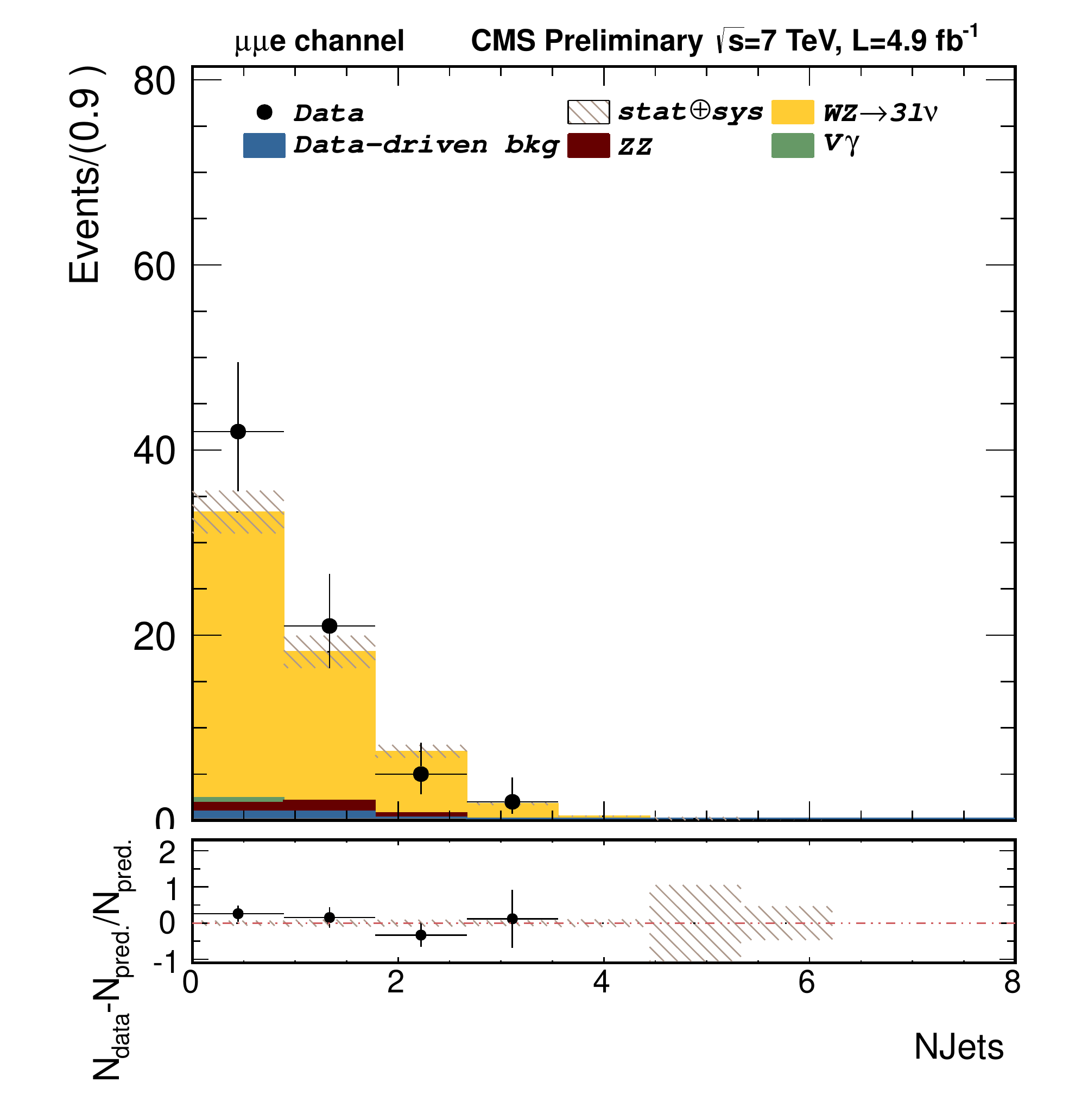}
	\end{subfigure}\quad
	\begin{subfigure}[b]{0.2\textwidth}
		\includegraphics[width=\textwidth]{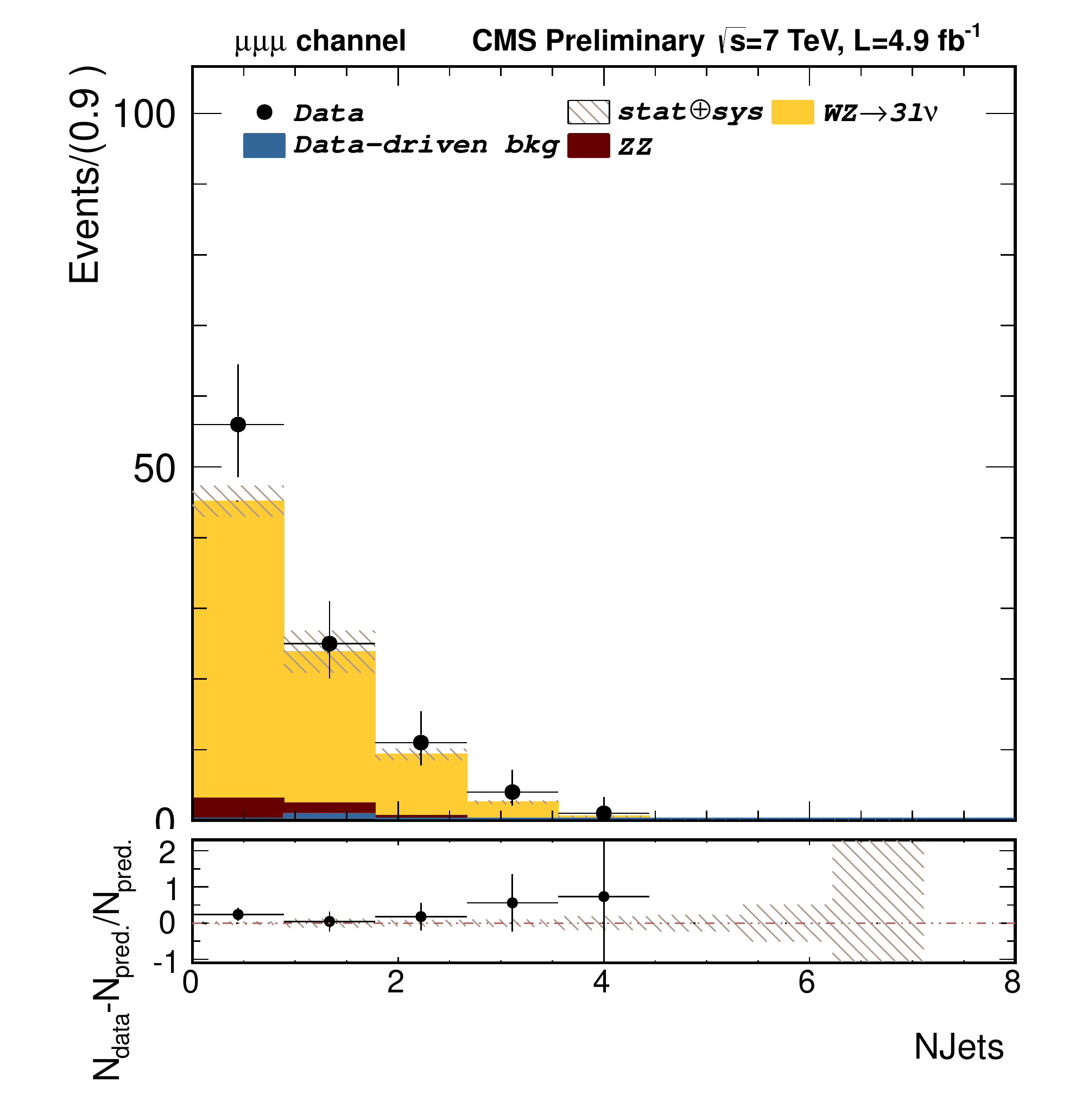}
	\end{subfigure}
	\caption[Number of jets at 7 TeV]{Number of 
	jets distribution at each event for the measured channels
	$eee$, $\mu ee$, $e\mu\mu$ and $\mu\mu\mu$ (from left to right) and
	after each analysis selection stage: after Z-candidate requirement (up row), after 
	W-candidate, without the \MET cut (middle row) and after W-candidate including \MET
	cut (bottom row).}
\end{sidewaysfigure}

\begin{sidewaysfigure}[!htpb]
	\centering
	\begin{subfigure}[b]{0.2\textwidth}
		\includegraphics[width=\textwidth]{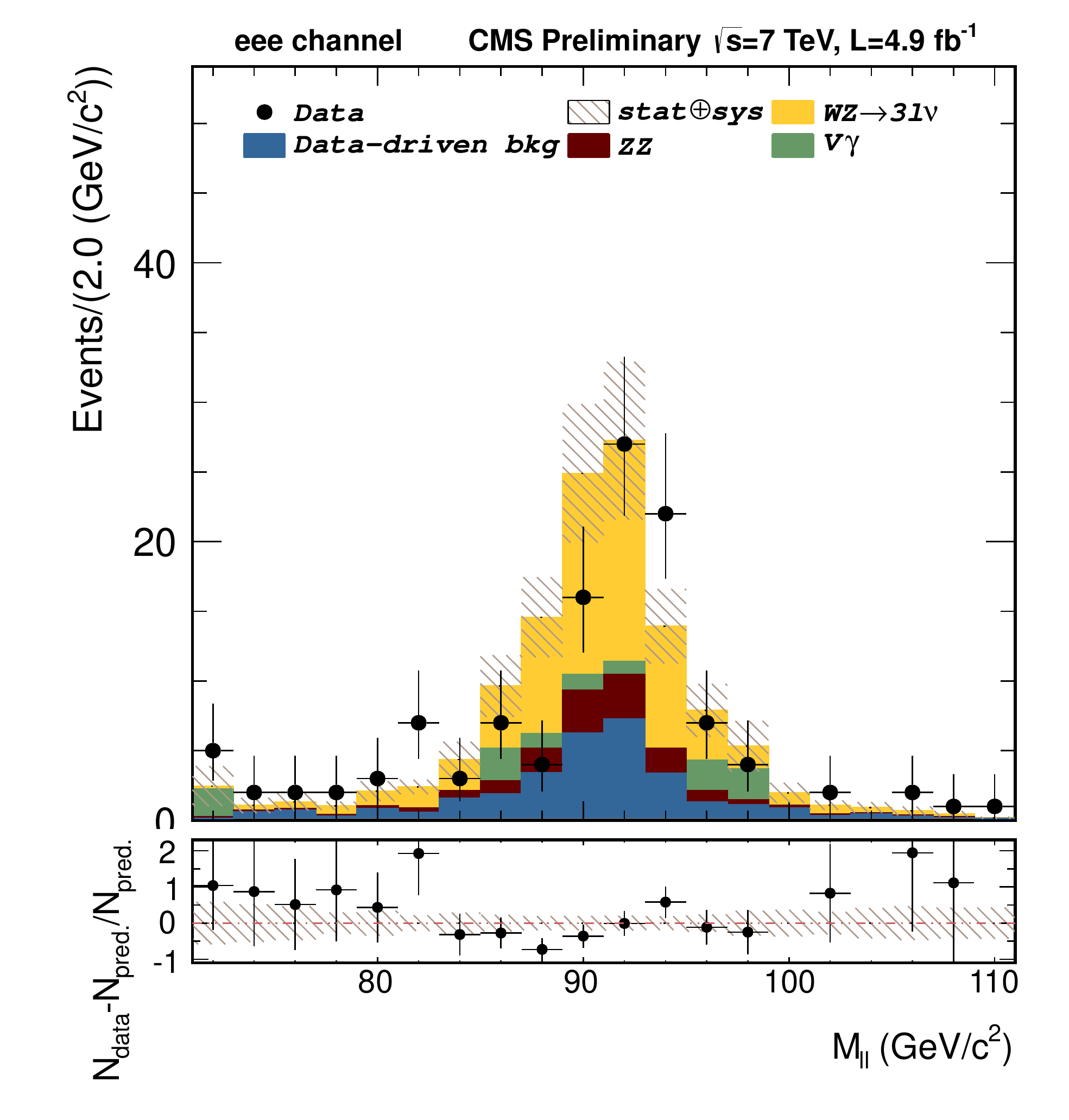}
	\end{subfigure}\quad
	\begin{subfigure}[b]{0.2\textwidth}
		\includegraphics[width=\textwidth]{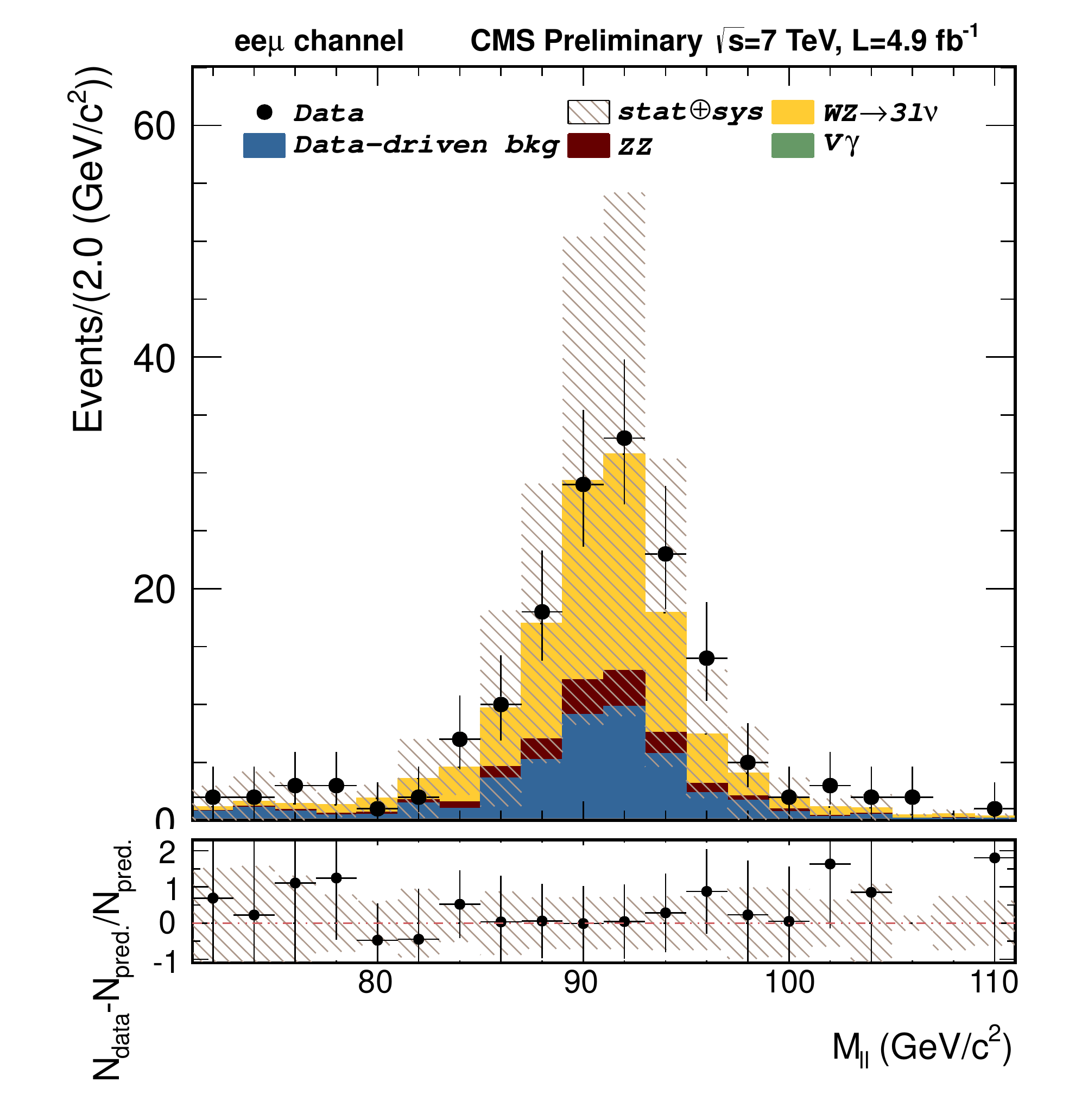}
	\end{subfigure}\quad
	\begin{subfigure}[b]{0.2\textwidth}
		\includegraphics[width=\textwidth]{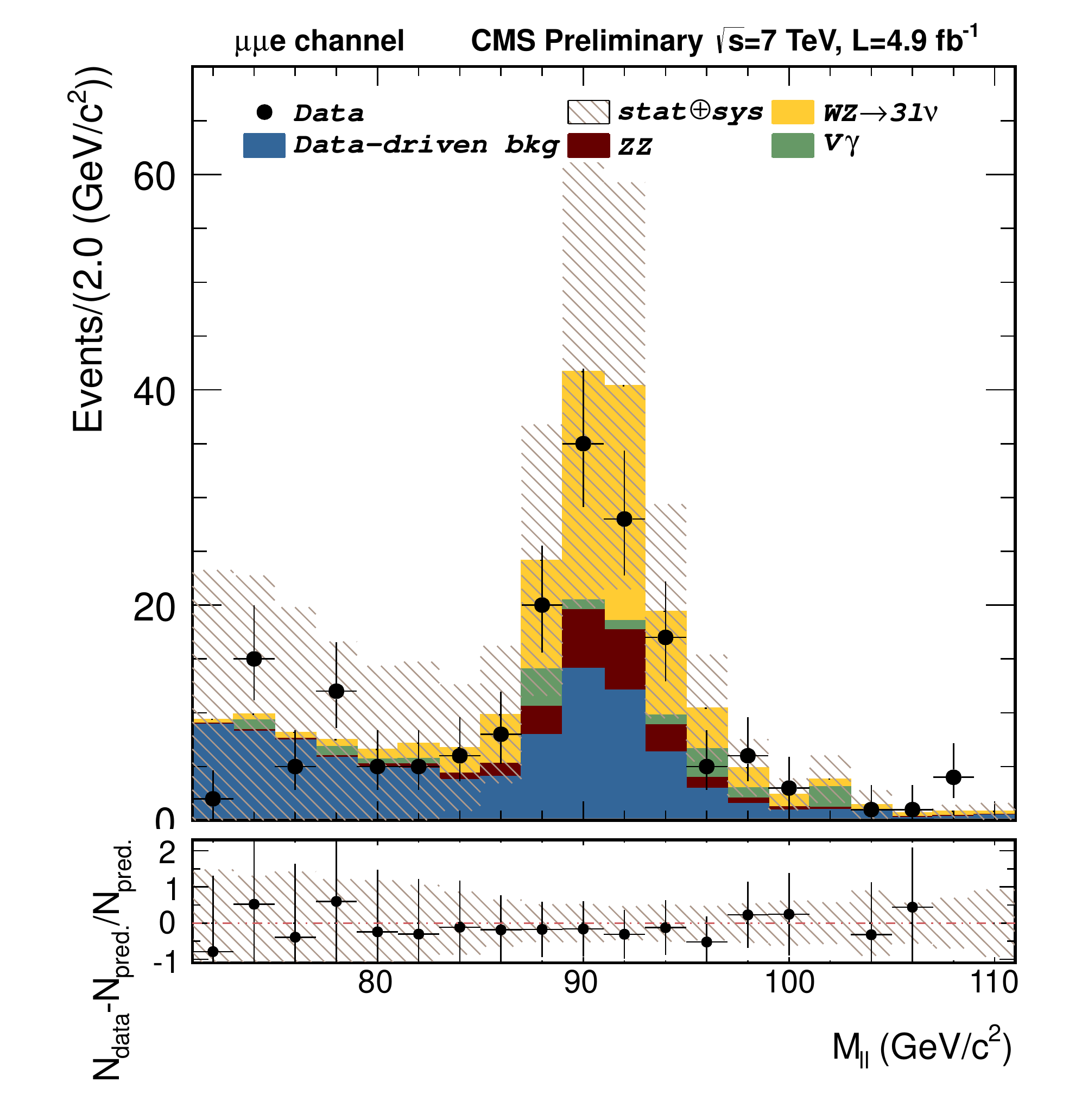}
	\end{subfigure}\quad
	\begin{subfigure}[b]{0.2\textwidth}
		\includegraphics[width=\textwidth]{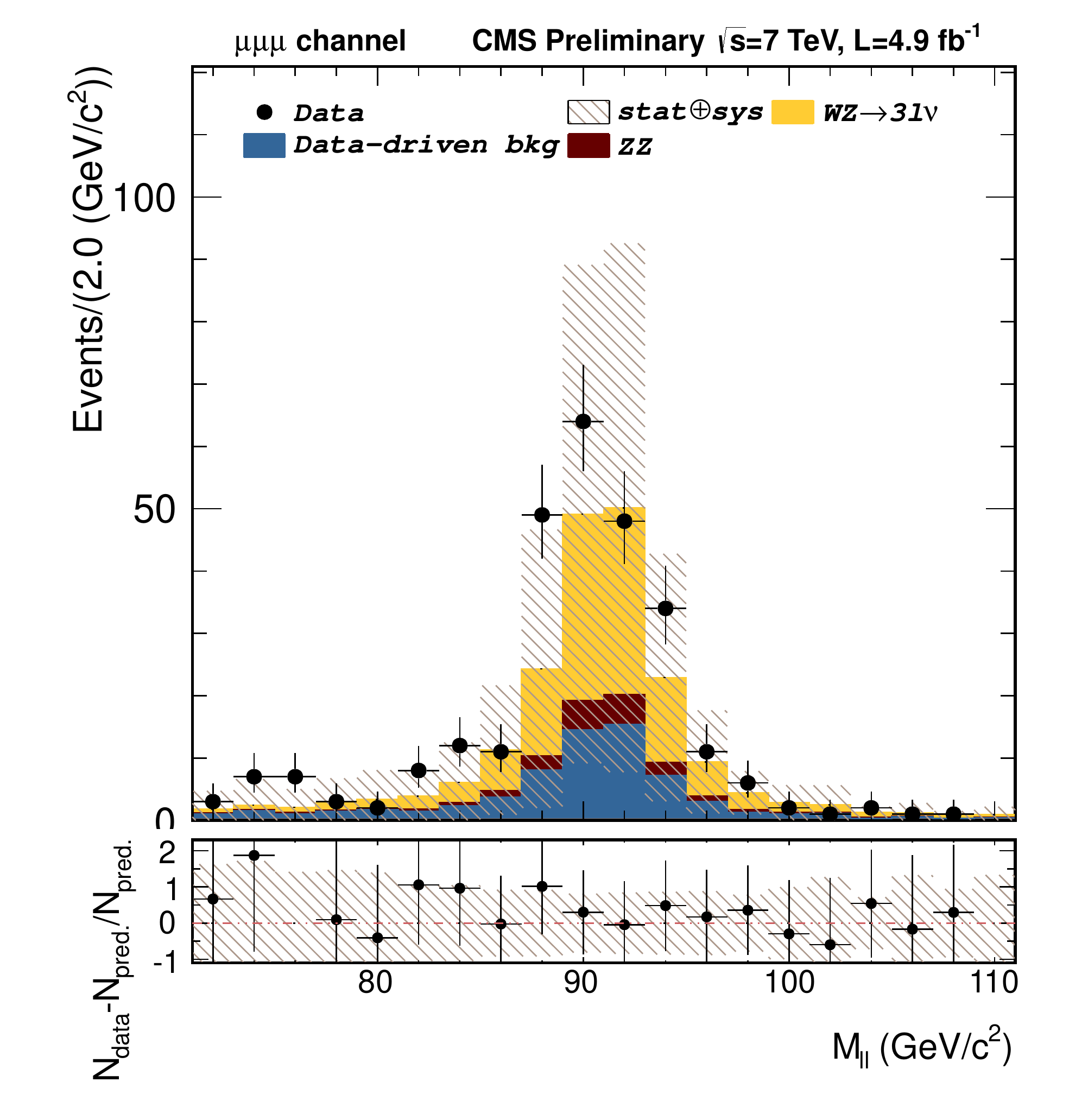}
	\end{subfigure}
	\vskip 1ex
	\centering
	\begin{subfigure}[b]{0.2\textwidth}
		\includegraphics[width=\textwidth]{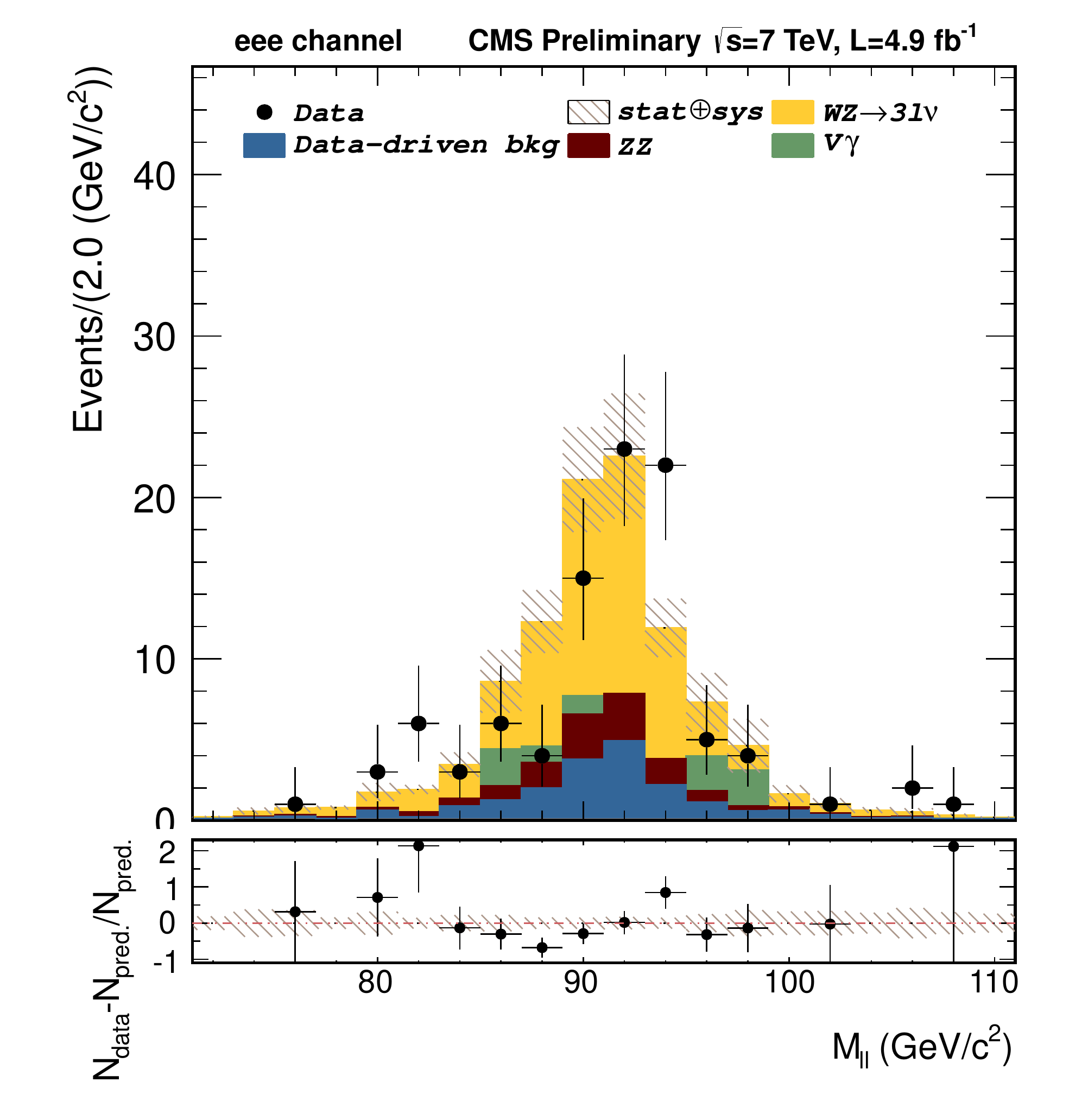}
	\end{subfigure}\quad
	\begin{subfigure}[b]{0.2\textwidth}
		\includegraphics[width=\textwidth]{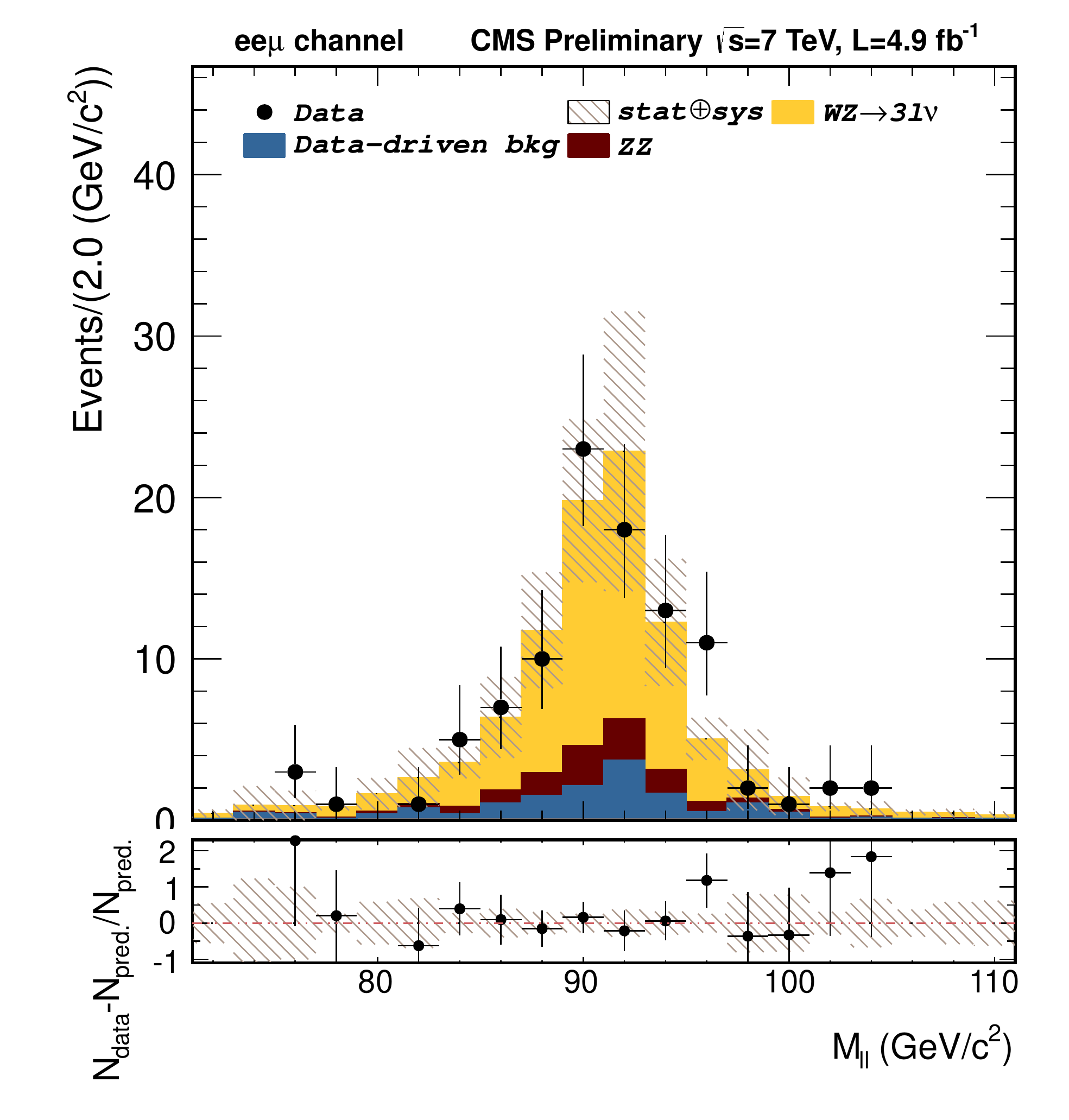}
	\end{subfigure}\quad
	\begin{subfigure}[b]{0.2\textwidth}
		\includegraphics[width=\textwidth]{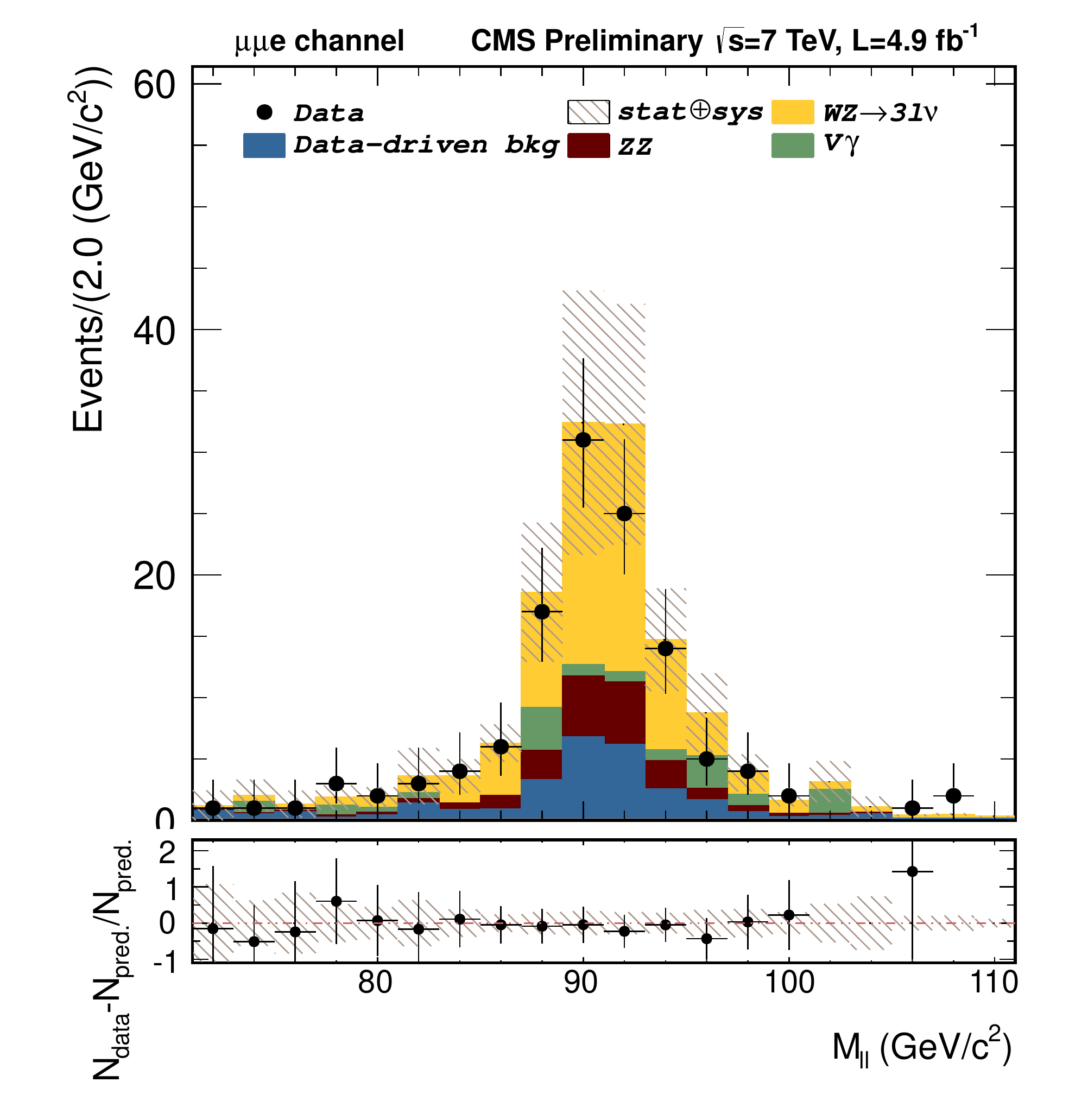}
	\end{subfigure}\quad
	\begin{subfigure}[b]{0.2\textwidth}
		\includegraphics[width=\textwidth]{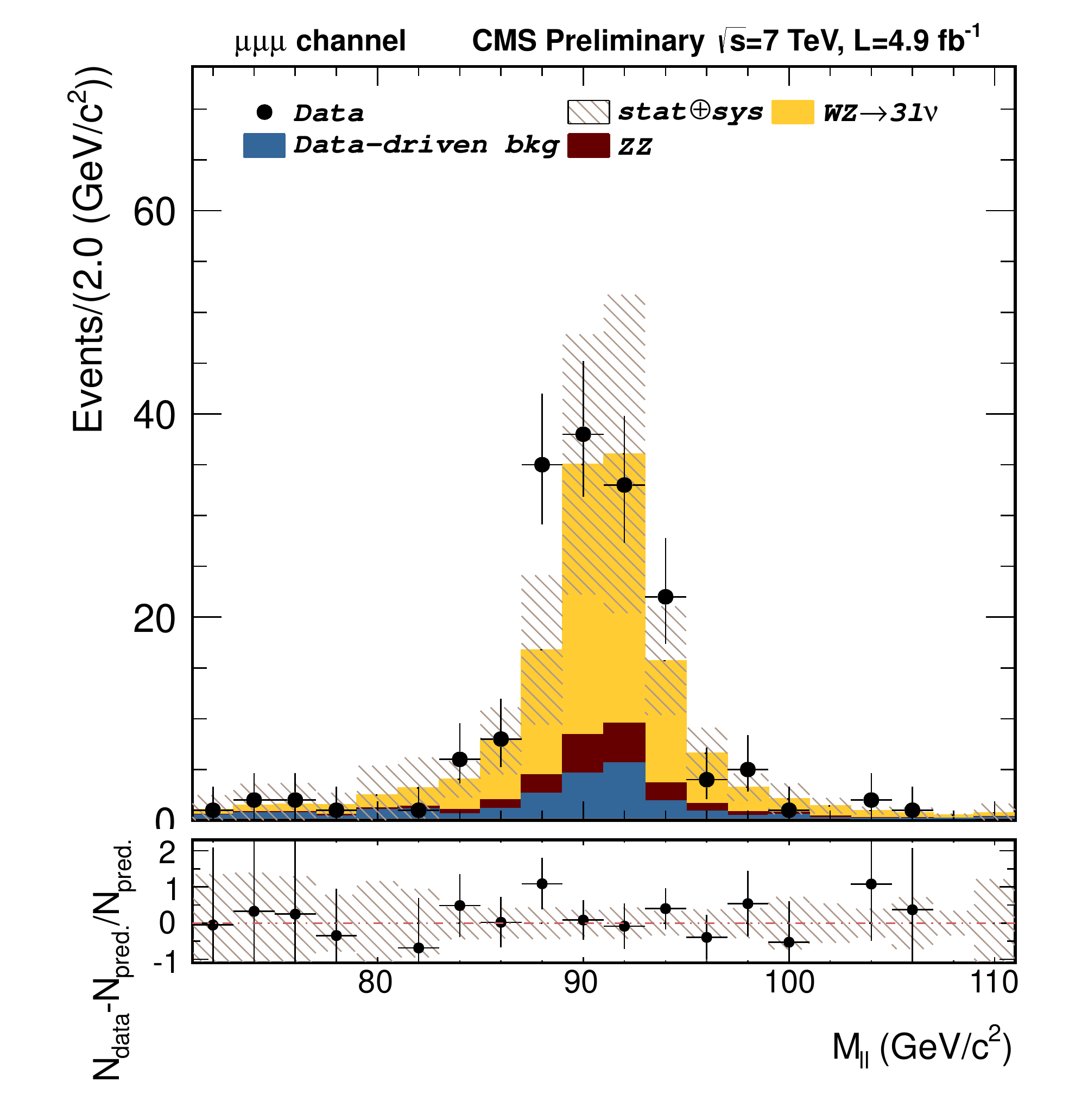}
	\end{subfigure}
	\vskip 1ex
	\begin{subfigure}[b]{0.2\textwidth}
		\includegraphics[width=\textwidth]{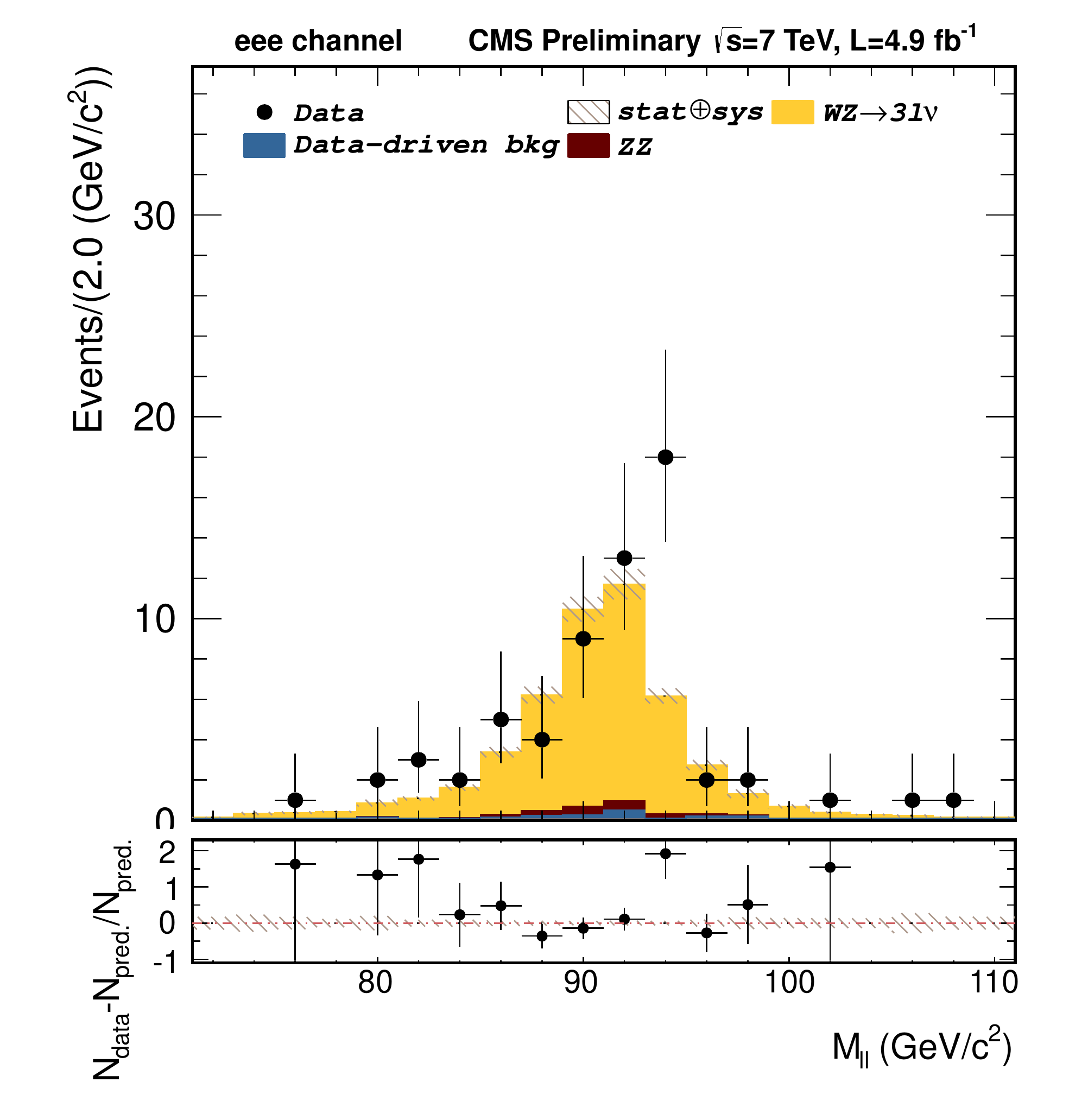}
	\end{subfigure}\quad
	\begin{subfigure}[b]{0.2\textwidth}
		\includegraphics[width=\textwidth]{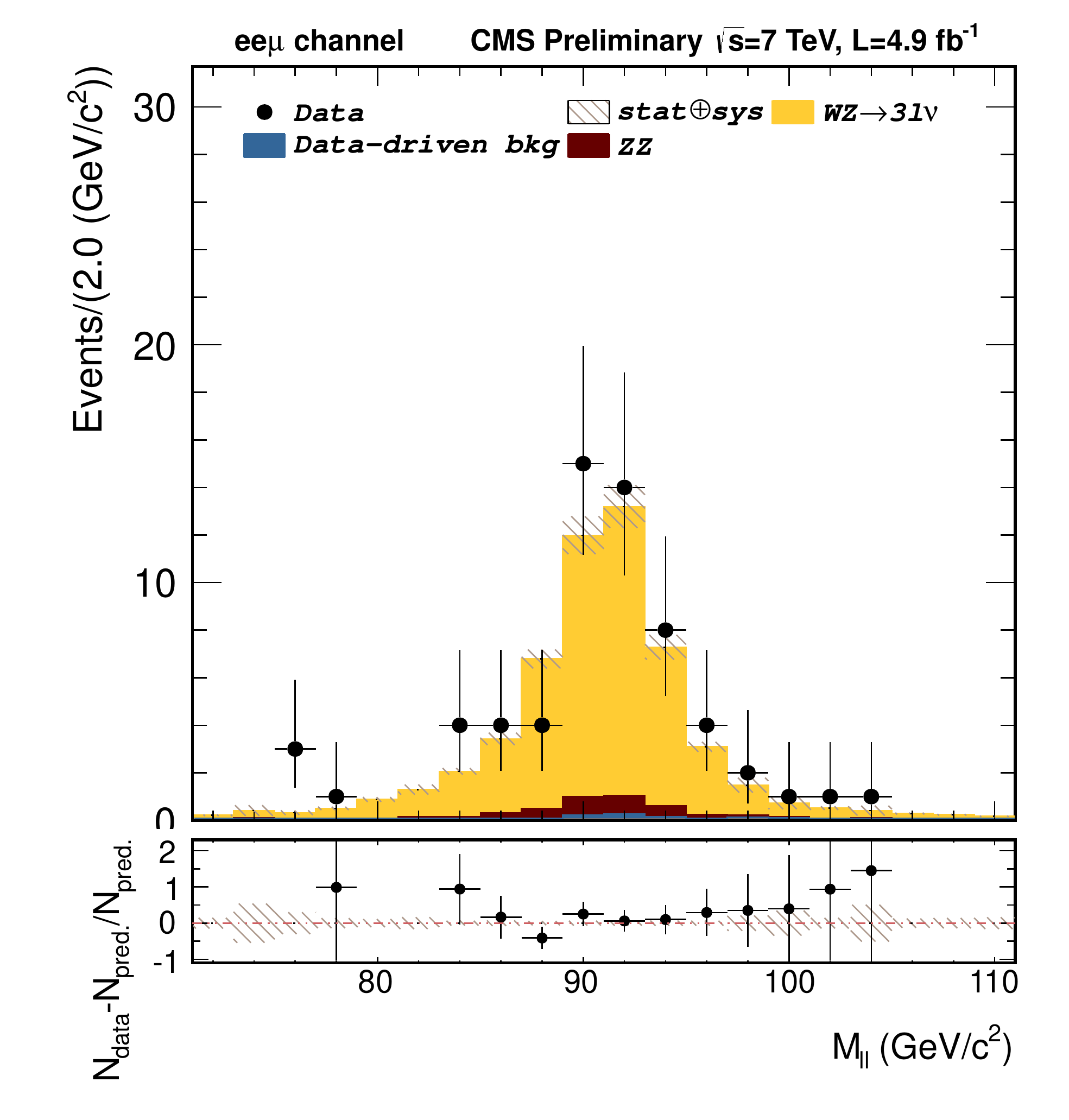}
	\end{subfigure}\quad
	\begin{subfigure}[b]{0.2\textwidth}
		\includegraphics[width=\textwidth]{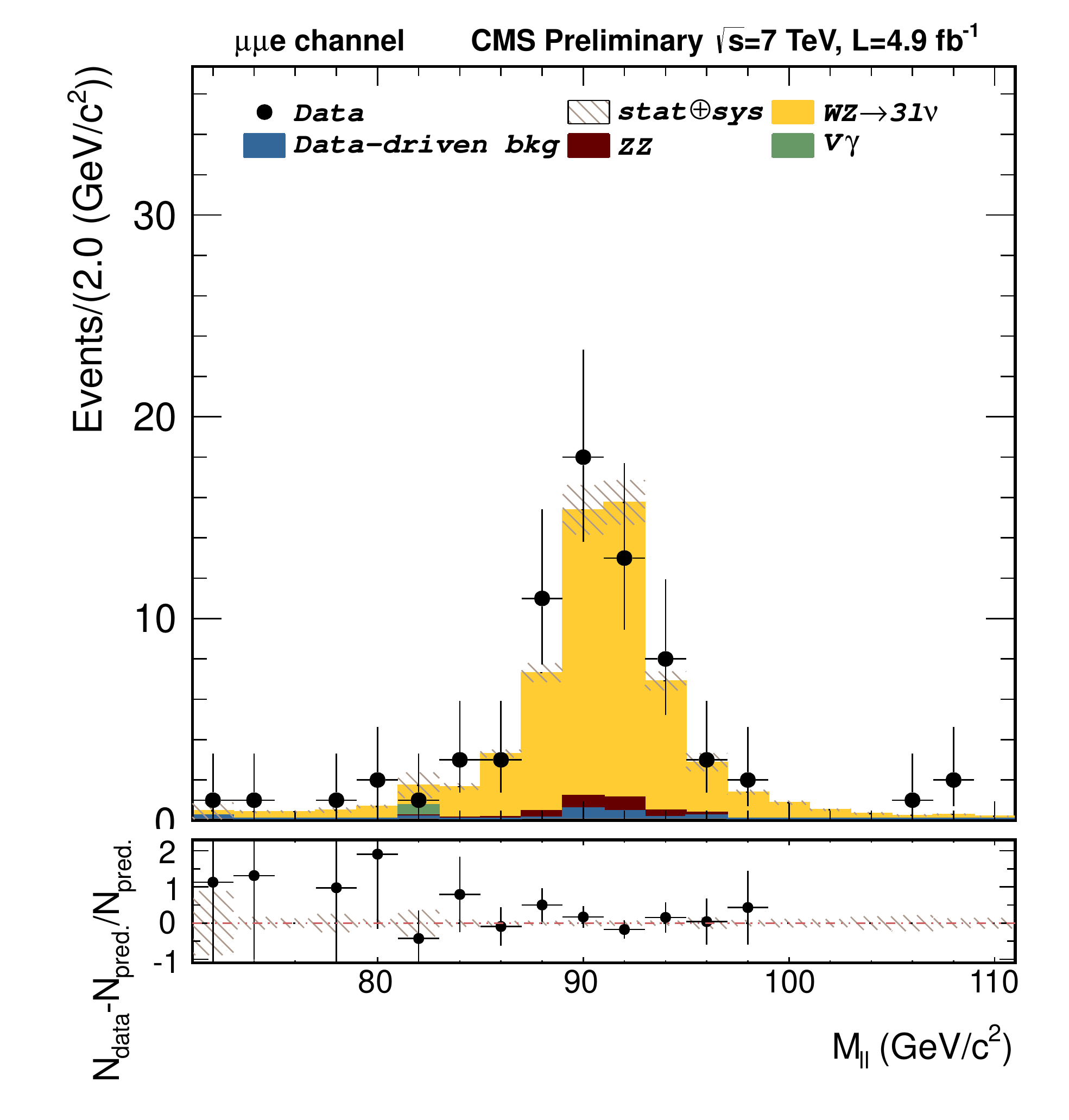}
	\end{subfigure}\quad
	\begin{subfigure}[b]{0.2\textwidth}
		\includegraphics[width=\textwidth]{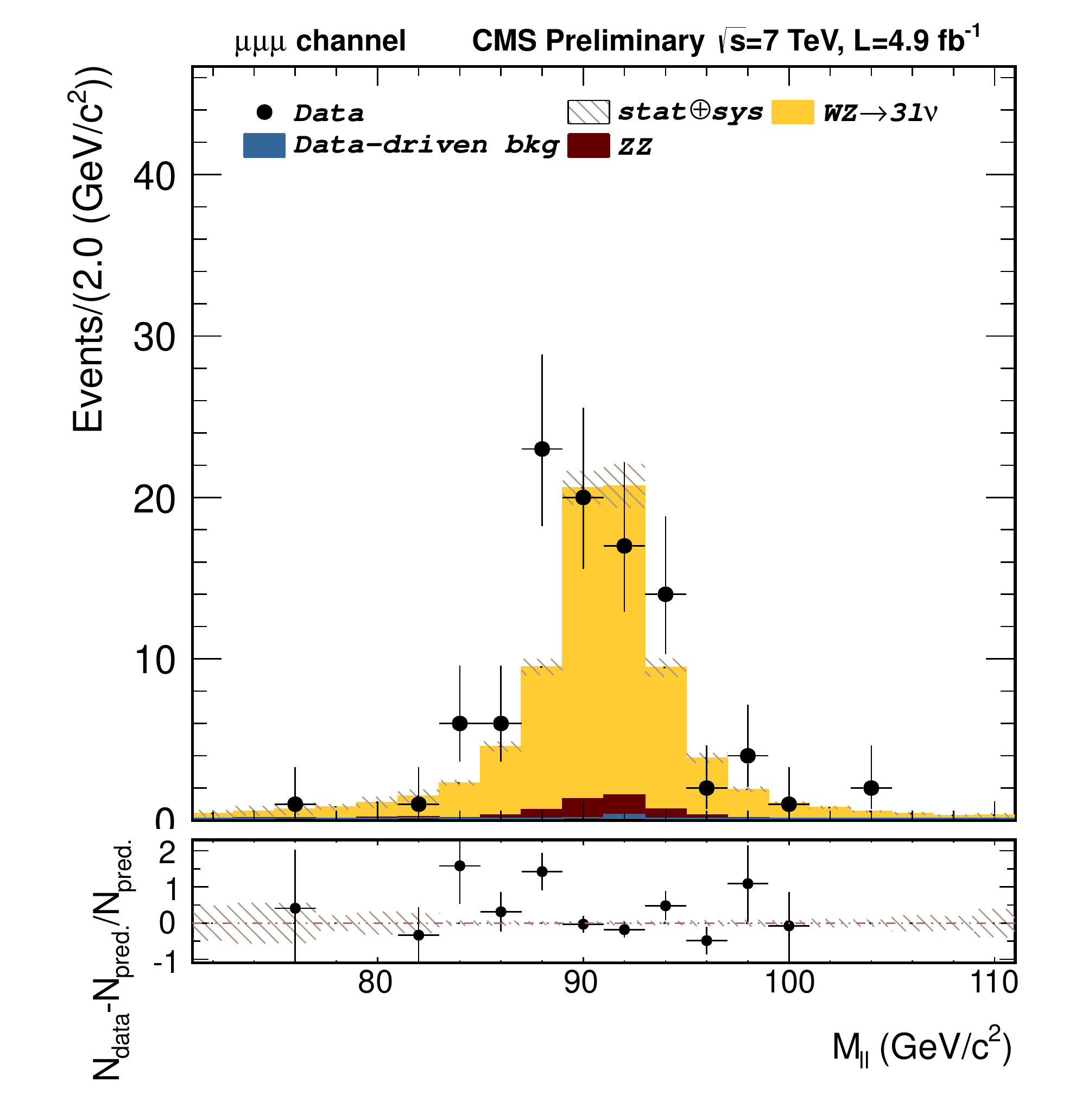}
	\end{subfigure}
	\caption[Invariant mass of the dilepton system at 7~\TeV]
	{Invariant mass of the Z-candidate dilepton system for the measured channels 
	$eee$, $\mu ee$, $e\mu\mu$ and $\mu\mu\mu$ (from left to right) and
	after each analysis selection stage: after Z-candidate requirement (up row), after 
	W-candidate, without the \MET cut (middle row) and after W-candidate including \MET
	cut (bottom row).}
\end{sidewaysfigure}

\begin{sidewaysfigure}[!htpb]
	\centering
	\begin{subfigure}[b]{0.2\textwidth}
		\includegraphics[width=\textwidth]{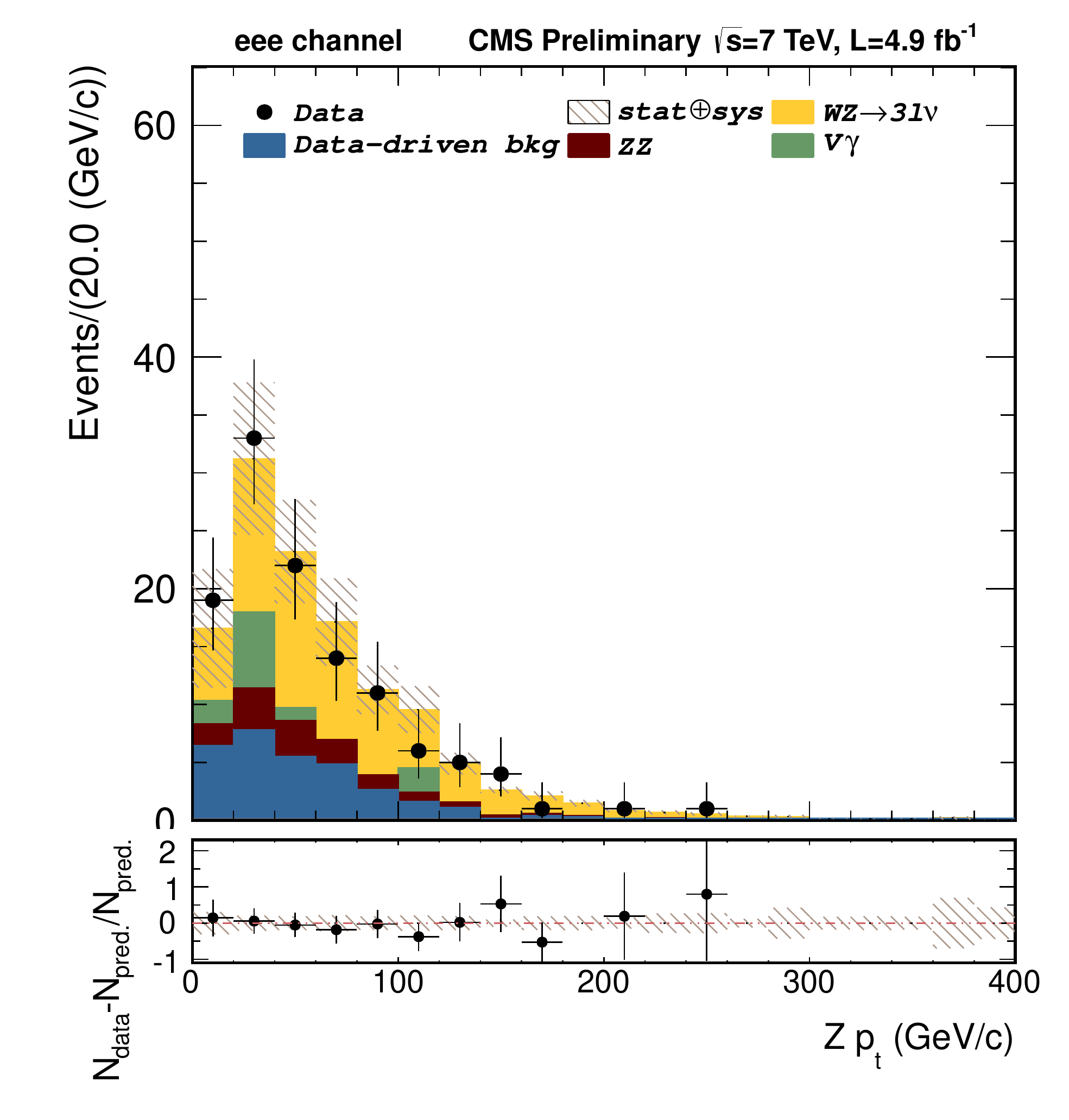}
	\end{subfigure}\quad
	\begin{subfigure}[b]{0.2\textwidth}
		\includegraphics[width=\textwidth]{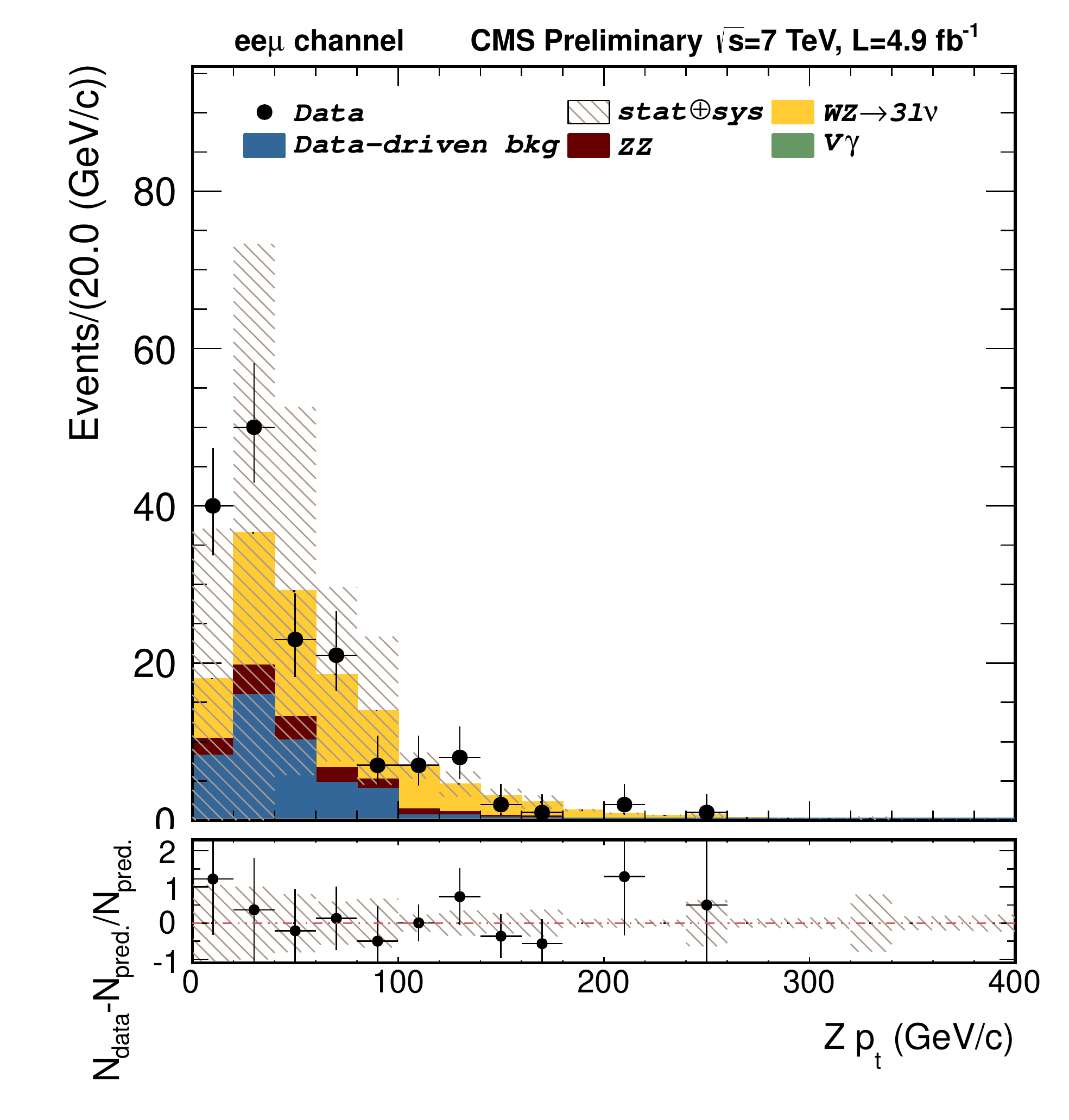}
	\end{subfigure}\quad
	\begin{subfigure}[b]{0.2\textwidth}
		\includegraphics[width=\textwidth]{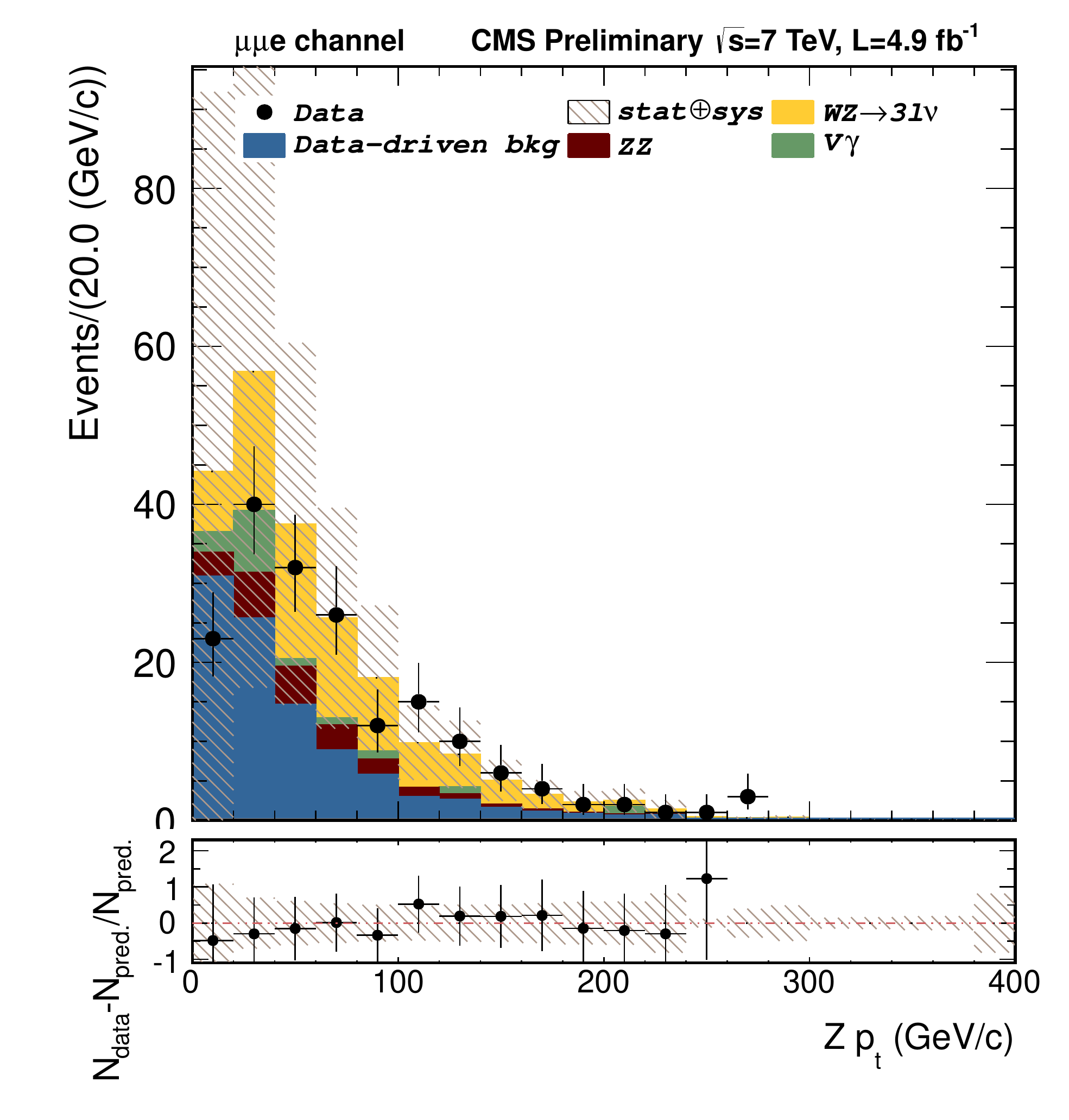}
	\end{subfigure}\quad
	\begin{subfigure}[b]{0.2\textwidth}
		\includegraphics[width=\textwidth]{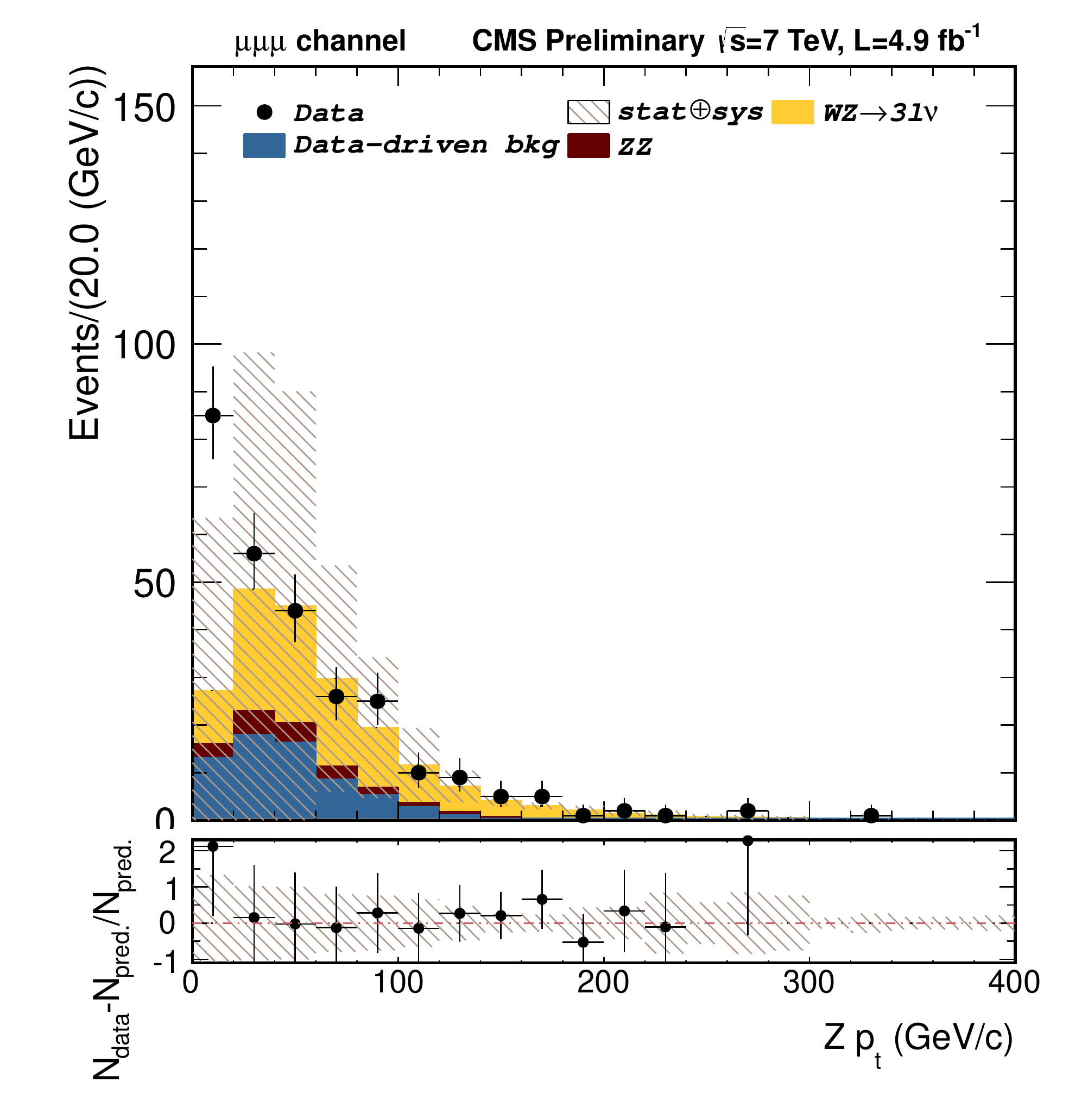}
	\end{subfigure}
	\vskip 1ex
	\centering
	\begin{subfigure}[b]{0.2\textwidth}
		\includegraphics[width=\textwidth]{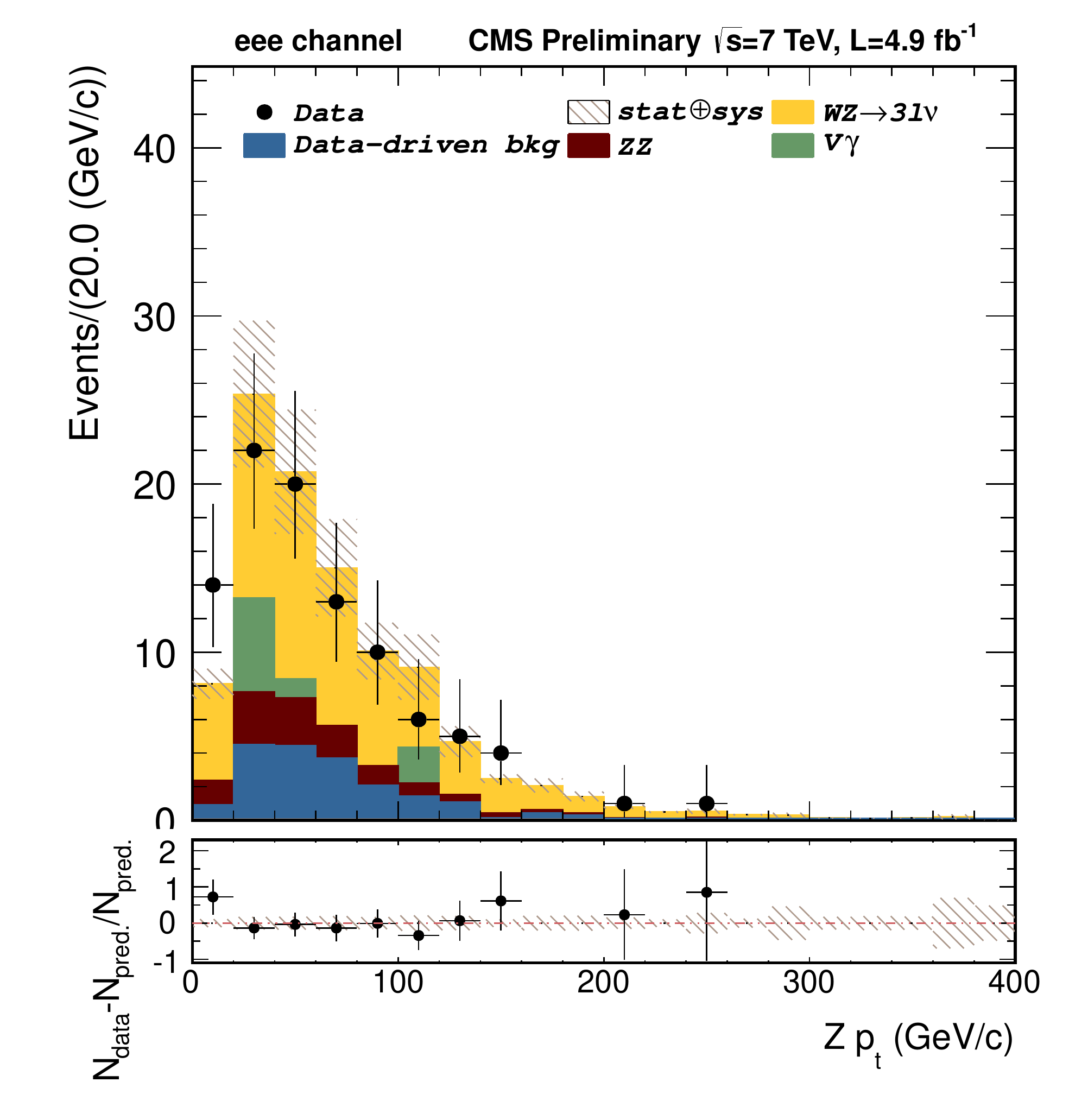}
	\end{subfigure}\quad
	\begin{subfigure}[b]{0.2\textwidth}
		\includegraphics[width=\textwidth]{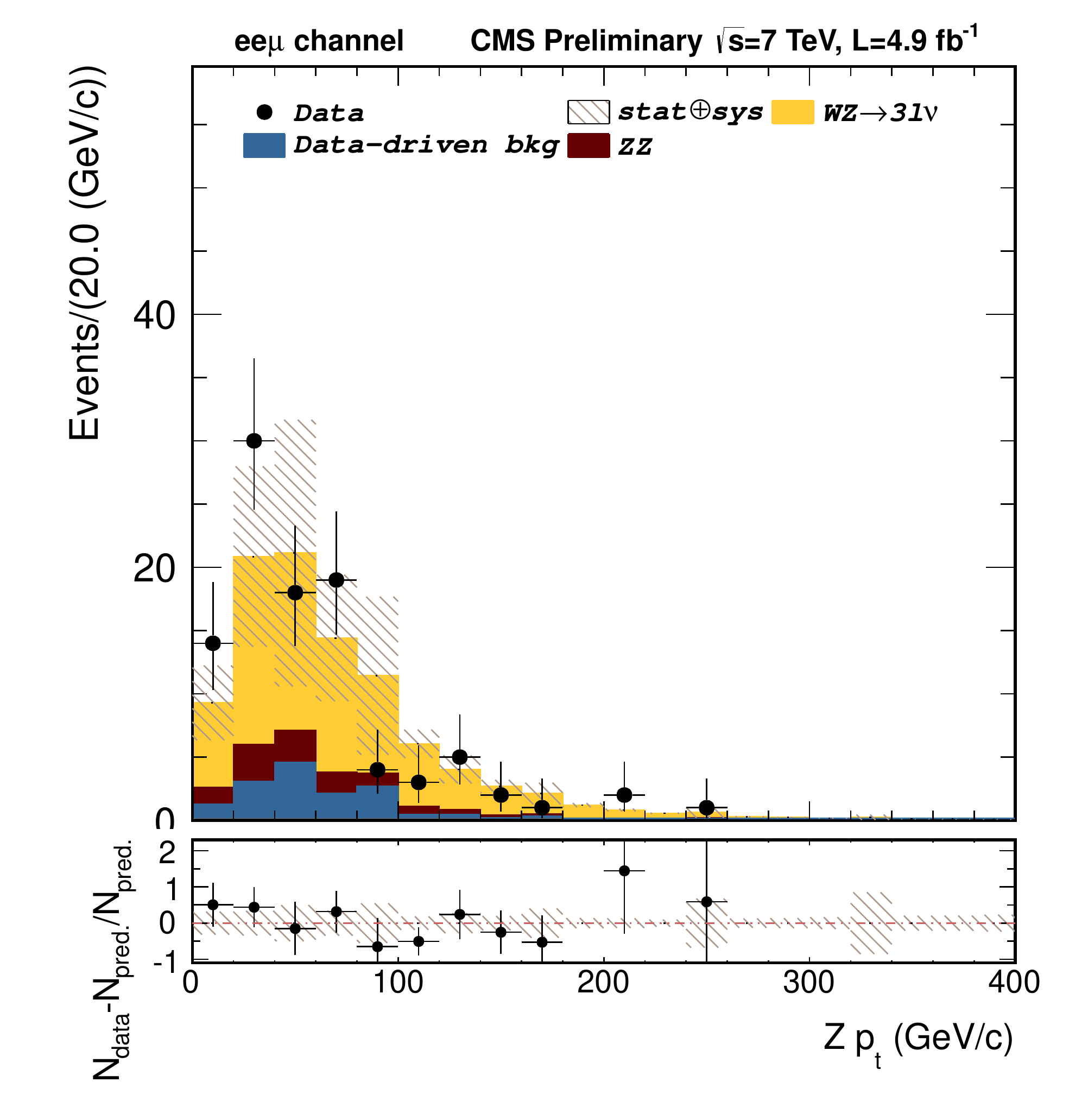}
	\end{subfigure}\quad
	\begin{subfigure}[b]{0.2\textwidth}
		\includegraphics[width=\textwidth]{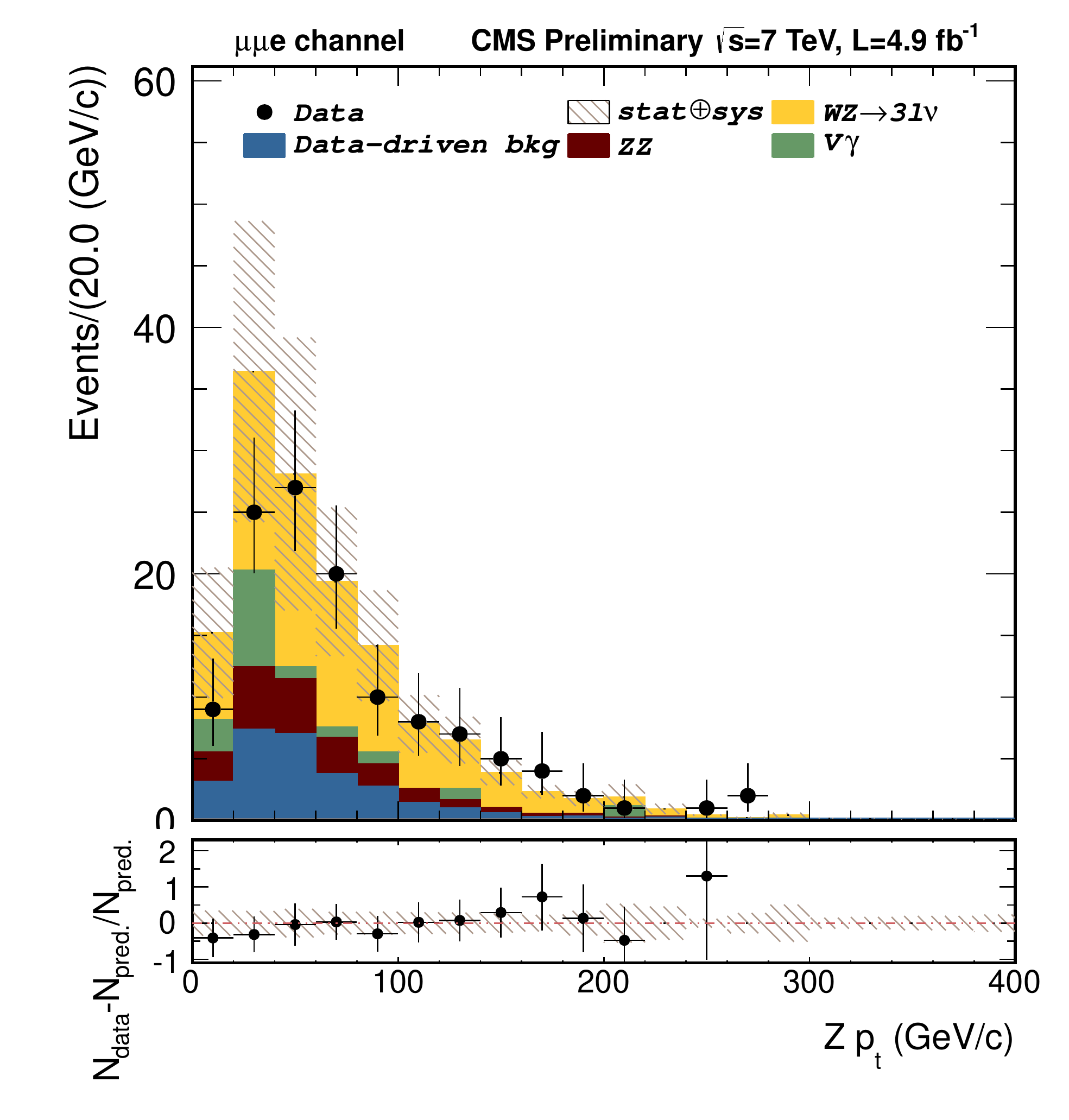}
	\end{subfigure}\quad
	\begin{subfigure}[b]{0.2\textwidth}
		\includegraphics[width=\textwidth]{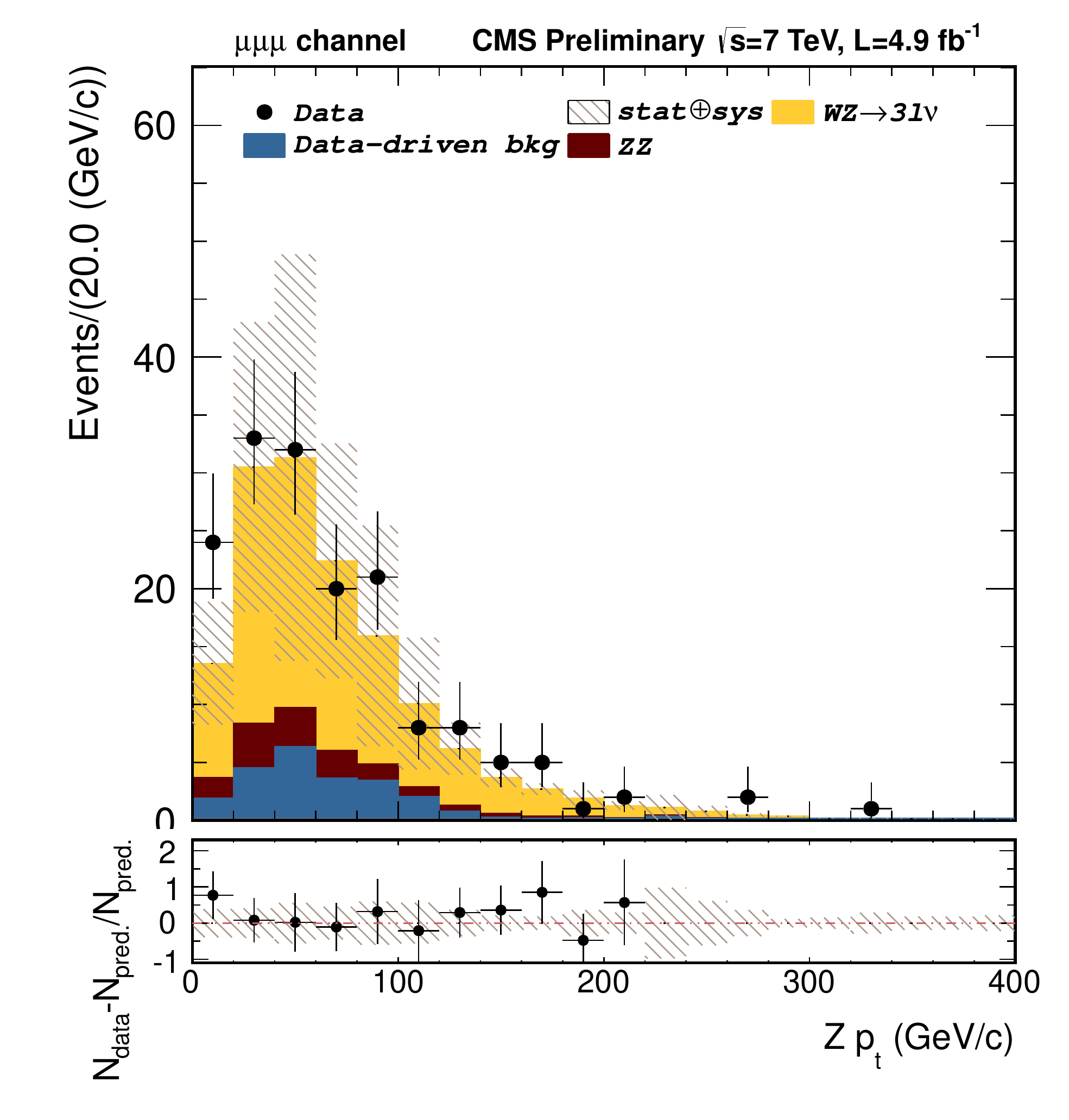}
	\end{subfigure}
	\vskip 1ex
	\begin{subfigure}[b]{0.2\textwidth}
		\includegraphics[width=\textwidth]{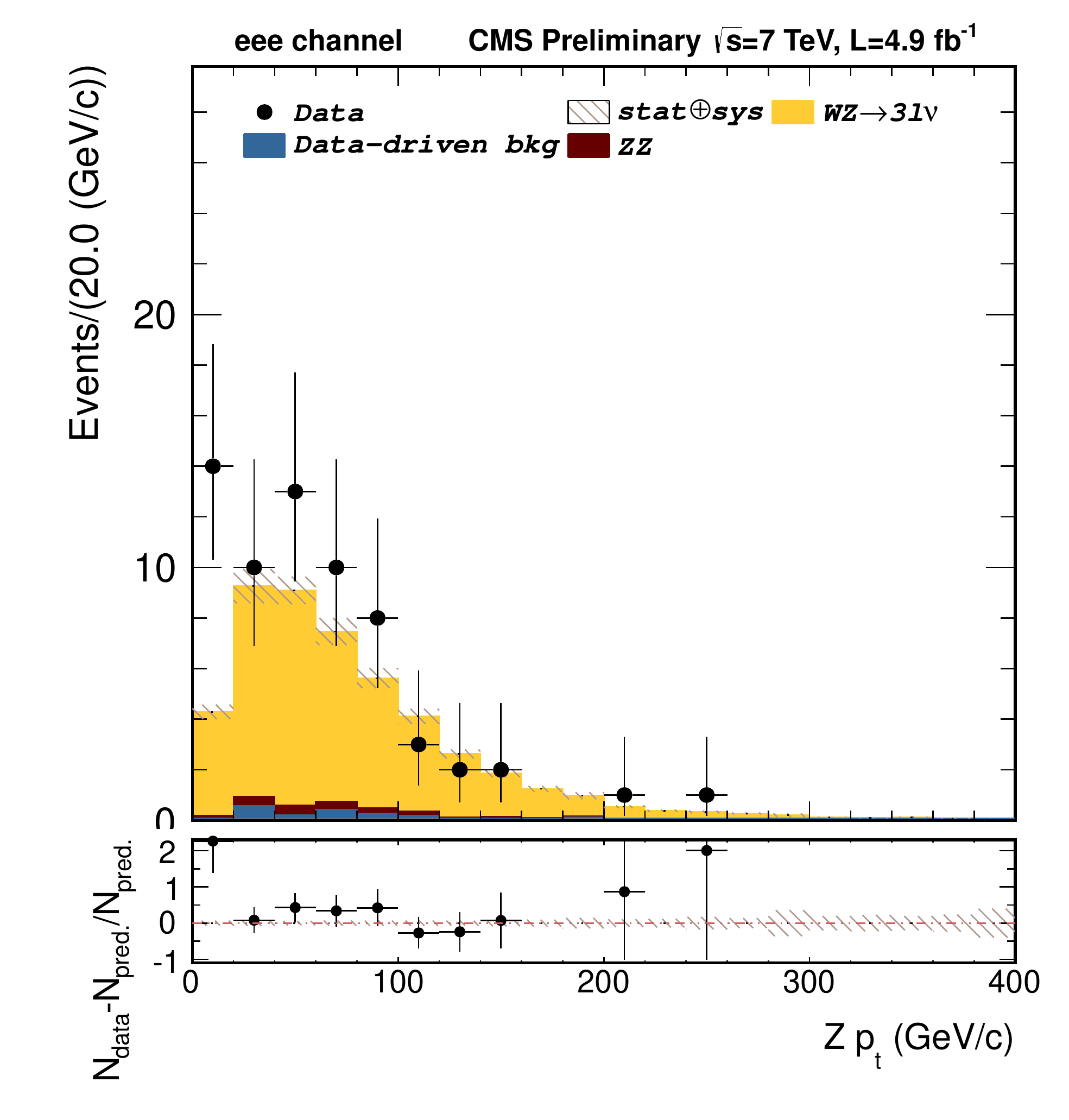}
	\end{subfigure}\quad
	\begin{subfigure}[b]{0.2\textwidth}
		\includegraphics[width=\textwidth]{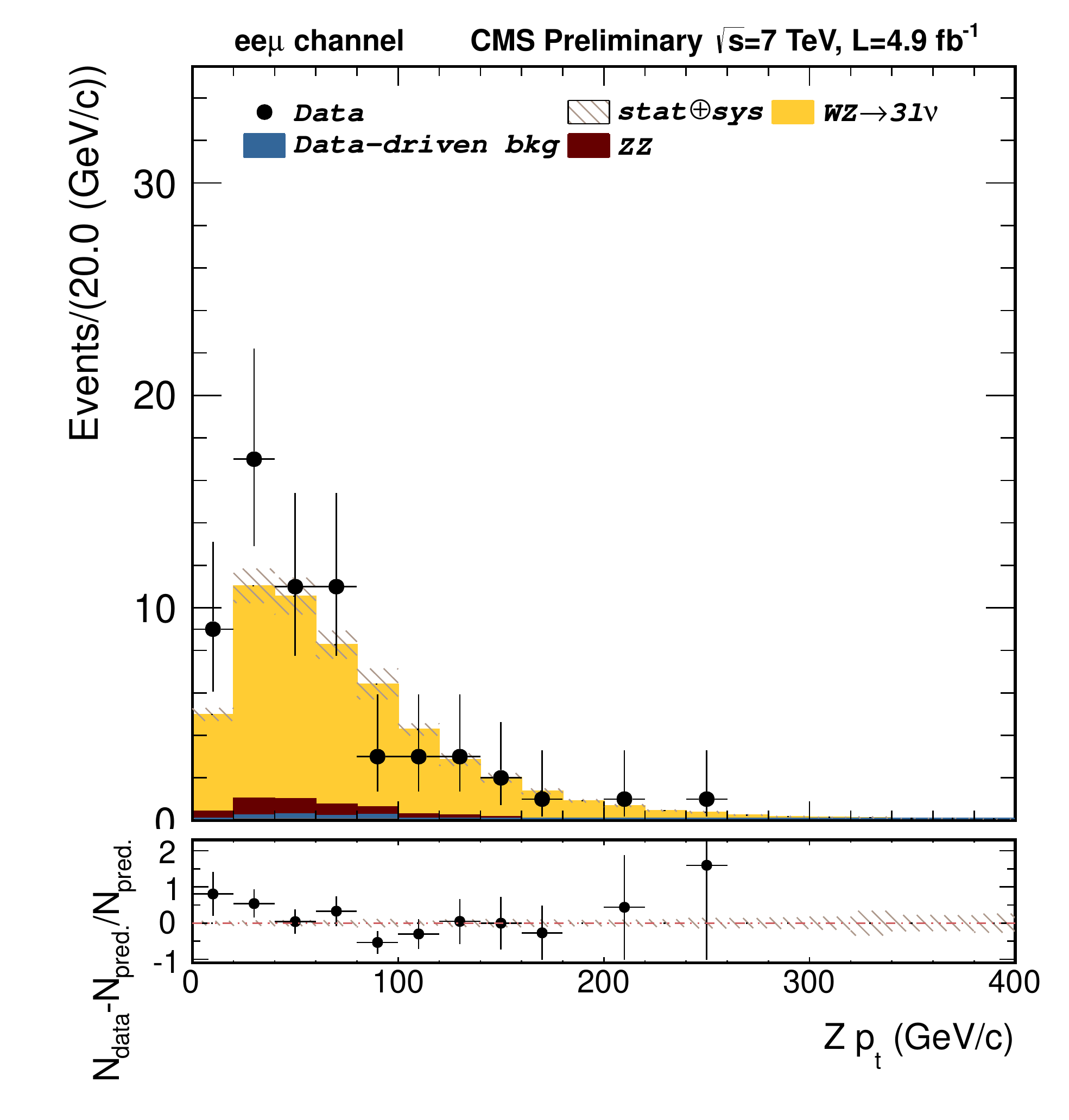}
	\end{subfigure}\quad
	\begin{subfigure}[b]{0.2\textwidth}
		\includegraphics[width=\textwidth]{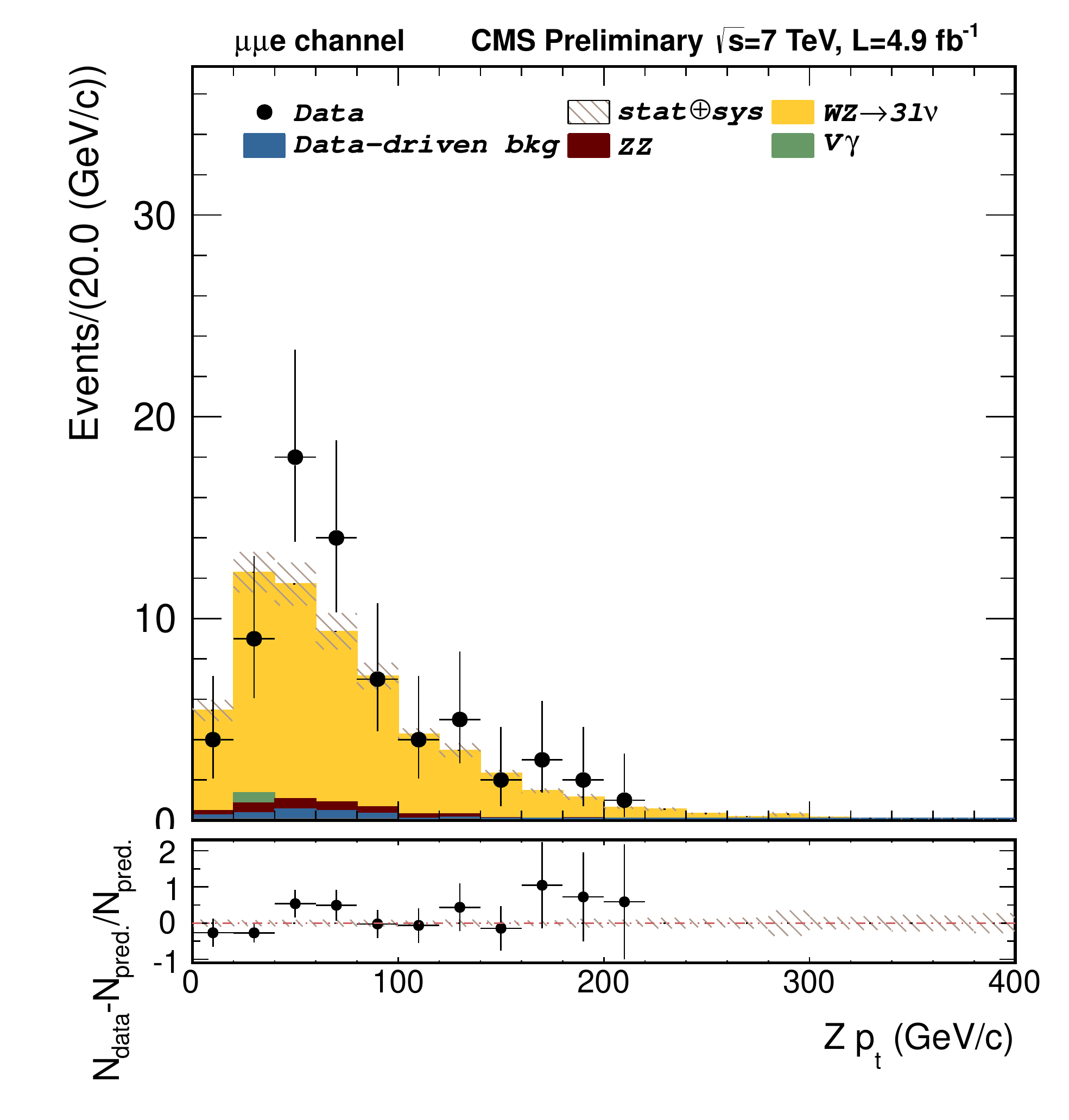}
	\end{subfigure}\quad
	\begin{subfigure}[b]{0.2\textwidth}
		\includegraphics[width=\textwidth]{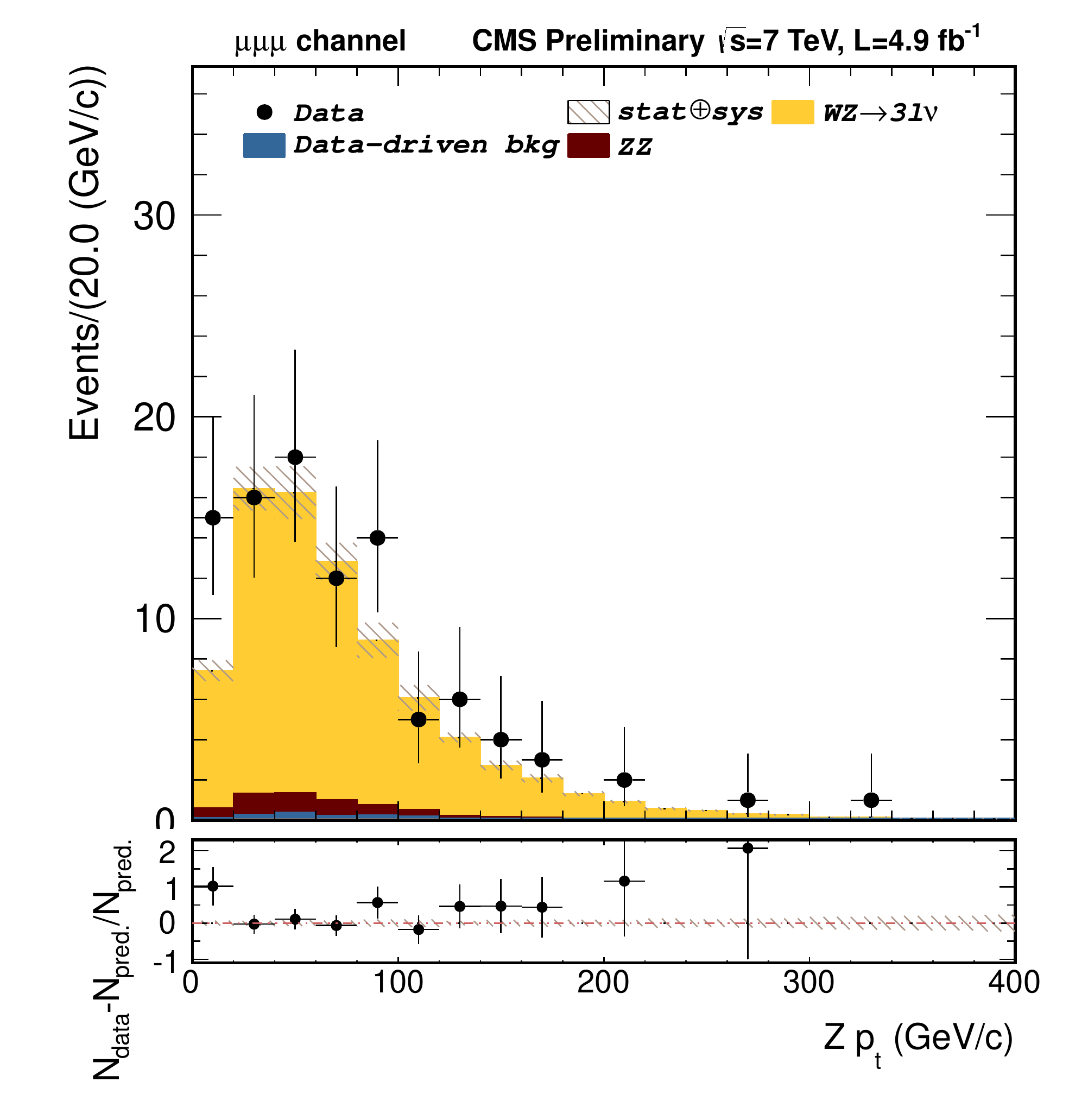}
	\end{subfigure}
	\caption[Transverse momentum of the dilepton system at 7~\TeV]
	{Transverse momentum of the Z-candidate dilepton system for the measured channels 
	$eee$, $\mu ee$, $e\mu\mu$ and $\mu\mu\mu$ (from left to right) and
	after each analysis selection stage: after Z-candidate requirement (up row), after 
	W-candidate, without the \MET cut (middle row) and after W-candidate including \MET
	cut (bottom row).}
\end{sidewaysfigure}

\begin{sidewaysfigure}[!htpb]
	\centering
	\begin{subfigure}[b]{0.2\textwidth}
		\includegraphics[width=\textwidth]{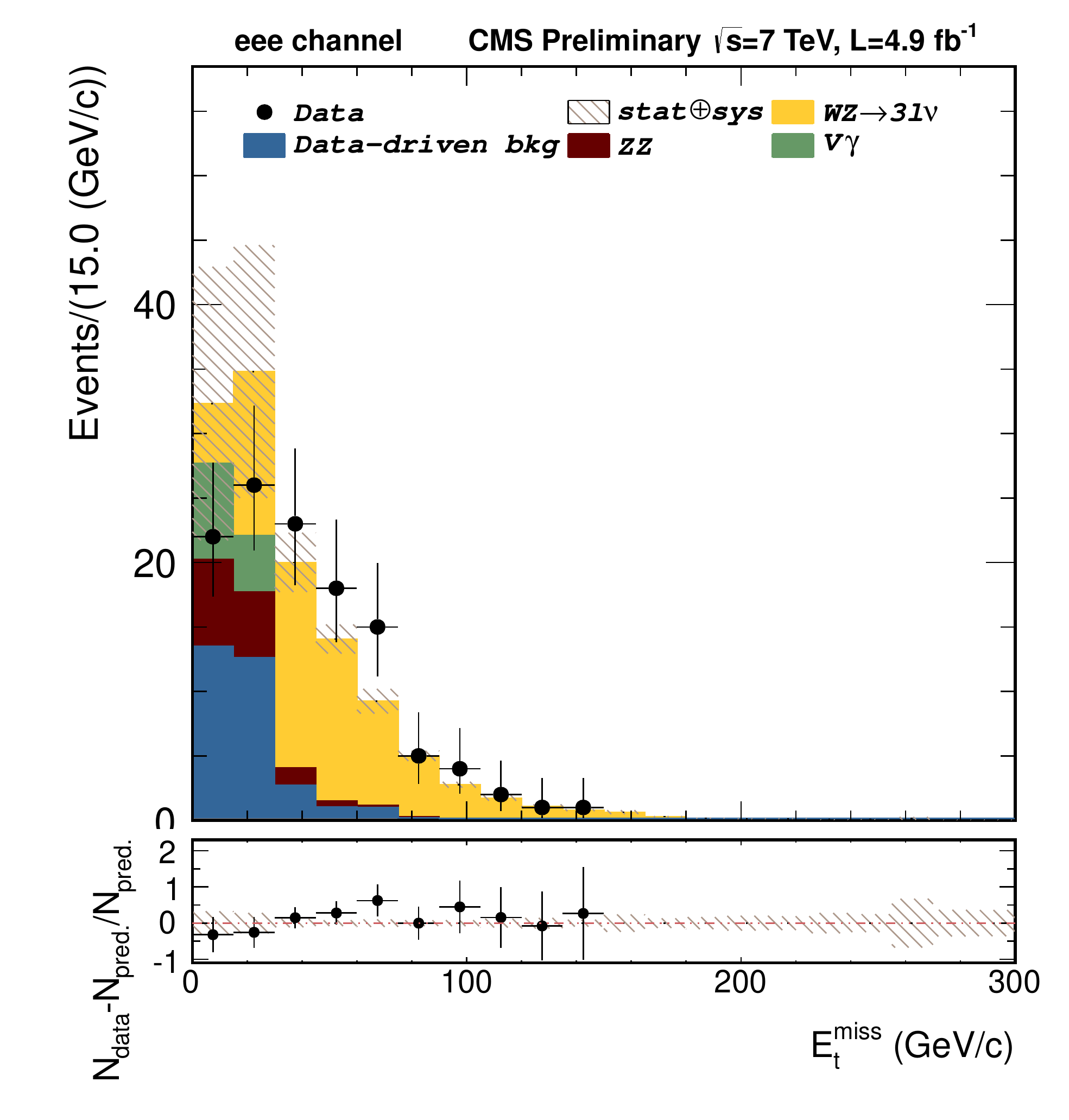}
	\end{subfigure}\quad
	\begin{subfigure}[b]{0.2\textwidth}
		\includegraphics[width=\textwidth]{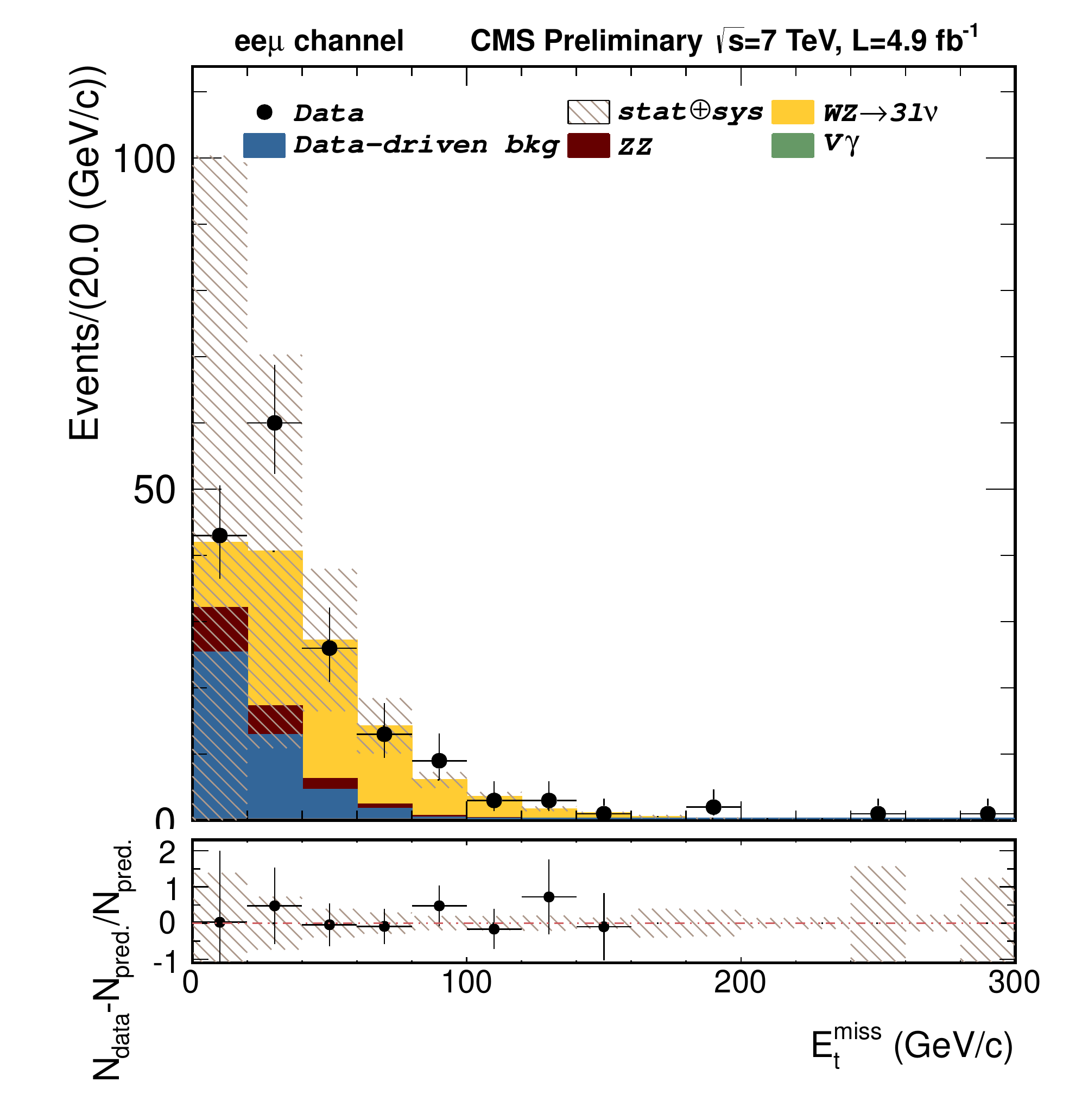}
	\end{subfigure}\quad
	\begin{subfigure}[b]{0.2\textwidth}
		\includegraphics[width=\textwidth]{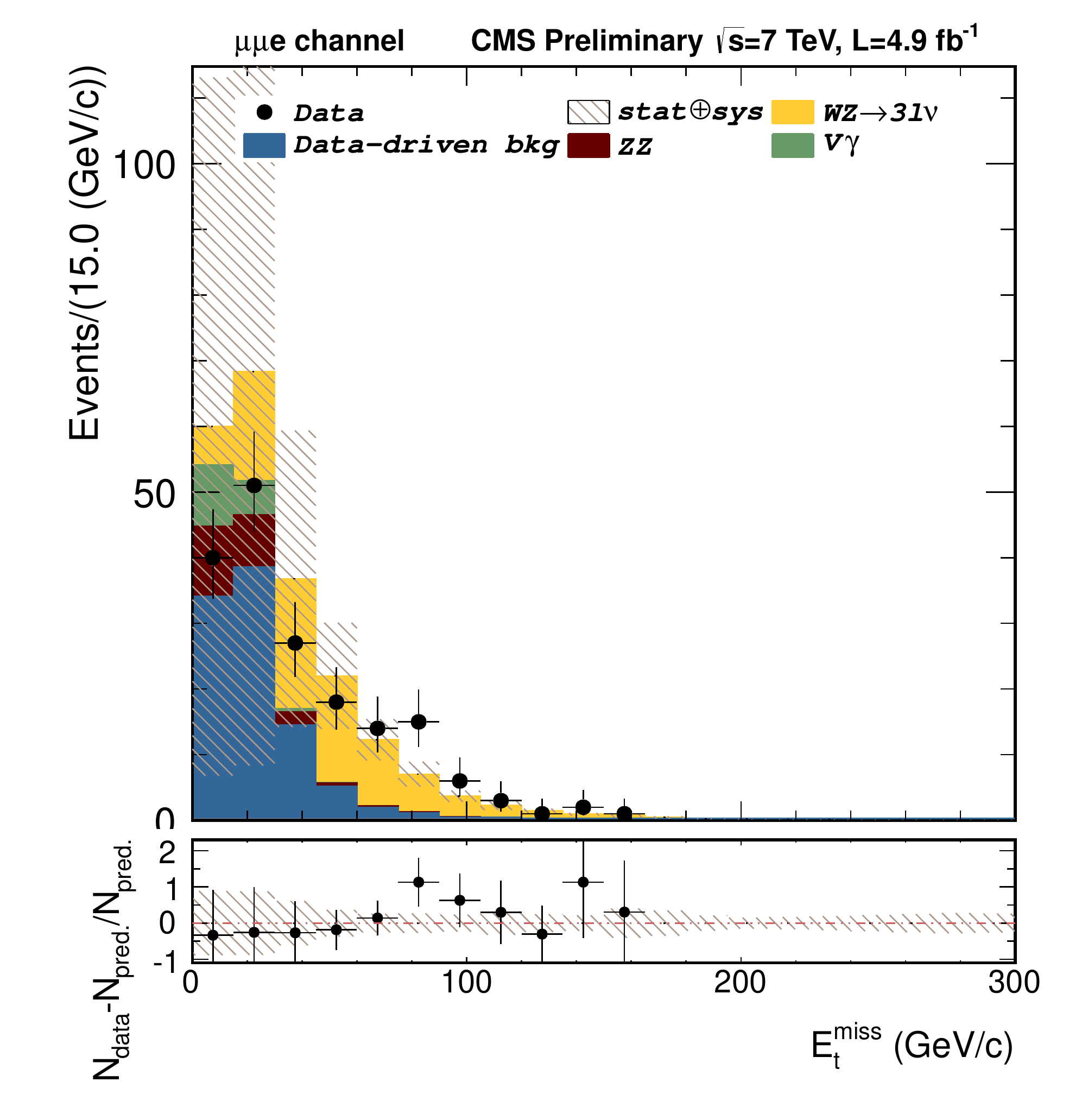}
	\end{subfigure}\quad
	\begin{subfigure}[b]{0.2\textwidth}
		\includegraphics[width=\textwidth]{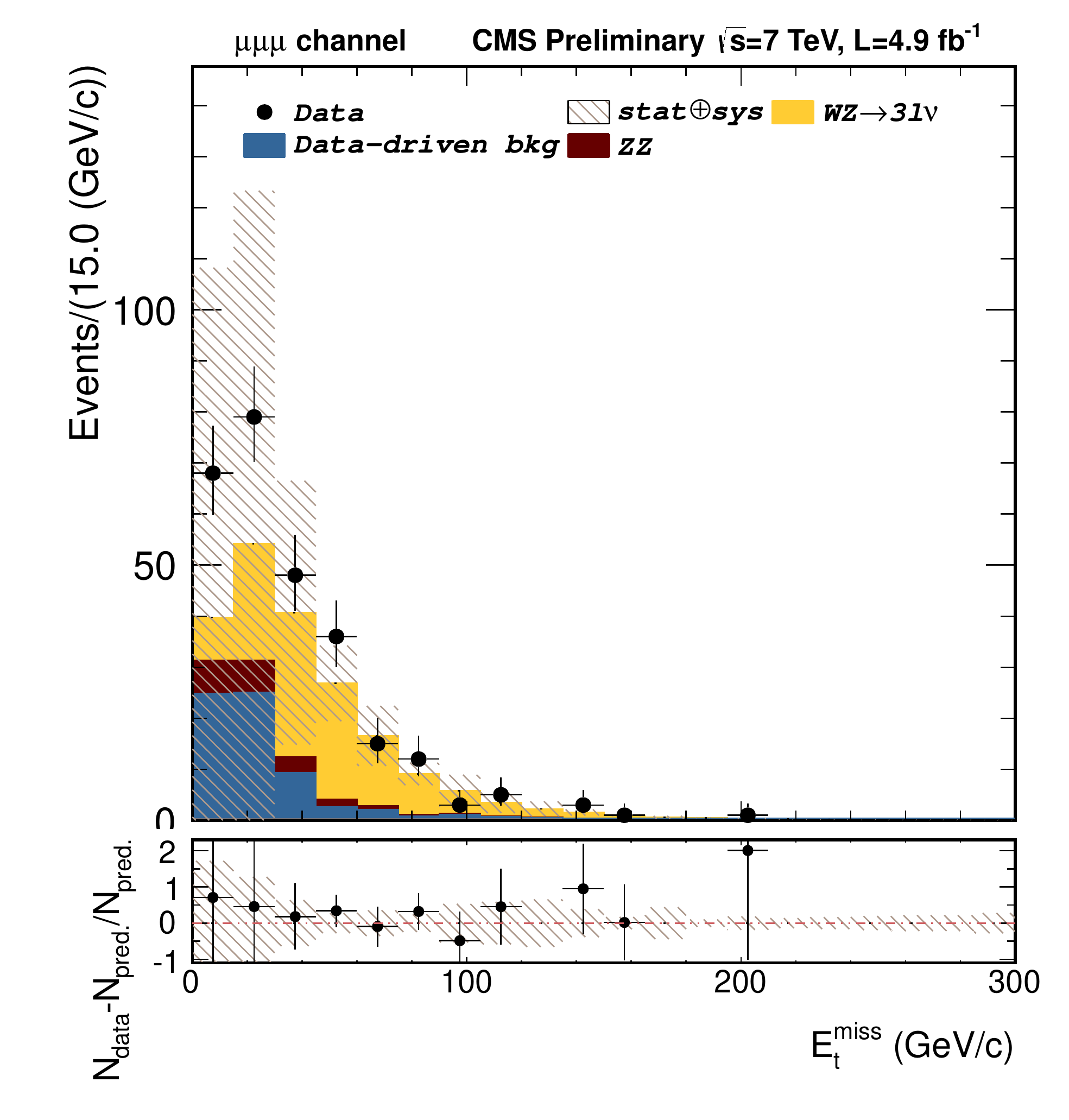}
	\end{subfigure}
	\vskip 1ex
	\centering
	\begin{subfigure}[b]{0.2\textwidth}
		\includegraphics[width=\textwidth]{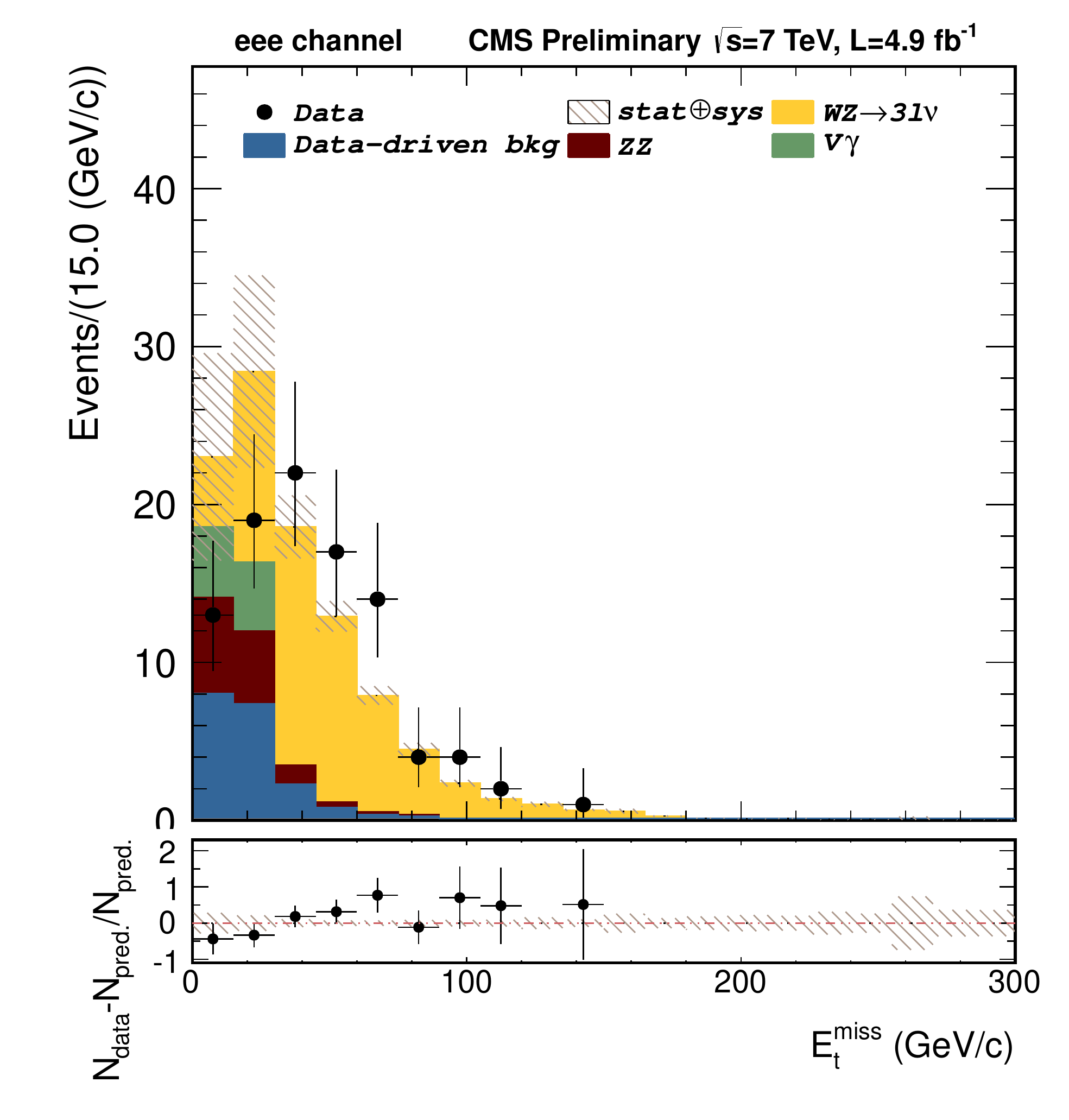}
	\end{subfigure}\quad
	\begin{subfigure}[b]{0.2\textwidth}
		\includegraphics[width=\textwidth]{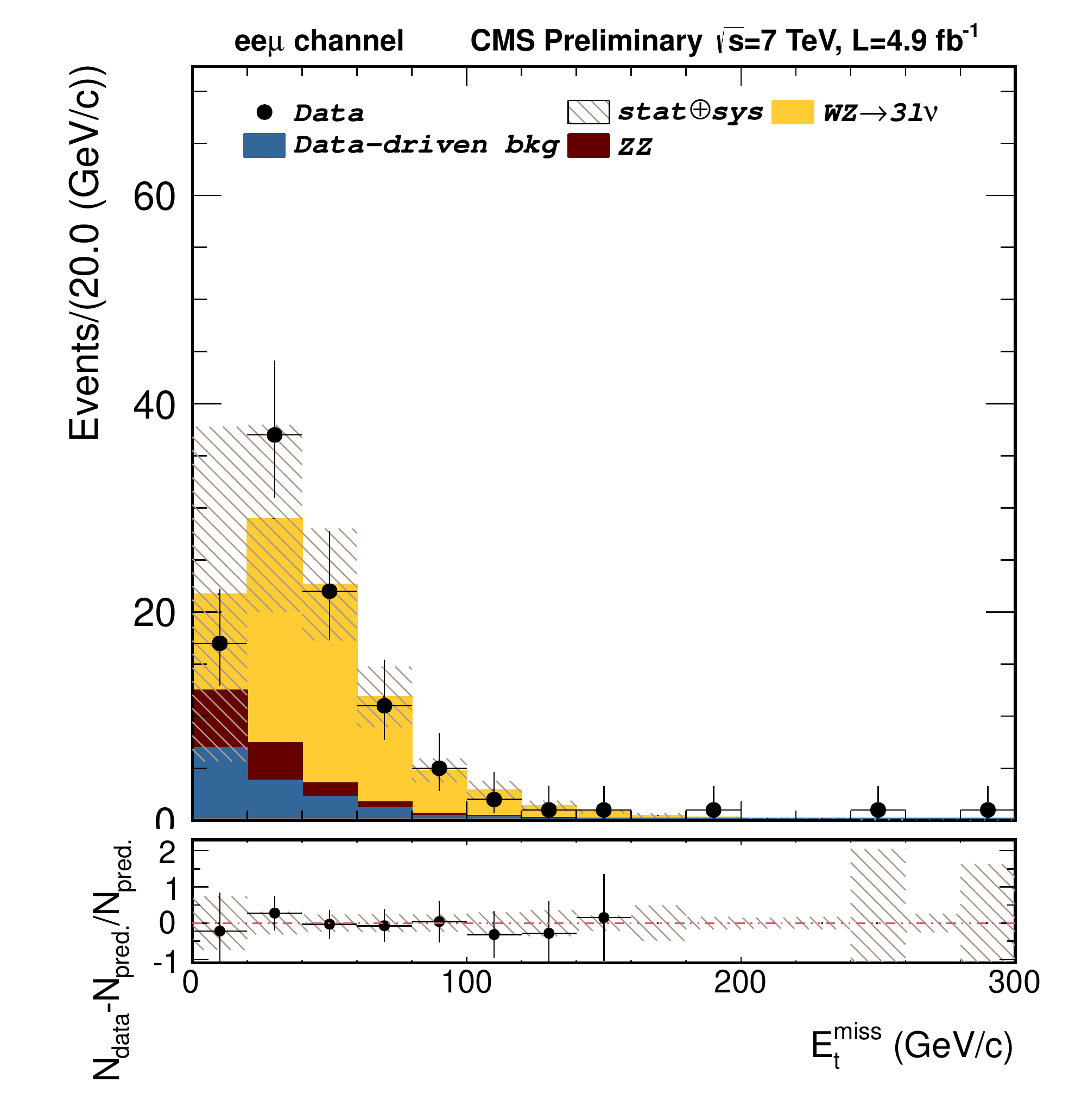}
	\end{subfigure}\quad
	\begin{subfigure}[b]{0.2\textwidth}
		\includegraphics[width=\textwidth]{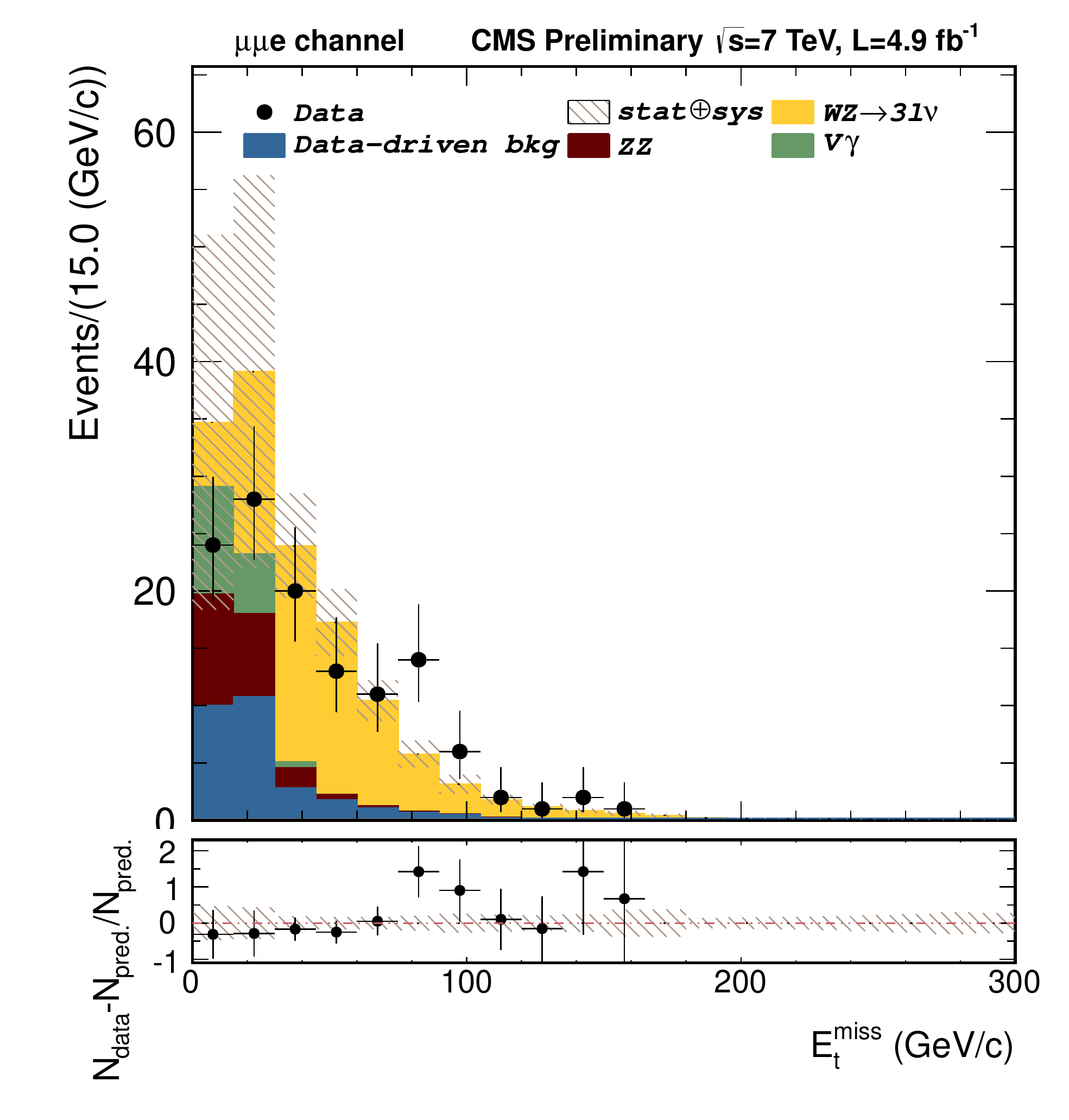}
	\end{subfigure}\quad
	\begin{subfigure}[b]{0.2\textwidth}
		\includegraphics[width=\textwidth]{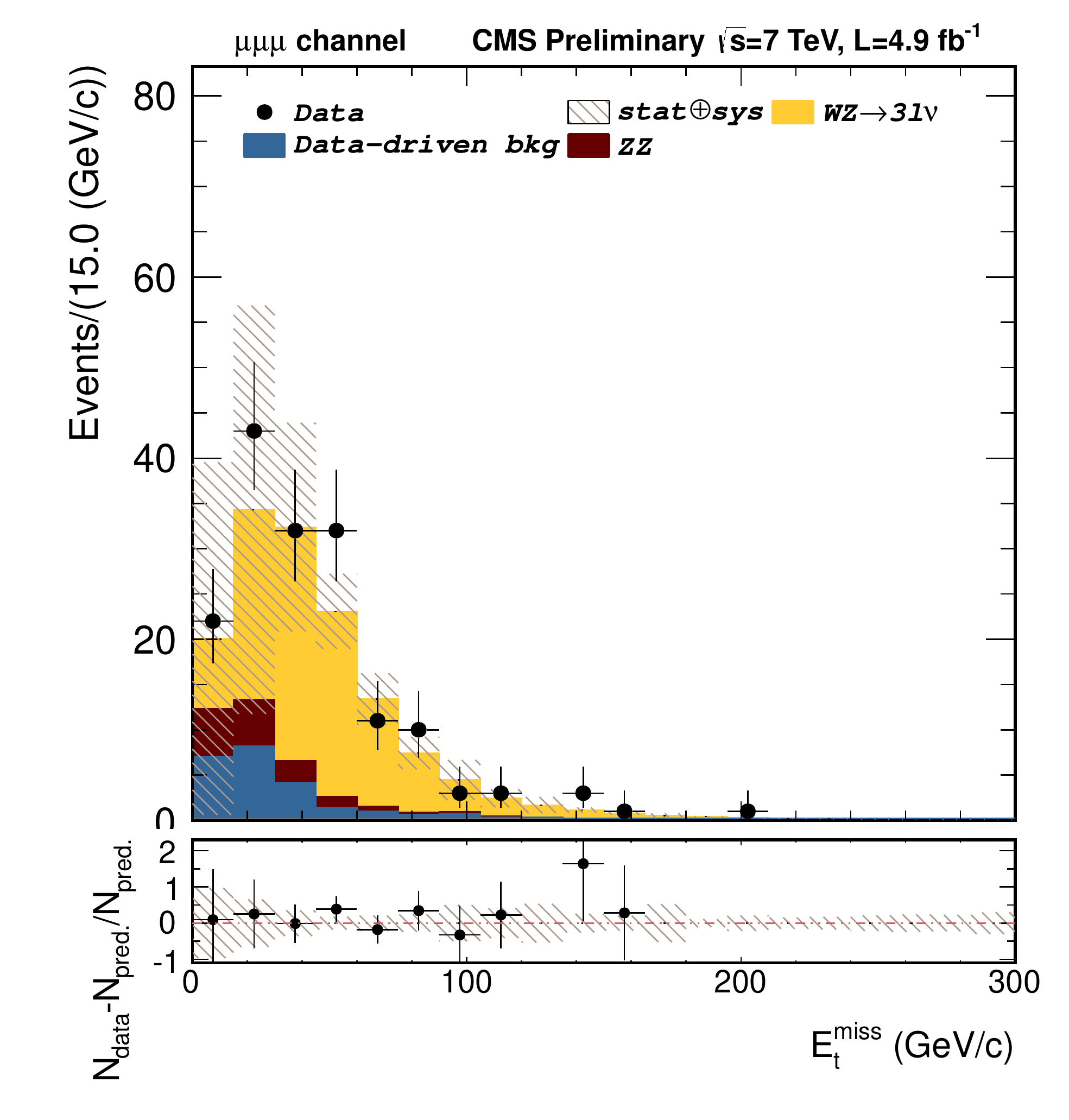}
	\end{subfigure}
	\vskip 1ex
	\begin{subfigure}[b]{0.2\textwidth}
		\includegraphics[width=\textwidth]{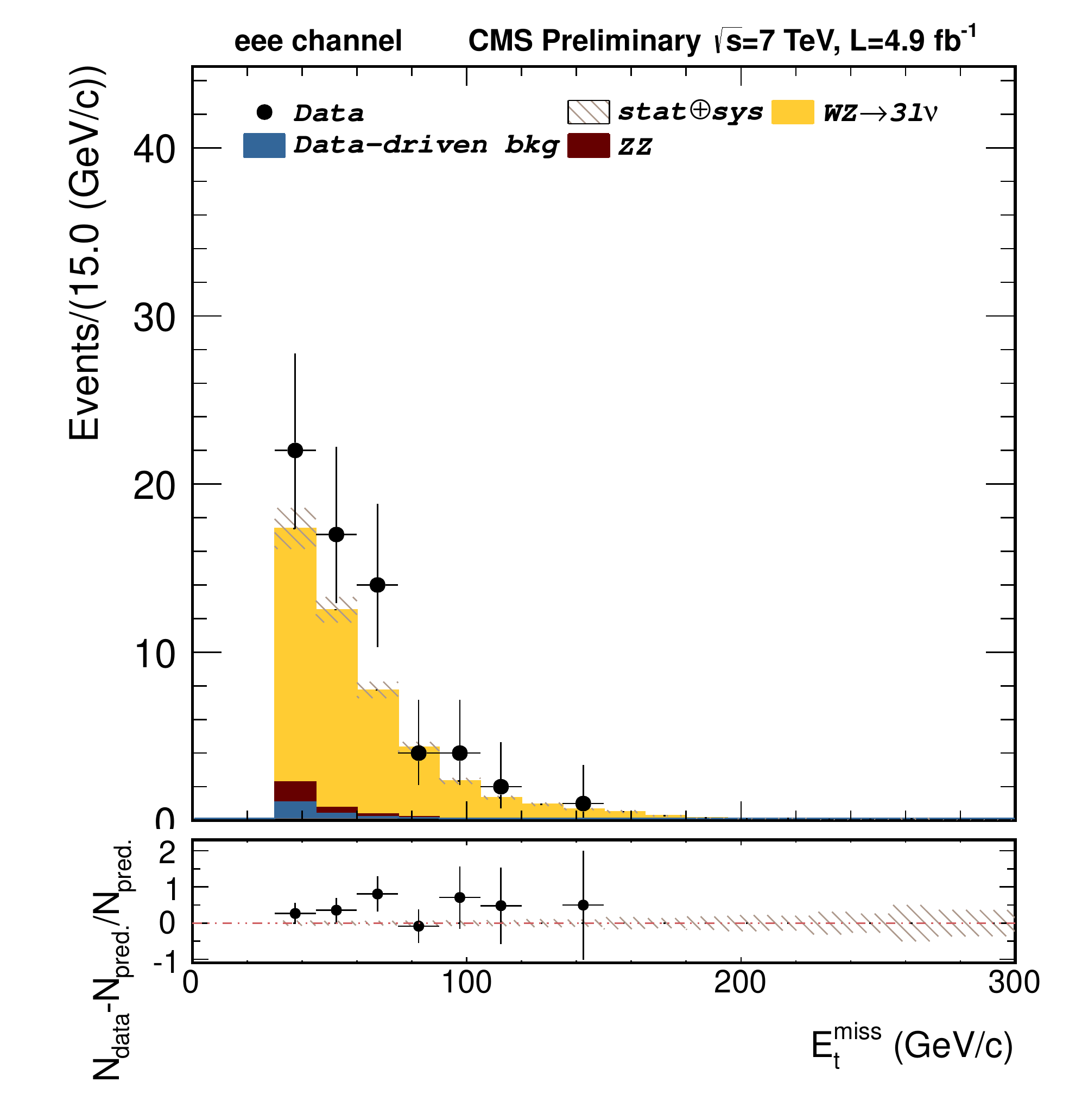}
	\end{subfigure}\quad
	\begin{subfigure}[b]{0.2\textwidth}
		\includegraphics[width=\textwidth]{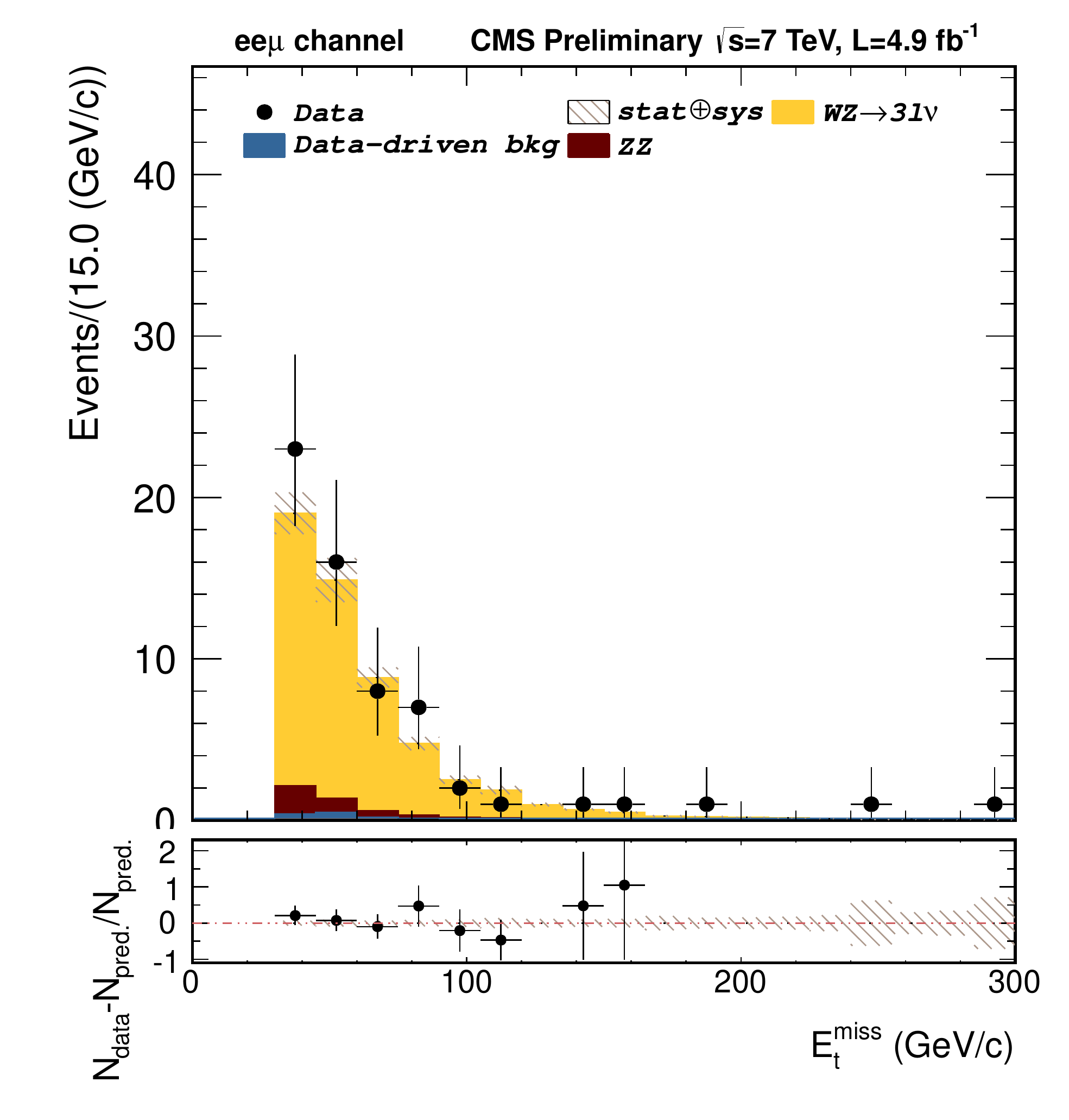}
	\end{subfigure}\quad
	\begin{subfigure}[b]{0.2\textwidth}
		\includegraphics[width=\textwidth]{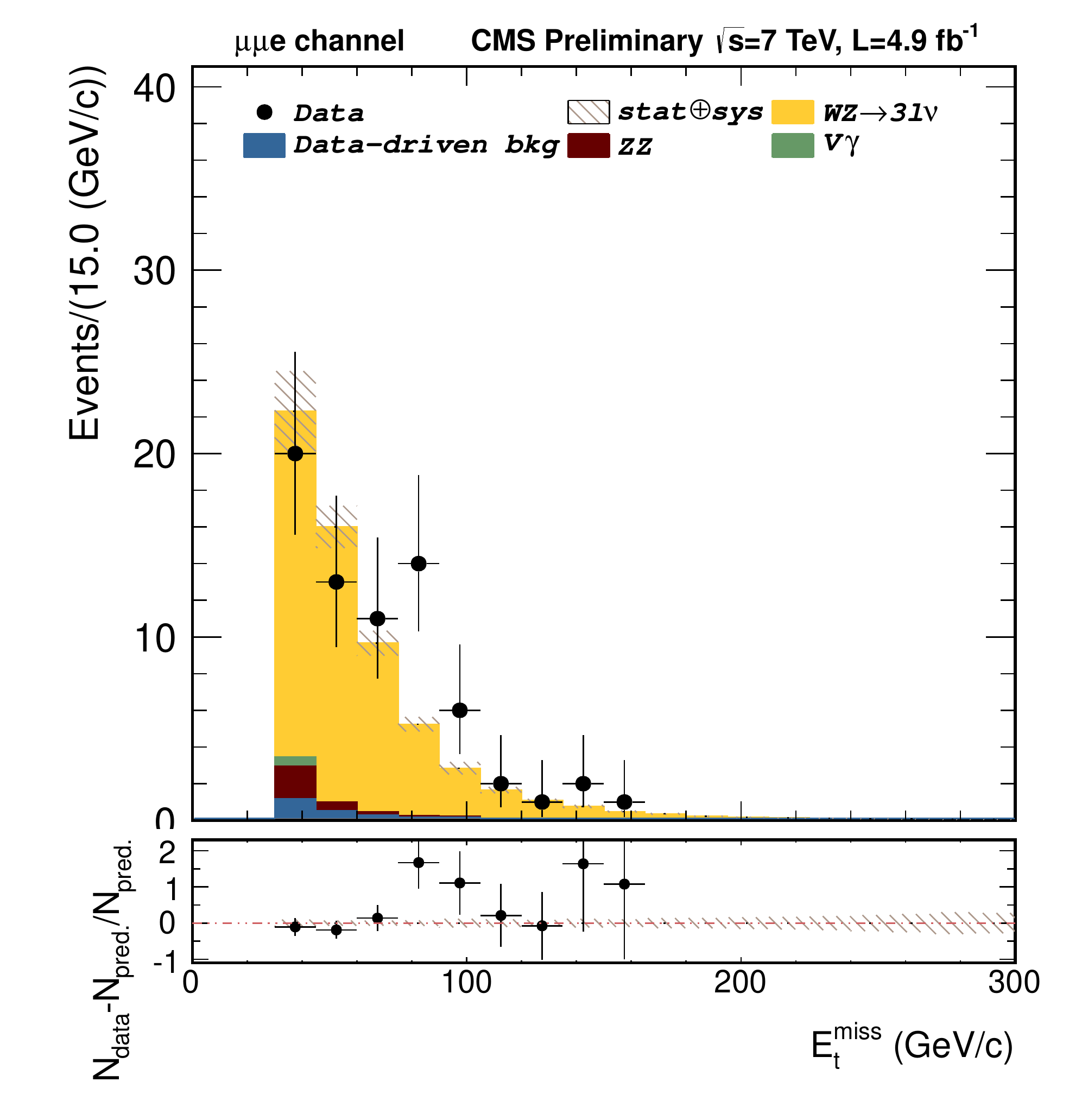}
	\end{subfigure}\quad
	\begin{subfigure}[b]{0.2\textwidth}
		\includegraphics[width=\textwidth]{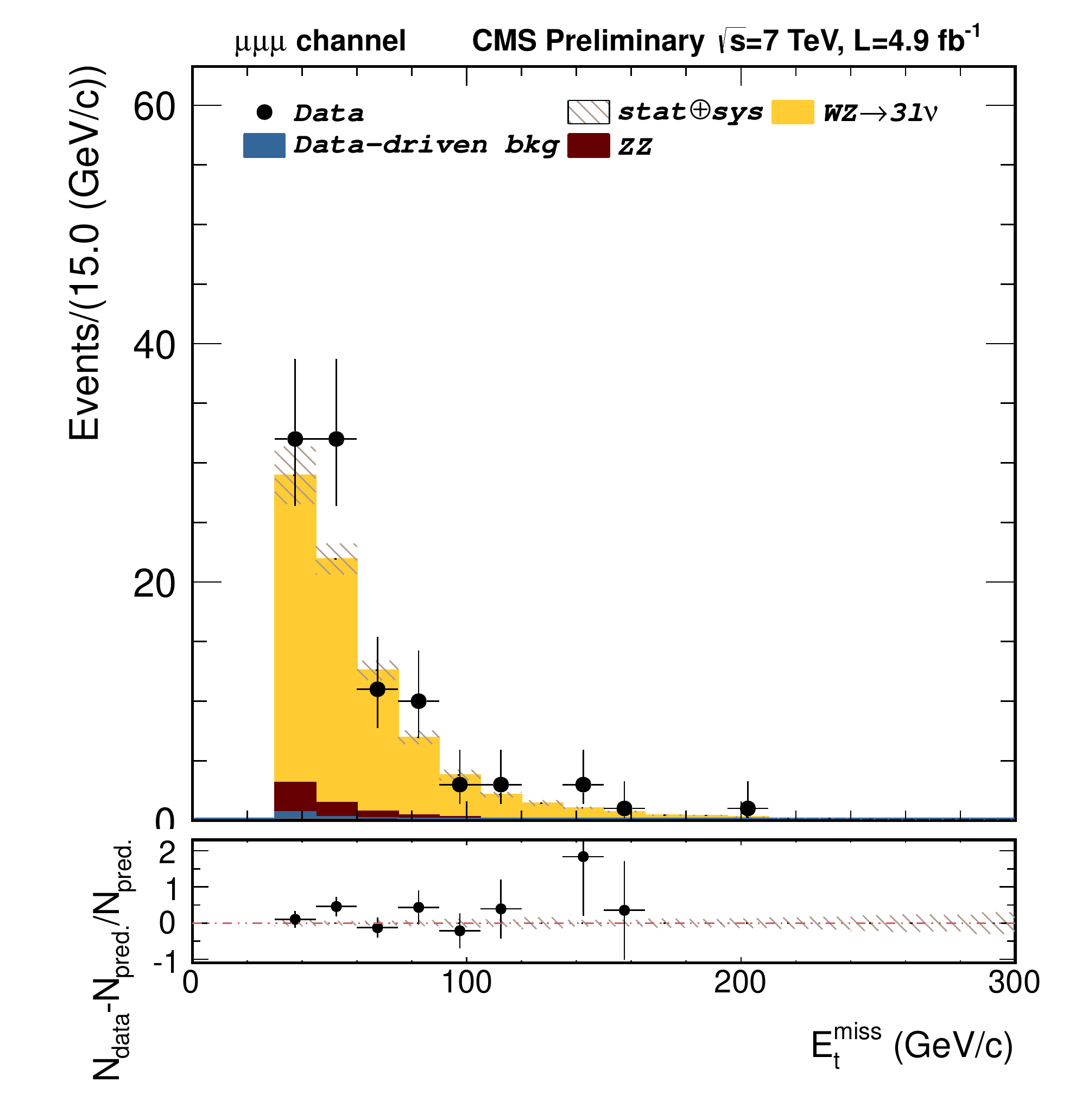}
	\end{subfigure}
	\caption[Missing transverse energy at 7~\TeV]{Missing 
	energy in the transverse plane at each event for the measured channels 
	$eee$, $\mu ee$, $e\mu\mu$ and $\mu\mu\mu$ (from left to right) and
	after each analysis selection stage: after Z-candidate requirement (up row), after 
	W-candidate, without the \MET cut (middle row) and after W-candidate including \MET
	cut (bottom row).}
\end{sidewaysfigure}

\begin{sidewaysfigure}[!htpb]
	\centering
	\begin{subfigure}[b]{0.2\textwidth}
		\includegraphics[width=\textwidth]{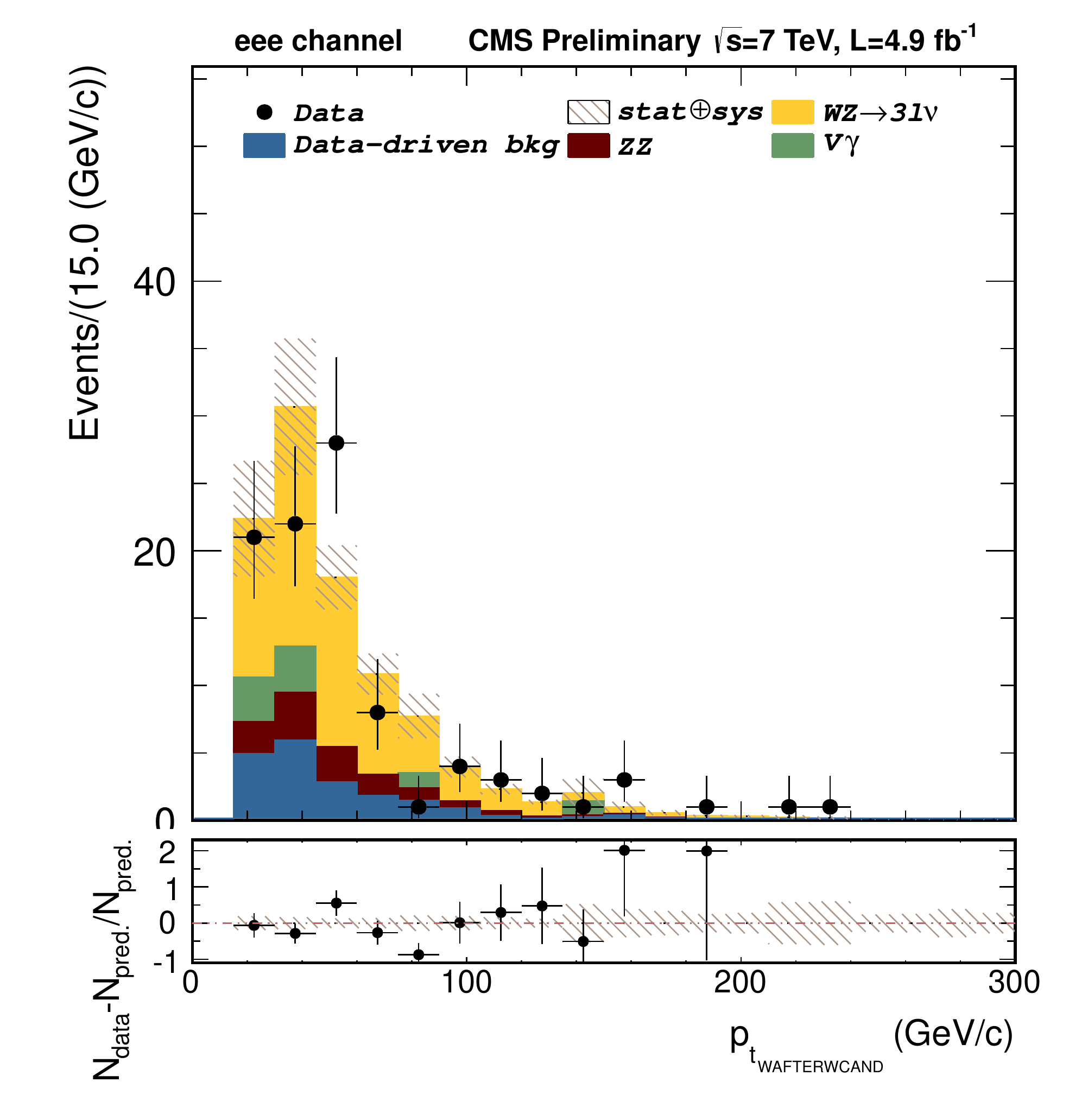}
	\end{subfigure}\quad
	\begin{subfigure}[b]{0.2\textwidth}
		\includegraphics[width=\textwidth]{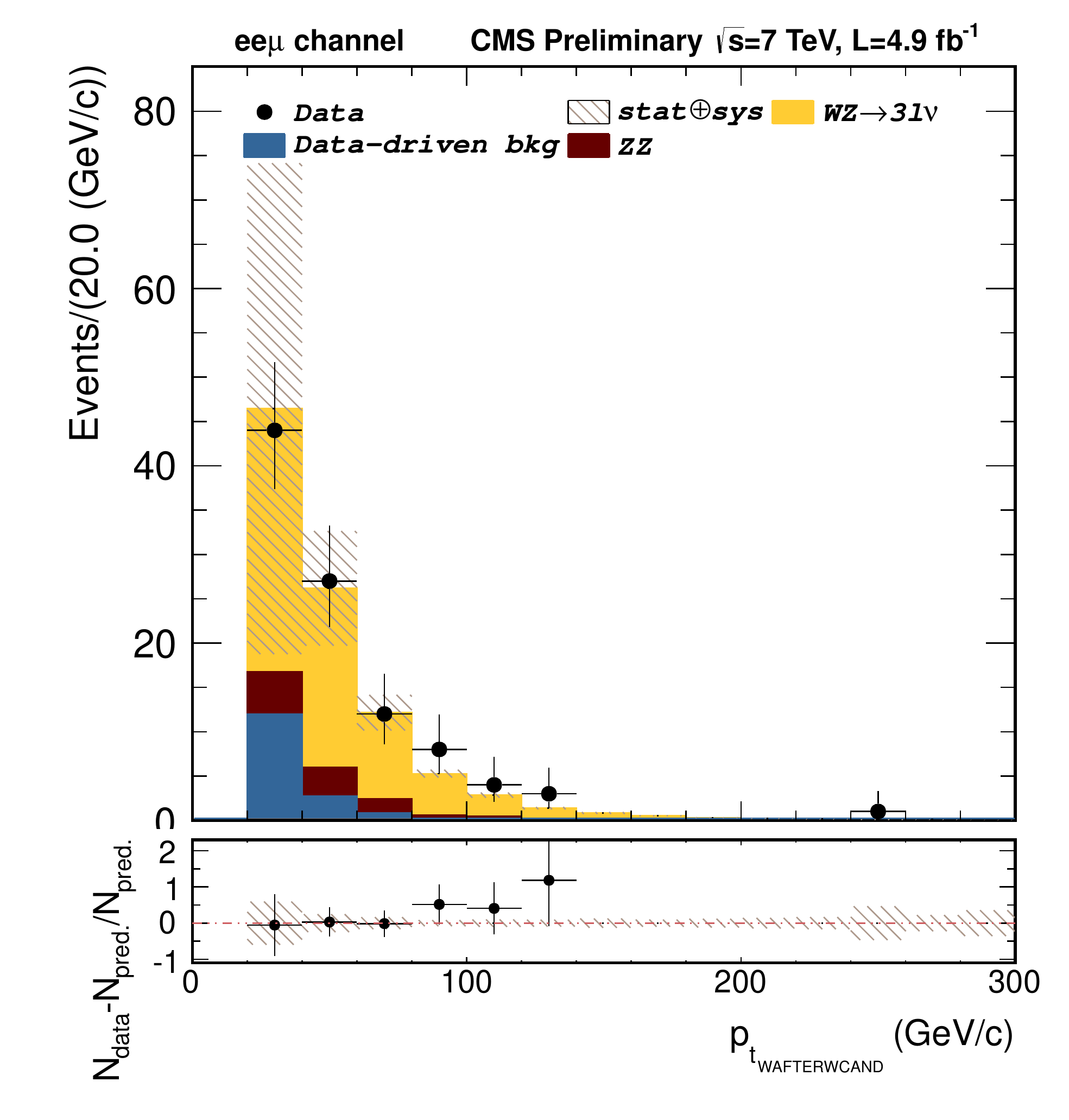}
	\end{subfigure}\quad
	\begin{subfigure}[b]{0.2\textwidth}
		\includegraphics[width=\textwidth]{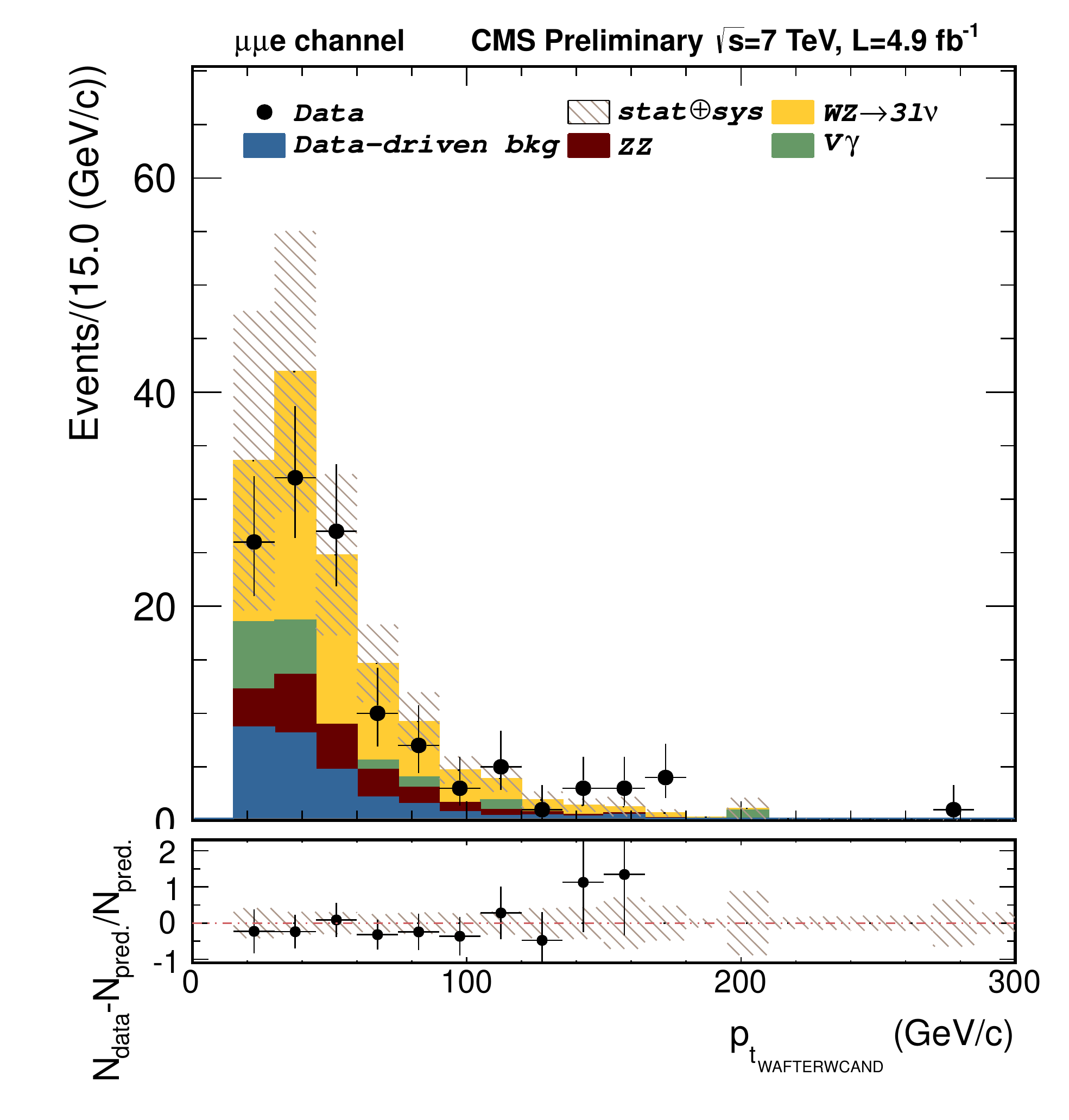}
	\end{subfigure}\quad
	\begin{subfigure}[b]{0.2\textwidth}
		\includegraphics[width=\textwidth]{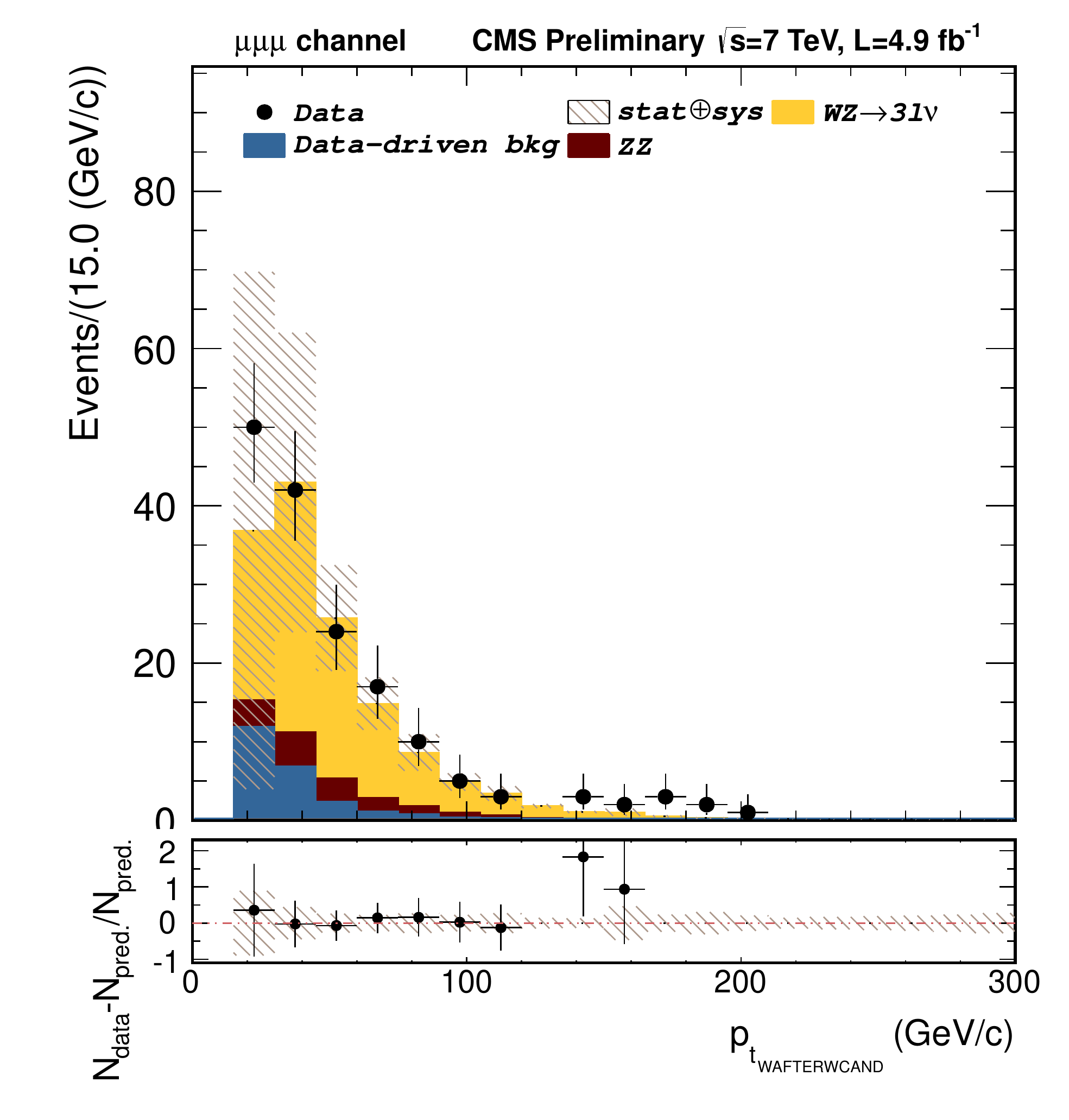}
	\end{subfigure}
	\vskip 1ex
	\begin{subfigure}[b]{0.2\textwidth}
		\includegraphics[width=\textwidth]{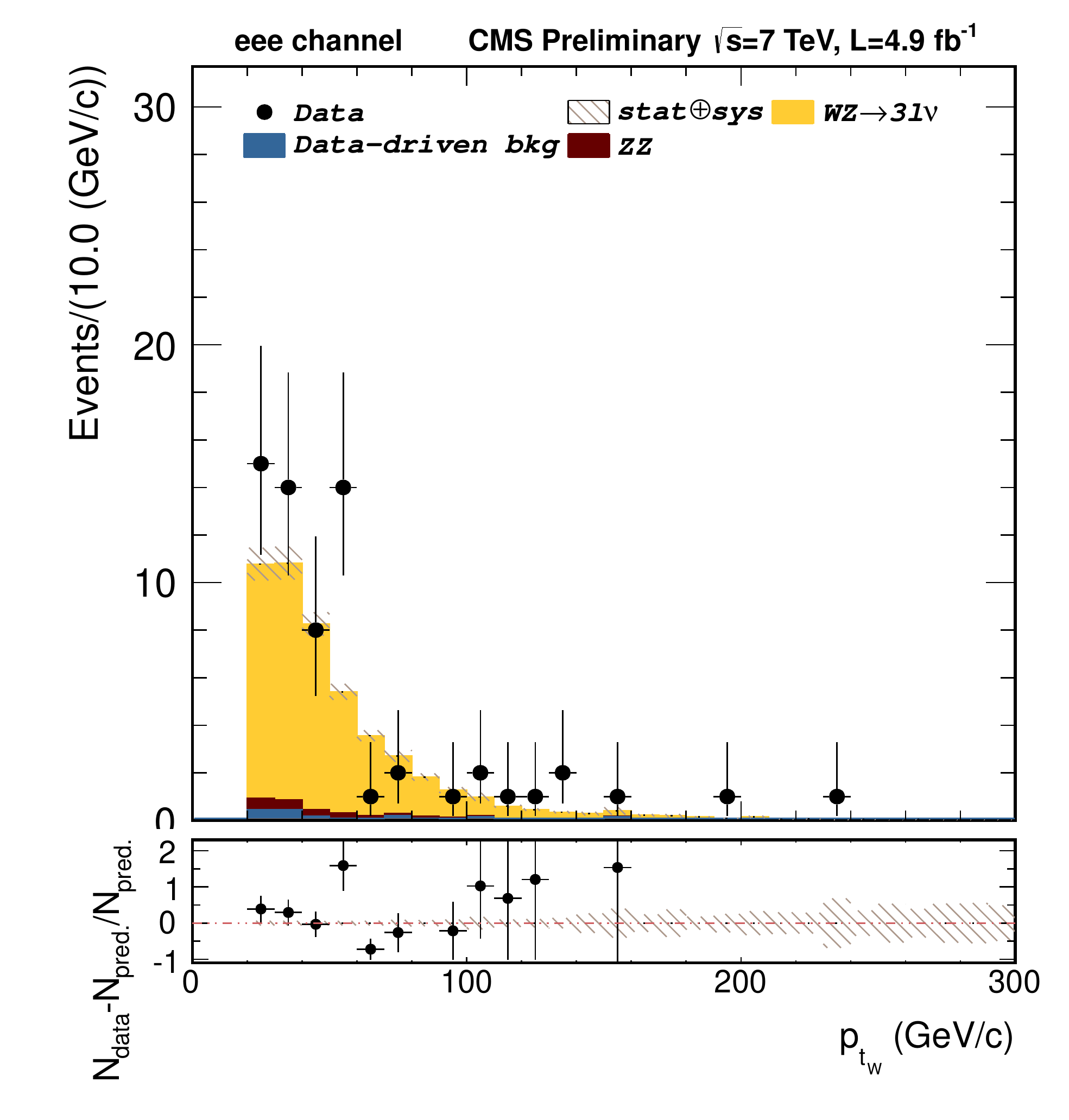}
	\end{subfigure}\quad
	\begin{subfigure}[b]{0.2\textwidth}
		\includegraphics[width=\textwidth]{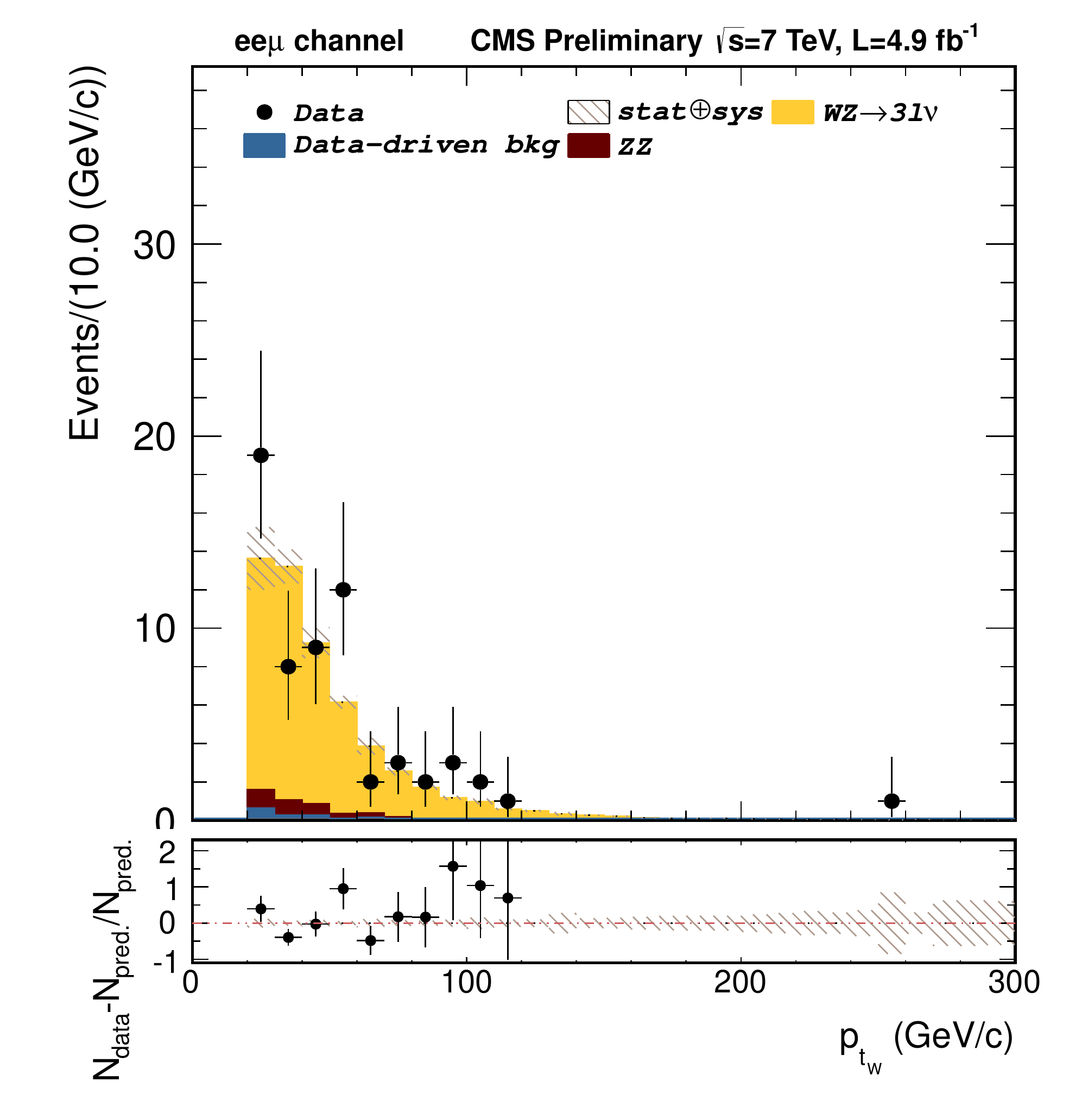}
	\end{subfigure}\quad
	\begin{subfigure}[b]{0.2\textwidth}
		\includegraphics[width=\textwidth]{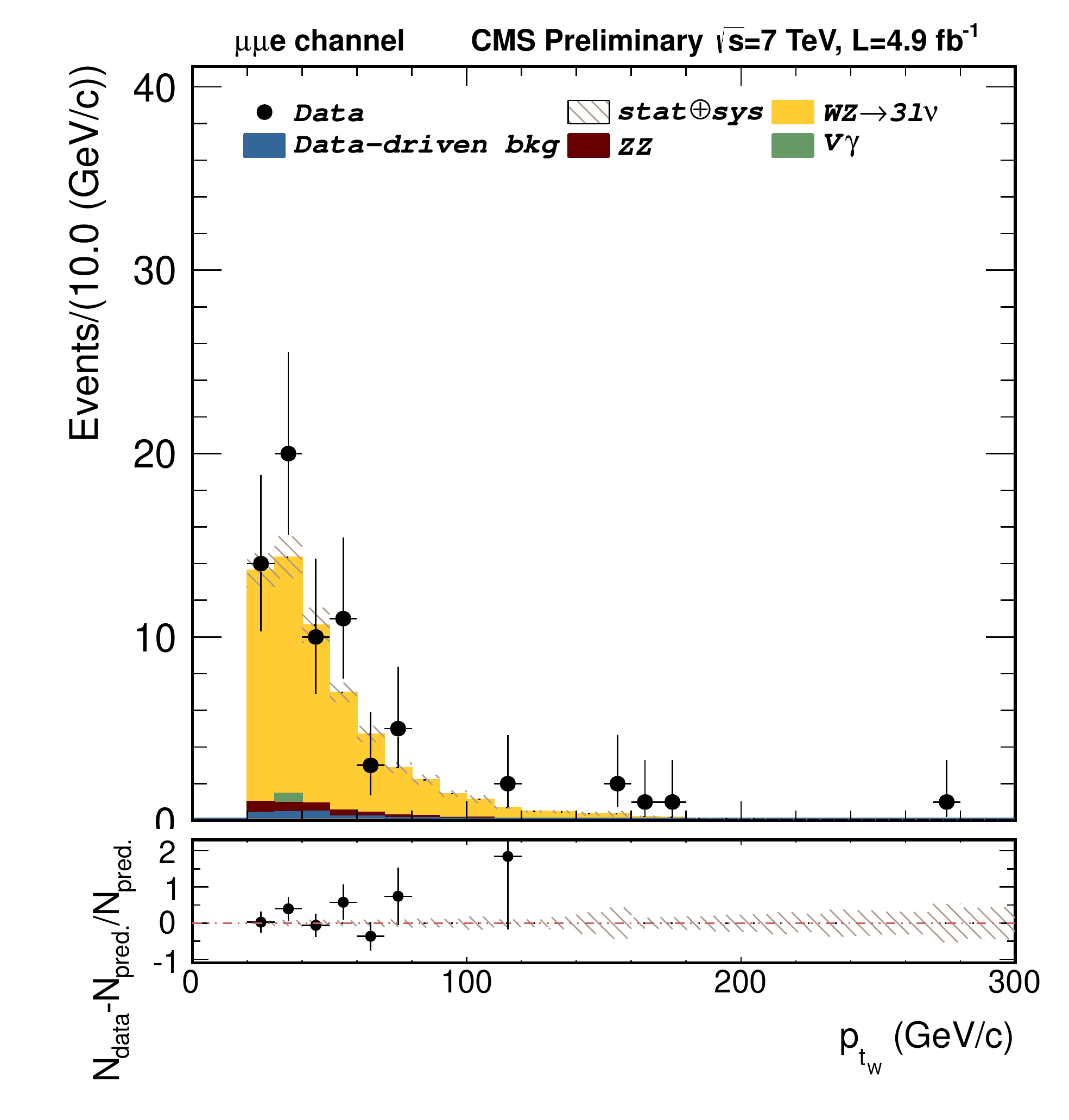}
	\end{subfigure}\quad
	\begin{subfigure}[b]{0.2\textwidth}
		\includegraphics[width=\textwidth]{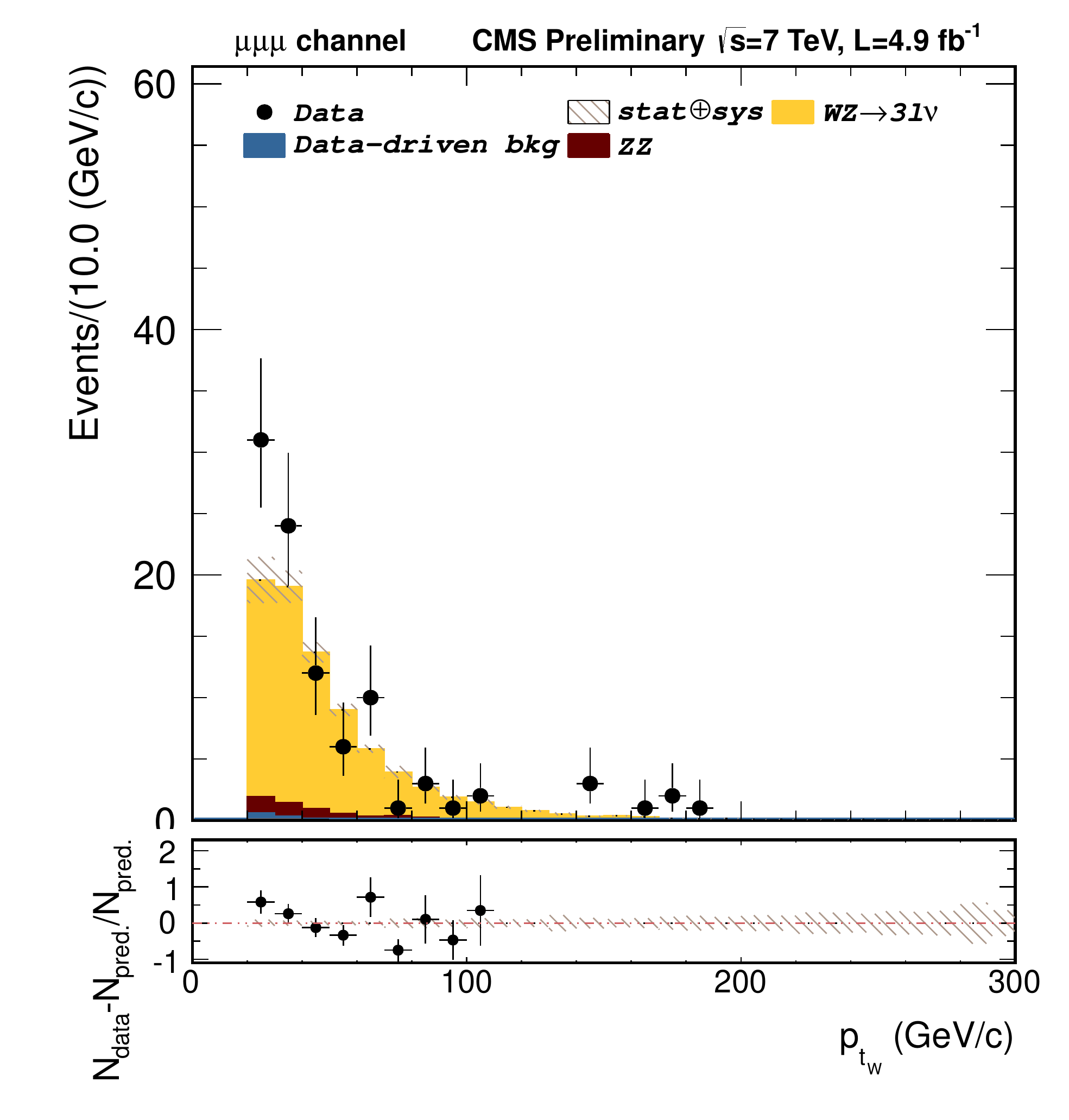}
	\end{subfigure}
	\caption[Transverse momentum of the W-candidate system at 7~\TeV]
	{Transverse momentum of the W-candidate system composed by
	the third selected lepton and \MET at each event for the measured channels $eee$,
	$\mu ee$, $e\mu\mu$ and $\mu\mu\mu$ 
	(from left to right) and after each analysis selection stage (once the W is selected): 
	after W-candidate requirement without the \MET cut (up row) and after W-candidate including
	\MET cut (bottom row).}
\end{sidewaysfigure}

\begin{sidewaysfigure}[!htpb]
	\centering
	\begin{subfigure}[b]{0.2\textwidth}
		\includegraphics[width=\textwidth]{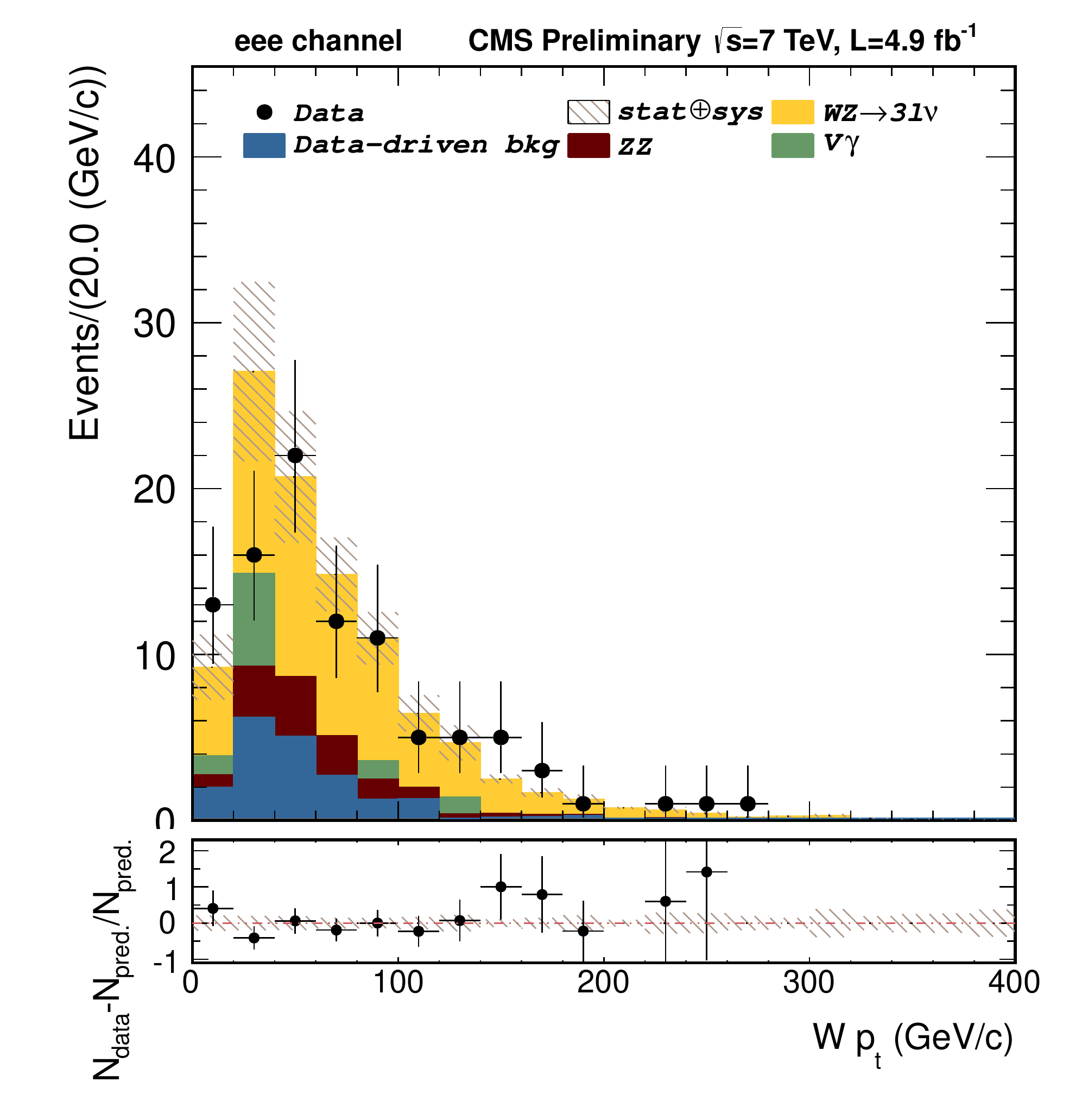}
	\end{subfigure}\quad
	\begin{subfigure}[b]{0.2\textwidth}
		\includegraphics[width=\textwidth]{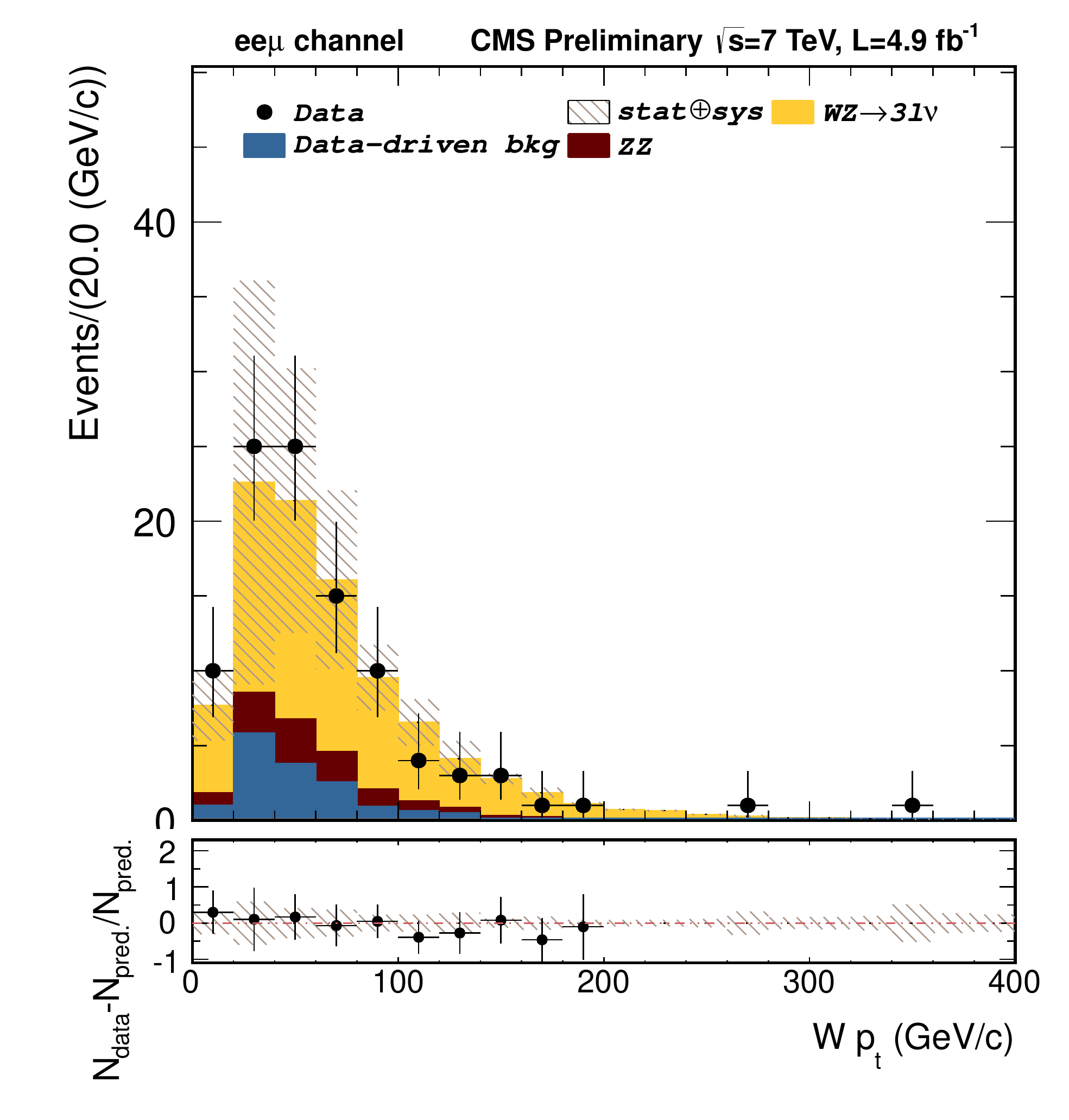}
	\end{subfigure}\quad
	\begin{subfigure}[b]{0.2\textwidth}
		\includegraphics[width=\textwidth]{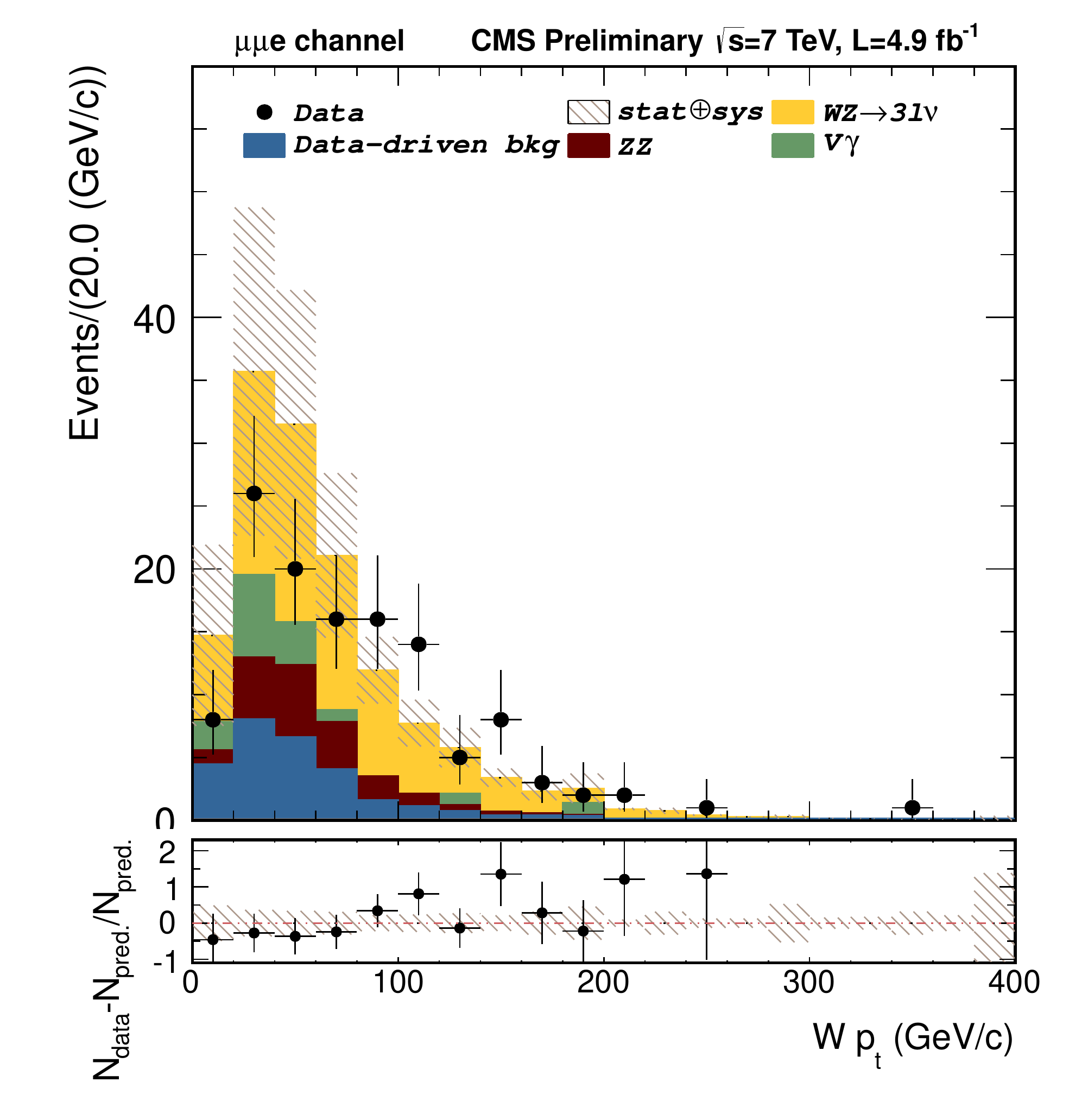}
	\end{subfigure}\quad
	\begin{subfigure}[b]{0.2\textwidth}
		\includegraphics[width=\textwidth]{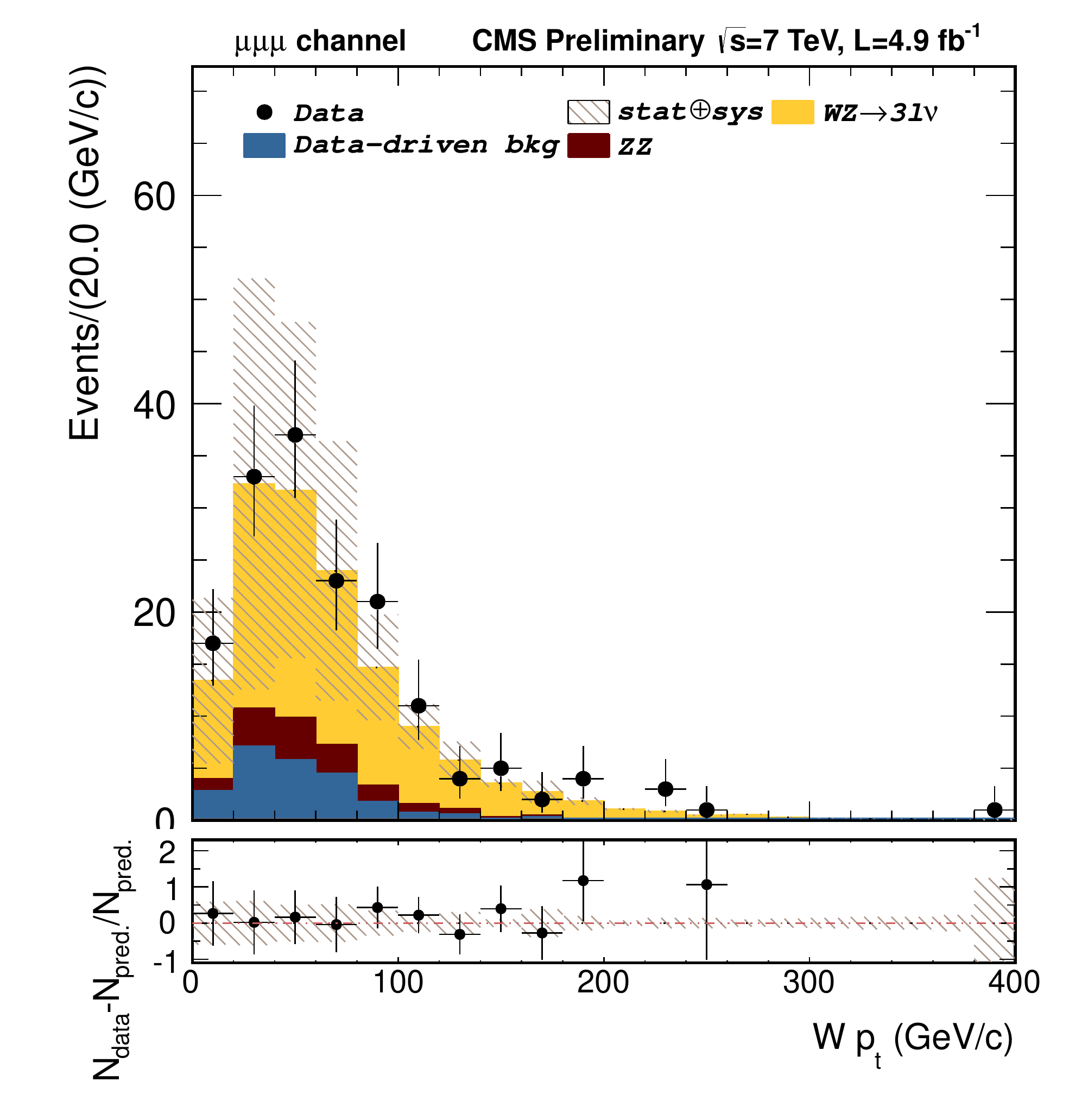}
	\end{subfigure}
	\vskip 1ex
	\begin{subfigure}[b]{0.2\textwidth}
		\includegraphics[width=\textwidth]{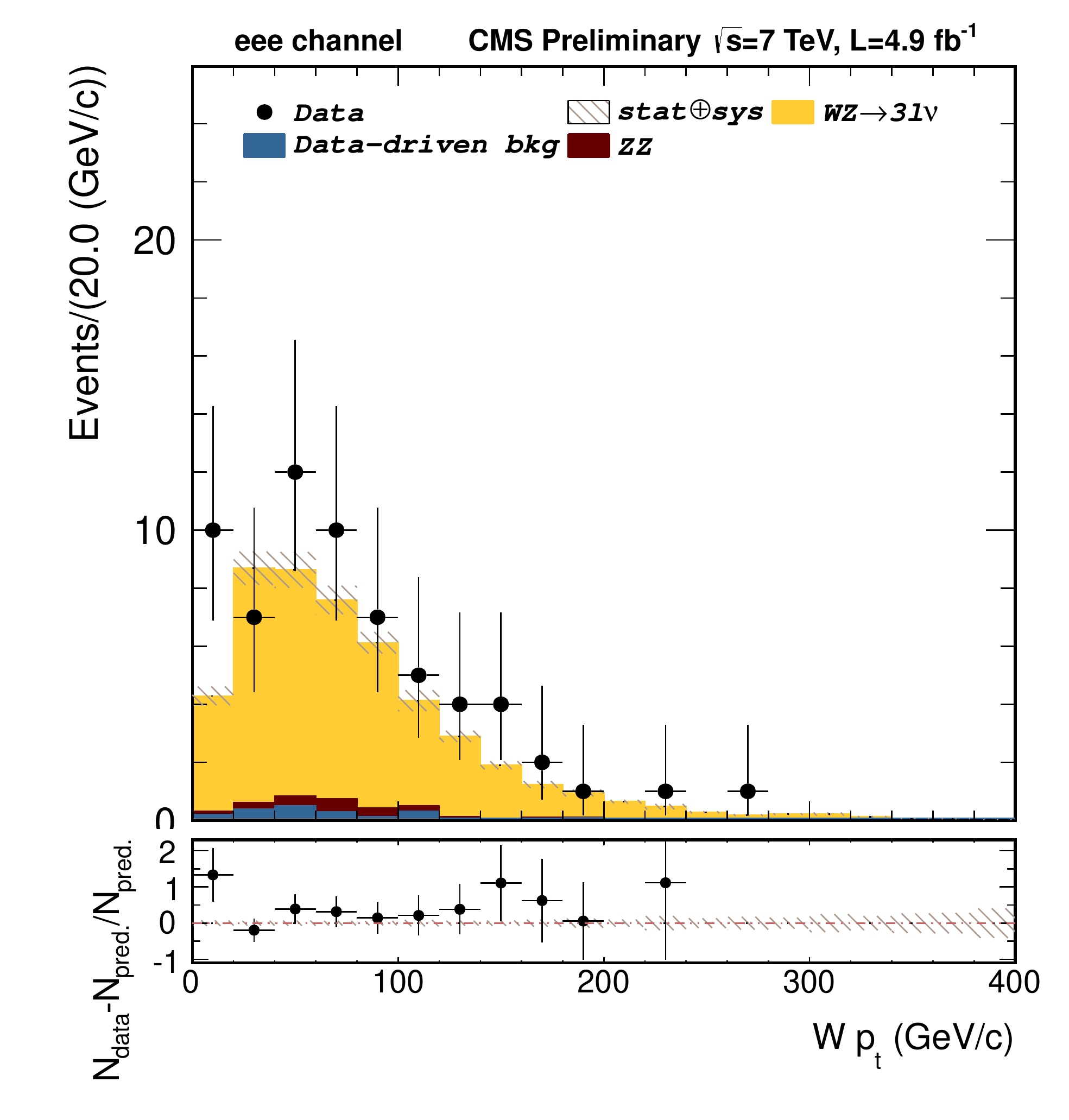}
	\end{subfigure}\quad
	\begin{subfigure}[b]{0.2\textwidth}
		\includegraphics[width=\textwidth]{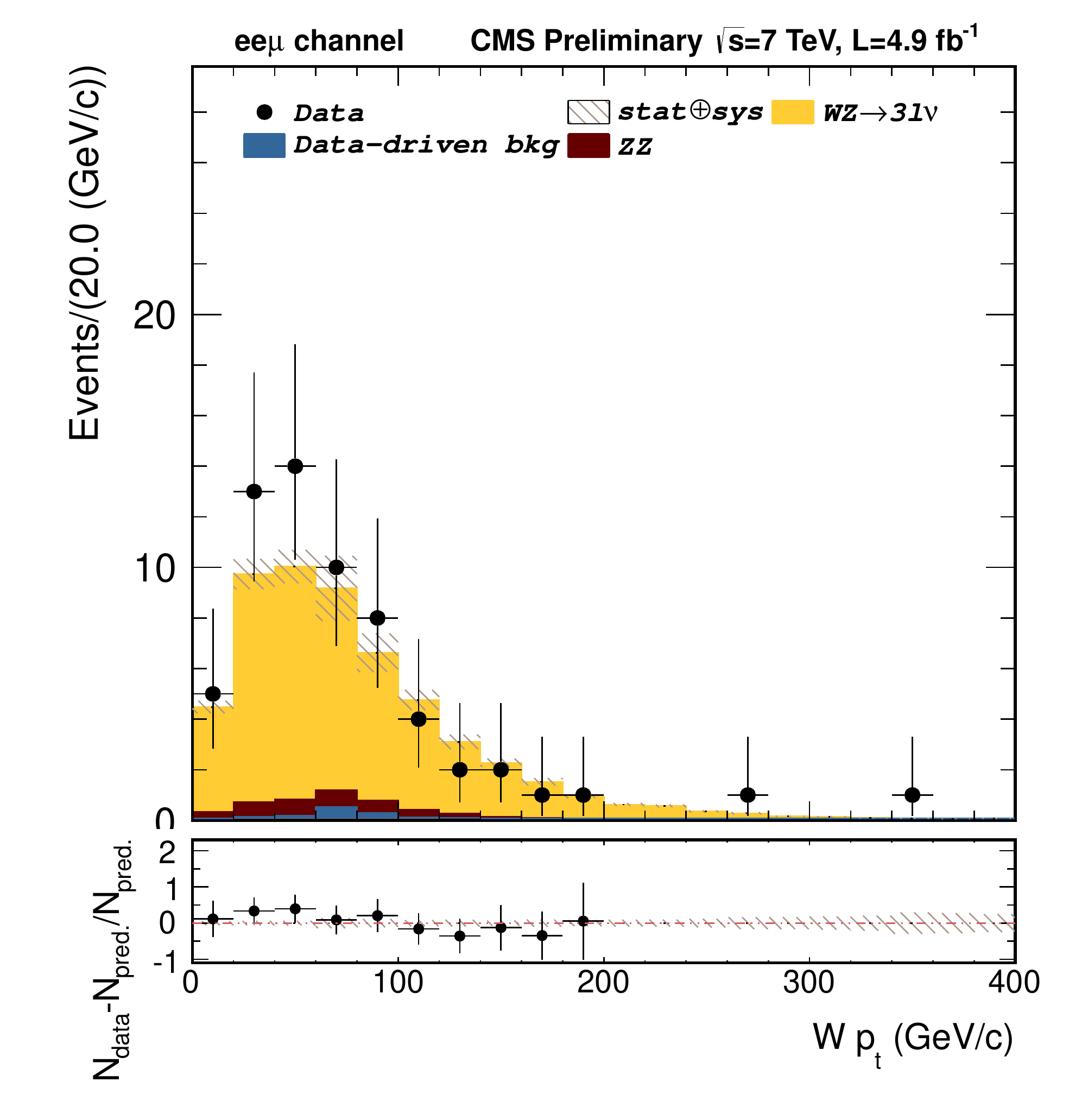}
	\end{subfigure}\quad
	\begin{subfigure}[b]{0.2\textwidth}
		\includegraphics[width=\textwidth]{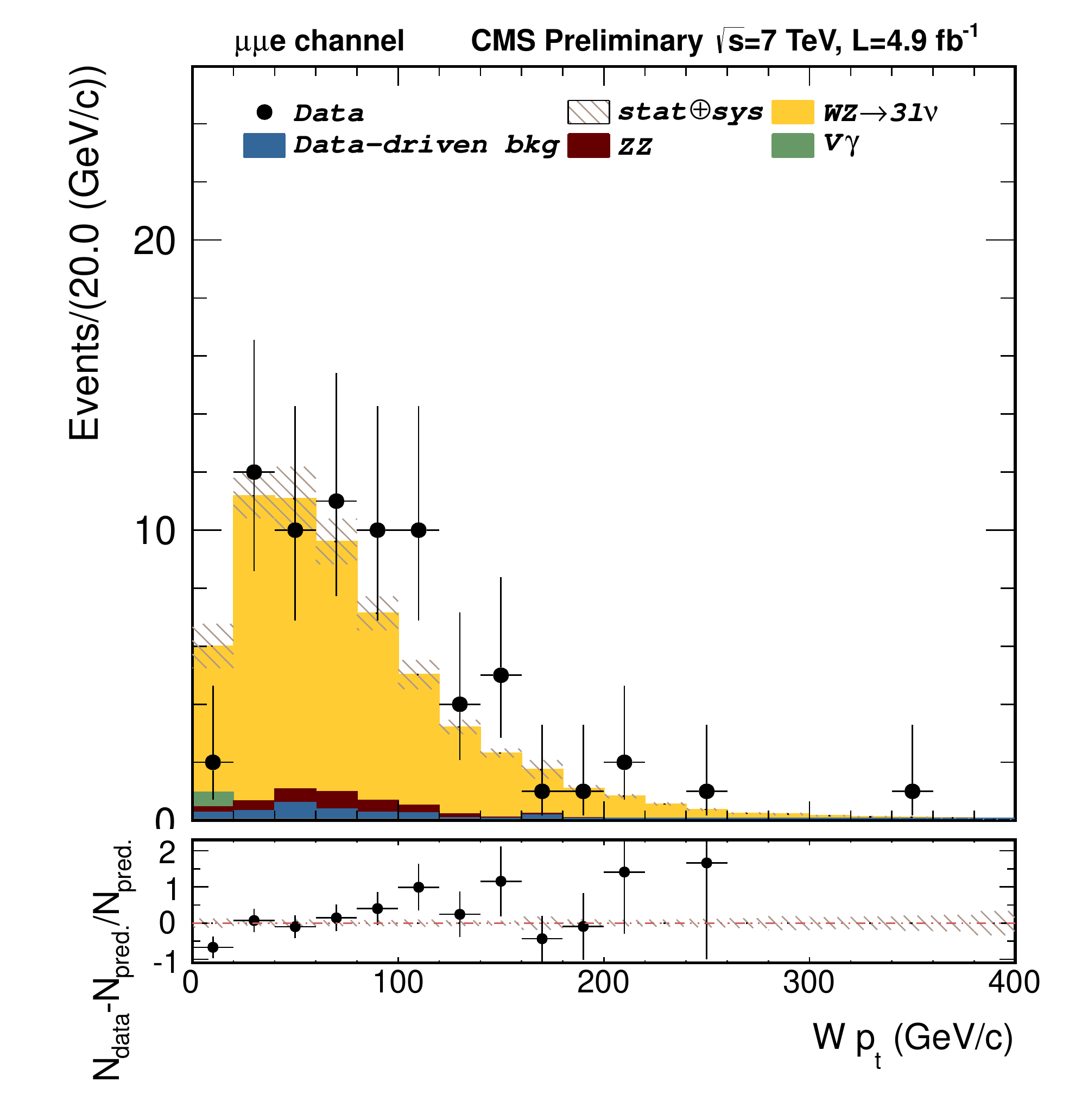}
	\end{subfigure}\quad
	\begin{subfigure}[b]{0.2\textwidth}
		\includegraphics[width=\textwidth]{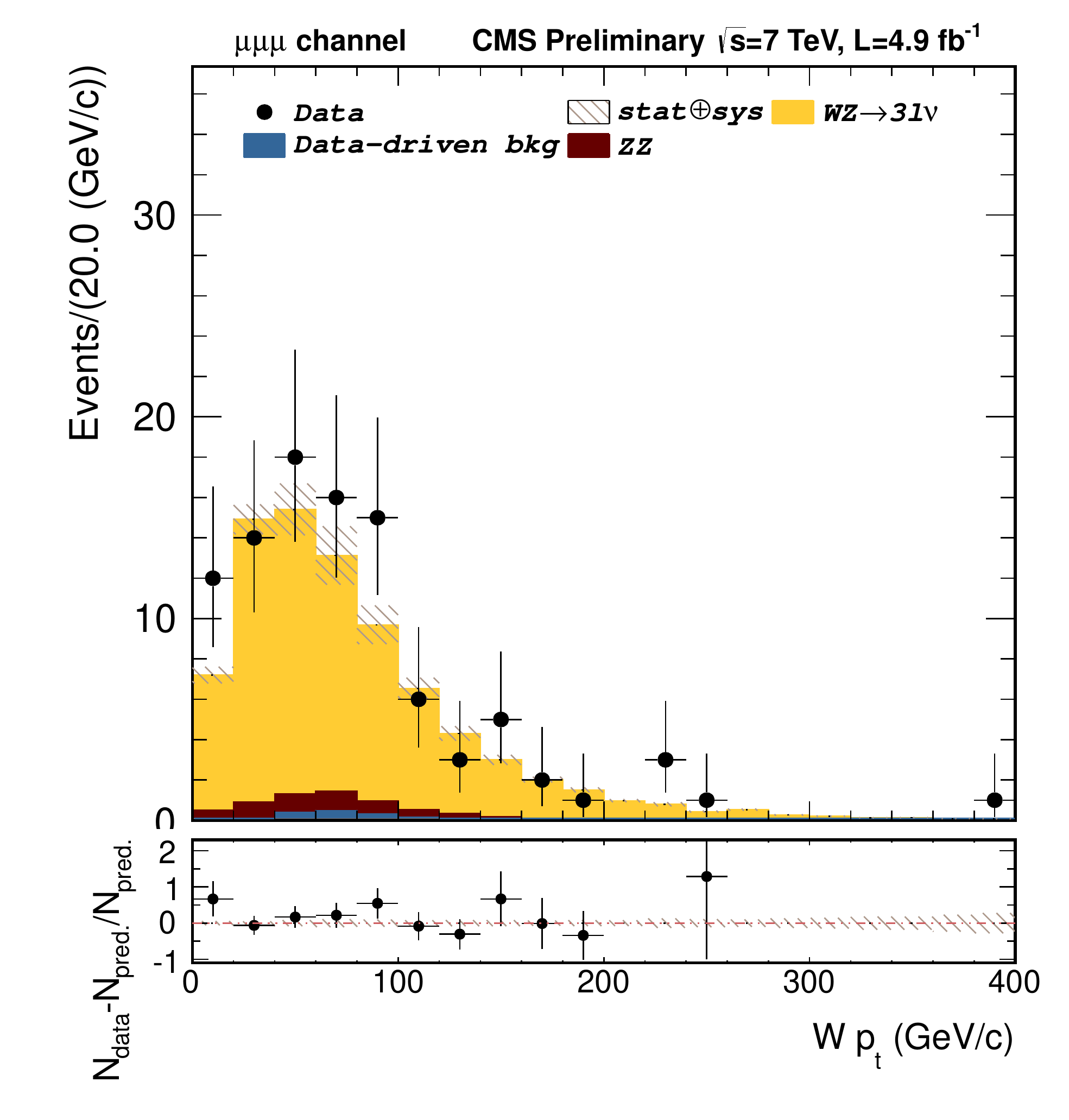}
	\end{subfigure}
	\caption[Transverse momentum of the W-candidate lepton at 7~\TeV]
	{Transverse momentum of the W-candidate lepton 
	at each event for the measured channels $eee$, $\mu ee$, $e\mu\mu$ and $\mu\mu\mu$ 
	(from left to right) and after each analysis selection stage (once the W is selected): 
	after W-candidate requirement without the \MET cut (up row) and after W-candidate including
	\MET cut (bottom row).}
\end{sidewaysfigure}

\begin{sidewaysfigure}[!htpb]
	\centering
	\begin{subfigure}[b]{0.2\textwidth}
		\includegraphics[width=\textwidth]{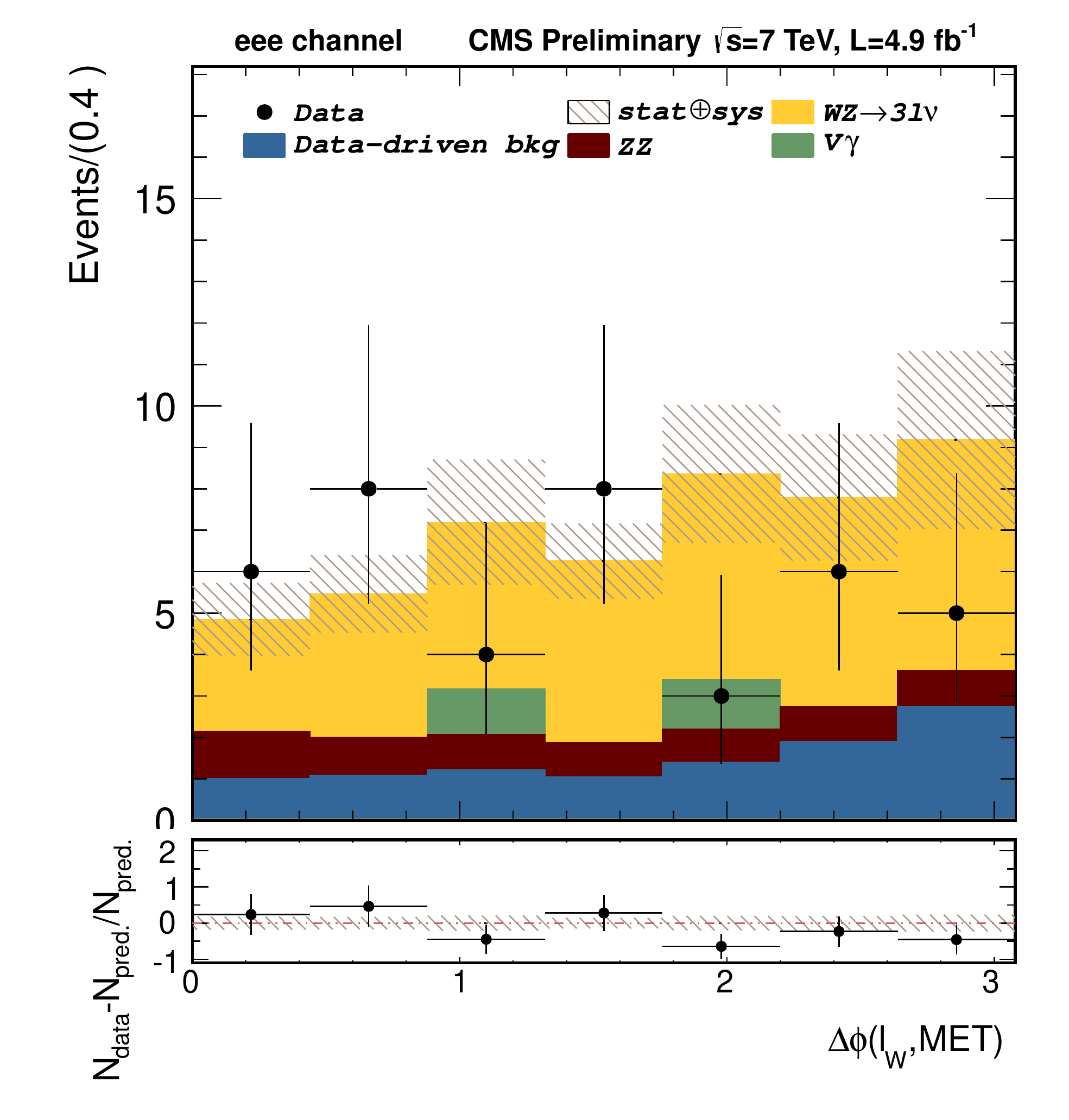}
	\end{subfigure}\quad
	\begin{subfigure}[b]{0.2\textwidth}
		\includegraphics[width=\textwidth]{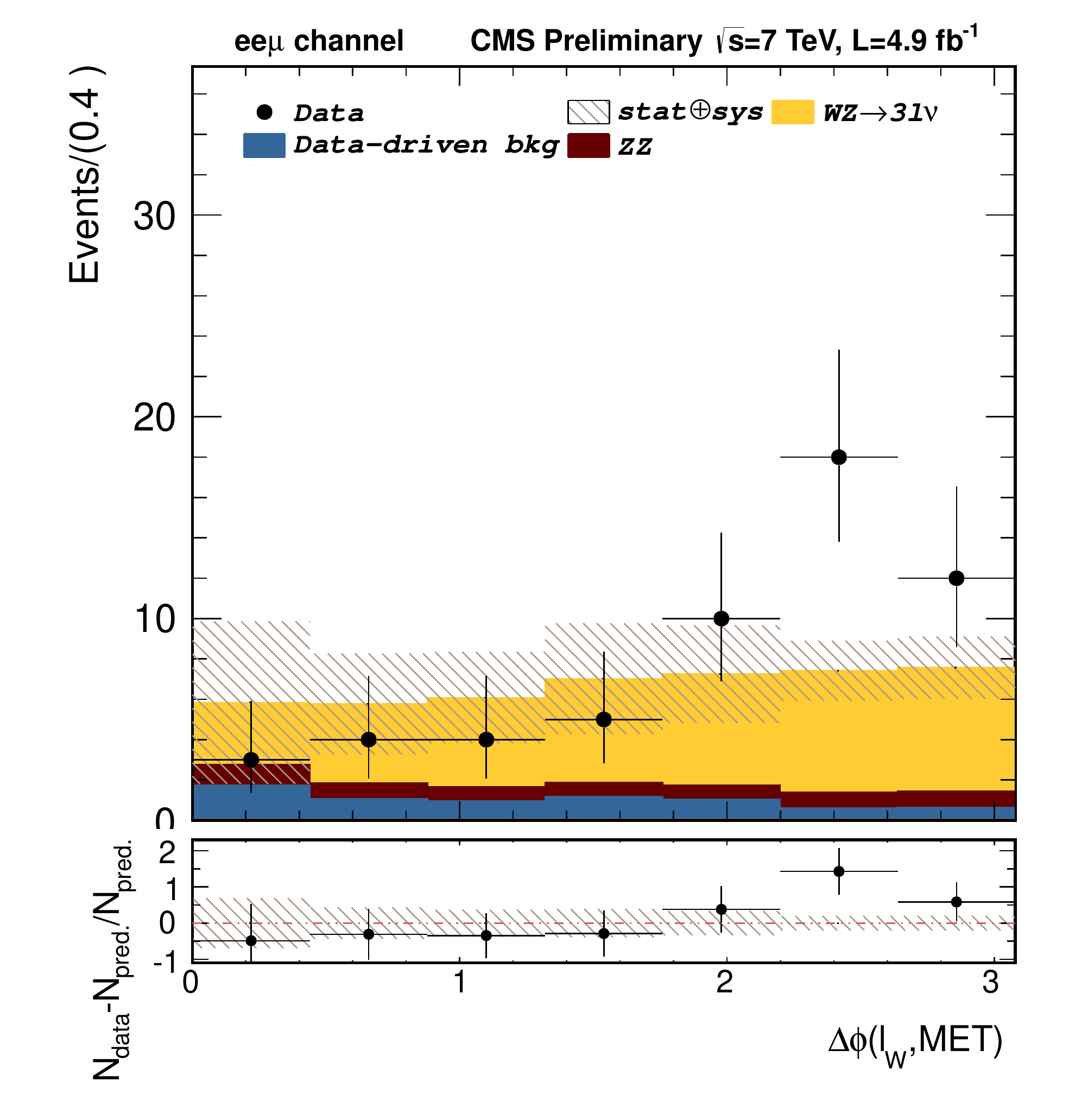}
	\end{subfigure}\quad
	\begin{subfigure}[b]{0.2\textwidth}
		\includegraphics[width=\textwidth]{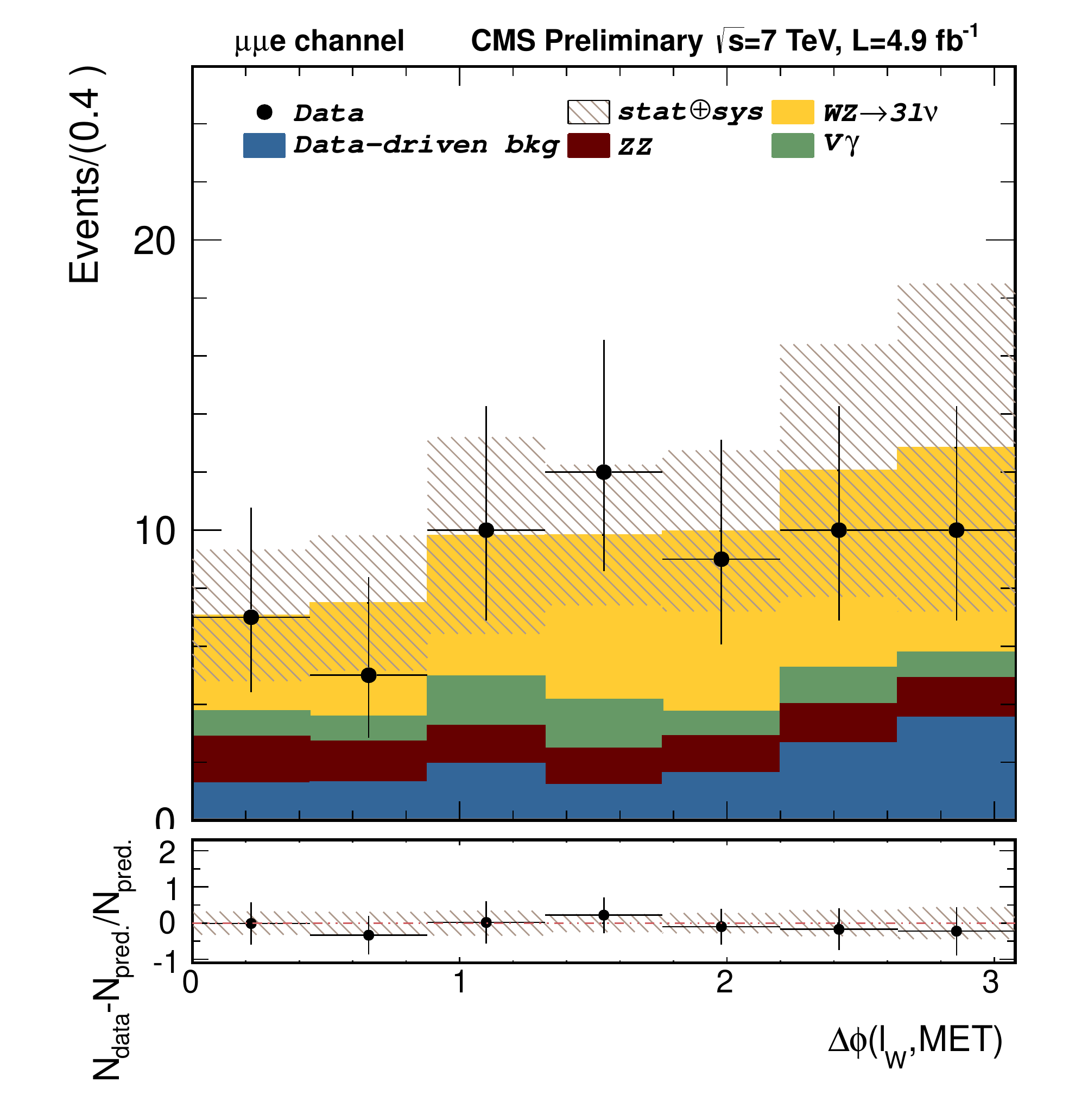}
	\end{subfigure}\quad
	\begin{subfigure}[b]{0.2\textwidth}
		\includegraphics[width=\textwidth]{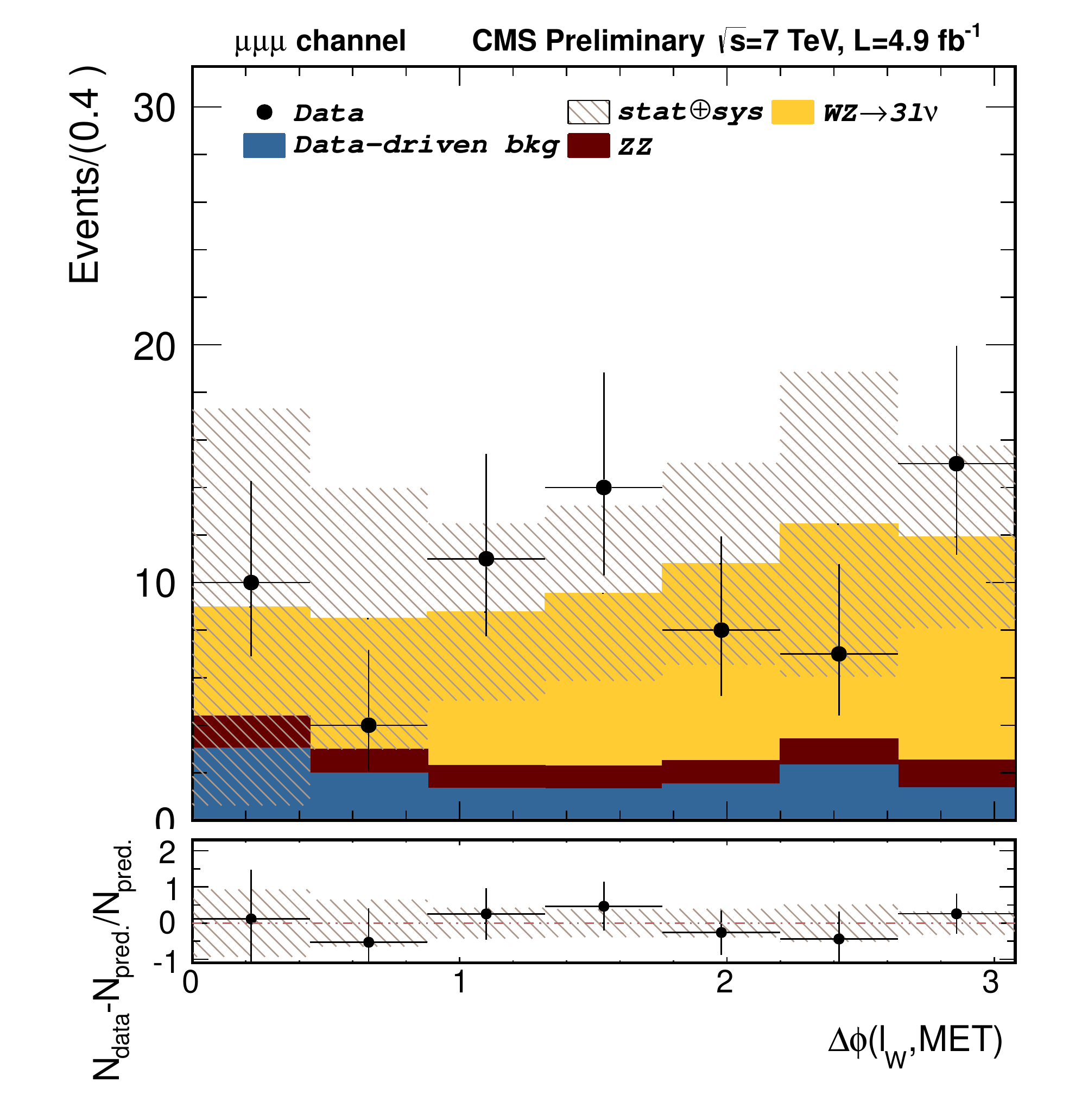}
	\end{subfigure}
	\vskip 1ex
	\begin{subfigure}[b]{0.2\textwidth}
		\includegraphics[width=\textwidth]{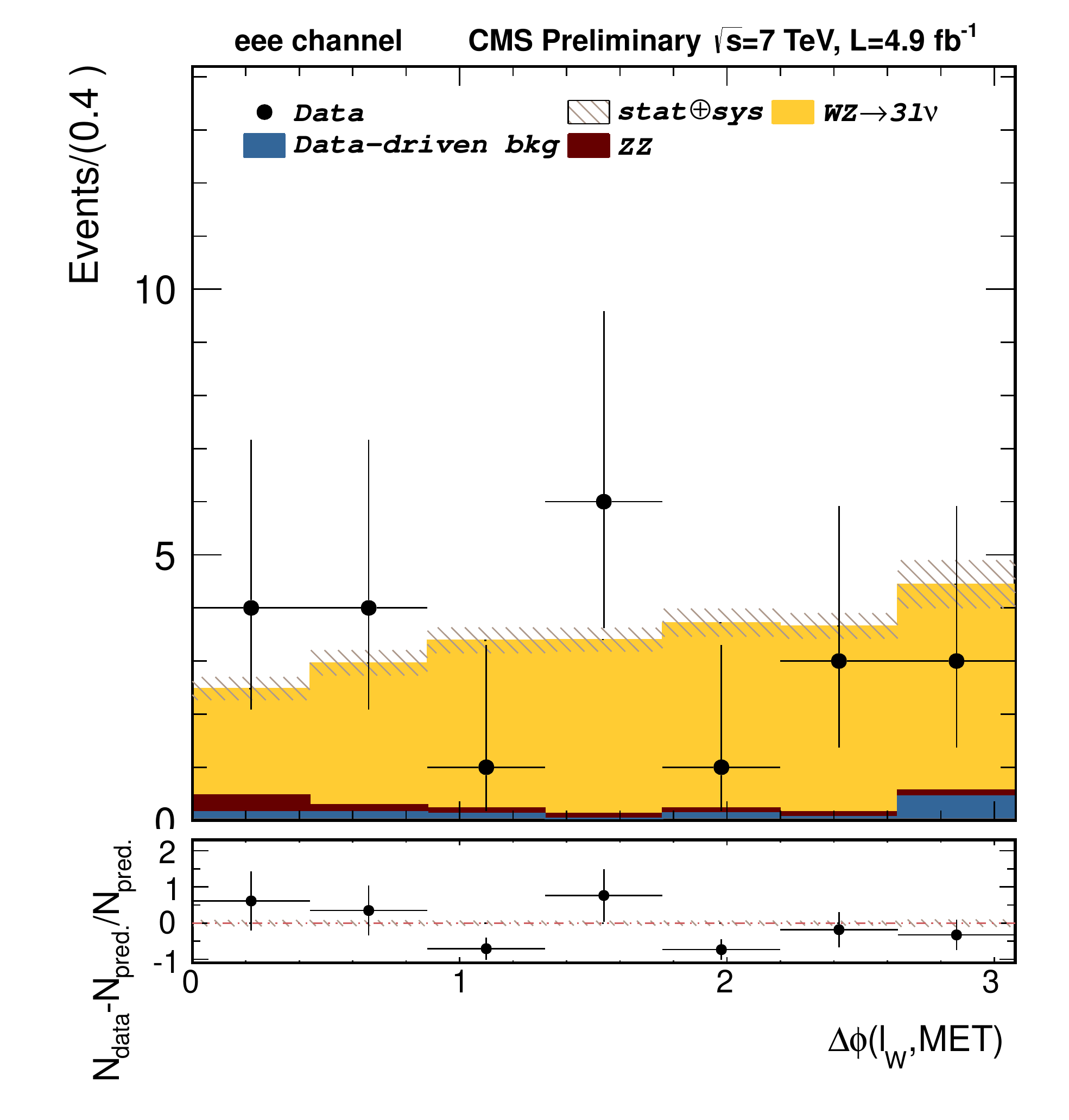}
	\end{subfigure}\quad
	\begin{subfigure}[b]{0.2\textwidth}
		\includegraphics[width=\textwidth]{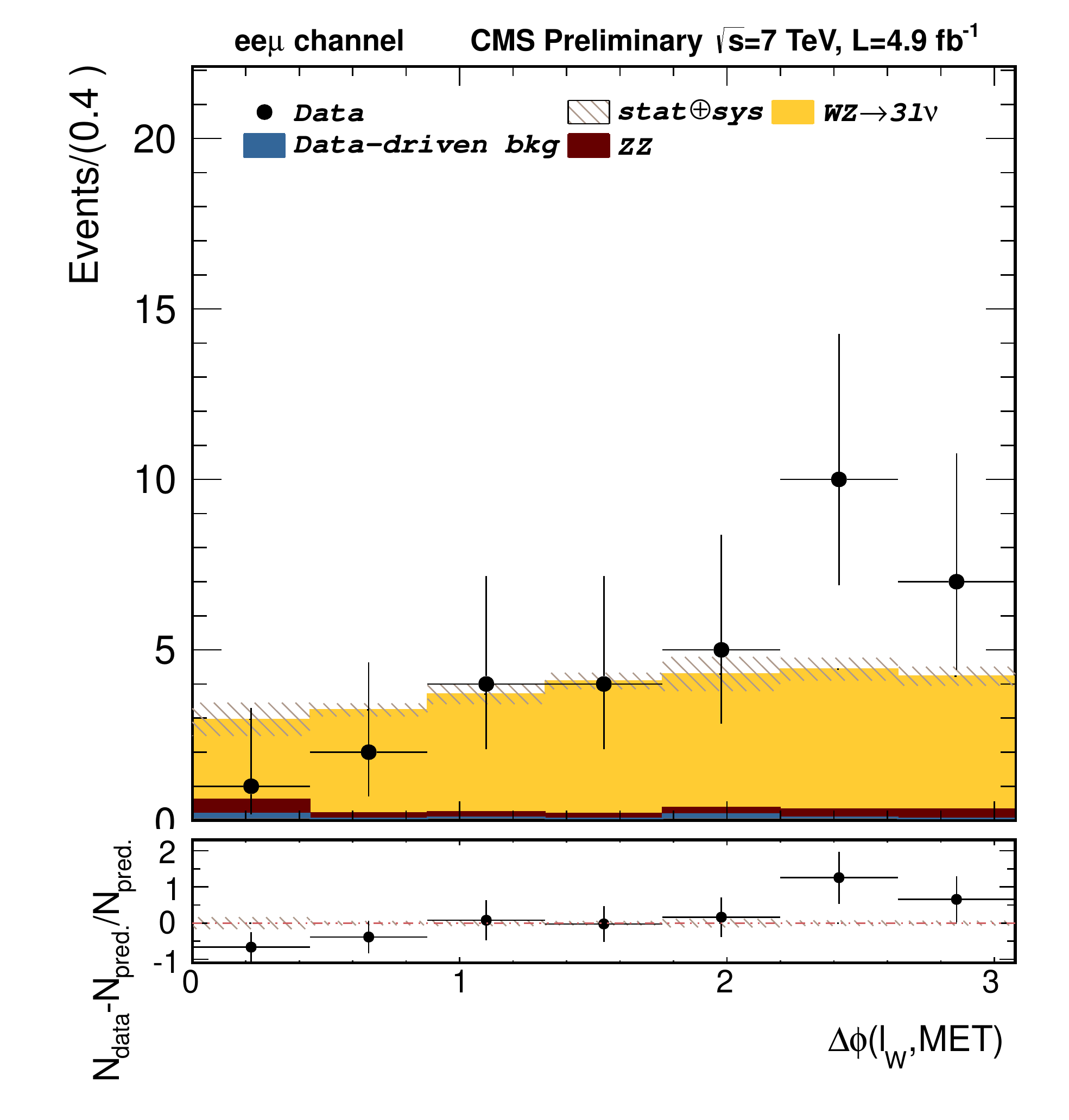}
	\end{subfigure}\quad
	\begin{subfigure}[b]{0.2\textwidth}
		\includegraphics[width=\textwidth]{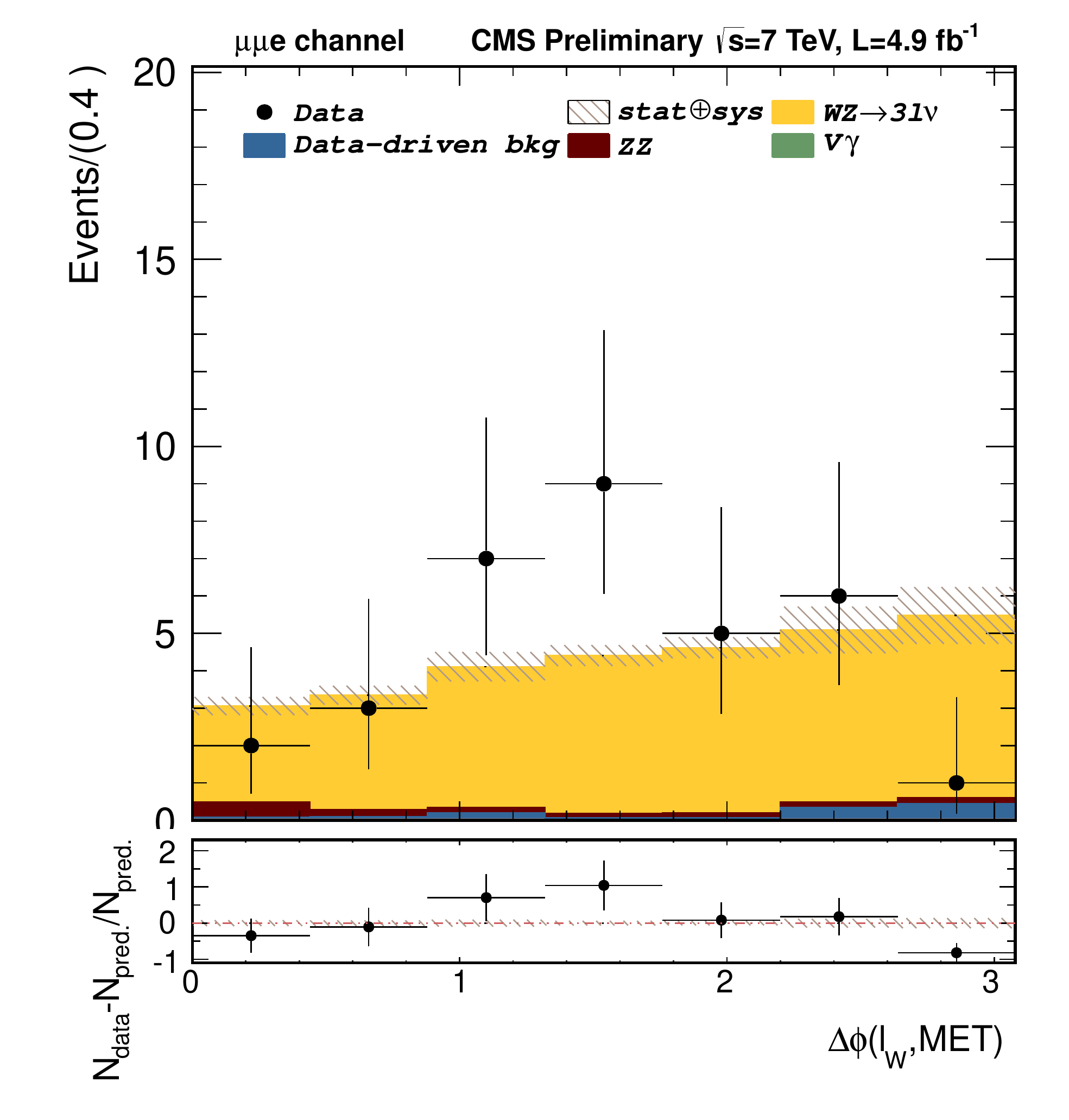}
	\end{subfigure}\quad
	\begin{subfigure}[b]{0.2\textwidth}
		\includegraphics[width=\textwidth]{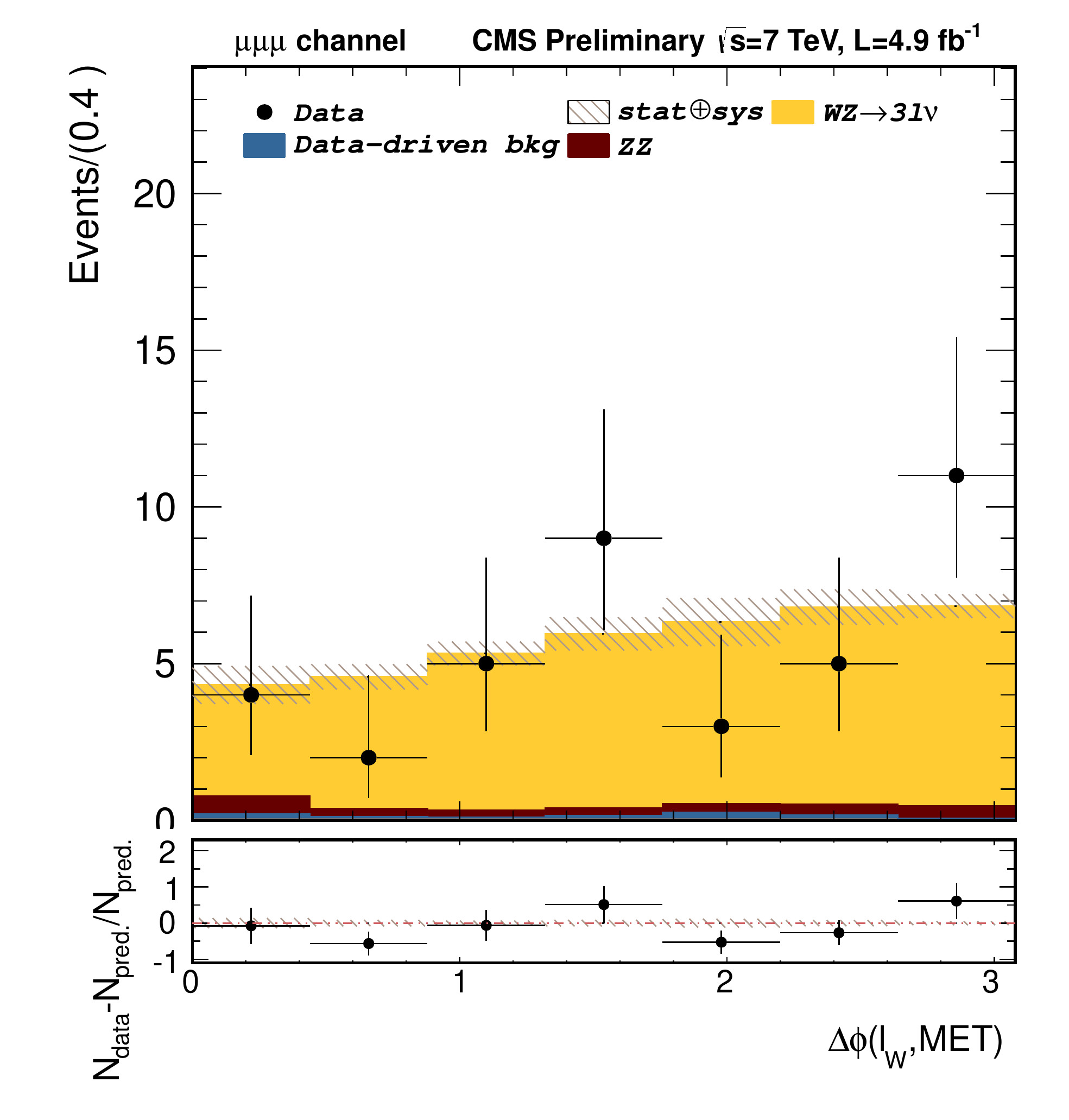}
	\end{subfigure}
	\caption[Azimuthal angle between W-candidate lepton and \MET at 7~\TeV]
	{Azimuthal angle between the W-candidate lepton and the \MET
	at each event for the measured channels $eee$, $\mu ee$, $e\mu\mu$ and $\mu\mu\mu$ 
	(from left to right) and after each analysis selection stage (once the W is selected): 
	after W-candidate requirement without the \MET cut (up row) and after W-candidate including
	\MET cut (bottom row).}
\end{sidewaysfigure}

\begin{sidewaysfigure}[!htpb]
	\centering
	\begin{subfigure}[b]{0.2\textwidth}
		\includegraphics[width=\textwidth]{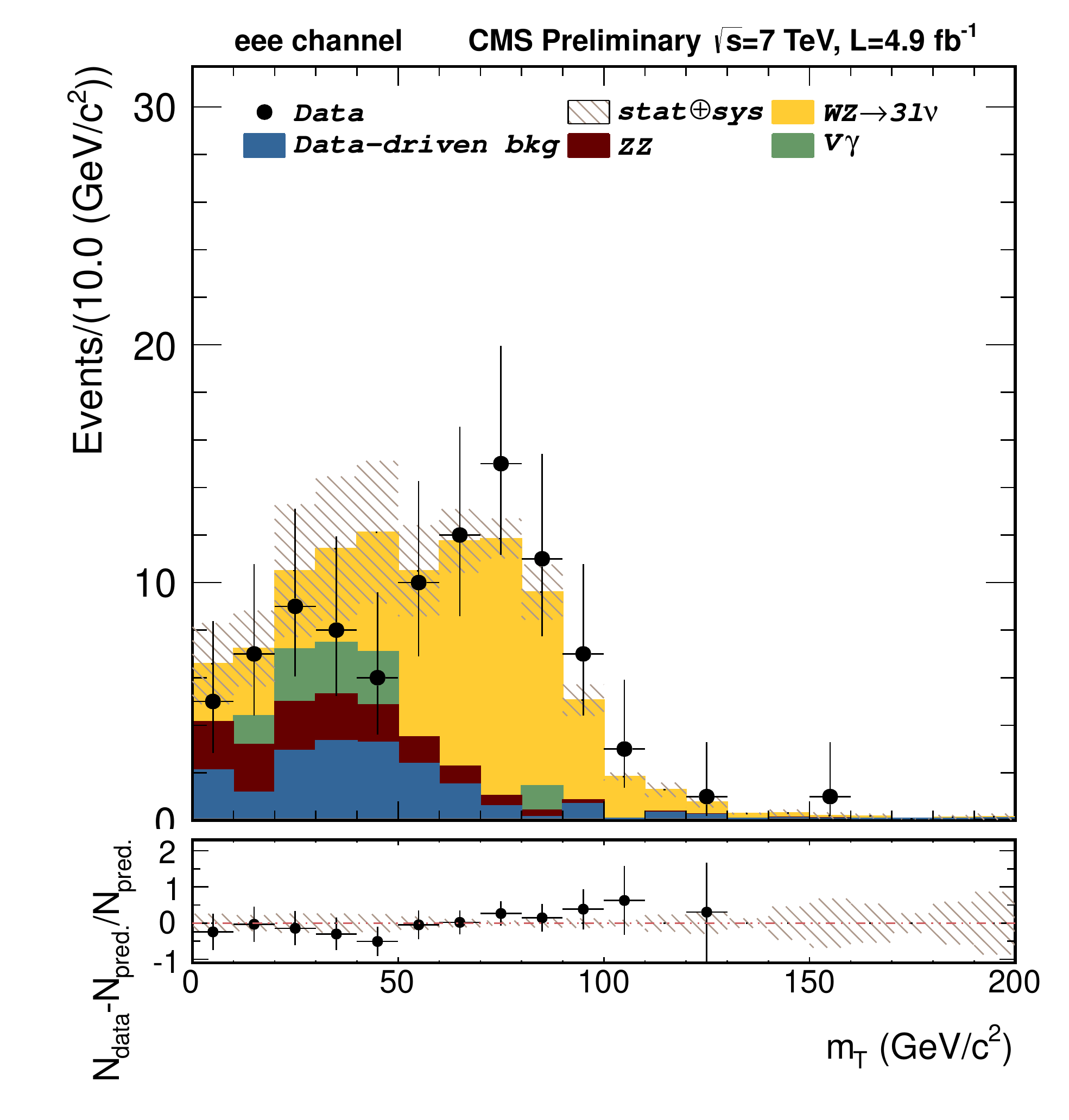}
	\end{subfigure}\quad
	\begin{subfigure}[b]{0.2\textwidth}
		\includegraphics[width=\textwidth]{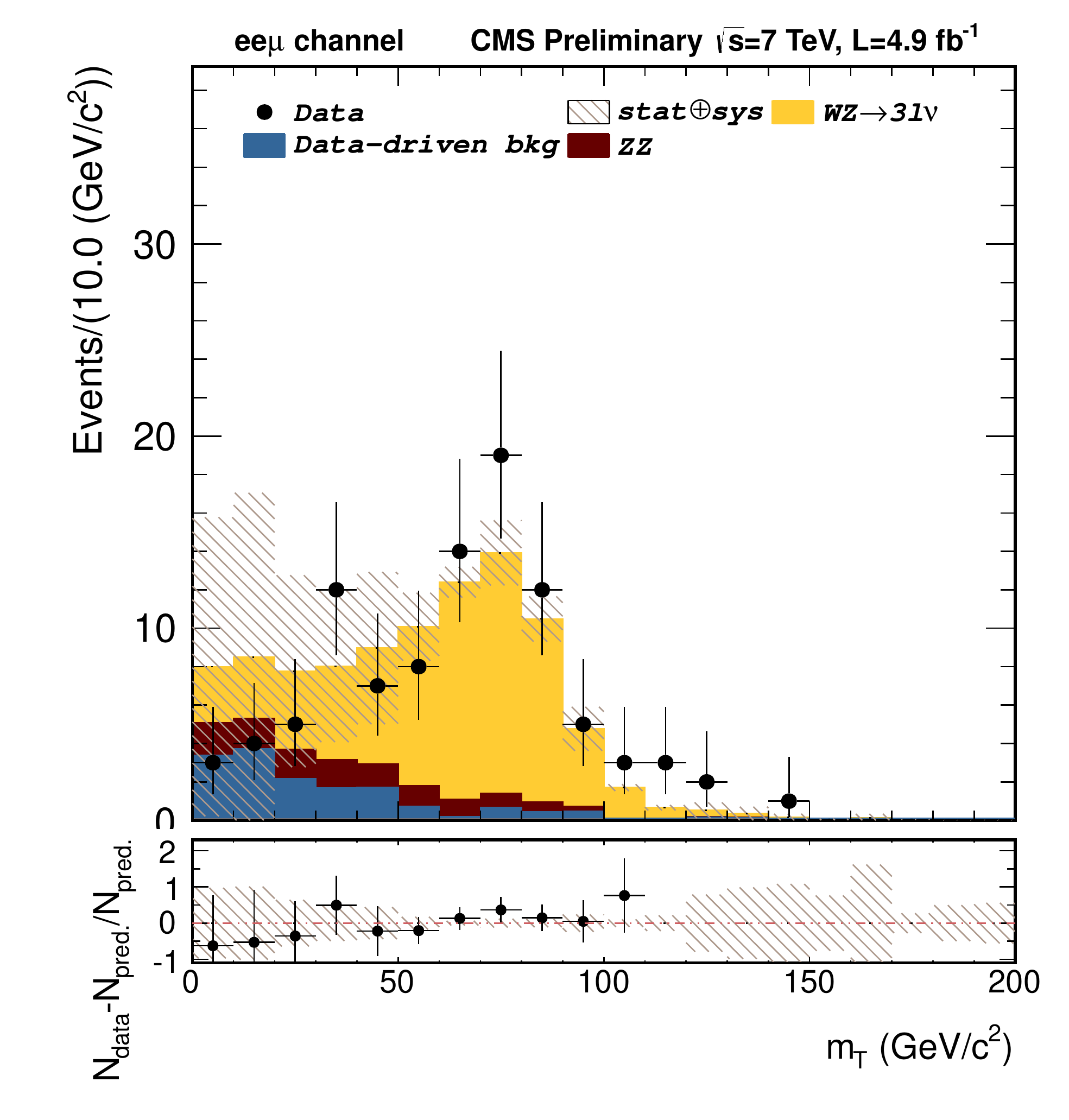}
	\end{subfigure}\quad
	\begin{subfigure}[b]{0.2\textwidth}
		\includegraphics[width=\textwidth]{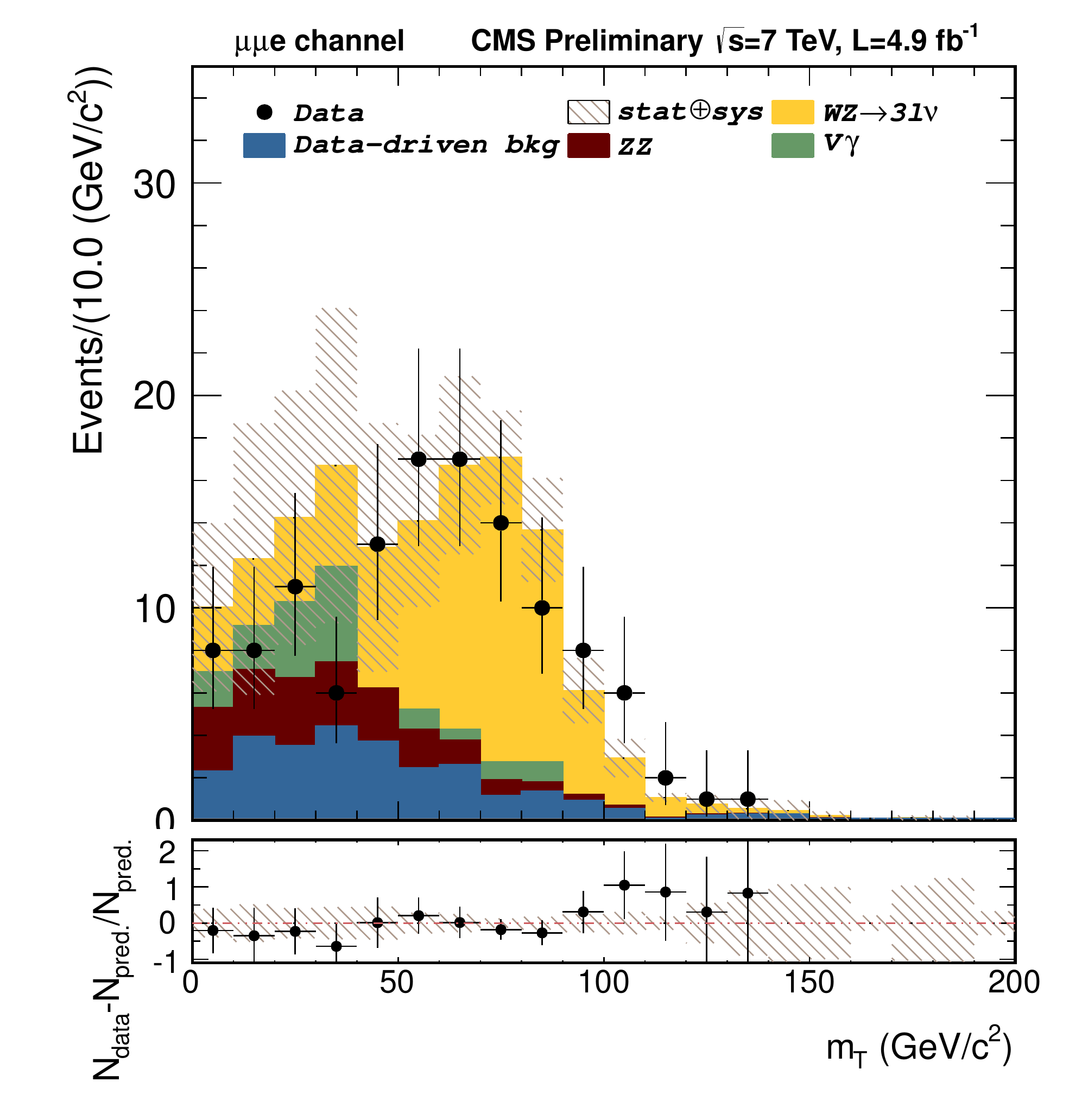}
	\end{subfigure}\quad
	\begin{subfigure}[b]{0.2\textwidth}
		\includegraphics[width=\textwidth]{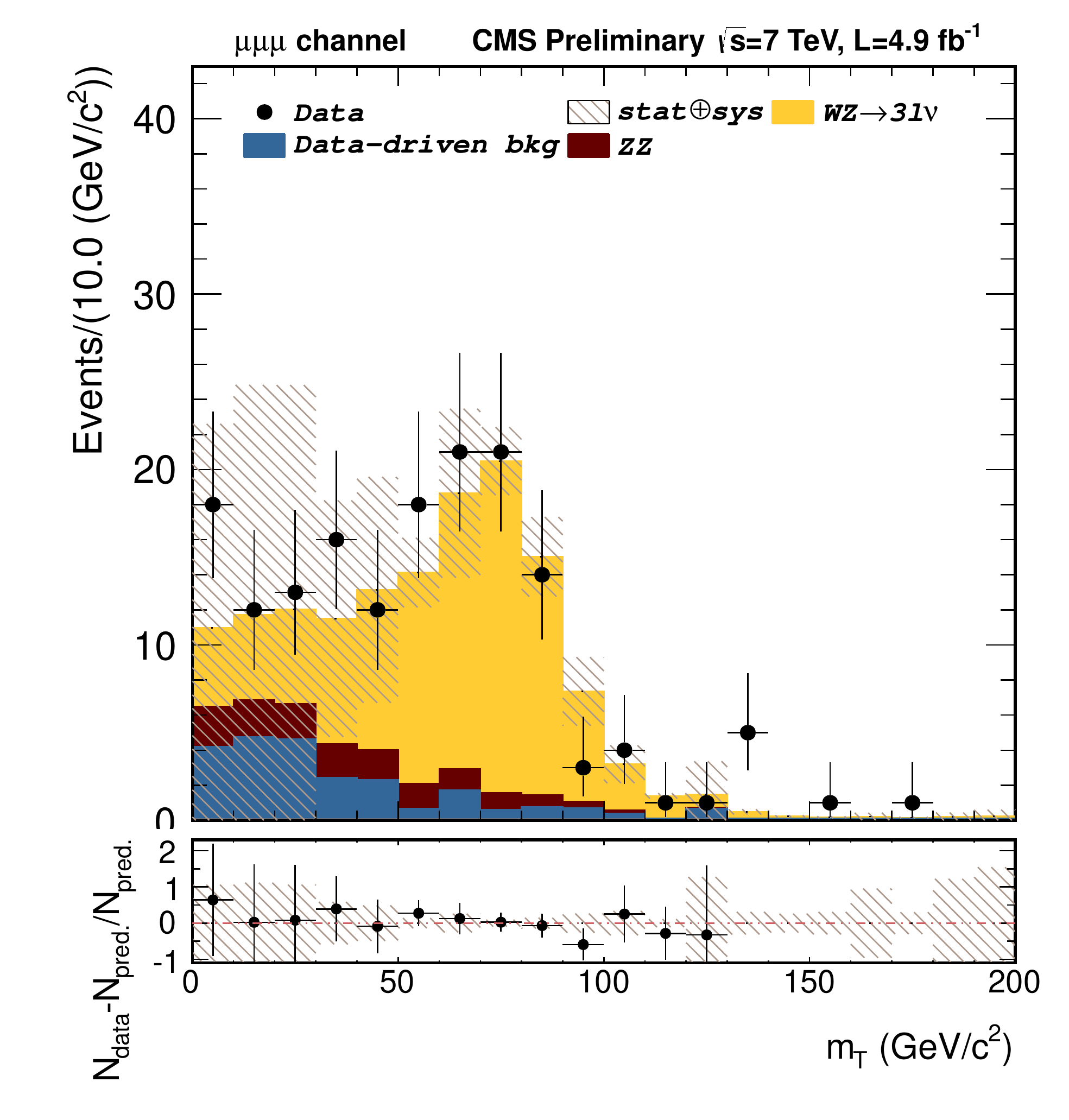}
	\end{subfigure}
	\vskip 1ex
	\begin{subfigure}[b]{0.2\textwidth}
		\includegraphics[width=\textwidth]{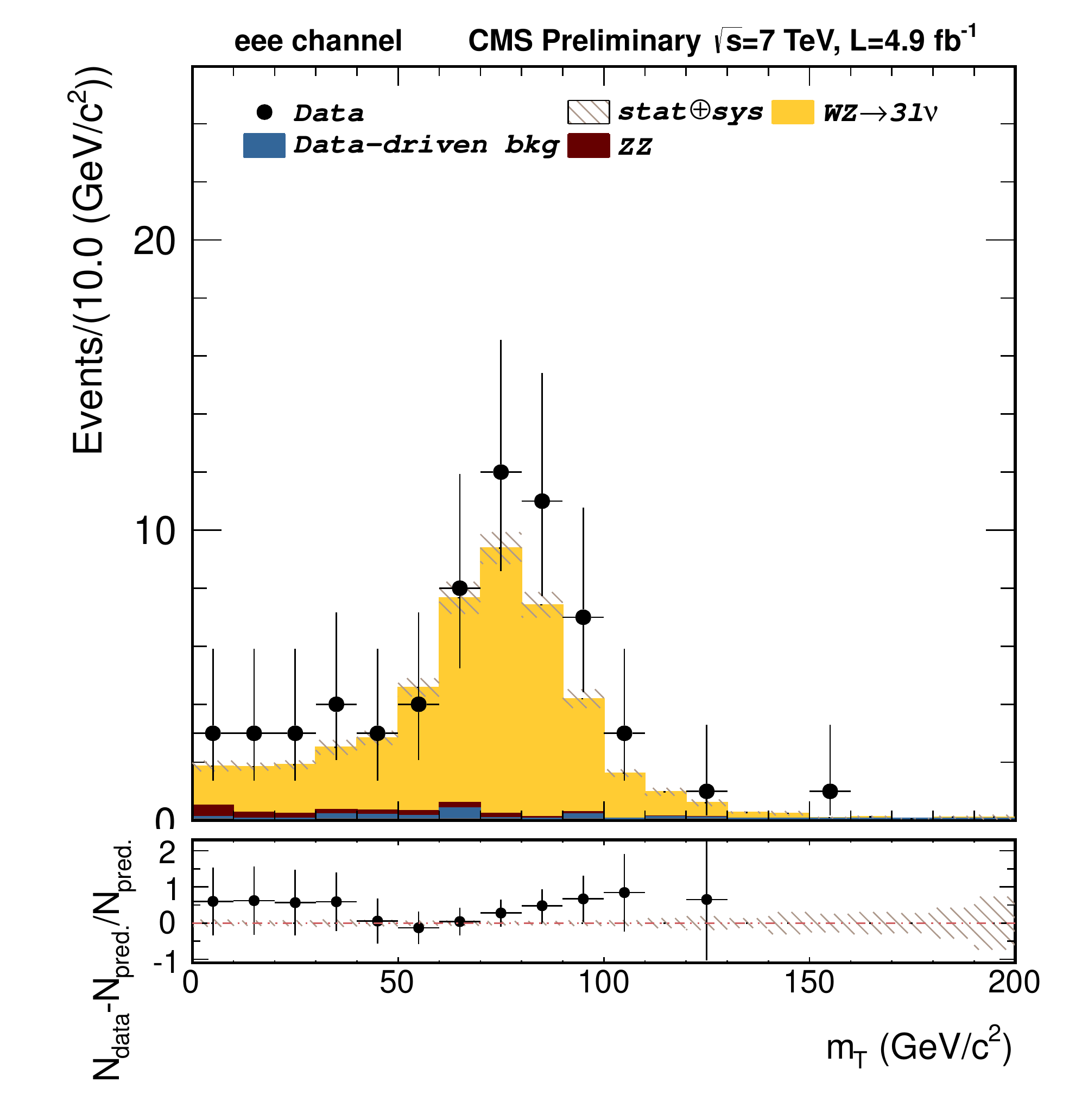}
	\end{subfigure}\quad
	\begin{subfigure}[b]{0.2\textwidth}
		\includegraphics[width=\textwidth]{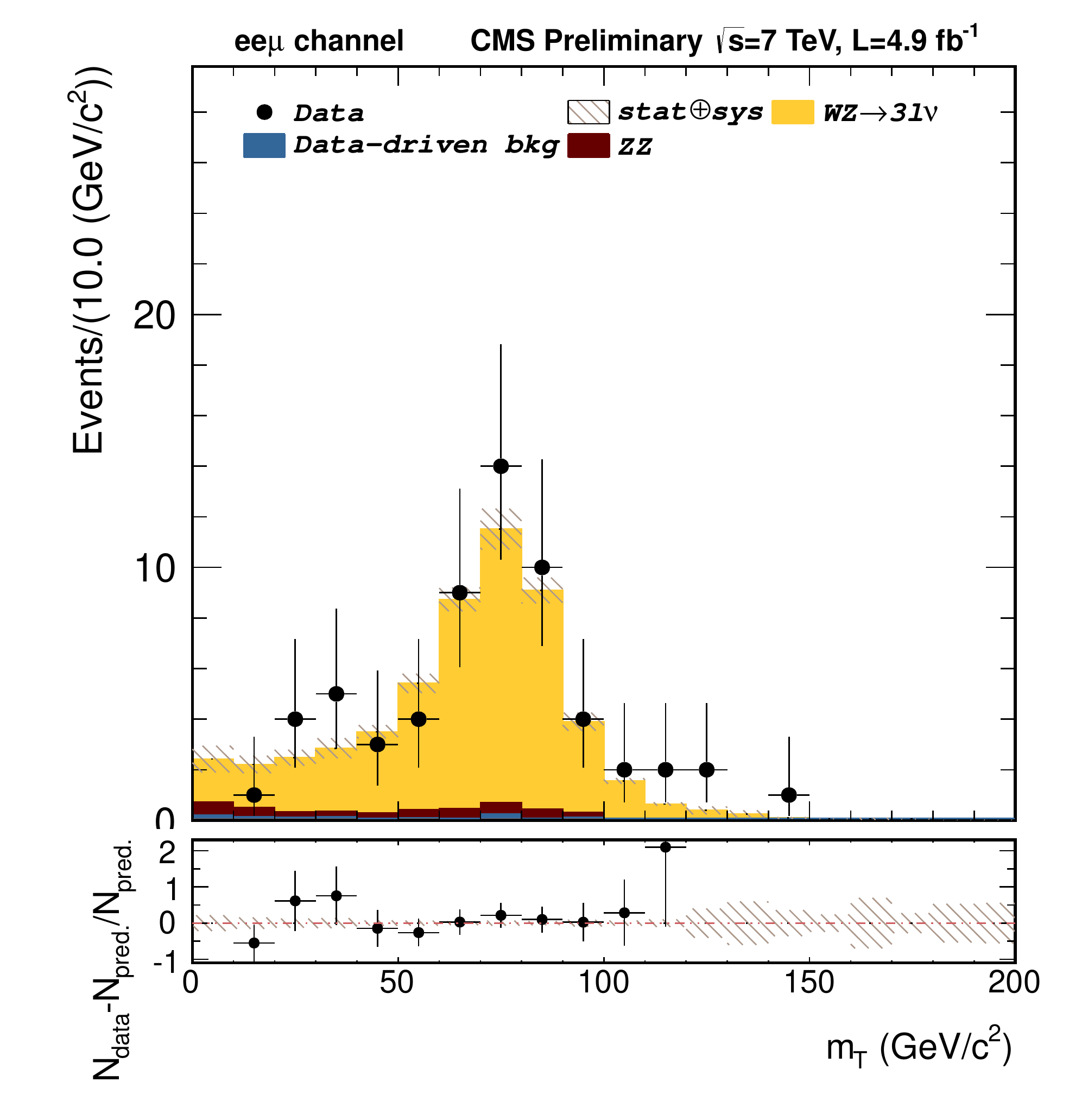}
	\end{subfigure}\quad
	\begin{subfigure}[b]{0.2\textwidth}
		\includegraphics[width=\textwidth]{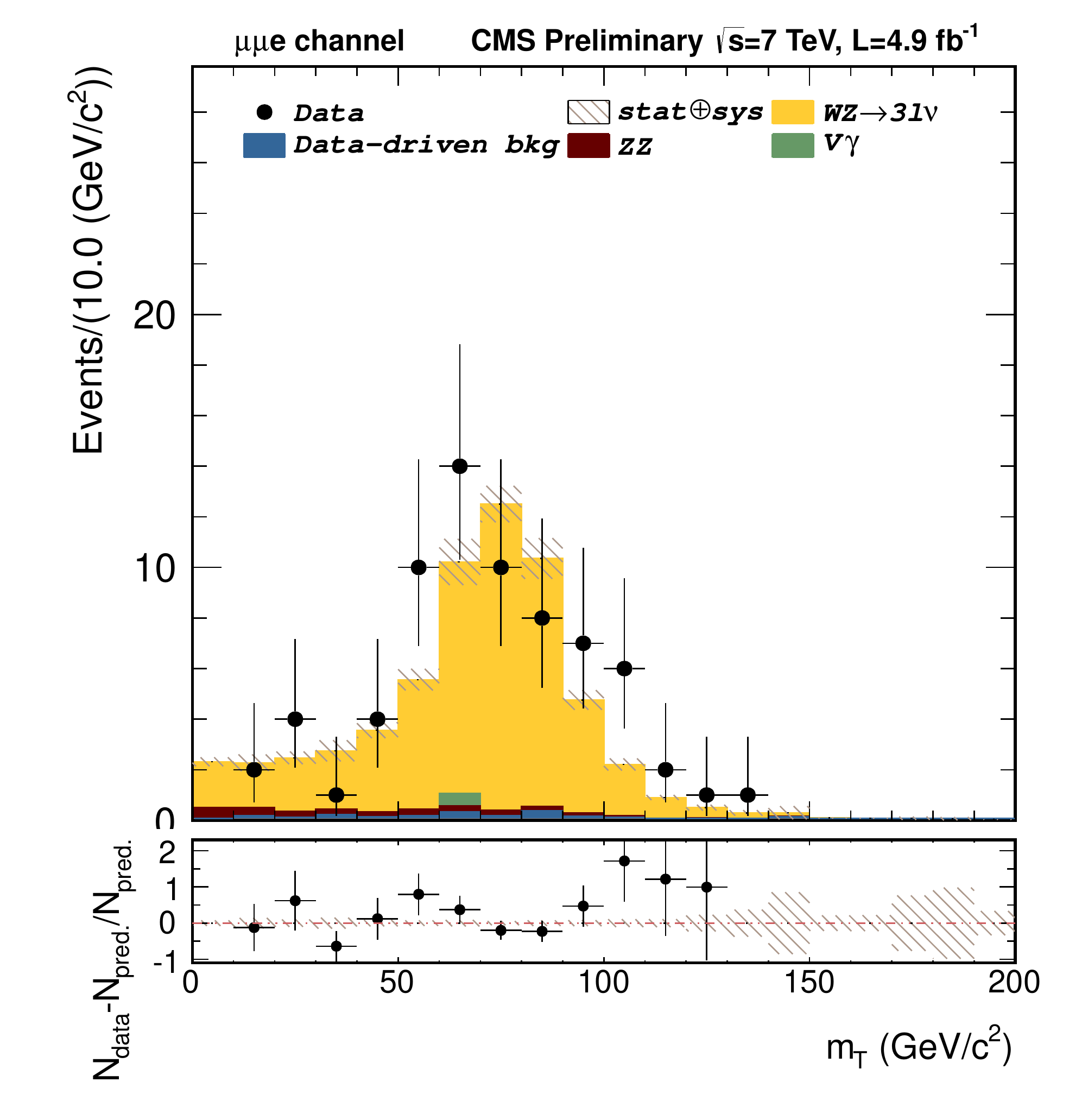}
	\end{subfigure}\quad
	\begin{subfigure}[b]{0.2\textwidth}
		\includegraphics[width=\textwidth]{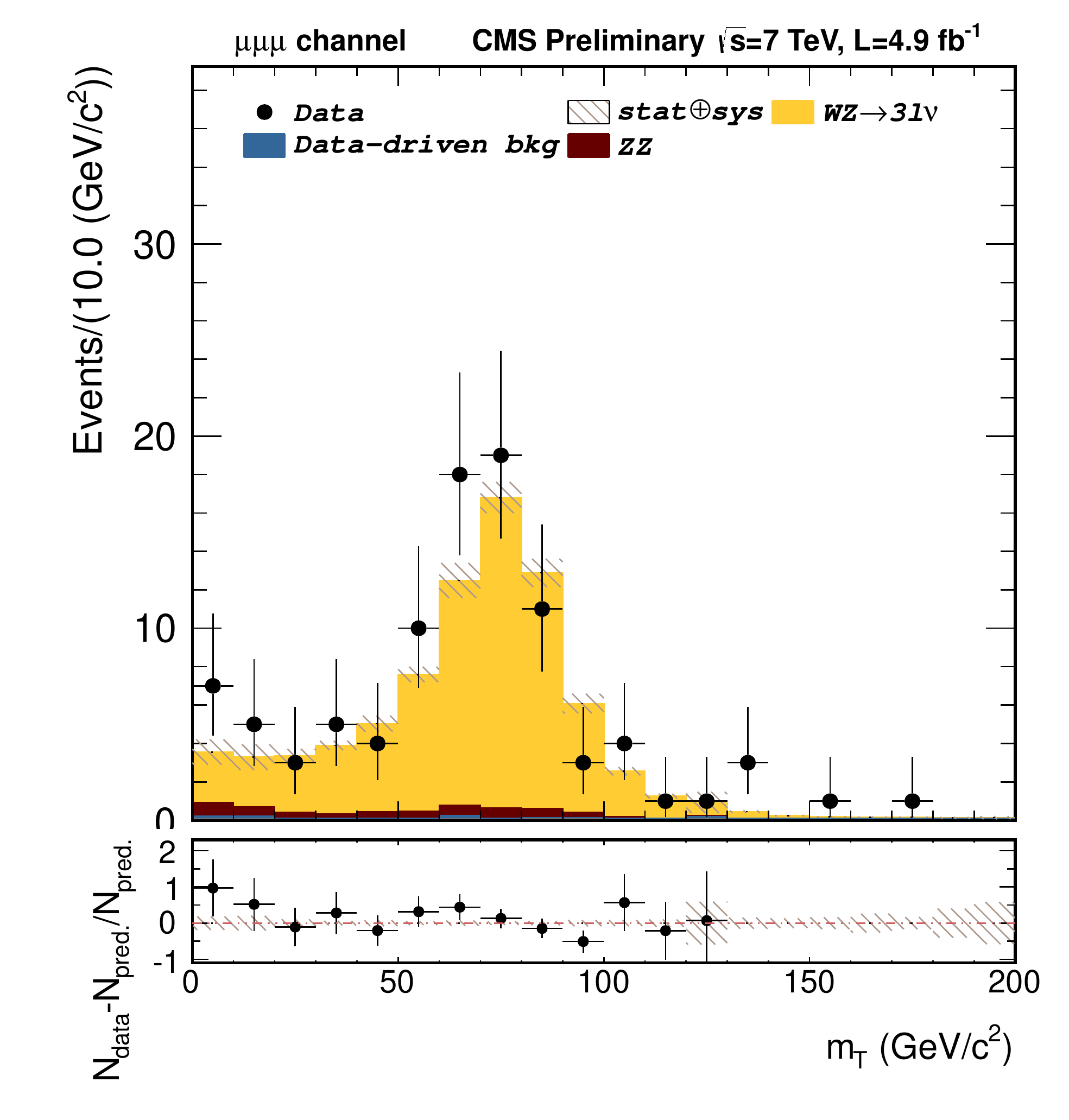}
	\end{subfigure}
	\caption[Transverse mass of the W-candidate lepton and \MET at 7~\TeV]
	{Transverse mass of the W-candidate lepton and the \MET
	at each event for the measured channels $eee$, $\mu ee$, $e\mu\mu$ and $\mu\mu\mu$ 
	(from left to right) and after each analysis selection stage (once the W is selected): 
	after W-candidate requirement without the \MET cut (up row) and after W-candidate including
	\MET cut (bottom row).}
\end{sidewaysfigure}

\begin{sidewaysfigure}[!htpb]
	\centering
	\begin{subfigure}[b]{0.2\textwidth}
		\includegraphics[width=\textwidth]{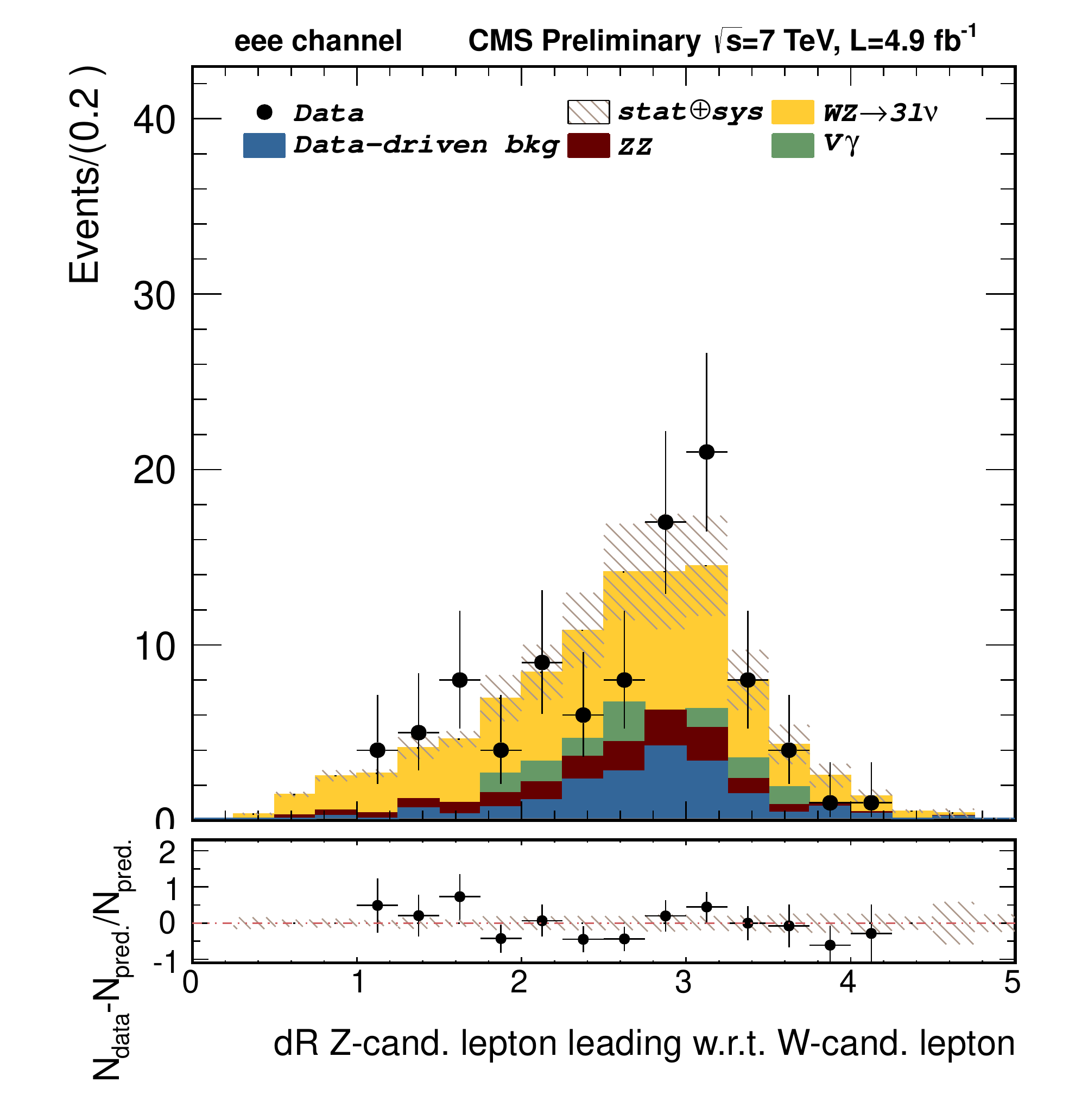}
	\end{subfigure}\quad
	\begin{subfigure}[b]{0.2\textwidth}
		\includegraphics[width=\textwidth]{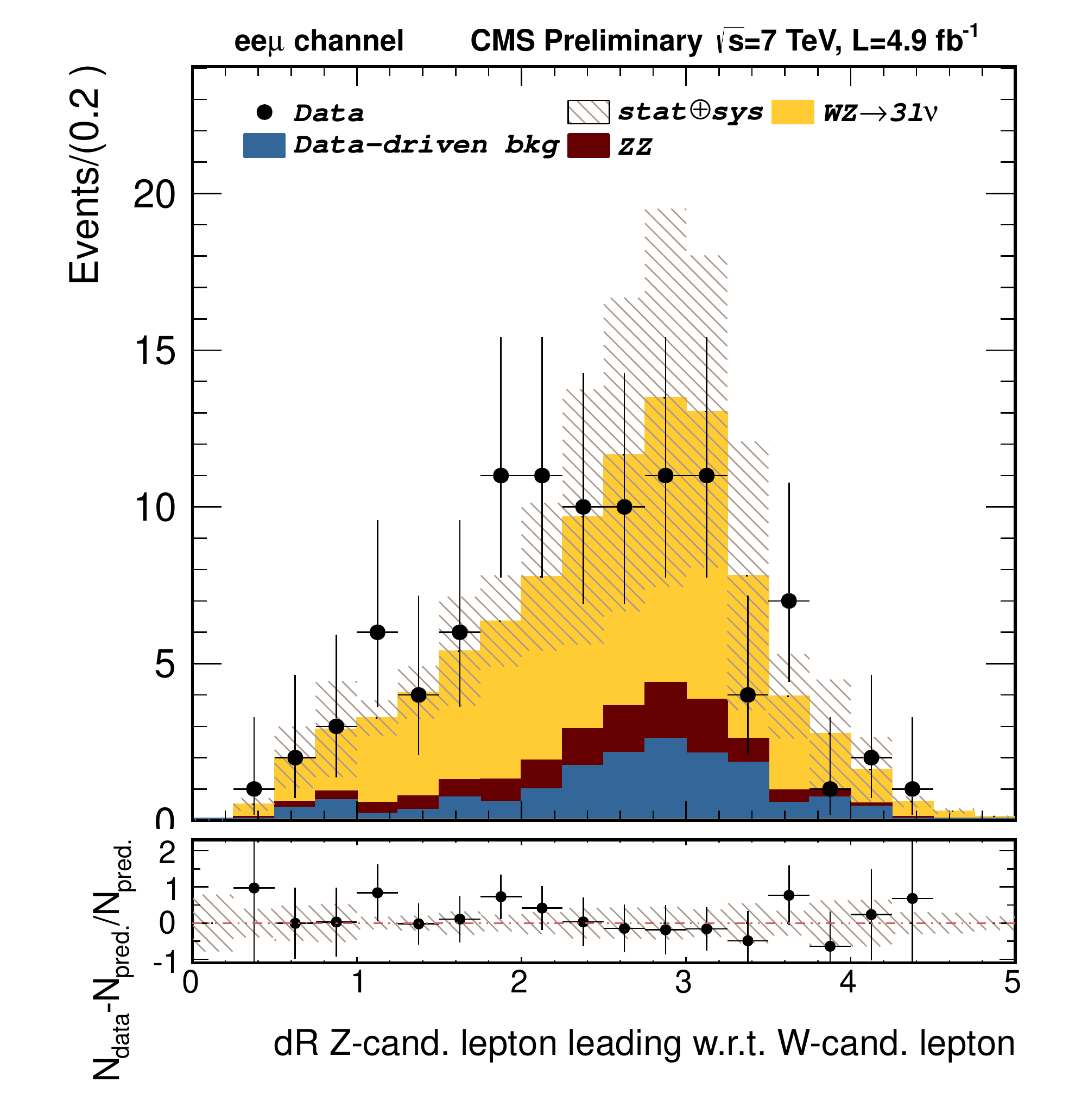}
	\end{subfigure}\quad
	\begin{subfigure}[b]{0.2\textwidth}
		\includegraphics[width=\textwidth]{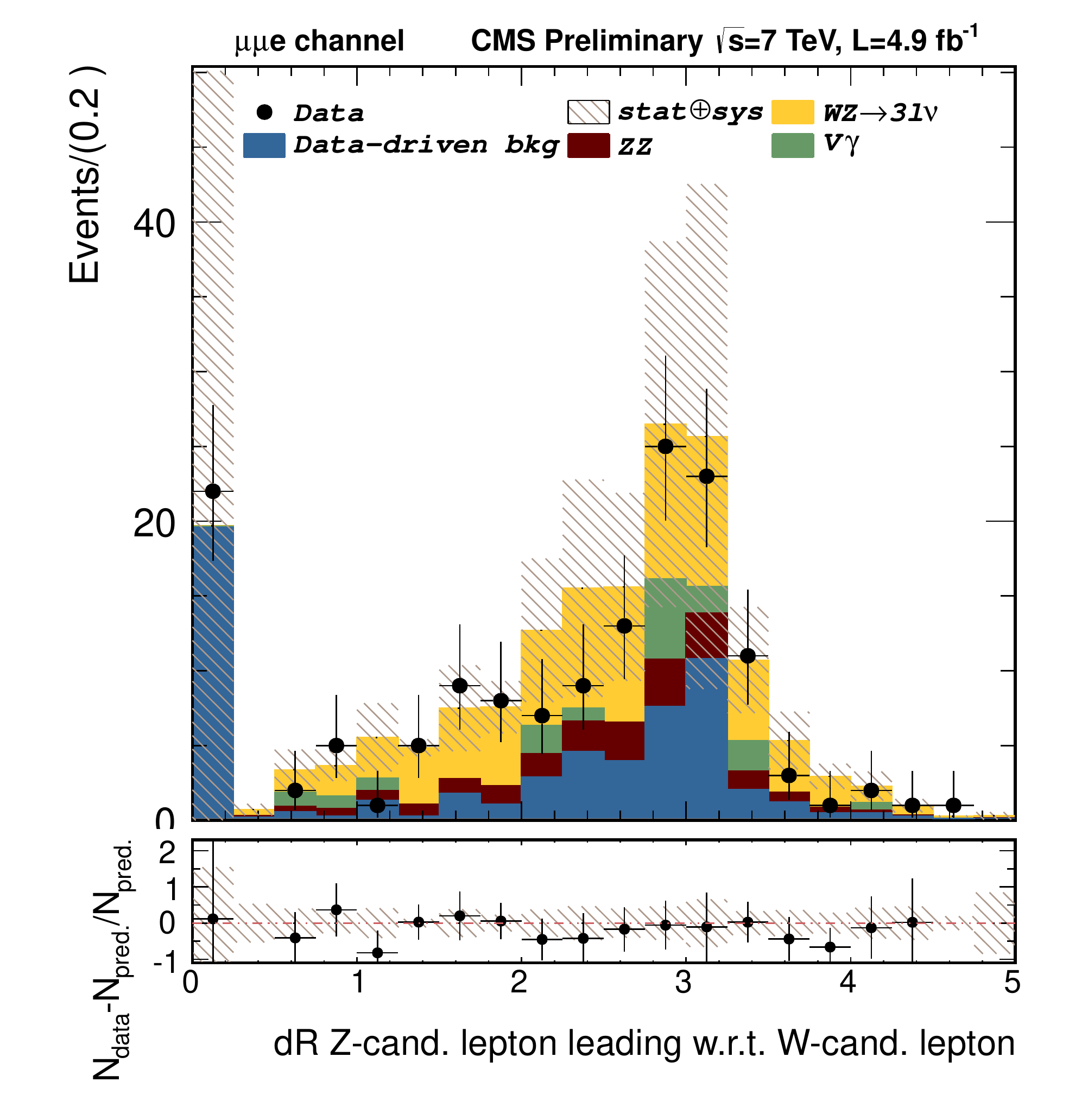}
	\end{subfigure}\quad
	\begin{subfigure}[b]{0.2\textwidth}
		\includegraphics[width=\textwidth]{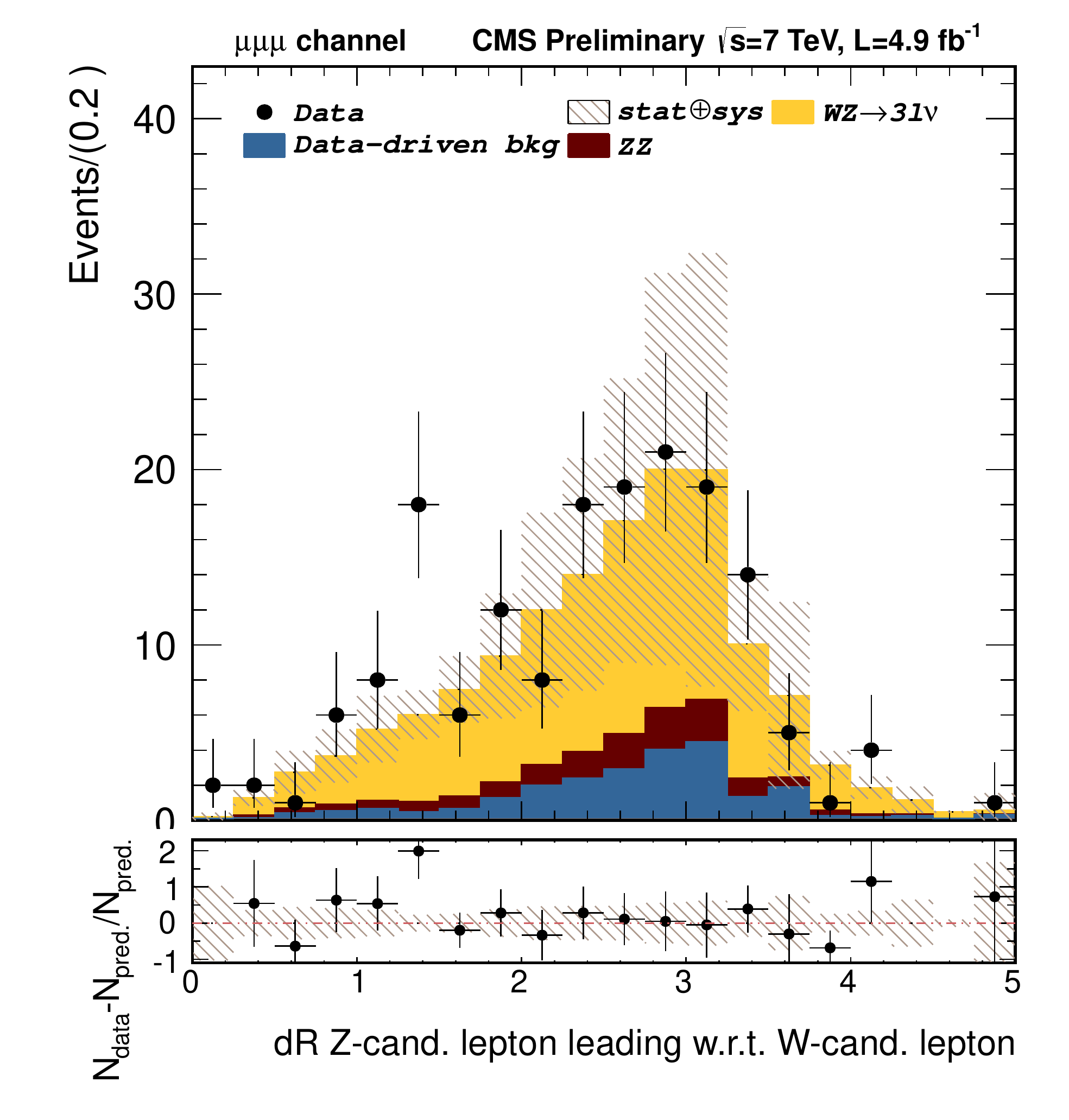}
	\end{subfigure}
	\caption[Angular distance between W-candidate lepton and Z-candidate leading lepton at 7~\TeV]  
	{Angular distance between the W-candidate lepton and the Z-candidate
	leading lepton at each event for the measured channels $eee$, $\mu ee$, $e\mu\mu$ and 
	$\mu\mu\mu$ (from left to right) after W-candidate requirement is applied.}
	\vskip 1em
	\centering
	\begin{subfigure}[b]{0.2\textwidth}
		\includegraphics[width=\textwidth]{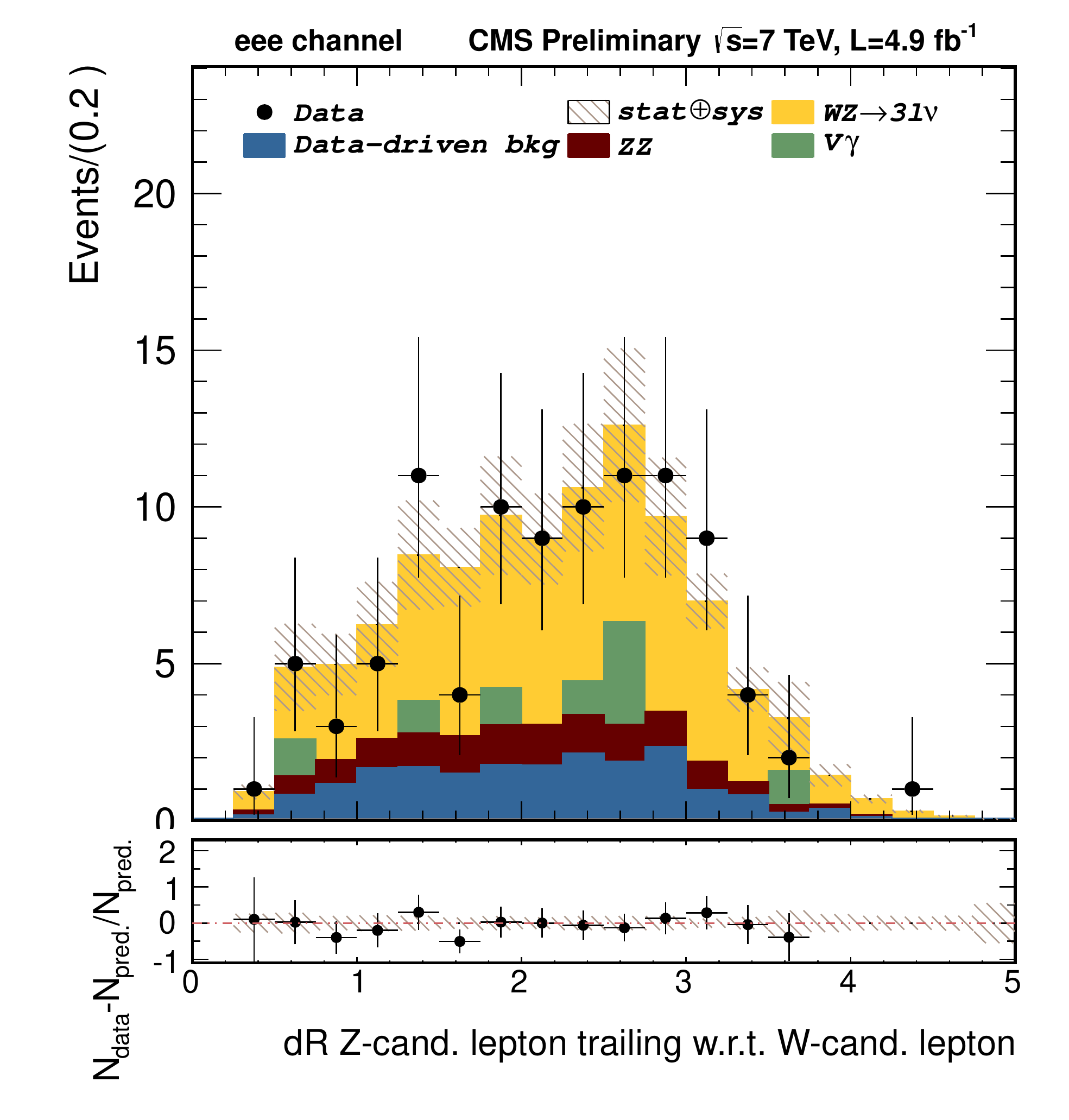}
	\end{subfigure}\quad
	\begin{subfigure}[b]{0.2\textwidth}
		\includegraphics[width=\textwidth]{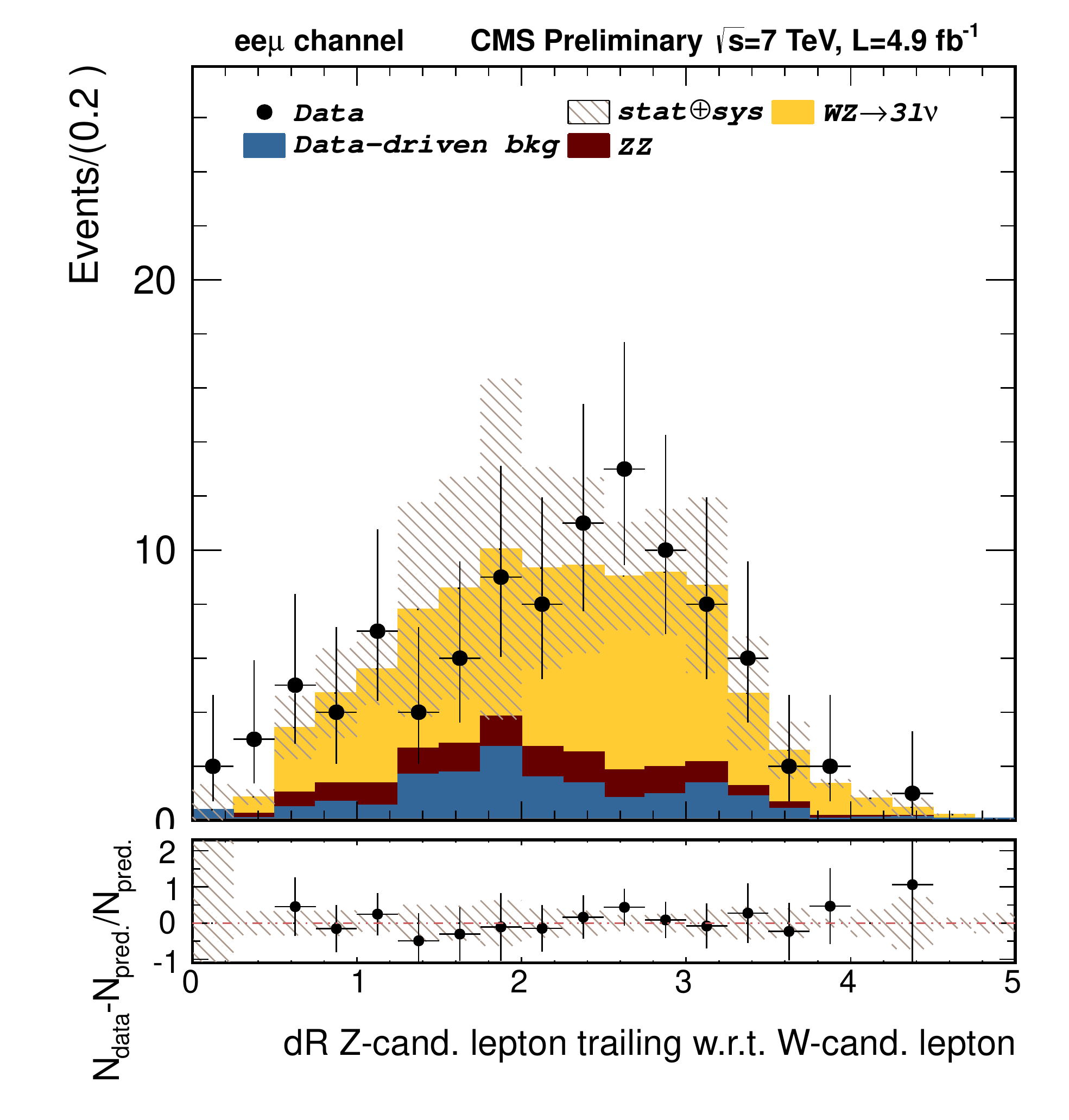}
	\end{subfigure}\quad
	\begin{subfigure}[b]{0.2\textwidth}
		\includegraphics[width=\textwidth]{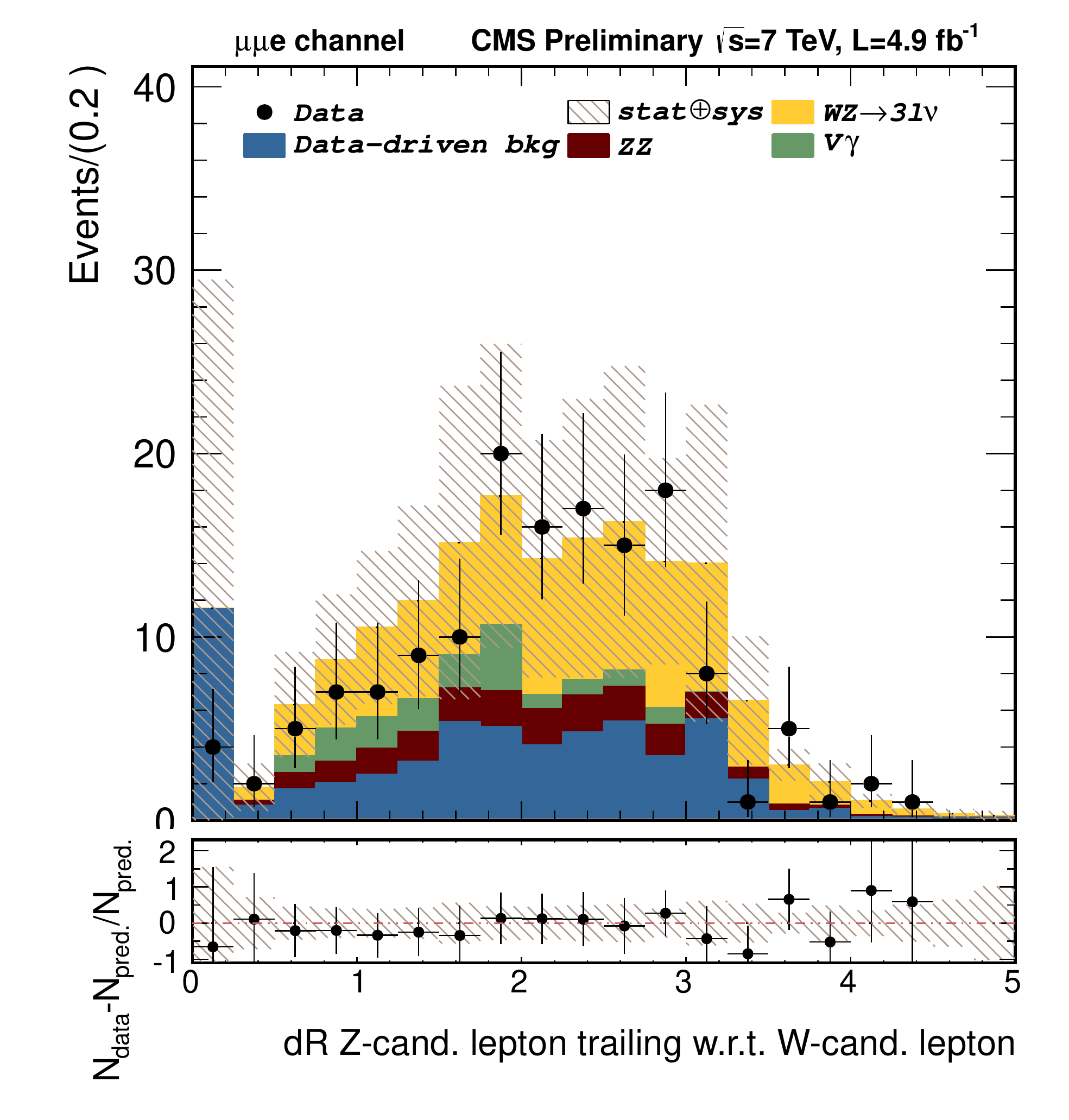}
	\end{subfigure}\quad
	\begin{subfigure}[b]{0.2\textwidth}
		\includegraphics[width=\textwidth]{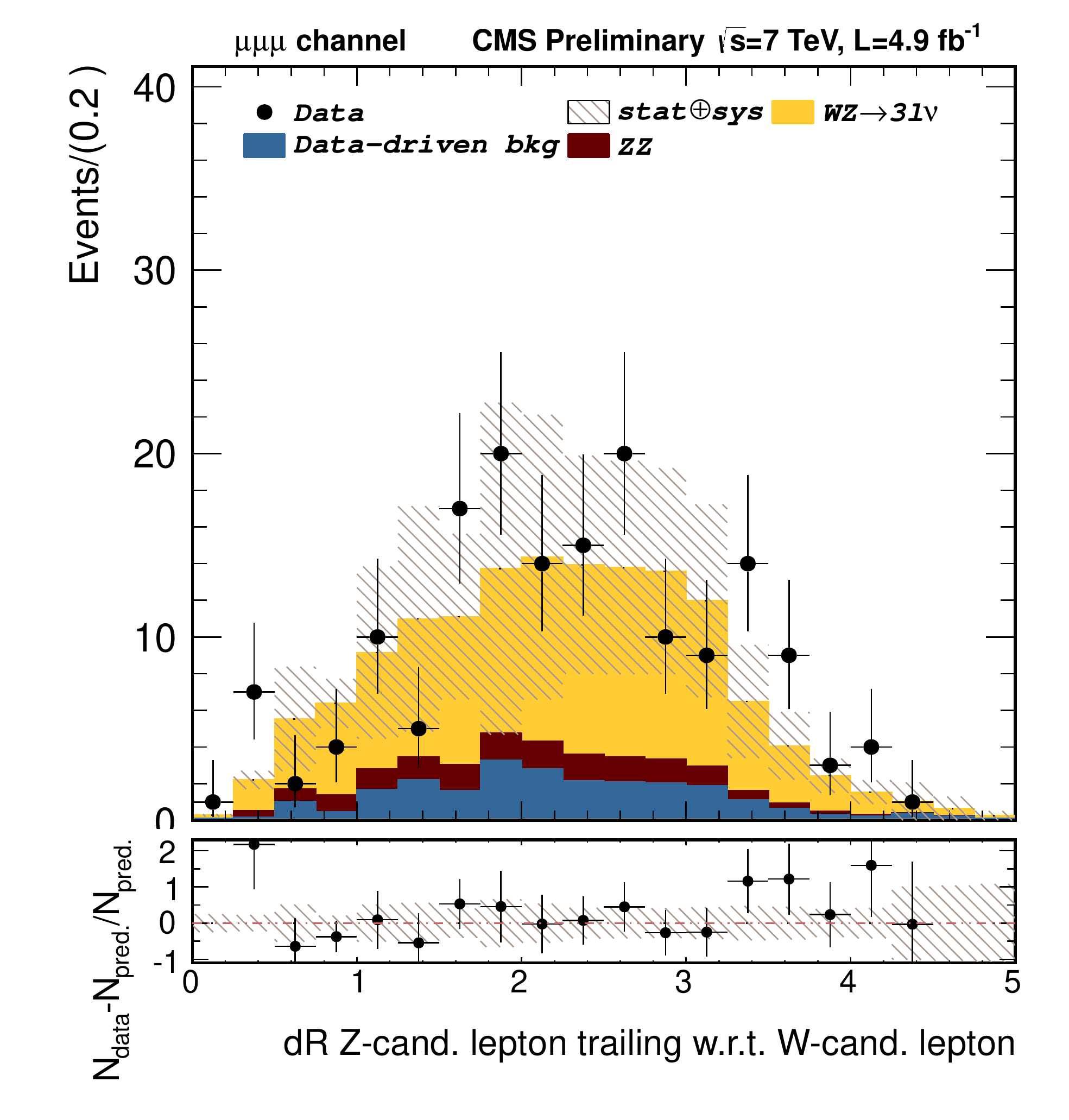}
	\end{subfigure}
	\caption[Angular distance between W-candidate lepton and Z-candidate trailing lepton at 7~\TeV]
	{Angular distance between the W-candidate lepton and the Z-candidate
	trailing lepton at each event for the measured channels $eee$, $\mu ee$, $e\mu\mu$ and 
	$\mu\mu\mu$ (from left to right) after W-candidate requirement is applied.}
\end{sidewaysfigure}

\begin{sidewaysfigure}[!htpb]
	\centering
	\begin{subfigure}[b]{0.2\textwidth}
		\includegraphics[width=\textwidth]{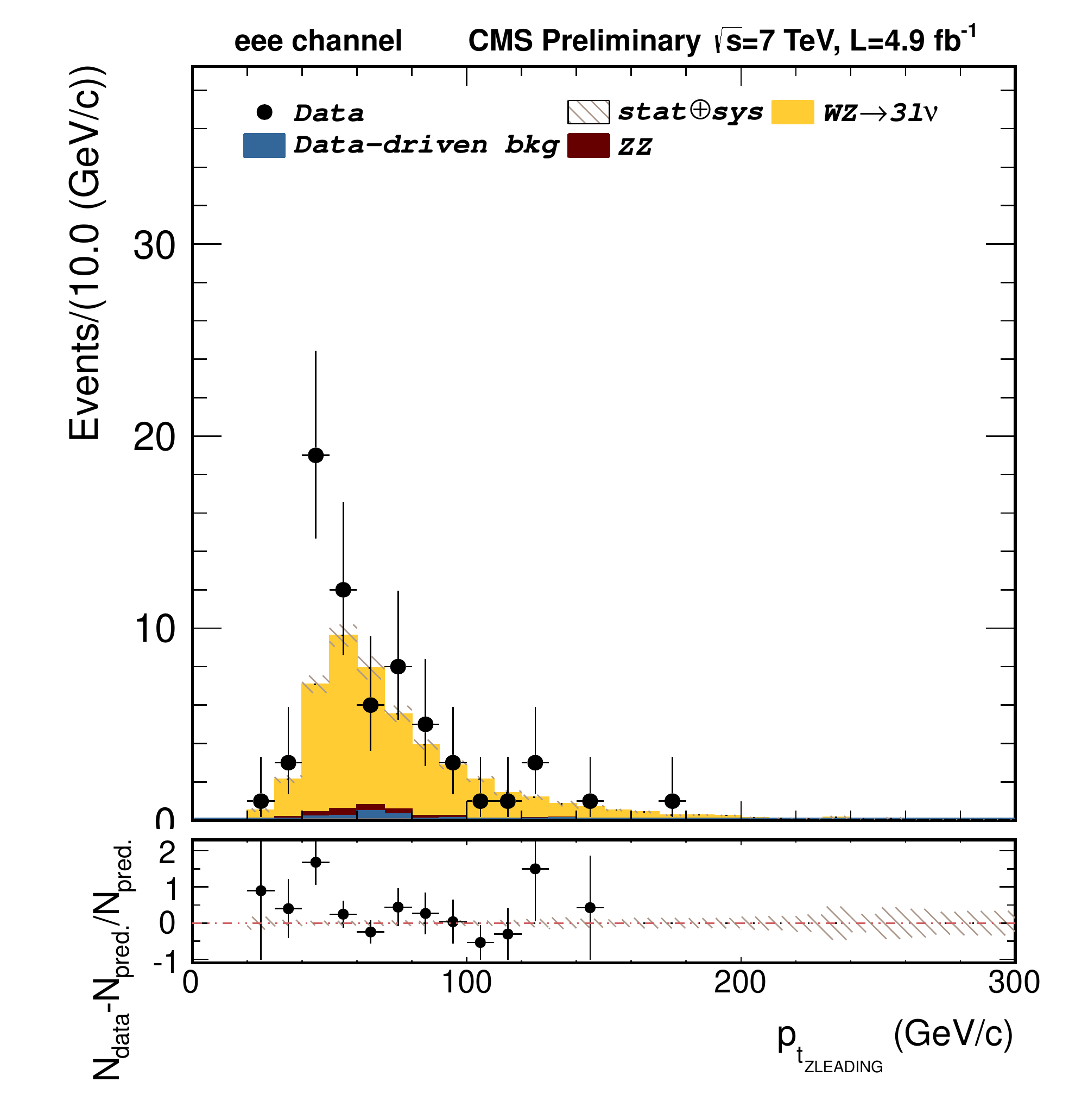}
	\end{subfigure}\quad
	\begin{subfigure}[b]{0.2\textwidth}
		\includegraphics[width=\textwidth]{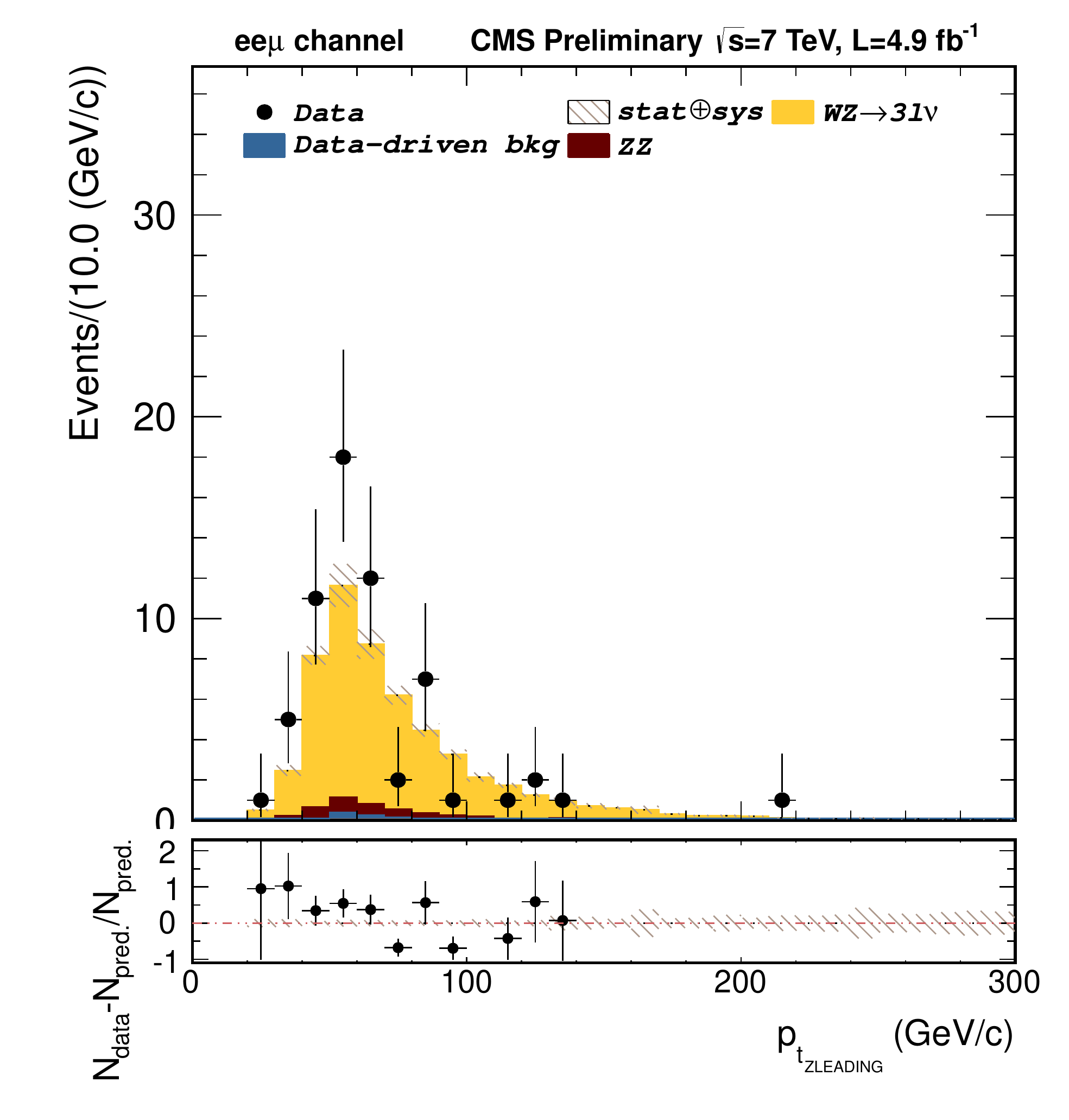}
	\end{subfigure}\quad
	\begin{subfigure}[b]{0.2\textwidth}
		\includegraphics[width=\textwidth]{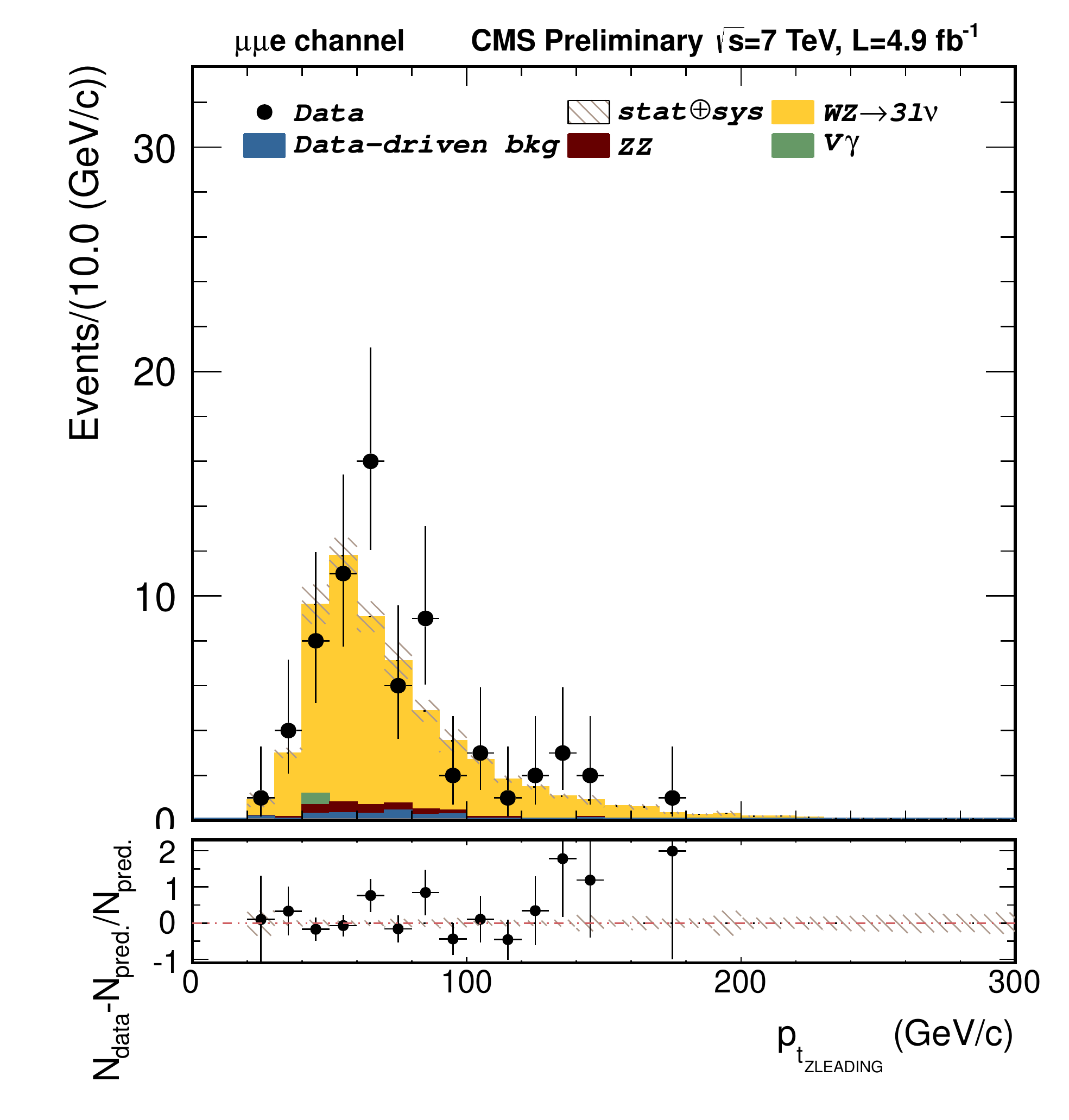}
	\end{subfigure}\quad
	\begin{subfigure}[b]{0.2\textwidth}
		\includegraphics[width=\textwidth]{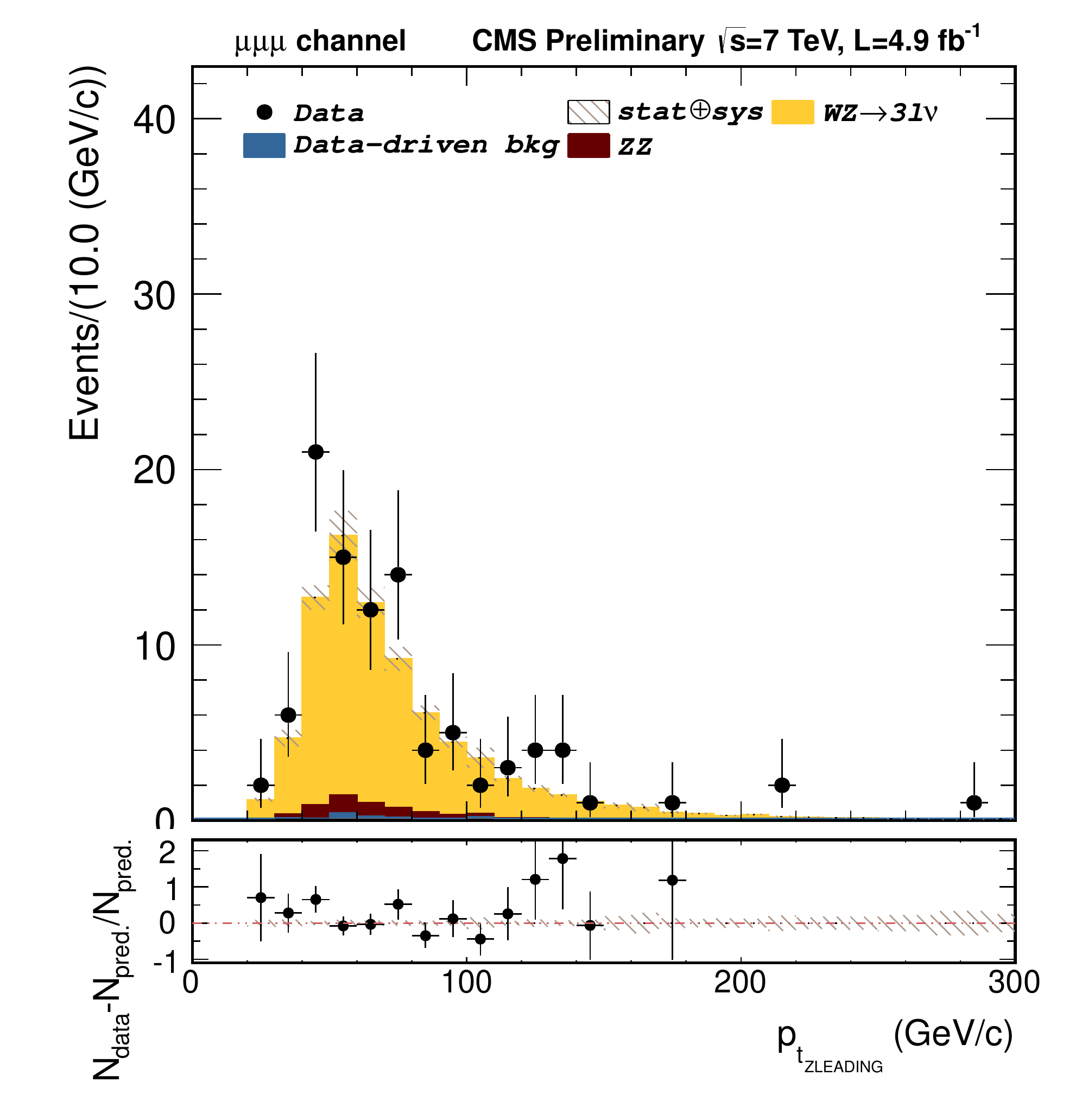}
	\end{subfigure}
	\caption[Transverse momentum of the Z-candidate leading lepton at 7~\TeV]
	{Transverse momentum of the Z-candidate
	leading lepton at each event for the measured channels $eee$, $\mu ee$, $e\mu\mu$ and 
	$\mu\mu\mu$ (from left to right) after W-candidate requirement is applied.}
	\vskip 1em
	\centering
	\begin{subfigure}[b]{0.2\textwidth}
		\includegraphics[width=\textwidth]{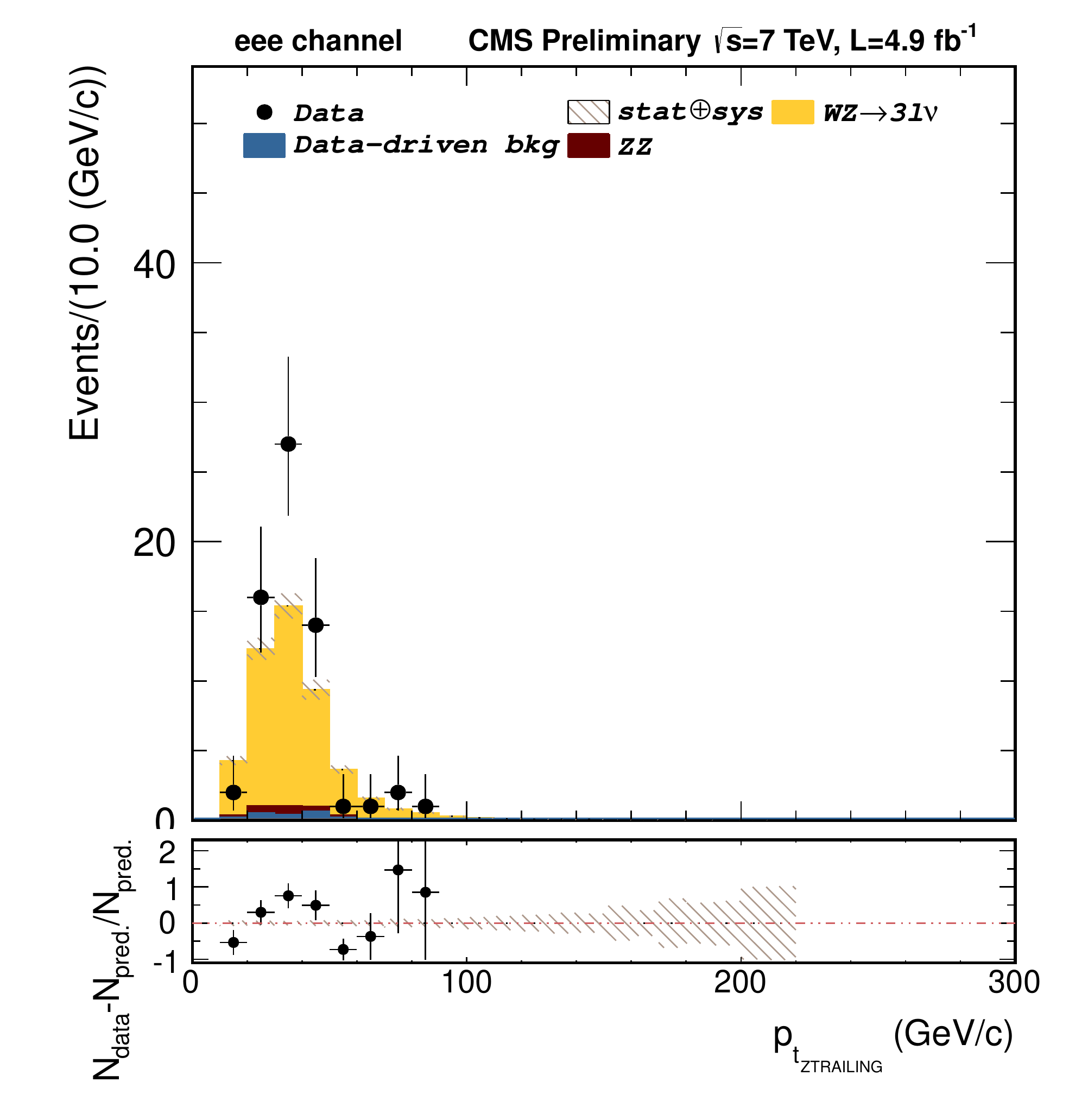}
	\end{subfigure}\quad
	\begin{subfigure}[b]{0.2\textwidth}
		\includegraphics[width=\textwidth]{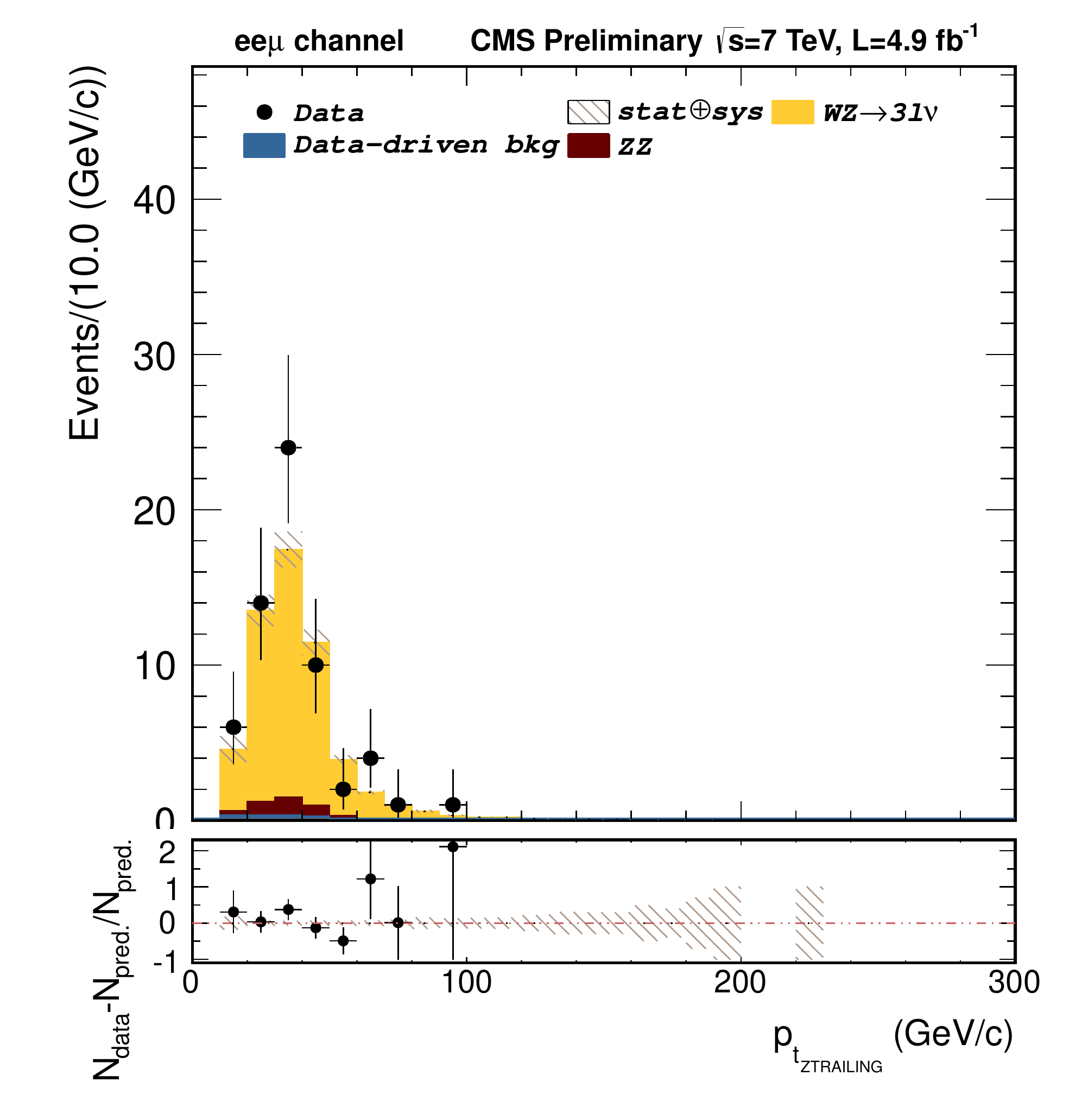}
	\end{subfigure}\quad
	\begin{subfigure}[b]{0.2\textwidth}
		\includegraphics[width=\textwidth]{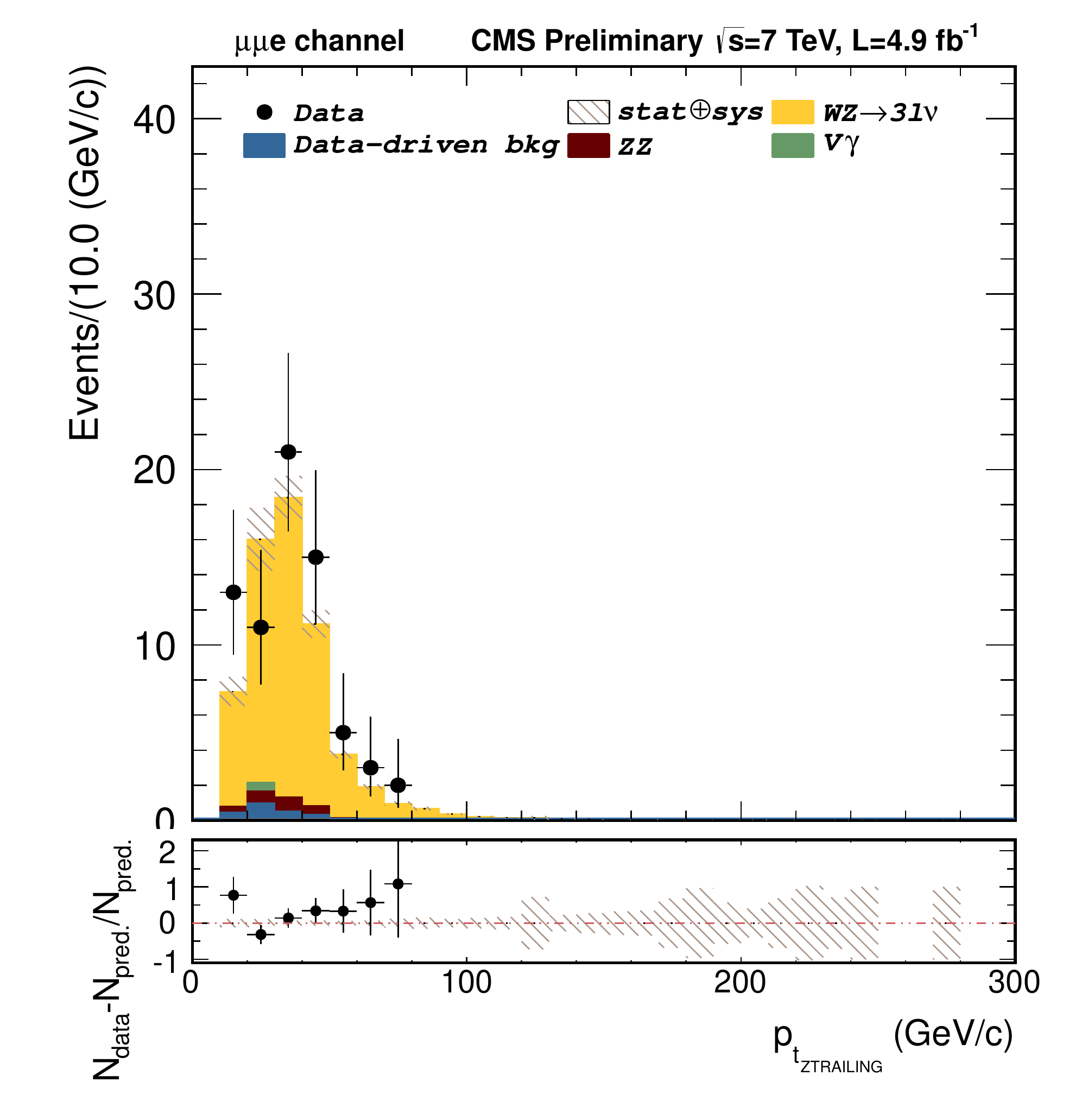}
	\end{subfigure}\quad
	\begin{subfigure}[b]{0.2\textwidth}
		\includegraphics[width=\textwidth]{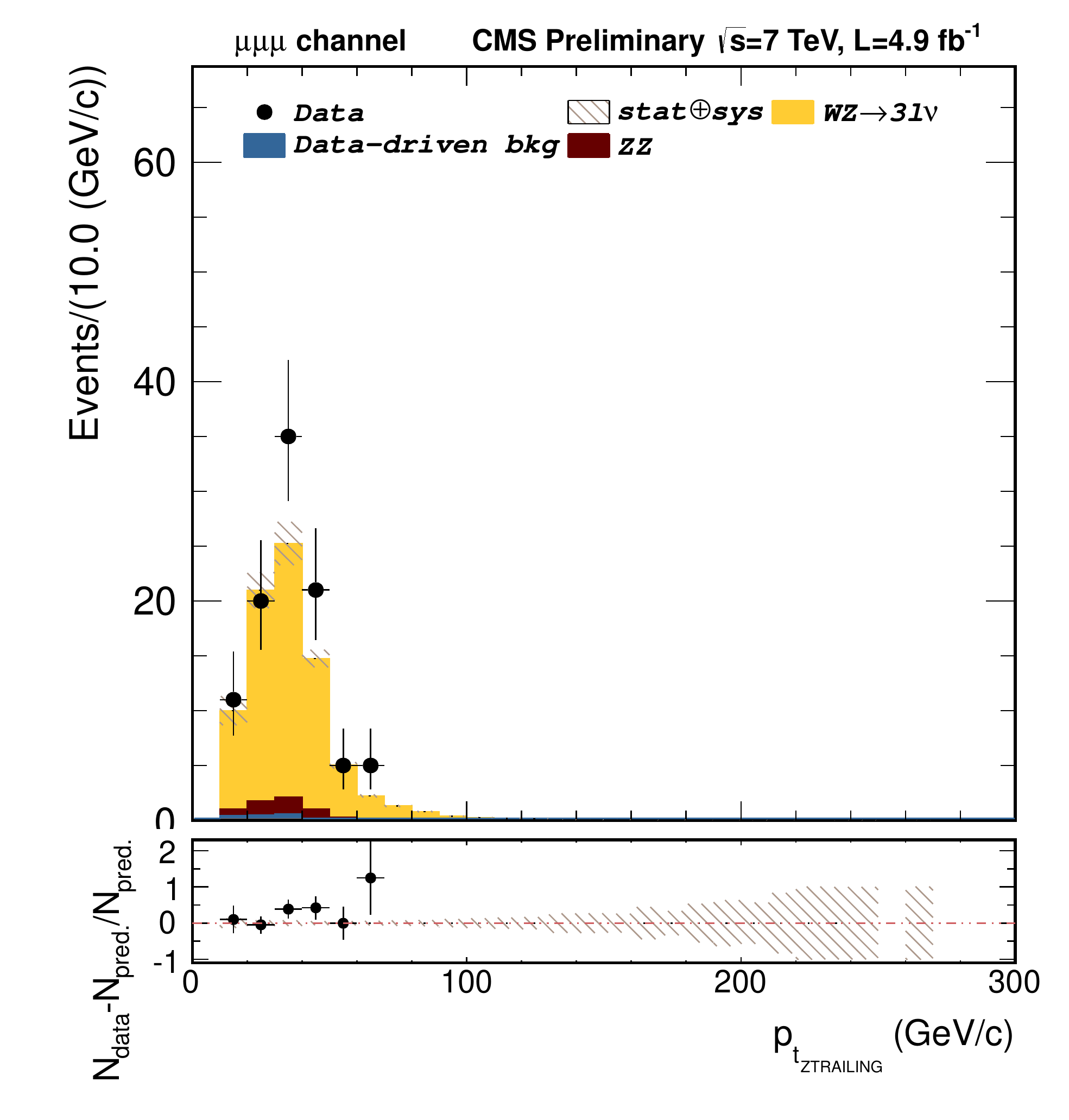}
	\end{subfigure}
	\caption[Transverse momentum of the Z-candidate trailing lepton at 7~\TeV]
	{Transverse momentum of the Z-candidate
	trailing lepton at each event for the measured channels $eee$, $\mu ee$, $e\mu\mu$ and 
	$\mu\mu\mu$ (from left to right) after W-candidate requirement is applied.}
\end{sidewaysfigure}
\clearpage

\section{Ratio analysis distributions at 7~\TeV}
The distributions shown in this subsection correspond to the 2011 analysis of the cross section
ratio between the \wzp and \wzm processes. The samples follows the colour conventions and the processing
explained in the previous section. The distributions are grouped by observable, showing in each figure 
two columns corresponding to the combined channel of the \wzm and \wzp and each row to a stage of the 
analysis. Note that since the sample splitting by charge lies in the \W-candidate, the analysis stages 
are shown from this requirement on.
\begin{figure}[!htpb]
	\centering
	\begin{subfigure}[b]{0.4\textwidth}
		\includegraphics[width=\textwidth]{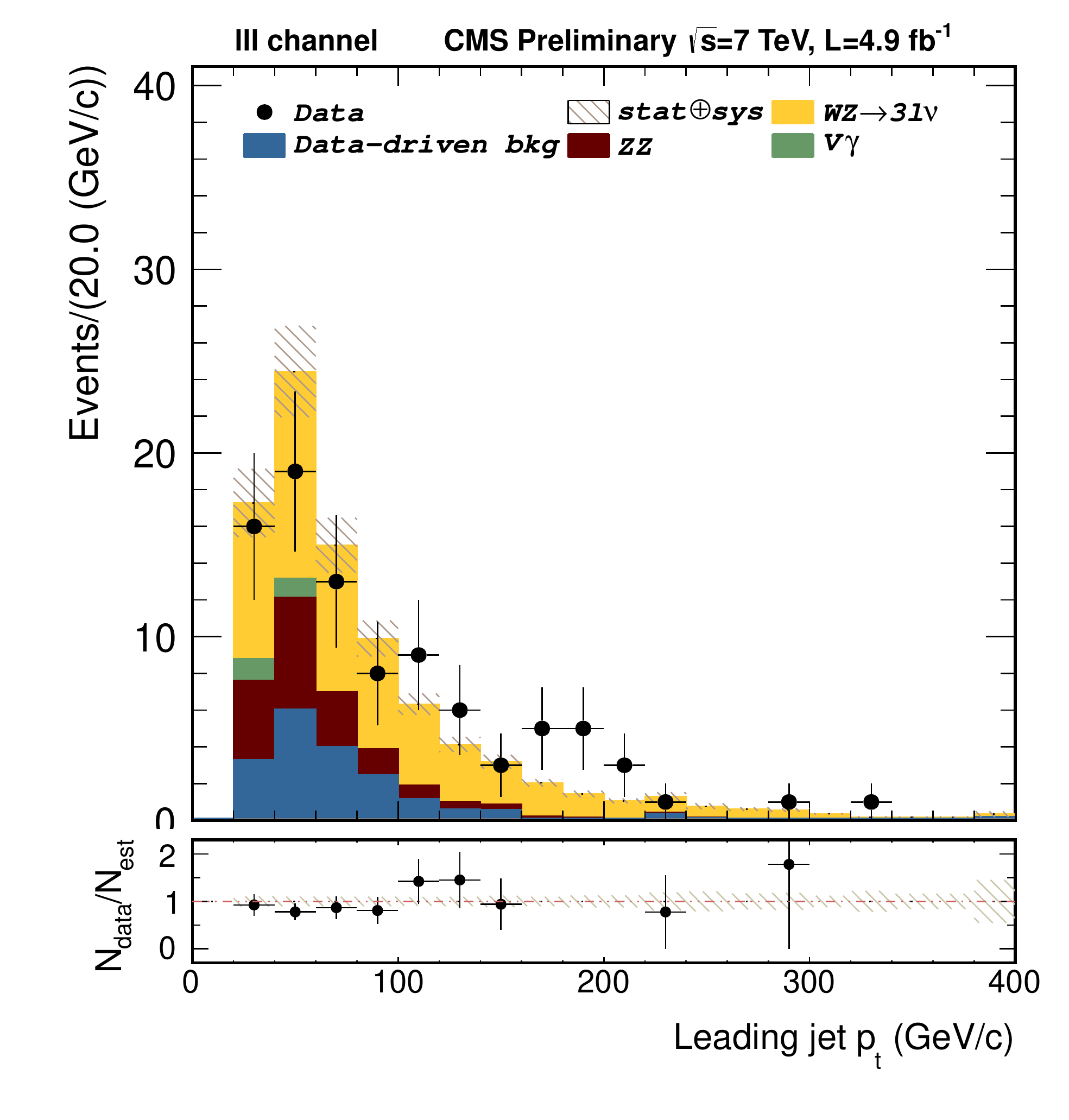}
	\end{subfigure}\quad
	\begin{subfigure}[b]{0.4\textwidth}
		\includegraphics[width=\textwidth]{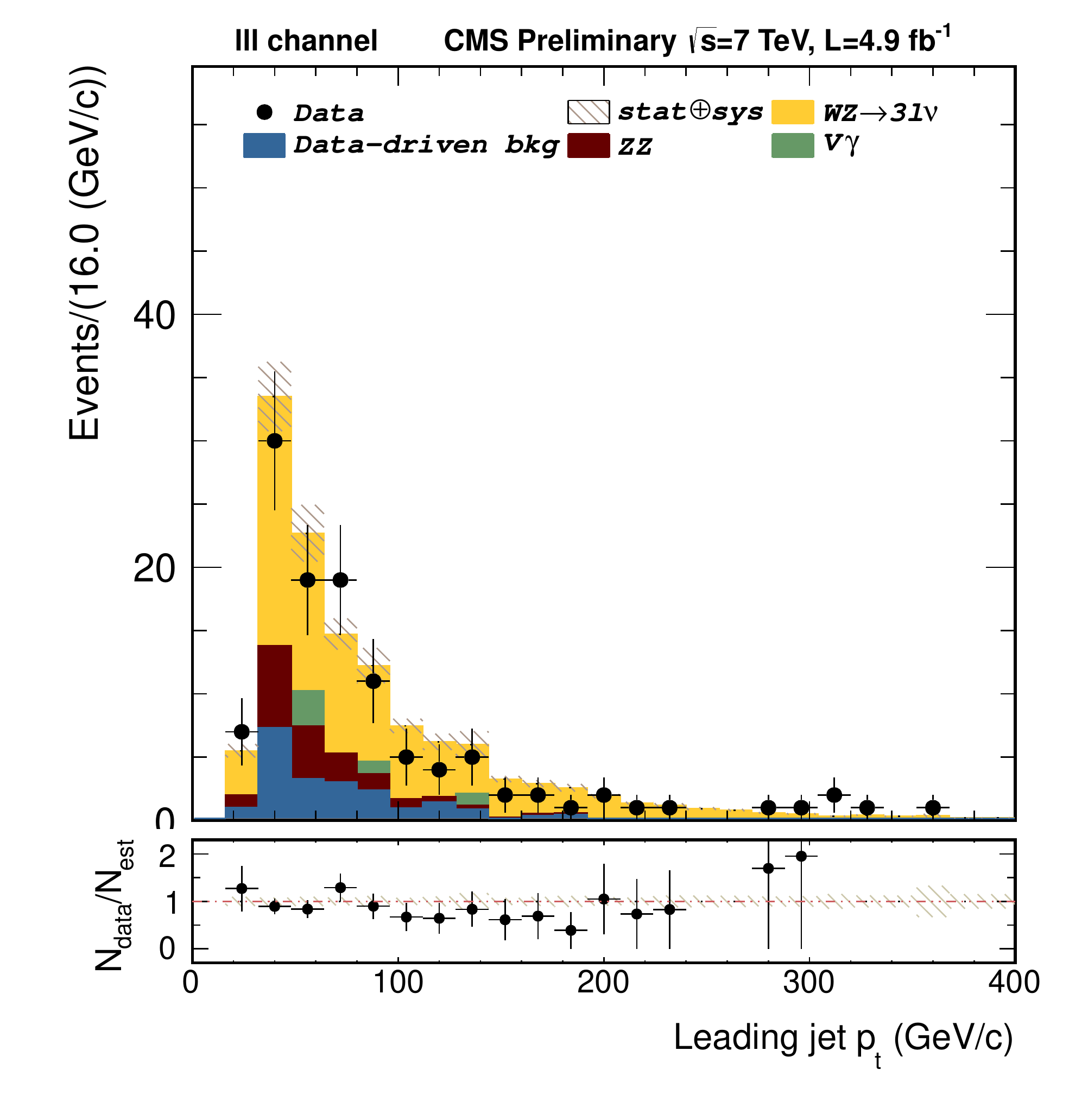}
	\end{subfigure}
	\vskip 1ex
	\begin{subfigure}[b]{0.4\textwidth}
		\includegraphics[width=\textwidth]{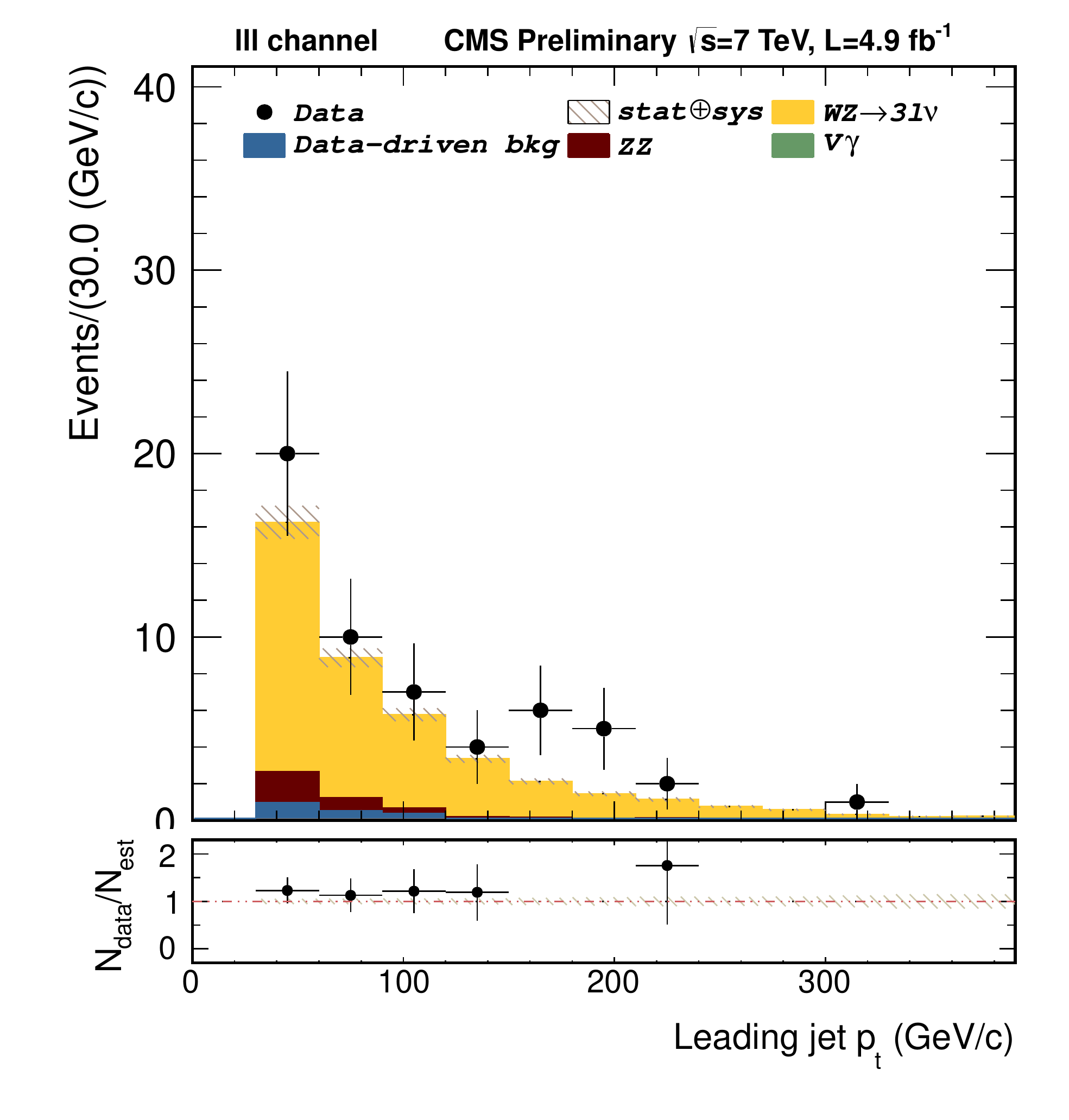}
	\end{subfigure}\quad
	\begin{subfigure}[b]{0.4\textwidth}
		\includegraphics[width=\textwidth]{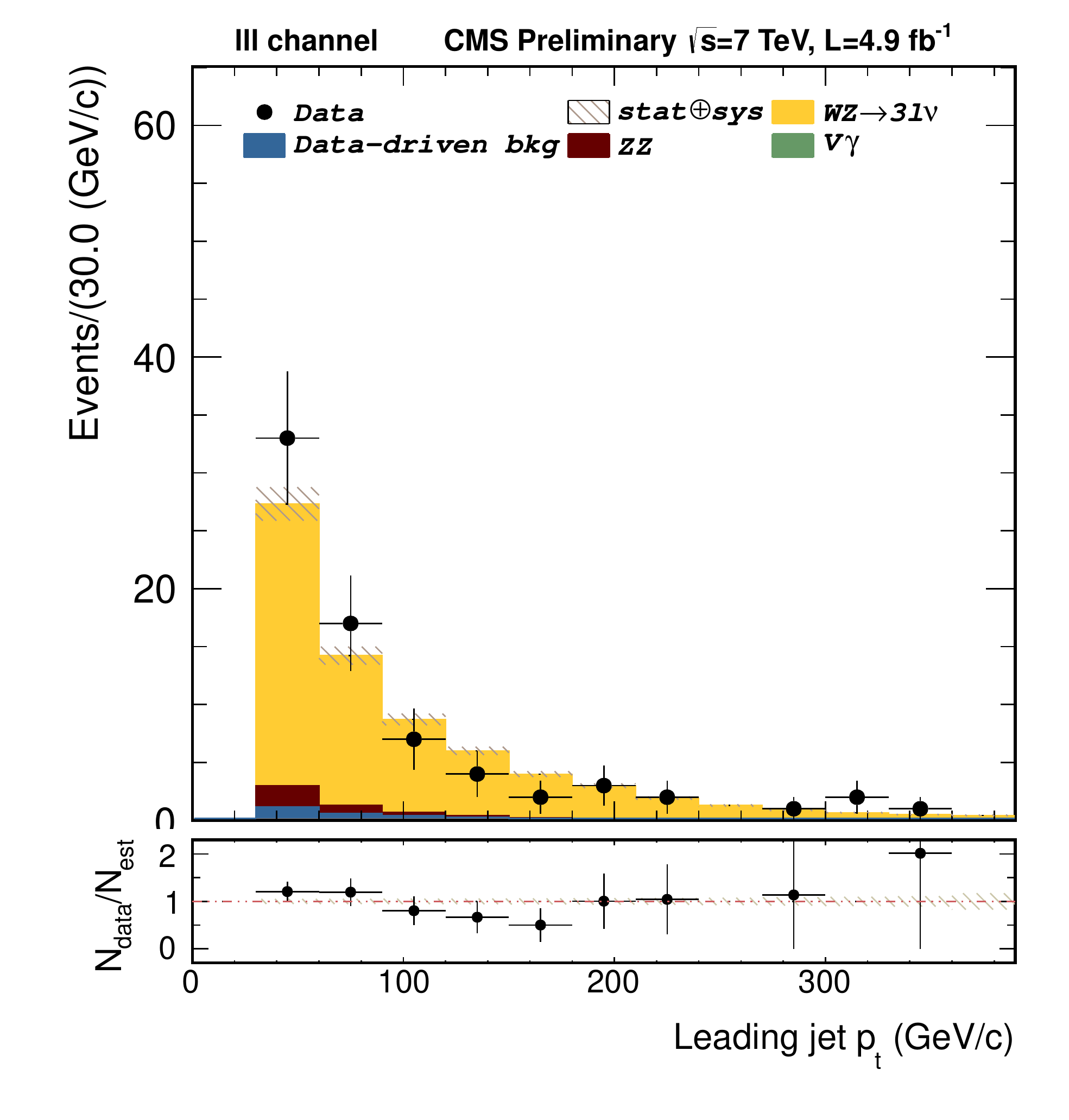}
	\end{subfigure}
	\caption[Transverse momentum of the leading jet at 7 TeV (ratio)]{Transverse 
	momentum distribution of the leading jet at each event for the \wzm (left column) and
	\wzp (right column) before the \MET cut (up row) and after (bottom row).}
\end{figure}

\begin{figure}
	\centering
	\begin{subfigure}[b]{0.3\textwidth}
		\includegraphics[width=\textwidth]{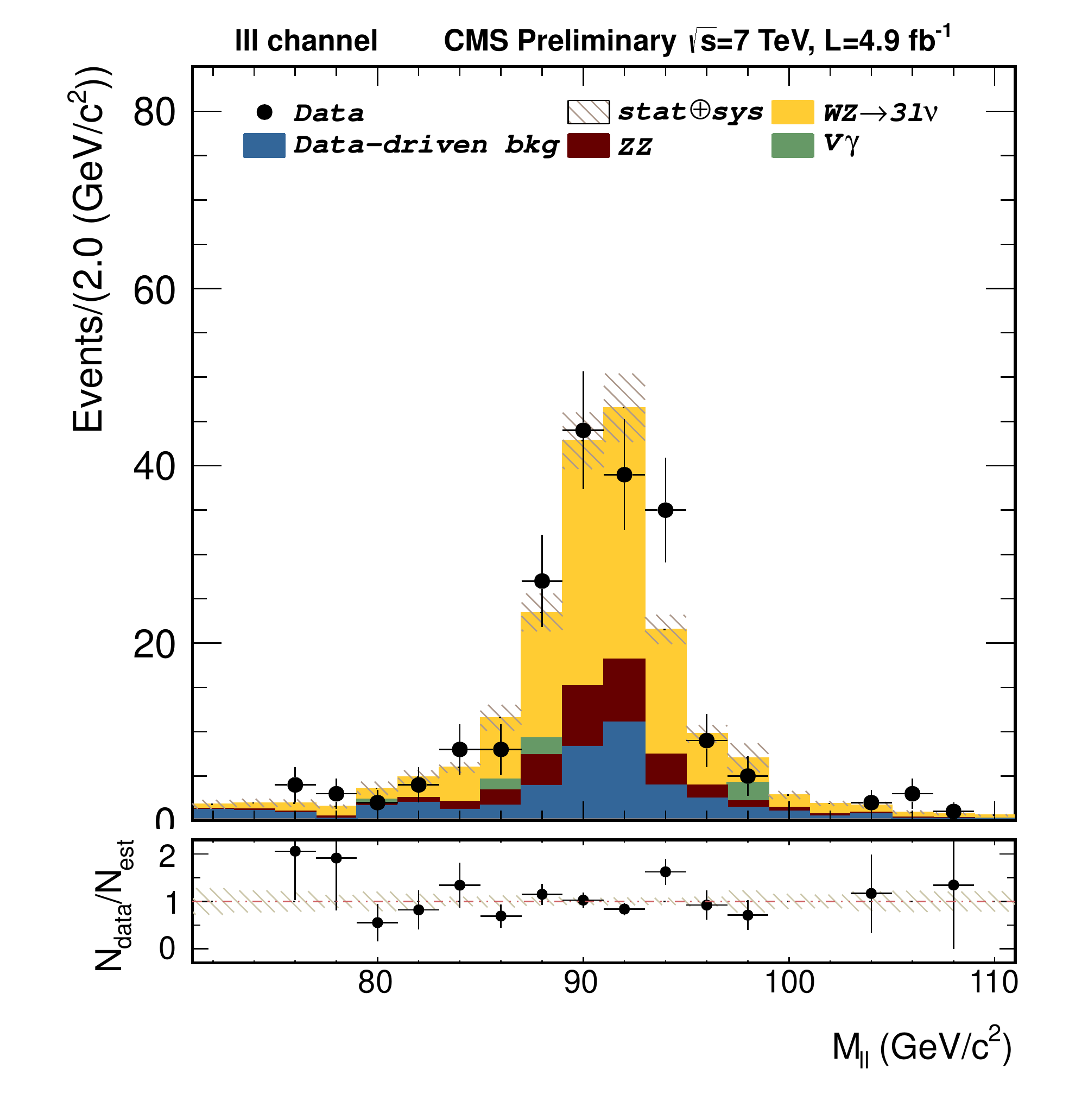}
	\end{subfigure}\quad
	\begin{subfigure}[b]{0.3\textwidth}
		\includegraphics[width=\textwidth]{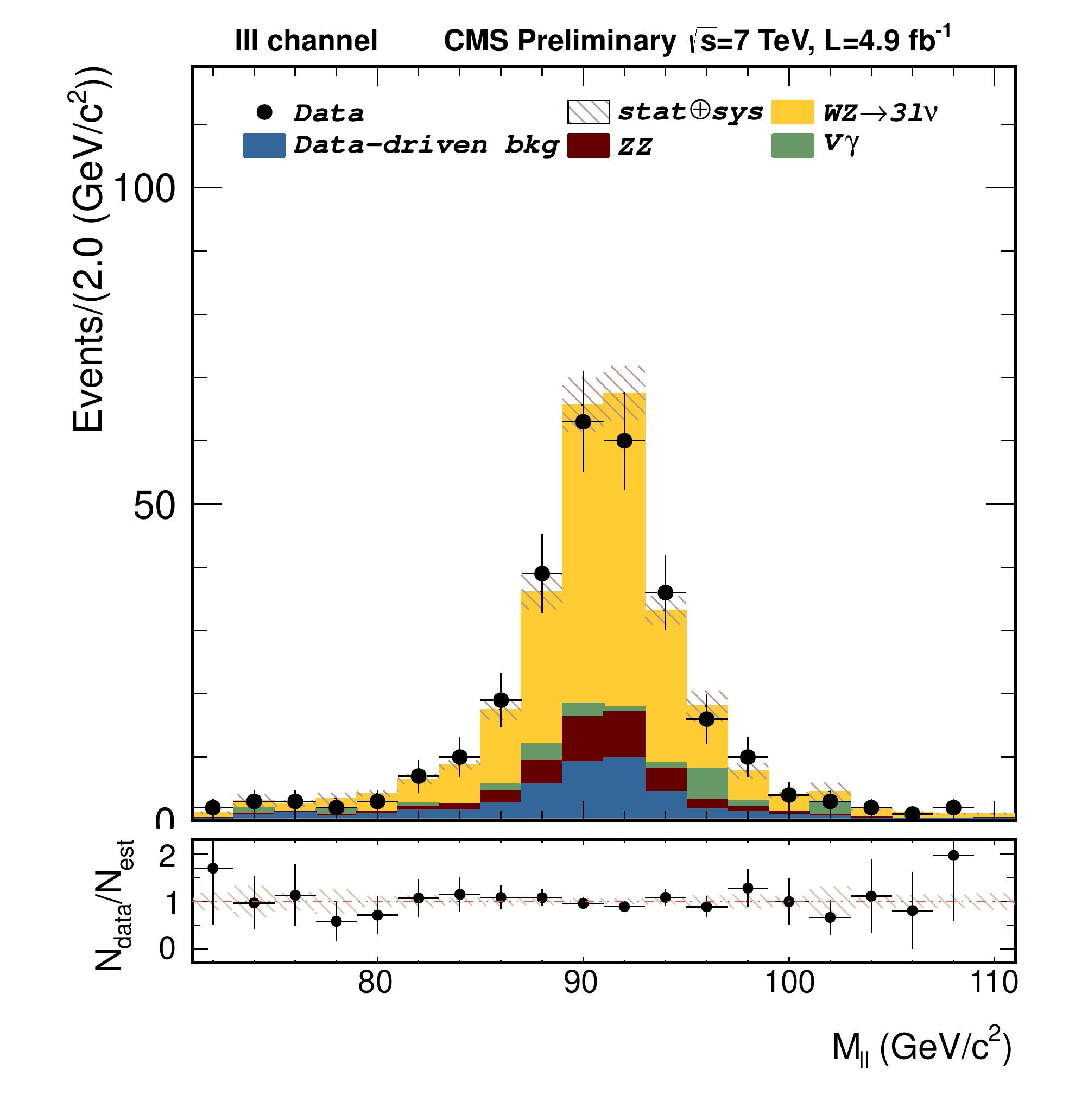}
	\end{subfigure}
	\vskip 1ex
	\begin{subfigure}[b]{0.3\textwidth}
		\includegraphics[width=\textwidth]{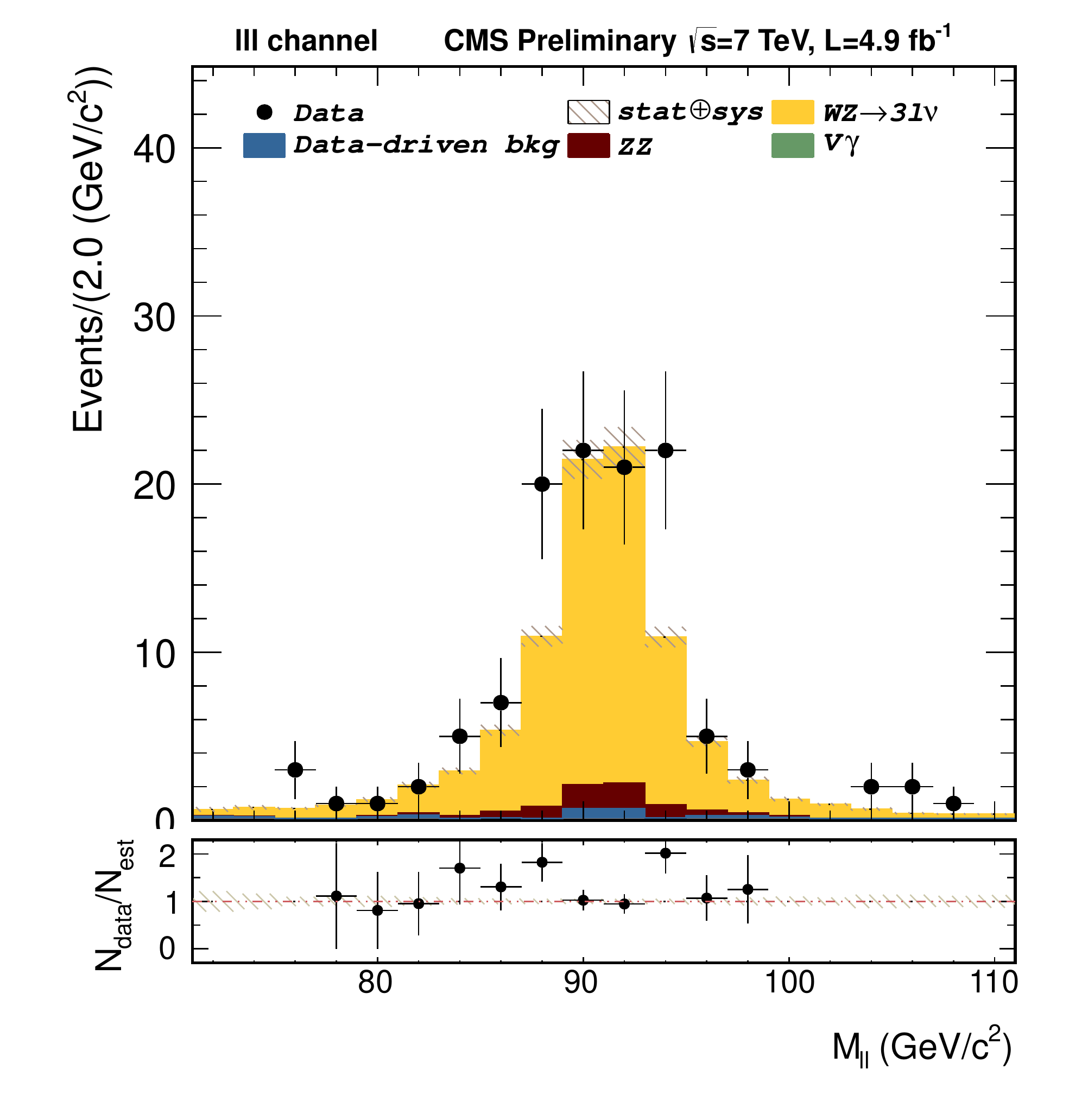}
	\end{subfigure}\quad
	\begin{subfigure}[b]{0.3\textwidth}
		\includegraphics[width=\textwidth]{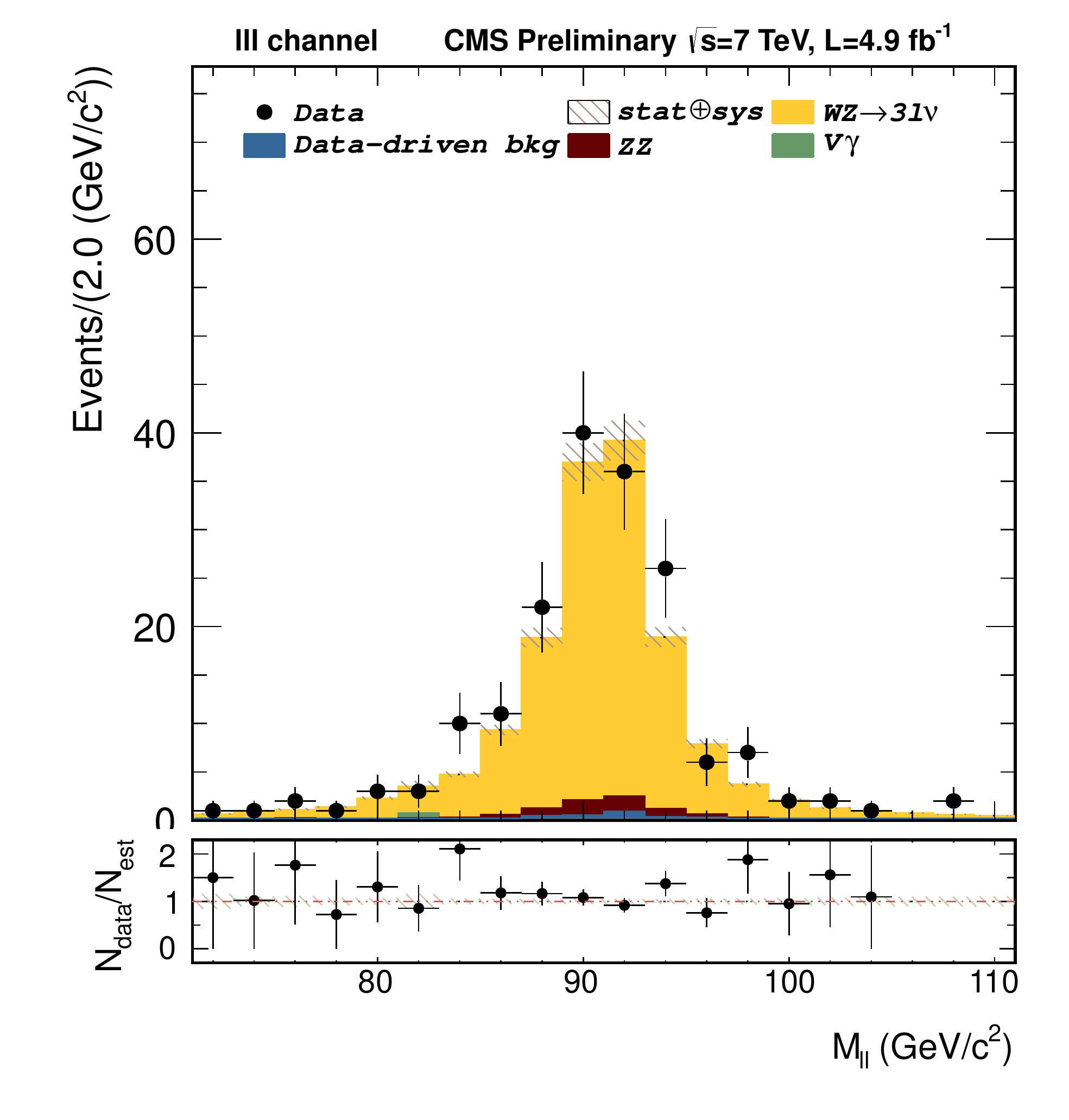}
	\end{subfigure}
	\caption[Invariant mass of the dilepton system at 7~\TeV (ratio)]
	{Invariant mass of the Z-candidate dilepton system for the \wzm (left column) and \wzp 
	(right column) before (up row) and after the \MET cut (bottom row).}
	\vskip 1em
	\begin{subfigure}[b]{0.3\textwidth}
		\includegraphics[width=\textwidth]{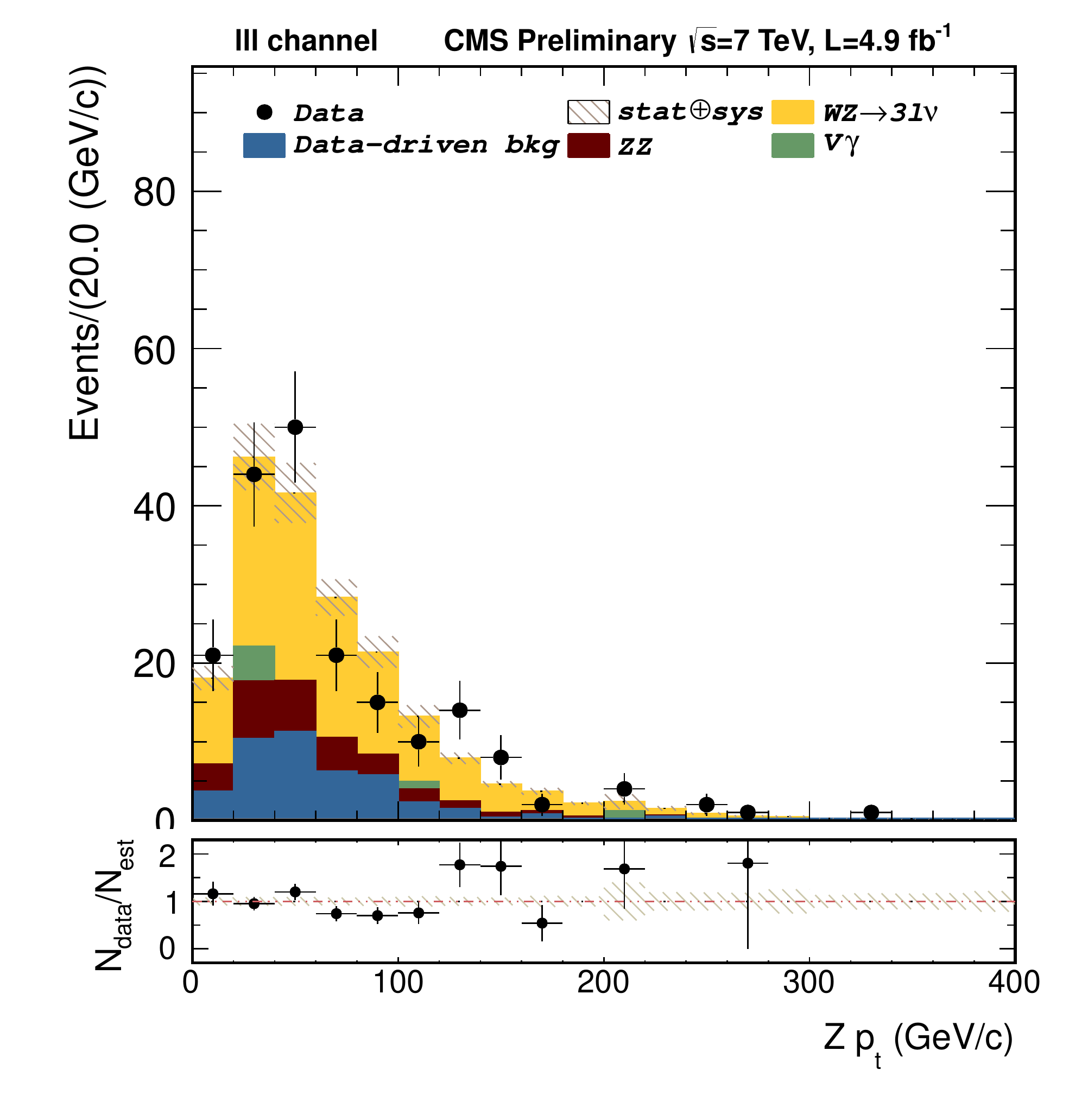}
	\end{subfigure}\quad
	\begin{subfigure}[b]{0.3\textwidth}
		\includegraphics[width=\textwidth]{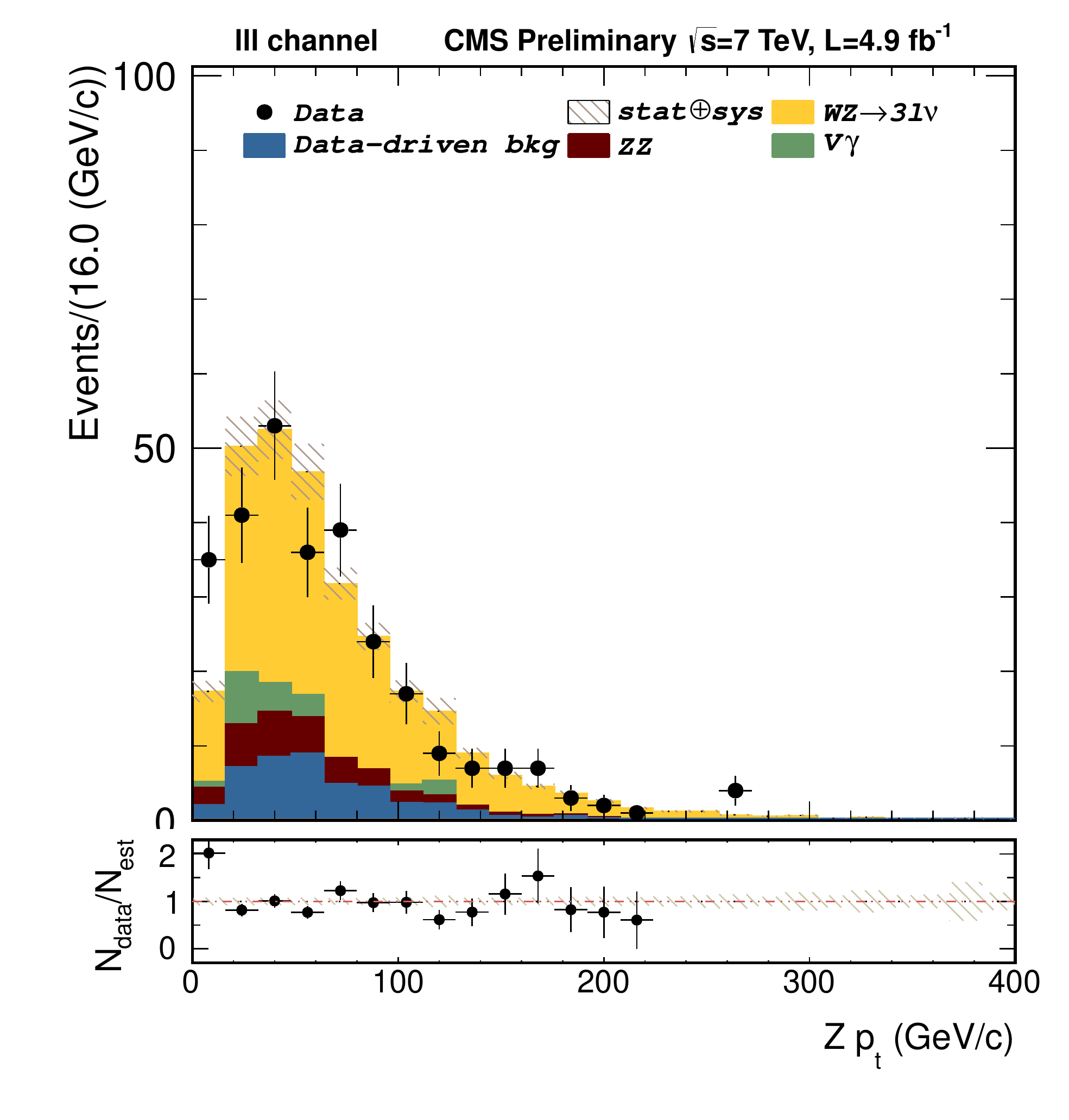}
	\end{subfigure}
	\vskip 1ex
	\begin{subfigure}[b]{0.3\textwidth}
		\includegraphics[width=\textwidth]{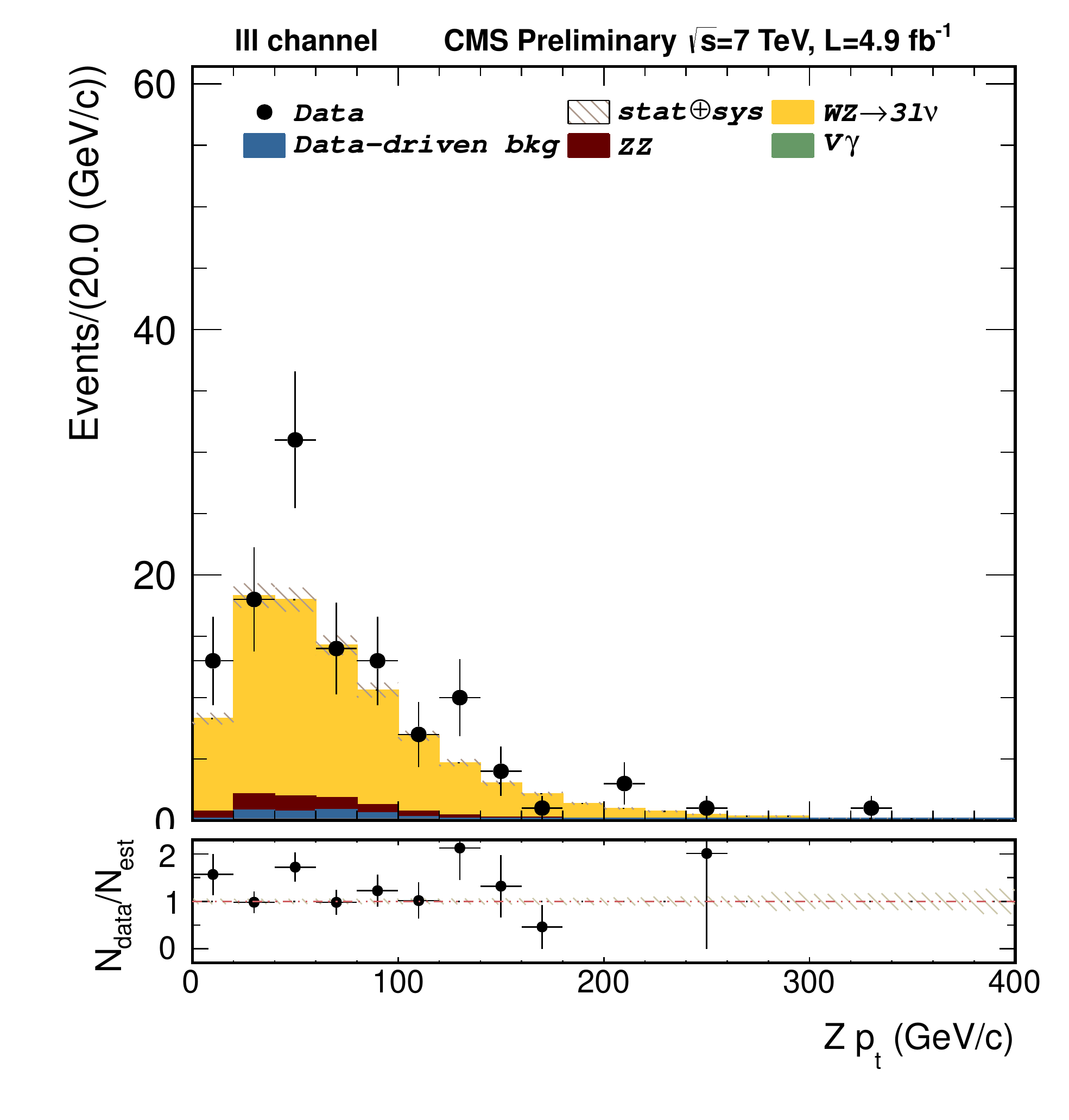}
	\end{subfigure}\quad
	\begin{subfigure}[b]{0.3\textwidth}
		\includegraphics[width=\textwidth]{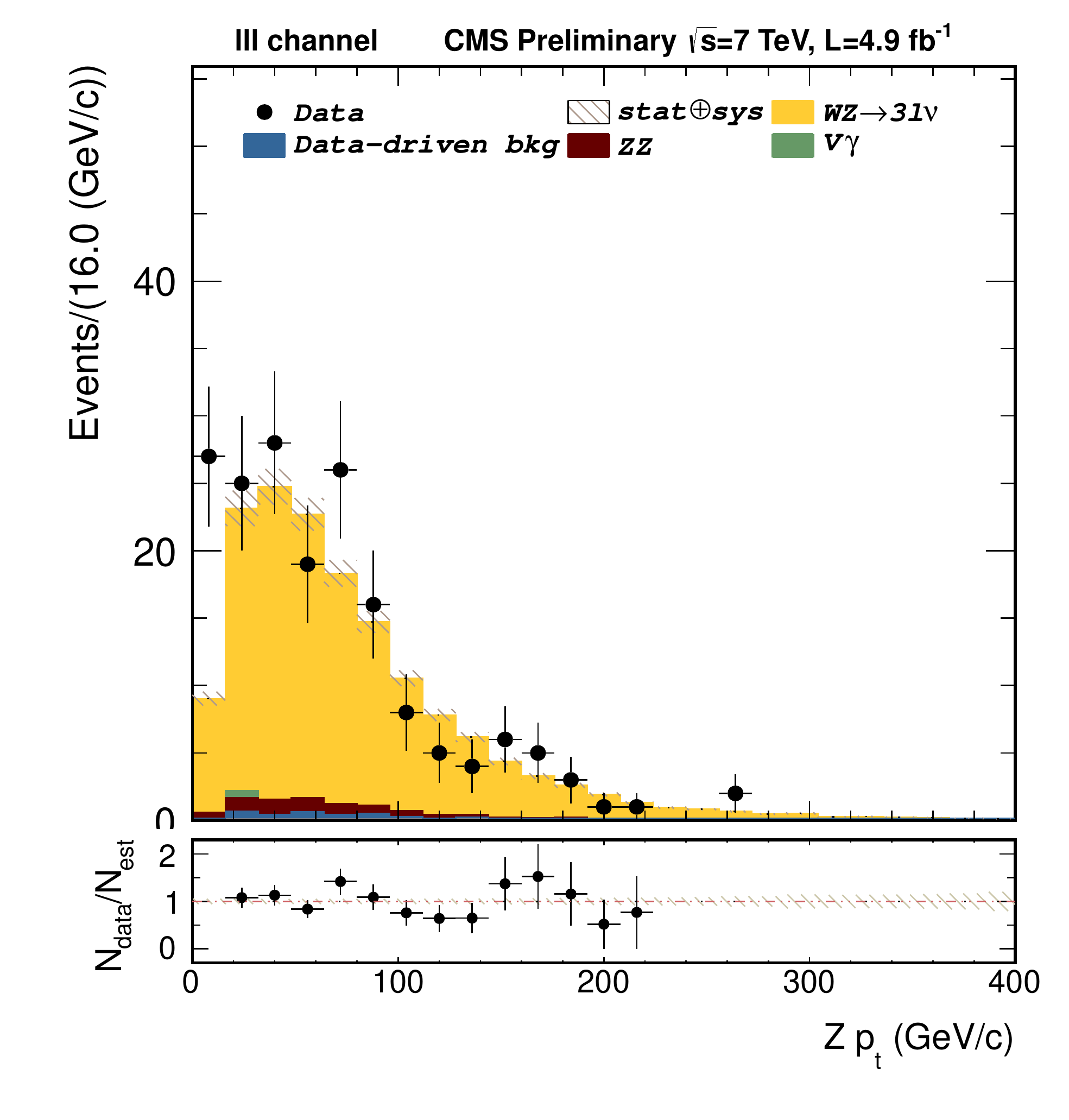}
	\end{subfigure}
	\caption[Transverse momentum of the dilepton system at 7~\TeV (ratio)]
	{Transverse momentum of the Z-candidate dilepton system for the \wzm (left column) and \wzp 
	(right column) before (up row) and after the \MET cut (bottom row).}
\end{figure}

\begin{figure}[!htpb]
	\centering
	\begin{subfigure}[b]{0.3\textwidth}
		\includegraphics[width=\textwidth]{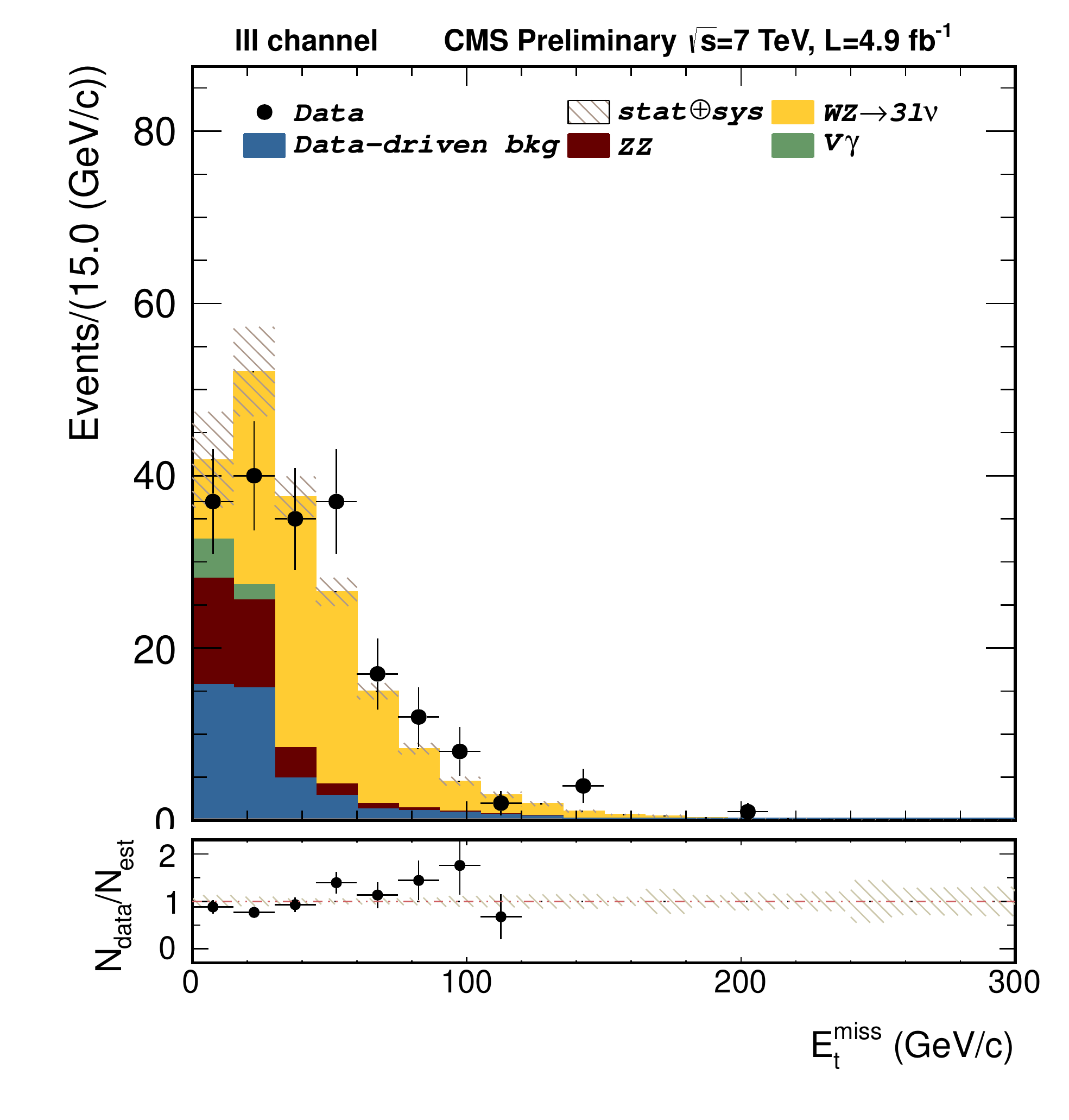}
	\end{subfigure}\quad
	\begin{subfigure}[b]{0.3\textwidth}
		\includegraphics[width=\textwidth]{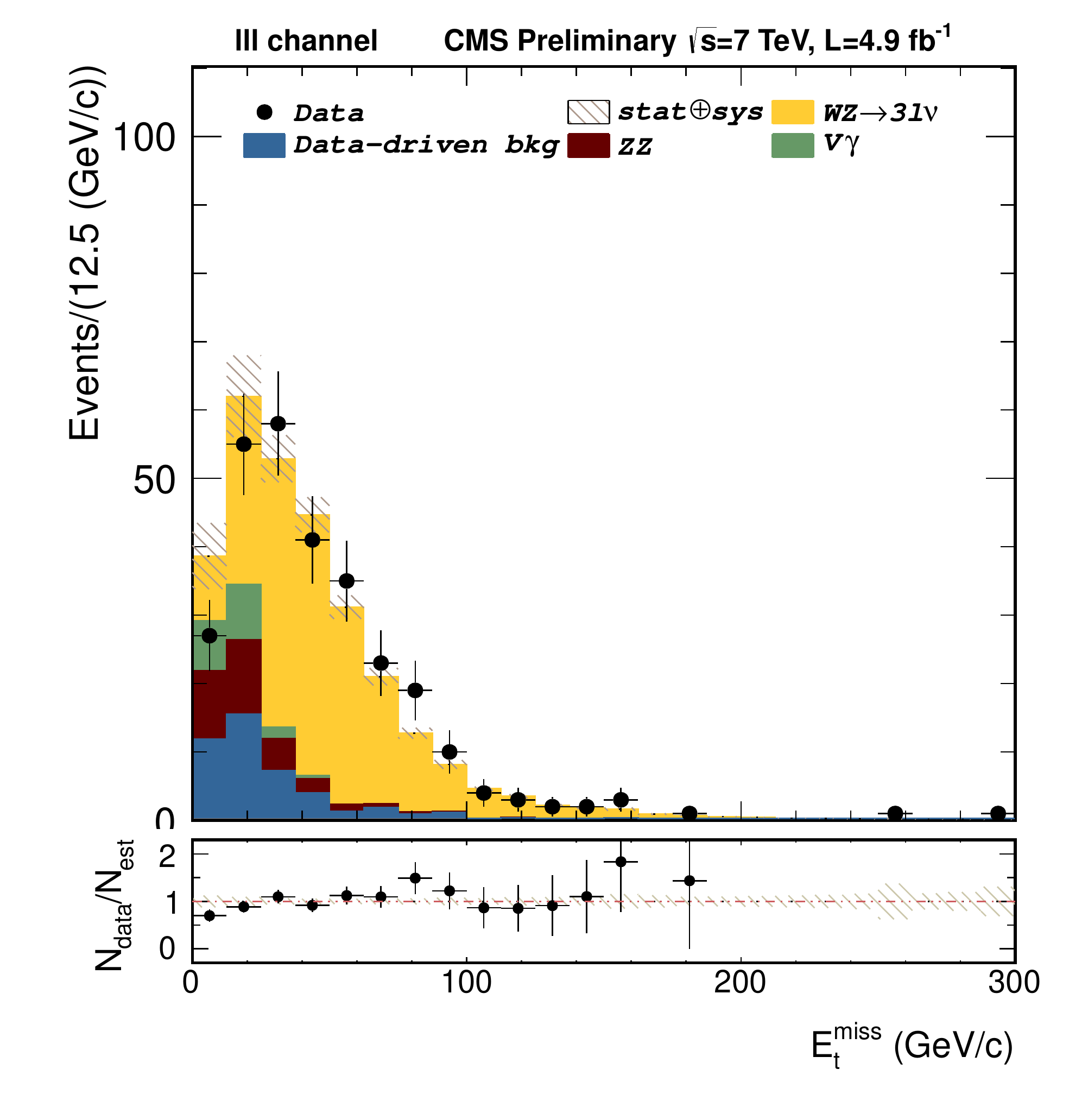}
	\end{subfigure}
	\vskip 1ex
	\begin{subfigure}[b]{0.3\textwidth}
		\includegraphics[width=\textwidth]{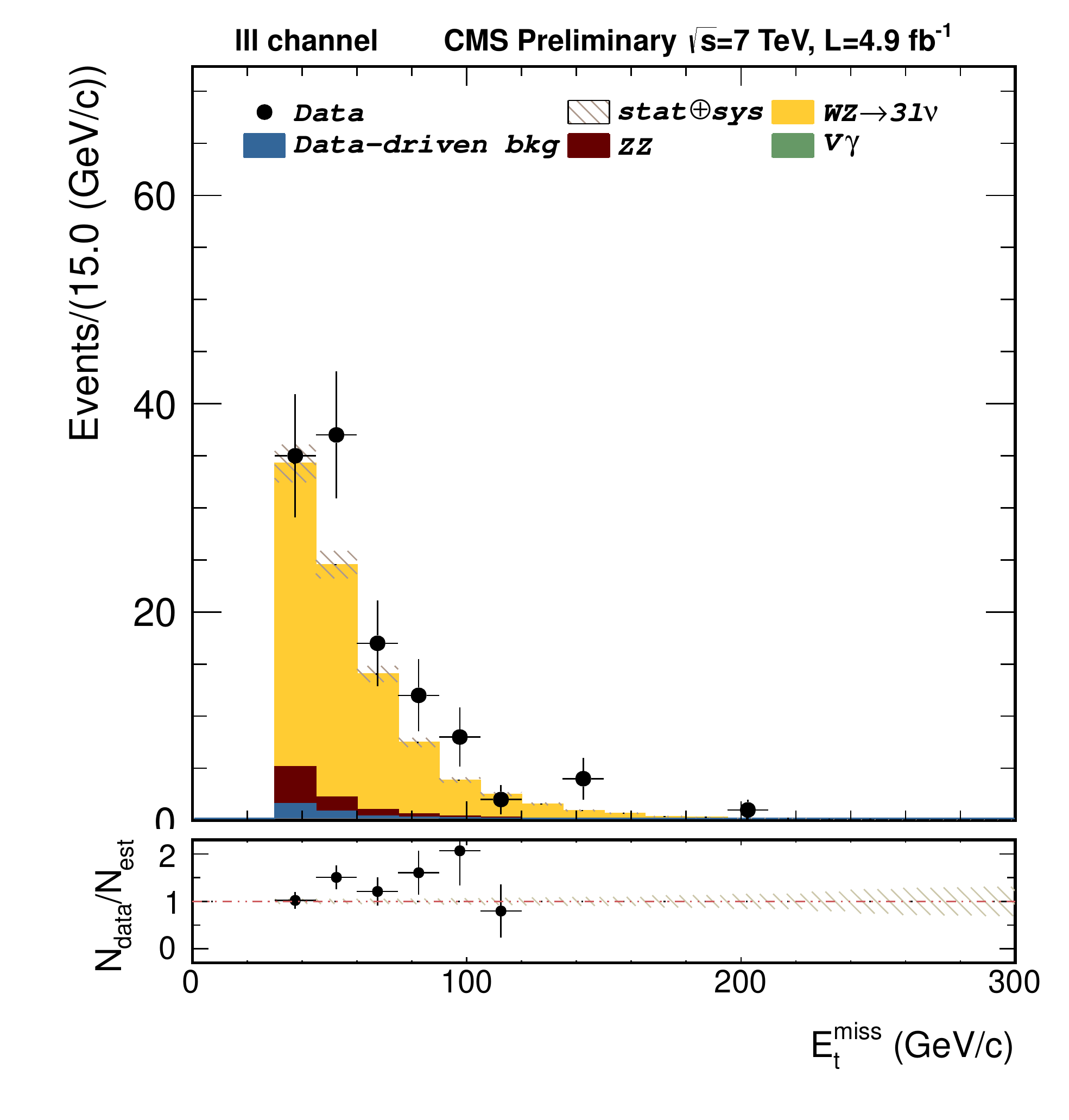}
	\end{subfigure}\quad
	\begin{subfigure}[b]{0.3\textwidth}
		\includegraphics[width=\textwidth]{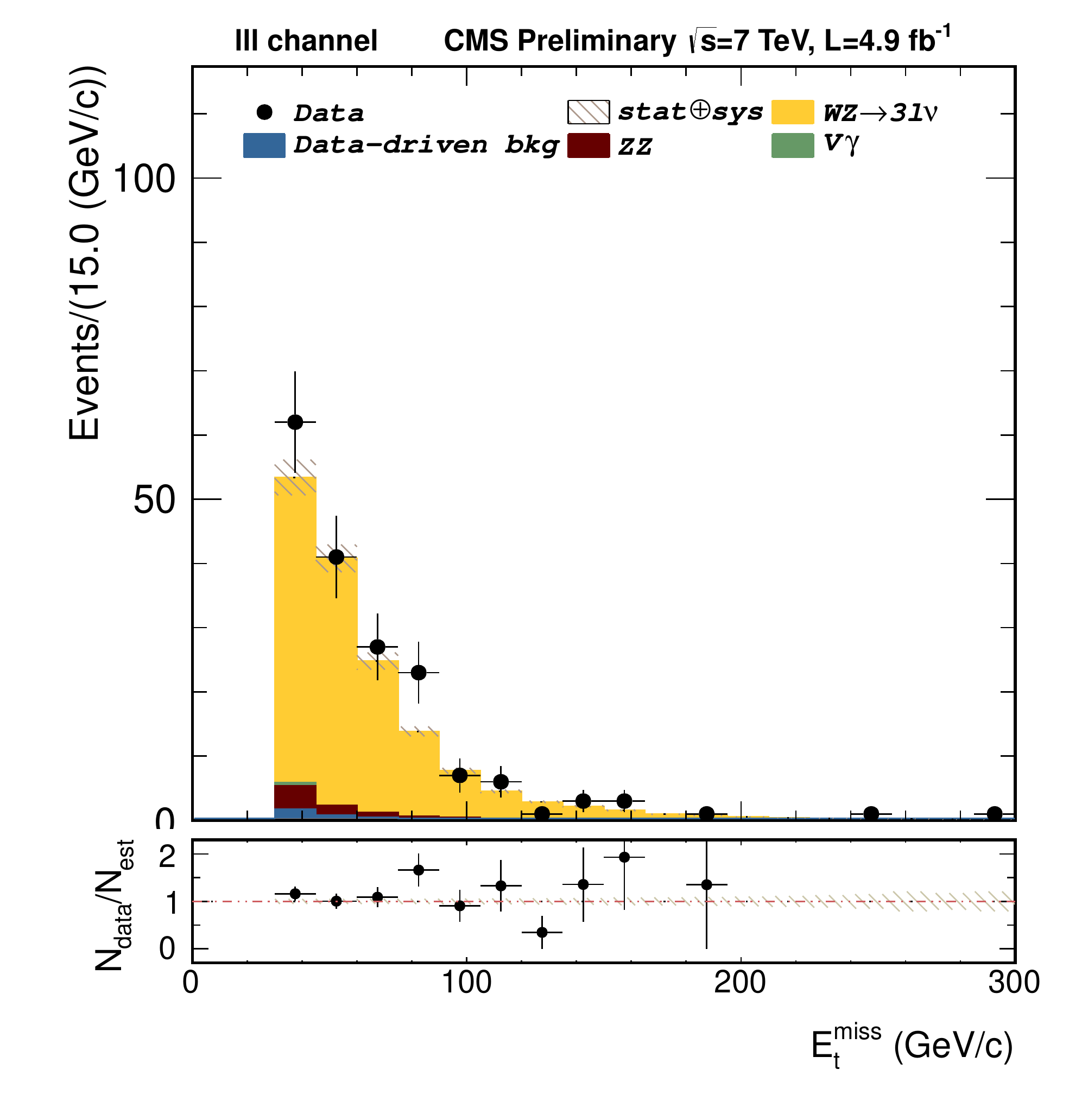}
	\end{subfigure}
	\caption[Missing transverse energy at 7~\TeV (ratio)]{Missing 
	energy in the transverse plane at each event for the \wzm (left column) and \wzp 
	(right column) before (up row) and after the \MET cut (bottom row).	}
	\vskip 1em
	\begin{subfigure}[b]{0.3\textwidth}
		\includegraphics[width=\textwidth]{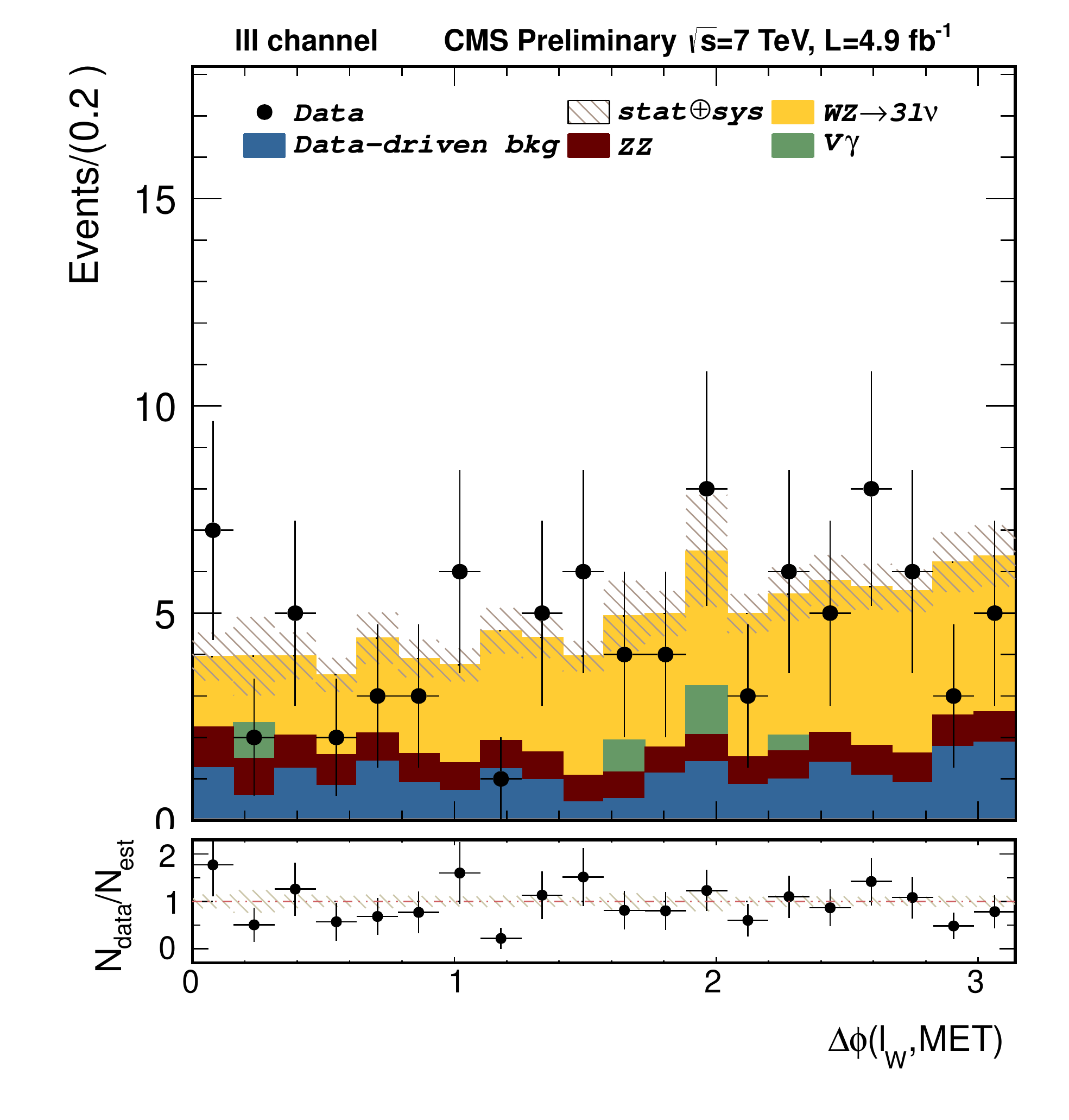}
	\end{subfigure}\quad
	\begin{subfigure}[b]{0.3\textwidth}
		\includegraphics[width=\textwidth]{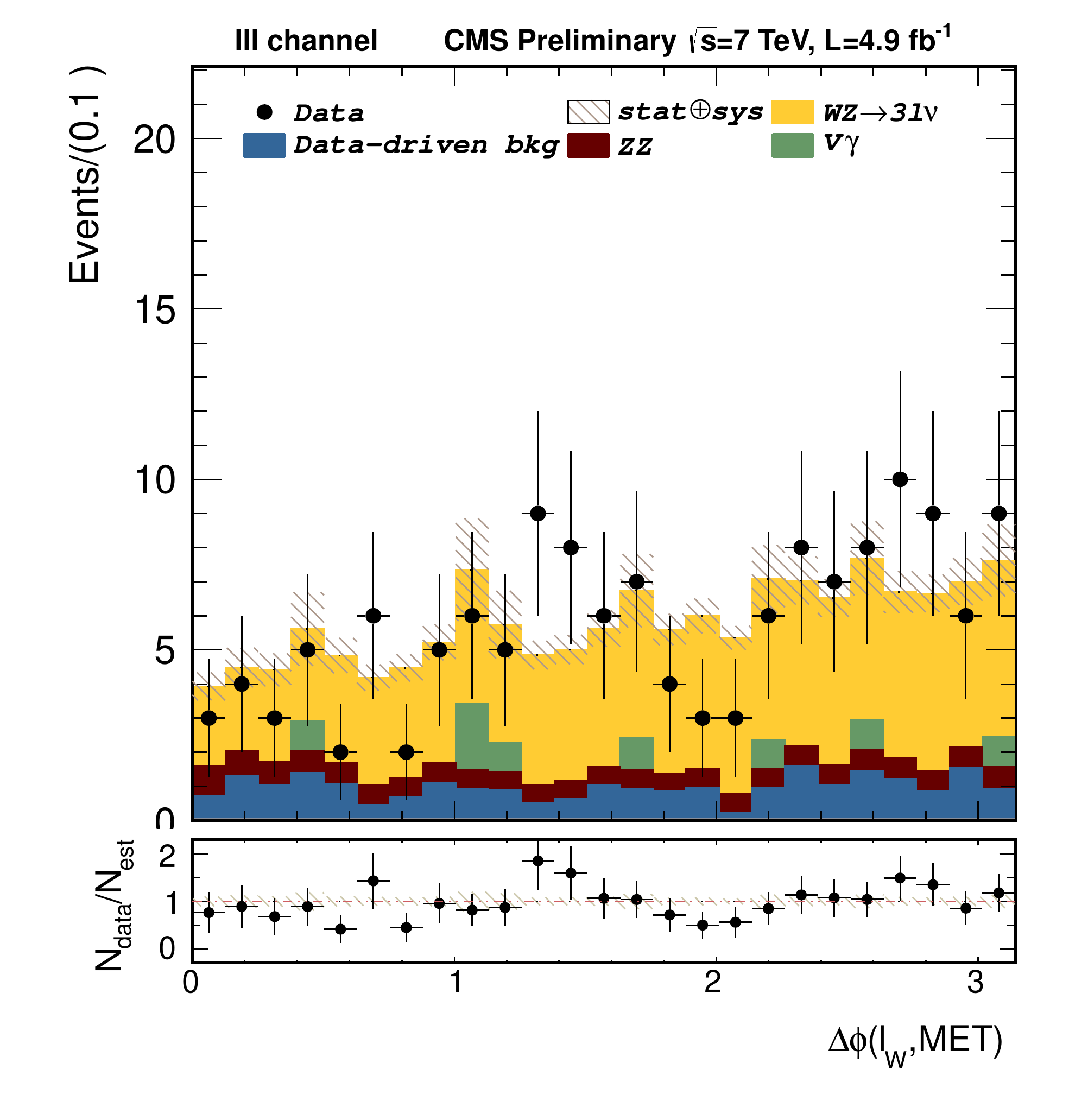}
	\end{subfigure}
	\vskip 1ex
	\begin{subfigure}[b]{0.3\textwidth}
		\includegraphics[width=\textwidth]{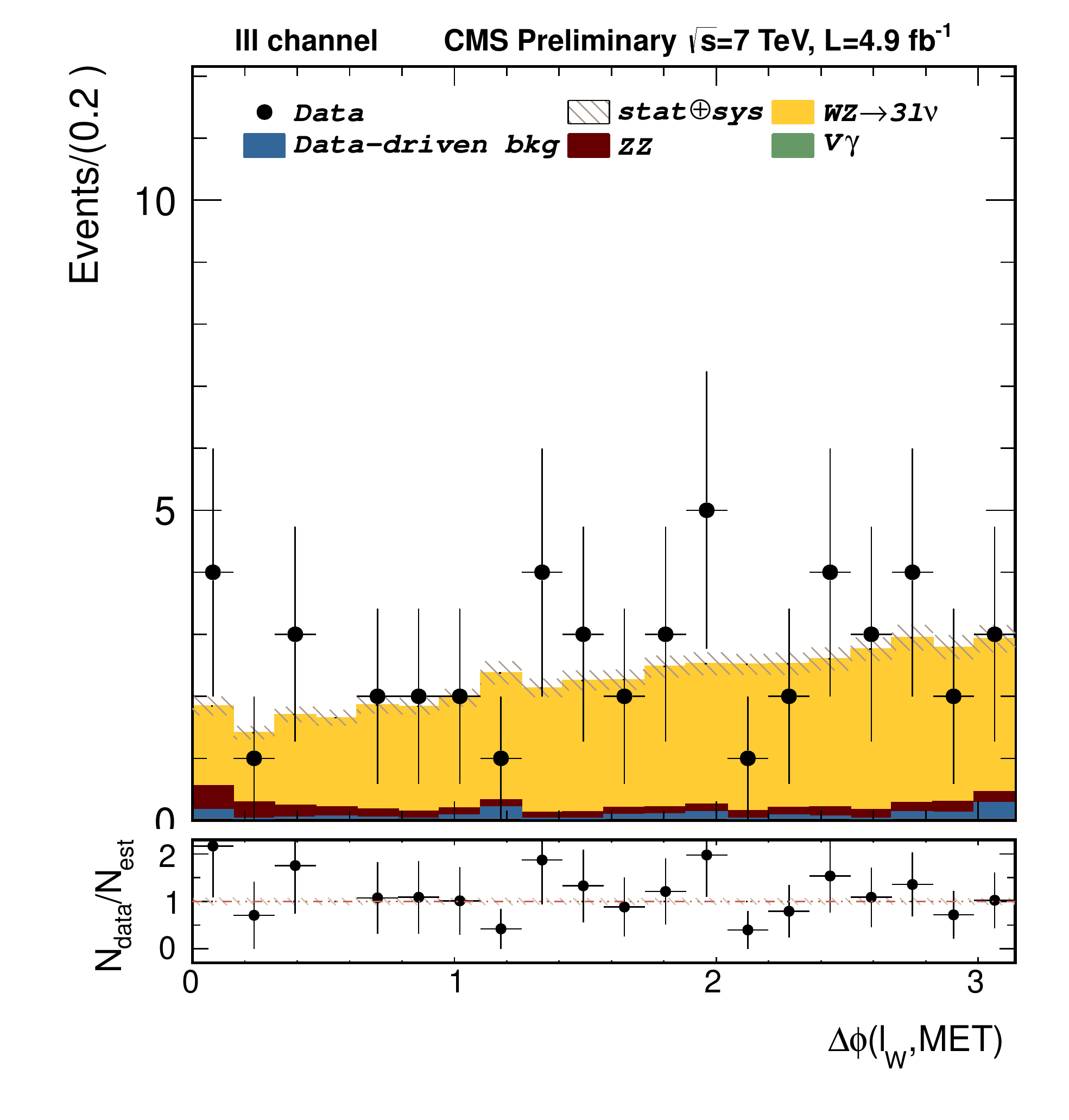}
	\end{subfigure}\quad
	\begin{subfigure}[b]{0.3\textwidth}
		\includegraphics[width=\textwidth]{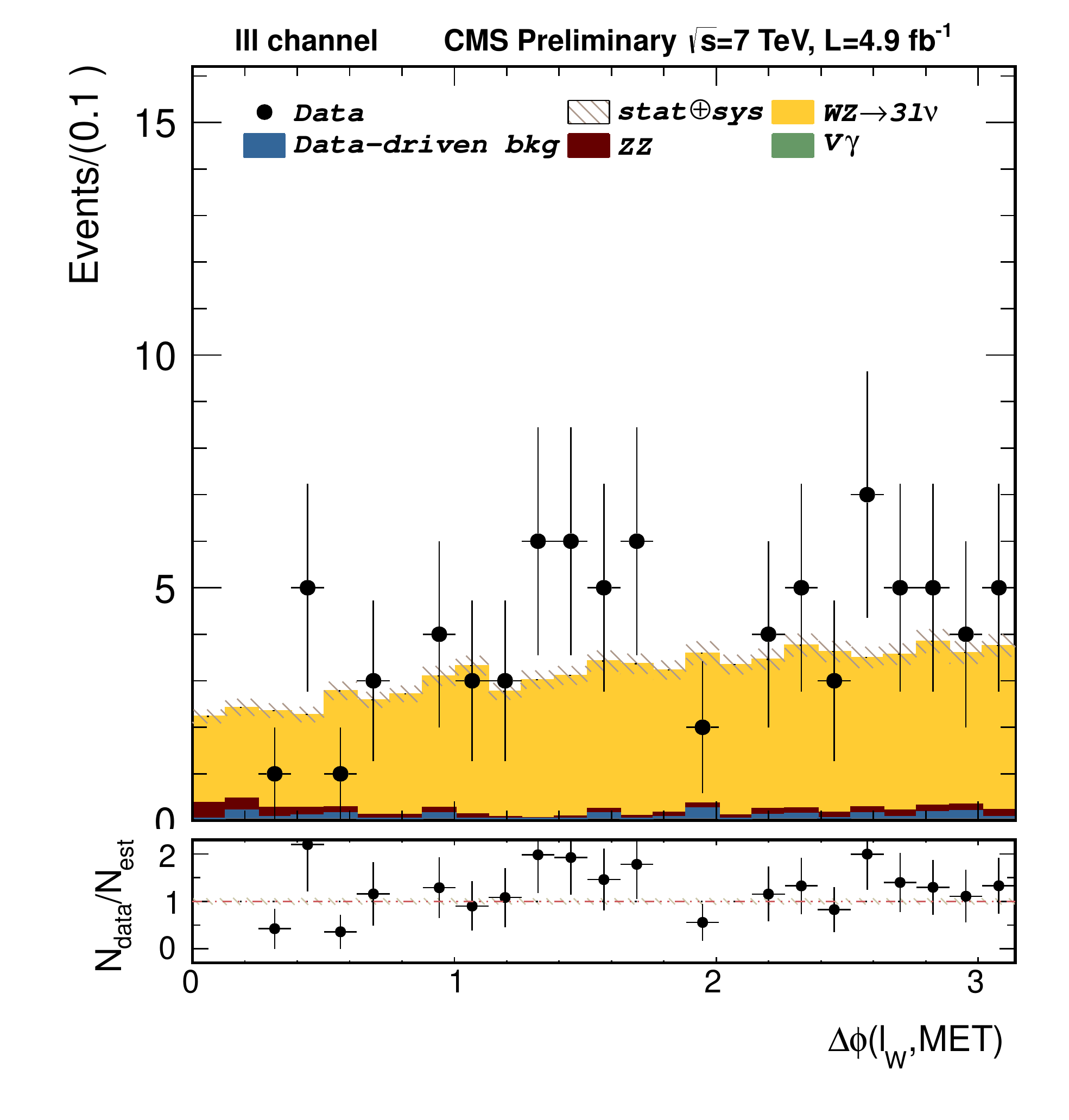}
	\end{subfigure}
	\caption[Azimuthal angle between W-candidate lepton and \MET at 7~\TeV (ratio)]
	{Azimuthal angle between the W-candidate lepton and the \MET
	at each event for the \wzm (left column) and \wzp (right column) before (up row) 
	and after the \MET cut (bottom row).}
\end{figure}

\begin{figure}[!htpb]
	\centering
	\begin{subfigure}[b]{0.3\textwidth}
		\includegraphics[width=\textwidth]{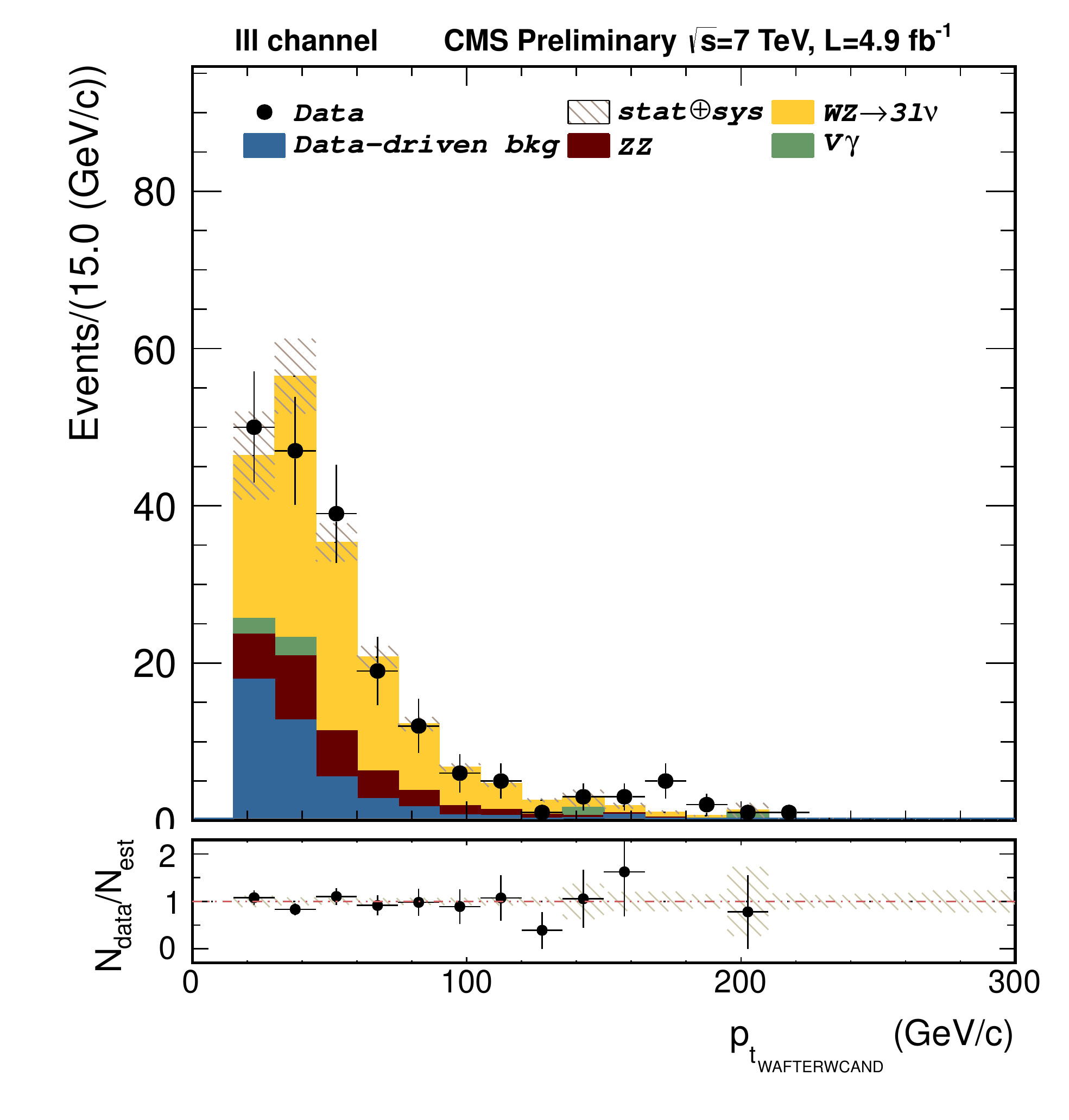}
	\end{subfigure}\quad
	\begin{subfigure}[b]{0.3\textwidth}
		\includegraphics[width=\textwidth]{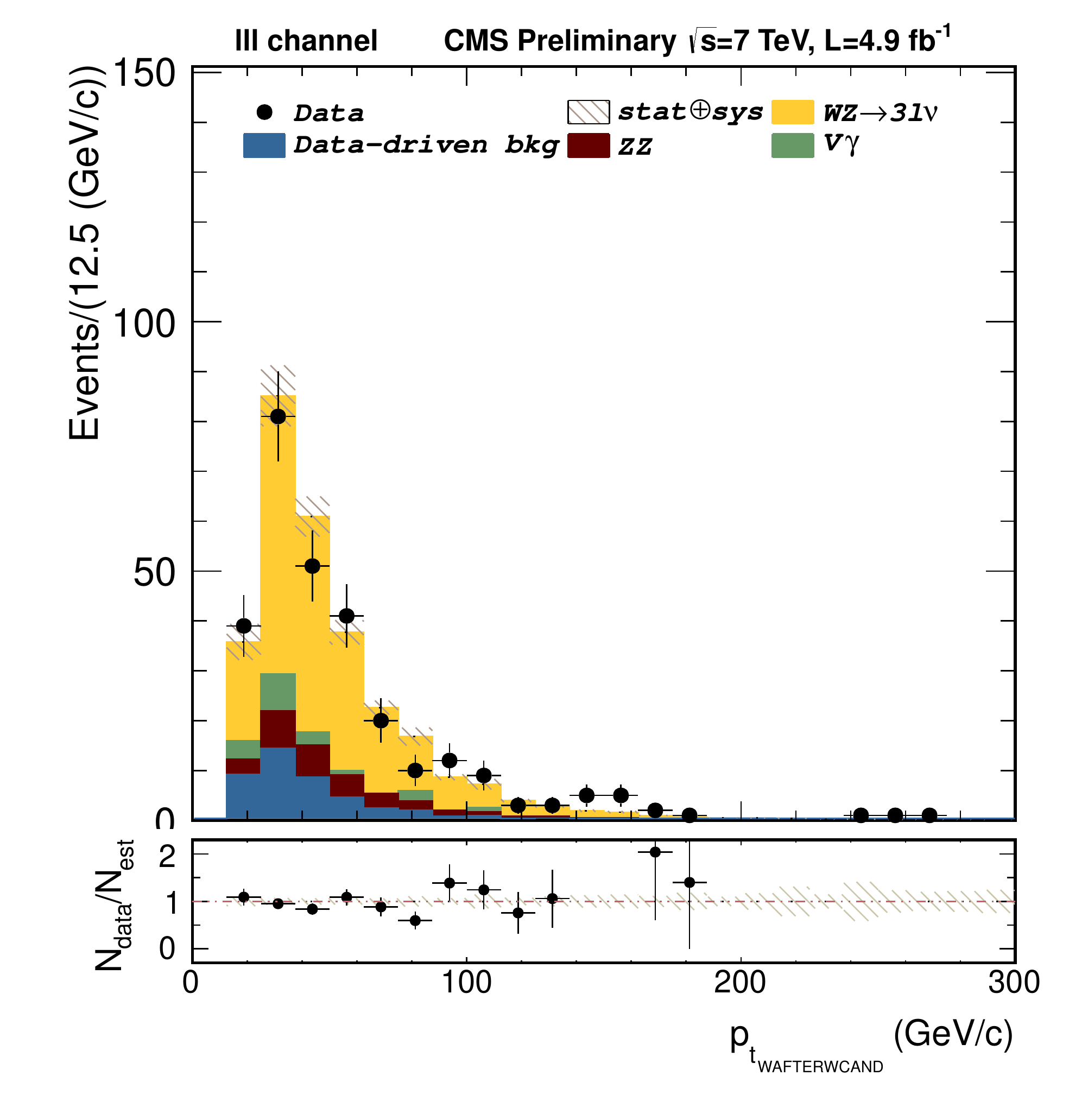}
	\end{subfigure}
	\vskip 1ex
	\begin{subfigure}[b]{0.3\textwidth}
		\includegraphics[width=\textwidth]{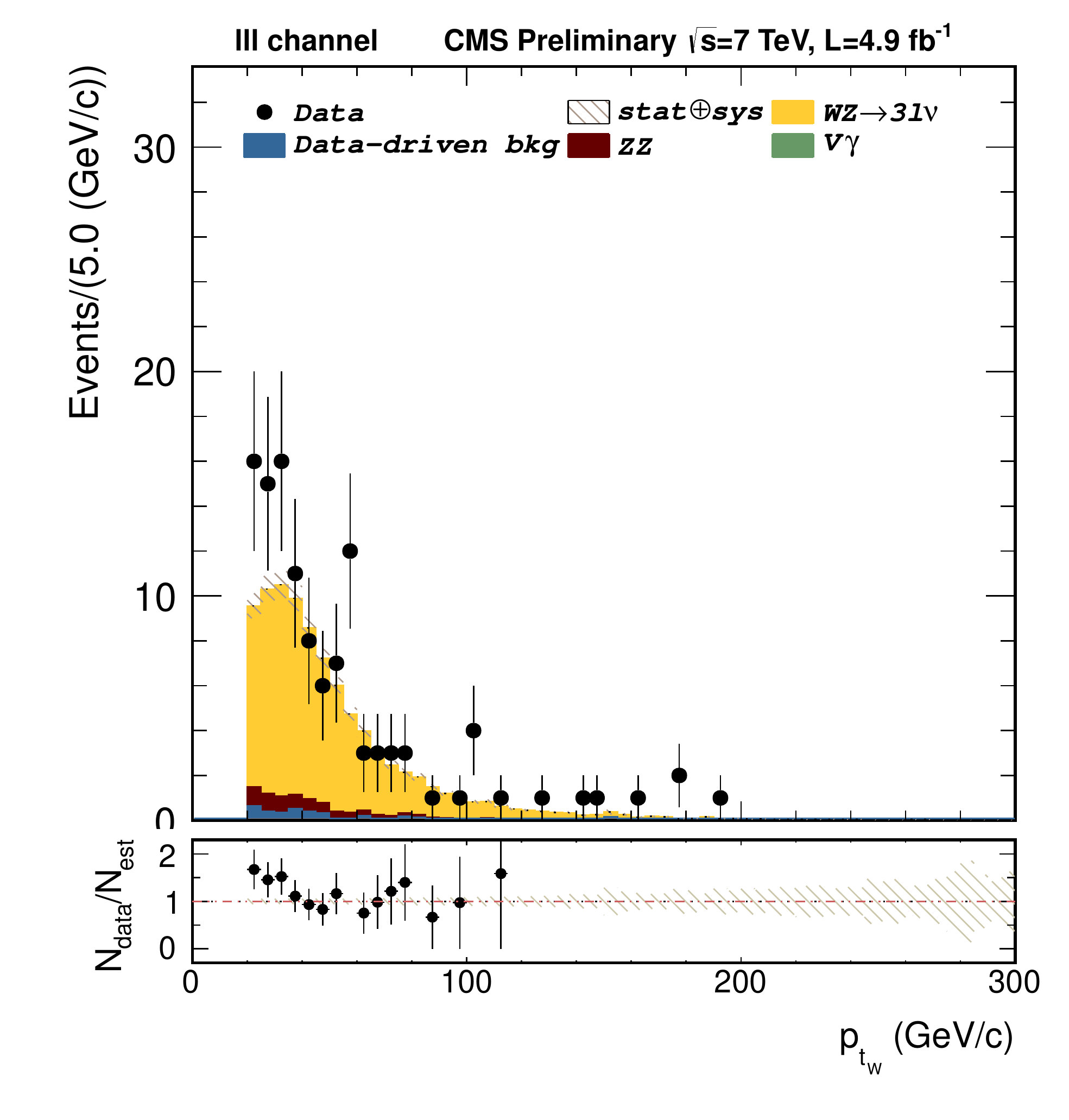}
	\end{subfigure}
	\begin{subfigure}[b]{0.3\textwidth}
		\includegraphics[width=\textwidth]{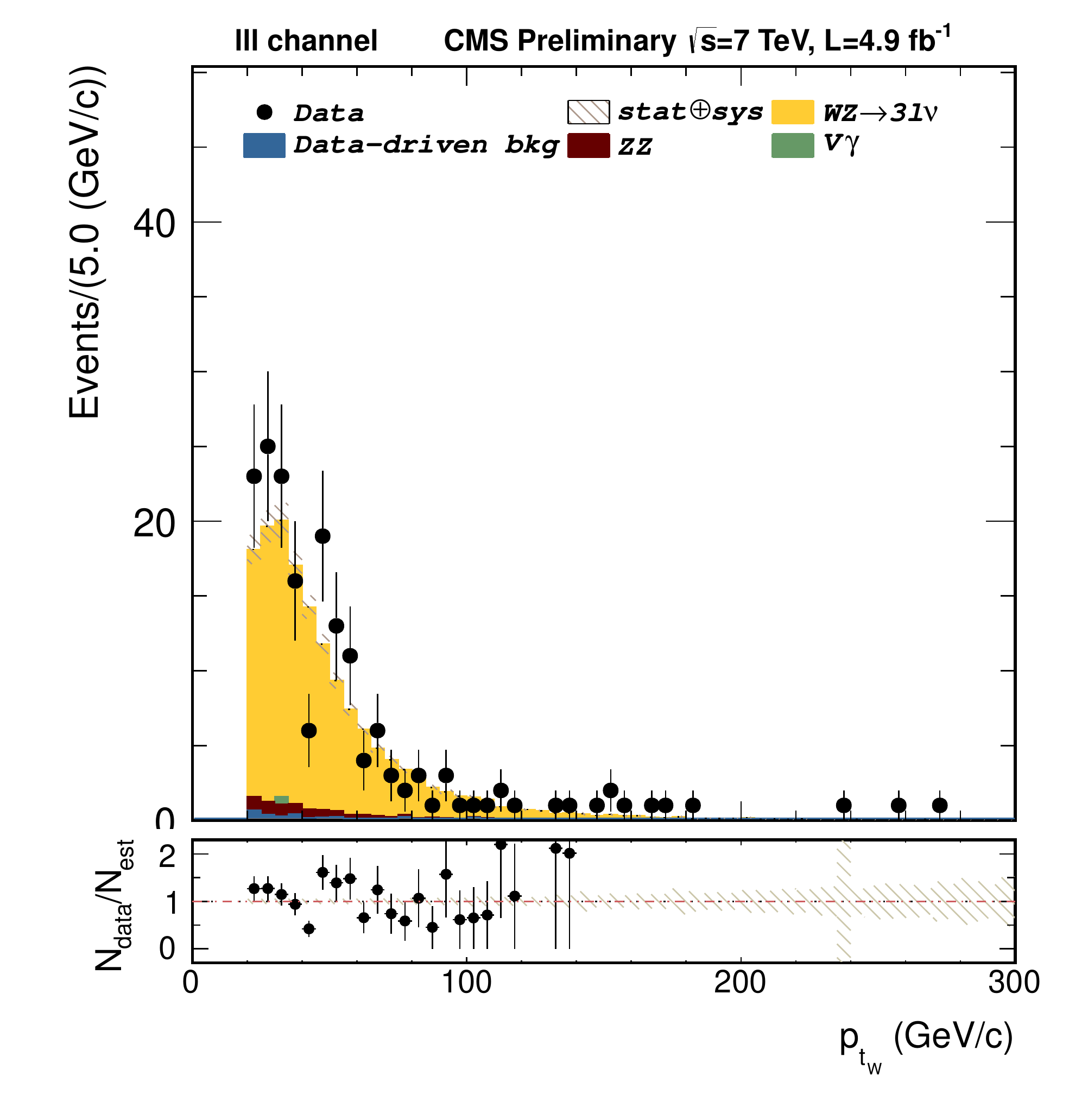}
	\end{subfigure}\quad
	\caption[Transverse momentum of the W-candidate system at 7~\TeV (ratio)]
	{Transverse momentum of the W-candidate system composed by
	the third selected lepton and \MET at each event for the \wzm (left column) and \wzp 
	(right column) before (up row) and after the \MET cut (bottom row).}
	\vskip 1em
	\begin{subfigure}[b]{0.3\textwidth}
		\includegraphics[width=\textwidth]{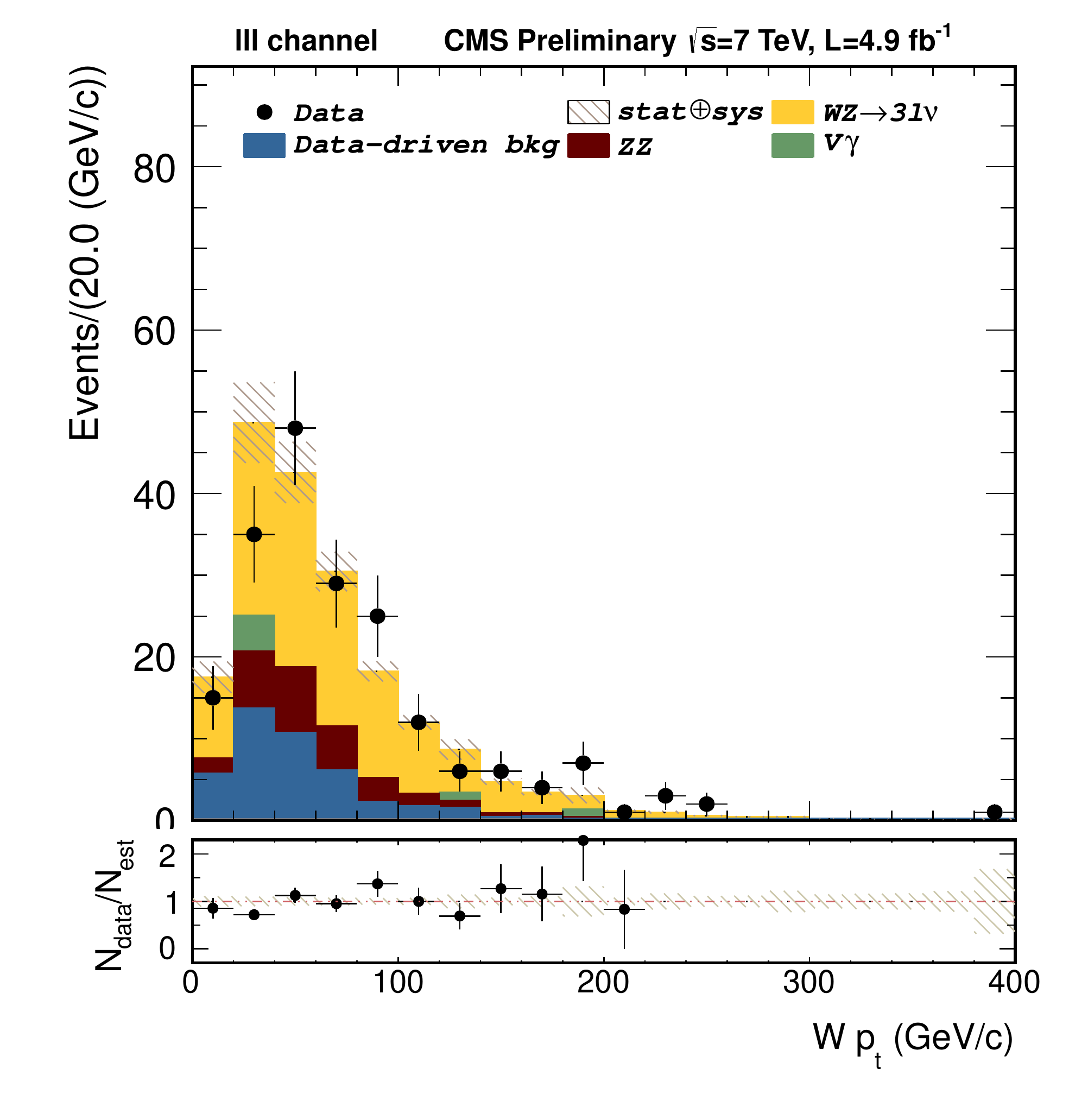}
	\end{subfigure}\quad
	\begin{subfigure}[b]{0.3\textwidth}
		\includegraphics[width=\textwidth]{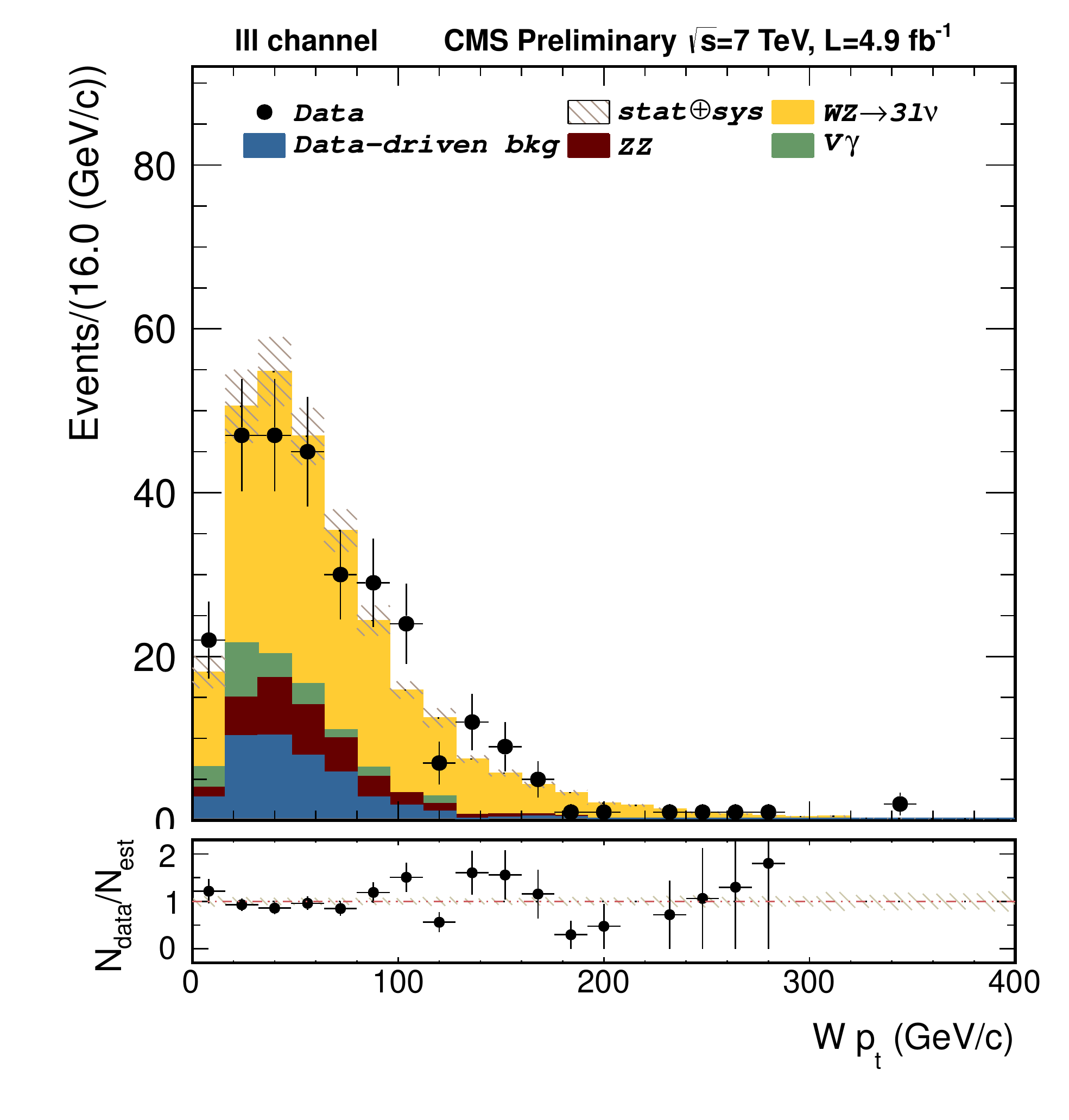}
	\end{subfigure}
	\vskip 1ex
	\begin{subfigure}[b]{0.3\textwidth}
		\includegraphics[width=\textwidth]{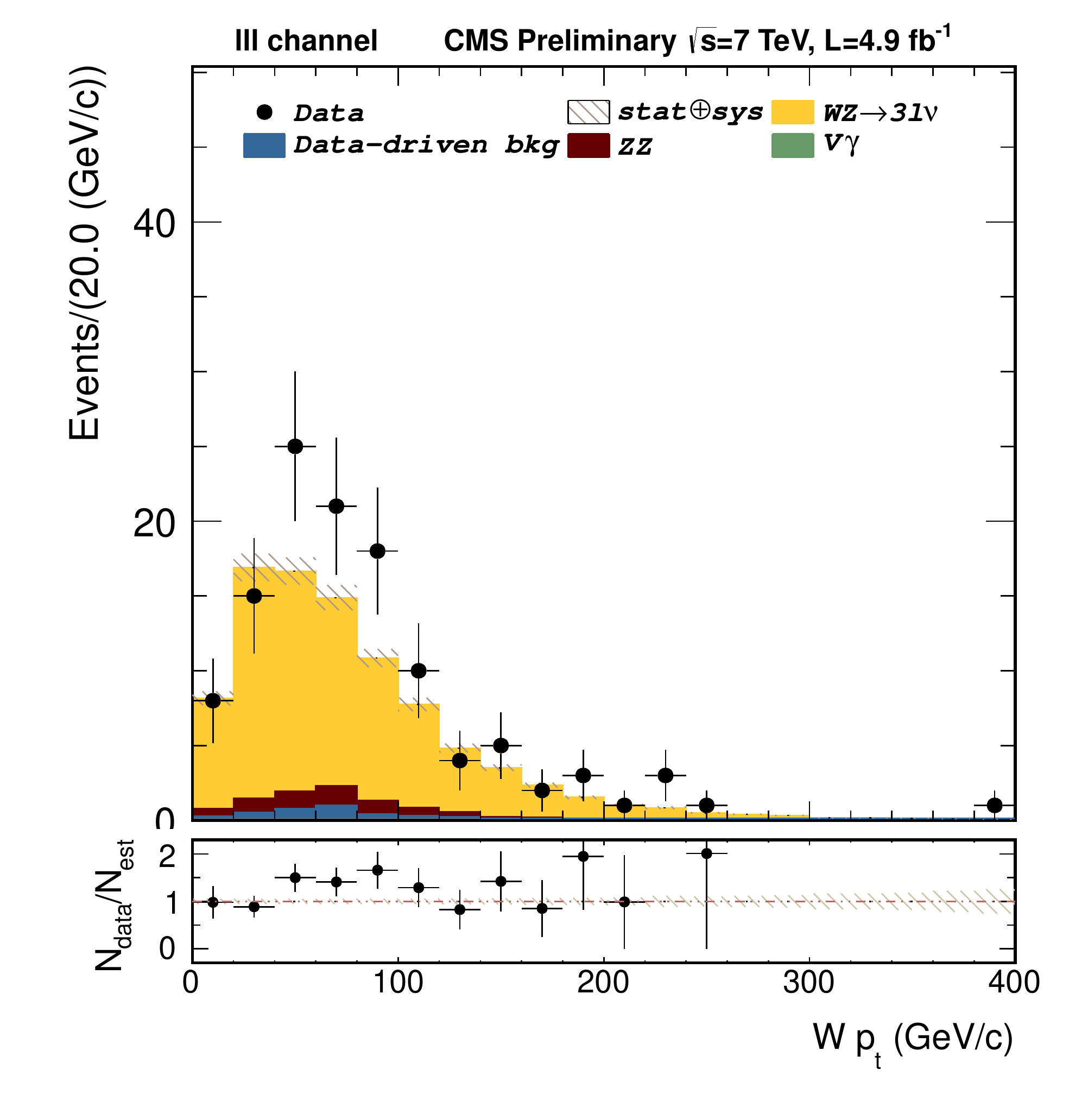}
	\end{subfigure}\quad
	\begin{subfigure}[b]{0.3\textwidth}
		\includegraphics[width=\textwidth]{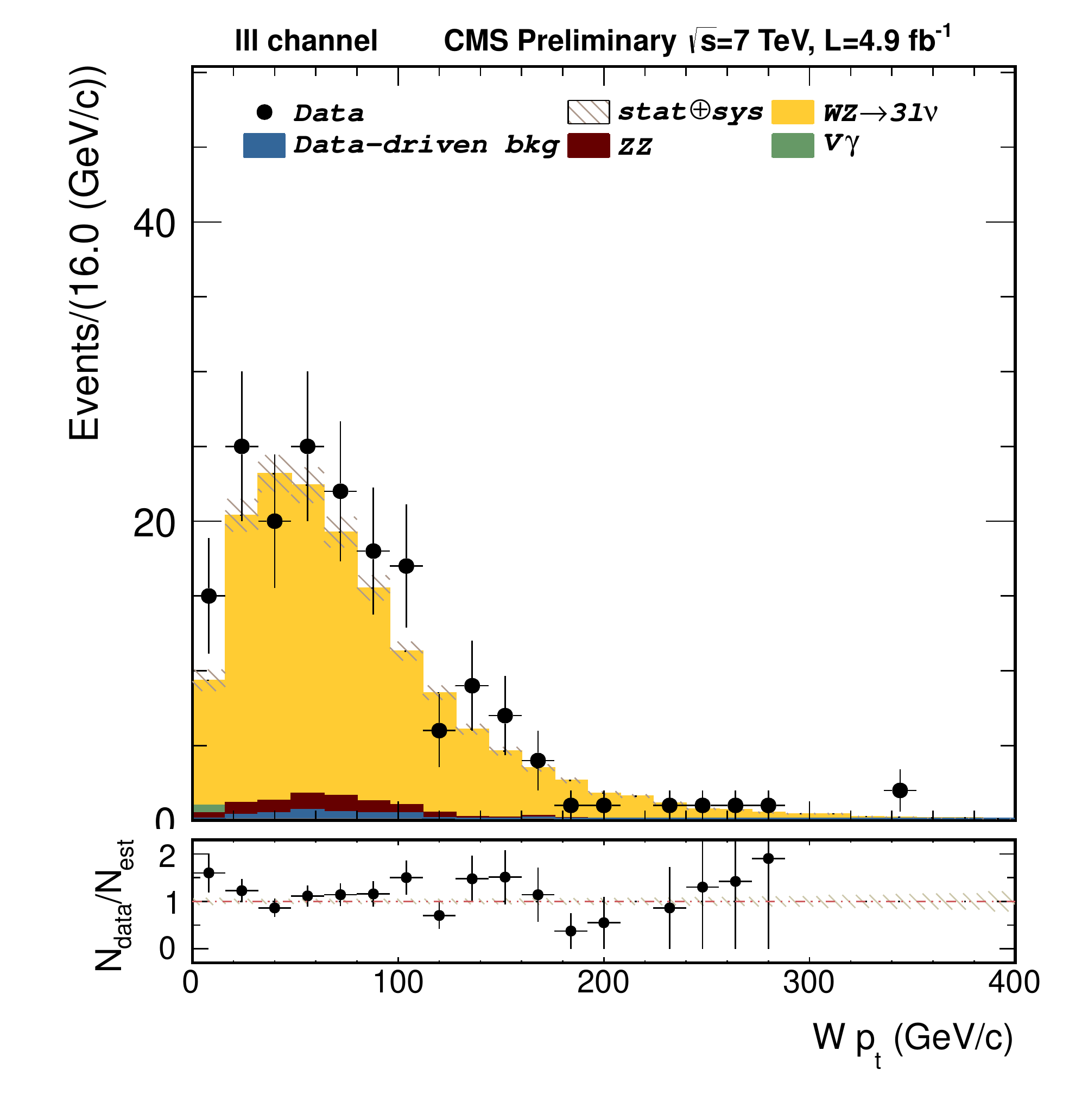}
	\end{subfigure}
	\caption[Transverse momentum of the W-candidate lepton at 7~\TeV (ratio)]
	{Transverse momentum of the W-candidate lepton 
	at each event for the \wzm (left column) and \wzp (right column) before (up row)
	and after the \MET cut (bottom row).}
\end{figure}

\begin{figure}[!htpb]
	\centering
	\begin{subfigure}[b]{0.3\textwidth}
		\includegraphics[width=\textwidth]{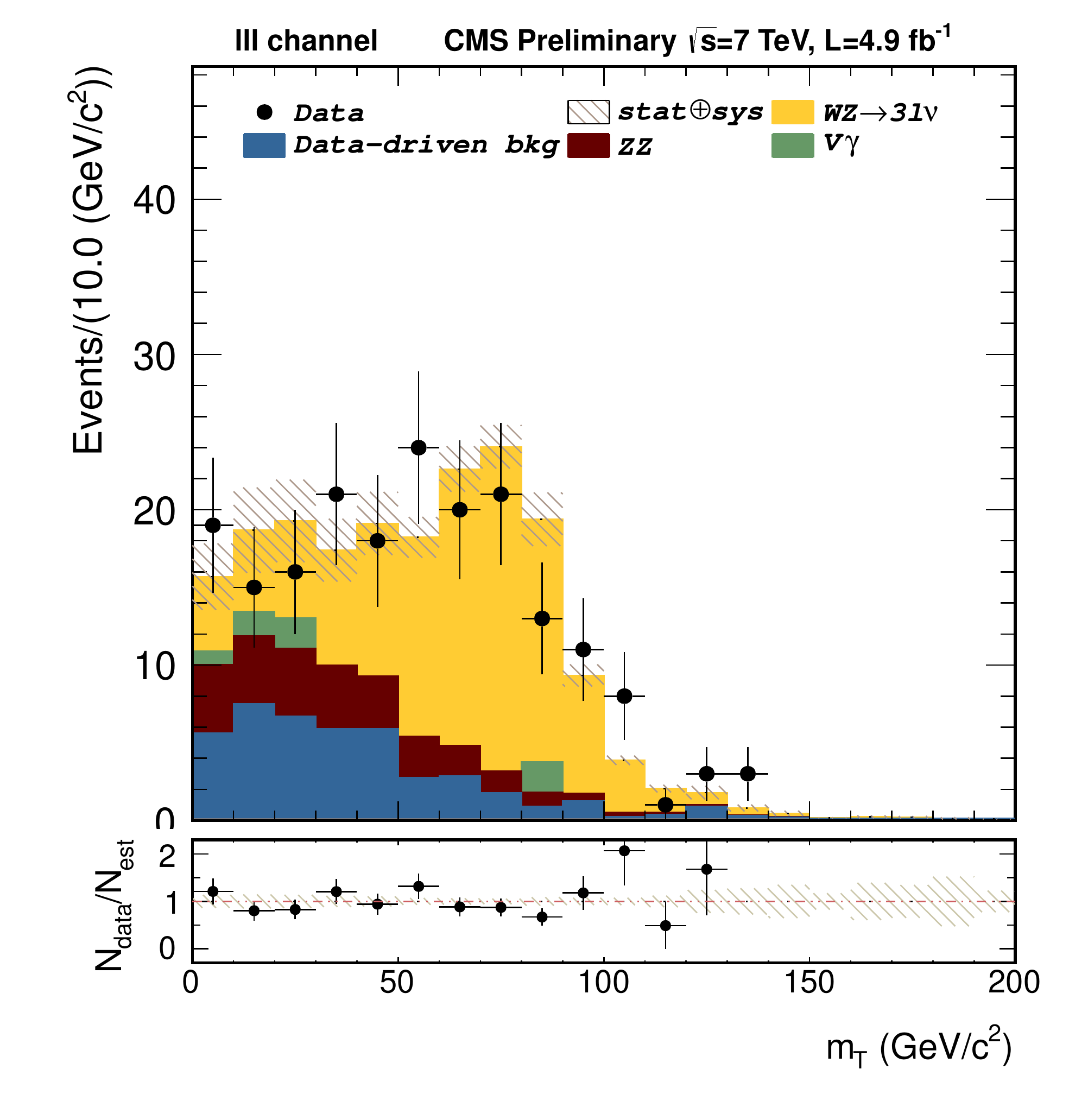}
	\end{subfigure}\quad
	\begin{subfigure}[b]{0.3\textwidth}
		\includegraphics[width=\textwidth]{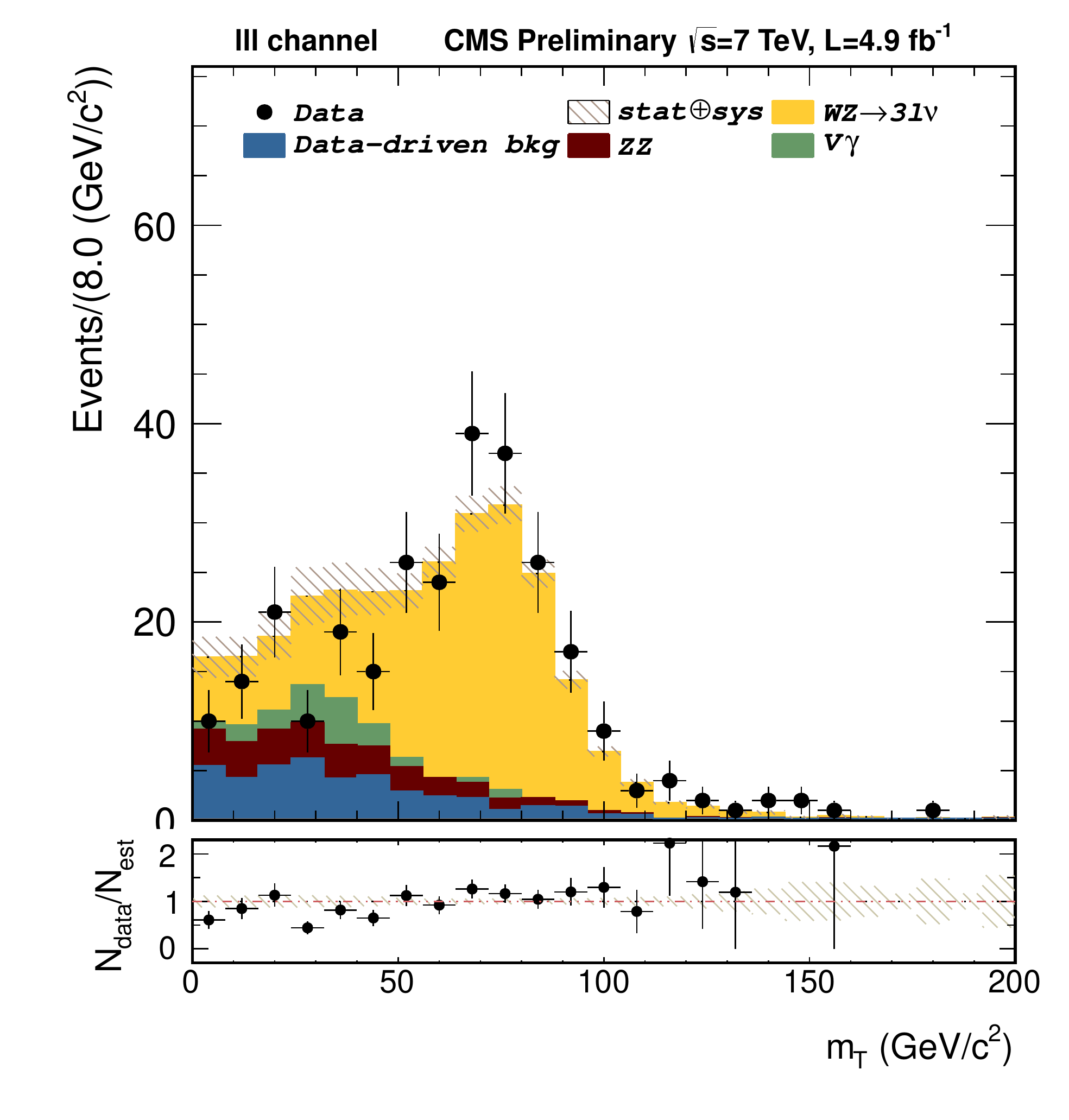}
	\end{subfigure}
	\vskip 1ex
	\begin{subfigure}[b]{0.3\textwidth}
		\includegraphics[width=\textwidth]{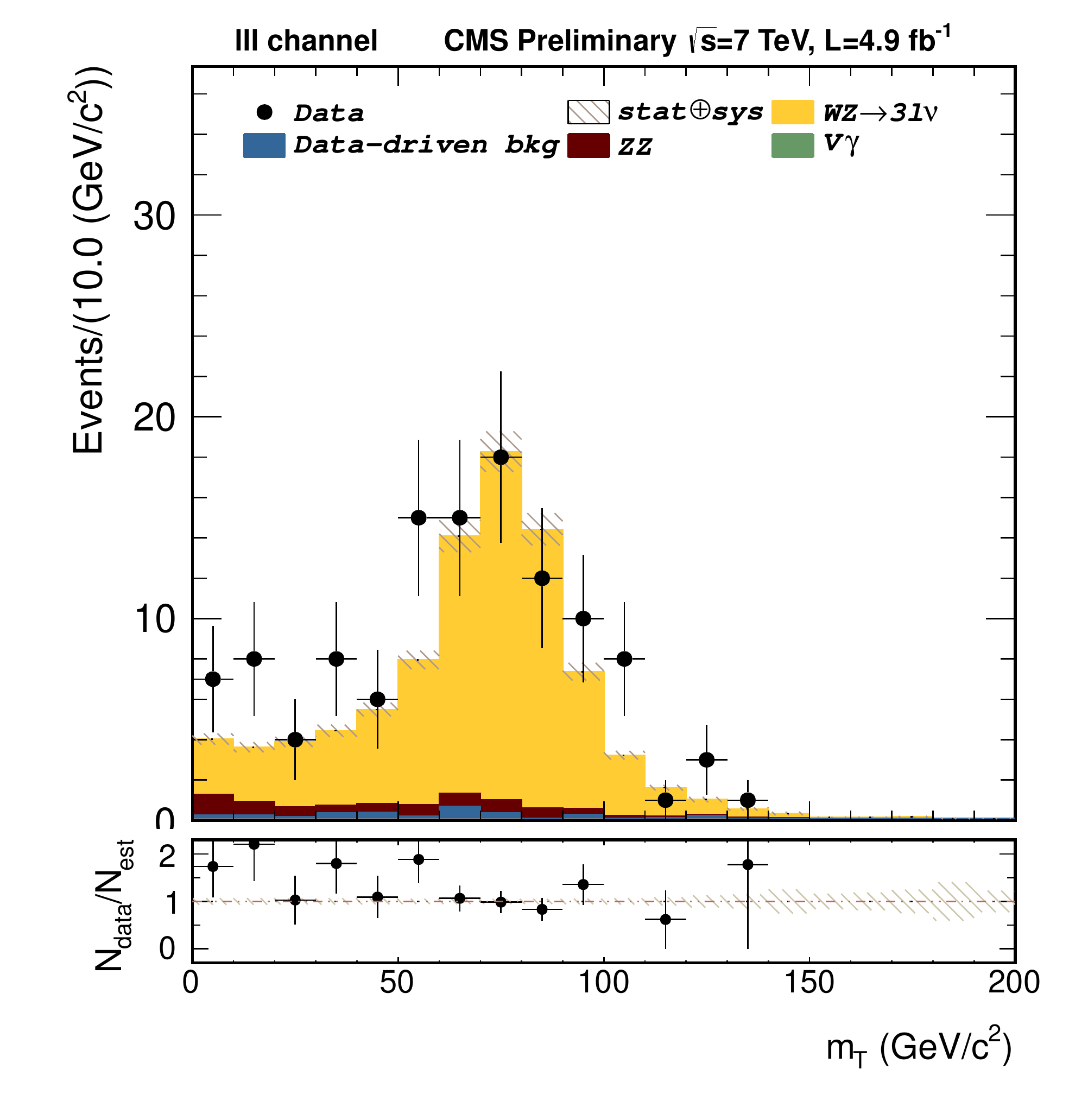}
	\end{subfigure}\quad
	\begin{subfigure}[b]{0.3\textwidth}
		\includegraphics[width=\textwidth]{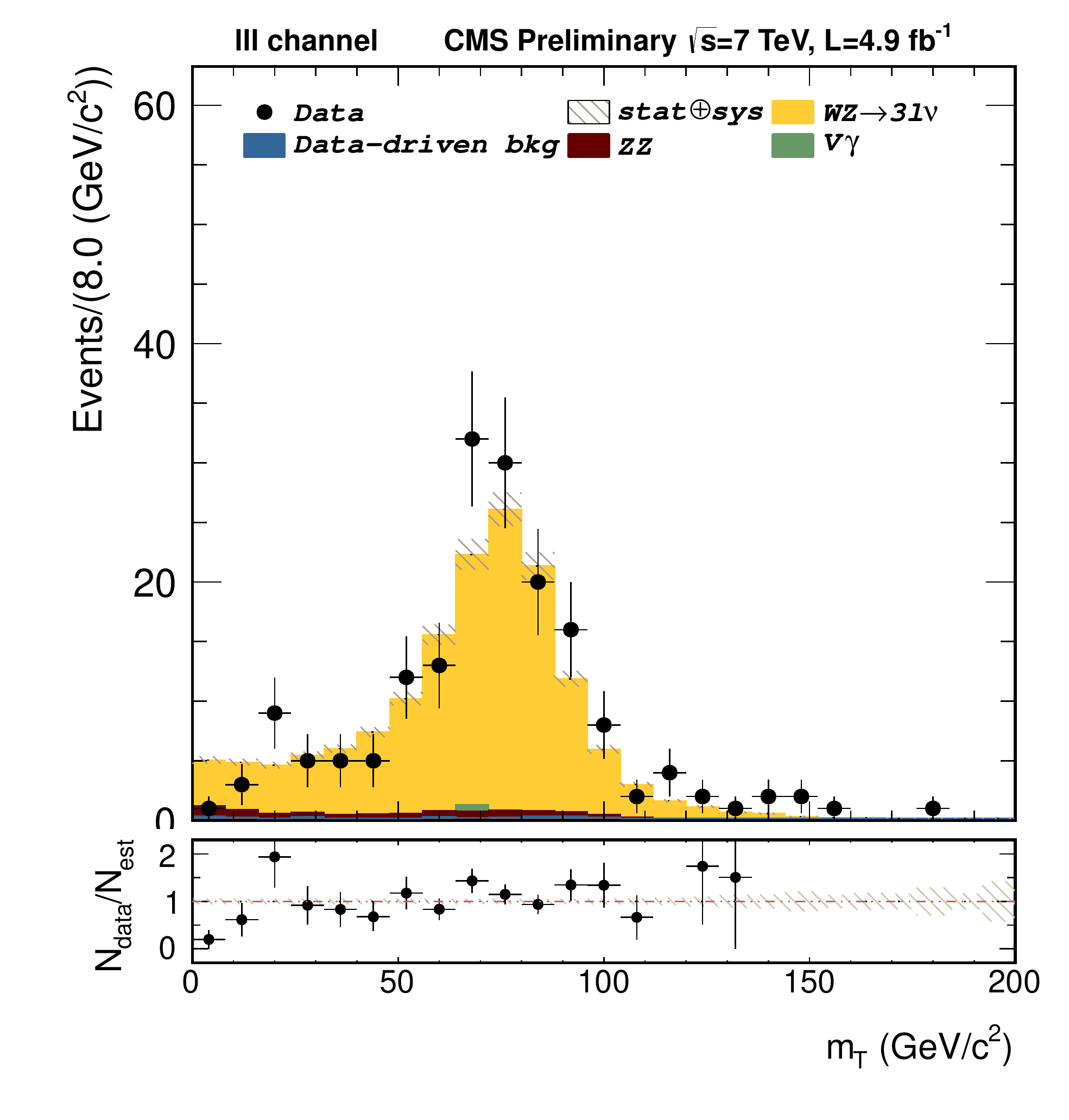}
	\end{subfigure}
	\caption[Transverse mass of the W-candidate lepton and \MET at 7~\TeV (ratio)]
	{Transverse mass of the W-candidate lepton and the \MET
	at each event for the \wzm (left column) and \wzp (right column) before (up row)
	and after the \MET cut (bottom row).}
	\vskip 1em
	\begin{subfigure}[b]{0.3\textwidth}
		\includegraphics[width=\textwidth]{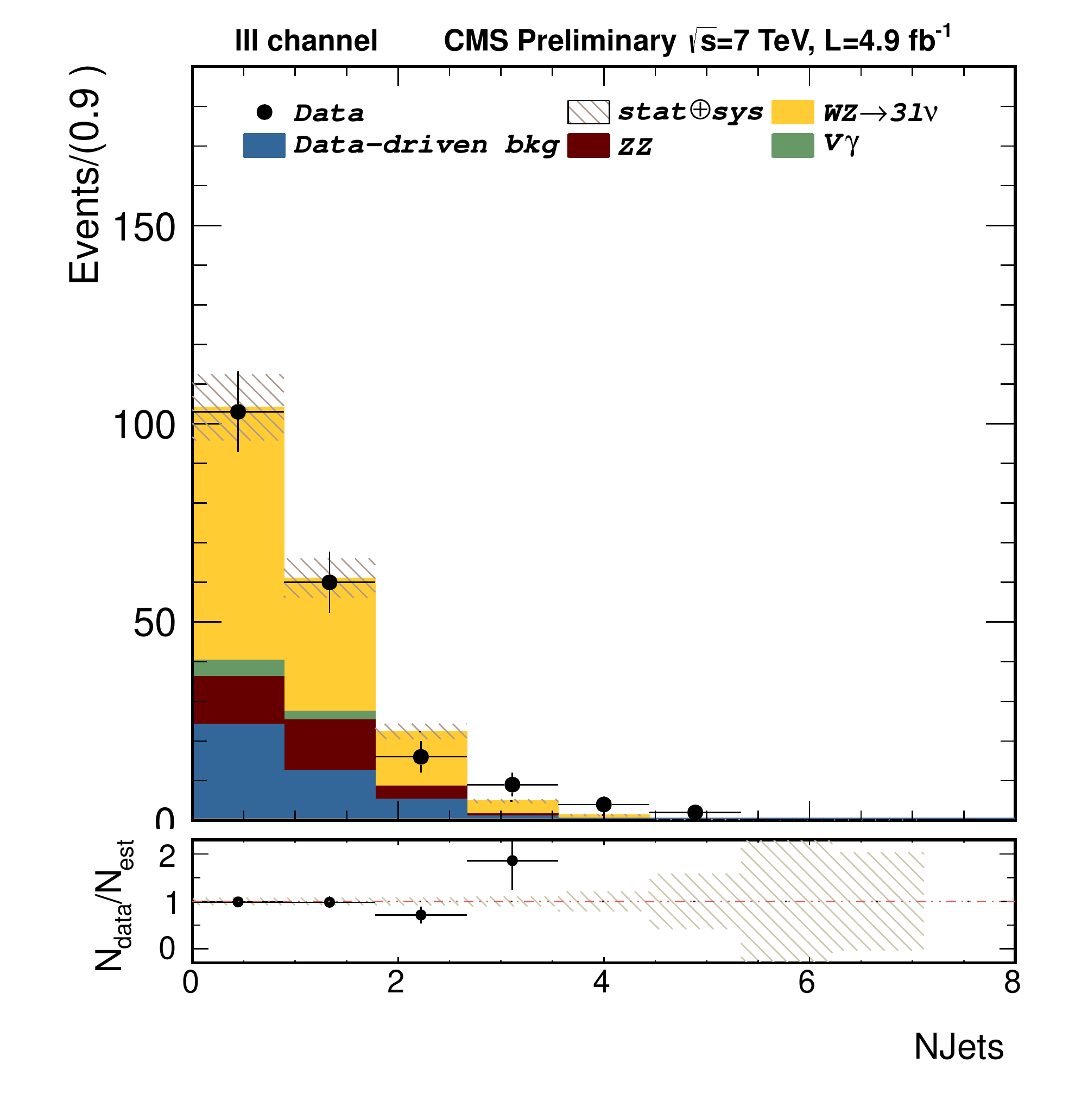}
	\end{subfigure}\quad
	\begin{subfigure}[b]{0.3\textwidth}
		\includegraphics[width=\textwidth]{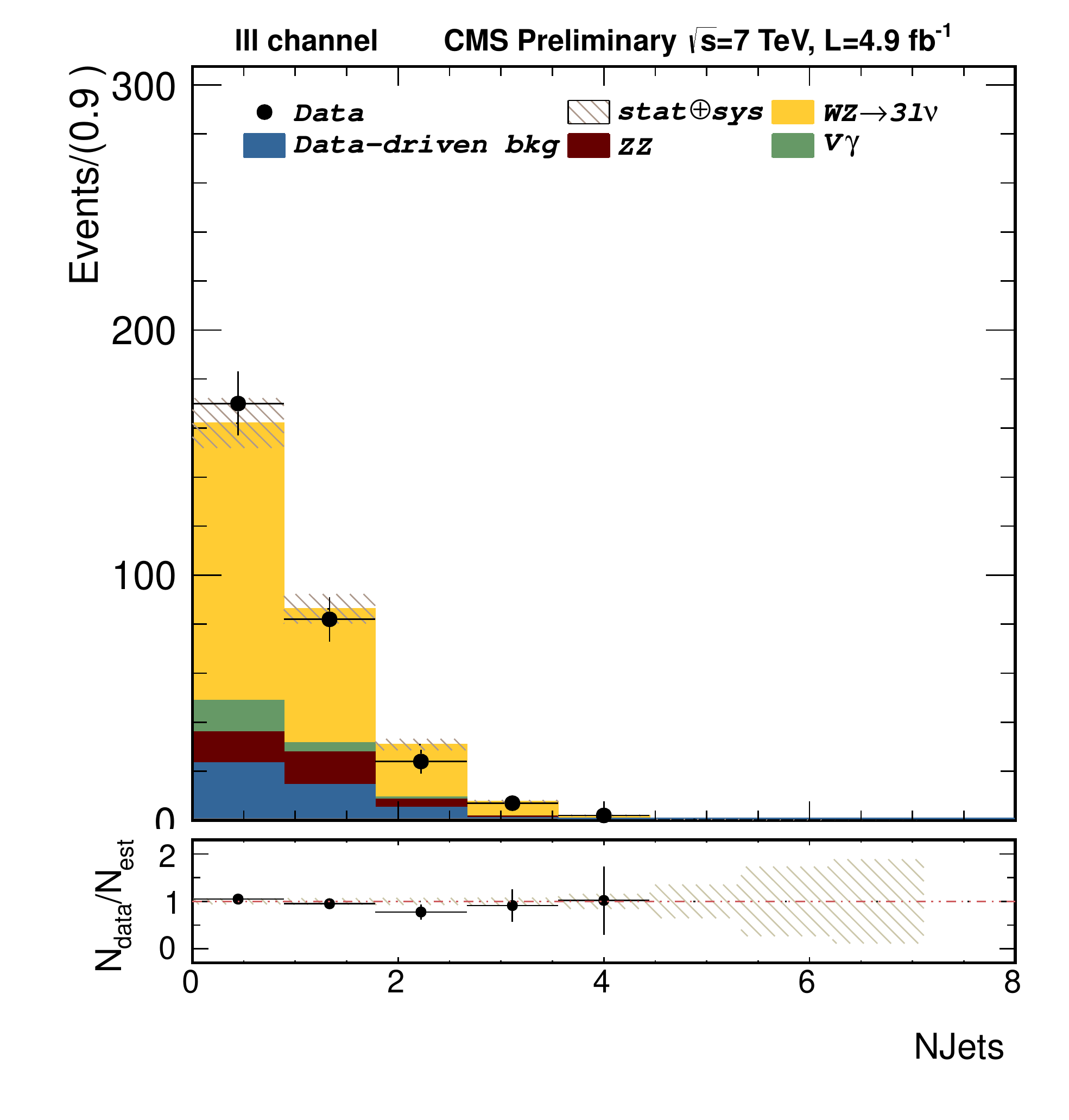}
	\end{subfigure}
	\vskip 1ex
	\begin{subfigure}[b]{0.3\textwidth}
		\includegraphics[width=\textwidth]{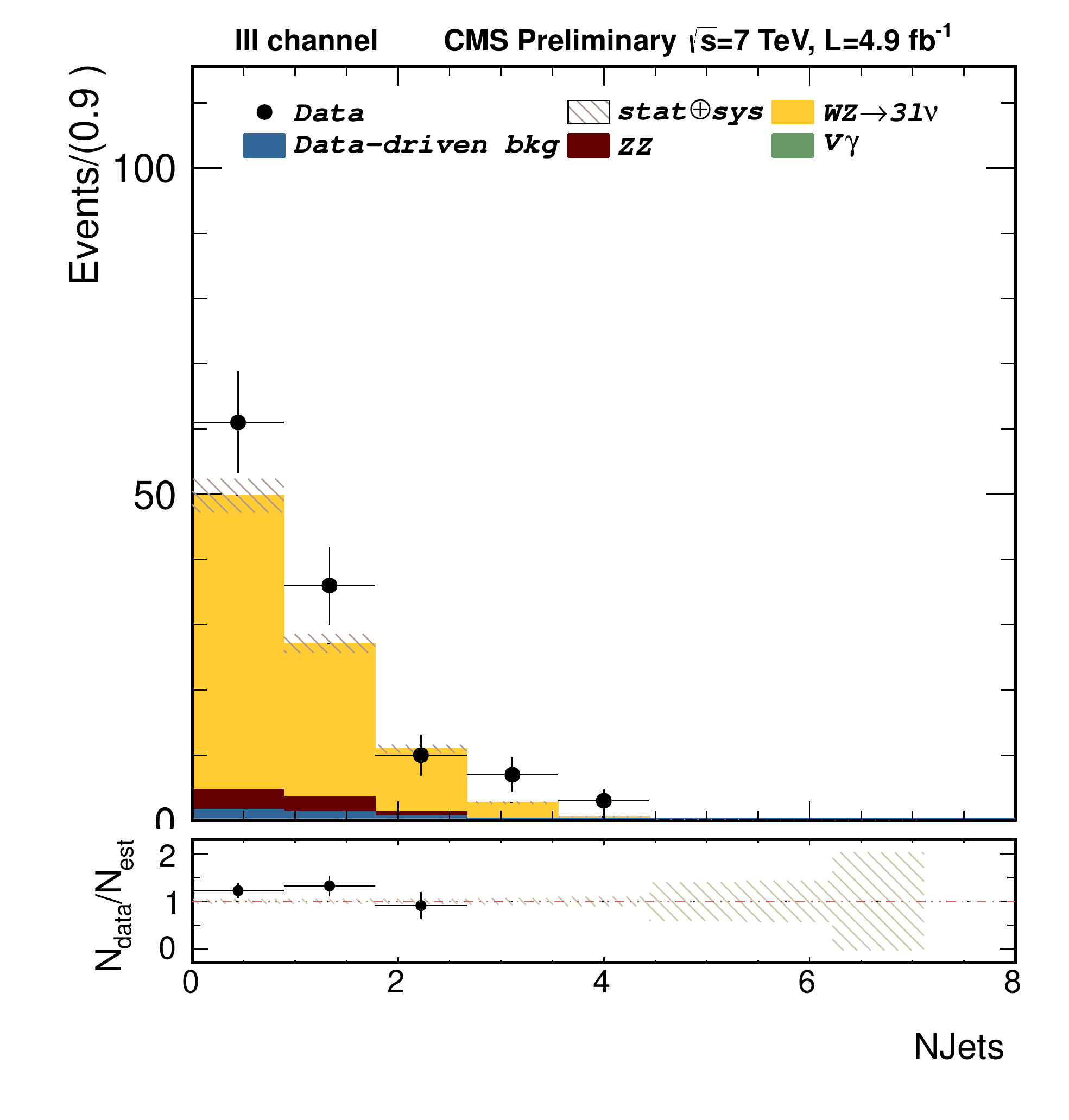}
	\end{subfigure}\quad
	\begin{subfigure}[b]{0.3\textwidth}
		\includegraphics[width=\textwidth]{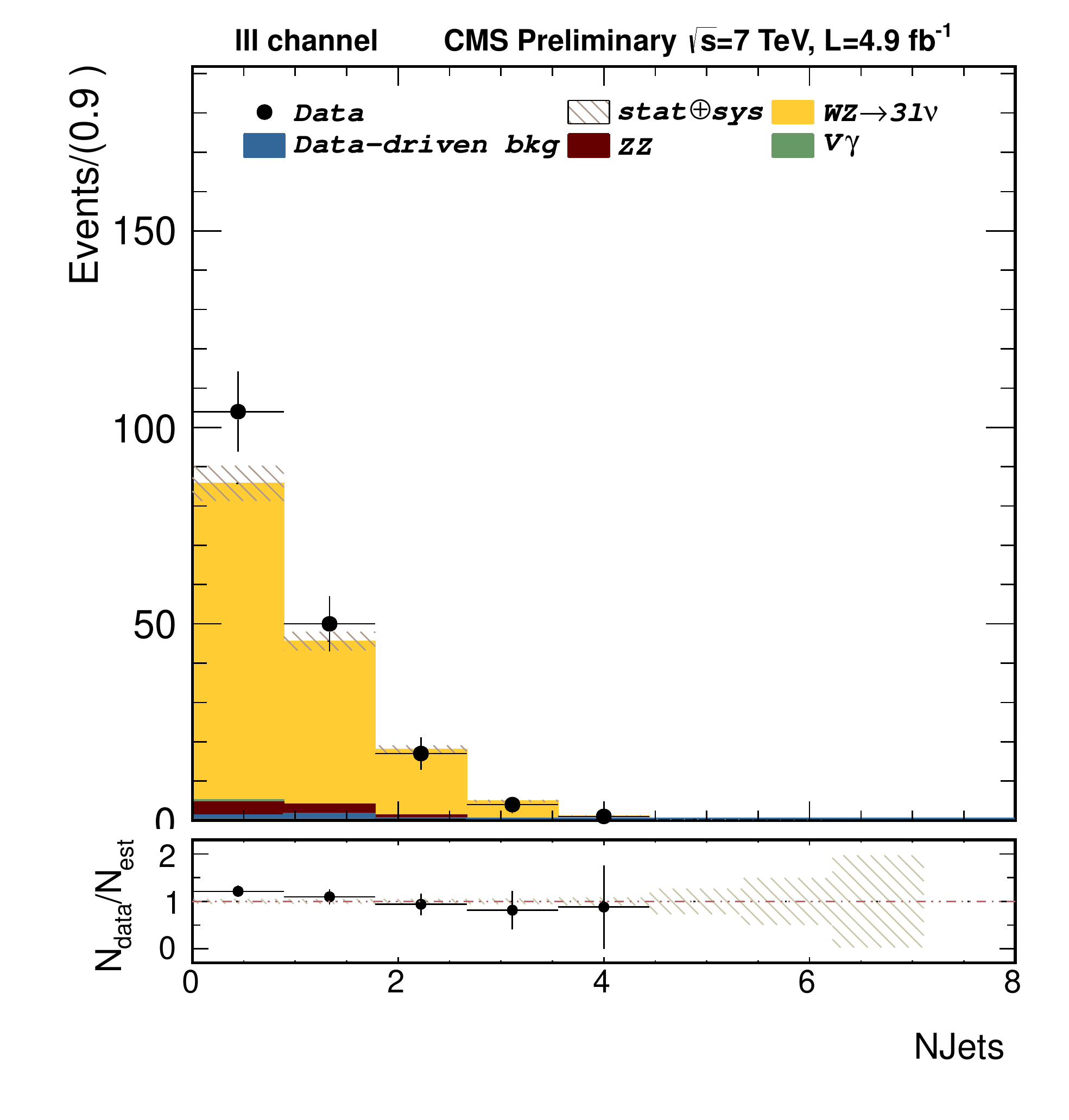}
	\end{subfigure}
	\caption[Number of jets at 7 TeV (ratio)]{Number of 
	jets distribution at each event for the \wzm (left column) and \wzp 
	(right column) before (up row) and after the \MET cut (bottom row).}
\end{figure}

\begin{figure}[!htpb]
	\centering
	\begin{subfigure}[b]{0.3\textwidth}
		\includegraphics[width=\textwidth]{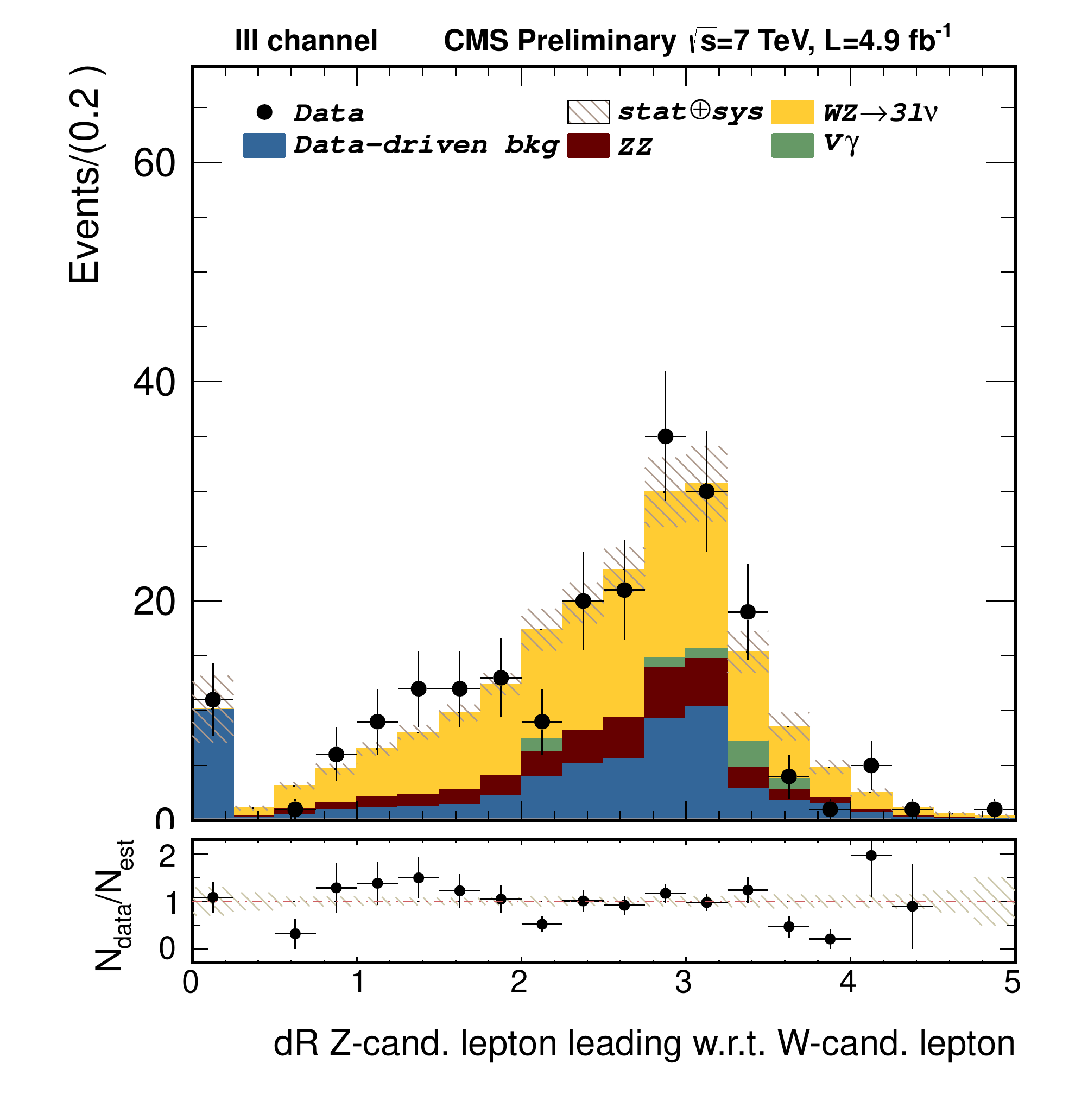}
	\end{subfigure}\quad
	\begin{subfigure}[b]{0.3\textwidth}
		\includegraphics[width=\textwidth]{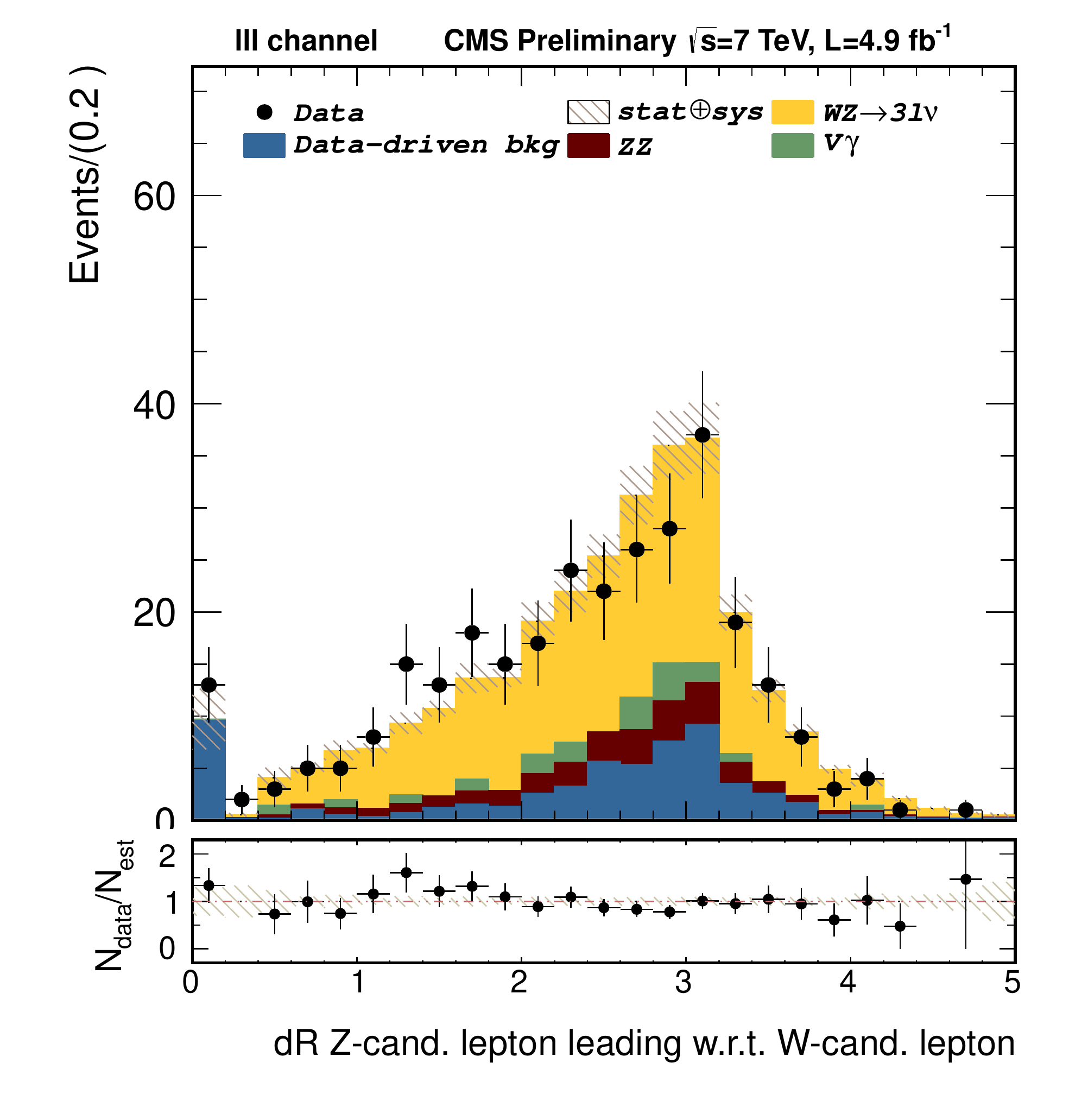}
	\end{subfigure}
	\vskip 1ex
	\begin{subfigure}[b]{0.3\textwidth}
		\includegraphics[width=\textwidth]{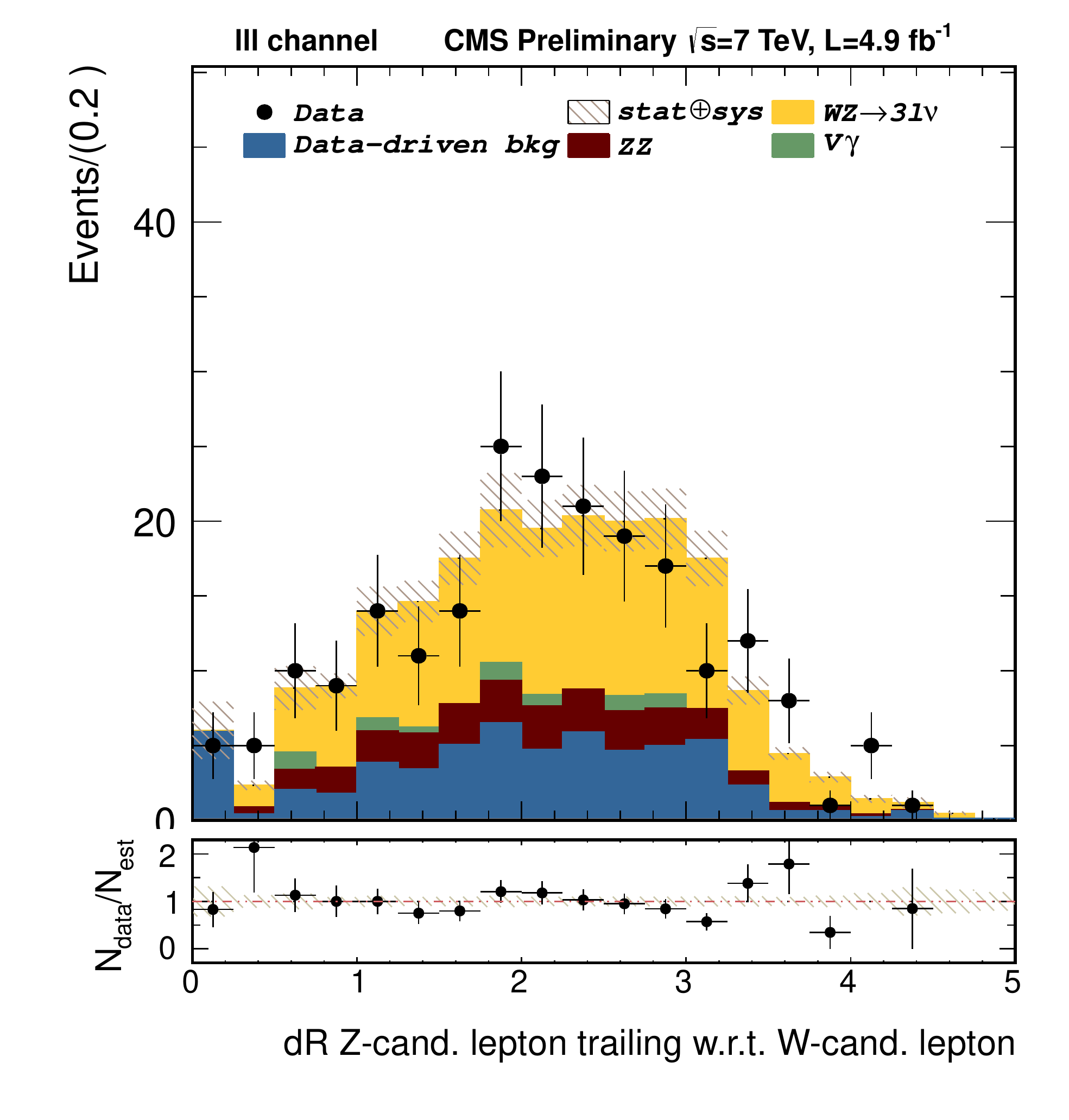}
	\end{subfigure}\quad
	\begin{subfigure}[b]{0.3\textwidth}
		\includegraphics[width=\textwidth]{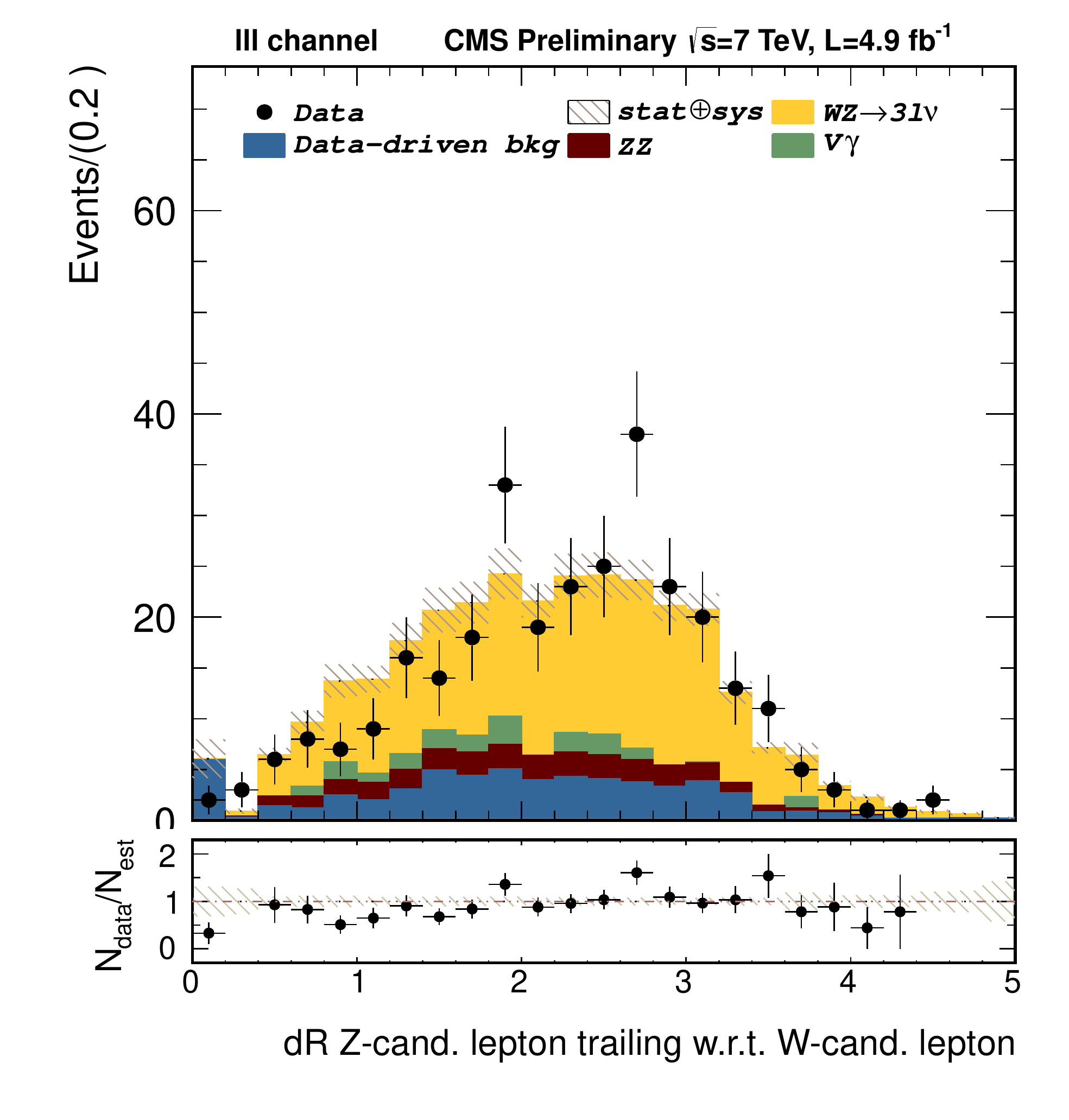}
	\end{subfigure}
	\caption[Angular distance between W-candidate lepton and Z-candidate leptons at 7~\TeV
	(ratio)]{Angular distance between the W-candidate lepton and the Z-candidate
	leading (up row) and trailing lepton (bottom row) at each event for the \wzm 
	(left column) and \wzp (right column) before \MET cut.}
	\vskip 1em
	\begin{subfigure}[b]{0.3\textwidth}
		\includegraphics[width=\textwidth]{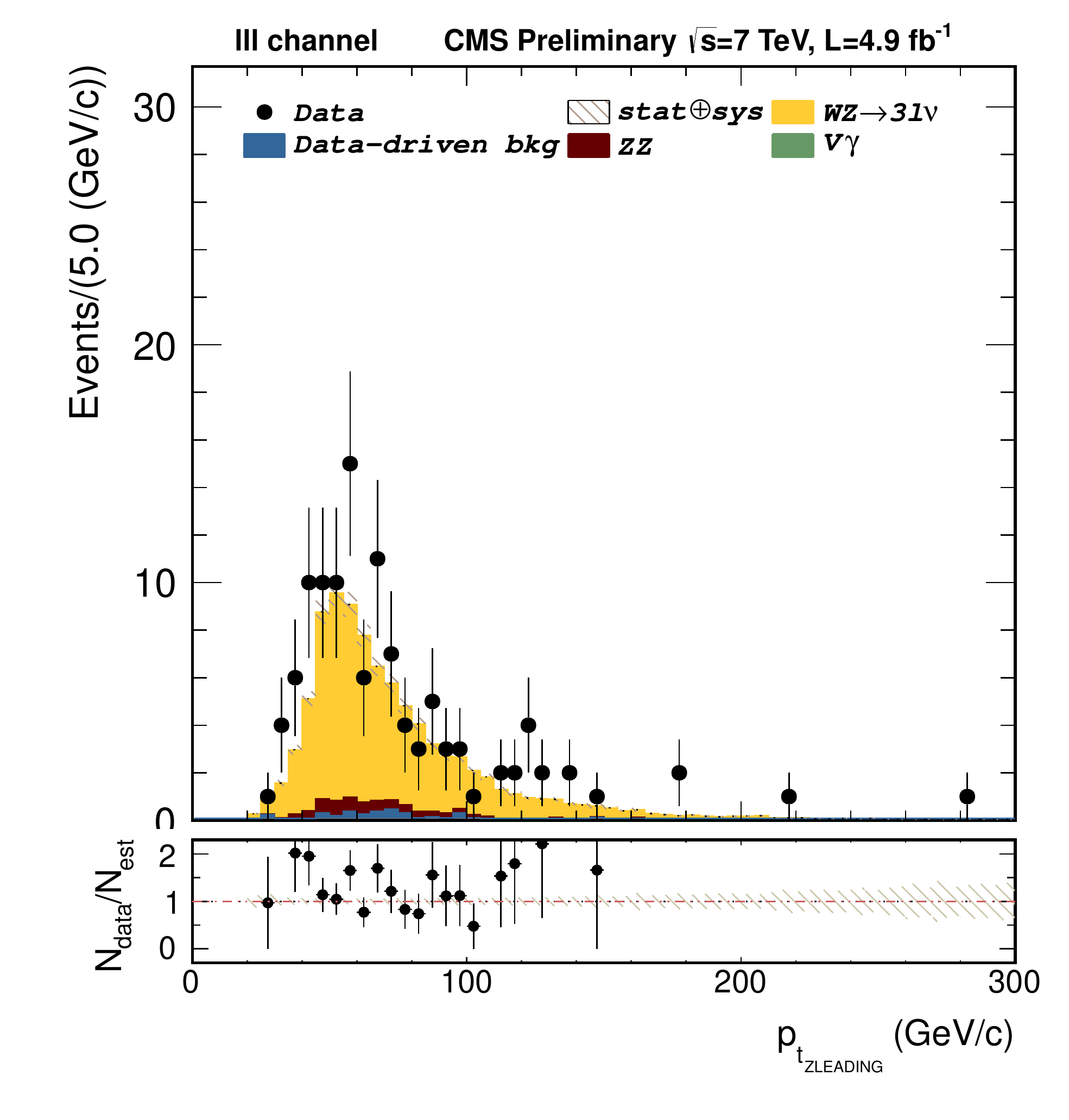}
	\end{subfigure}\quad
	\begin{subfigure}[b]{0.3\textwidth}
		\includegraphics[width=\textwidth]{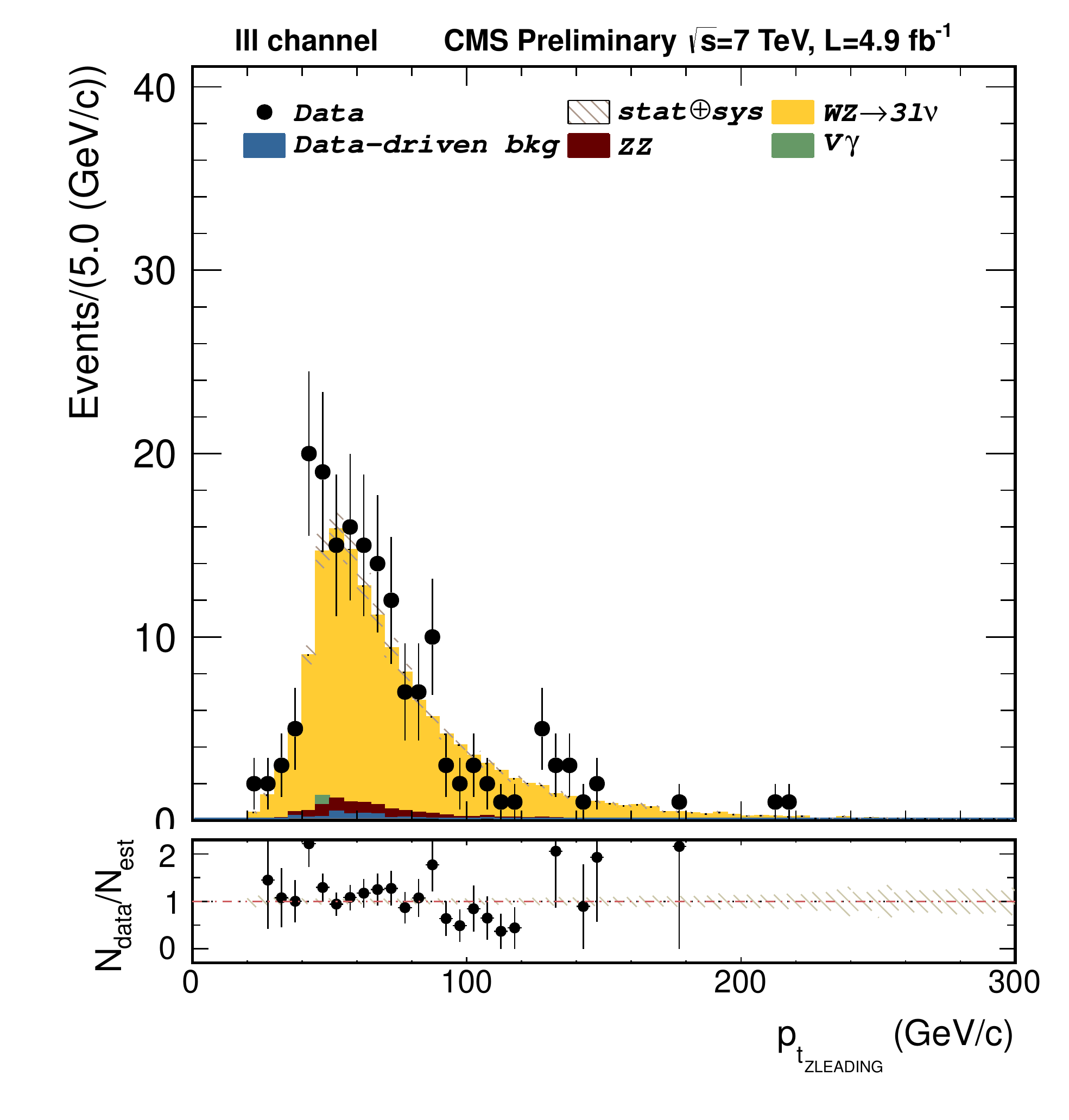}
	\end{subfigure}
	\vskip 1ex
	\begin{subfigure}[b]{0.3\textwidth}
		\includegraphics[width=\textwidth]{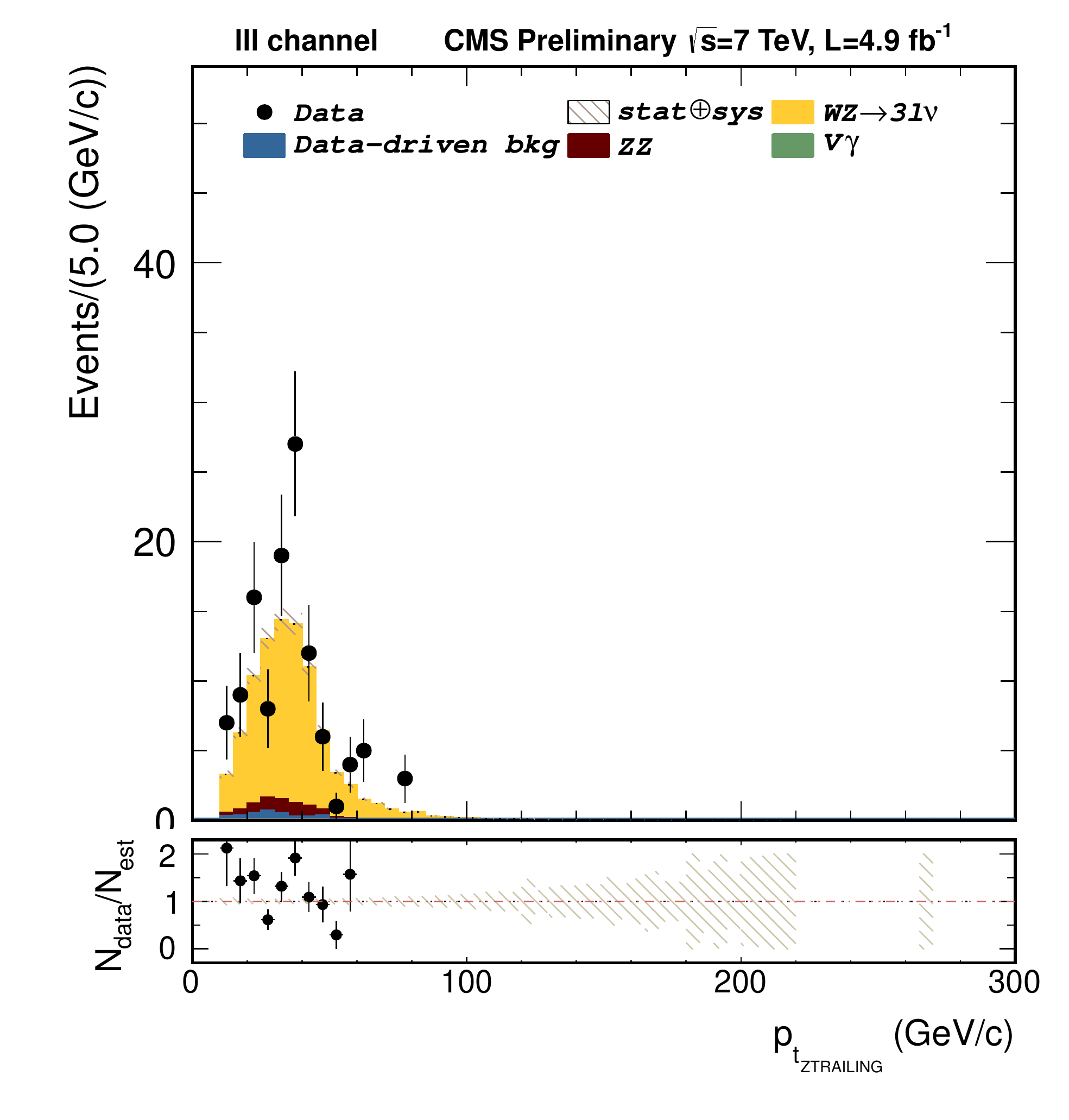}
	\end{subfigure}
	\begin{subfigure}[b]{0.3\textwidth}
		\includegraphics[width=\textwidth]{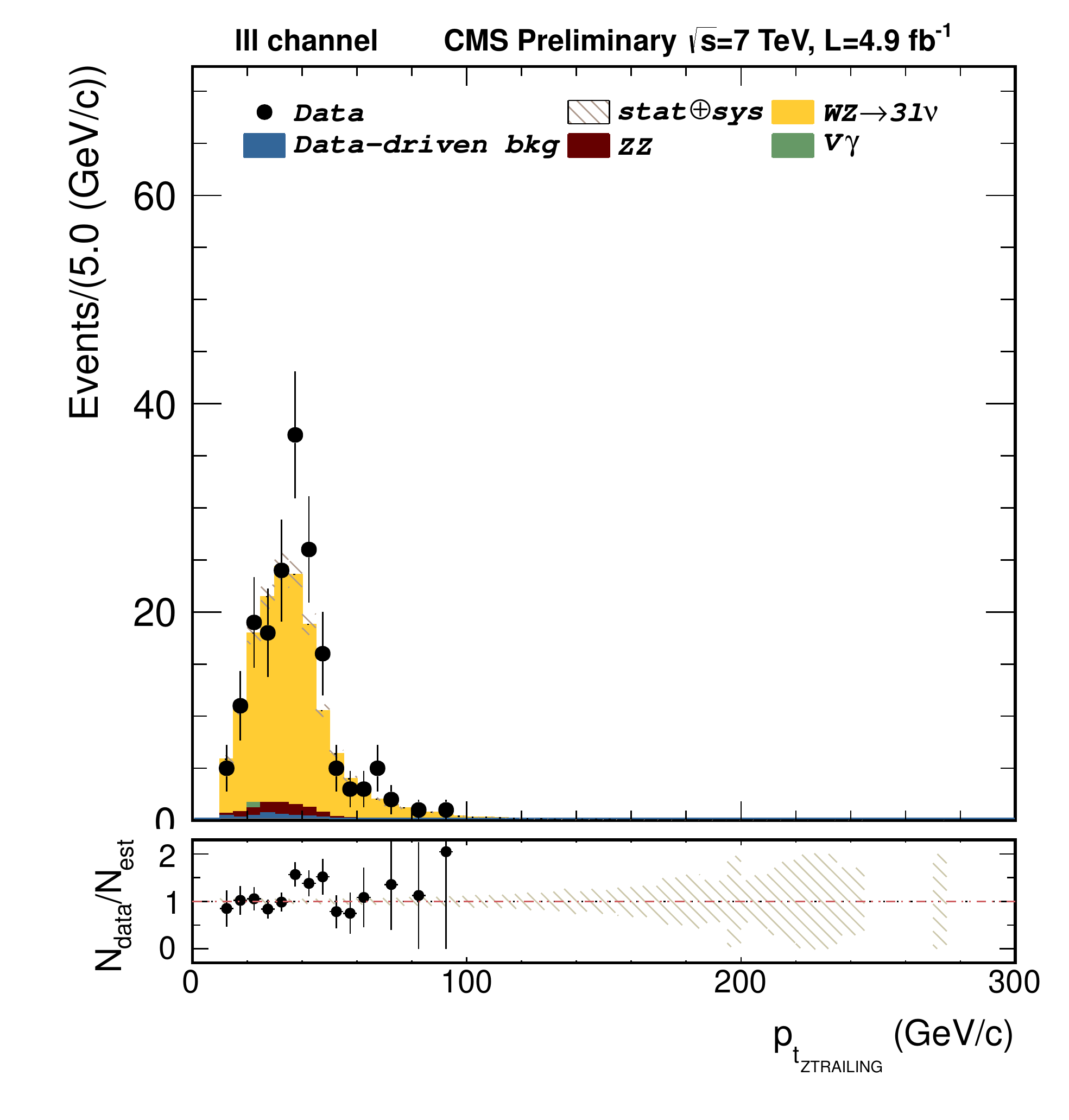}
	\end{subfigure}
	\caption[Transverse momentum of the Z-candidate leptons at 7~\TeV (ratio)]
	{Transverse momentum of the Z-candidate leading (up row) and trailing lepton 
	(bottom row) at each event for the \wzm (left column) and \wzp (right column) after the 
	\MET cut.}
\end{figure}

\clearpage
\section{Cross section analysis distributions at 8~\TeV}
The distributions shown in this subsection correspond to the 2012 analysis of the inclusive
\WZ cross section using data corresponding to an integrated luminosity of 19.6~\fbinv. In addition
to the same processes used in the 7~\TeV analysis estimated with simulated Monte Carlo samples, 
the VVV (V=W,Z) processes are also considered (with light blue in figures) along with the 
W$\gamma$ and W$\gamma^*$ denoted as WV (grey in figures).
\begin{sidewaysfigure}[!htpb]
	\centering
	\begin{subfigure}[b]{0.2\textwidth}
		\includegraphics[width=\textwidth,height=\textwidth]{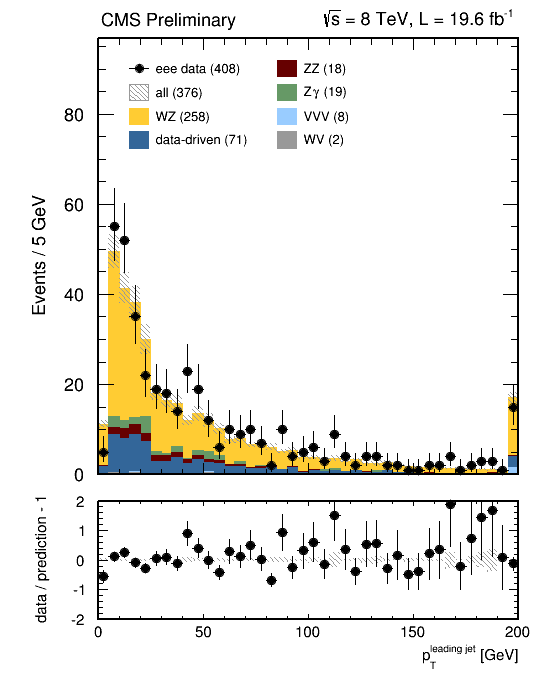}
	\end{subfigure}\quad
	\begin{subfigure}[b]{0.2\textwidth}
		\includegraphics[width=\textwidth,height=\textwidth]{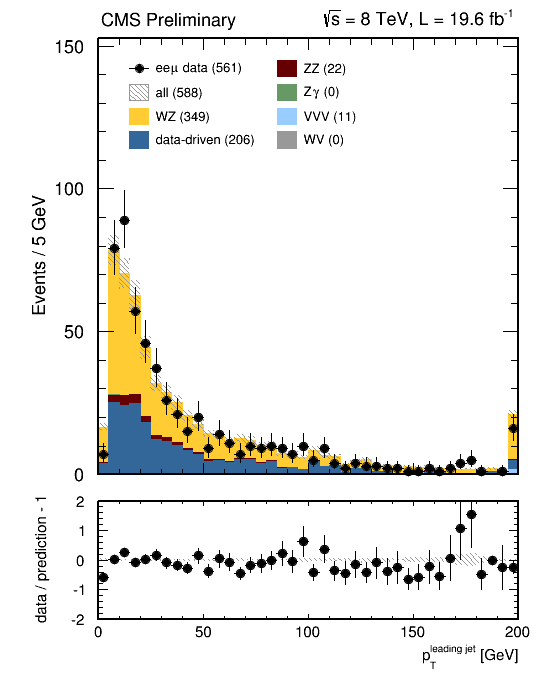}
	\end{subfigure}\quad
	\begin{subfigure}[b]{0.2\textwidth}
		\includegraphics[width=\textwidth,height=\textwidth]{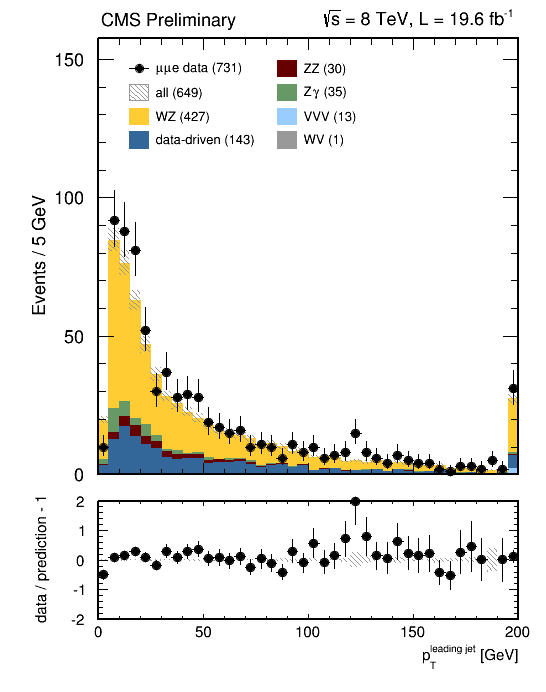}
	\end{subfigure}\quad
	\begin{subfigure}[b]{0.2\textwidth}
		\includegraphics[width=\textwidth,height=\textwidth]{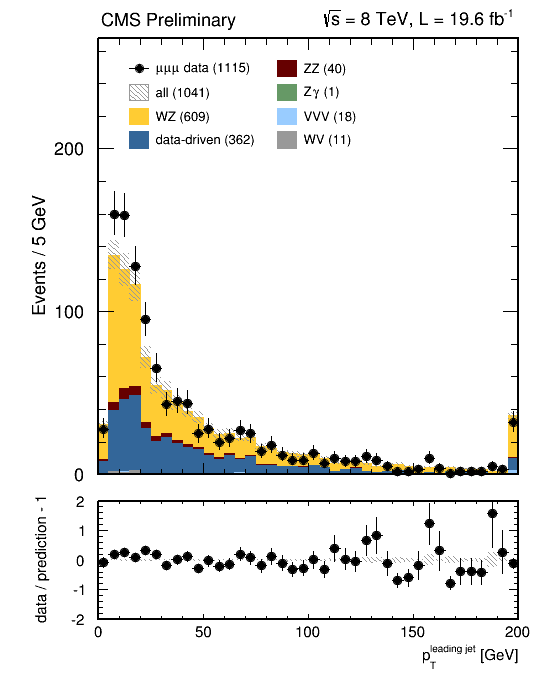}
	\end{subfigure}
	\vskip 1ex
	\centering
	\begin{subfigure}[b]{0.2\textwidth}
		\includegraphics[width=\textwidth,height=\textwidth]{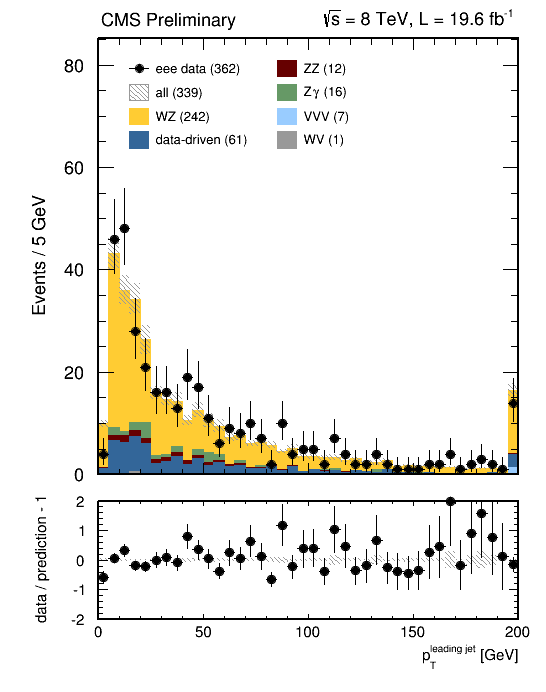}
	\end{subfigure}\quad
	\begin{subfigure}[b]{0.2\textwidth}
		\includegraphics[width=\textwidth,height=\textwidth]{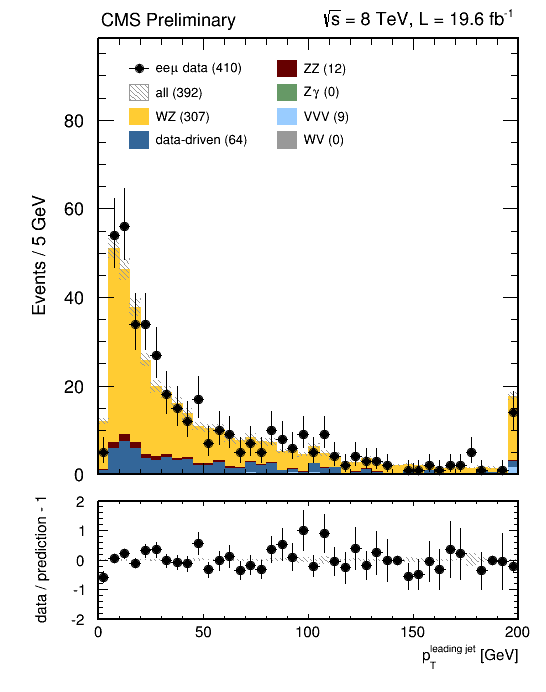}
	\end{subfigure}\quad
	\begin{subfigure}[b]{0.2\textwidth}
		\includegraphics[width=\textwidth,height=\textwidth]{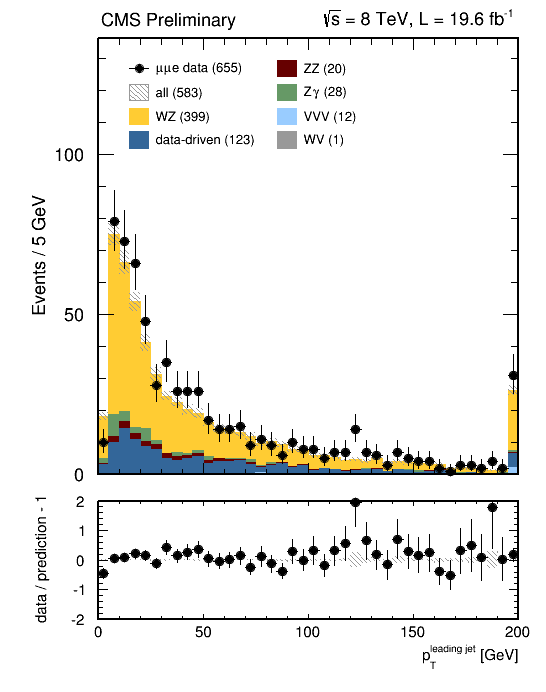}
	\end{subfigure}\quad
	\begin{subfigure}[b]{0.2\textwidth}
		\includegraphics[width=\textwidth,height=\textwidth]{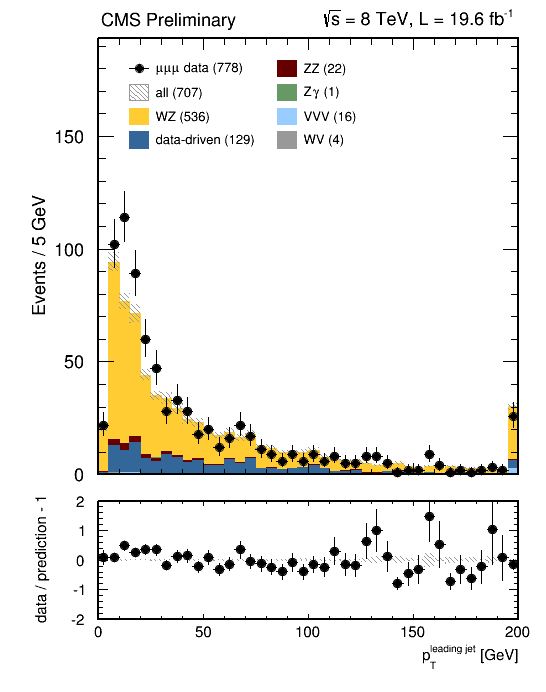}
	\end{subfigure}
	\vskip 1ex
	\begin{subfigure}[b]{0.2\textwidth}
		\includegraphics[width=\textwidth,height=\textwidth]{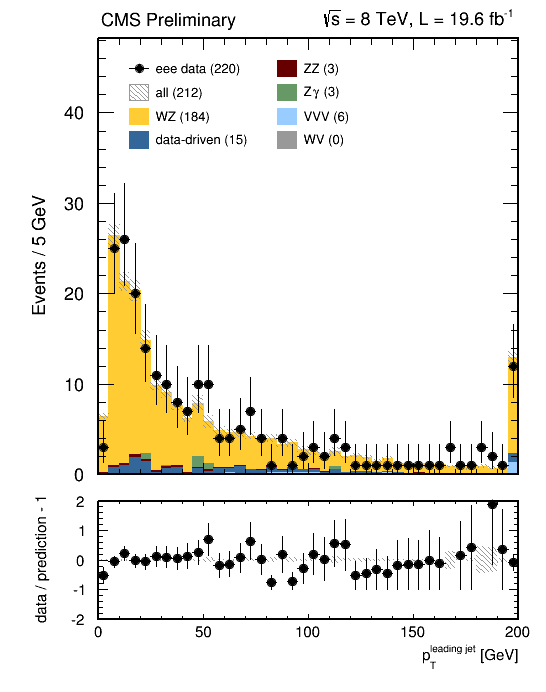}
	\end{subfigure}\quad
	\begin{subfigure}[b]{0.2\textwidth}
		\includegraphics[width=\textwidth,height=\textwidth]{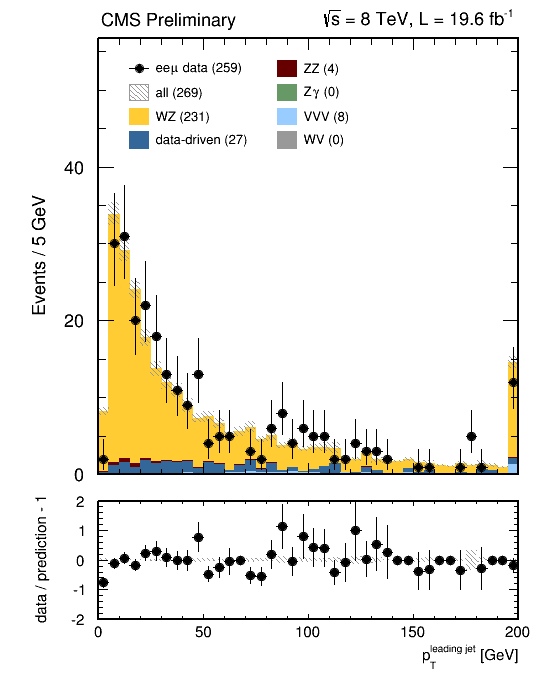}
	\end{subfigure}\quad
	\begin{subfigure}[b]{0.2\textwidth}
		\includegraphics[width=\textwidth,height=\textwidth]{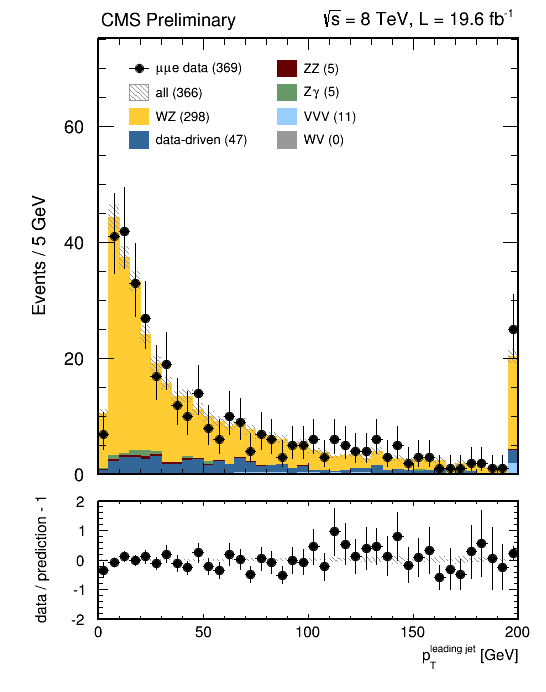}
	\end{subfigure}\quad
	\begin{subfigure}[b]{0.2\textwidth}
		\includegraphics[width=\textwidth,height=\textwidth]{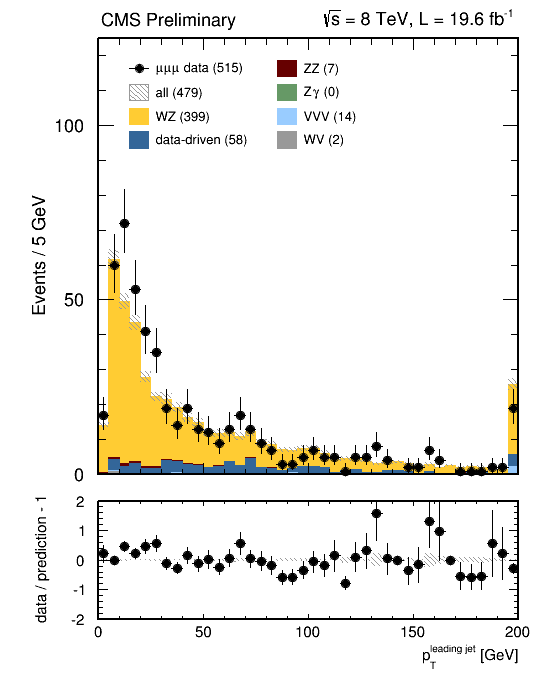}
	\end{subfigure}
	\caption[Transverse momentum of the leading jet at 8 TeV]{Transverse 
	momentum distribution of the leading jet at each event for the
	measured channels $eee$, $\mu ee$, $e\mu\mu$ and $\mu\mu\mu$ (from left to right) and
	after each analysis selection stage: after Z-candidate requirement (up row), after 
	W-candidate, without the \MET cut (middle row) and after W-candidate including \MET
	cut (bottom row).}
\end{sidewaysfigure}

\begin{sidewaysfigure}[!htpb]
	\centering
	\begin{subfigure}[b]{0.2\textwidth}
		\includegraphics[width=\textwidth,height=\textwidth]{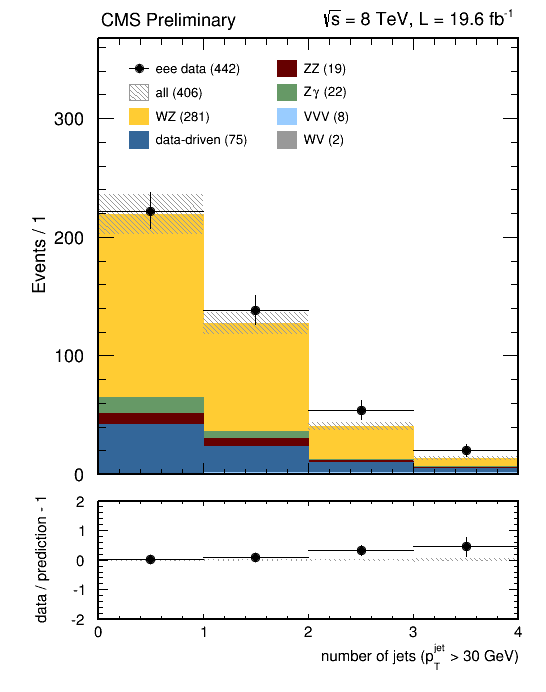}
	\end{subfigure}\quad
	\begin{subfigure}[b]{0.2\textwidth}
		\includegraphics[width=\textwidth,height=\textwidth]{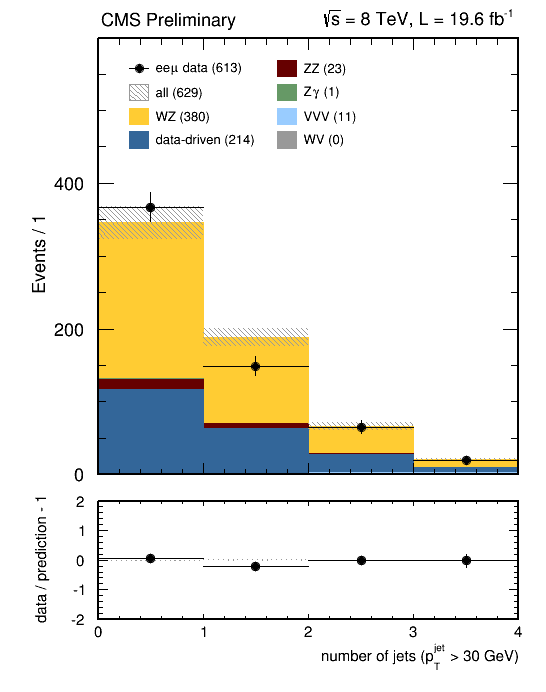}
	\end{subfigure}\quad
	\begin{subfigure}[b]{0.2\textwidth}
		\includegraphics[width=\textwidth,height=\textwidth]{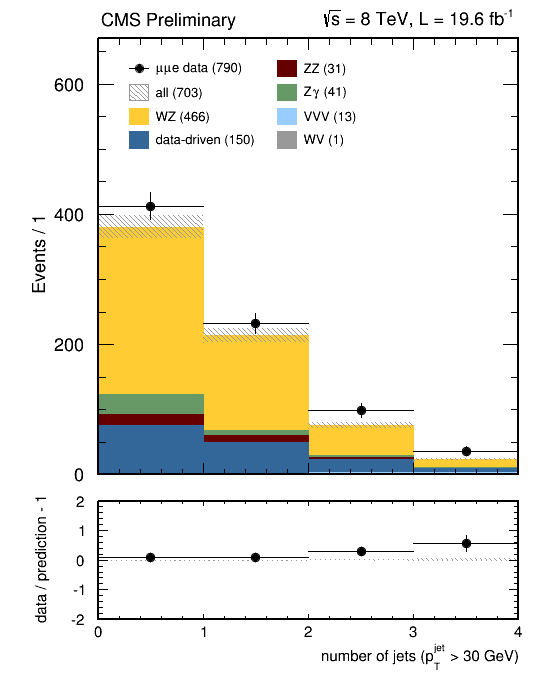}
	\end{subfigure}\quad
	\begin{subfigure}[b]{0.2\textwidth}
		\includegraphics[width=\textwidth,height=\textwidth]{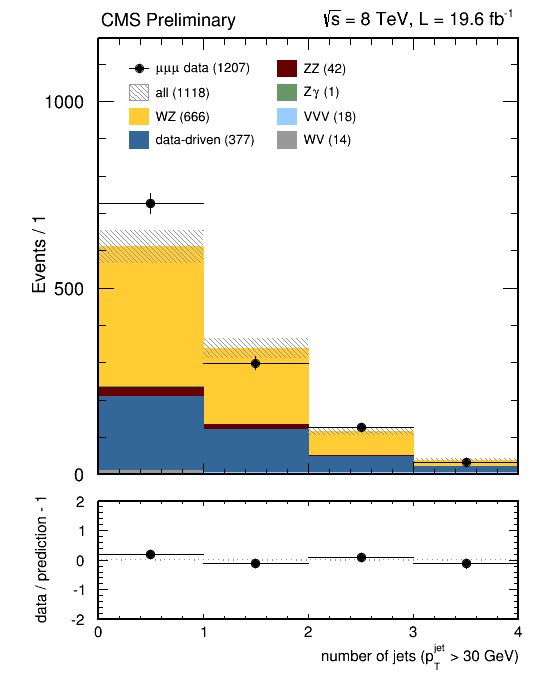}
	\end{subfigure}
	\vskip 1ex
	\centering
	\begin{subfigure}[b]{0.2\textwidth}
		\includegraphics[width=\textwidth,height=\textwidth]{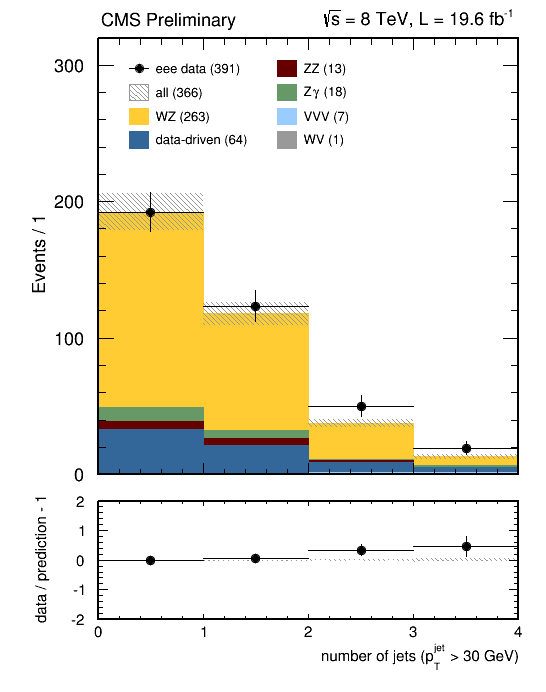}
	\end{subfigure}\quad
	\begin{subfigure}[b]{0.2\textwidth}
		\includegraphics[width=\textwidth,height=\textwidth]{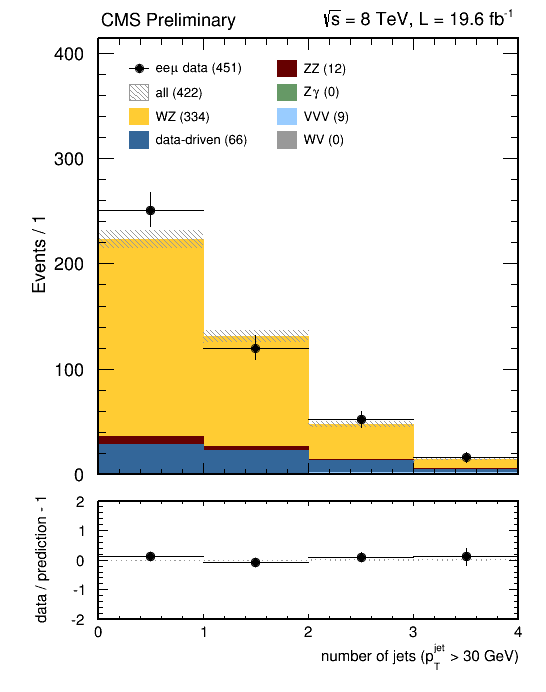}
	\end{subfigure}\quad
	\begin{subfigure}[b]{0.2\textwidth}
		\includegraphics[width=\textwidth,height=\textwidth]{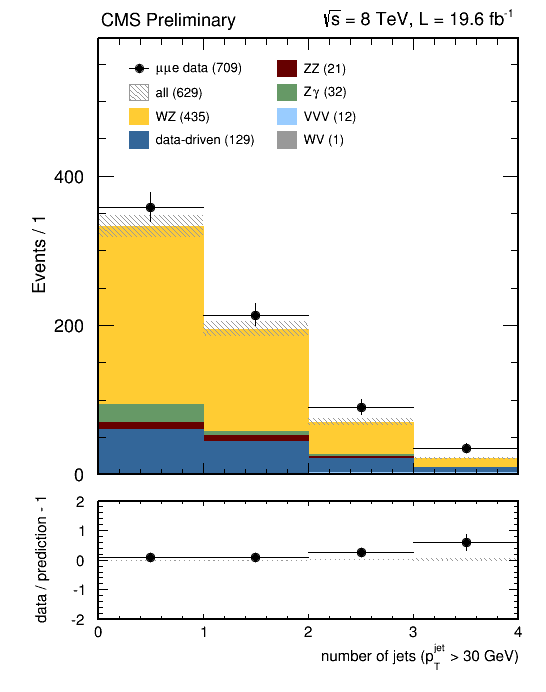}
	\end{subfigure}\quad
	\begin{subfigure}[b]{0.2\textwidth}
		\includegraphics[width=\textwidth,height=\textwidth]{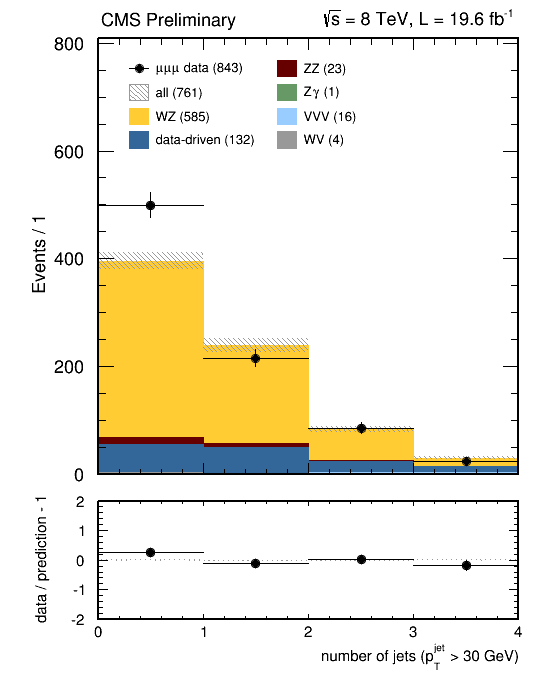}
	\end{subfigure}
	\vskip 1ex
	\begin{subfigure}[b]{0.2\textwidth}
		\includegraphics[width=\textwidth,height=\textwidth]{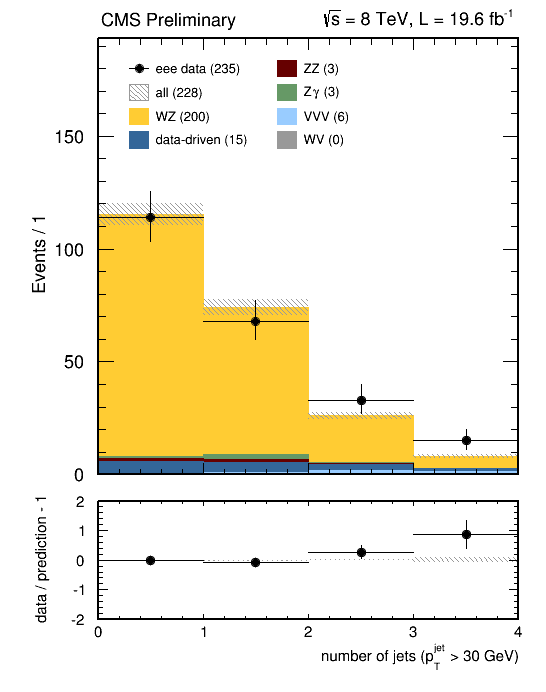}
	\end{subfigure}\quad
	\begin{subfigure}[b]{0.2\textwidth}
		\includegraphics[width=\textwidth,height=\textwidth]{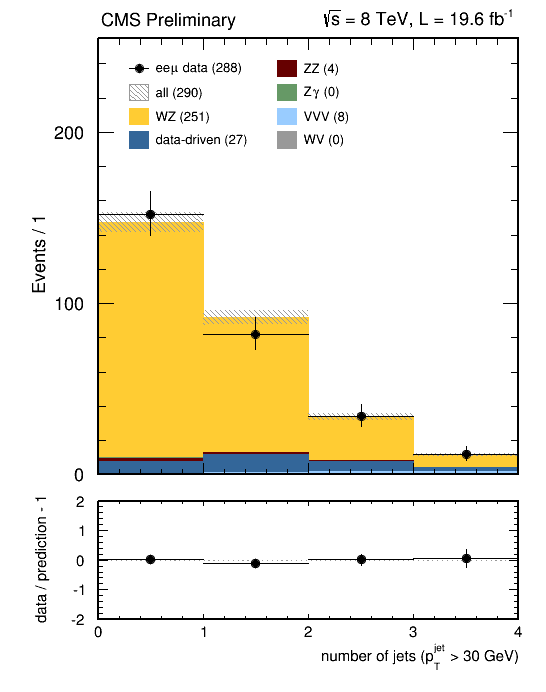}
	\end{subfigure}\quad
	\begin{subfigure}[b]{0.2\textwidth}
		\includegraphics[width=\textwidth,height=\textwidth]{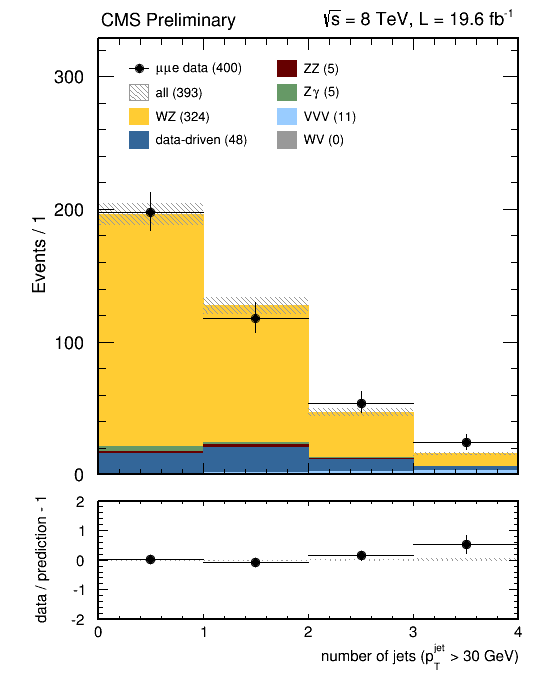}
	\end{subfigure}\quad
	\begin{subfigure}[b]{0.2\textwidth}
		\includegraphics[width=\textwidth,height=\textwidth]{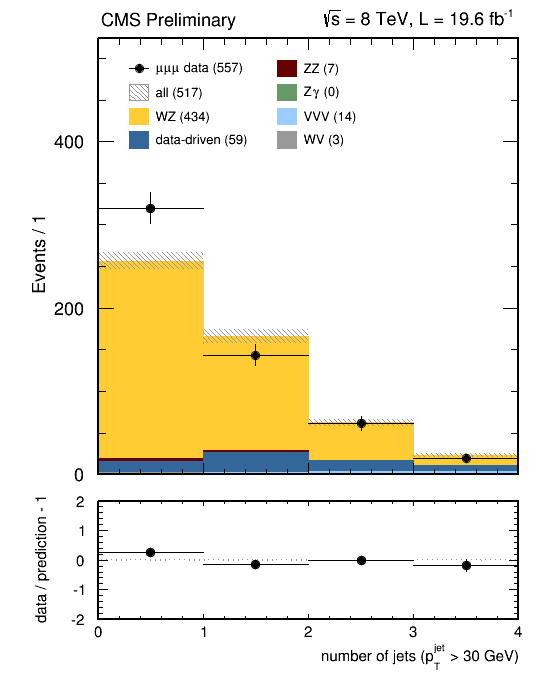}
	\end{subfigure}
	\caption[Number of jets at 8 TeV]{Number of 
	jets distribution at each event for the measured channels
	$eee$, $\mu ee$, $e\mu\mu$ and $\mu\mu\mu$ (from left to right) and
	after each analysis selection stage: after Z-candidate requirement (up row), after 
	W-candidate, without the \MET cut (middle row) and after W-candidate including \MET
	cut (bottom row).}
\end{sidewaysfigure}

\begin{sidewaysfigure}[!htpb]
	\centering
	\begin{subfigure}[b]{0.2\textwidth}
		\includegraphics[width=\textwidth,height=\textwidth]{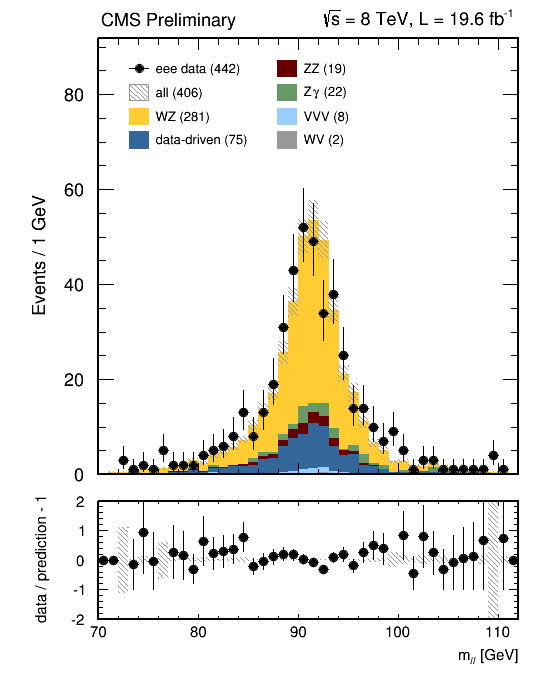}
	\end{subfigure}\quad
	\begin{subfigure}[b]{0.2\textwidth}
		\includegraphics[width=\textwidth,height=\textwidth]{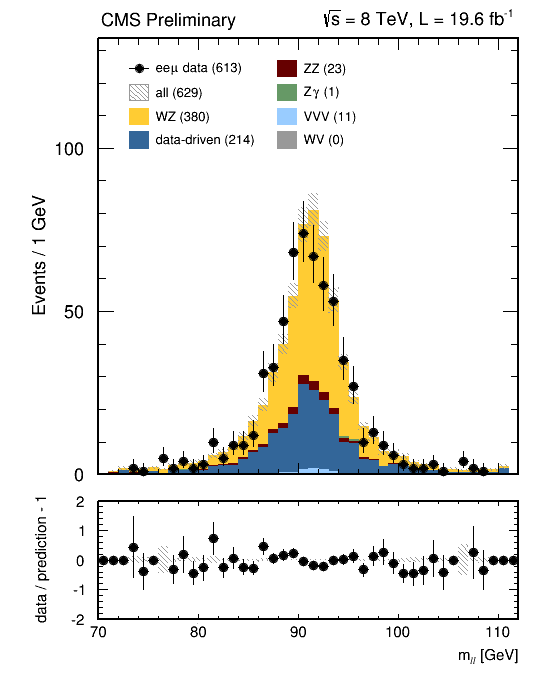}
	\end{subfigure}\quad
	\begin{subfigure}[b]{0.2\textwidth}
		\includegraphics[width=\textwidth,height=\textwidth]{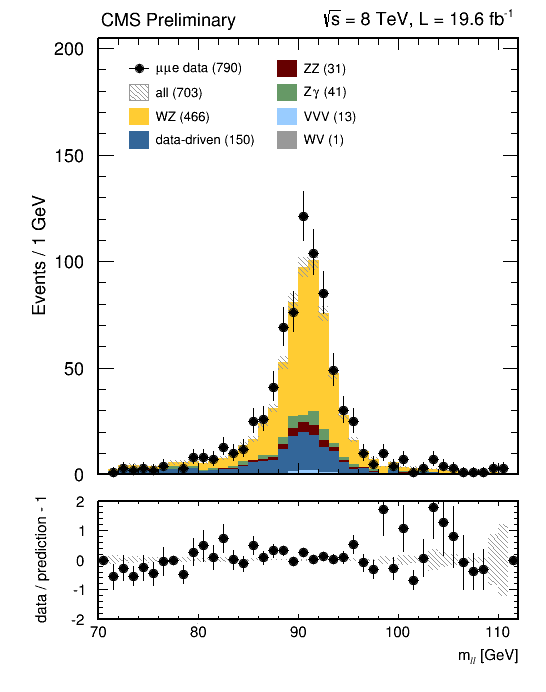}
	\end{subfigure}\quad
	\begin{subfigure}[b]{0.2\textwidth}
		\includegraphics[width=\textwidth,height=\textwidth]{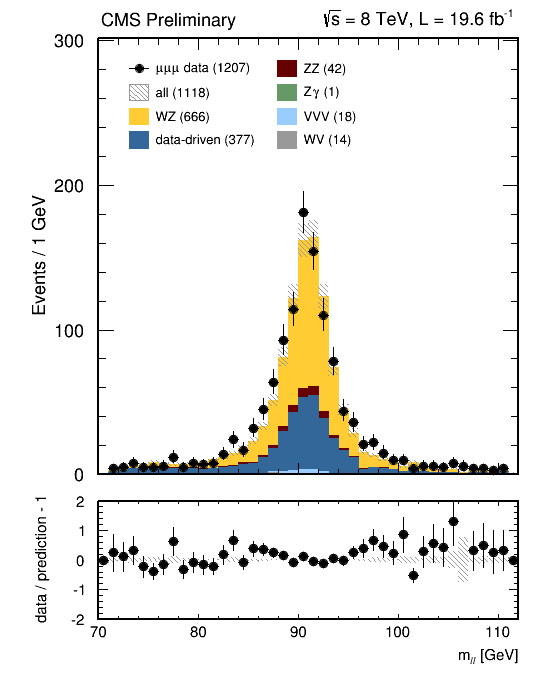}
	\end{subfigure}
	\vskip 1ex
	\centering
	\begin{subfigure}[b]{0.2\textwidth}
		\includegraphics[width=\textwidth,height=\textwidth]{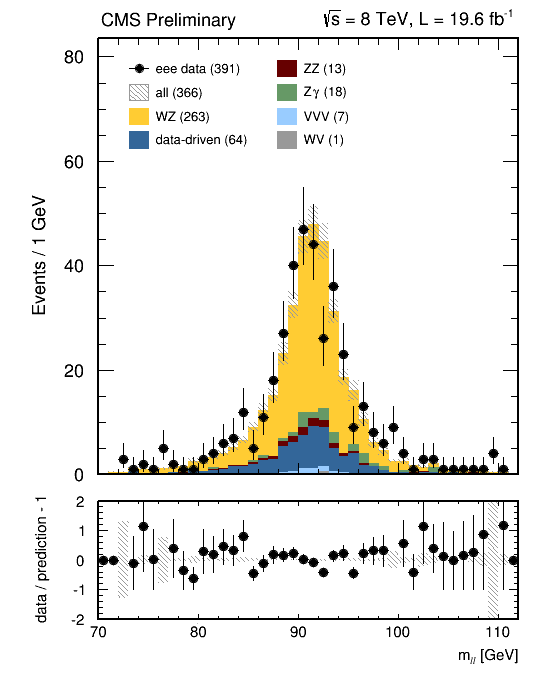}
	\end{subfigure}\quad
	\begin{subfigure}[b]{0.2\textwidth}
		\includegraphics[width=\textwidth,height=\textwidth]{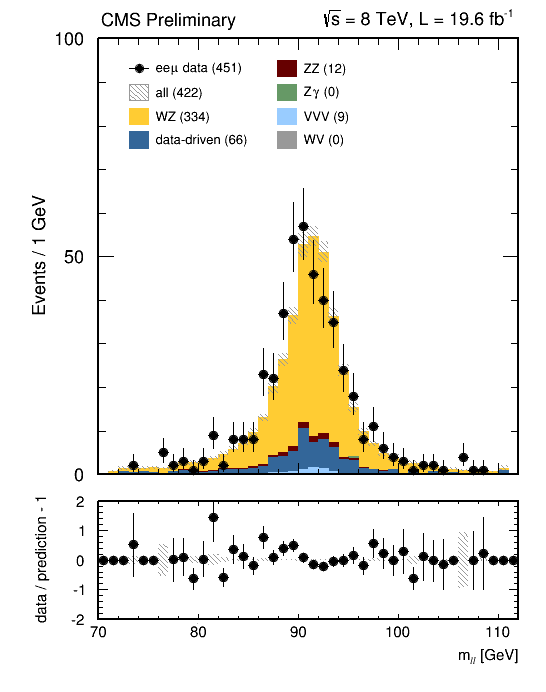}
	\end{subfigure}\quad
	\begin{subfigure}[b]{0.2\textwidth}
		\includegraphics[width=\textwidth,height=\textwidth]{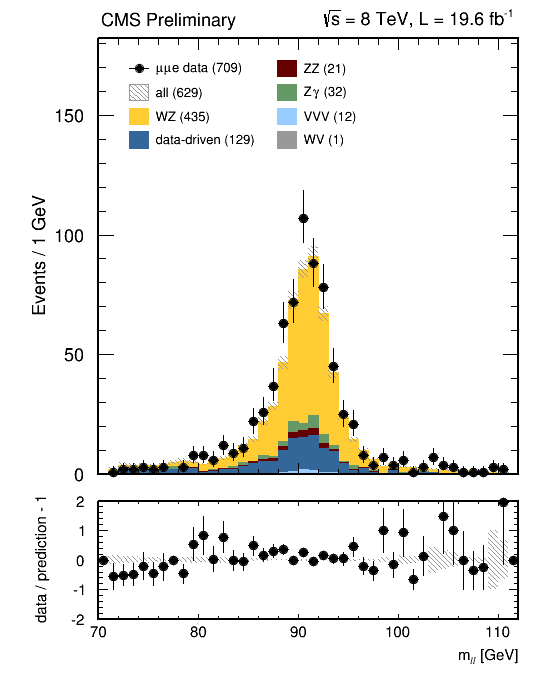}
	\end{subfigure}\quad
	\begin{subfigure}[b]{0.2\textwidth}
		\includegraphics[width=\textwidth,height=\textwidth]{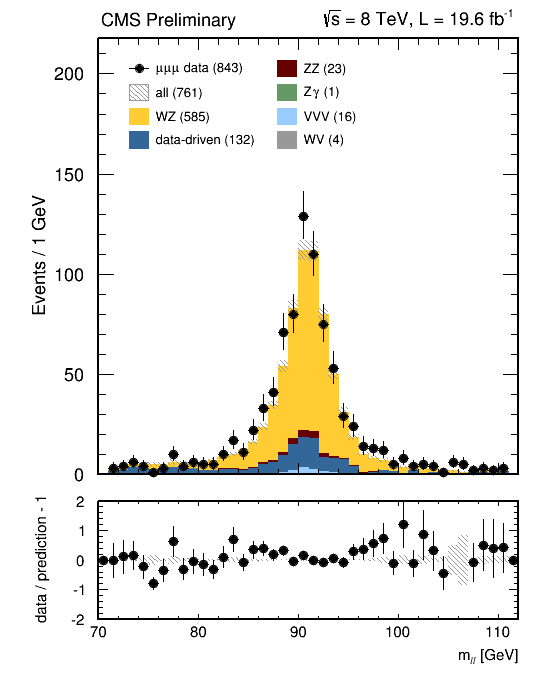}
	\end{subfigure}
	\vskip 1ex
	\begin{subfigure}[b]{0.2\textwidth}
		\includegraphics[width=\textwidth,height=\textwidth]{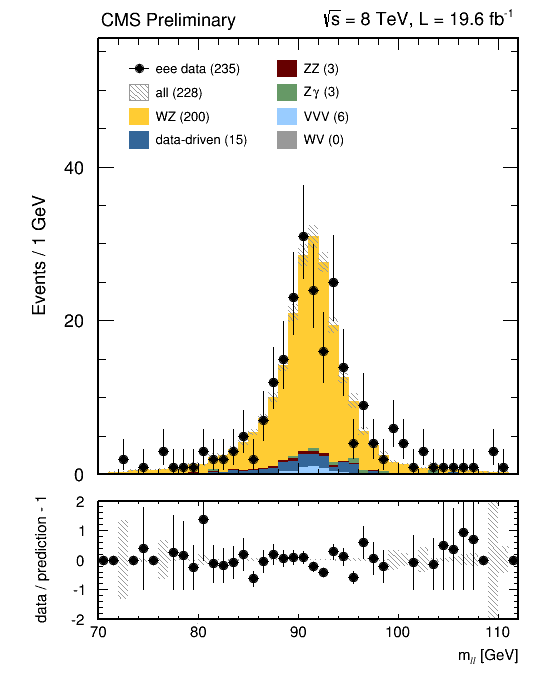}
	\end{subfigure}\quad
	\begin{subfigure}[b]{0.2\textwidth}
		\includegraphics[width=\textwidth,height=\textwidth]{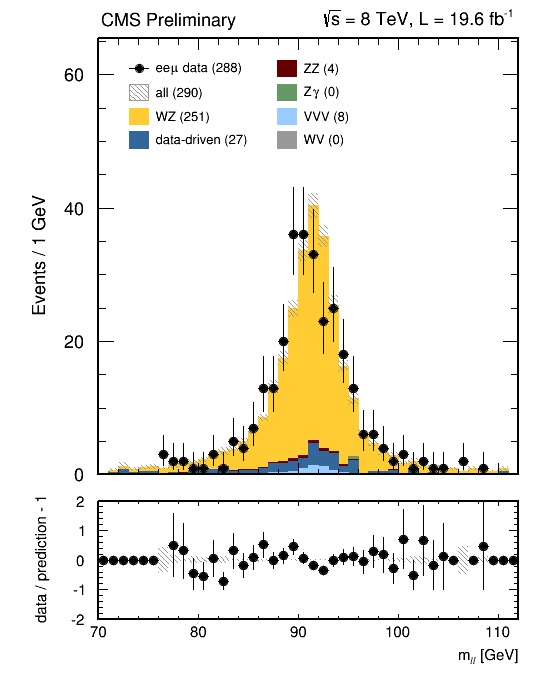}
	\end{subfigure}\quad
	\begin{subfigure}[b]{0.2\textwidth}
		\includegraphics[width=\textwidth,height=\textwidth]{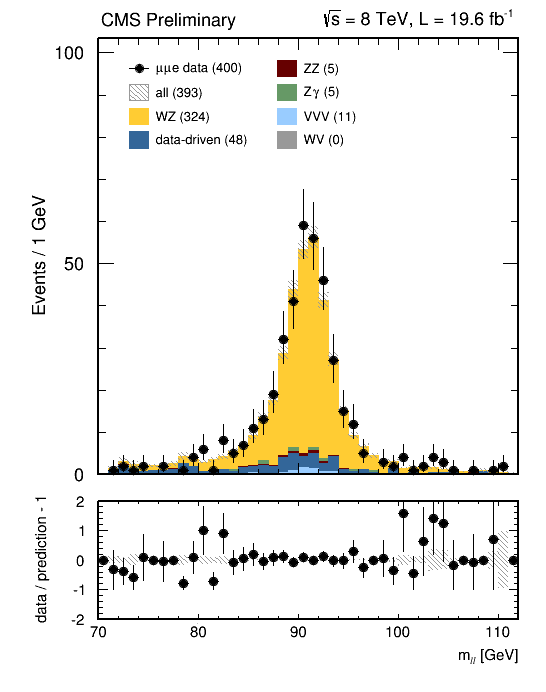}
	\end{subfigure}\quad
	\begin{subfigure}[b]{0.2\textwidth}
		\includegraphics[width=\textwidth,height=\textwidth]{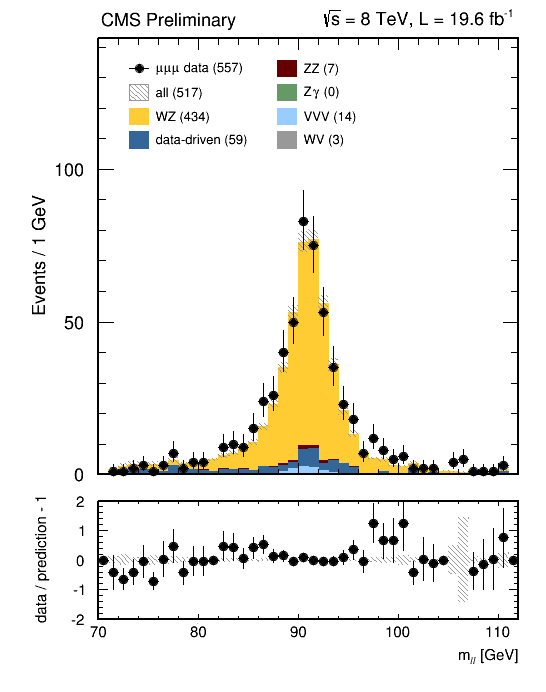}
	\end{subfigure}
	\caption[Invariant mass of the dilepton system at 8~\TeV]
	{Invariant mass of the Z-candidate dilepton system for the measured channels 
	$eee$, $\mu ee$, $e\mu\mu$ and $\mu\mu\mu$ (from left to right) and
	after each analysis selection stage: after Z-candidate requirement (up row), after 
	W-candidate, without the \MET cut (middle row) and after W-candidate including \MET
	cut (bottom row).}
\end{sidewaysfigure}

\begin{sidewaysfigure}[!htpb]
	\centering
	\begin{subfigure}[b]{0.2\textwidth}
		\includegraphics[width=\textwidth,height=\textwidth]{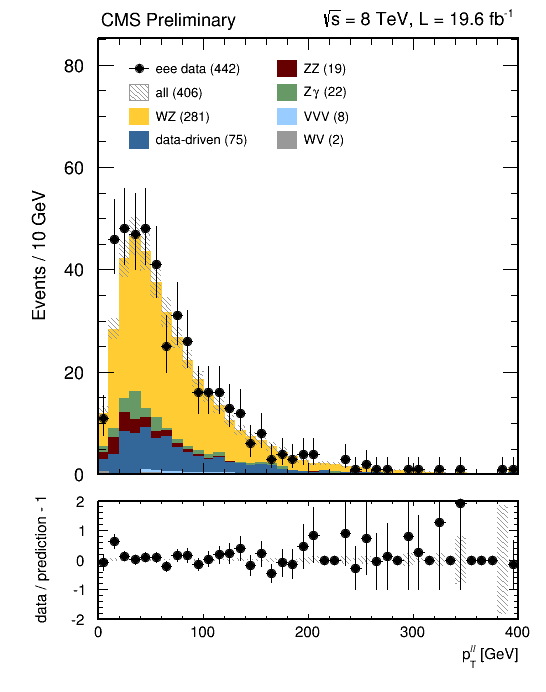}
	\end{subfigure}\quad
	\begin{subfigure}[b]{0.2\textwidth}
		\includegraphics[width=\textwidth,height=\textwidth]{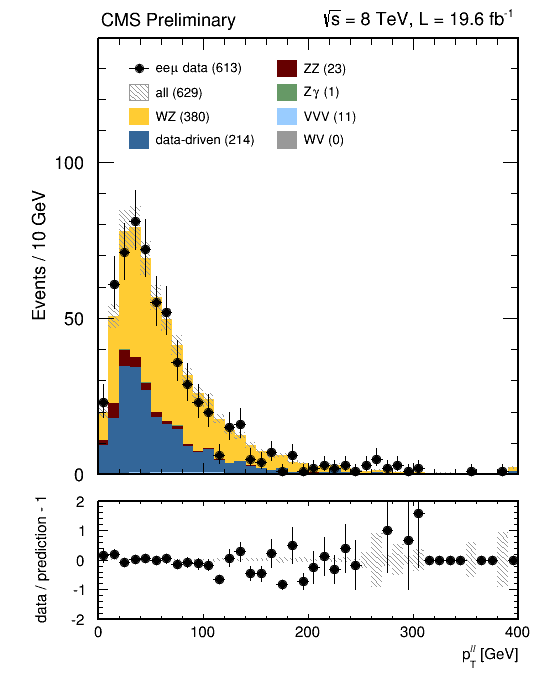}
	\end{subfigure}\quad
	\begin{subfigure}[b]{0.2\textwidth}
		\includegraphics[width=\textwidth,height=\textwidth]{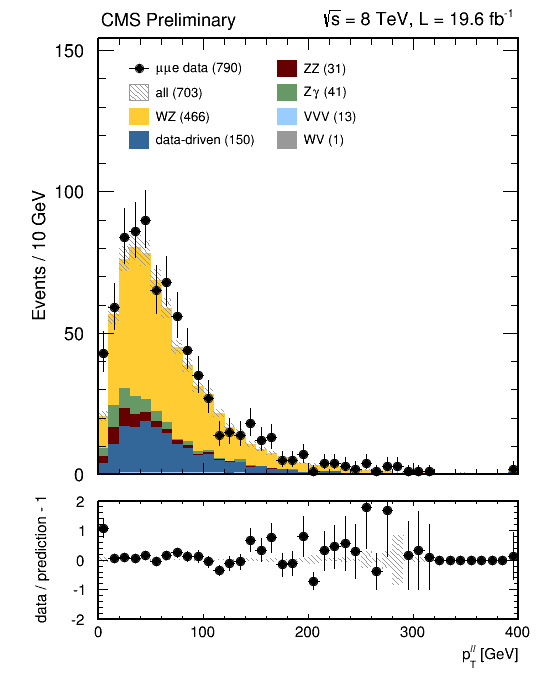}
	\end{subfigure}\quad
	\begin{subfigure}[b]{0.2\textwidth}
		\includegraphics[width=\textwidth,height=\textwidth]{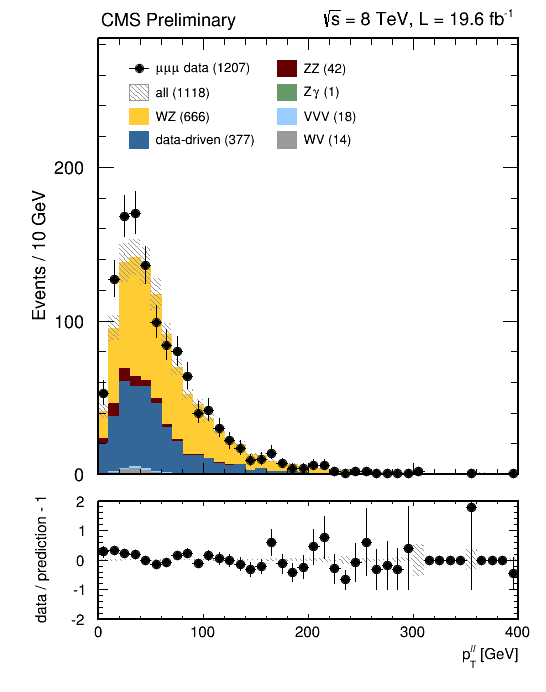}
	\end{subfigure}
	\vskip 1ex
	\centering
	\begin{subfigure}[b]{0.2\textwidth}
		\includegraphics[width=\textwidth,height=\textwidth]{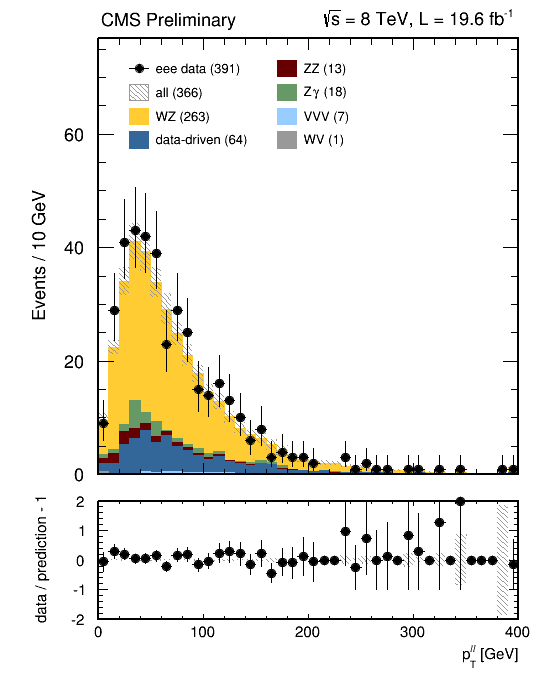}
	\end{subfigure}\quad
	\begin{subfigure}[b]{0.2\textwidth}
		\includegraphics[width=\textwidth,height=\textwidth]{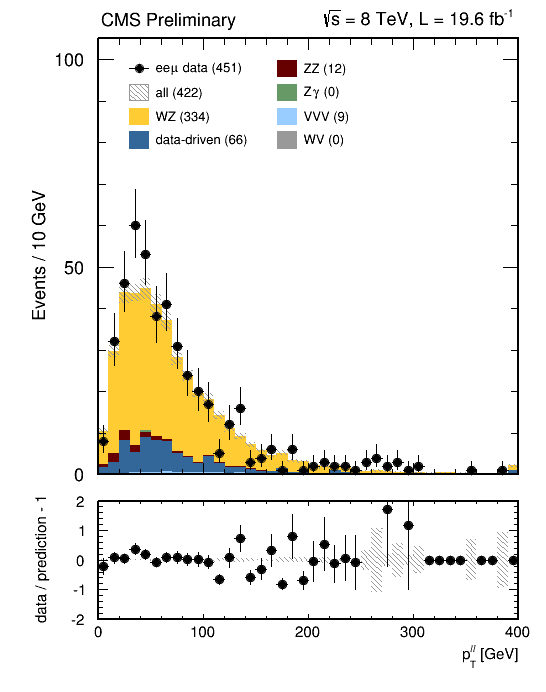}
	\end{subfigure}\quad
	\begin{subfigure}[b]{0.2\textwidth}
		\includegraphics[width=\textwidth,height=\textwidth]{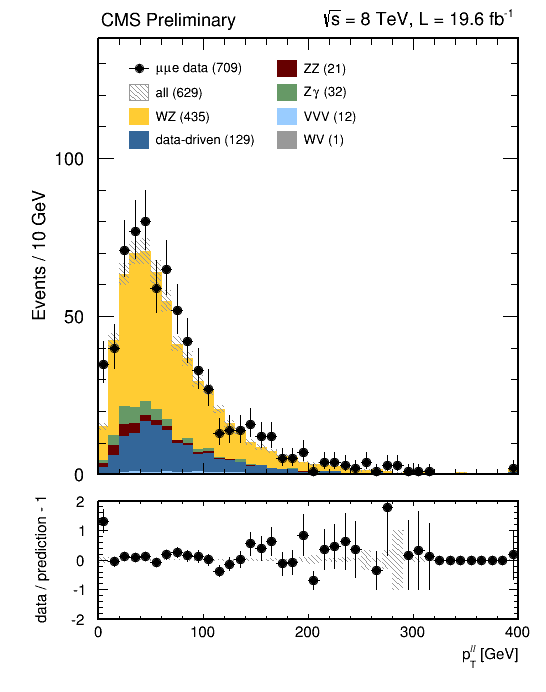}
	\end{subfigure}\quad
	\begin{subfigure}[b]{0.2\textwidth}
		\includegraphics[width=\textwidth,height=\textwidth]{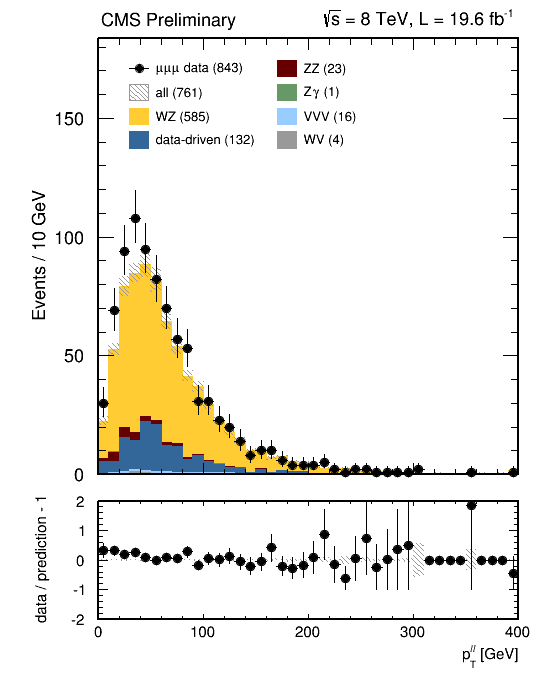}
	\end{subfigure}
	\vskip 1ex
	\begin{subfigure}[b]{0.2\textwidth}
		\includegraphics[width=\textwidth,height=\textwidth]{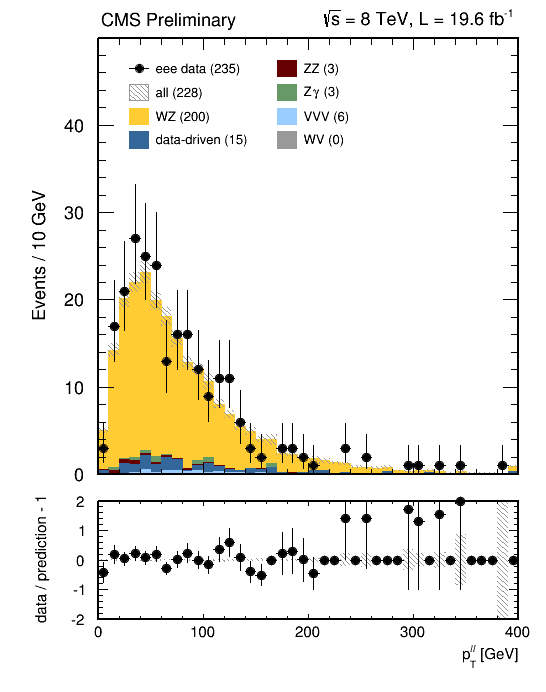}
	\end{subfigure}\quad
	\begin{subfigure}[b]{0.2\textwidth}
		\includegraphics[width=\textwidth,height=\textwidth]{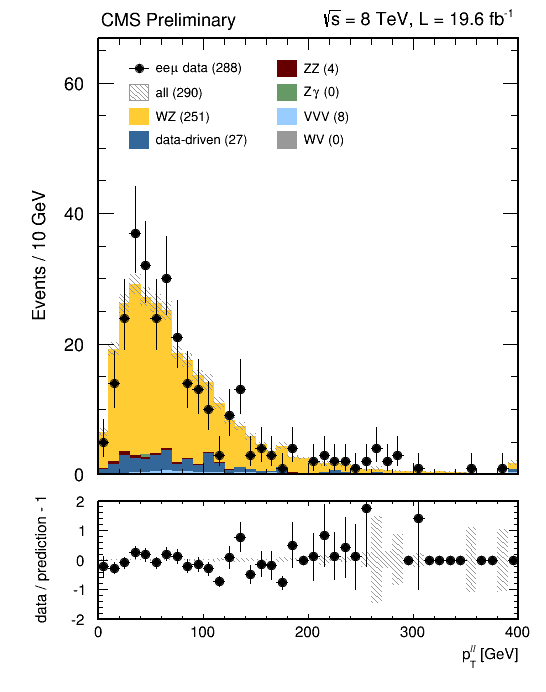}
	\end{subfigure}\quad
	\begin{subfigure}[b]{0.2\textwidth}
		\includegraphics[width=\textwidth,height=\textwidth]{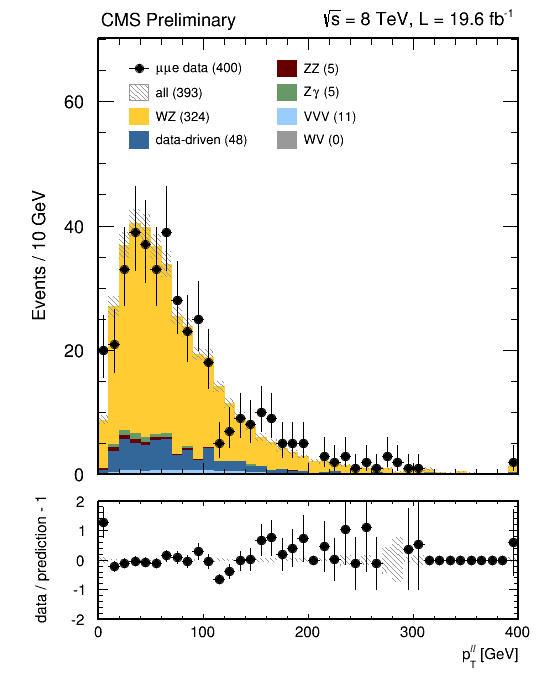}
	\end{subfigure}\quad
	\begin{subfigure}[b]{0.2\textwidth}
		\includegraphics[width=\textwidth,height=\textwidth]{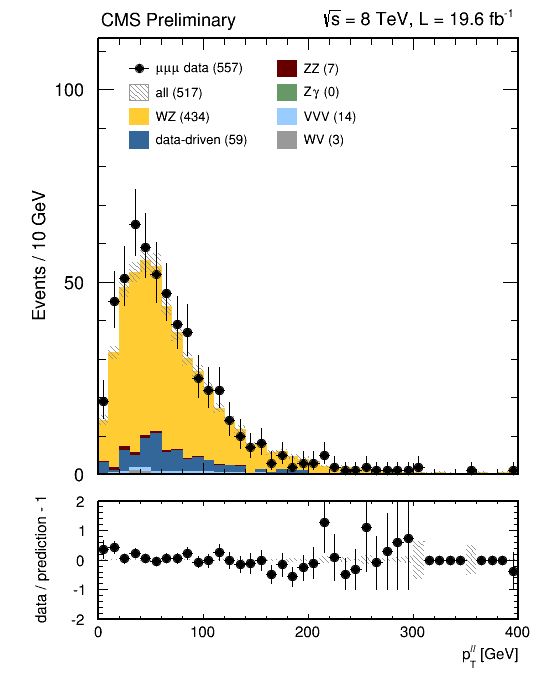}
	\end{subfigure}
	\caption[Transverse momentum of the dilepton system at 8~\TeV]
	{Transverse momentum of the Z-candidate dilepton system for the measured channels 
	$eee$, $\mu ee$, $e\mu\mu$ and $\mu\mu\mu$ (from left to right) and
	after each analysis selection stage: after Z-candidate requirement (up row), after 
	W-candidate, without the \MET cut (middle row) and after W-candidate including \MET
	cut (bottom row).}
\end{sidewaysfigure}

\begin{sidewaysfigure}[!htpb]
	\centering
	\begin{subfigure}[b]{0.2\textwidth}
		\includegraphics[width=\textwidth,height=\textwidth]{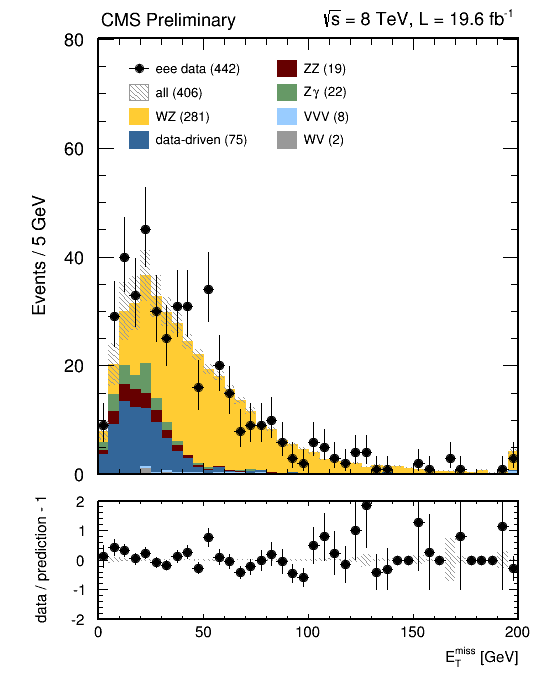}
	\end{subfigure}\quad
	\begin{subfigure}[b]{0.2\textwidth}
		\includegraphics[width=\textwidth,height=\textwidth]{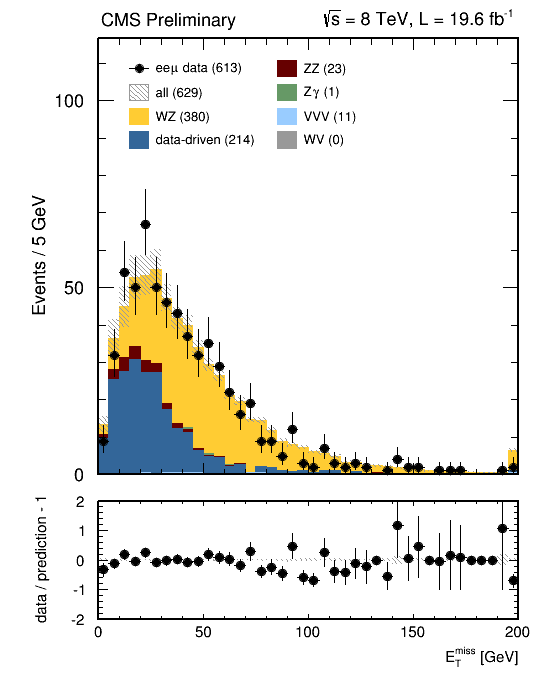}
	\end{subfigure}\quad
	\begin{subfigure}[b]{0.2\textwidth}
		\includegraphics[width=\textwidth,height=\textwidth]{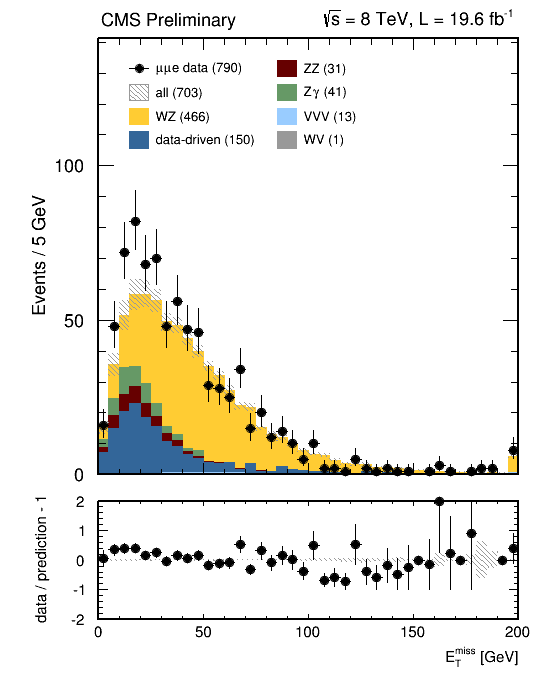}
	\end{subfigure}\quad
	\begin{subfigure}[b]{0.2\textwidth}
		\includegraphics[width=\textwidth,height=\textwidth]{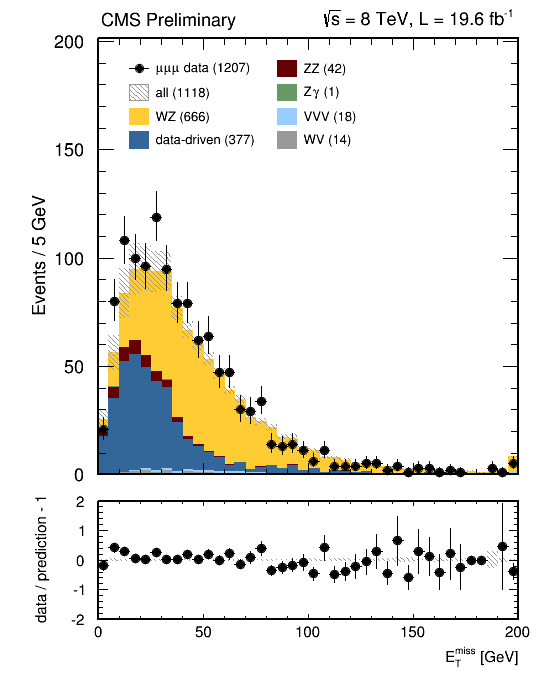}
	\end{subfigure}
	\vskip 1ex
	\centering
	\begin{subfigure}[b]{0.2\textwidth}
		\includegraphics[width=\textwidth,height=\textwidth]{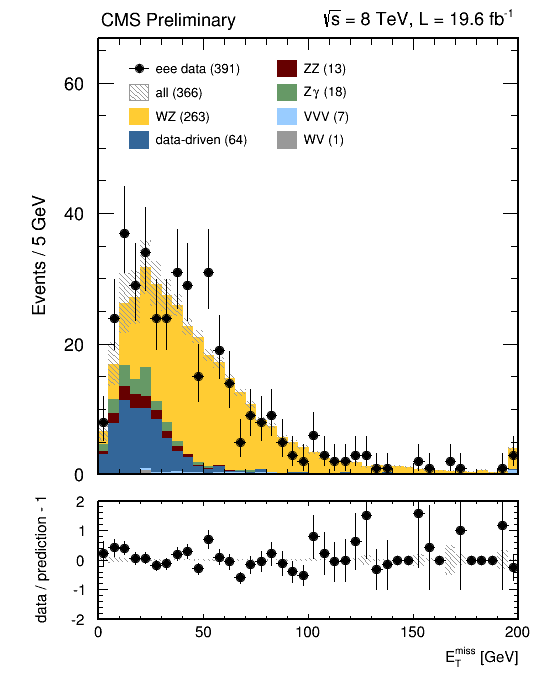}
	\end{subfigure}\quad
	\begin{subfigure}[b]{0.2\textwidth}
		\includegraphics[width=\textwidth,height=\textwidth]{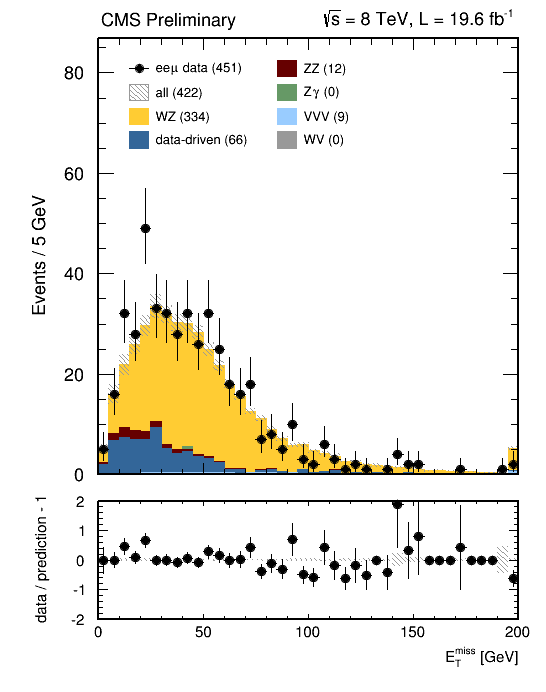}
	\end{subfigure}\quad
	\begin{subfigure}[b]{0.2\textwidth}
		\includegraphics[width=\textwidth,height=\textwidth]{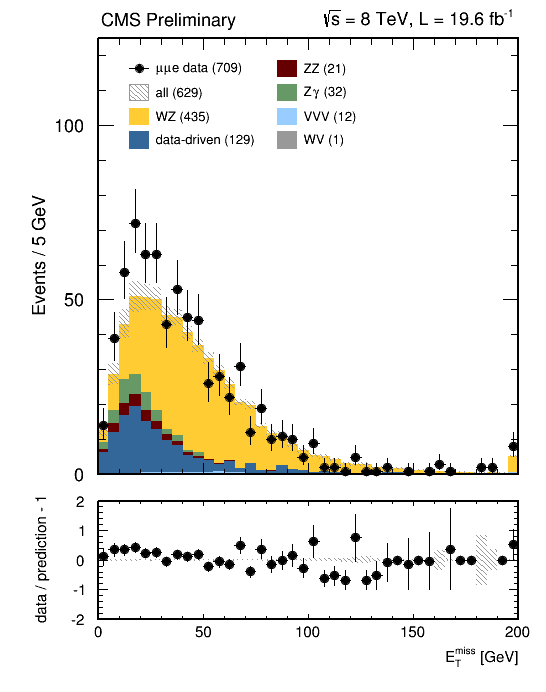}
	\end{subfigure}\quad
	\begin{subfigure}[b]{0.2\textwidth}
		\includegraphics[width=\textwidth,height=\textwidth]{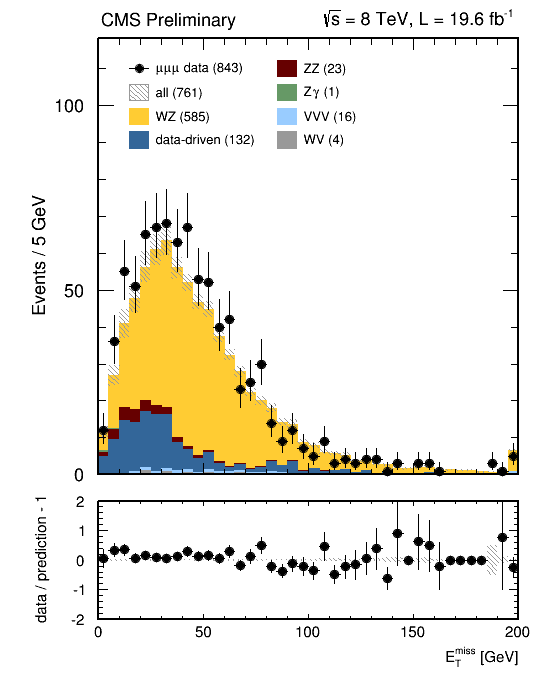}
	\end{subfigure}
	\vskip 1ex
	\begin{subfigure}[b]{0.2\textwidth}
		\includegraphics[width=\textwidth,height=\textwidth]{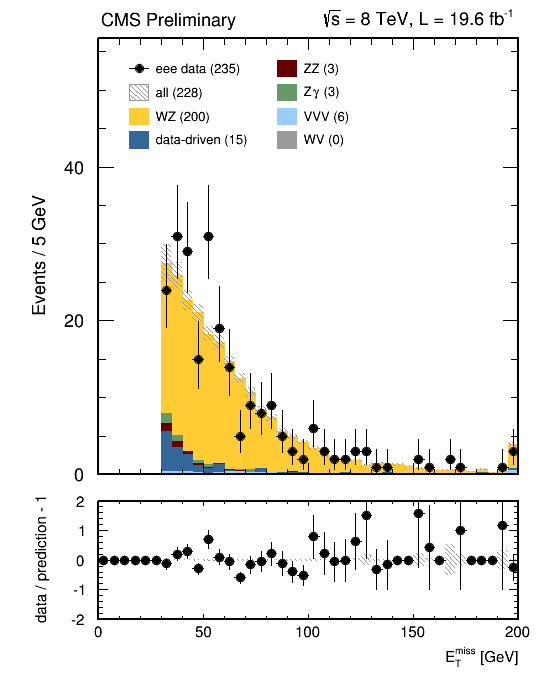}
	\end{subfigure}\quad
	\begin{subfigure}[b]{0.2\textwidth}
		\includegraphics[width=\textwidth,height=\textwidth]{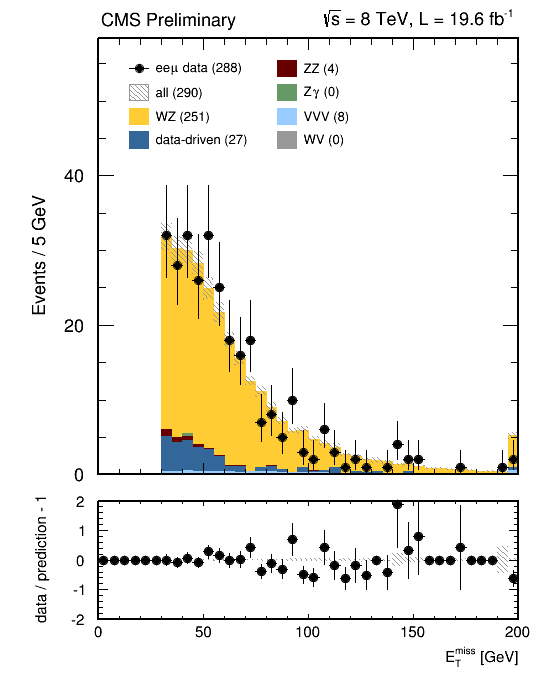}
	\end{subfigure}\quad
	\begin{subfigure}[b]{0.2\textwidth}
		\includegraphics[width=\textwidth,height=\textwidth]{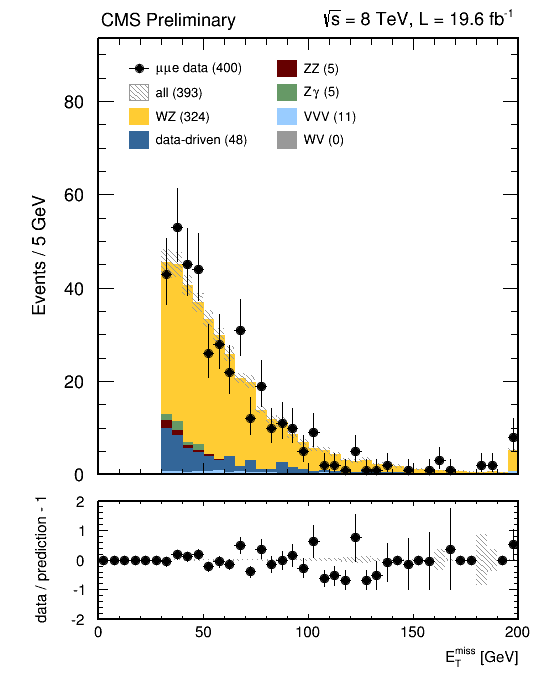}
	\end{subfigure}\quad
	\begin{subfigure}[b]{0.2\textwidth}
		\includegraphics[width=\textwidth,height=\textwidth]{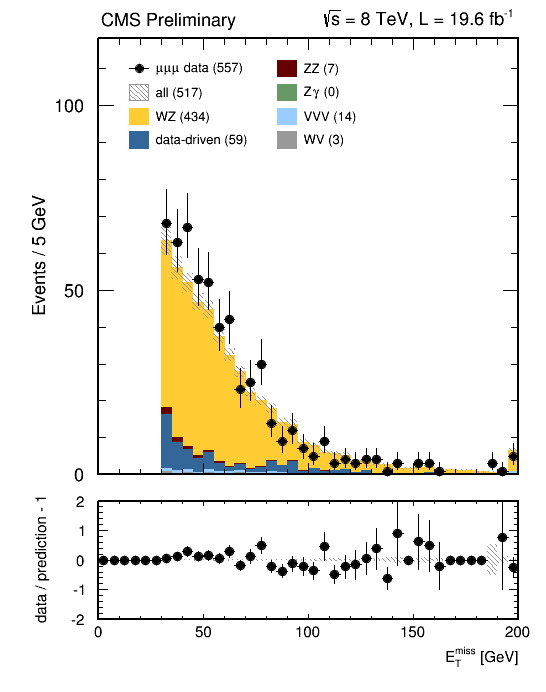}
	\end{subfigure}
	\caption[Missing transverse energy at 8~\TeV]{Missing 
	energy in the transverse plane at each event for the measured channels 
	$eee$, $\mu ee$, $e\mu\mu$ and $\mu\mu\mu$ (from left to right) and
	after each analysis selection stage: after Z-candidate requirement (up row), after 
	W-candidate, without the \MET cut (middle row) and after W-candidate including \MET
	cut (bottom row).}
\end{sidewaysfigure}

\begin{sidewaysfigure}[!htpb]
	\centering
	\begin{subfigure}[b]{0.2\textwidth}
		\includegraphics[width=\textwidth,height=\textwidth]{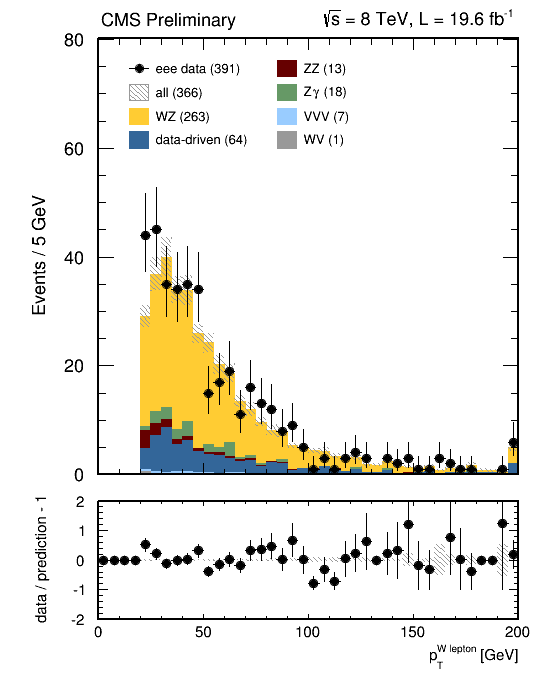}
	\end{subfigure}\quad
	\begin{subfigure}[b]{0.2\textwidth}
		\includegraphics[width=\textwidth,height=\textwidth]{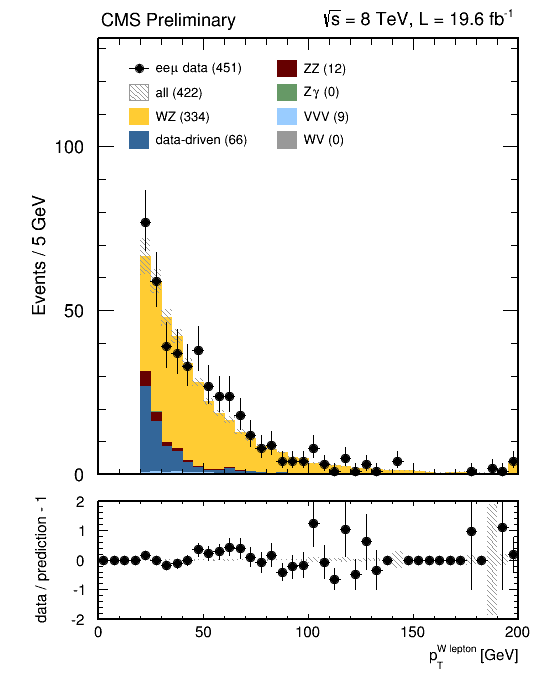}
	\end{subfigure}\quad
	\begin{subfigure}[b]{0.2\textwidth}
		\includegraphics[width=\textwidth,height=\textwidth]{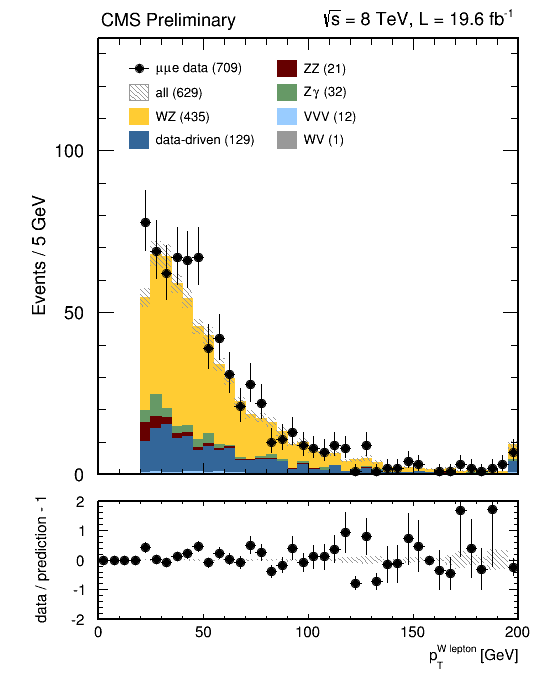}
	\end{subfigure}\quad
	\begin{subfigure}[b]{0.2\textwidth}
		\includegraphics[width=\textwidth,height=\textwidth]{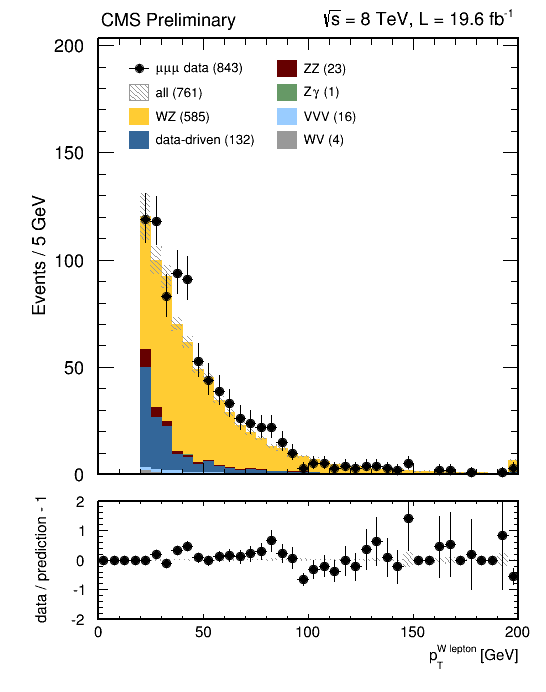}
	\end{subfigure}
	\vskip 1ex
	\begin{subfigure}[b]{0.2\textwidth}
		\includegraphics[width=\textwidth,height=\textwidth]{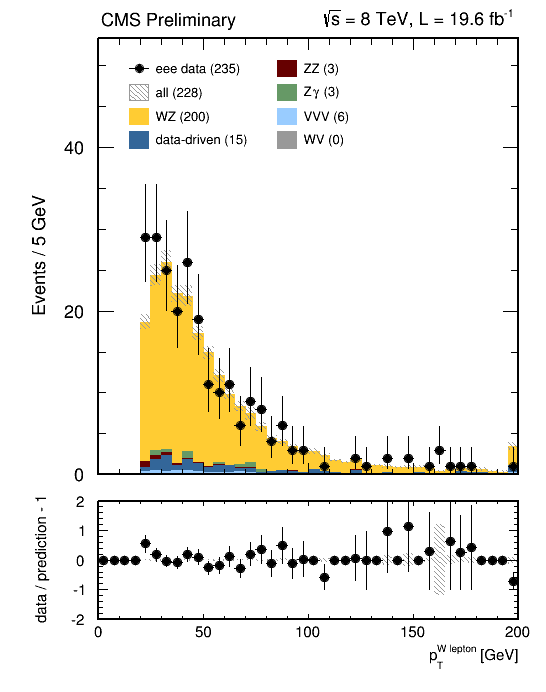}
	\end{subfigure}\quad
	\begin{subfigure}[b]{0.2\textwidth}
		\includegraphics[width=\textwidth,height=\textwidth]{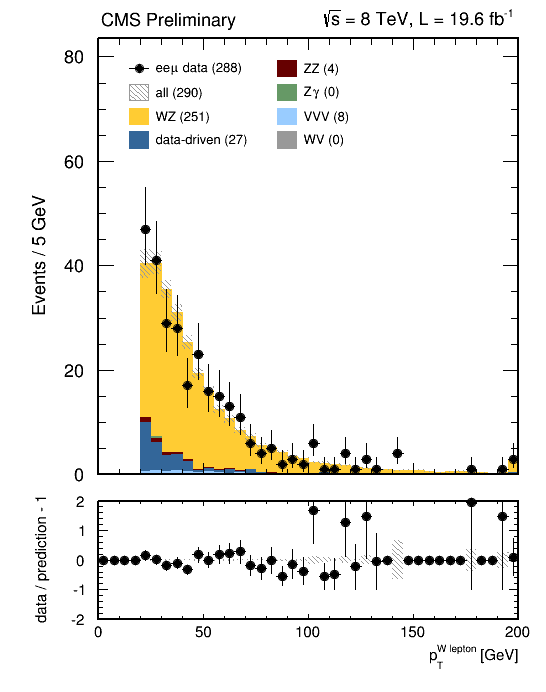}
	\end{subfigure}\quad
	\begin{subfigure}[b]{0.2\textwidth}
		\includegraphics[width=\textwidth,height=\textwidth]{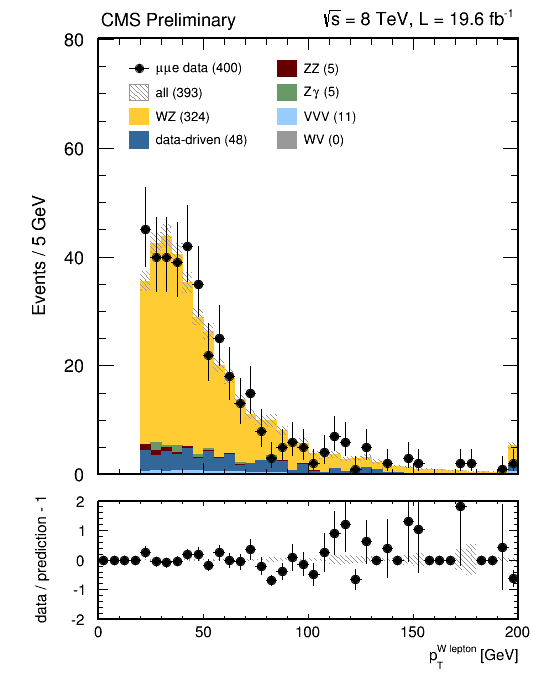}
	\end{subfigure}\quad
	\begin{subfigure}[b]{0.2\textwidth}
		\includegraphics[width=\textwidth,height=\textwidth]{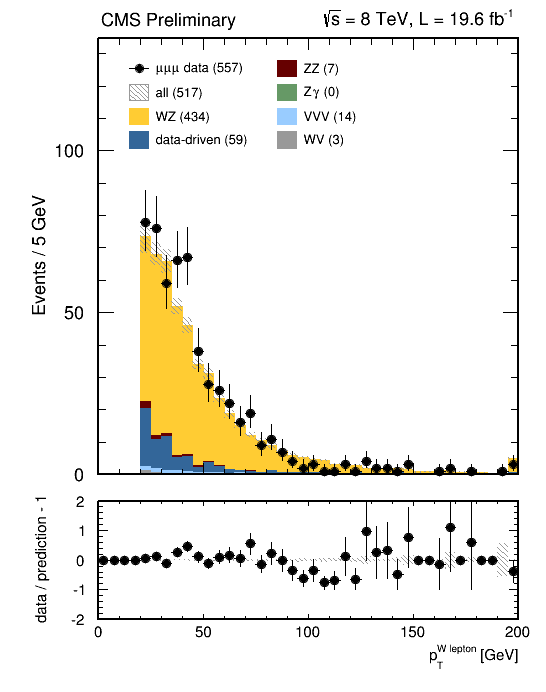}
	\end{subfigure}
	\caption[Transverse momentum of the W-candidate system at 8~\TeV]
	{Transverse momentum of the W-candidate system composed by
	the third selected lepton and \MET at each event for the measured channels $eee$,
	$\mu ee$, $e\mu\mu$ and $\mu\mu\mu$ 
	(from left to right) and after each analysis selection stage (once the W is selected): 
	after W-candidate requirement without the \MET cut (up row) and after W-candidate including
	\MET cut (bottom row).}
\end{sidewaysfigure}

\begin{sidewaysfigure}[!htpb]
	\centering
	\begin{subfigure}[b]{0.2\textwidth}
		\includegraphics[width=\textwidth,height=\textwidth]{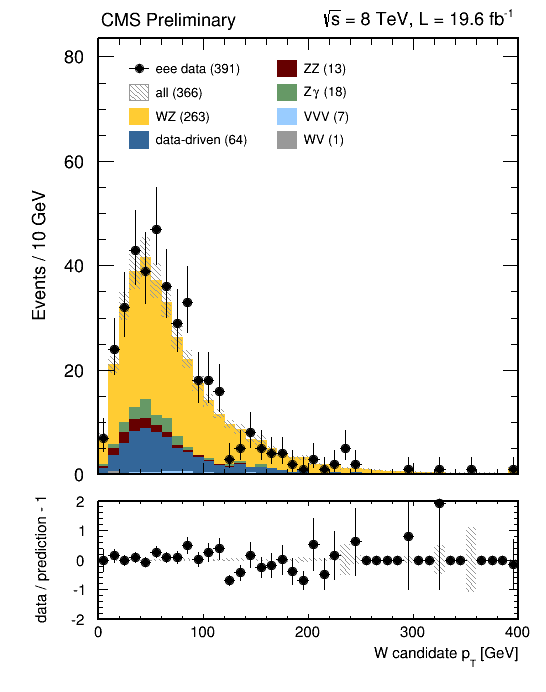}
	\end{subfigure}\quad
	\begin{subfigure}[b]{0.2\textwidth}
		\includegraphics[width=\textwidth,height=\textwidth]{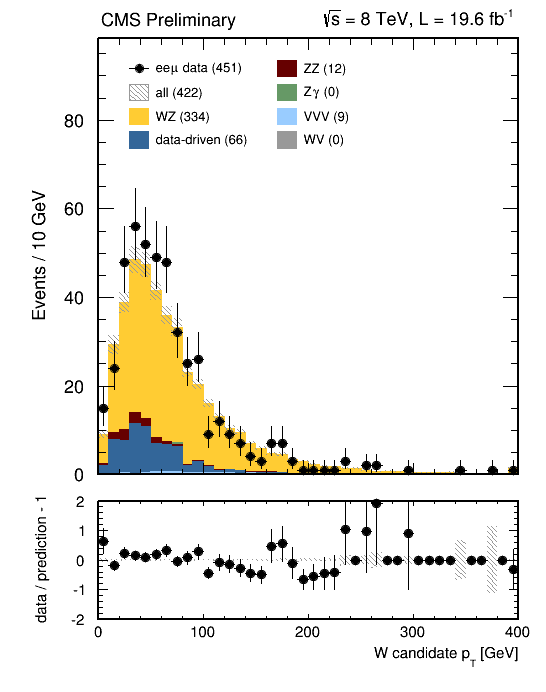}
	\end{subfigure}\quad
	\begin{subfigure}[b]{0.2\textwidth}
		\includegraphics[width=\textwidth,height=\textwidth]{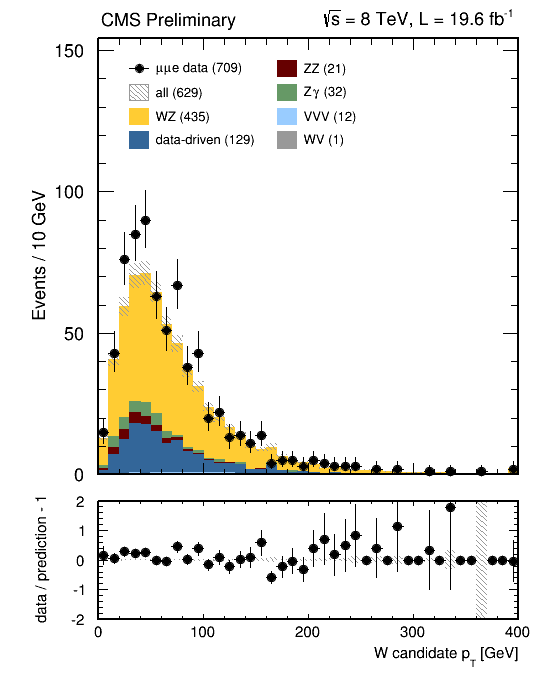}
	\end{subfigure}\quad
	\begin{subfigure}[b]{0.2\textwidth}
		\includegraphics[width=\textwidth,height=\textwidth]{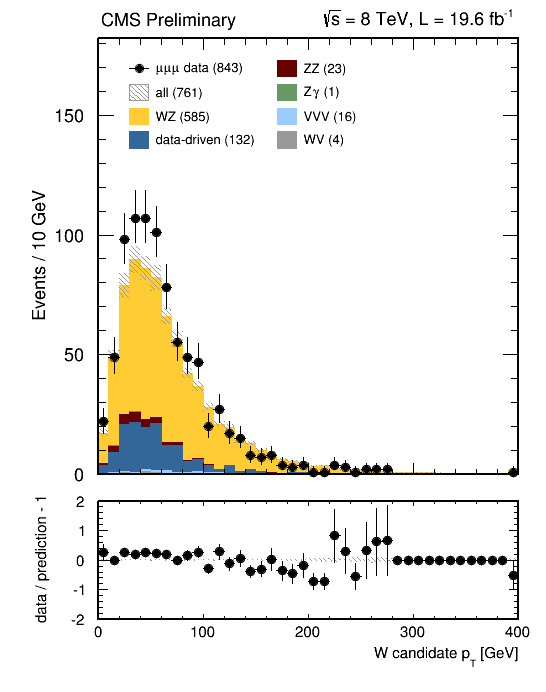}
	\end{subfigure}
	\vskip 1ex
	\begin{subfigure}[b]{0.2\textwidth}
		\includegraphics[width=\textwidth,height=\textwidth]{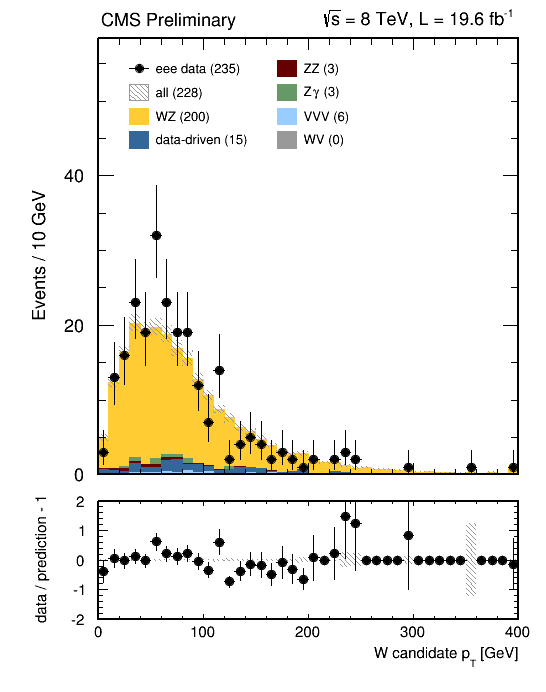}
	\end{subfigure}\quad
	\begin{subfigure}[b]{0.2\textwidth}
		\includegraphics[width=\textwidth,height=\textwidth]{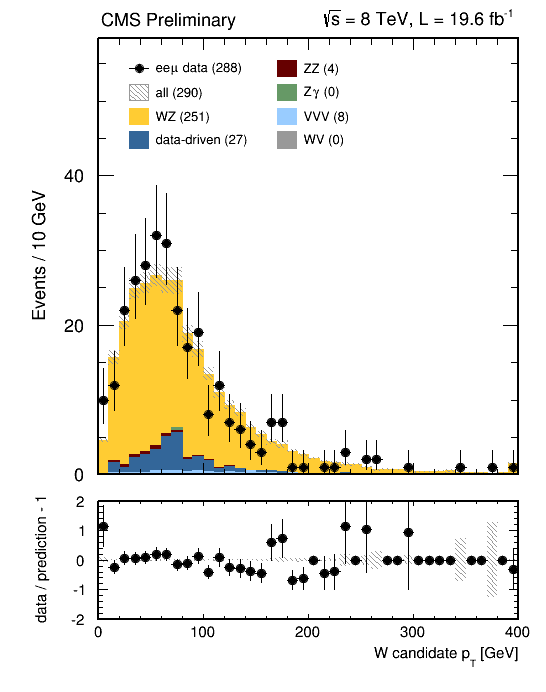}
	\end{subfigure}\quad
	\begin{subfigure}[b]{0.2\textwidth}
		\includegraphics[width=\textwidth,height=\textwidth]{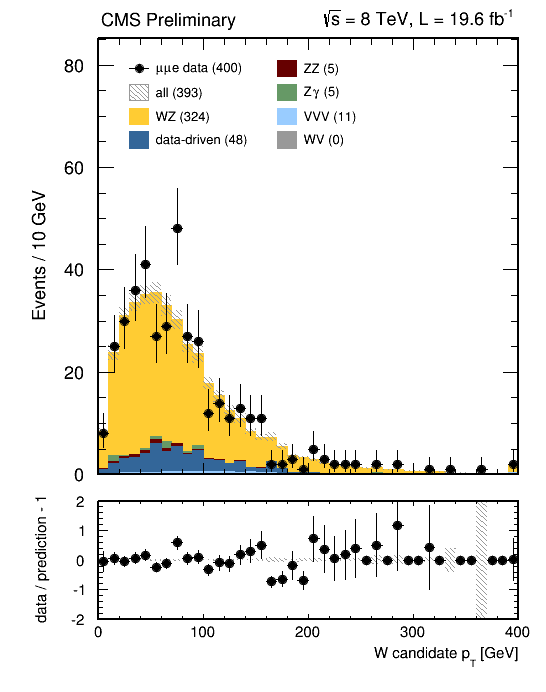}
	\end{subfigure}\quad
	\begin{subfigure}[b]{0.2\textwidth}
		\includegraphics[width=\textwidth,height=\textwidth]{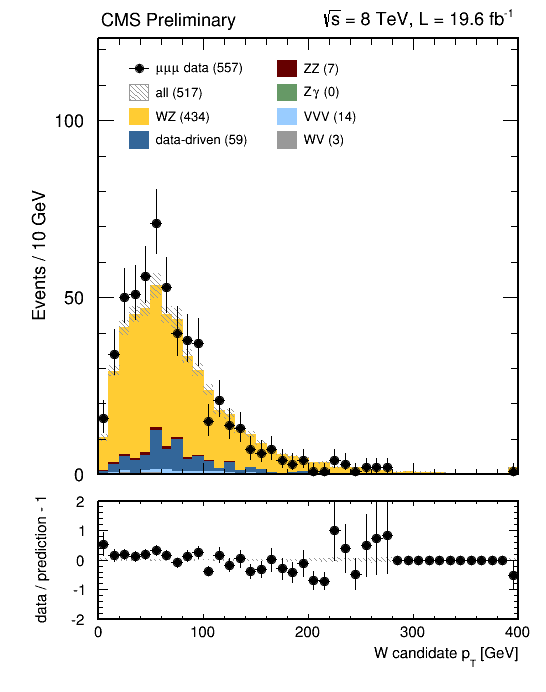}
	\end{subfigure}
	\caption[Transverse momentum of the W-candidate lepton at 8~\TeV]
	{Transverse momentum of the W-candidate lepton 
	at each event for the measured channels $eee$, $\mu ee$, $e\mu\mu$ and $\mu\mu\mu$ 
	(from left to right) and after each analysis selection stage (once the W is selected): 
	after W-candidate requirement without the \MET cut (up row) and after W-candidate including
	\MET cut (bottom row).}
\end{sidewaysfigure}

\begin{sidewaysfigure}[!htpb]
	\centering
	\begin{subfigure}[b]{0.2\textwidth}
		\includegraphics[width=\textwidth,height=\textwidth]{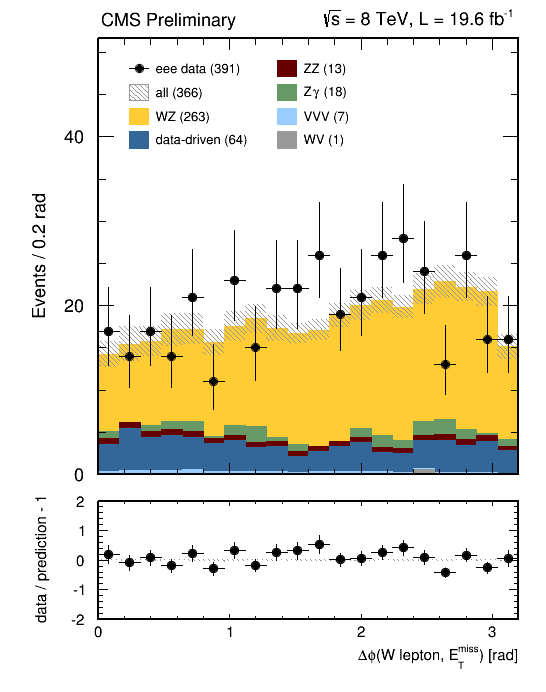}
	\end{subfigure}\quad
	\begin{subfigure}[b]{0.2\textwidth}
		\includegraphics[width=\textwidth,height=\textwidth]{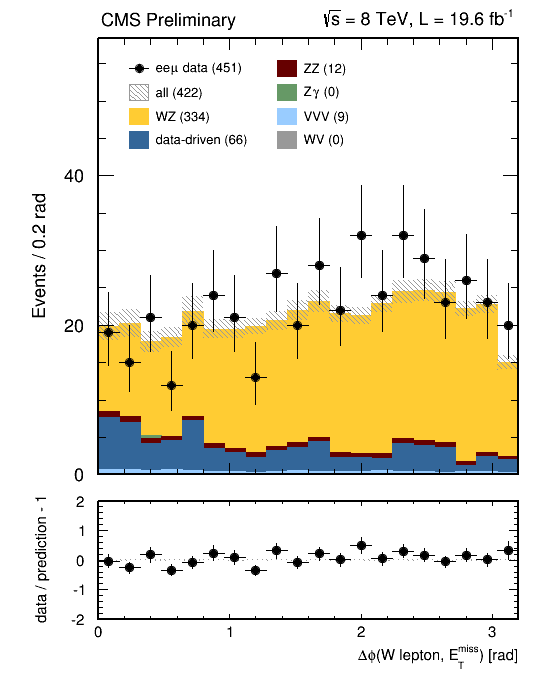}
	\end{subfigure}\quad
	\begin{subfigure}[b]{0.2\textwidth}
		\includegraphics[width=\textwidth,height=\textwidth]{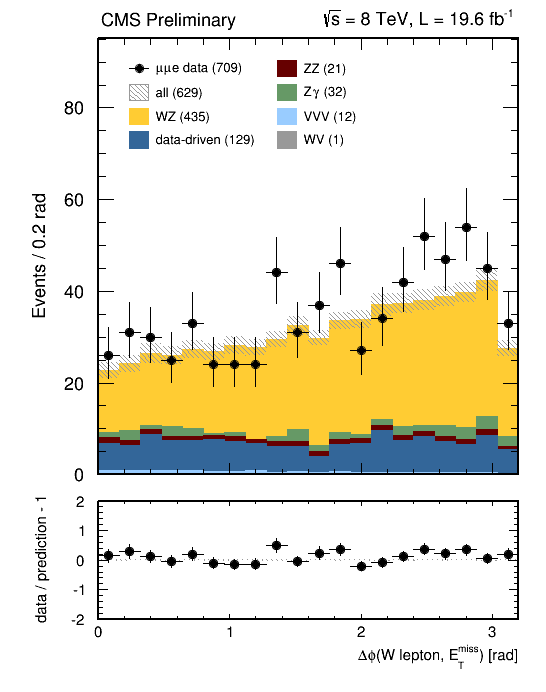}
	\end{subfigure}\quad
	\begin{subfigure}[b]{0.2\textwidth}
		\includegraphics[width=\textwidth,height=\textwidth]{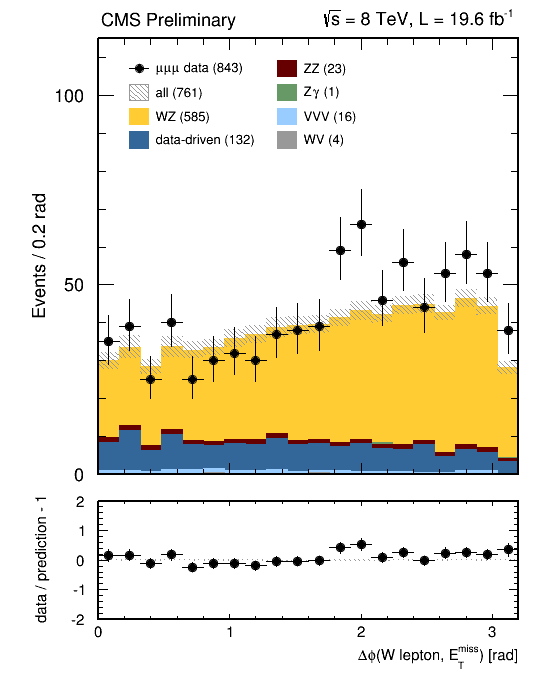}
	\end{subfigure}
	\vskip 1ex
	\begin{subfigure}[b]{0.2\textwidth}
		\includegraphics[width=\textwidth,height=\textwidth]{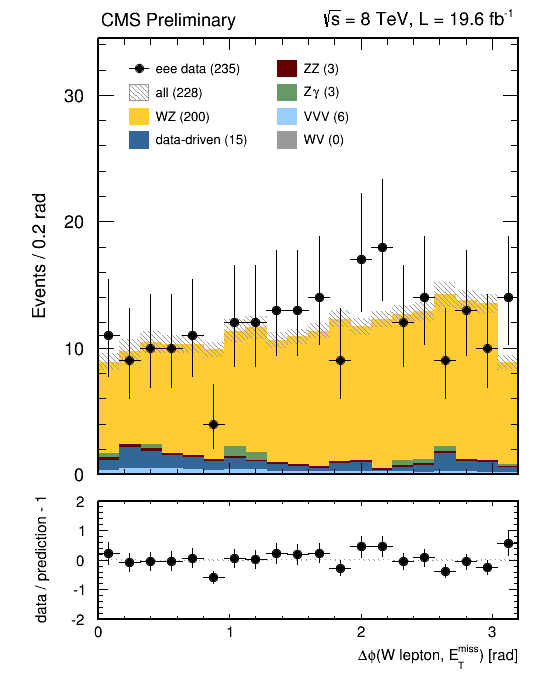}
	\end{subfigure}\quad
	\begin{subfigure}[b]{0.2\textwidth}
		\includegraphics[width=\textwidth,height=\textwidth]{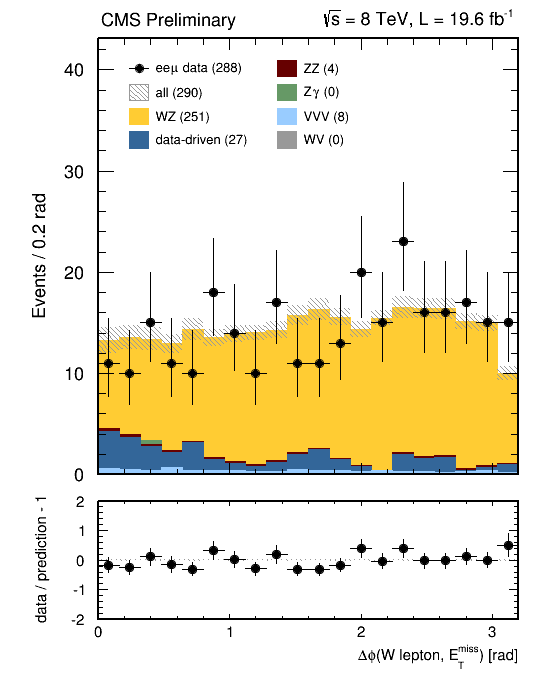}
	\end{subfigure}\quad
	\begin{subfigure}[b]{0.2\textwidth}
		\includegraphics[width=\textwidth,height=\textwidth]{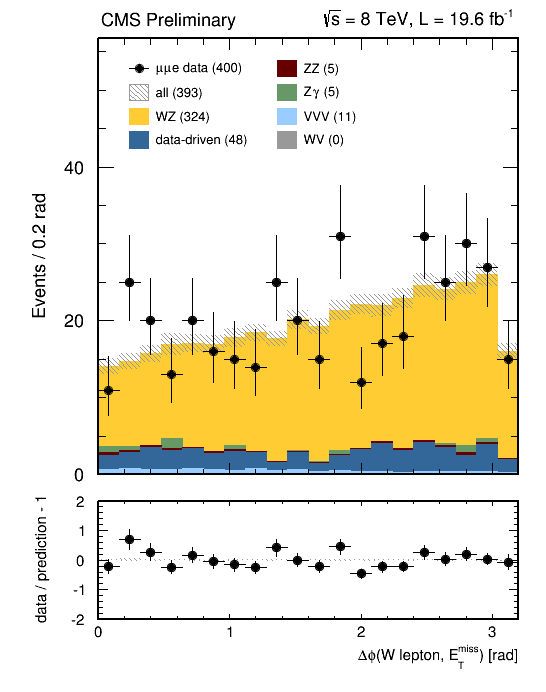}
	\end{subfigure}\quad
	\begin{subfigure}[b]{0.2\textwidth}
		\includegraphics[width=\textwidth,height=\textwidth]{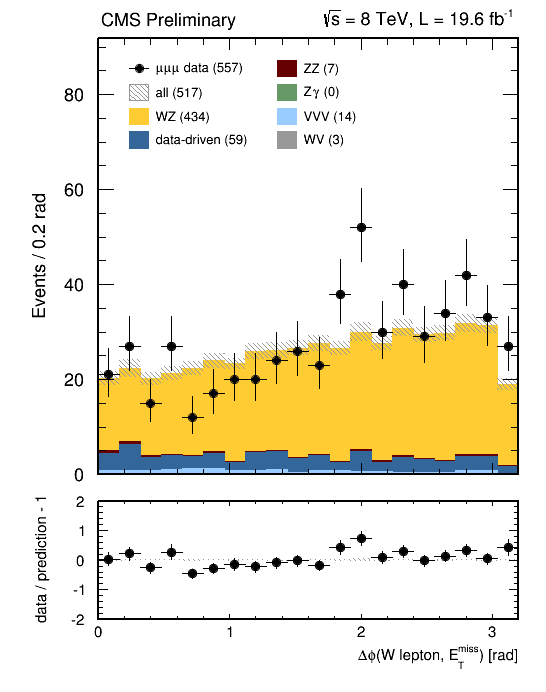}
	\end{subfigure}
	\caption[Azimuthal angle between W-candidate lepton and \MET at 8~\TeV]
	{Azimuthal angle between the W-candidate lepton and the \MET
	at each event for the measured channels $eee$, $\mu ee$, $e\mu\mu$ and $\mu\mu\mu$ 
	(from left to right) and after each analysis selection stage (once the W is selected): 
	after W-candidate requirement without the \MET cut (up row) and after W-candidate including
	\MET cut (bottom row).}
\end{sidewaysfigure}

\begin{sidewaysfigure}[!htpb]
	\centering
	\begin{subfigure}[b]{0.2\textwidth}
		\includegraphics[width=\textwidth,height=\textwidth]{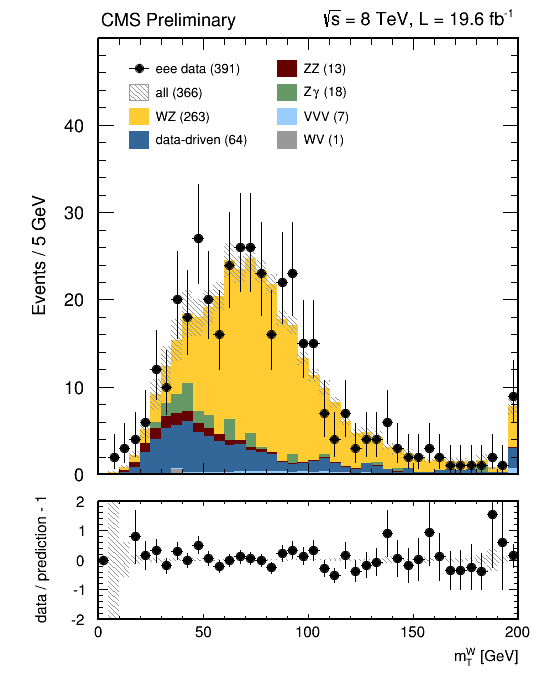}
	\end{subfigure}\quad
	\begin{subfigure}[b]{0.2\textwidth}
		\includegraphics[width=\textwidth,height=\textwidth]{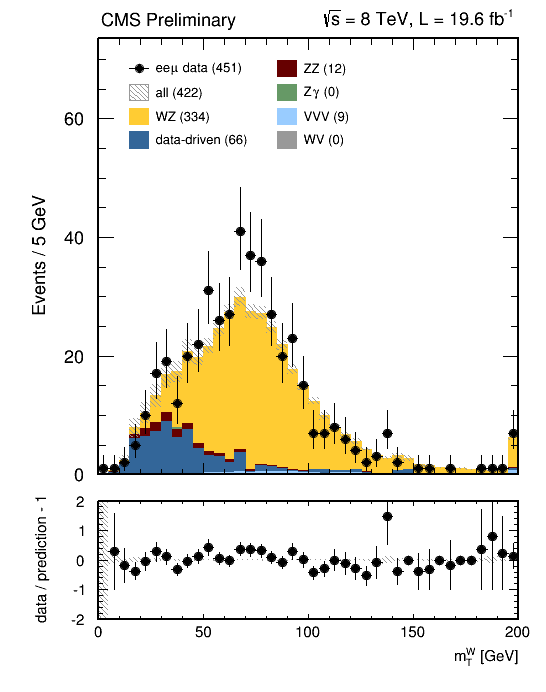}
	\end{subfigure}\quad
	\begin{subfigure}[b]{0.2\textwidth}
		\includegraphics[width=\textwidth,height=\textwidth]{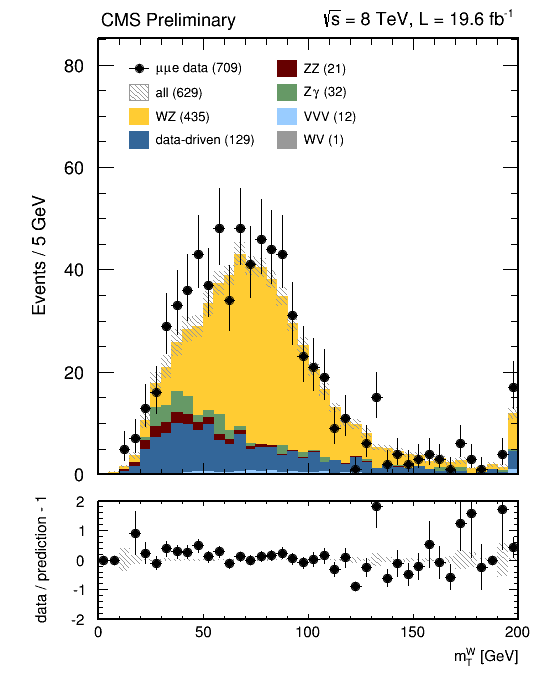}
	\end{subfigure}\quad
	\begin{subfigure}[b]{0.2\textwidth}
		\includegraphics[width=\textwidth,height=\textwidth]{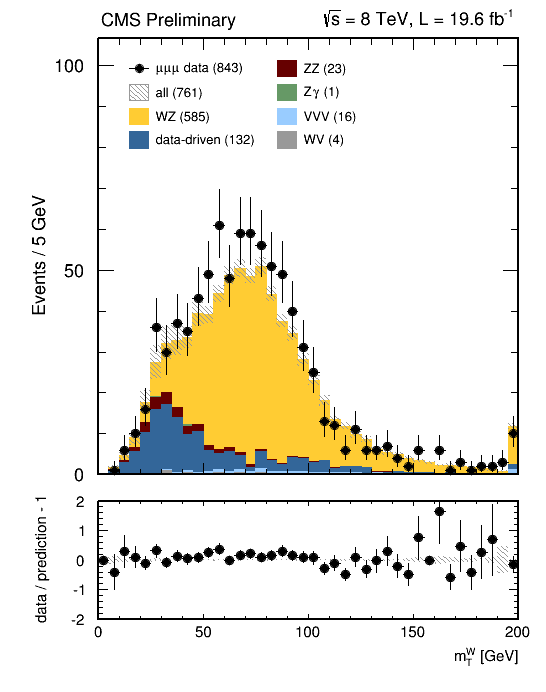}
	\end{subfigure}
	\vskip 1ex
	\begin{subfigure}[b]{0.2\textwidth}
		\includegraphics[width=\textwidth,height=\textwidth]{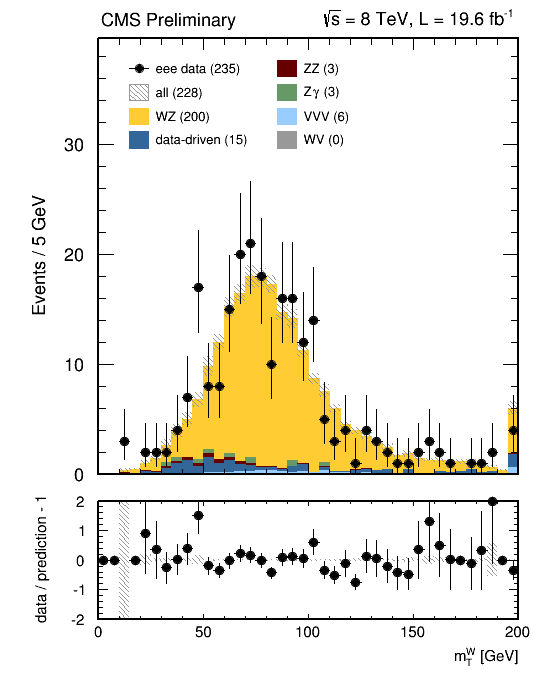}
	\end{subfigure}\quad
	\begin{subfigure}[b]{0.2\textwidth}
		\includegraphics[width=\textwidth,height=\textwidth]{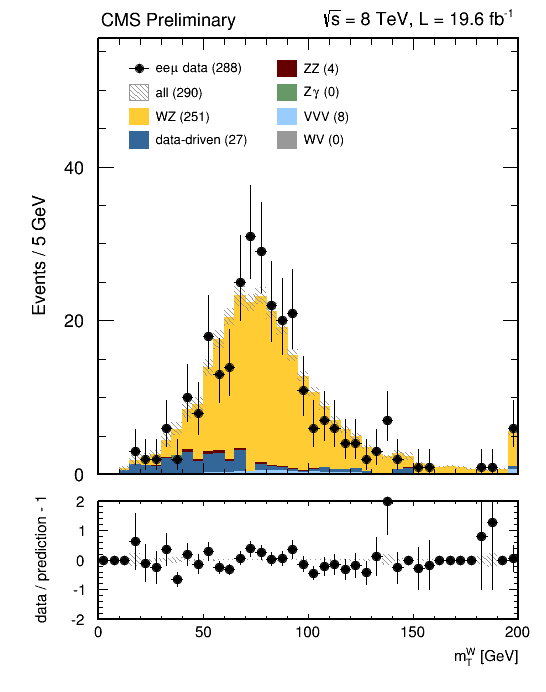}
	\end{subfigure}\quad
	\begin{subfigure}[b]{0.2\textwidth}
		\includegraphics[width=\textwidth,height=\textwidth]{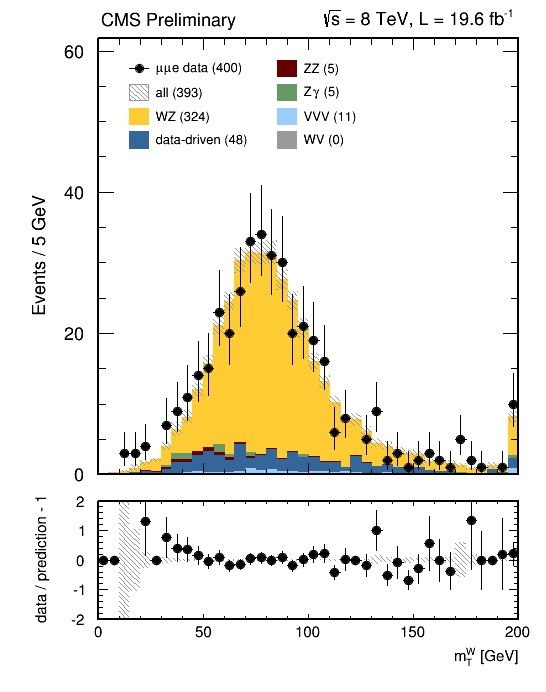}
	\end{subfigure}\quad
	\begin{subfigure}[b]{0.2\textwidth}
		\includegraphics[width=\textwidth,height=\textwidth]{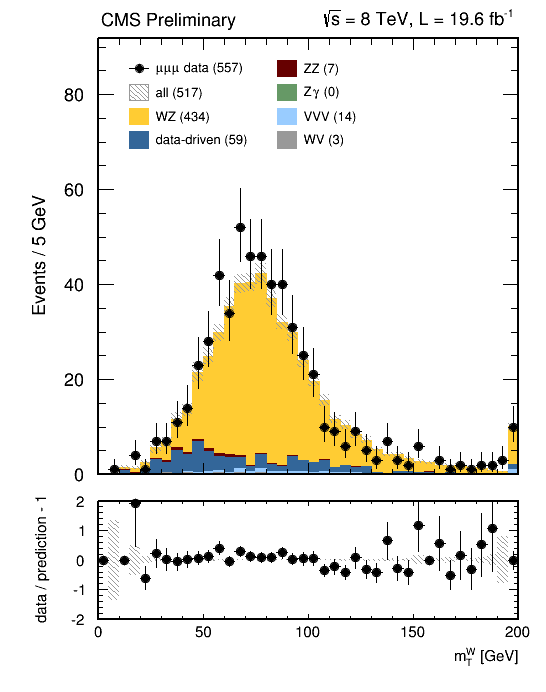}
	\end{subfigure}
	\caption[Transverse mass of the W-candidate lepton and \MET at 8~\TeV]
	{Transverse mass of the W-candidate lepton and the \MET
	at each event for the measured channels $eee$, $\mu ee$, $e\mu\mu$ and $\mu\mu\mu$ 
	(from left to right) and after each analysis selection stage (once the W is selected): 
	after W-candidate requirement without the \MET cut (up row) and after W-candidate including
	\MET cut (bottom row).}
\end{sidewaysfigure}

\begin{sidewaysfigure}[!htpb]
	\centering
	\begin{subfigure}[b]{0.2\textwidth}
		\includegraphics[width=\textwidth,height=\textwidth]{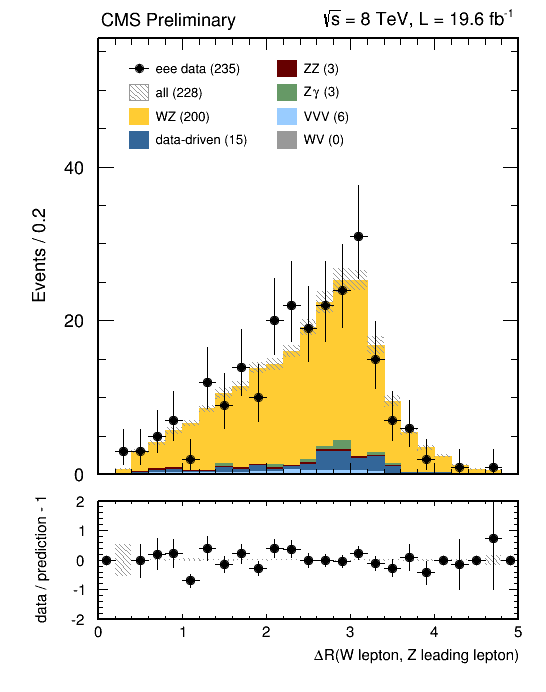}
	\end{subfigure}\quad
	\begin{subfigure}[b]{0.2\textwidth}
		\includegraphics[width=\textwidth,height=\textwidth]{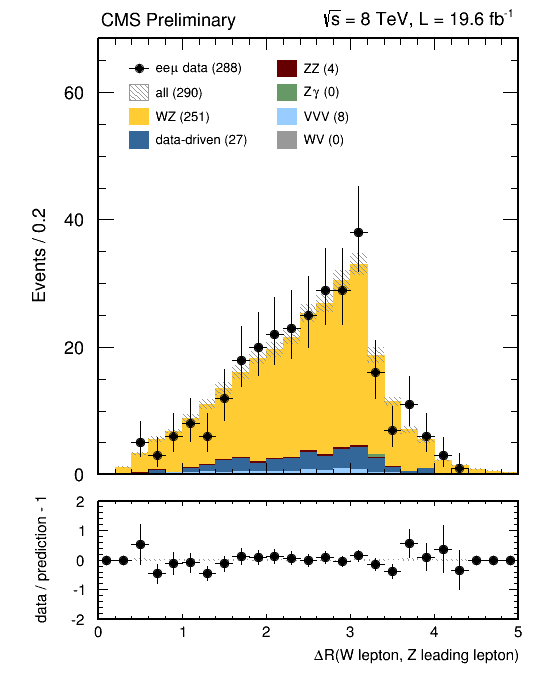}
	\end{subfigure}\quad
	\begin{subfigure}[b]{0.2\textwidth}
		\includegraphics[width=\textwidth,height=\textwidth]{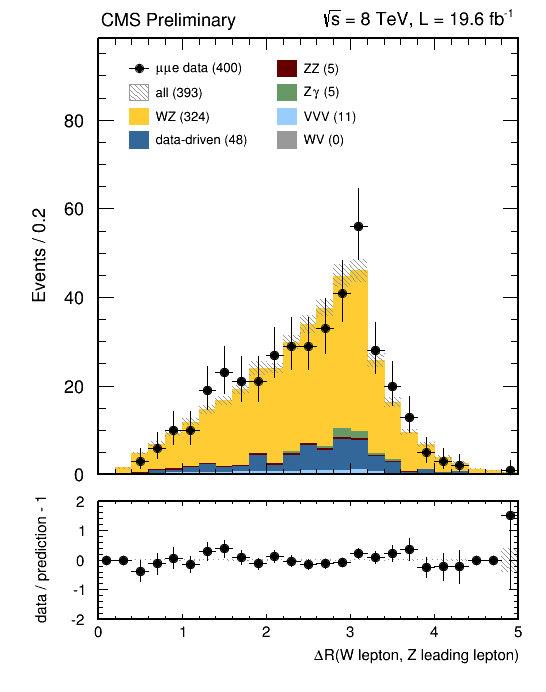}
	\end{subfigure}\quad
	\begin{subfigure}[b]{0.2\textwidth}
		\includegraphics[width=\textwidth,height=\textwidth]{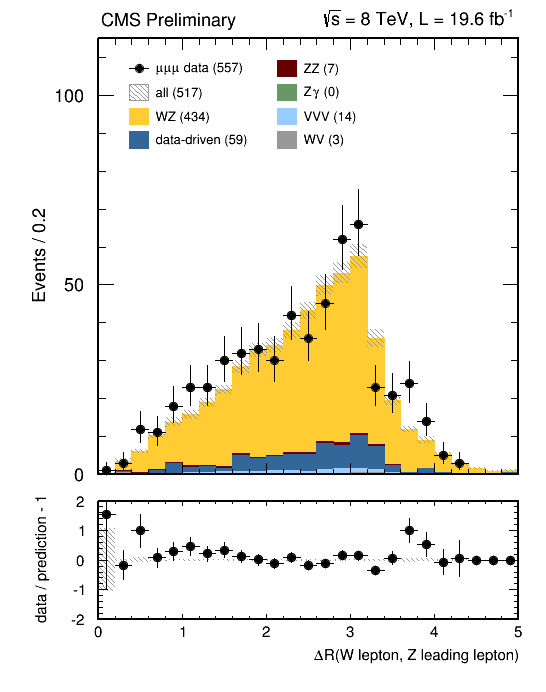}
	\end{subfigure}
	\caption[Angular distance between W-candidate lepton and Z-candidate leading lepton at 8~\TeV]
	{Angular distance between the W-candidate lepton and the Z-candidate
	leading lepton at each event for the measured channels $eee$, $\mu ee$, $e\mu\mu$ and 
	$\mu\mu\mu$ (from left to right) after W-candidate requirement is applied.}
	\vskip 1em
	\centering
	\begin{subfigure}[b]{0.2\textwidth}
		\includegraphics[width=\textwidth,height=\textwidth]{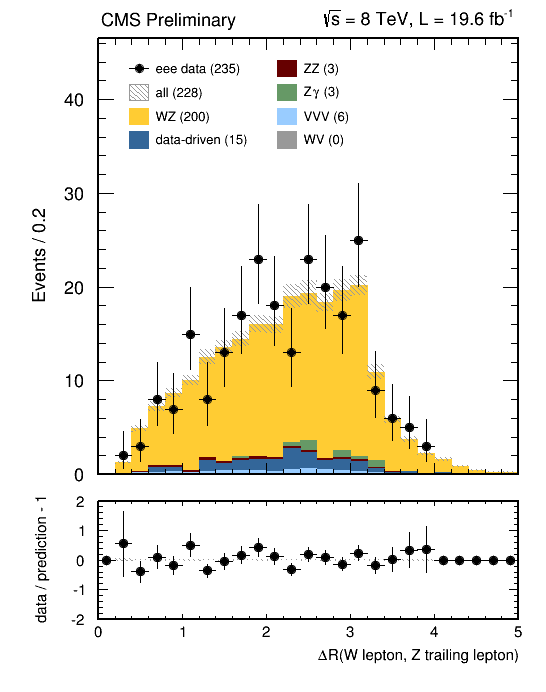}
	\end{subfigure}\quad
	\begin{subfigure}[b]{0.2\textwidth}
		\includegraphics[width=\textwidth,height=\textwidth]{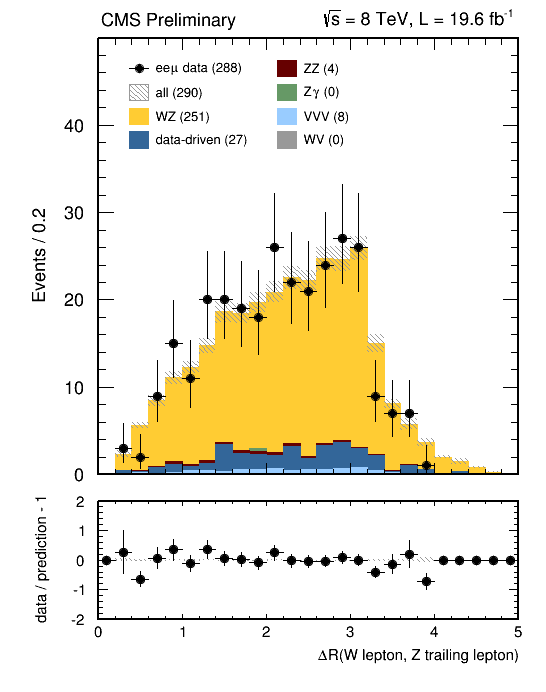}
	\end{subfigure}\quad
	\begin{subfigure}[b]{0.2\textwidth}
		\includegraphics[width=\textwidth,height=\textwidth]{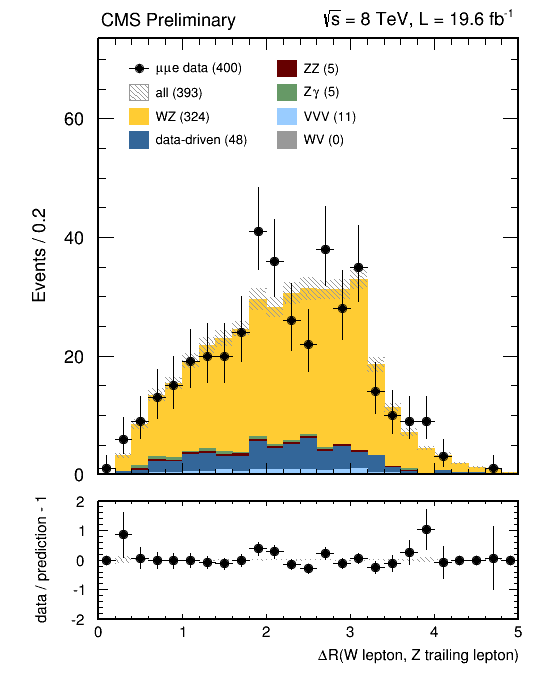}
	\end{subfigure}\quad
	\begin{subfigure}[b]{0.2\textwidth}
		\includegraphics[width=\textwidth,height=\textwidth]{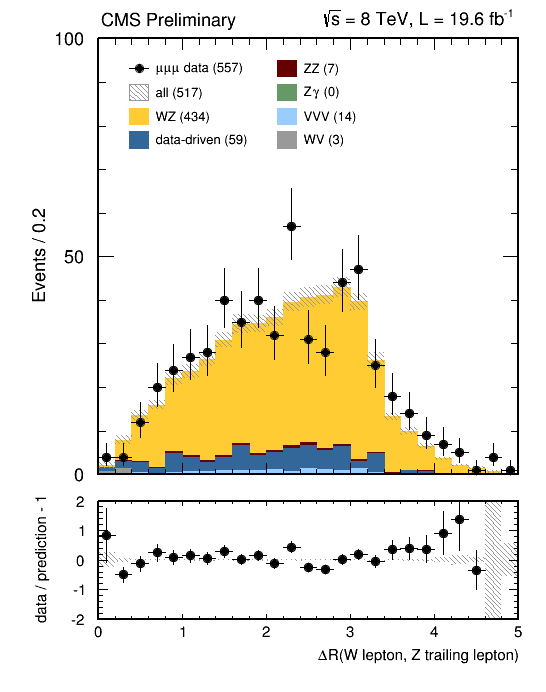}
	\end{subfigure}
	\caption[Angular distance between W-candidate lepton and Z-candidate trailing lepton at 8~\TeV]
	{Angular distance between the W-candidate lepton and the Z-candidate
	trailing lepton at each event for the measured channels $eee$, $\mu ee$, $e\mu\mu$ and 
	$\mu\mu\mu$ (from left to right) after W-candidate requirement is applied.}
\end{sidewaysfigure}

\begin{sidewaysfigure}[!htpb]
	\centering
	\begin{subfigure}[b]{0.2\textwidth}
		\includegraphics[width=\textwidth,height=\textwidth]{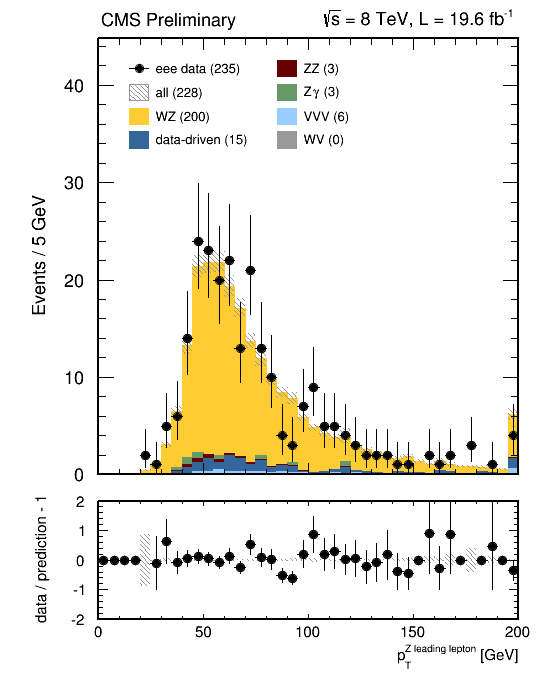}
	\end{subfigure}\quad
	\begin{subfigure}[b]{0.2\textwidth}
		\includegraphics[width=\textwidth,height=\textwidth]{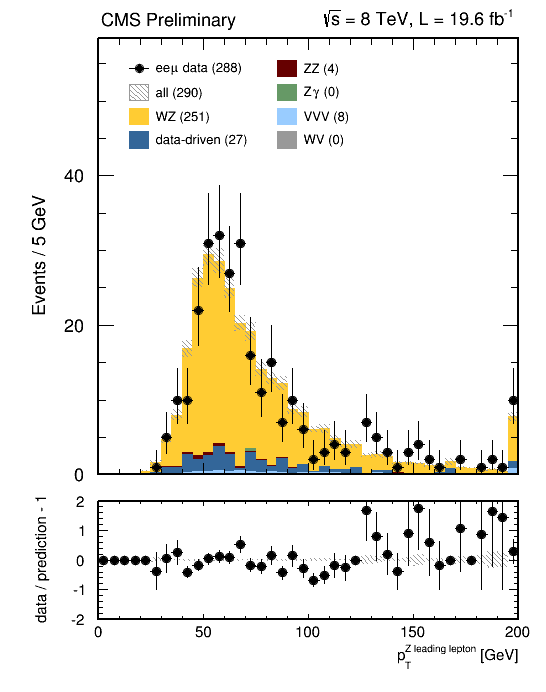}
	\end{subfigure}\quad
	\begin{subfigure}[b]{0.2\textwidth}
		\includegraphics[width=\textwidth,height=\textwidth]{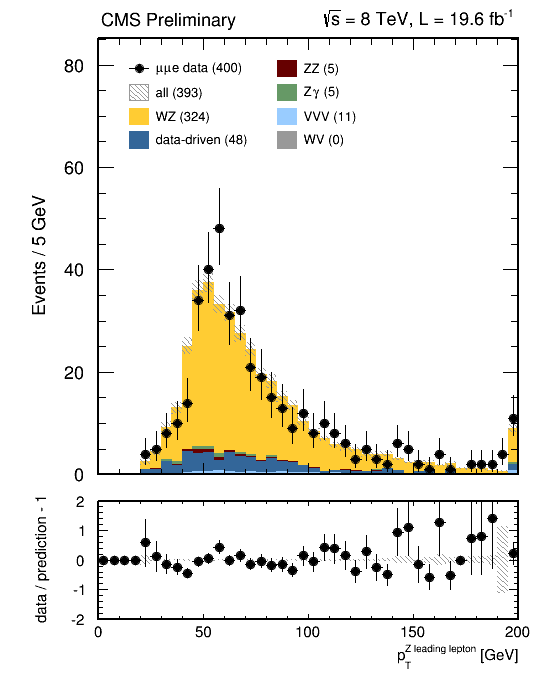}
	\end{subfigure}\quad
	\begin{subfigure}[b]{0.2\textwidth}
		\includegraphics[width=\textwidth,height=\textwidth]{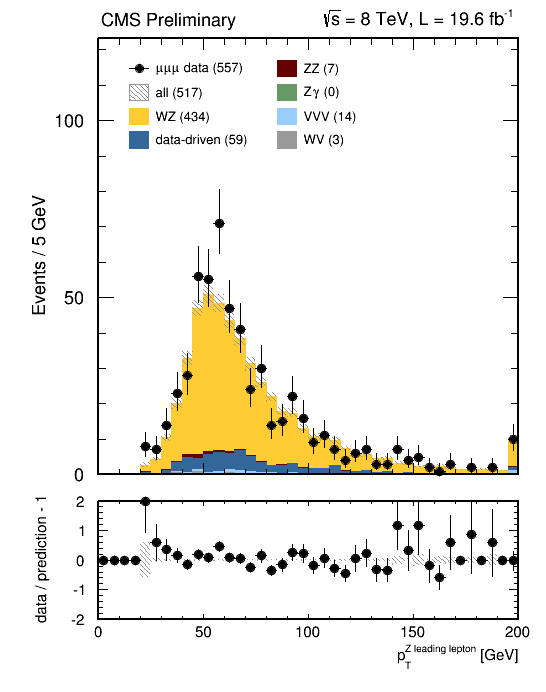}
	\end{subfigure}
	\caption[Transverse momentum of the Z-candidate leading lepton at 8~\TeV]
	{Transverse momentum of the Z-candidate
	leading lepton at each event for the measured channels $eee$, $\mu ee$, $e\mu\mu$ and 
	$\mu\mu\mu$ (from left to right) after W-candidate requirement is applied.}
	\vskip 1em
	\centering
	\begin{subfigure}[b]{0.2\textwidth}
		\includegraphics[width=\textwidth,height=\textwidth]{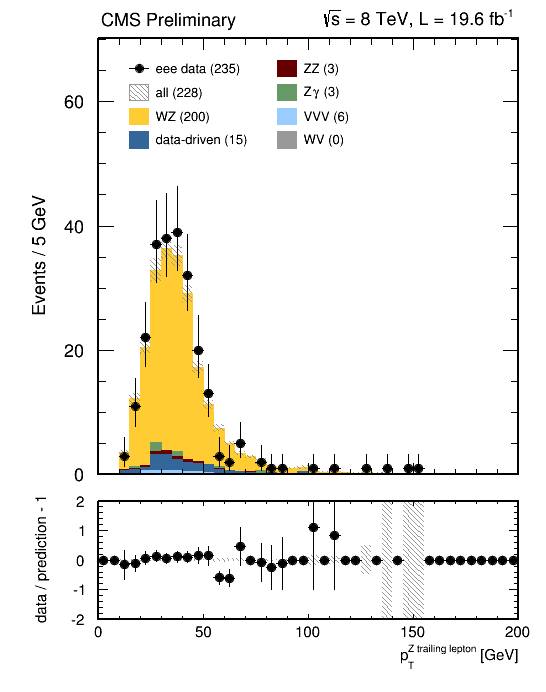}
	\end{subfigure}\quad
	\begin{subfigure}[b]{0.2\textwidth}
		\includegraphics[width=\textwidth,height=\textwidth]{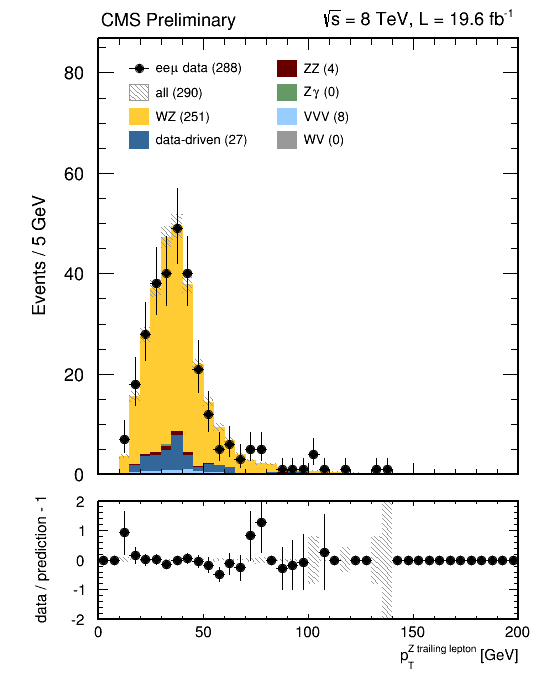}
	\end{subfigure}\quad
	\begin{subfigure}[b]{0.2\textwidth}
		\includegraphics[width=\textwidth,height=\textwidth]{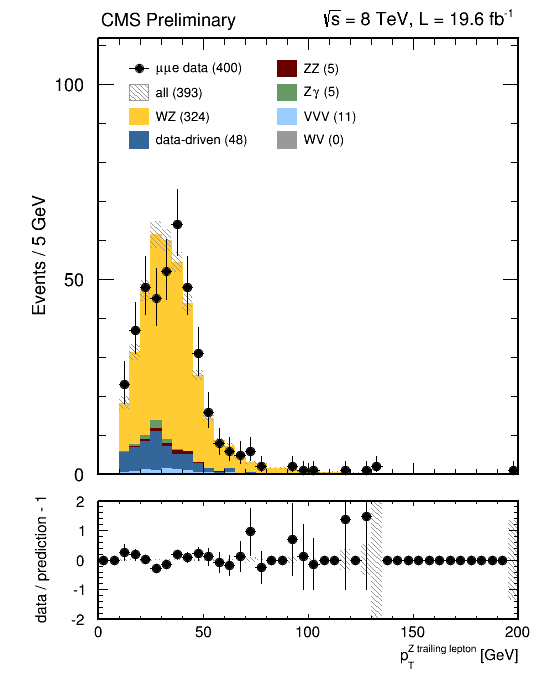}
	\end{subfigure}\quad
	\begin{subfigure}[b]{0.2\textwidth}
		\includegraphics[width=\textwidth,height=\textwidth]{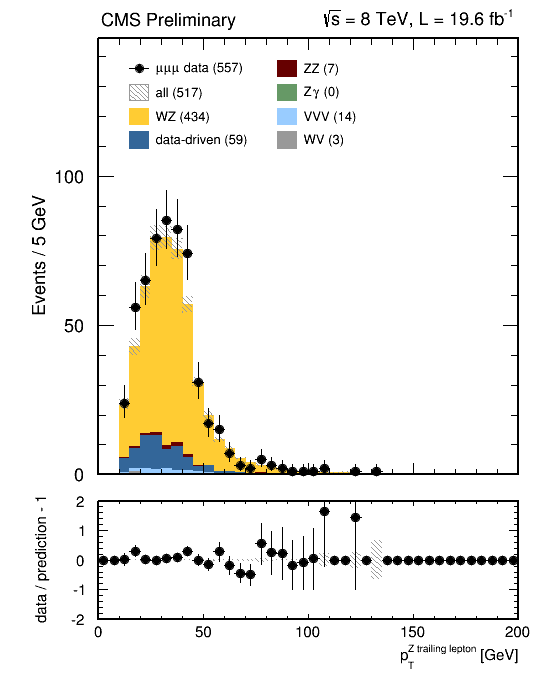}
	\end{subfigure}
	\caption[Transverse momentum of the Z-candidate trailing lepton at 8~\TeV]
	{Transverse momentum of the Z-candidate
	trailing lepton at each event for the measured channels $eee$, $\mu ee$, $e\mu\mu$ and 
	$\mu\mu\mu$ (from left to right) after W-candidate requirement is applied.}
\end{sidewaysfigure}
\clearpage

\section{Ratio analysis distributions at 8~\TeV}
The distributions shown in this subsection correspond to the 2012 analysis of the cross section
ratio between the \wzp and \wzm processes. The samples follows the colour conventions and the processing
explained in the previous sections. The distributions are grouped by observable, showing in each figure 
two columns corresponding to the combined channel of the \wzm and \wzp and each row to a stage of the 
analysis. Note that since the sample splitting by charge lies in the \W-candidate, the analysis stages 
are shown from this requirement on.
\begin{figure}[!htpb]
	\centering
	\begin{subfigure}[b]{0.4\textwidth}
		\includegraphics[width=\textwidth,height=\textwidth]{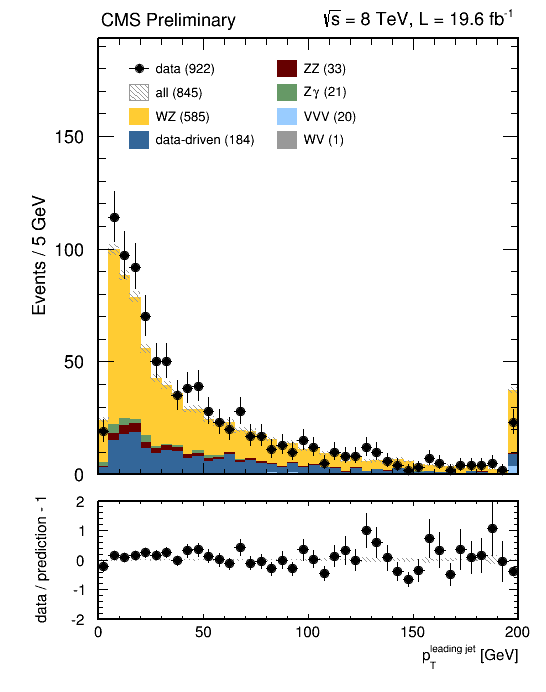}
	\end{subfigure}\quad
	\begin{subfigure}[b]{0.4\textwidth}
		\includegraphics[width=\textwidth,height=\textwidth]{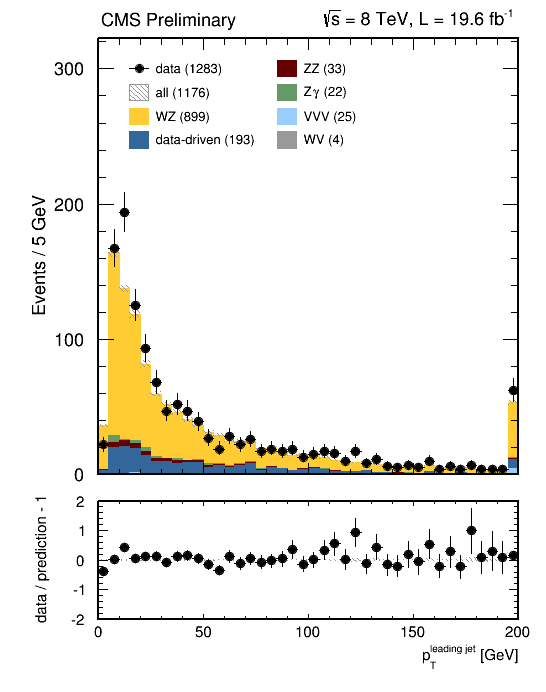}
	\end{subfigure}
	\vskip 1ex
	\begin{subfigure}[b]{0.4\textwidth}
		\includegraphics[width=\textwidth,height=\textwidth]{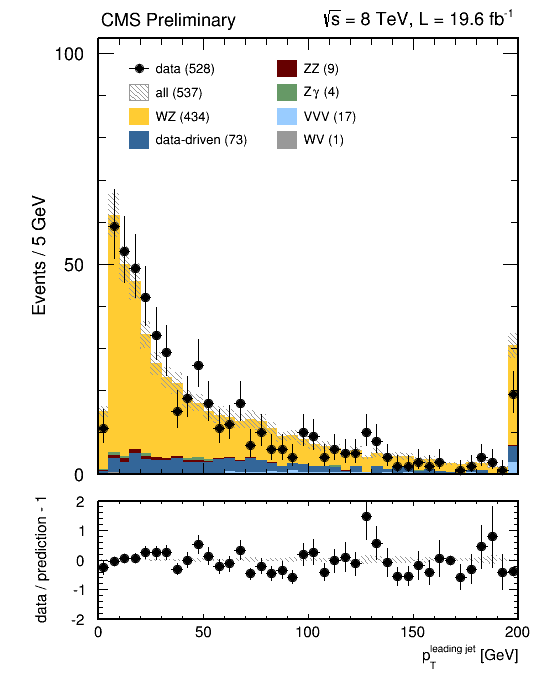}
	\end{subfigure}\quad
	\begin{subfigure}[b]{0.4\textwidth}
		\includegraphics[width=\textwidth,height=\textwidth]{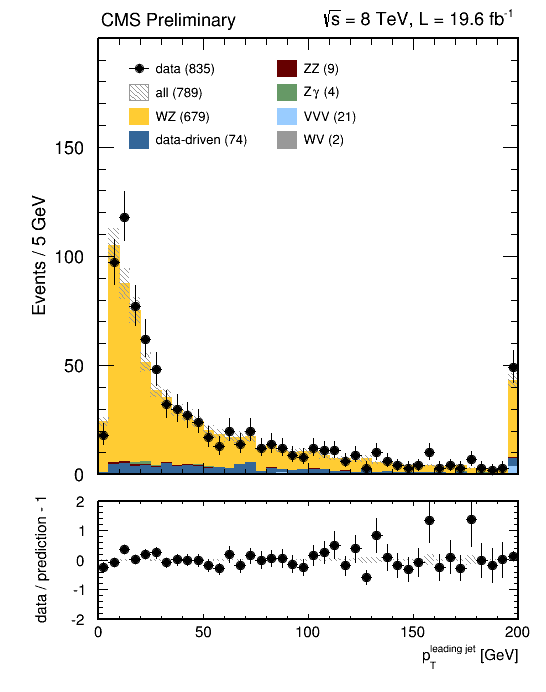}
	\end{subfigure}
	\caption[Transverse momentum of the leading jet at 8 TeV (ratio)]{Transverse 
	momentum distribution of the leading jet at each event for the \wzm (left column) and
	\wzp (right column) before the \MET cut (up row) and after (bottom row).}
\end{figure}

\begin{figure}
	\centering
	\begin{subfigure}[b]{0.3\textwidth}
		\includegraphics[width=\textwidth,height=\textwidth]{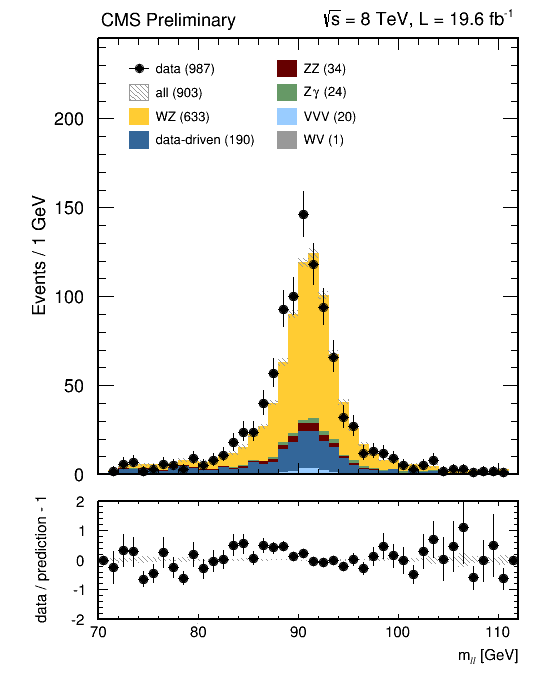}
	\end{subfigure}\quad
	\begin{subfigure}[b]{0.3\textwidth}
		\includegraphics[width=\textwidth,height=\textwidth]{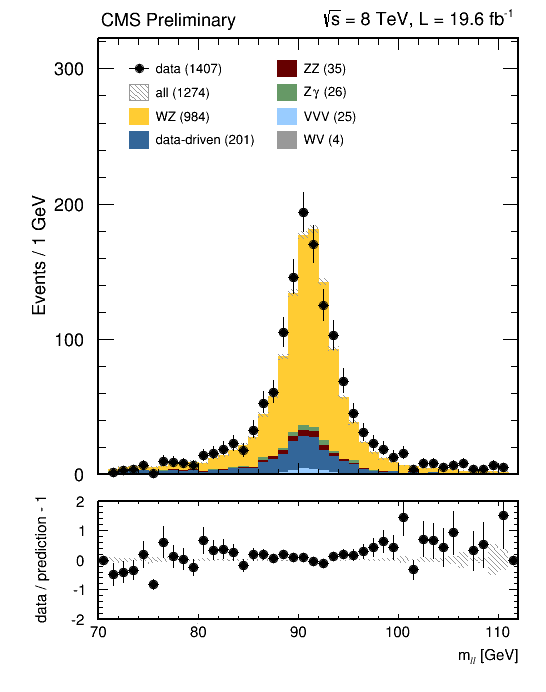}
	\end{subfigure}
	\vskip 1ex
	\begin{subfigure}[b]{0.3\textwidth}
		\includegraphics[width=\textwidth,height=\textwidth]{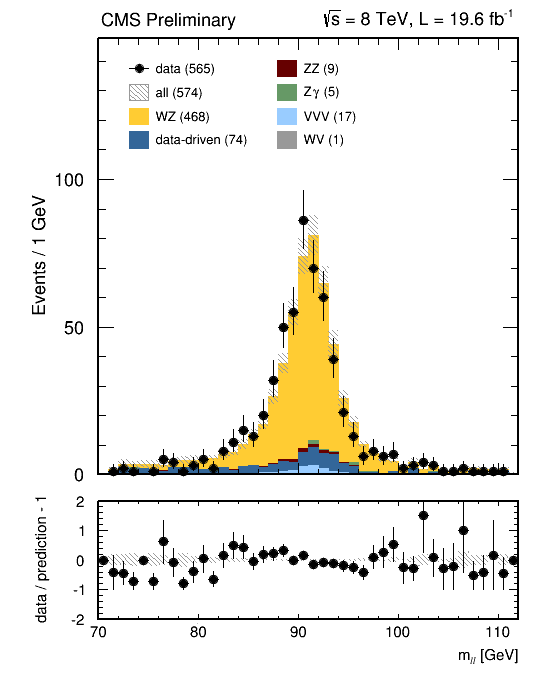}
	\end{subfigure}\quad
	\begin{subfigure}[b]{0.3\textwidth}
		\includegraphics[width=\textwidth,height=\textwidth]{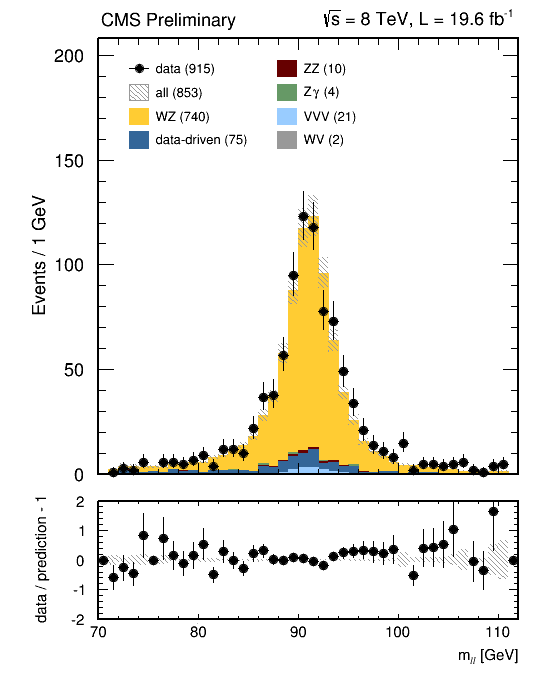}
	\end{subfigure}
	\caption[Invariant mass of the dilepton system at 8~\TeV (ratio)]
	{Invariant mass of the Z-candidate dilepton system for the \wzm (left column) and \wzp 
	(right column) before (up row) and after the \MET cut (bottom row).}
	\vskip 1em
	\begin{subfigure}[b]{0.3\textwidth}
		\includegraphics[width=\textwidth,height=\textwidth]{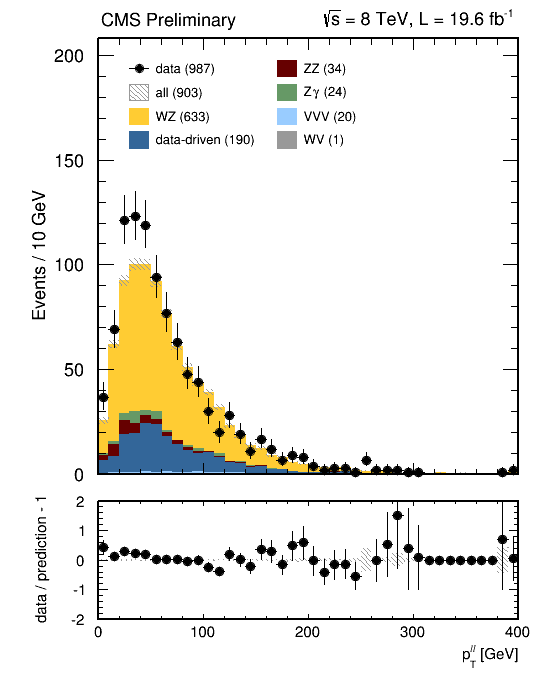}
	\end{subfigure}\quad
	\begin{subfigure}[b]{0.3\textwidth}
		\includegraphics[width=\textwidth,height=\textwidth]{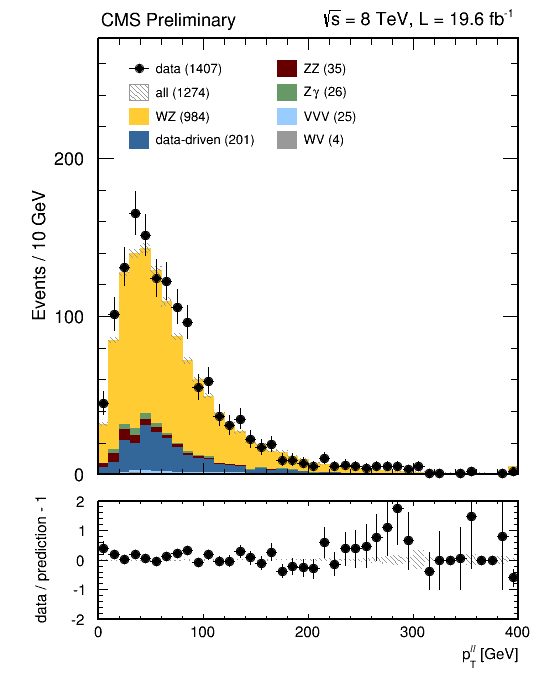}
	\end{subfigure}
	\vskip 1ex
	\begin{subfigure}[b]{0.3\textwidth}
		\includegraphics[width=\textwidth,height=\textwidth]{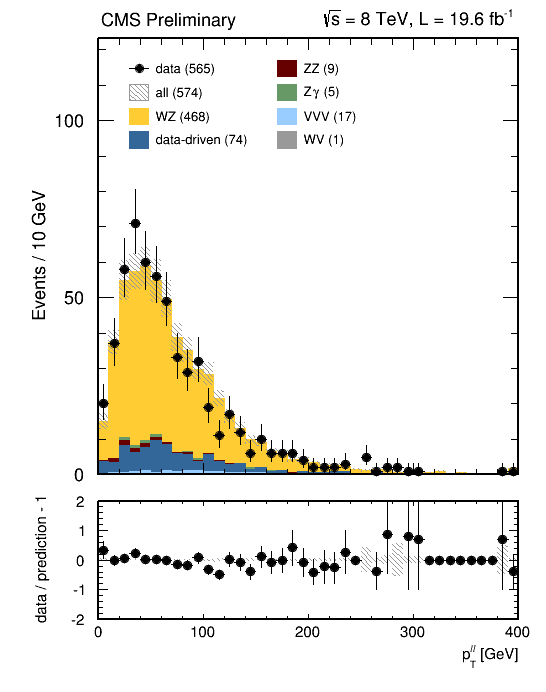}
	\end{subfigure}\quad
	\begin{subfigure}[b]{0.3\textwidth}
		\includegraphics[width=\textwidth,height=\textwidth]{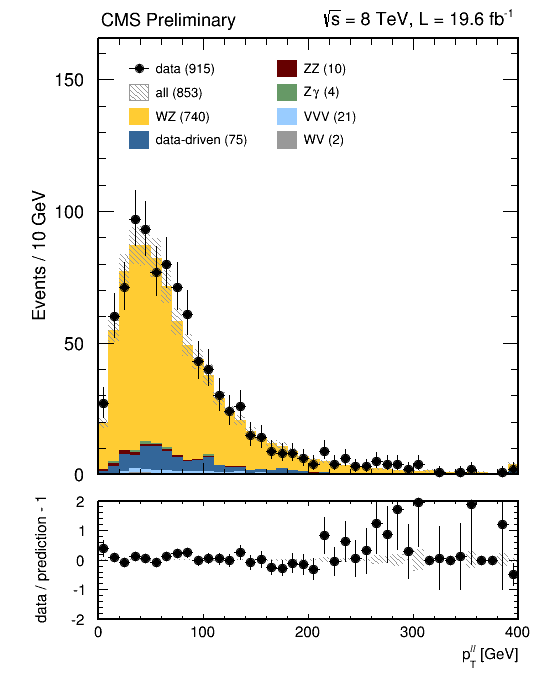}
	\end{subfigure}
	\caption[Transverse momentum of the dilepton system at 8~\TeV (ratio)]
	{Transverse momentum of the Z-candidate dilepton system for the \wzm (left column) and \wzp 
	(right column) before (up row) and after the \MET cut (bottom row).}
\end{figure}

\begin{figure}[!htpb]
	\centering
	\begin{subfigure}[b]{0.3\textwidth}
		\includegraphics[width=\textwidth,height=\textwidth]{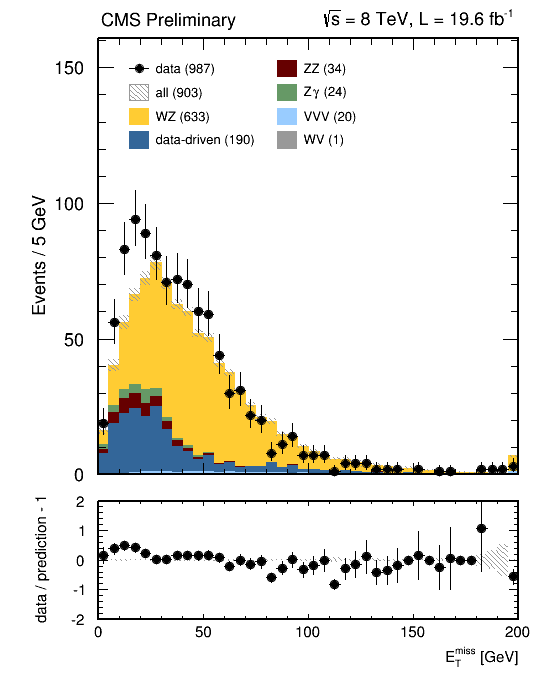}
	\end{subfigure}\quad
	\begin{subfigure}[b]{0.3\textwidth}
		\includegraphics[width=\textwidth,height=\textwidth]{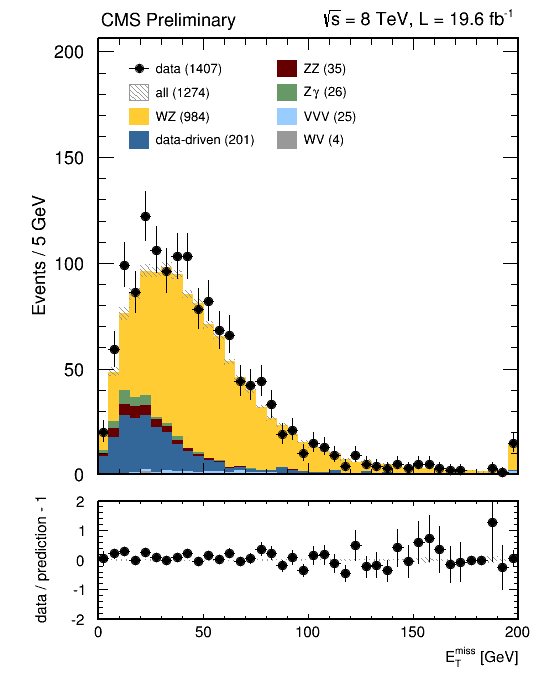}
	\end{subfigure}
	\vskip 1ex
	\begin{subfigure}[b]{0.3\textwidth}
		\includegraphics[width=\textwidth,height=\textwidth]{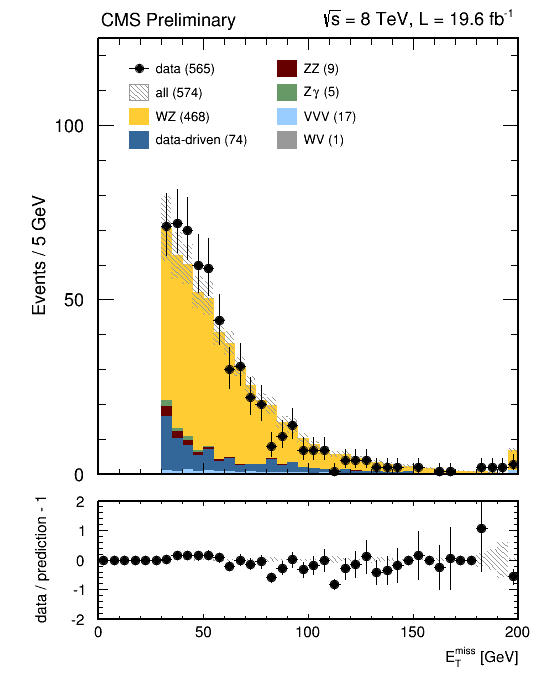}
	\end{subfigure}\quad
	\begin{subfigure}[b]{0.3\textwidth}
		\includegraphics[width=\textwidth,height=\textwidth]{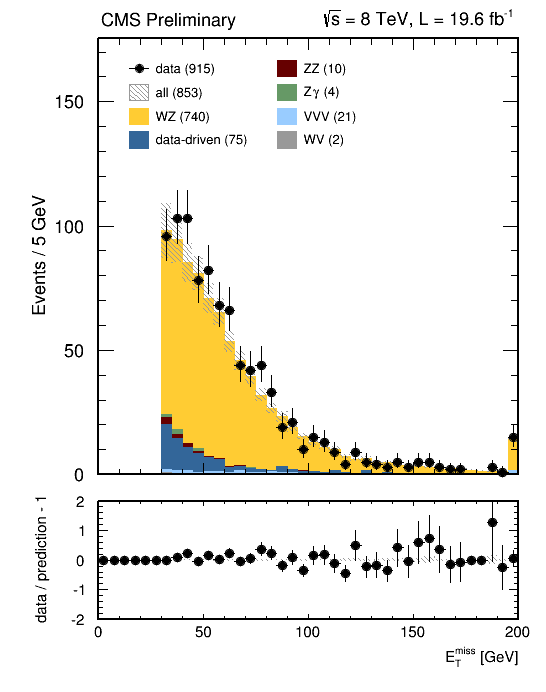}
	\end{subfigure}
	\caption[Missing transverse energy at 8~\TeV (ratio)]{Missing 
	energy in the transverse plane at each event for the \wzm (left column) and \wzp 
	(right column) before (up row) and after the \MET cut (bottom row).	}
	\vskip 1em
	\begin{subfigure}[b]{0.3\textwidth}
		\includegraphics[width=\textwidth,height=\textwidth]{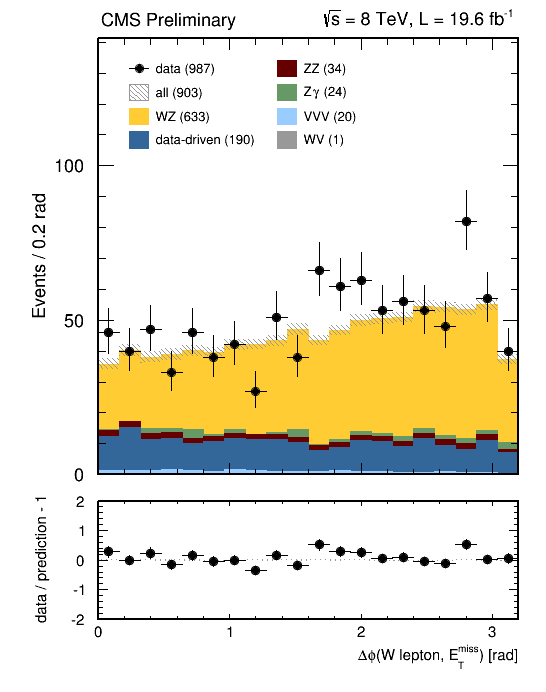}
	\end{subfigure}\quad
	\begin{subfigure}[b]{0.3\textwidth}
		\includegraphics[width=\textwidth,height=\textwidth]{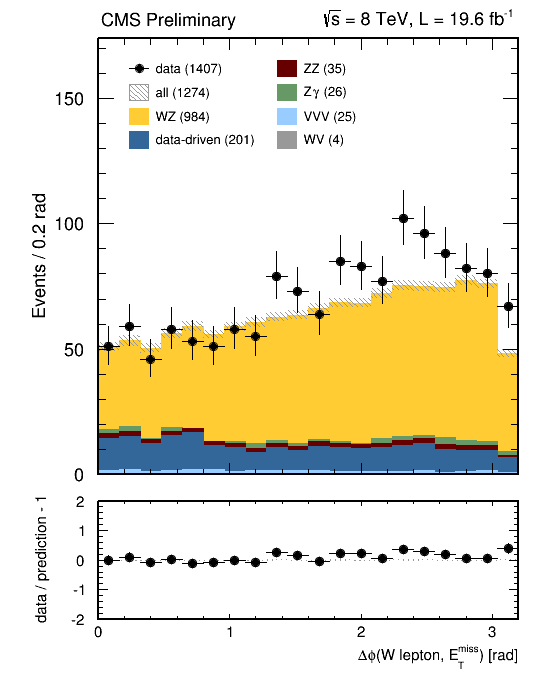}
	\end{subfigure}
	\vskip 1ex
	\begin{subfigure}[b]{0.3\textwidth}
		\includegraphics[width=\textwidth,height=\textwidth]{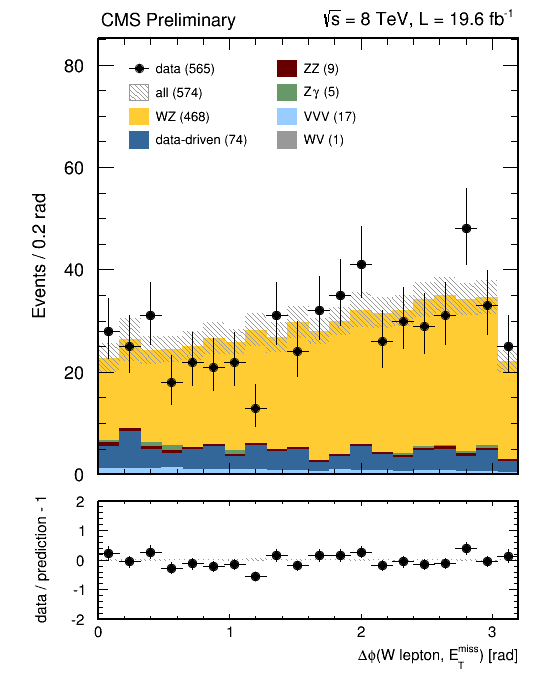}
	\end{subfigure}\quad
	\begin{subfigure}[b]{0.3\textwidth}
		\includegraphics[width=\textwidth,height=\textwidth]{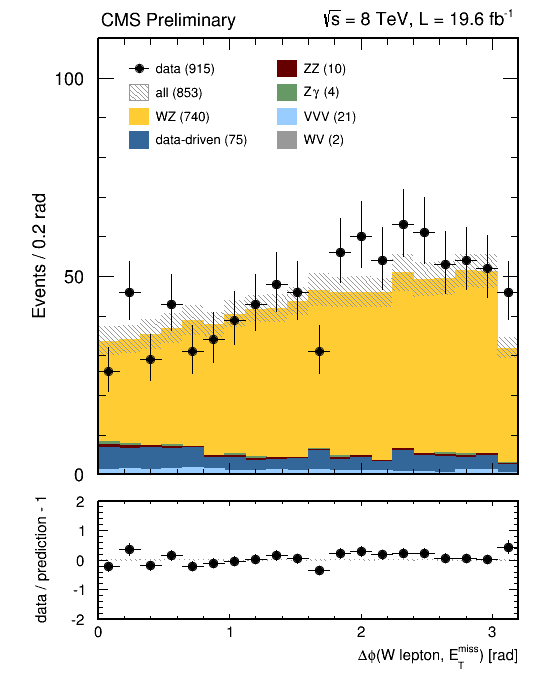}
	\end{subfigure}
	\caption[Azimuthal angle between W-candidate lepton and \MET at 8~\TeV (ratio)]
	{Azimuthal angle between the W-candidate lepton and the \MET
	at each event for the \wzm (left column) and \wzp (right column) before (up row) 
	and after the \MET cut (bottom row).}
\end{figure}

\begin{figure}[!htpb]
	\centering
	\begin{subfigure}[b]{0.3\textwidth}
		\includegraphics[width=\textwidth,height=\textwidth]{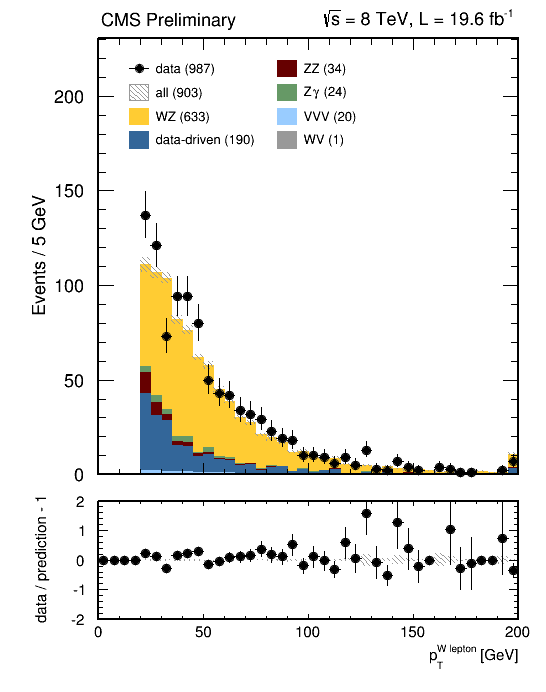}
	\end{subfigure}\quad
	\begin{subfigure}[b]{0.3\textwidth}
		\includegraphics[width=\textwidth,height=\textwidth]{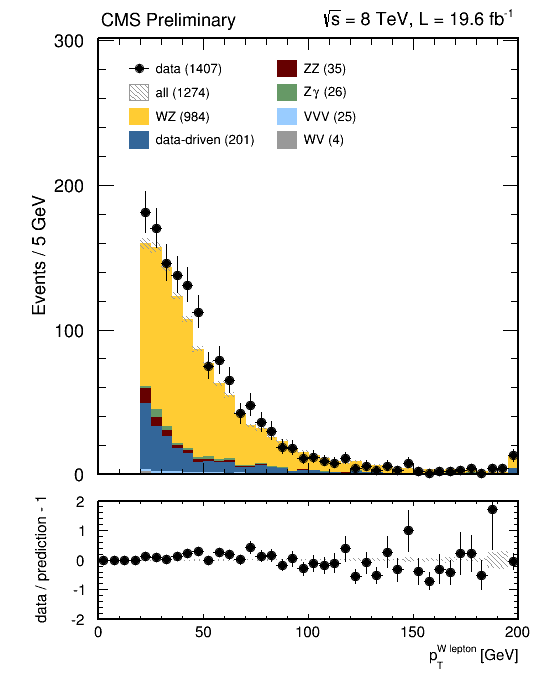}
	\end{subfigure}
	\vskip 1ex
	\begin{subfigure}[b]{0.3\textwidth}
		\includegraphics[width=\textwidth,height=\textwidth]{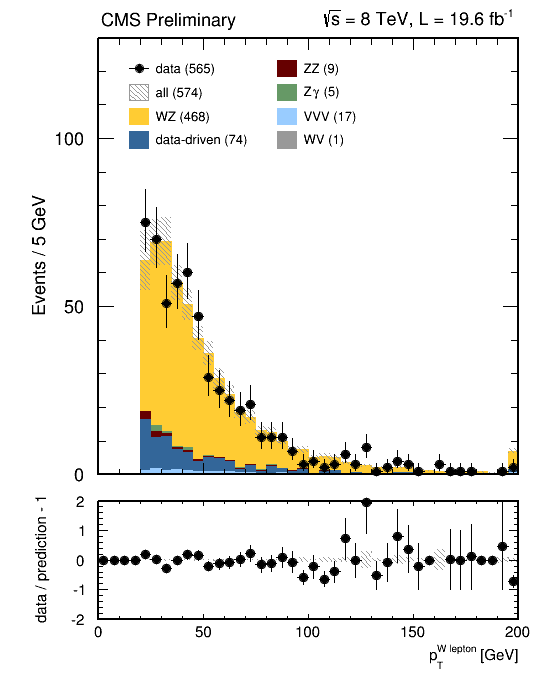}
	\end{subfigure}
	\begin{subfigure}[b]{0.3\textwidth}
		\includegraphics[width=\textwidth,height=\textwidth]{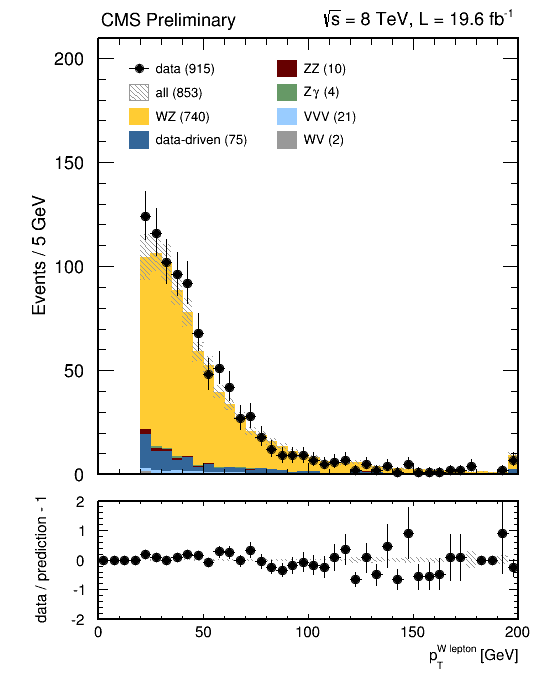}
	\end{subfigure}\quad
	\caption[Transverse momentum of the W-candidate system at 8~\TeV (ratio)]
	{Transverse momentum of the W-candidate system composed by
	the third selected lepton and \MET at each event for the \wzm (left column) and \wzp 
	(right column) before (up row) and after the \MET cut (bottom row).}
	\vskip 1em
	\begin{subfigure}[b]{0.3\textwidth}
		\includegraphics[width=\textwidth,height=\textwidth]{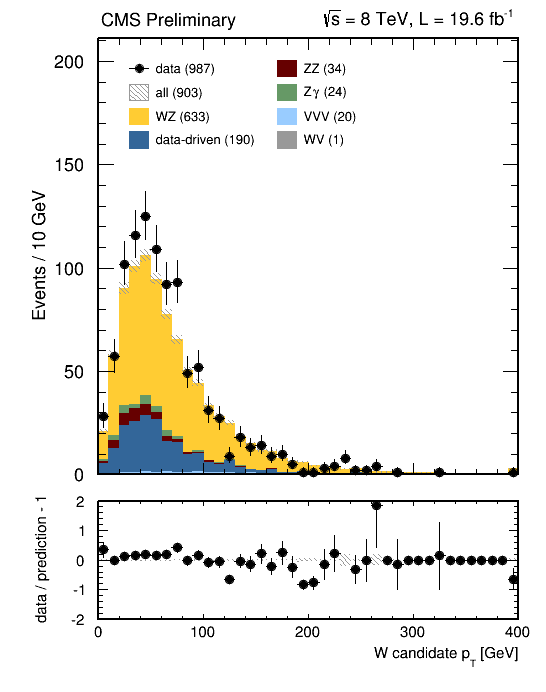}
	\end{subfigure}\quad
	\begin{subfigure}[b]{0.3\textwidth}
		\includegraphics[width=\textwidth,height=\textwidth]{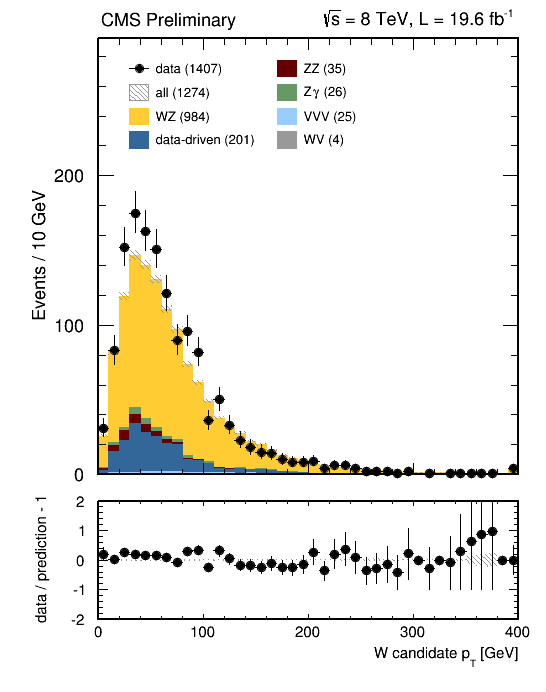}
	\end{subfigure}
	\vskip 1ex
	\begin{subfigure}[b]{0.3\textwidth}
		\includegraphics[width=\textwidth,height=\textwidth]{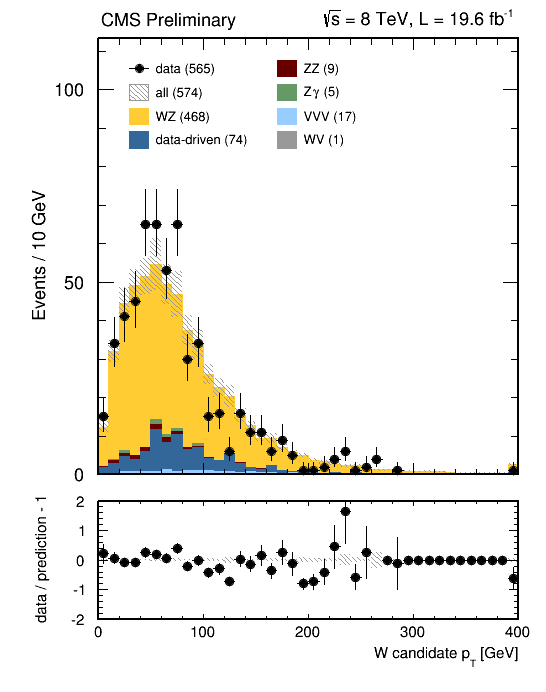}
	\end{subfigure}\quad
	\begin{subfigure}[b]{0.3\textwidth}
		\includegraphics[width=\textwidth,height=\textwidth]{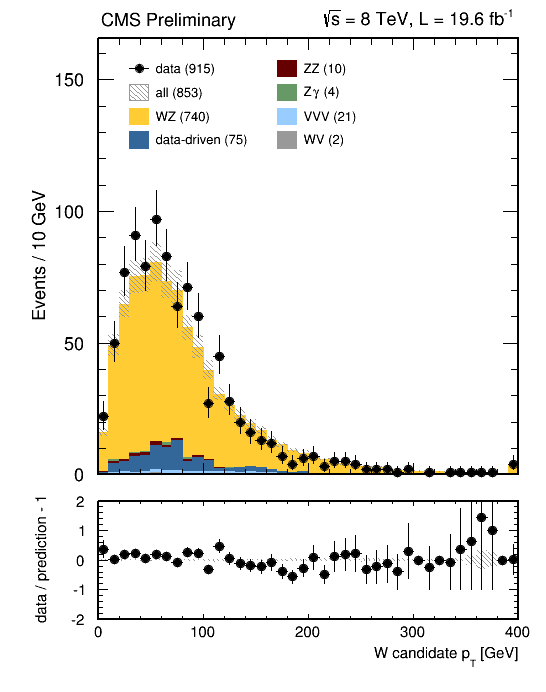}
	\end{subfigure}
	\caption[Transverse momentum of the W-candidate lepton at 8~\TeV (ratio)]
	{Transverse momentum of the W-candidate lepton 
	at each event for the \wzm (left column) and \wzp (right column) before (up row)
	and after the \MET cut (bottom row).}
\end{figure}

\begin{figure}[!htpb]
	\centering
	\begin{subfigure}[b]{0.3\textwidth}
		\includegraphics[width=\textwidth,height=\textwidth]{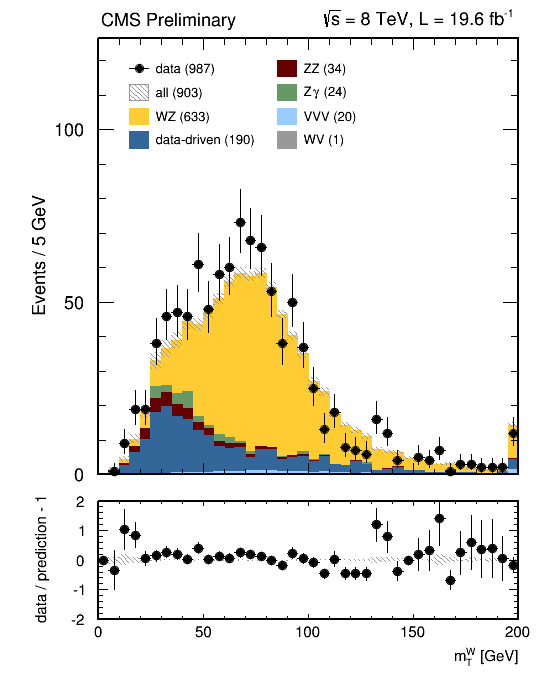}
	\end{subfigure}\quad
	\begin{subfigure}[b]{0.3\textwidth}
		\includegraphics[width=\textwidth,height=\textwidth]{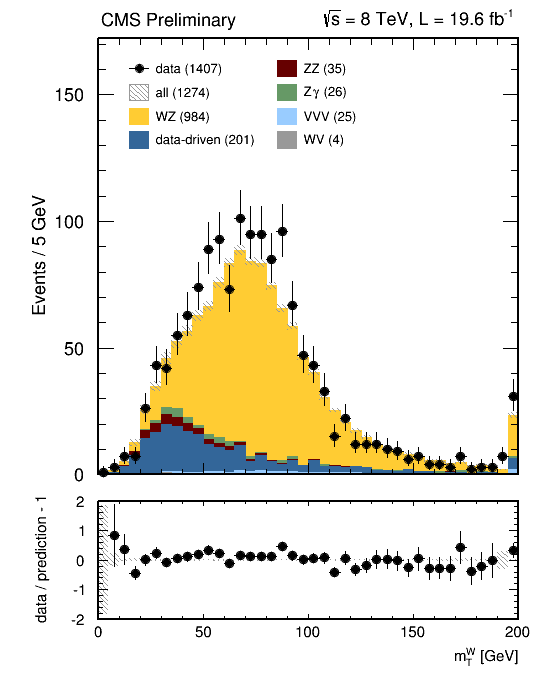}
	\end{subfigure}
	\vskip 1ex
	\begin{subfigure}[b]{0.3\textwidth}
		\includegraphics[width=\textwidth,height=\textwidth]{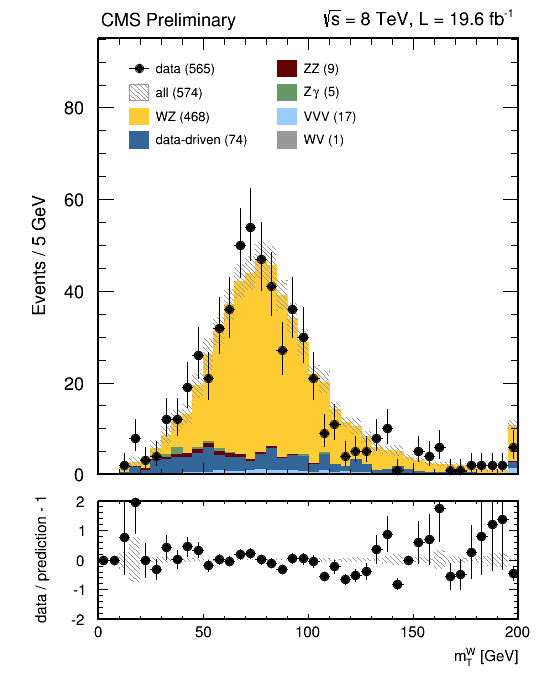}
	\end{subfigure}\quad
	\begin{subfigure}[b]{0.3\textwidth}
		\includegraphics[width=\textwidth,height=\textwidth]{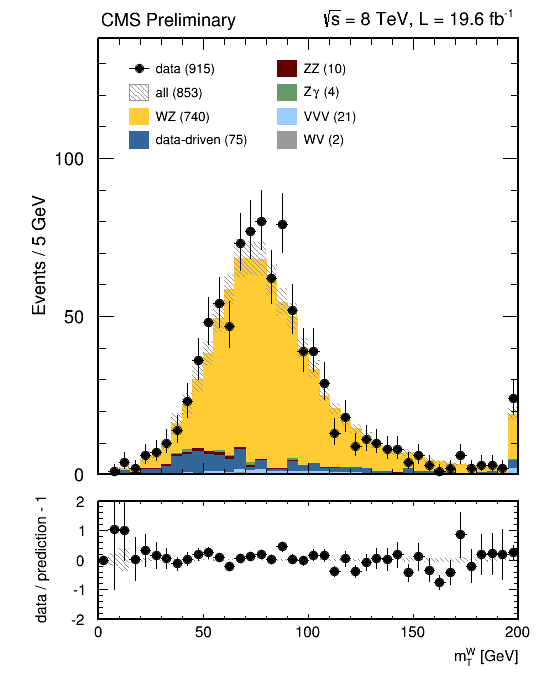}
	\end{subfigure}
	\caption[Transverse mass of the W-candidate lepton and \MET at 8~\TeV (ratio)]
	{Transverse mass of the W-candidate lepton and the \MET
	at each event for the \wzm (left column) and \wzp (right column) before (up row)
	and after the \MET cut (bottom row).}
	\vskip 1em
	\begin{subfigure}[b]{0.3\textwidth}
		\includegraphics[width=\textwidth,height=\textwidth]{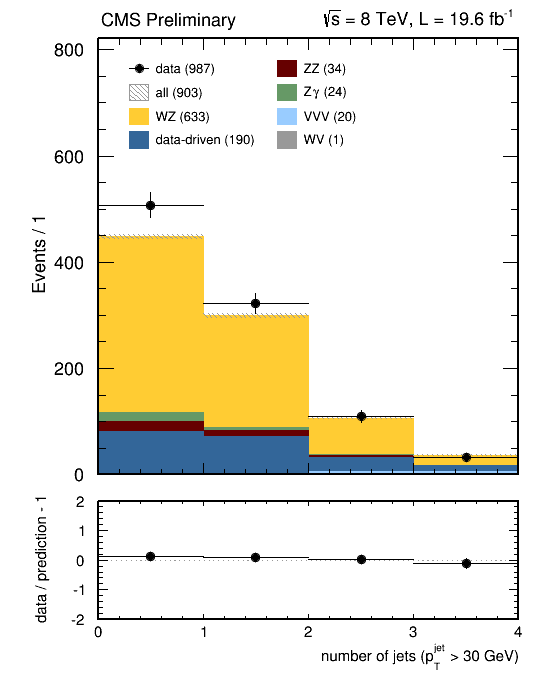}
	\end{subfigure}\quad
	\begin{subfigure}[b]{0.3\textwidth}
		\includegraphics[width=\textwidth,height=\textwidth]{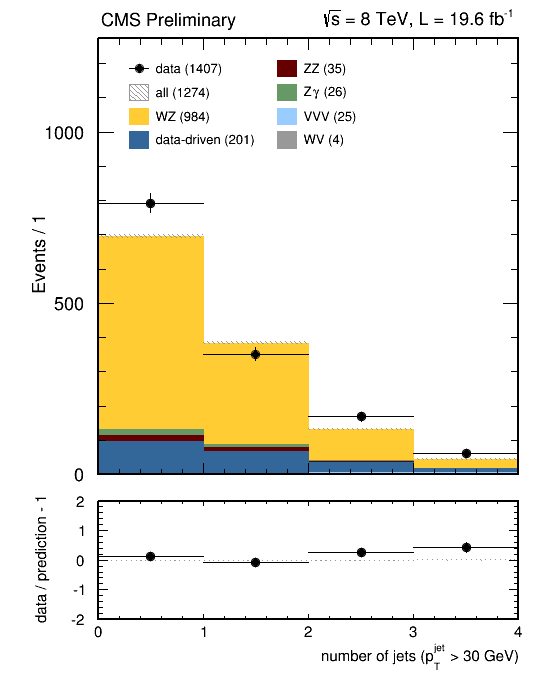}
	\end{subfigure}
	\vskip 1ex
	\begin{subfigure}[b]{0.3\textwidth}
		\includegraphics[width=\textwidth,height=\textwidth]{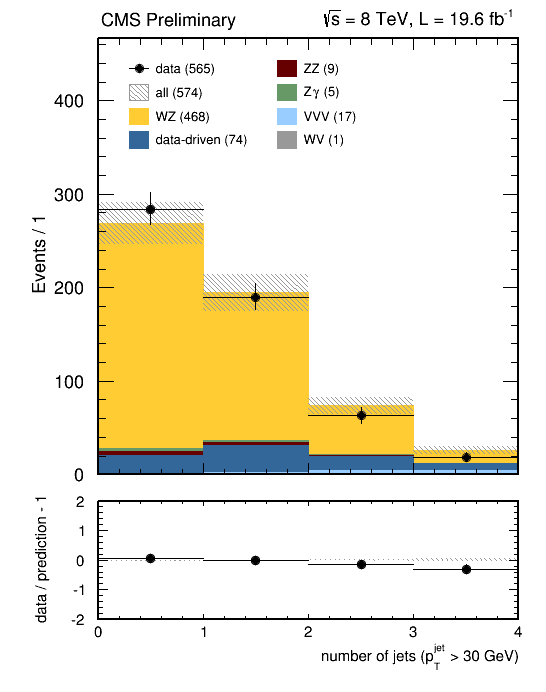}
	\end{subfigure}\quad
	\begin{subfigure}[b]{0.3\textwidth}
		\includegraphics[width=\textwidth,height=\textwidth]{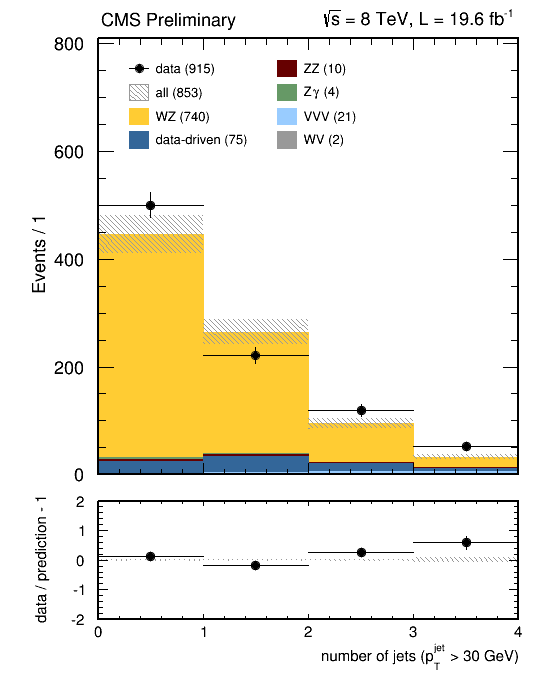}
	\end{subfigure}
	\caption[Number of jets at 8 TeV (ratio)]{Number of 
	jets distribution at each event for the \wzm (left column) and \wzp 
	(right column) before (up row) and after the \MET cut (bottom row).}
\end{figure}

\begin{figure}[!htpb]
	\centering
	\begin{subfigure}[b]{0.3\textwidth}
		\includegraphics[width=\textwidth,height=\textwidth]{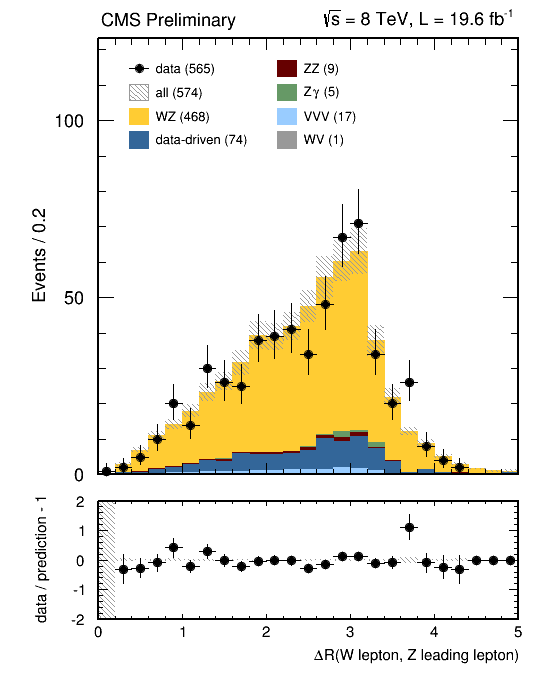}
	\end{subfigure}\quad
	\begin{subfigure}[b]{0.3\textwidth}
		\includegraphics[width=\textwidth,height=\textwidth]{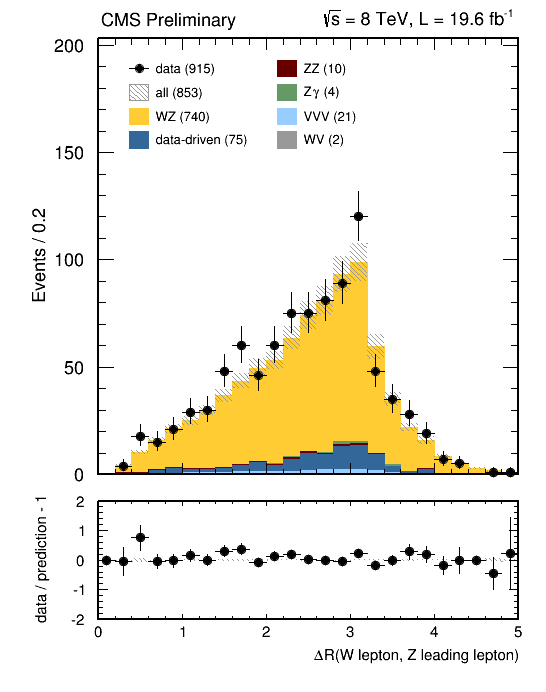}
	\end{subfigure}
	\vskip 1ex
	\begin{subfigure}[b]{0.3\textwidth}
		\includegraphics[width=\textwidth,height=\textwidth]{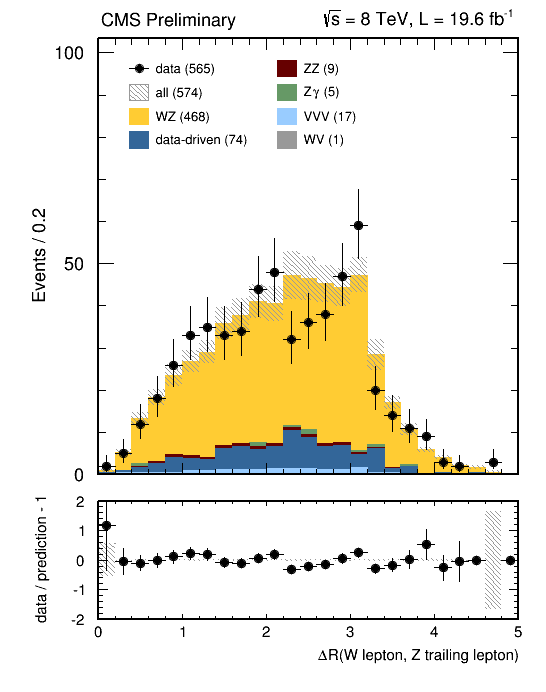}
	\end{subfigure}\quad
	\begin{subfigure}[b]{0.3\textwidth}
		\includegraphics[width=\textwidth,height=\textwidth]{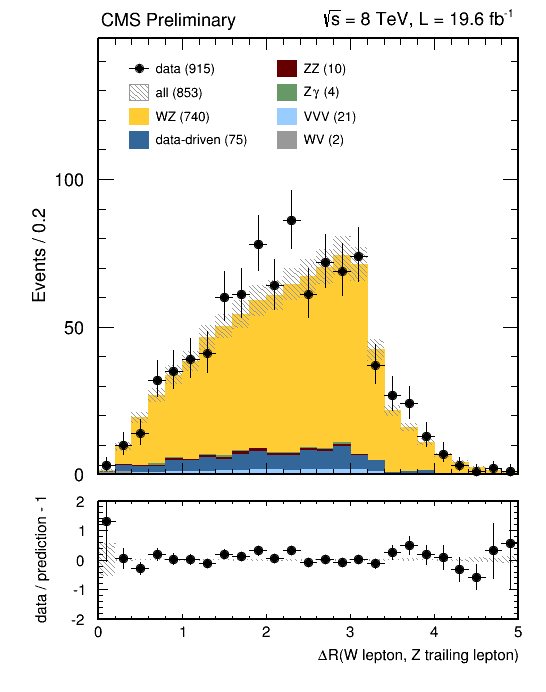}
	\end{subfigure}
	\caption[Angular distance between W-candidate lepton and Z-candidate leptons at 8~\TeV (ratio)]
	{Angular distance between the W-candidate lepton and the Z-candidate
	leading (up row) and trailing lepton (bottom row) at each event for the \wzm 
	(left column) and \wzp (right column) before \MET cut.}
	\vskip 1em
	\begin{subfigure}[b]{0.3\textwidth}
		\includegraphics[width=\textwidth,height=\textwidth]{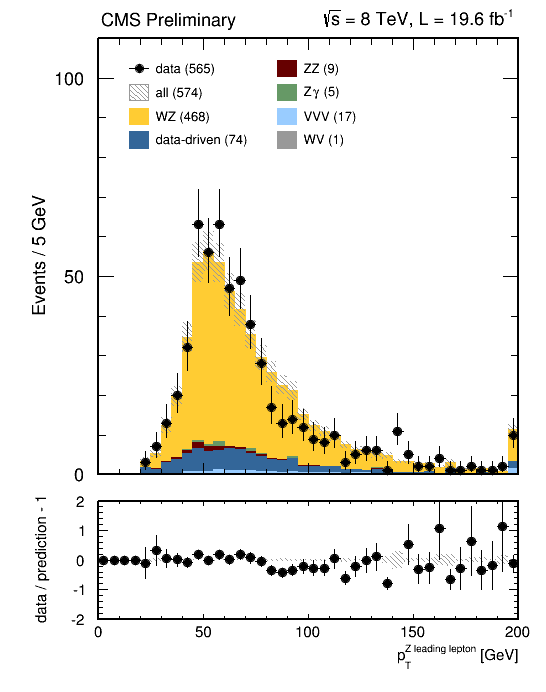}
	\end{subfigure}\quad
	\begin{subfigure}[b]{0.3\textwidth}
		\includegraphics[width=\textwidth,height=\textwidth]{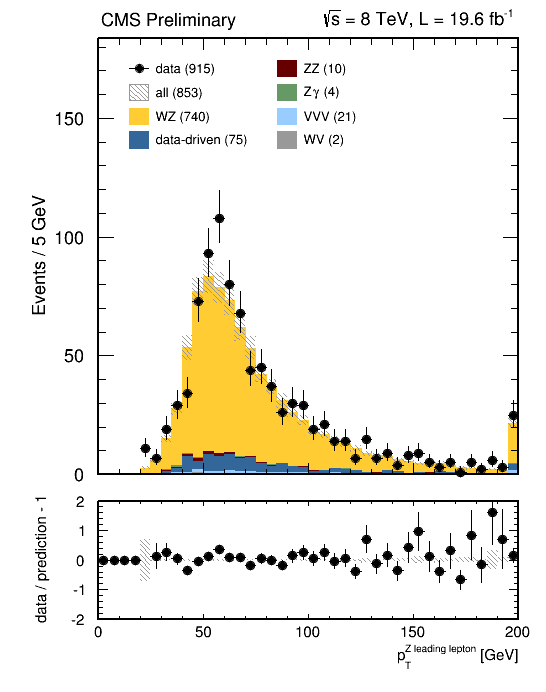}
	\end{subfigure}
	\vskip 1ex
	\begin{subfigure}[b]{0.3\textwidth}
		\includegraphics[width=\textwidth,height=\textwidth]{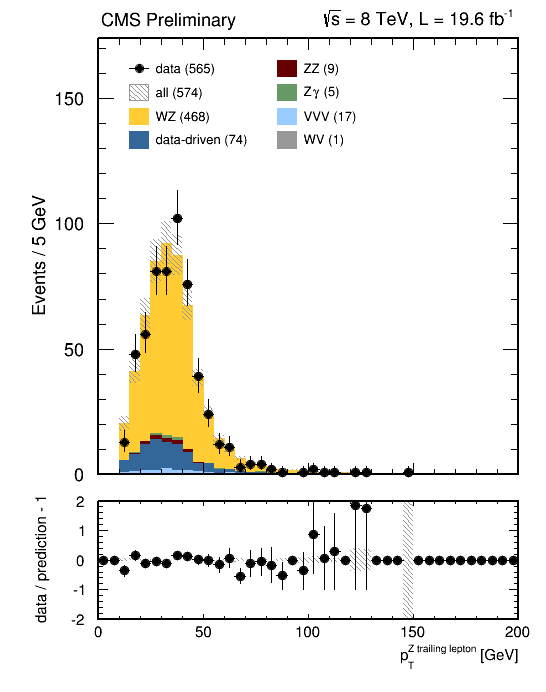}
	\end{subfigure}
	\begin{subfigure}[b]{0.3\textwidth}
		\includegraphics[width=\textwidth,height=\textwidth]{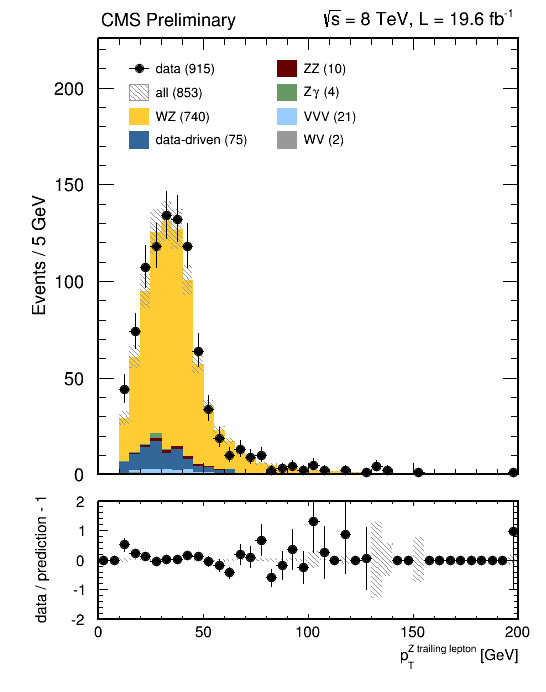}
	\end{subfigure}\quad
	\caption[Transverse momentum of the Z-candidate leptons at 8~\TeV (ratio)]
	{Transverse momentum of the Z-candidate leading (up row) and trailing lepton 
	(bottom row) at each event for the \wzm (left column) and \wzp (right column) after the 
	\MET cut.}
\end{figure}


\backmatter 

\selectlanguage{spanish}
\let\oldchaptermark\chaptermark
\renewcommand{\chaptermark}[1]{\markboth{\MakeUppercase{#1 }}{}}
\addtocounter{chapter}{1}
\renewcommand*{\theequation}{R.\arabic{equation}}
\renewcommand*{\thetable}{R.\arabic{table}}
\renewcommand*{\thesection}{\arabic{section}}
\setcounter{section}{0}
\chapter{Resumen}
\epigraph{\qquad\qquad Hab\'ia una vez, un circo...}{--- \textup{Los payasos de la tele}}
El arranque del acelerador de part\'iculas m\'as potente del mundo, el LHC (de sus siglas en 
ingl\'es \emph{Large Hadron Collider}), en 2009 y sus tres a\~nos de impresionante rendimiento
han permitido a todos los experimentos situados en el anillo del acelerador, entre ellos CMS 
(del ingl\'es \emph{Compact Muon Solenoid}), almacenar millones de datos de colisiones 
prot\'on--prot\'on, accediendo por primera vez a la escala de energ\'ias del teraelectronvoltio
(\TeV). El an\'alisis de estos datos ha reforzado las predicciones del Modelo Est\'andar
de part\'iculas y ha posibilitado una serie de importantes resultados destancando entre ellos 
el descubrimiento de una nueva part\'icula compuesta, el mes\'on
$\chi_b(3P)$, la creaci\'on de plasma de quarks y gluones, la primera observaci\'on de la 
desintegraci\'on del mes\'on $B_s^0$ a dos muones cuyos resultados son consistentes con el Modelo 
Est\'andar y el descubrimiento de una nueva part\'icula elemental, un bos\'on cuyas propiedades 
medidas hasta la fecha son consistentes con el bos\'on de Higgs predicho por el Modelo Est\'andar.
La obtenci\'on de tan notables resultados ha necesitado de un previo re-descubrimiento del Modelo 
Est\'andar de part\'iculas. 

El Modelo Est\'andar de part\'iculas, establecido como modelo ortodoxo de las interacciones 
entre las part\'iculas a mediados de los a\~nos 70 del siglo pasado, ha sido verificado experimentalmente
a lo largo de estos a\~nos, confirmando las predicciones, en muchos casos con notable precisi\'on.
Este conocimiento experimental del modelo ha sido explotado para calibrar los datos, permitiendo
una mejor comprensi\'on de los complejos aparatos de medida usados para la 
detecci\'on de part\'iculas. Los procesos predichos por el Modelo Est\'andar han sido observados
de nuevo y se han realizado nuevas medidas en el nuevo rango de energ\'ias que ha alcanzado el 
acelerador: 7 y 8~\TeV. Estas nuevas medidas, de destacado valor cient\'ifico por s\'i mismas, 
representan el primer paso hacia los descubrimientos mencionados previamente, puesto que estos 
procesos, ya medidos y estudiados con el Modelo Est\'andar, introducir\'an ruido a la se\~nal del
nuevo proceso que quiere medirse y, por tanto, deben entenderse y controlarse perfectamente para 
poder ser sustra\'idos o estimados de los datos observados. En particular, la producci\'on de 
dibosones \WZ aparece recurrentemente como fondo de diversas b\'usquedas del bos\'on de Higgs y en 
modelos de nueva f\'isica, siendo, por consiguiente, importante un conocimiento preciso y minucioso
del proceso para controlar los an\'alisis de b\'usqueda. Asimismo, la secci\'on eficaz de 
producci\'on del proceso nunca ha sido medido en los rangos de energ\'ia que ha alcanzado el LHC,
proporcionando una nueva medida a confirmar por las predicciones del Modelo Est\'andar. 

Puesto que el bos\'on \W est\'a 
cargado, el estudio del proceso de producci\'on de \WZ pasa por estudiar la producci\'on de \wzm y
\wzp, y en particular, un observable interesante a medir es el cociente de la secci\'on eficaz de
ambos procesos, $\sigma_{\wzm}/\sigma_{\wzp}$. Este cociente puede ser medido con mayor precisi\'on 
que las secciones eficaces debido a que es posible cancelar en algunos casos, o reducir de forma 
importante en otros, algunas de las incertidumbres experimentales. Adem\'as, el cociente de 
secciones eficaces resulta ser m\'as sensible a las funciones de distribuci\'on part\'onicas 
(PDFs, del ingl\'es \emph{Parton Distribution Functions}), unas funciones fenomenol\'ogicas que se 
utilizan como \emph{inputs} en los c\'alculos te\'oricos de secci\'on eficaz. En consecuencia, este
observable puede utilizarse para validar o constre\~nir los diferentes conjuntos de PDFs que 
existen.

En esta tesis doctoral se presenta y desarrolla el trabajo realizado para medir la secci\'on eficaz
de producci\'on de dibosones \WZ en colisiones prot\'on-prot\'on con una energ\'ia de centro de 
masas de 7 y 8~\TeV, junto con la medida del cociente de producci\'on de \wzm y \wzp. Los datos 
analizados se obtuvieron durante los a\~nos 2011 (\comene=7~\TeV) y 2012 (\comene=8~\TeV) con el 
detector CMS, equivalente a 4.9~\fbinv y 19.6~\fbinv de luminosidad integrada para 2011 y 2012 
respectivamente. La secci\'on eficaz de producci\'on de \WZ ha sido medida con anterioridad a 
menor energ\'ia, \comene=1.96~\TeV, en los experimentos CDF~\cite{Aaltonen:2012vu} y 
D0~\cite{Aaltonen:2012vu} del acelerador americano \emph{Tevatron}, y recientemente en 
\emph{ATLAS}, el otro gran experimento de prop\'osito general del LHC, a 7 y 8~\TeV de energ\'ia
de centro de masas~\cite{Aad2012341,Aad:2012twa,ATLAS:2013fma}. CMS ha presentado 
resultados~\cite{CMS:2011dqa} a 7~\TeV utilizando datos que corresponden a 1.1~\fbinv de 
luminosidad integrada, y que constituyen parte de varias tesis 
doctorales~\cite{Martelli:2012zva,Klukas:2012era}. Los \'ultimos resultados actualizados a 
7~\TeV utilizando todos los datos disponibles, as\'i como los nueva medida que ha presentado 
CMS a 8~\TeV, y junto con la primera medida experimental realizada del cociente de producci\'on de 
\wzm y \wzp~\cite{CMS-PAS-SMP-12-006}, se han extraido fundamentalmente de esta tesis doctoral. El
proceso de an\'alisis se comparti\'o con otro grupo de trabajo dentro de la 
colaboraci\'on\footnote{Cuyo esfuerzo quedar\'a plasmado en otra tesis doctoral que est\'a fase de
en preparaci\'on}, de forma que de manera independiente, aunque utilizando la misma metodolog\'ia y
datos iniciales, se obtuvieron los mismos resultados. Este proceso proporcion\'o rubustez al 
an\'alisis y lo protegi\'o contra posibles (y probables) errores de c\'odigo, a\~nadiendo solidez a 
las medidas. Los resultados de esta tesis constituyen, pues, los resultados que la colaboraci\'on 
CMS ha presentado en conferencias y cuyo art\'iculo correspondiente est\'a en fase de preparaci\'on. 

\section{Marco te\'orico}
El Modelo Est\'andar de part\'iculas~\cite{Halzen:1984mc} (ME) se ha establecido, a lo largo de 
estos cuarenta a\~nos desde que a mitades de los a\~nos 70 del siglo XX se finaliz\'o su 
formulaci\'on con la confirmaci\'on experimental de los \emph{quarks}, como la teor\'ia que mejor 
describe experimentalmente las interacciones de las part\'iculas subat\'omicas. El modelo 
caracteriza tres de las cuatro interacciones fundamentales conocidas: la electromagn\'etica, la 
d\'ebil y la fuerte. A pesar de los esfuerzos realizados para incluir la cuarta interacci\'on, 
la gravitatoria, hasta la fecha no ha sido posible acomodarla a la teor\'ia; este hecho, junto con 
otras cuestiones fundamentales sin resolver, evita que el ME sea una teor\'ia completa de las 
interacciones fundamentales aunque s\'i es un modelo efectivo que proporciona predicciones 
te\'oricas consistentes con los resultados experimentales en los rangos energ\'eticos alcanzados 
hasta la fecha. 

El modelo describe la materia y antimateria a trav\'es de campos fermi\'onicos y las interacciones 
mediante campos bos\'onicos. As\'i, el ME incluye 61 part\'iculas elementales, 48 fermiones de spin
$1/2$ y 13 bosones mediadores de fuerza. Los fermiones, clasificados de acuerdo a c\'omo 
interaccionan, es decir qu\'e \emph{cargas} llevan, se dividen en \emph{quarks} y \emph{leptones}. 
Hay seis quarks\footnote{En realidad, hay 18 quarks, cada uno de los seis con diferente carga de
color: roja, verde y azul.} (\emph{up}, \emph{down}, \emph{charm}, \emph{strange}, \emph{top} y
\emph{bottom}) y seis leptones (\emph{electr\'on}, \emph{neutrino electr\'onico}, \emph{mu\'on}, 
\emph{neutrino mu\'onico}, \emph{tau\'on} y \emph{neutrino tau\'onico}), cada uno con su 
correspondiente antipart\'icula. Cada tipo de quark y lept\'on es designado como 
\emph{sabor}. A su vez, los leptones y quarks se agrupan en \emph{generaciones}
formadas por part\'iculas que exhiben un comportamiento f\'isico similar. La propiedad distintiva 
de los quarks es que son portadores de carga de color y, por tanto, interaccionan a trav\'es de
la fuerza fuerte. Los quarks, adem\'as, llevan carga el\'ectrica y carga de isosp\'in d\'ebil, 
interactuando tambi\'en mediante la fuerza electromagn\'etica y d\'ebil. Los leptones, en
cambio, no llevan carga de color, interact\'uan a trav\'es de la fuerza d\'ebil aunque, el 
electr\'on, mu\'on y tau\'on lo hacen tambi\'en a trav\'es de la electromagn\'etica. As\'i, los 
neutrinos, al no llevar carga el\'ectrica e interactuar \'unicamente mediante la interacci\'on
d\'ebil, son extremadamente complicados de detectar.
\begin{table}[!htpb]
	\centering
	\begin{tabular}{lc c c c}\hline\hline
		\multicolumn{4}{c}{}  & Interacciones\\ \hline
		\multirow{2}{*}{LEPTONES} 
		 &  $\nu_e$ & $\nu_{\mu}$ & $\nu_{\tau}$ & d\'ebil\\
		 & $e$ & $\mu$ & $\tau$ & d\'ebil, EM\\\hline
		\multirow{2}{*}{QUARKS}   
		 & $u$ & $c$ & $t$ & d\'ebil, EM, fuerte\\
		 & $d$ & $s$ & $b$ & d\'ebil, EM, fuerte\\\hline
	\end{tabular}
	\caption[]{Taxonom\'ia de los fermiones del ME, mostrando
		sus respectivos s\'imbolos: $\nu$ (neutrino), $e$ (electron), $\mu$ (mu\'on),
		$\tau$ (tau\'on) y la inicial del nombre de cada quark. Cada quark se encuentra
		con tres cargas diferentes de color (carga de la interacci\'on fuerte), siendo en
		total $3\times6=18$ quarks. Adem\'as, cada fermi\'on tiene su correspondiente 
		antiparticula (de carga el\'ectrica opuesta). La agrupaci\'on en filas responde a la 
		asociaci\'on por generaciones. La \'ultima columna muestra la interacci\'on a 
		la que son sensibles cada fermi\'on de la misma generaci\'on 
		(EM=electromagn\'etica).}\label{res:table:fermions}
\end{table}

Los bosones, por su parte, son las part\'iculas de spin $1$ que el modelo utiliza para mediar las
interacciones. Son los portadores de las fuerzas electromagn\'etica (fot\'on, $\gamma$), 
d\'ebil (bosones \Z, $W^+$ y $W^-$) y fuerte (ocho gluones, $g$). Tanto el fot\'on como los gluones
son bosones sin masa, y, en el caso de los gluones, adem\'as tambi\'en pueden interaccionar consigo 
mismos. Los bosones \Z, $W^+$ y $W^-$, responsables de mediar las interacciones d\'ebiles 
entre part\'iculas de distinto sabor, son, por el contrario, bosones masivos. El bos\'on \W, al 
portar carga el\'ectrica, tambi\'en se acopla con la interacci\'on electromagn\'etica. Los tres 
bosones masivos junto con el fot\'on se agrupan de forma que colectivamente son los mediadores
de la interacci\'on electrod\'ebil (EWK, del ingl\'es electroweak), la unificaci\'on en el ME de
la teor\'ia electromagn\'etica y d\'ebil.

El ME se formula matem\'aticamente a trav\'es de teor\'ia cu\'antica de campos~\cite{Peskin:1995qft}, 
donde un \emph{lagrangiano} controla la din\'amica y la cinem\'atica de la teor\'ia. La 
construcci\'on del modelo sigue los procedimientos habituales para construir la mayor\'ia de 
teor\'ias de campo, postulando un conjunto de simetr\'ias del sistema, que a su vez definen las 
interacciones del mismo. Adem\'as de la simetr\'ia global de Poincar\'e, puesto que el ME es una 
teor\'ia relativista, el ME viene definido por la simetr\'ia interna del sistema
$SU(3)_C\otimes SU(2)_L\otimes U(1)_Y$. La formulaci\'on del lagrangiano m\'as general con la
simetria interna mencionada, predice part\'iculas sin masa, predicci\'on inconsistente con los 
datos experimentales. As\'i, se introduce un mecanismo \emph{ad-hoc} para proporcionar masa a las 
part\'iculas y que respeta la invarianza del sistema a la simetr\'ia local, completando el ME. Este
mecanismo es conocido como \emph{mecanismo de Higgs}, cuya consecuencia directa es la 
introducci\'on de un nuevo bos\'on en la teor\'ia, el bos\'on de Higgs, encargado de proporcionar 
masa a los fermiones y a los bosones \W y \Z, mientras que permite al fot\'on y a los gluones no 
tener masa.

\section{El experimento}
El LHC es un acelerador de part\'iculas que permite colisionar protones a 14~\TeV de energ\'ia en
el centro de masas en su dise\~no nominal. Durante los tres primeros a\~nos de funcionamiento ha
alcanzado las energ\'ias 7~\TeV y 8~\TeV, proporcionando por primera vez acceso experimental a esos
rangos energ\'eticos y con una luminosidad instant\'anea capaz de producir millones de procesos
de baja tasa de producci\'on. En el anillo principal del acelerador, de unos 27 kil\'ometros de 
circunferencia instalado cerca de Ginebra, se sit\'uan cuatro grandes detectores en el centro de
otros tantos puntos de colisi\'on: ALICE, ATLAS, CMS y LHCb; estos aparatos van a detectar y 
almacenar las colisiones de part\'iculas de alta energ\'ia que se produzcan.

El detector CMS~\cite{Chatrchyan:2008zzk} es un detector de prop\'osito general, compacto y 
herm\'etico de unos 21 metros de largo y 14 de di\'ametro, cuyo peso aproximado es de 12500 toneladas.
El dise\~no del detector comprende un solenoide superconductor que proporciona un campo 
magn\'etico uniforme de 3.8~T, en cuyo interior se emplazan diferentes sistemas de 
detecci\'on de part\'iculas. En su parte m\'as interna, rodeando al punto de colisi\'on, se
encuentra el sistema interno de detecci\'on de trazas, compuesto por un detector de p\'ixeles de
tres capas cil\'indricas de radios comprendidos entre 4.4 y 10.2~\cm, y un detector de 
bandas de silicio compuesto por diez capas cil\'indricas de detecci\'on que se extienden hacia
el exterior alcanzando un radio de 1.1~m. Cada sistema cil\'indrico se completa con dos tapas, 
permitiendo extender la aceptancia de detecci\'on hasta $|\eta|<2.5$. Los detectores de trazas
se rodean de un calorimetro electromagn\'etico de cristal de tungstato de plomo (ECAL, del 
ingl\'es electromagnetic calorimeter) de fina granularidad en el plano transverso, y de un 
calor\'imetro hadr\'onico basado en detectores de centelleo (HCAL, del ingl\'es hadron calorimeter)
que cubren la regi\'on $|\eta| < 3$. El hierro de retorno, en la parte externa del solenoide, est\'a 
instrumentado con detectores gaseosos que se utilizan para identificar muones en el rango 
$|\eta|<2.4$. El barril del cilindro del sistema de detecci\'on de muones est\'a formado por 
c\'amaras de deriva (DT, del ingl\'es drift tube) mientras que las tapas del cilindro montan 
c\'amaras de bandas cat\'odicas (CSC, del ingl\'es cathode strip chamber) complementadas por
c\'amaras de l\'aminas resistivas (RPC, resistive plate chamber).

Los sucesos de colisiones se seleccionan utilizando el \emph{sistema de selecci\'on de datos},
llamado \emph{trigger}. Debido a limitaciones de almacenamiento y velocidad de procesamiento, 
no todas las colisiones producidas en CMS puede almacenarse. De hecho, dada la frecuencia de cruce
de los haces de protones proporcionados por el LHC en el punto de interacci\'on, unas 100 millones
de colisiones por segundo se est\'an produciendo en CMS, de las cuales s\'olo un porcentaje muy 
peque\~no ser\'an colisiones de altas energ\'ias e interesantes desde el punto de vista de 
an\'alisis. El sistema de \emph{trigger} es el encargado de seleccionar dichas colisiones 
\emph{interesantes} a trav\'es de parte de los sistemas de detecci\'on de CMS, siendo el sistema lo
suficientemente flexible para poder seleccionar sucesos dependiendo de sus propiedades medidas y 
organizarlos en diferentes subconjuntos de datos seg\'un el contenido del suceso.

\section{An\'alisis del proceso \WZ}
La producci\'on de dibosones \WZ en un colisionador prot\'onico se produce principalmente por la
aniquilaci\'on de quarks $u$ ($d$) y antiquarks $\bar{d}$ ($\bar{u}$) de los protones que 
colisionan para producir un boson $\W^+$ ($W^-$) que a su vez pierde energ\'ia al interaccionar
d\'ebilmente produciendo un boson \Z. La desintegraci\'on lept\'onica de los bosones es, entre
todas las desintegraciones posibles, la m\'as l\'impia experimentalmente. El \W produce un lept\'on
con su misma carga acompa\~nado de un neutrino, indetectable en CMS pero que puede inferirse 
aplicando conservaci\'on de energ\'ia en el plano transverso. Por su parte, el bos\'on \Z se 
desintegra lept\'onicamente en dos leptones del mismo sabor y carga 
opuesta. As\'i, experimentalmente la signatura del proceso se caracteriza por tres leptones,
dos de ellos del mismo sabor y carga opuesta, y una cantidad apreciable de energ\'ia perdida en
el plano tranverso (designada como \MET).

La secci\'on eficaz de producci\'on de un proceso $X$ puede estimarse a partir de
\begin{equation}
	\sigma(pp\to X) = \frac{N_S}{\mathcal{A}\cdot\varepsilon\cdot\lumi_{int}}\;,
	\label{res:eq:xs}
\end{equation}
siendo $N_S$ el n\'umero de sucesos observados del proceso $X$, i.e. la se\~nal; $\mathcal{A}$ 
designa la aceptancia del detector; $\varepsilon$ es la eficiencia de detectar el proceso; y 
$\lumi_{int}$ es la luminosidad integrada. As\'i, medir una secci\'on eficaz equivale a seleccionar
sucesos de se\~nal evaluando las eficiencias de detecci\'on, que en l\'ineas generales define la
metodolog\'ia utilizada en la medida de la secci\'on eficaz de cualquier proceso. 

Espec\'ificamente, la metodolog\'ia utilizada en la medida de la tasa de producci\'on del proceso 
\WZ y del cociente $\sigma_{\wzm}/\sigma_{wzp}$ de este trabajo de tesis sigue las siguientes 
l\'ineas argumentales. Los datos proporcionados
por el detector CMS se seleccionan mediante una serie de cortes de calidad sobre los objetos
fundamentales del an\'alisis, esto es, los leptones, y posteriormente se aplican una sucesi\'on de
\emph{cortes secuenciales} optimizados para extraer sucesos de se\~nal. Puesto que entre los 
sucesos seleccionados van a encontrarse contaminaci\'on de otros procesos, se utilizan m\'etodos
basados en datos (\emph{data-driven}) y m\'etodos basados en simulaci\'on (Monte Carlo) para 
estimar estos fondos. Las eficiencias de reconstrucci\'on de objectos (donde se incluyen trigger,
identificaci\'on y aislamiento) se eval\'uan con m\'etodos basados en datos experimentales llamados
\emph{tag-and-probe}. Finalmente, utilizando una muestra simulada de \WZ se obtiene la aceptancia
y la eficiencia de selecci\'on.

\subsection{Reconstrucci\'on de objetos}
Los datos recolectados por CMS se procesan de forma centralizada, proporcionando una
reconstrucci\'on gen\'erica de los sucesos a todo an\'alisis realizado en la colaboraci\'on.  
Cada an\'alisis puede refinar el contenido de los sucesos, adecu\'andolos a las necesidades del 
an\'alisis particular. Para la selecci\'on de los estados finales del proceso \WZ, los sucesos
reconstruidos son inicialmente filtrados con cortes de calidad sobre los leptones finales del 
suceso; en particular se requiere una buena identificaci\'on de los candidatos a 
muones~\cite{Chatrchyan:2011tz} y de los candidatos a electrones~\cite{CMS-PAS-EGM-10-004}. 
Adem\'as, se comprueba que cada candidato a lept\'on sea compatible con el v\'ertice primario del 
suceso, que se escoge como aquel cuyas trayectorias asociadas suman un mayor $\sum \pt^2$. Este 
criterio proporciona una asignaci\'on correcta del v\'ertice primario en m\'as del 99\% de los casos. 

Los leptones que provienen de una desintegraci\'on de bosones \W o \Z tienden a estar
aislados de otras part\'iculas del suceso, mientras que hadrones que se han identificado
incorrectamente como leptones, o leptones que provienen de desintegraciones de quarks pesados
generalmente est\'an cerca de un jet. As\'i, una forma de discriminar entre estos dos tipos de
leptones es mediante el uso del \emph{aislamiento}. El aislamiento puede definirse a trav\'es de
un cono construido alrededor de la direcci\'on de la trayectoria del lept\'on en el v\'ertice del 
suceso; se calcula la suma escalar de la energ\'ia transversa de cada part\'icula reconstruida en 
el suceso que es compatible con el v\'ertice primario escogido y que est\'a contenida en el cono
(sin utilizar la del propio lept\'on). Si dicha suma excede aproximadamente el 10\% del momento 
transverso del lept\'on, se rechaza, consider\'andolo no aislado (el porcentaje exacto
de \pt depende del sabor del lept\'on, de su momento transverso y de su pseudorapidez). 

La \MET, debida al neutrino que proviene de la desintegraci\'on del \W que no interact\'ua con el 
detector, se calcula mediante un m\'etodo llamado \emph{particle-flow}~\cite{CMS:2010eua}. El 
algoritmo de \emph{particle-flow} combina la informaci\'on del detector de trazas, del 
espectr\'ometro de muones y de todos los calor\'imetros de CMS para reconstruir objectos de acuerdo
con el tipo de part\'iculas: electrones, muones, fotones y hadrones neutros o cargados. Esta 
reconstrucci\'on, utilizando todos los subsistemas de detecci\'on disponibles, permite conseguir correcciones
precisas en las energ\'ias de las part\'iculas, adem\'as de proporcionar un grado significativo
de redundancia que se traduce en una medida de la \MET menos sensible a los errores en las 
calibraciones calorim\'etricas. La \MET se calcula como la magnitud del vector suma, cambiado de 
signo, de las energ\'ias transversas de todos los objetos \emph{particle-flow}.

\subsection{Selecci\'on de sucesos}
La desintegraci\'on lept\'onica de los dibosones \WZ define el estado final de la se\~nal como
tres leptones de alto momento y aislados, dos de carga opuesta y mismo sabor, adem\'as de una 
cantidad apreciable de \MET debido al paso del neutrino por el detector sin interactuar con sus 
sistemas. La desintegraci\'on del \W o \Z en tauones tambi\'en es considerada se\~nal, siempre y 
cuando el tau\'on se desintegre lept\'onicamente. Dependiendo del sabor de los leptones finales, 
podemos realizar la medida utilizando cuatro estados finales, i.e. \emph{canales}, diferentes y 
excluyentes: $eee$, $\mu ee$, $e\mu\mu$ y $\mu\mu\mu$ (sobreentendiendo la presencia de \MET en 
todos ellos). 

Los sucesos que sobreviven a los cortes de calidad de los leptones son evaluados mediante cortes
secuenciales que rechazan todo aquel suceso que no cumpla:
\begin{enumerate}
	\item El suceso tiene que haber sido seleccionado con un trigger que requiera la presencia
		de dos leptones, ya sean dos muones o dos electrones.
	\item El suceso tiene que contener s\'olo tres leptones, electrones o muones, que satisfagan
		los criterios de calidad descritos previamente.
	\item El suceso contiene un candidato a \Z, construido a partir de dos leptones de signo
		opuesto e igual sabor. Adem\'as el lept\'on de mayor \pt tiene que cumplir 
		$\pt>20~\GeV$ y el de menor, $\pt>10~\GeV$. La masa invariante del sistema de los
		dos leptones debe estar en una ventana de $20\GeV$ alrededor de la masa nominal del
		\Z.
	\item El tercer lept\'on, asociado al \W, se require que cumpla $\pt>20\GeV$.
	\item La \MET del sucesos tiene que ser mayor de $30~\GeV$.
\end{enumerate}
Los requisitos (1) y (2) se designan a lo largo de este trabajo como corte de \emph{preselecci\'on},
el requisito (3) como \emph{candidato-Z} y los requisitos (4) y (5) como \emph{candidato-W}. En el 
caso de la medida del cociente de secciones eficaces entre \wzm y \wzp, el requisito (4) es 
ampliado de tal forma que incluya la carga del lept\'on separando as\'i la muestra en dos submuestras, 
cada una utilizada para medir \wzm y \wzp.

\subsection{Estudio de fondos}
El uso de los cortes secuenciales en los sucesos permite seleccionar con gran pureza una muestra 
con candidatos a \WZ, pero quedan remanentes de sucesos que provienen de otros procesos, 
contaminando la se\~nal. Estos sucesos, llamados \emph{fondos} o \emph{ruido}, se pueden 
clasificar, dependiendo de su origen, en fondos instrumentales o fondos f\'isicos. 

Los fondos f\'isicos son aquellos procesos de producci\'on cuyo estado final es e\-xac\-ta\-men\-te
igual al 
estado final del \WZ, esto es tres leptones aislados de alto momento y una cantidad apreciable de 
\MET. El proceso ZZ desintegr\'andose ambos bosones \Z lept\'onicamente constituye un fondo 
f\'isico del \WZ cuando uno de los cuatro leptones no sea detectado, ya sea por aceptancia o por 
ineficiencias de detecci\'on, presentando un estado final de tres leptones, dos de carga opuesta e 
igual sabor, y una cantidad apreciable de \MET debido al lept\'on perdido. As\'i, tambi\'en 
constituyen fondos f\'isicos los procesos VVV (siendo V=W,Z), aunque estos procesos tiene una baja 
tasa de producci\'on en comparaci\'on con el \WZ o el ZZ, y solamente van a ser considerados en el
an\'alisis de 8~\TeV, donde la cantidad de datos acumulados es lo suficientemente grande como para 
que aparezcan algunos de dichos procesos. Estos fondos f\'isicos son irreducibles en el sentido 
que no es posible distinguirlos en su estado final de la se\~nal; para estimarlos se genera toda
la cadena de desintegraci\'on del proceso considerado a partir de c\'alculos te\'oricos y 
t\'ecnicas Monte Carlo, reproduciendo las posibles part\'iculas finales del proceso y su 
cinem\'atica. Se simula el paso de dichas part\'iculas finales a trav\'es del detector 
obteniendo la emulaci\'on de la respuesta del detector en forma de datos equivalentes a 
los datos que se obtendr\'ian experimentalmente. El proceso es generado tantas veces como sea 
posible de tal forma que se consiga un n\'umero significativo de sucesos. La simulaci\'on de 
procesos y su desintegraci\'on se realiza a trav\'es de software especializado, llamados 
\emph{generadores de sucesos}, que son utilizandos ampliamente por toda la comunidad para estimar y 
obtener predicciones te\'oricas en colisiones de part\'iculas de altas energ\'ias. As\'i, todos
los procesos involucrados considerados, tanto se\~nal como fondos del an\'alisis del \WZ han 
sido simulados y comparados, donde procediera, con los datos experimentales. Los procesos de 
se\~nal y fondo han sido generados utilizando principalmente \MADGRAPH~\cite{Alwall:2011uj}, 
\PYTHIA~\cite{Sjostrand:2006za} y \textsc{powheg}~\cite{Alioli:2010xd}. El conjunto de PDFs 
utilizado para producir las muestras fueron \textsc{cteq6l}~\cite{Lai:2010nw} y 
\textsc{ct10}~\cite{Lai:2010vv} dependiendo del generador. En todos los procesos, la respuesta del
detector se simula utilizando una descripci\'on detallada de CMS basada en el paquete 
\GEANTfour~\cite{Allison:2006ve}. Finalmente, las muestras simuladas fueron repesadas para producir
la distribuci\'on del n\'umero de interacciones prot\'on-prot\'on en cada cruce de haces 
(\emph{pileup}) observados en los datos experimentales, dado que esa informaci\'on no estaba 
disponible en el momento de procesar las muestras de Monte Carlo.

Los fondos instrumentales son procesos cuyo estado final queda distorsionado por la medida 
err\'onea de la desintegraci\'on de una o varias part\'iculas del proceso, siendo identificadas como 
leptones aislados cuando originalmente no lo eran. Estos leptones, llamados 
\emph{fakes}, pueden ser jets hadr\'onicos que se identificaron err\'oneamente como leptones, o
verdaderos leptones pero que se desintegraron a partir de quarks pesados. Este tipo de fondo, a su 
vez, pueden clasificarse seg\'un el n\'umero de leptones \emph{fakes} creados.
\begin{itemize}
	\item[$\bullet$] Tres \emph{fakes} creados. Por ejemplo, en procesos de multijets en QCD, o en procesos 
		W+jets, donde el \W se desintegra hadr\'onicamente.
	\item[$\bullet$] Dos \emph{fakes} creados. El proceso W+jets, donde el \W se desintegra lept\'onicamente
		es un ejemplo de ello.
	\item[$\bullet$] Un \emph{fake} creado. El proceso Z+jets, con dos leptones de alto momento, mismo sabor
		y signo opuesto, si el \Z se desintegr\'o lept\'onicamente. Tambi\'en el proceso 
		\ttbar es un ejemplo de este caso cuando los quark top se desintegran a \W y estos
		a su vez se desintegran lept\'onicamente.
\end{itemize}
Los dos primeros casos, la creaci\'on de tres y dos \emph{fakes}, es bastante improbable y, de hecho, se 
comprueba que en este an\'alisis pueden negligirse. As\'i, los fondos instrumentales se reducen al
caso de creaci\'on de un \emph{fake}, cuya contribuci\'on principal la constituyen los procesos Z+jets y 
\ttbar. 

La estimaci\'on de los fondos instrumentales se ha realizado utilizando un m\'etodo basado en 
datos, llamado \emph{feakable object method} (FOM). El FOM es un m\'etodo que estima la 
contribuci\'on de leptones \emph{fakes} explotando las, a priori, distintas eficiencias de 
identificaci\'on e aislamiento que tienen los leptones \emph{fakes} y los leptones desintegrados 
de un \W o \Z, llamados \emph{prompt}. Para ello, se relajan los criterios de calidad de los leptones del 
an\'alisis, obteniendo un nuevo conjunto de leptones llamados \emph{loose} o \emph{fakeables}
que son utilizados para comprobar las eficiencias de identificaci\'on e aislamiento, relacionadas
directamente con la probabilidad de que sean leptones \emph{fake} o \emph{prompt}. El m\'etodo 
extrae en datos experimentales las eficiencias de los leptones \emph{prompt} y \emph{fake} y las
utiliza para estimar las diferentes contribuciones en el estado final de tres leptones, es decir,
estima las contribuciones en el an\'alisis, de fondos generados con tres leptones \emph{fake}
($N_{\fr\fr\fr}^{N_{t3}}$), con dos leptones \emph{fake} y uno \emph{prompt} 
($N_{\fr\fr\pr}^{N_{t3}}$), con un lept\'on \emph{fake} y dos \emph{prompts} 
($N_{\pr\pr\fr}^{N_{t3}}$), y con tres leptones \emph{prompt} ($N_{\pr\pr\pr}^{N_{t3}}$). En 
particular, la estimaci\'on $N_{\pr\pr\pr}^{N_{t3}}$ est\'a definiendo el proceso \WZ, aunque 
tambi\'en incorpora las contribuciones de los procesos irreducibles, as\'i sustrayendo a 
$N_{\pr\pr\pr}^{N_{t3}}$ las estimaciones basadas en Monte Carlo de los fondos irreducibles, 
obtenemos directamente la estimaci\'on del n\'umero de sucesos de se\~nal $N_S$.

\subsection{Sistem\'aticos de las medidas}
La medida de cada observable utilizado en la ec.~\eqref{res:eq:xs} tendr\'a asociadas incertidumbres
debidas a efectos sistem\'aticos que han sido identificados, adem\'as de los errores asociados a la
estad\'istica utilizada para realizar la medida. Los efectos sistem\'aticos identificados en las
medidas de secci\'on eficaz pueden separarse en varios grupos. El primer grupo combina sistem\'aticos
que afectan a la aceptancia y a las eficiencias de reconstrucci\'on e identificaci\'on de los 
objetos de estado final, y son determinados a trav\'es de simulaciones Monte Carlo. En este grupo
se incluyen las incertidumbres asociadas a la escala y resoluci\'on de energ\'ia de los leptones y 
de la \MET, incertidumbres asociadas al rescalado del pileup en las muestras simuladas, y tambi\'en
las incertidumbres te\'oricas de las PDFs y de las escalas de renormalizaci\'on y factorizaci\'on
utilizadas en el c\'alculo de la secci\'on eficaz de la se\~nal. 
El segundo grupo incluye las incertidumbres introducidas al utilizar los factores de correcci\'on
de eficiencias, factores que corrigen las eficiencias calculadas en muestras Monte Carlo con las 
eficiencias calculadas en datos experimentales. Las eficiencias de reconstrucci\'on e 
identificaci\'on de leptones se calculan con m\'etodos \emph{tag-and-probe} tanto en datos
experimentales como en simulaciones Monte Carlo; la diferencia en el cociente de eficiencias 
entre datos y Monte Carlo se toma como incertidumbre. 
El tercer grupo de incertidumbres sistem\'aticas implica al n\'umero de sucesos de fondo 
estimados. En particular, al n\'umero estimado por el m\'etodo \emph{data-driven} se le asigna
una incertidumbre debida a la elecci\'on de la muestra experimental utilizada para estimar la
tasa de \emph{fakes}. A los fondos estimados con simulaciones Monte Carlo, se les asignan un error
sistem\'atico debido a la secci\'on eficaz del proceso utilizada para normalizar el n\'umero de 
sucesos; cuando est\'a disponible una medida experimental, se utiliza el error de la medida como
error sistem\'atico de la secci\'on eficaz del proceso, en caso contrario se utiliza el error
te\'orico asociado al obtener dicha secci\'on eficaz.
Finalmente, tambi\'en se incluye la incertidumbre asociada a la medida de la luminosidad.

La medida del cociente de secciones eficaces entre los procesos \wzm y \wzp sufrir\'an de las 
mismas fuentas de errores sistem\'aticos descritos en el p\'arrafo anterior, pero al realizar
el cociente, la incertidumbre de la luminosidad se cancela y el resto de sistem\'aticos 
aten\'uan su efecto debido a que la direcci\'on de dichos efectos sistem\'aticos es id\'entico en
ambos procesos. La \'unica fuente de error sistem\'atico que no se ver\'a reducida es la debida a 
las diferencias en las eficiencias entre los leptones positivos y negativos asociados al \W, 
sistem\'atico no considerado en la medida de secci\'on eficaz inclusiva debido a que dicha 
diferencia ha sido medida como compatible a la unidad dentro de errores. En el caso de la medida
de $\sigma_{\wzm}/\sigma_{\wzp}$, los errores de la medida del cociente de eficiencias entre 
leptones positivos y negativos han sido propagados como sistem\'aticos, siendo esta fuente la que 
contribuye en mayor proporci\'on a los errores sistem\'aticos.

\section{Discusi\'on de resultados}
La aplicaci\'on de la metodolog\'ia desarrollada a lo largo de este trabajo de tesis, y sintetizada
en este cap\'itulo resumen, ha permitido obtener las medidas de la secci\'on eficaz del proceso 
$pp\to\WZ+X$ y del cociente entre procesos $\sigma(pp\to\wzm+X)/\sigma(pp\to\wzp+X)$ con energ\'ias 
de centro de masas de 7~\TeV y 8~\TeV, en el espacio de fases definido en una ventana de 20~\GeV 
alrededor de la masa nominal del bos\'on \Z. Las medidas han sido realizadas en cuatro canales de
desintegraci\'on con estados finales definidos por los leptones de los bosones \W y \Z: $eee$, 
$\mu ee$, $e\mu\mu$ y $\mu\mu\mu$. Las cuatro medidas han sido combinadas utilizando un m\'etodo
linear que tiene en cuenta las correlaciones entre los errores de la medida para obtener la mejor
estimaci\'on. 

El resultado obtenido para $\sqrt{s}=7~\TeV$ utilizando los datos detectados por CMS durante el 
a\~no 2011, que corresponden a una luminosidad integrada de 4.9~\fbinv es
\begin{equation*}
	\sigma_{7\TeV}(pp\to\WZ+X)= 20.8\pm1.3_{\text{stat}}\pm1.1_{\text{sys}}\pm0.5_{\text{lumi}}\, \text{pb}
\end{equation*}
donde los errores han sido separados dependiendo de la fuente que los produce, estad\'istico (stat),
sistem\'atico (sys) o sistem\'atico de la luminosidad (lumi). La medida es compatible, dentro de errores, con la 
predicci\'on te\'orica del Modelo Est\'andar para este proceso, $17.8^{+0.7}_{-0.5}$~pb.

La secci\'on eficaz medida con una energ\'ia de centro de masas de 8~\TeV, utilizando los datos 
detectados por CMS durante el a\~no 2012 y que corresponden a una luminosidad integrada de 
19.6~\fbinv es
\begin{equation*}
	\sigma_{8\TeV}(pp\to\WZ+X)= 24.6\pm0.8_{\text{stat}}\pm1.1_{\text{sys}}\pm1.1_{\text{lumi}}\, \text{pb}
\end{equation*}
siendo tambi\'en compatible, dentro de errores, con la predicci\'on te\'orica de $21.9^{+0.9}_{-0.5}$~pb.

El medida del cociente de secciones eficaces a 7~\TeV 
\begin{equation*}
	        \left(\frac{\sigma_{\wzm}}{\sigma_{\wzp}}\right)_{7~\TeV}= 0.547\pm0.075_{\text{stat}}\pm0.011_{\text{sys}}
\end{equation*}
se compara perfectamente con la predicci\'on te\'orica, dentro de errores, de 
$0.563^{+0.002}_{-0.001}$ donde se utiliza el conjunto de PDFs \textsc{mstw8} para calcular la 
predicci\'on. Utilizando otro conjunto de PDFs, \textsc{ct-10}, se obtiene una predicci\'on con un
valor central mucho m\'as cercano al valor medido $0.546^{+0.002}_{-0.001}$.

En el caso del cociente a 8~\TeV, se obtiene una medida de
\begin{equation*}
	\left(\frac{\sigma_{\wzm}}{\sigma_{\wzp}}\right)_{8~\TeV}= 0.551\pm0.035_{\text{stat}}\pm0.010_{\text{sys}}\,,
\end{equation*}
que tambi\'en presenta un acuerdo excelente con las predicciones te\'oricas, \linebreak
$0.580\pm0.001$ 
utilizando el conjunto de PDFs \textsc{mstw8} y $0.563\pm0.001$ utilizando \textsc{ct-10}.

Los datos almacenados en CMS de colisiones prot\'on-prot\'on a \comene=7~\TeV han permitido
medir la secci\'on eficaz de producci\'on asociada del proceso \WZ, y por primera vez, el
cociente de secciones eficaces entre \wzm y \wzp, obteniendo resultados dominados por errores
estad\'isticos. Los cantidad de datos de colisiones a \comene=8~\TeV, que se incrementaron
sustancialmente respecto a 7~\TeV, ha posibilitado unas medidas de mayor precisi\'on, a punto
de alcanzar la sensibilidad de las predicciones te\'oricas. Una interesante mejora del presente
an\'alisis incluir\'ia un estudio detallado de las fuentes de error sistem\'aticas para intentar 
minimizarlas y reducir su tama\~no, alcanzando as\'i la sensibilidad te\'orica. Adem\'as, los datos
disponibles para 8~\TeV permiten extender las medidas a secciones eficaces diferenciales en \emph{bins}
de variables cinem\'aticas sensibles, proporcionando una comparaci\'on m\'as detallada de la medida
con la teor\'ia. El mismo an\'alisis est\'a capacitado para comprobar la presencia de 
\emph{acoplamientos an\'omalos de tres bosones gauge} a trav\'es del v\'ertice WWZ, donde los 
l\'imites, en caso de no encontrar dichos acoplamientos, podr\'ian ser mejorados sustancialmente
respecto a la literatura.

Las medidas realizadas en este trabajo de tesis son resultados originales que ser\'an publicados 
por la colaboraci\'on CMS en un art\'iculo, en fase de redacci\'on, y que han sido presentadas en
distintas conferencias especializadas. En particular, el cociente de secciones eficaces no hab\'ia 
sido medido previamente, siendo \'este el primer texto, junto con las presentaciones en 
conferencias, que lo contiene.

\renewcommand{\thesection}{\arabic{chapter}.\arabic{section}}
\selectlanguage{english}
\renewcommand{\chaptermark}[1]{\oldchaptermark{1}}

\bibliographystyle{Classes/tdrbibstyle.bst}
\renewcommand{\bibname}{References} 
\bibliography{XRF-references} 
\glsaddall[types={acronym,main}]
\printglossary[style=altlist]  
\renewcommand{\glsnamefont}[1]{\textbf{ #1}}
\printglossary[type=acronym,style=long] 
\renewcommand{\glsnamefont}[1]{\small{\textsc{\normalfont #1}}}
\printglossary[type=index,style=mcolindex] 
\end{document}